\documentclass[a4paper, twoside, 12pt]{article}

\usepackage[utf8]{inputenc}  
\usepackage[T1]{fontenc}      
\usepackage[english]{babel}  
\usepackage{layout}						
\usepackage[top=2.5cm, bottom=2.5cm, left=2.5cm, right=2.5cm]{geometry} 
\usepackage{setspace} 				
\usepackage{color}
\usepackage{graphicx}
\usepackage{geometry}
\usepackage{amsmath}
\usepackage{graphicx}
\usepackage{txfonts}
\usepackage{amsmath}
\usepackage{bm}
\usepackage{natbib}
\usepackage[backref=page, hidelinks,colorlinks = true,citecolor = blue, linkcolor = blue, urlcolor = blue, draft=false]{hyperref}
\usepackage[figuresright, quiet]{rotating}
\usepackage{tablefootnote}
\usepackage[all]{hypcap}
\usepackage{tabularx,booktabs}
\usepackage{multirow}
\usepackage[labelfont=bf]{caption}
\usepackage{subcaption}
\usepackage{csquotes}
\usepackage{fancyhdr}
\usepackage{etoc}
\usepackage{color}
\usepackage{chngcntr}
\usepackage{enumitem}
\usepackage{afterpage}
\usepackage{ifoddpage}
\usepackage{mathtools}
\usepackage{ulem}
\usepackage{tocloft}
\usepackage[]{algorithm2e}
\usepackage{listings}
\usepackage{makecell}

\renewcommand*{\backref}[1]{}
\renewcommand*{\backrefalt}[4]
{%
    \ifcase #1 []%
        \or        [#2]
        \else      [#2]
    \fi
}

\DeclareMathOperator\arctanh{arctanh}

\makeatletter
     \renewcommand*\l@figure{\vspace{0.07cm}\@dottedtocline{1}{0.7em}{3.0em}}
     \renewcommand*\l@table{\vspace{0.1cm}\@dottedtocline{1}{0.7em}{2.7em}}
\makeatother

\counterwithin{figure}{section}
\counterwithin{table}{section}
\counterwithin{equation}{section}

\definecolor{bleu}{rgb}{0.0, 0.3, 1.0}
\definecolor{orange}{rgb}{1.0, 0.5, 0.3}
\definecolor{vert}{rgb}{0.196,0.804,0.196}

\title{THESE DE DOCTORAT DE L'ETABLISSEMENT UNIVERSITE BOURGOGNE FRANCHE COMTE}
\author{David \bsc{Cornu}}
\date{EMPTY}

\newcolumntype{Y}{>{\arraybackslash}X}
\newcolumntype{x}{>{\arraybackslash}p}
\newcolumntype{m}{>{\centering\arraybackslash}X}

\newcommand{\Uline}{\rule{0.85\linewidth}{0.6mm}}
\newcommand{\arcsec}{^{\prime\prime}}

\begin{document}

\def\aj{AJ}                   
\def\actaa{Acta Astronomica}      
\def\araa{ARA\&A}             
\def\apj{ApJ}                 
\def\apjl{ApJL}                
\def\apjs{ApJs}               
\def\ao{Applied Optics}          
\def\apss{Astrophysics and Space Science}             
\def\aap{A\&A}                
\def\aapr{A\&Ar}          
\def\aaps{A\&As}              
\def\azh{Astronomicheskii Zhurnal}                 
\def\baas{Bulletin of the AAS}               
\def\bac{Bulletin of the Astronomical Institutes of Czechoslovakia }
\def\caa{Chinese Astronomy and Astrophysics}
\def\cjaa{Chinese Journal of Astronomy and Astrophysics}
\def\icarus{Icarus}           
\def\jcap{Journal of Cosmology and Astroparticle Physics}
\def\jrasc{Journal of the RAS of Canada}             
\def\memras{Memoirs of the RAS}            
\def\mnras{MNRAS}             
\def\na{New Astronomy}                
\def\nar{New Astronomy Review}          
\def\pra{Physical Review A: General Physics}        
\def\prb{Physical Review B: Solid State}        
\def\prc{Physical Review C}        
\def\prd{Physical Review D}        
\def\pre{Physical Review E}        
\def\prl{Physical Review Letters}    
\def\pasa{Publications of the Astron. Soc. of Australia}               
\def\pasp{PASP}               
\def\pasj{PASJ}               
\def\rmxaa{Revista Mexicana de Astronomia y Astrofisica}%
\def\qjras{Quarterly Journal of the RAS}             
\def\skytel{Sky and Telescope}             
\def\solphys{Solar Physics}      
\def\sovast{Soviet Astronomy}      
\def\ssr{Space Science Reviews}     
\def\zap{eitschrift fuer Astrophysik}                 
\def\nat{Nature}              
\def\iaucirc{IAU Cirulars}       
\def\aplett{Astrophysics Letters} 
\def\apspr{Astrophysics Space Physics Research}
\def\bain{Bulletin Astronomical Institute of the Netherlands} 
\def\fcp{Fundamental Cosmic Physics}  
\def\gca{Geochimica Cosmochimica Acta}   
\def\grl{Geophysics Research Letters} 
\def\jcp{Journal of Chemical Physics}      
\def\jgr{Journal of Geophysics Research}    
\def\jqsrt{Journal of Quantitiative Spectroscopy and Radiative Transfer}
\def\memsai{Mem. Societa Astronomica Italiana}
\def\nphysa{Nuclear Physics A}   
\def\physrep{Physics Reports}   
\def\physscr{Physica Scripta}   
\def\planss{Planetary Space Science}   
\def\procspie{Proceedings of the SPIE}   

\pagenumbering{gobble}

\begin{titlepage}
\newgeometry{left=1.0cm,right=1.0cm,top=1cm,bottom=1cm}
	\begin{center}
		\begin{minipage}[t]{\textwidth}
			\begin{minipage}[!t]{0.24\textwidth}
			\centering
				\includegraphics[width=\textwidth]{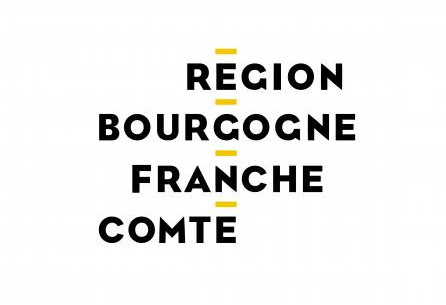}
			\end{minipage}
			\hfill
			\begin{minipage}[t]{0.47\textwidth}
			\centering
			\vspace{-1.5cm}
			\textsc{\textbf{ {\underline{\Huge PhD Thesis}} \vspace{0.6cm}\\ \small{Thèse de doctorat en vue de l'obtention du grade de}}}
			\end{minipage}
			\hfill
			\begin{minipage}[!t]{0.24\textwidth}
			\centering
				\vspace{-0.3cm}
				\includegraphics[width=0.7\textwidth]{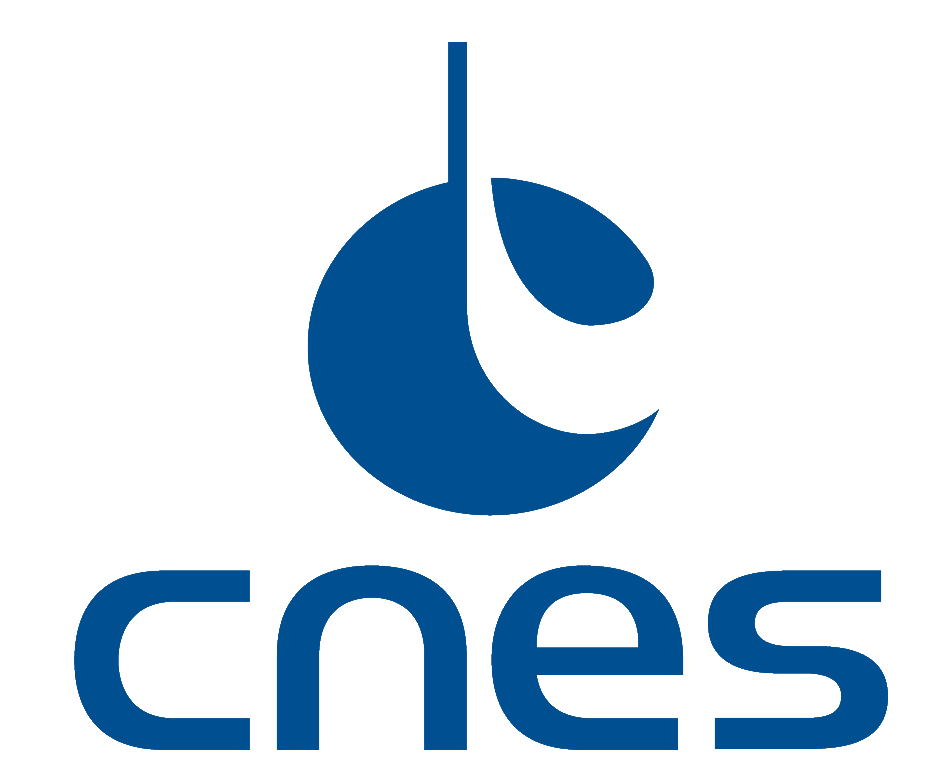}
			\end{minipage}
		\end{minipage}
		\vspace{0.4cm}\\
		\textsc{\textbf{\Large Docteur de l'Université Bourgogne Franche-Comté }}\\
		\vspace{1.2cm}
		
		\Uline \\[0.8cm]
		{\Huge \textbf{Modeling the 3D Milky Way \vspace{-0.0cm}\\ using \textsc{Machine Learning} \vspace{0.2cm}\\ with Gaia and infrared surveys}} \\[0.8cm]
		\Uline \\[1.0cm]
		
		\centering
		
		\textbf{Specialization : \textsc{Astrophysics}}\vspace{0.3cm}\\
		presented and defended by \vspace{0.3cm}\\
		\textbf{\textsc{\Large David Cornu}} \vspace{0.3cm}\\
		Besançon, September 29, 2020\vspace{1.2cm}\\
		
		Under the supervision of: \textbf{\textsc{Annie Robin}} and \textbf{\textsc{Julien Montillaud}} \vspace{0.6cm}\\
		
		\rule{0.9\linewidth}{0.4mm} \vspace{0.5cm}\\
		
		\begin{minipage}[!ht]{0.8\textwidth}
		\flushleft
		
		\hspace{0.1cm} \textbf{\large{Composition du jury :}} \vspace{0.3cm}\\
		
		\small
		\begin{tabularx}{\hsize}{l l l}
		\textbf{Rosine, Lallement} & Directrice de recherche - Université PSL (GEPI) & \textbf{Rapportrice} \\
		\textbf{Luis M., Sarro Baro} & Associate professor - UNED (AI Dept.) & \textbf{Rapporteur} \\
		\textbf{Anne S.M., Buckner} & Research Fellow - University of Exeter & Examinatrice \\
		\textbf{Douglas J., Marshall} & Maître de conférences - Université Toulouse III, P. Sabatier & Examinateur \\
		\textbf{Raphael, Couturier} & Professeur - IUT Belfort-Montbéliard & \textbf{President} \\
		\textbf{Sylvain, Bontemps} & Directeur de recherche - Université de Bordeaux  & Examinateur \\
		\textbf{Annie, Robin} & Directrice de recherche - Université de Franche-Comté  & Directrice de thèse \\
		\textbf{Julien, Montillaud} & Maître de conférences - Université de Franche-Comté & Codirecteur de thèse \\
		\end{tabularx}
		
		\end{minipage}

		\vspace{1.5cm}
		\begin{minipage}[!ht]{\textwidth}
			\begin{minipage}[b]{0.20\textwidth}
			\centering
				\includegraphics[width=0.6\textwidth]{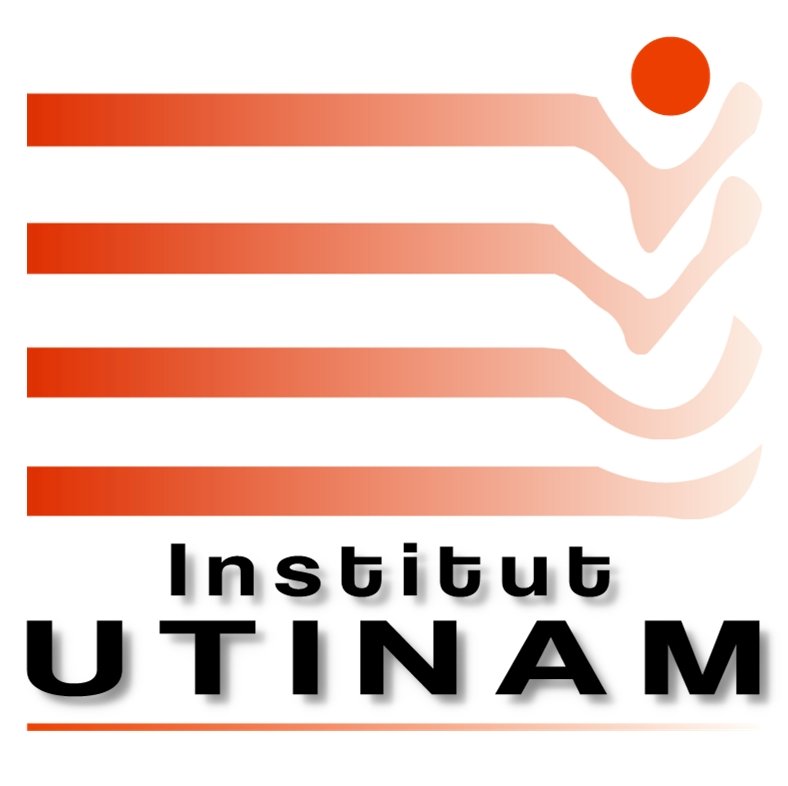}
			\end{minipage}
			\hfill
			\begin{minipage}[!ht]{0.58\textwidth}
			\centering
			\vspace{-2.2cm}
			\textbf{Ecole Doctorale : Carnot-Pasteur}, ED 553\\
			\textbf{Laboratoire : Institut UTINAM, Observatoire de Besançon} \\
			UMR CNRS 6213, Besançon, France 
			\end{minipage}
			\hfill
			\begin{minipage}[!t]{0.20\textwidth}
			\centering
			\vspace{-2.5cm}
				\includegraphics[width=1.0\textwidth]{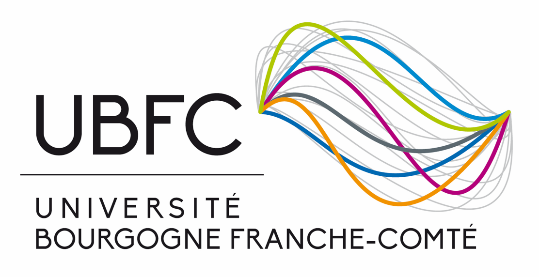}
			\end{minipage}
		\end{minipage}

	\end{center}
\end{titlepage}

\restoregeometry
\newpage
\null
\thispagestyle{empty}

\newpage
\addcontentsline{toc}{section}{Acknowledgments}

\newgeometry{top=1.3cm,bottom=1.3cm}

\textbf{\large Remerciements formels}\vspace{-0.3cm} \\

Cette thèse a été co-financée par le Centre National d’études Spatiales (CNES) et par la région Bourgogne Franche-Comté. Le CNES, dans son rôle d’employeur, nous a assuré son soutien administratif et financier à plusieurs reprises. Le laboratoire UTINAM et l’Observatoire des Sciences de l’Univers THETA de l’université de Bourgogne Franche-Comté ont été les lieux d’accueil quotidien pour nos travaux, et nous tenons à remercier les différentes instances (laboratoire, équipe, OSU) qui nous ont soutenu financièrement et administrativement à de nombreuses reprises. Nos remerciements se portent également vers le Mésocentre de l’Université de Franche-Comté et le Master CompuPhys qui nous ont permis d’accéder à d’importantes ressources de calculs indispensables au travail réalisé. Enfin, nous remercions différents projets et collaborations auxquels nous avons eu la chance de participer : le projet GALETTE financé par PCMI, le projet Besançon-Budapest-Collaboration (BBC), le groupe Galactic Cold Cores (GCC), et l’International Space Science Institute (ISSI) qui a hébergé un groupe de travail autour du modèle de la Galaxie de Besançon.\\

\textbf{\large Remerciements informels} \vspace{-0.3cm}\\

Tout d’abord je tiens profondément à remercier Julien Montillaud, envers qui toute ma gratitude ne saurait s’exprimer en seulement quelques lignes. Merci à Julien d’abord pour sa disponibilité, sa bienveillance, et surtout son éternelle patience, dans l’encadrement de cette thèse. Sa motivation débordante et communicative fut un véritable moteur pour moi au quotidien. Merci aussi pour tout ce qu’il m’a appris, tant sur le plan scientifique qu’humain. Merci de m’avoir aidé à donner le meilleur de moi-même et de m’avoir poussé à élargir mes compétences et intérêts. Ce serait un immense plaisir pour moi que nous puissions continuer à travailler ensemble dans les années à venir. \vspace{-0.3cm}\\

Merci également à Annie Robin pour son encadrement. Annie a su me faire profiter de sa très grande expérience et a toujours été disponible pour m’aider à avancer dans mes travaux. Merci notamment pour son regard critique très pertinent qui nous a permis de souvent dépasser nos objectifs et d’étendre nos champs d’expertise. Je tiens à chaleureusement remercier les autres collègues (passés ou actuels) du bâtiment des horloges : Jean-Baptiste, Nadège, Céline et Guillaume. Merci à chacun d’entre eux pour leur regard scientifique sur nos progrès, pour leurs suggestions de travaux, et pour tous leurs conseils qui m’ont permis de progresser dans de nombreux domaines. J’espère avoir la chance de pouvoir continuer à travailler avec eux, et que nous aurons l’occasion de concrétiser beaucoup des idées que nous avons pu avoir ensemble. Merci également à eux pour les moments de partages, au laboratoire ou en mission, et d’une manière plus générale pour m’avoir intégré aussi chaleureusement à l’équipe. \vspace{-0.3cm}\\

Merci à tous les membres de l’observatoire et du laboratoire UTINAM pour leur accueil. Je pense en particulier à tous mes collègues avec qui j’ai partagé "journal club", réunions, séminaires, mais aussi de nombreux moments de convivialité. J’adresse tous mes voeux de réussite à mes collègues doctorants et jeunes docteurs qui ont été une incroyable source de soutien et d’échange. Merci à tous mes collègues pour la place qu’ils m’ont laissée prendre dans ce groupe; je ne crois pas retrouver facilement une ambiance aussi bienveillante au sein d’une équipe.\\

\textbf{\large Special thanks:}\vspace{-0.3cm}\\

I want to warmly thank all the jury members for their clear interest in my work and for all their comments and suggestions that helped me to improve the quality of the present manuscript and to consider new research ideas. I also want to express a great thank you to Anne Buckner for her help to improve the manuscript language quality.

\newpage
\restoregeometry
\thispagestyle{empty}
\section*{}
\addcontentsline{toc}{section}{Abstract}
\begin{center}
\vspace{-1.5cm}
\textbf{\Large Abstract}\\
\end{center}

Large-scale structure in the Milky Way (MW) is, observationally, not well constrained. Studying the morphology of other galaxies is straightforward but the observation of our home galaxy is made difficult by our internal viewpoint. Stellar confusion and screening by interstellar matter are strong observational limitations to assess the underlying 3D structure of the MW. At the same time, very large-scale astronomical surveys are made available and are expected to allow new studies to overcome the previous limitations. The Gaia survey that contains around 1.6 billion star distances is the new flagship of MW structure and stellar population analyses, and can be combined with other large-scale infrared (IR) surveys to provide unprecedented long distance measurements inside the Galactic Plane. Concurrently, the past two decades have seen an explosion of the use of Machine Learning (ML) methods that are also increasingly employed in astronomy. With these methods it is possible to automate complex problem solving and efficient extraction of statistical information from very large datasets.\\

\vspace{-0.2cm}
In the present work we first describe our construction of a ML classifier to improve a widely adopted classification scheme for Young Stellar Object (YSO) candidates. Born in dense interstellar environments, these young stars have not yet had time to significantly move away from their formation site and therefore can be used as a probe of the densest structures in the interstellar medium. The combination of YSO identification and Gaia distance measurements enables the reconstruction of dense cloud structures in 3D. Our ML classifier is based on Artificial Neural Networks (ANN) and uses IR data from the Spitzer Space Telescope to reconstruct the YSO classification automatically from given examples. We extensively explore dataset constructions and the effect of imbalanced classes in order to optimize our ANN prediction and to provide reliable estimates of its accuracy for each class. Our method is suitable for large-scale YSO candidate identification and provides a membership probability for each object. This probability can be used to select the most reliable objects for subsequent applications like cloud structure reconstruction.\\

\vspace{-0.2cm}
In the second part, we present a new method for reconstructing the 3D extinction distribution of the MW and that is based on Convolutional Neural Networks (CNN). With this approach it is possible to efficiently predict individual line of sight extinction profiles using IR data from the 2MASS survey. The CNN is trained using a large-scale Galactic model, the Besançon Galaxy Model, and learns to infer the extinction distance distribution by comparing results of the model with observed data. This method has been employed to reconstruct a large Galactic Plane portion toward the Carina arm and has demonstrated competitive predictions with other state-of-the-art 3D extinction maps. Our results are noticeably predicting spatially coherent structures and significantly reduced artifacts that are frequent in maps using similar datasets. We show that this method is able to resolve distant structures up to 10 kpc with a formal resolution of 100 pc. Our CNN was found to be capable of combining 2MASS and Gaia datasets without the necessity of a cross match. This allows the network to use relevant information from each dataset depending on the distance in an automated fashion. The results from this combined prediction are encouraging and open the possibility for future full Galactic Plane prediction using a larger combination of various datasets.

\newpage

\thispagestyle{empty}
\section*{}
\addcontentsline{toc}{section}{Résumé en Français}
\begin{center}
\vspace{-1.5cm}
\textbf{\Large Résumé en Français}\\
\end{center}
\textbf{\large Titre en français : Modélisation de la Voie Lactée en 3D par \textsc{machine learning} avec les données infrarouges et Gaia}\\

La structure à grande échelle de la Voie-Lactée (VL) n'est actuellement toujours pas parfaitement contrainte. Contrairement aux autres galaxies, il est difficile d'observer directement sa structure du fait de notre appartenance à celle-ci. La confusion entre les étoiles et l'occultation de la lumière par le milieu interstellaire (MIS) sont les principales sources de difficulté qui empêchent la reconstruction de la structure sous-jacente de la VL. Par ailleurs, de plus en plus de relevés astronomiques de grande ampleur sont disponibles et permettent de surmonter ces difficultés. Le relevé Gaia et ses 1.6 milliards mesures de distances aux étoiles est le nouvel outil de prédilection pour l’étude de la structure de la VL et l’analyse des populations stellaires. Ces nouvelles données peuvent être combinées avec d’autres grands relevés infrarouges (IR) afin d’effectuer des mesures à des distances jusque-là inégalées. Par ailleurs, le nombre d’applications reposant sur des méthodes d’apprentissage machine (AM) s’est envolé ces vingt dernières années et celles-ci sont de plus en plus employées en astronomie. Ces méthodes sont capables d’automatiser la résolution de problèmes complexes ou encore d’extraire efficacement des statistiques sur de grands jeux de données.\\

\vspace{-0.2cm}
Dans cette étude, nous commençons par décrire la construction d’un outil de classification par AM utilisé pour améliorer les méthodes classiques de classification des Jeunes Objets Stellaires (JOS). Comme les étoiles naissent dans un environnement interstellaire dense, il est possible d’utiliser les plus jeunes d’entre elles, qui n’ont pas encore eu le temps de s’éloigner de leur lieux de formation, afin d’identifier les structures denses du MIS. La combinaison des JOS et des distances mesurées par Gaia permet alors de reconstruire la structure 3D des nuages denses. Notre méthode de classification par AM est basée sur les réseaux de neurones artificiels et se sert des données du télescope spatial Spitzer pour reconstruire automatiquement la classification des JOS sur la base d’une liste d’exemples. Nous détaillons la construction des jeux de données associés ainsi que l’effet du déséquilibre entre les classes, ce qui permet d’optimiser les prédictions du réseau et d’estimer la précision associée. Cette méthode est capable d’identifier des JOS dans de très grands relevés tout en fournissant une probabilité d’appartenance pour chacun des objets testés. Celle-ci peut alors être utilisée pour retenir les objets les plus fiables afin de reconstruire la structure des nuages.\\

\vspace{-0.2cm}
Dans une seconde partie, nous présentons une méthode permettant de reconstruire la distribution 3D de l’extinction dans la VL et reposant sur des réseaux de neurones convolutifs. Cette approche permet de prédire des profils d’extinction sur la base de données IR provenant du relevé 2MASS. Ce réseau est entraîné à l’aide du modèle de la Galaxie de Besançon afin de reproduire la distribution en distance de l’extinction à grande échelle en s’appuyant sur la comparaison entre le modèle et les données observées. Nous avons ainsi reconstruit une grande portion du plan Galactique dans la région du bras de la Carène, et avons montré que notre prédiction est compétitive avec d’autres cartes d’extinction 3D qui font référence. Nos résultats sont notamment capables de prédire des structures spatialement cohérentes, et parviennent à réduire les artefacts fréquents dits ``doigts de Dieu''. Cette méthode est parvenue à résoudre des structures distantes jusqu’à 10 kpc avec une résolution formelle de 100 pc. Notre réseau est également capable de combiner les données 2MASS et Gaia sans avoir recours à une identification croisée. Cela permet d’utiliser automatiquement le jeu de données le plus pertinent en fonction de la distance. Les résultats de cette prédiction combinée sont encourageants et ouvrent la voie à de nouvelles reconstructions du plan Galactique en combinant davantage de jeux de données.

\newpage
\null
\thispagestyle{empty}

\newpage
\newgeometry{left=2.7cm, right=2.7cm, top=2.9cm,bottom=2.9cm}
\renewcommand{\contentsname}{\vspace{-2cm}}
\tableofcontents

\newpage
\restoregeometry

\pagestyle{fancy}
\fancyhead[LO]{}
\fancyhead[RE]{}
\fancyhead[LE]{\bf \rightmark}
\fancyhead[RO]{\bf \nouppercase{\leftmark}}

\pagenumbering{arabic}
\setcounter{page}{0}



\thispagestyle{empty}
\null
\newpage
\thispagestyle{empty}
\hfill
\vspace{0.3\textheight}
\part{Context}

\newpage
\null
\thispagestyle{empty}
\newpage
\etocsetnexttocdepth{4}
\etocsettocstyle{\subsection*{Part I: Context}}{}
\localtableofcontents{}

\newpage
\null
\thispagestyle{empty}
\newpage
\section{Milky Way 3D structure}

In this section we introduce astronomical knowledge that is relevant to understand the context of the present study. We noticeably describe the presently admitted view of our galaxy the Milky Way along with a few order of magnitude for the useful astronomical objects and quantity. We describe the expected structural information of the Milky Way and highlight its support from observational constrains. In a second time we expose some properties of the interstellar medium, summarizing its link with the stars in the galaxy. We end by describing the extinction from the ISM and its link with the structural information of the Milky Way. 

\etocsettocstyle{\subsubsection*{\vspace{-1cm}}}{}
\localtableofcontents

\subsection{Review of useful properties of the Milky Way}

	\subsubsection{The only galaxy that can be observed from the inside}

A natural beginning of a work on the Milky Way galaxy structure would be to define what a galaxy exactly is. Still, the presently accepted definition is not that old. In the 1920's, two visions of our place in the universe was opposed in what was latter called the "Great debate". The argument was mostly opposing the two astronomers Harlow Shapley and Heber D. Curtis. The former defended the thesis that every astronomical object observed and especially what they called distant spiral nebulae was part of our Milky Way, so these nebulae must be close and small. The second one in contrast, argued that these objects were very distant and very large and were likely to be external galaxies that look very alike our own Milky Way of billions of stars. They published a common paper containing two parts where each of them exposed their arguments and that was titled \textit{The Scale of the Universe} \citet{shapley_curtis_1921}. The difference in physical scale between the two points of view was of several orders of magnitude, illustrating how much was remaining to understand just 100 years ago.\\

A few years later another famous astronomer, Edwin Hubble, published a paper that estimated the distance of these spiral nebulae based on the known absolute magnitude of Cepheid variable stars \citep{hubble_1926}. The distances he found are today known to be significantly underestimated, still they were already large enough distances to support the thesis defended by Curtis that these nebulae were very large, very massive, distant structures. Later, he published the study that gave birth to the Hubble law and that correlates the distance and radial velocity of other galaxies with their reddening \citep{Hubble_1929}, which is now understood as a cosmological effect of the universe expansion. The known scale of the universe had drastically changed in a few years.\\

	\begin{figure*}[!t]
	\hspace{-1.4cm}
	\begin{minipage}{1.18\textwidth}
	\centering
	\begin{subfigure}[!t]{0.32\textwidth}
	\includegraphics[width=1.0\hsize, height=1.0\hsize]{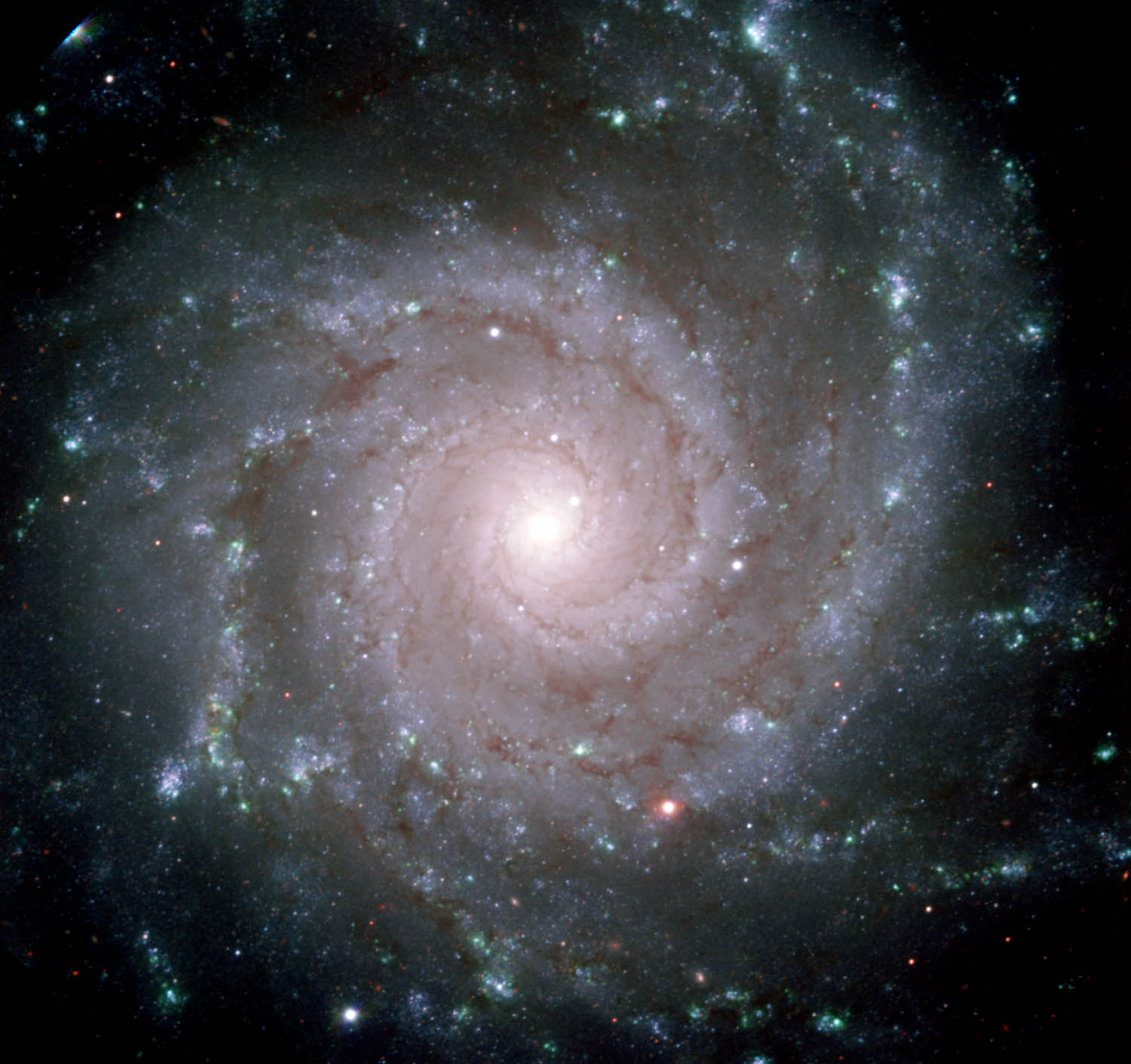}
	\end{subfigure}
	\begin{subfigure}[!t]{0.32\textwidth}
	\includegraphics[width=1.0\hsize, height=1.0\hsize]{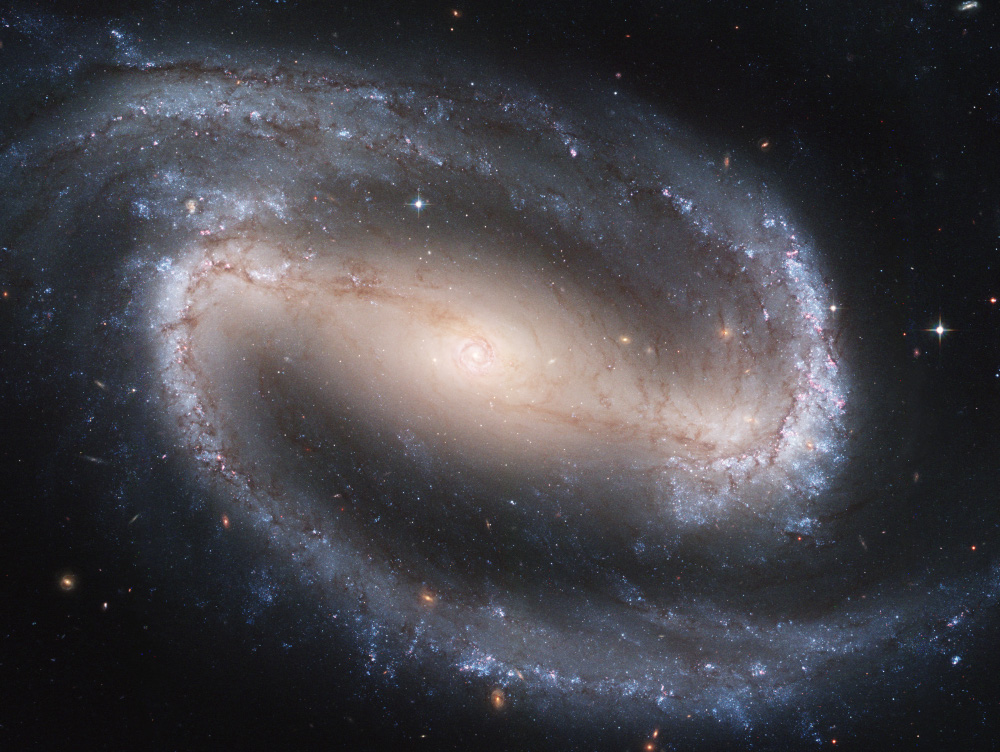}
	\end{subfigure}
	\begin{subfigure}[!t]{0.32\textwidth}
	\includegraphics[width=1.0\hsize, height=1.0\hsize]{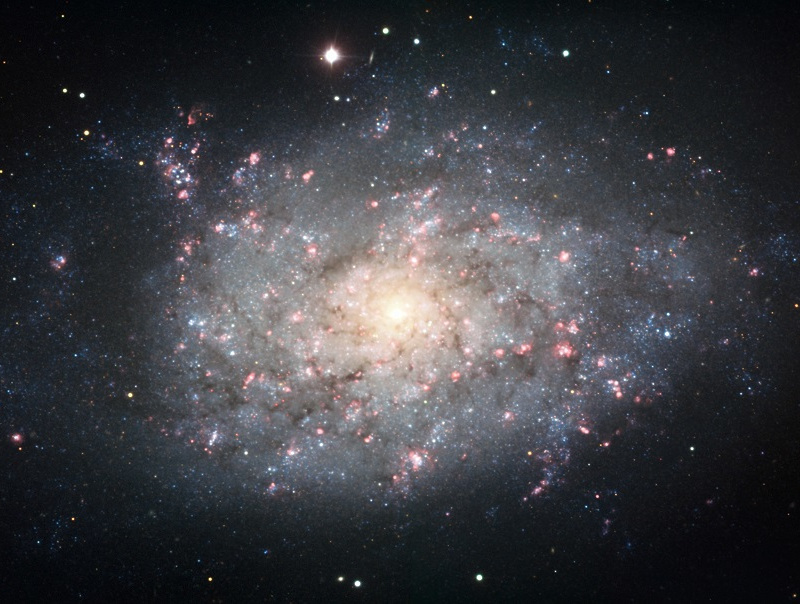}
	\end{subfigure}
	\end{minipage}
	\caption[Spiral galaxy examples]{Examples of spiral galaxies that present different detailed morphology. From left to right, \href{https://www.gemini.edu/gallery/media/perfect-spiral-m74}{NGC 628 (M74)}, a grand design spiral galaxy (SA(s)c) observed by the 8.1-meter North Telescope of the Gemini Observatory , \href{https://apod.nasa.gov/apod/ap200611.html}{NGC 1300} a barred spiral galaxy (SB(s)bc) observed by the Hubble space telescope, and \href{http://annesastronomynews.com/annes-image-of-the-day-spiral-galaxy-ngc-7793/}{NGC 7793} a flocculent galaxy (SA(s)d) observed by the ESO Very Large Telescope (VLT).}
    \label{spiral_gal_morphology_comparison}
\end{figure*}

We do not aim at making an historical overview of the astronomical knowledge about galaxies here, but this story illustrates the fact that knowing the physical scale and boundary of our own galaxy was a tricky question at this time \citep[e.g][]{Kapteyn_1922}. The currently accepted view of a galaxy is a system of stars, dust, gas, and dark matter that is gravitationally bound. Their size can vary from a few kpc to more than 100 kpc and their mass estimates are mostly between $10^5$ and $10^{13}$ $\mathrm{M_\odot}$ based on rotation curves. Galaxies have many different forms, as described by \citet{Hubble_1936} and successively refined in \citet[][,...]{DeVaucouleurs1959, DeVaucouleurs1991, Lintott2008}, and the most noticeable ones for the present study are spiral galaxies. Figure~\ref{spiral_gal_morphology_comparison} shows three typical observed galaxies of this type (NGC 628 - M74, NGC 1300 and NGC 7793) with a face-on view of their plane spanned by spiral-shaped arms that start from the bulge and coil out progressively. This figure also illustrates the large variety of possible spiral structures, with a variable number of arms and a center that can be a roughly-spherical bulge or in other cases an elongated bar. There is also an opposition between two views of galaxy structures: (i) the grand-design view that corresponds to very well resolved narrow spiral arms at large scale which could be the case of M74 in the left frame, and (ii) the flocculent view of more sub-structured galaxies with sub-arms, arm discontinuities, bridges between them, and that does not always follow the expected spiral shape as illustrated in the left frame with NGC 7793. Most of the star formation is believed to occur in the arms \citep{Solomon_1989, Salim_2010} even if the galaxy mass is mostly evenly sprayed over the whole disk \citep{McMillan_2017}. This disk is rotating around the central bulge or bar region that hosts a supermassive black hole for most galaxies \citep{Heckman_2014} and concentrates an important fraction of its mass. This global structure of a galaxy is expected to come from their formation process that started quickly in the early stages of the universe and that is still going on today \citep{Freeman_2002}. While there are multiple views on which process dominates the galaxies formation, it is mostly accepted that there is a gravitational collapse of matter at large scale  \citep{Cooper_2010} and that the rotating accreted matter speeds up with the decrease in the structure size creating a flatten disk shape structure \citep{Brook_2004}. Interestingly, there are still to this day a lot of unknowns about our home galaxy structure, size, mass, detailed 3D distribution of star, etc \citep{Bland-Hawthorn_2016}. We are presently in an uncomfortable situation, where we know more about other galaxies large-scale structures that are far way from us, than about our own galaxy structure. This is due exactly to the fact the we are part of this galaxy. While it is possible to see other galaxies face-on, our own galaxy obscures itself since we are inside the galactic plane and relatively far away from the center. \\

	\begin{figure}[!t]
	\centering
	\includegraphics[width=0.70\hsize]{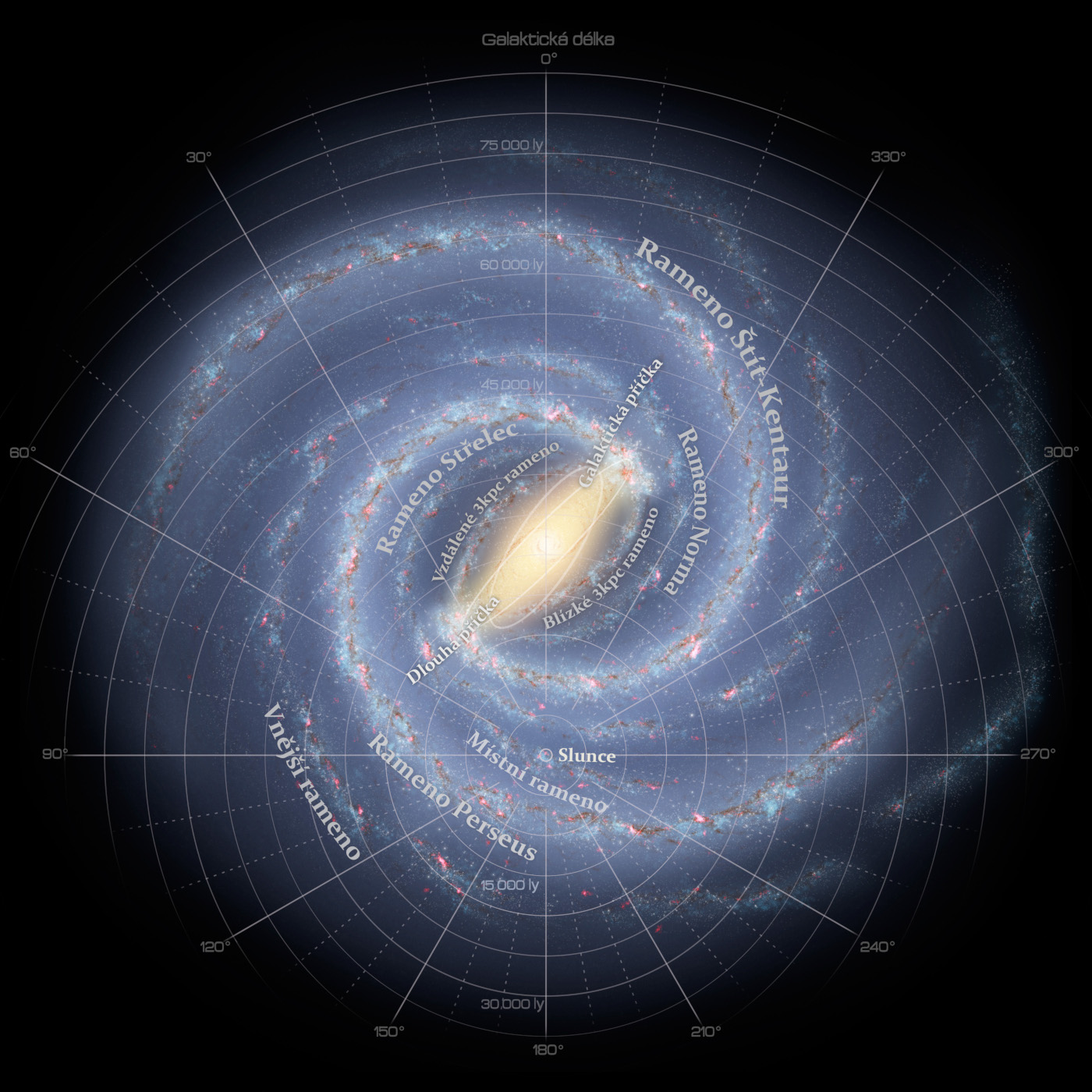}
	\caption[Artistic view of the Milky Way]{Most common artistic face-on view of the Milky Way that illustrates the expected Milky Way bulge and arms. \textit{ From} \citet{Hurt_2008}.}
	\label{milky_way_hurt}
	\end{figure}

	\begin{figure}[!t]
	\centering
	\includegraphics[width=1.0\hsize]{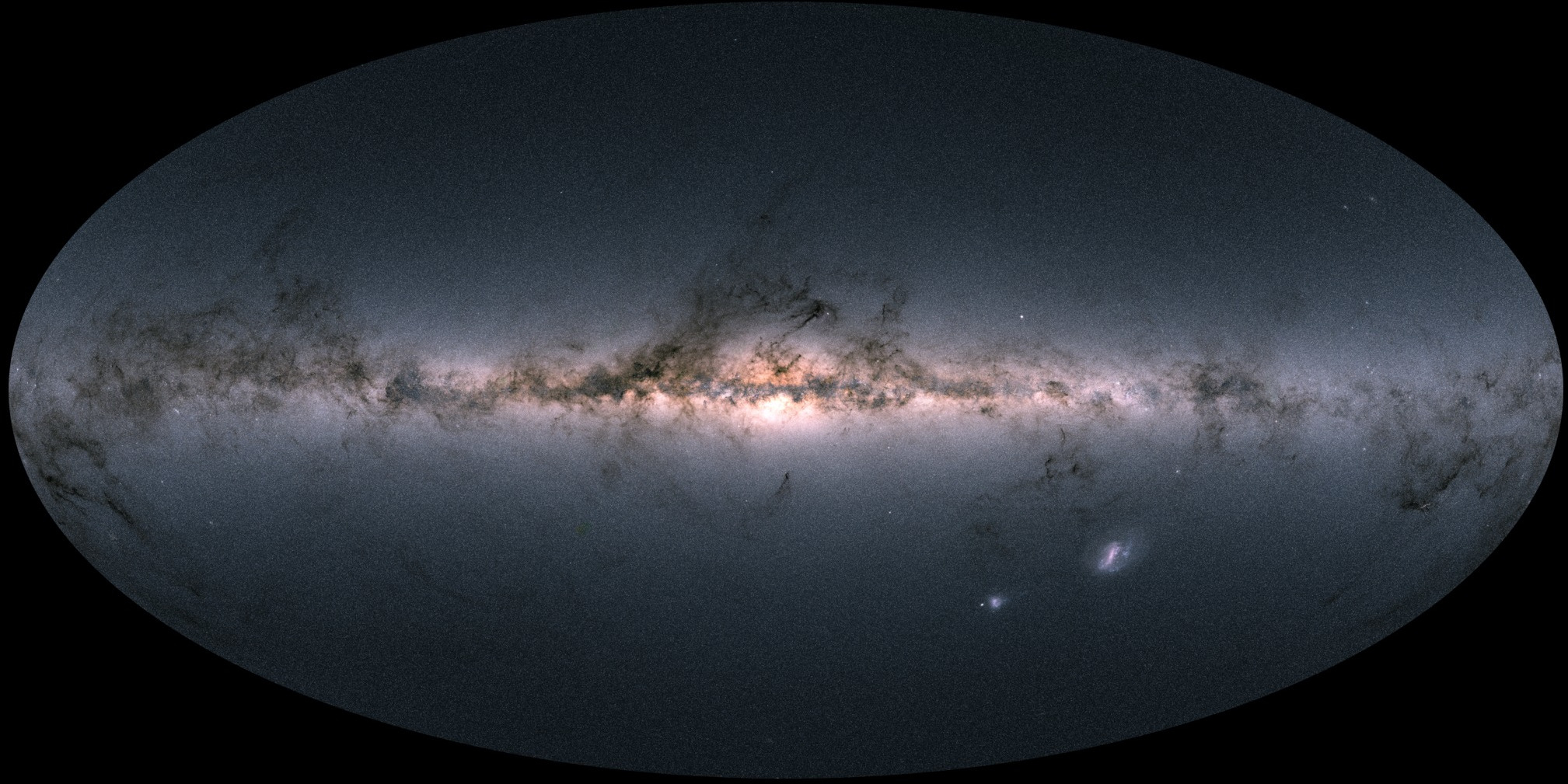}
	 \caption[Gaia DR2 view of the Milky Way in the plane of the sky]{Gaia DR2 view of the Milky Way in the plane of the sky. The image is not a photograph, but a map of the 1.6 billion star brightness in the survey. The image is encoded using the Gaia magnitude bands following, Red: $\mathrm{G_{RP}}$, Green: G and Blue: $\mathrm{G_{BP}}$. {\it From } \citet{Gaia_Collaboration_2018_global}}
    \label{milky_way_Gaia_DR2}

	\end{figure} 

\newpage
The most used Milky Way representation is the one presented in Figure~\ref{milky_way_hurt} from \citet{Hurt_2008}. Despite the fact that this view was constructed based on some observations and on strong theoretical knowledge from the observation of other galaxy structures, it remains mostly an artistic representation that is strongly underconstrained. This view conveys the idea that this is the present state-of-the-art astronomical knowledge of our galaxy structure, even though only sparse and heterogeneous observational evidences are available to this day. What can truly be observed from our standpoint looks like the Figure \ref{milky_way_Gaia_DR2} that contains all the observed stars from the Gaia DR2 mission that we will describe in Section~\ref{modern_large_scale_surveys}. This view illustrates the difficulty caused by our position inside the Milky Way. A great thought experiment that we got from a colleague, is to picture the Milky Way as an expanded forest. Once inside, it is possible to see through a tenth of meter depending on the tree density, but at some point the accumulation of trees and vegetation with the distance makes the view opaque. It is therefore impossible to properly assess the size of the forest from the inside. In this view the trees correspond to the stars and the most diffuse vegetation to the gas and dust distributed in the Milky Way disk. From this it is more clear why it is difficult to reconstruct the Milky Way large-scale structure. Still, it remains a favored position to study the interstellar medium and the stars themselves.\\

\vspace{-0.7cm}
	\subsubsection{Expected structural information}
	\label{milky_way_structure}
	
In the present section we summarize some of the Milky Way (hereafter MW) properties based on the present knowledge \citep[mostly following][]{Bland-Hawthorn_2016} in order to contextualize the present study. The MW is a rather evolved galaxy that has a decreasing star formation and does not present traces of important merger history. It is usually classified as a Spiral Sb-Sbc galaxy, and most of the representations account for 4 spiral arms. Most measurements predict a stellar mass around $5 \times 10^{10}\, \mathrm{M_\odot}$ and total galactic mass from the large dark matter halo around  $\sim 1.5 \times 10^{12}\, \mathrm{M_\odot}$. The stellar disk radius is often estimated at 10 kpc and is often separated in two stellar populations, one from a thin disk with a scale height estimated around $\mathrm{z^t} \simeq 300$ pc and an older one from a thick disk with a scale height $\mathrm{z^T}\simeq 900 pc$ depending on the study, both presenting a flaring, i.e. an increase in height scale with galactic radius. The Sun position is estimated at around 8 kpc from the center of the MW and roughly positioned at an elevation of $\mathrm{z_0}\simeq 25 pc$. At the center of the MW is a super massive black hole named Sagittarius $\mathrm{A^*}$ for which the mass is estimated at approximately $4 \times 10^6\, \mathrm{M_\odot}$, around which a nuclear star cluster is found. They are themselves embedded in an X-shaped (or peanut-shaped) bulge structure that is $\sim 3$ kpc long and with a scale height of $\sim 0.5$ kpc \citep{Robin_2012}. There is then a ``long bar'' or ``thin bar'' region that extend after the bulge up to a 5 kpc half-length \citep{Wegg_2015} but with a quickly decreasing height profile with a mean $180$ pc scale height.\\

The previous elements are considered to be the main components of the Milky Way and provide an accurate global representation based on relatively well-constrained observations. Then the arm structures are more difficult to constrain since they are mostly defined by their higher luminosity or peculiar stellar population and do not represent strong star over-density in stellar mass \citep{Salim_2010,McMillan_2017}. They are proposed to be self-propagated compression waves created by the differential rotation of the galaxy. In this model an arm is a spiral-shaped local compression that triggers star formation and propagates through the galactic plane, which explains that the star velocities do not match those of the arms \citep{Shu_2016}. This process triggers intense star formation episodes, where massive stars are more likely to form than in other regions of the galaxy, highlighting the spiral arm shape. Since these massive stars have a very short life-time, they are gone soon after the passage of the wave, inducing relatively narrow spiral arm structures. As we will expose in Section~\ref{stellar_formation_dense_environment}, stars form from dense cloud compression, therefore the arm structures are also traced by dense interstellar environment which will be at the center of the present study. Overall, in contrast with what Figure~\ref{milky_way_hurt} support, the MW arms are mostly under relatively weak observational constrains at these day which is discussed in Section~\ref{obs_constraints_MW}\\ 
	
\vspace{-0.5cm}
\subsection{The interstellar medium}

	\subsubsection{The bridge between stellar population and interstellar medium}
	\label{stellar_formation_dense_environment}
	
	The main components of galaxies are stars, but they evolve in a more diffuse matter environment, the Inter Stellar Medium (ISM), to which they are bound through a complex interplay. Mainly, the ISM is a mixture of gas and dust with a huge diversity of states and detailed composition. Overall the mass of the ISM is divided as $70.4\%$ Hydrogen, $28.1\%$ Helium, and $1.5\%$ of heavier atoms, almost all of it being in a gas state with less than $1\%$ of the mass of this matter being in the form of solid dust grains \citep{Ferriere_2001}. This matter is distributed very heterogeneously in the galactic environment, from very warm diffuse ($T_K > 10^5\, \mathrm{K}$ and $n < 0.01\, \mathrm{cm^{-3}}  $) and almost transparent large-scale structures to very dense and cold structures ($T_K \sim 10\, \mathrm{K}$ and $n > 10^3\, \mathrm{cm^{-3}}  $) at much smaller scales with a continuum of structures between the two, including for example interstellar molecular filaments.\\
	
	The ISM evolution is determined by the complex interplay between the magneto-hydrodynamics laws, which describe how the gas flows in the galaxy, gravity and self-gravity, which contribute to shaping and compressing the gas at all scales, as well as a number of processes related to stellar evolution, like the propagation of supernova shock waves, the gas heating by photoelectric effect on dust grains or gas ionization by stellar ultra-violet (UV) flux. The ISM represents a significant portion of the mass of the Milky way, equivalent to around 10 to 15\% of the total stellar mass. This is known to be the matter from which stars form as explained in detail, for example, by \citet{McKee_2007}, \citet{Kennicutt_2012}, and references therein, and described briefly here. From the proportions reported above, it is visible that the MW has already converted most of its gas into stars. Under the combined effects of dynamics and gravity the interstellar medium will contract hierarchically creating dense clouds. At some point they will become optically thick and their inside will get cooler by preventing the ambient UV light from stars to penetrate the cloud deeply. The low temperatures enable a complex chemistry catalyzed by dust grains that allows the creation of larger molecules and lets dust grain themselves grow in size, which changes the optical properties of the densest structures. The definition threshold of a dense cloud is a tricky question that is often solved using a certain amount of CO emission or by the dust reddening amount that is much higher in dense clouds. If the cloud is massive enough so that gravity is stronger than the gas support (kinetic energy, turbulence, magnetic field), it will collapse gravitationally, starting the formation of a protostar. It ultimately leads to the formation of a star that is supported by nuclear fusion in its core, i.e. a main-sequence star. The steps of star-formation, from the gravitational collapse to the main-sequence star are described where they are useful in Section~\ref{yso_def_and_use}.\\
	
	The important point here is that stars are formed through the collapse of the dense interstellar medium. Once formed, the stars will progressively get away from their original structure, so that identifying Young Stellar Objects (YSO) that did not have time to move too much is a suitable way to reconstruct large-scale dense-cloud structures that are massive enough to form stars (Sect.~\ref{3d_yso_gaia}). A large part of the present study, namely the part II, is dedicated to the construction of a YSO identification method that is a prerequisite of the previous approach.\\
	
	\vspace{-0.5cm}
	More generally the link between the stars and the ISM is not one-sided since there are a lot of feedback from the stars on the ISM, of which we give some examples. First the star light warms and ionizes the ISM but also breaks any complex molecule or even evaporate dust grains. In addition, any element other than hydrogen and helium originate from the stellar nucleosynthesis and are dispersed after the star end of life. The supernova explosions that ends the life of massive stars ($M > 8 \mathrm{M_\odot}$) play a major and ambiguous role in the ISM evolution. This phenomenon is known to inject a very significant mechanical energy to the ISM, which can blow away dense structures and enrich it with new elements that are only formed during such energetic events. It can also have the opposite effect and induce compression waves on the ISM, leading to triggered star formation \citep[e.g][]{padoan_2017}.\\
	
	At large scales, the ISM is shaped by the global dynamic of the Milky Way. Indeed larger ISM structure have been observed to follow the spiral arms in other galaxies \citep[e.g][]{Elmegreen_2003} and in simulations \citep{Bournaud_2002} with the more local structures getting their energy mostly from gravitational instabilities from the spiral arms that drive the turbulent regime and by inward mass accretion \citep{Bournaud_2010}. This explains why the large-scale structures of galaxies can be traced using the distribution of the dense ISM. Still, smaller scales of the local ISM distribution are made more complex by various feedback effects like supernovae \citep{Hennebelle_2014} that are important to account for the flocculent substructures in galaxies, as it is illustrated by the Figure~\ref{milky_way_Gaia_DR2} to explain higher latitude dense clouds, or again in the right frame of Figure~\ref{spiral_gal_morphology_comparison}.

	\subsubsection{Interstellar medium extinction and emission}
	\label{intro_extinction}

\begin{figure}[!t]
	\centering
	\includegraphics[width=0.86\hsize]{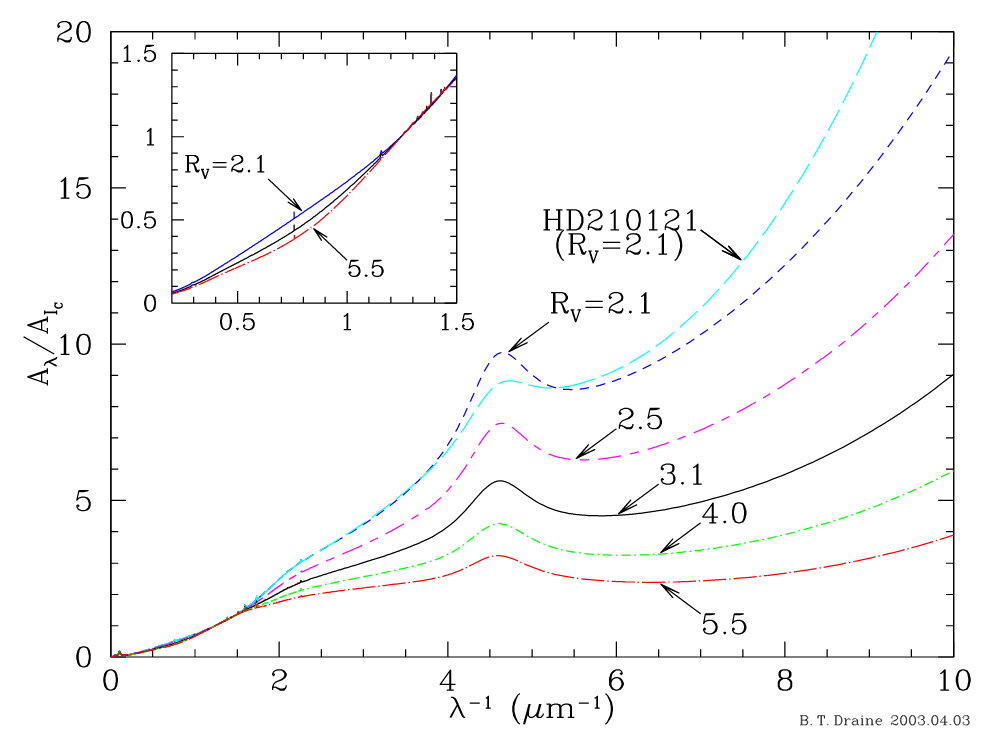}
	\caption[Extinction curve variation as a function of $R_V$.]{Extinction curves based on the prescriptions from \citet{fitzpatrick_correcting_1999} for different values of $R_V$. {\it From} \citet{Draine_2003}.}
    \label{extinction_curve_fig}
	\end{figure}

	While the ISM can be studied through its several interactions with stars, it is possible to perform more direct detection of the ISM structures. The first observable effect, even by the human eye, is the screening effect of the background stars by the interstellar clouds as can be observed in the Milky Way plane in Figure~\ref{milky_way_Gaia_DR2}. This is due to an effect of the ISM called extinction and that sums two physical effects: the absorption and the scattering of the light by the matter in the light path. For astronomical considerations this effect induced predominantly by interstellar dust grains \citep[][and reference therein]{Draine_2003}. This extinction is usually characterized by the quantity $A_\lambda$:
	\begin{equation}
	A_{\lambda} = 2.5 \log \Bigg(\, \frac{F^0_{\lambda}}{F_{\lambda}}\, \Bigg)
	\label{basic_ext_eq}
	\end{equation} 
where $F^0$ is the luminosity flux before the clouds, $F$ is the flux after the cloud and $A_\lambda$ is the total extinction at the wavelength $\lambda$. An important point is that all these quantities depend of the wavelength of the light. This is due to the dependence of scattering to the ratio between the wavelength and the grain size, while the dust absorption spectrum also depends on the grain composition. Therefore, the extinction strongly correlates with the dust grain size distribution in the ISM defining what is called an extinction curve, or extinction law (Fig.~\ref{extinction_curve_fig}). It was exposed by \citet{Cardelli_1989} and refined by \citet{fitzpatrick_correcting_1999} that it is possible to parametrize this law using a single dimensionless parameter $R_V$ that is expressed as:
\begin{equation}
R_V = \frac{A_V}{E(B-V)}
\label{extinction_curve_eq}
\end{equation} 
where $A_V$ is the extinction in the V band ($\lambda = 550$\, nm, $\Delta \lambda = 88$\, nm), and $E(B-V)$ is the reddening (or selective extinction) between the B ($\lambda = 445$\, nm, $\Delta \lambda = 94$\, nm) and the V bands, defined as $E(B-V) = A_B - A_V$.
\newpage

This reddening is an important aspect of the process since it corresponds to an effective shift of the apparent color of the observed stars under the effect of extinction. We show the shape of the typical extinction curves for different values of $R_V$ in Figure~\ref{extinction_curve_fig}. In first approximation, the dust grain composition and size distribution in the diffuse interstellar medium is globally constant across the Milky Way. This leads to a rather constant extinction law in this medium, although significant variations are observed, mostly toward dense molecular gas \citep[e.g.][and references therein]{Schirmer_2020}. These variations are generally well parameterized by a single parameter \citep[$R_V$, Eq.~\ref{basic_ext_eq}, Fig.~\ref{extinction_curve_fig}][]{Cardelli_1989}, although more complex variations were reported toward the Galactic Center \citep{Nataf_2016}, and $R_V$ appears to vary even on large galactic scales for the diffuse ISM \citep{Schlafly_2016}. From this observationally constrained law, we see that shorter wavelengths are much more affected by extinction than the longer ones. This is the cause of the reddening of the observed light.\\

One of the most important aspects of extinction is that it is an integrated quantity over the full light path from the emitting astronomical object down to the observer. However, since the extinction quantity is characteristic of the amount of dust it can be used as a probe of the dense regions of the Milky Way. There are noticeable relations between the extinction and the column density of atomic or molecular gas. The main difficulty is then to reconstruct the distance distribution along a given line of sight, called an extinction profile. The reconstruction of the 3D distribution of the extinction in the MW is a powerful approach as it would directly map the 3D distribution of the dense ISM. This is theoretically a suitable method to provide constraints on the spiral arms of the MW that we described as underconstrained in Section~\ref{milky_way_structure}. The second half of the present study, namely Part III, is devoted to a new approach to reconstruct large-scale 3D extinction maps with a large distance range based on multiple observational surveys. More details on existing studies about this approach and their implications are given in the corresponding part introduction section \ref{ext_map_first_section}.\\

Finally, another ISM observable that is used in this work is dust emission. The heating of dust grains via the absorption of star light is balanced, in average, by their cooling due to a continuous thermal or stochastic emission at infrared (IR) wavelengths. We note that in very dense environments, collisions can also become a heating process. The typical wavelength range of dust emission is between $ 1 < \lambda < 10^3 \, \mathrm{\mu m}$ with various contributions induced from the diverse populations of dust grains. Figure~\ref{dust_emission_fig} shows the typical dust emission as a function of the wavelength, separating different grain population contributions as modeled by \citep{Compiegne_2011} along with observational constraints. We illustrate the use of dust emission in Figure~\ref{planck_optdepth_skyplane} that shows the reconstructed dust optical depth at 353 GHz based on a modified black body fitting of the dust emission observed by the Planck space telescope \citep{Planck_2016}. This map will noticeably be used to perform morphology comparison of the dust distribution over the plane of the sky in Sections~\ref{2mass_maps_section} and~\ref{gaia_2mass_ext_section}. The dust emission can also be used to distinguish different early-protostar stages. Indeed, stars begin their formation embedded into dense envelopes that are heated by the protostar that is then visible in the spectral energy distribution (SED) of the object. In subsequent stages, the envelope is evaporated and a dusty emitting disk remains. These emission properties are used in Section~\ref{intro_yso} as a tracer for YSO classification in the infrared. \\

	\begin{figure}[!t]
	\vspace{2.50cm}
	\begin{minipage}{1.0\textwidth}
	\centering
	\includegraphics[width=0.94\hsize]{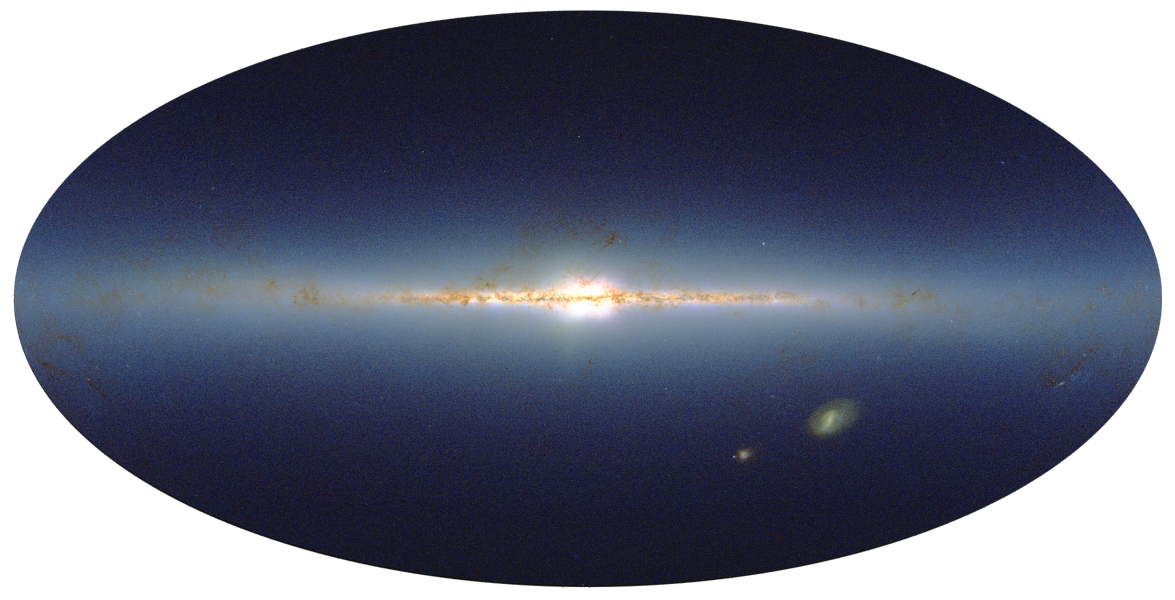}
	\end{minipage}
	\caption[2MASS view of the Milky Way in the plane of the sky]{2MASS view of the Milky Way in the plane of the sky. The image is encoded using the 2MASS magnitude bands following, Red: J, Green: H and Blue: $\mathrm{K_s}$. {\it From } \citet{skrutskie_two_2006}}
	\label{2MASS_skyplane}
	\end{figure}

\begin{figure}[!t]
	\centering
	\includegraphics[width=0.84\hsize]{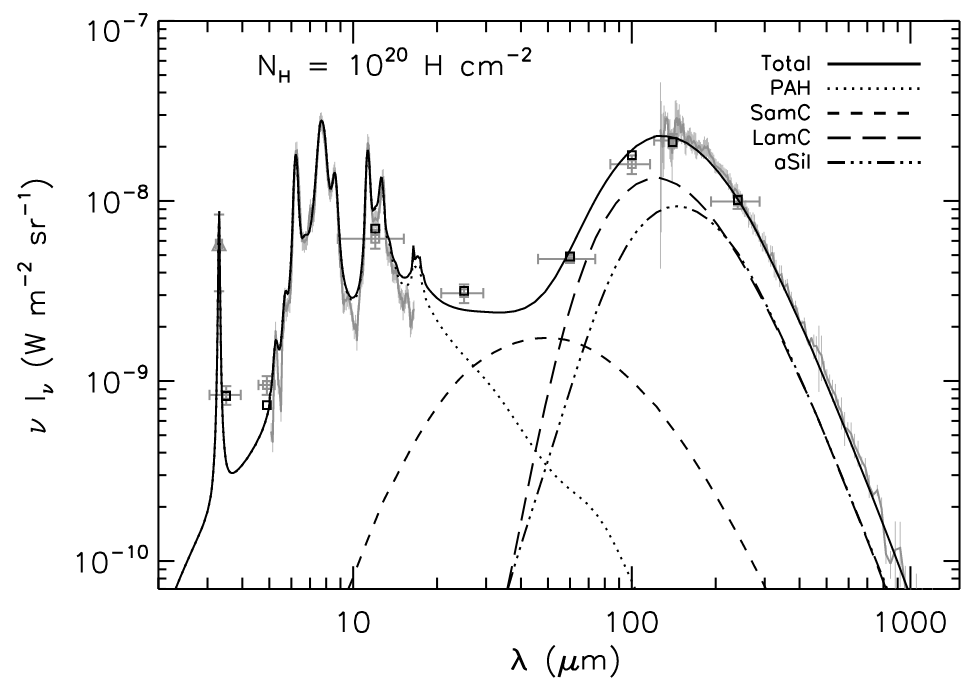}
	\caption[Dust emission as a function of wavelength]{Typical observed dust emission for the diffuse interstellar medium at high-galactic latitude for a given $N_H = 10^{20} \mathrm{H\,cm^{-2}}$. The mid-IR ($\sim 5-15\, \mathrm{\mu m}$) and far-IR ($\sim 100-1000\, \mathrm{\mu m}$) spectra are from ISOCAM/CVF (ISO satellite) and FIRAS (COBE satellite), respectively. Squares are the photometric measurements from DIRBE (COBE). The continuous line is the DustEm model prediction. {\it From} \citet{Compiegne_2011}.}
	\label{dust_emission_fig}
	\end{figure}
	
	\begin{figure}[!t]
	\centering
	\vspace{-0.5cm}
	\includegraphics[width=1.0\hsize]{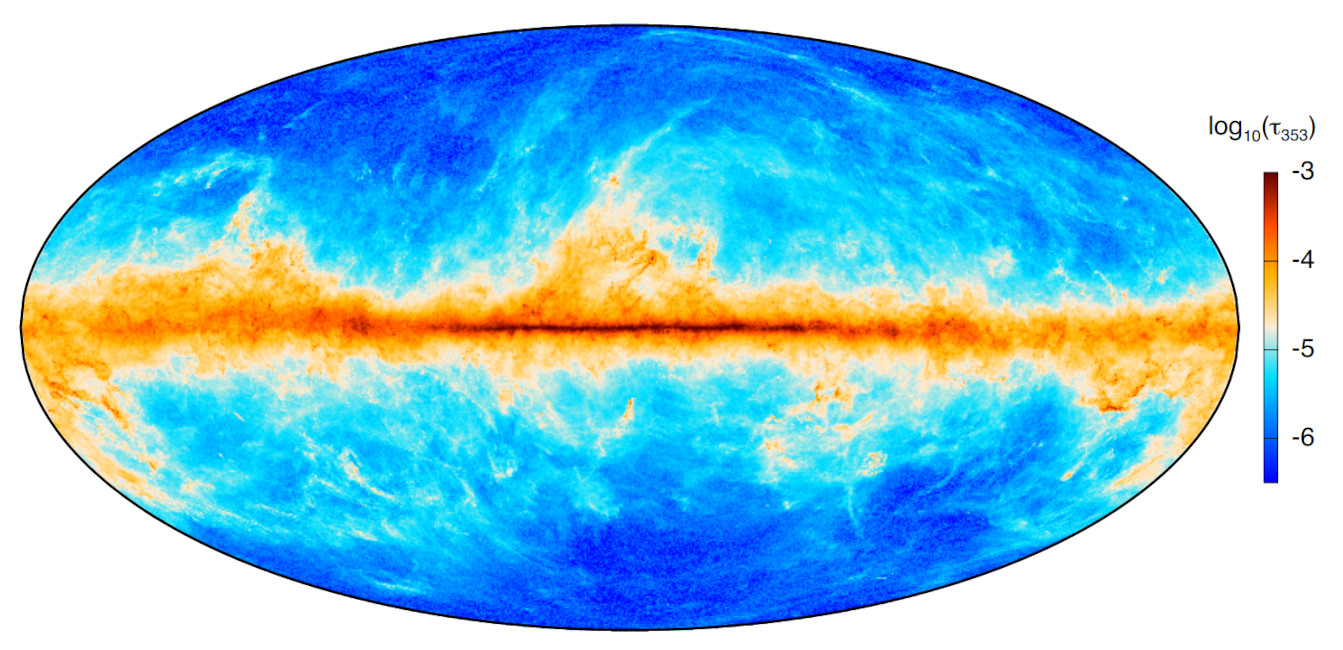}
	\caption[Planck dust opacity view of the Milky Way in the plane of the sky]{Planck dust opacity at 353 GHz of the Milky Way in the plane of the sky, as fitted from a modified black body based on dust emission. {\it From } \citet{Planck_2014}.}
    \label{planck_optdepth_skyplane}
\end{figure}

\clearpage
\subsection{Observational constraints on the Milky Way structure}
\label{obs_constraints_MW}

	\begin{figure}[!t]
	\centering
	\includegraphics[width=0.65\hsize]{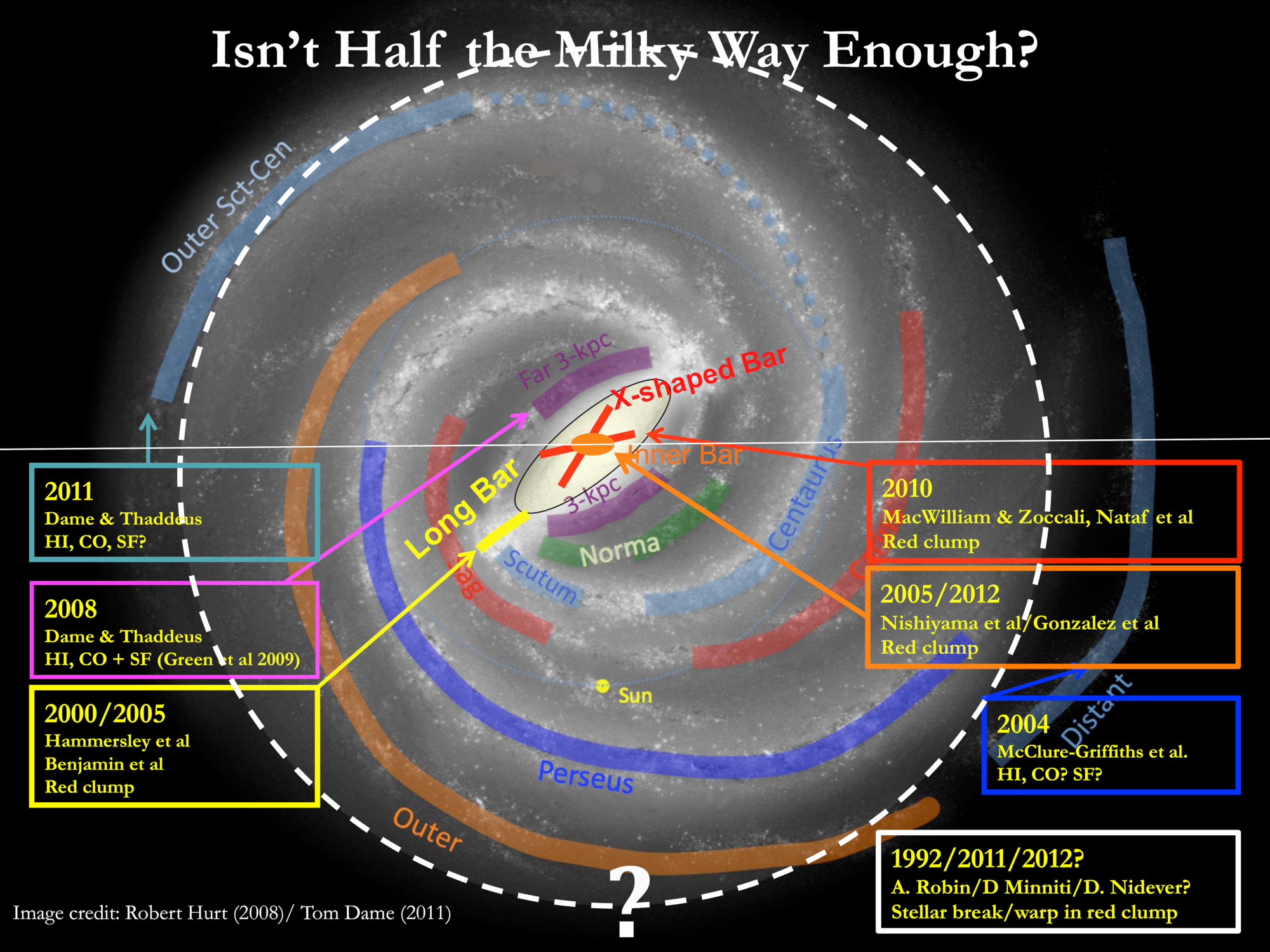}
	\caption[Milky Way illustration with observational constraints]{Artistic representation of the Milky Way annotated with compiled effective knowledge from 2014 on the spiral arms based on observational constraints. Each structure is associated with a reference publication. \textit{This image is taken from }\citet{Benjamin_2014} \textit{which adapted it from} \citet{Hurt_2008}.}
    \label{milky_way_annoted}
	\end{figure}

We show in Figure~\ref{milky_way_annoted} a carefully annotated version of the artistic face-on view made by \citet{Benjamin_2014} and that represents a census of the existing published constraints on each expected spiral arm structure of the MW a few years ago. This figure puts the emphasis on the fact that there is a significant portion of the Milky Way structure that is not constrained due to its position behind the Galactic Center. Here we describe some of the present existing work that have added constrains on the Milky Way structure. One of the oldest method to infer the galaxy spiral arms existence and position has been to measure the atomic hydrogen HI emission \citep{van_de_Hulst_1954}. Since HI is already mainly presents in ISM clouds that follows the global Milky Way structure, it is possible to use it to reconstruct roughly the galactic structure \citep{Kalberla_2009}. HI is observed through its hyperfine transition that emits at a 21 cm wavelength at which the interstellar medium is mostly transparent, granting the possibility of high distance measurements. The main approach is then to use the Doppler frequency shifting to reconstruct the velocity of coherent structures in the spectra. With some assumptions on the Milk Way circular geometry it is noticeably possible to reconstruct the arms tangent position in order to reconstruct the galactic rotation curve. HI data were also used to infer the position of some galactic arms or substructure \citep{McClure-Griffiths_2004}, but it remains too diffuse to highlight very strongly a global spiral structures.\\

\begin{figure*}[!t]
\hspace{-1.8cm}
\begin{minipage}{1.25\textwidth}
	\centering
	\begin{subfigure}[t]{\textwidth}
	\includegraphics[width=1.0\hsize]{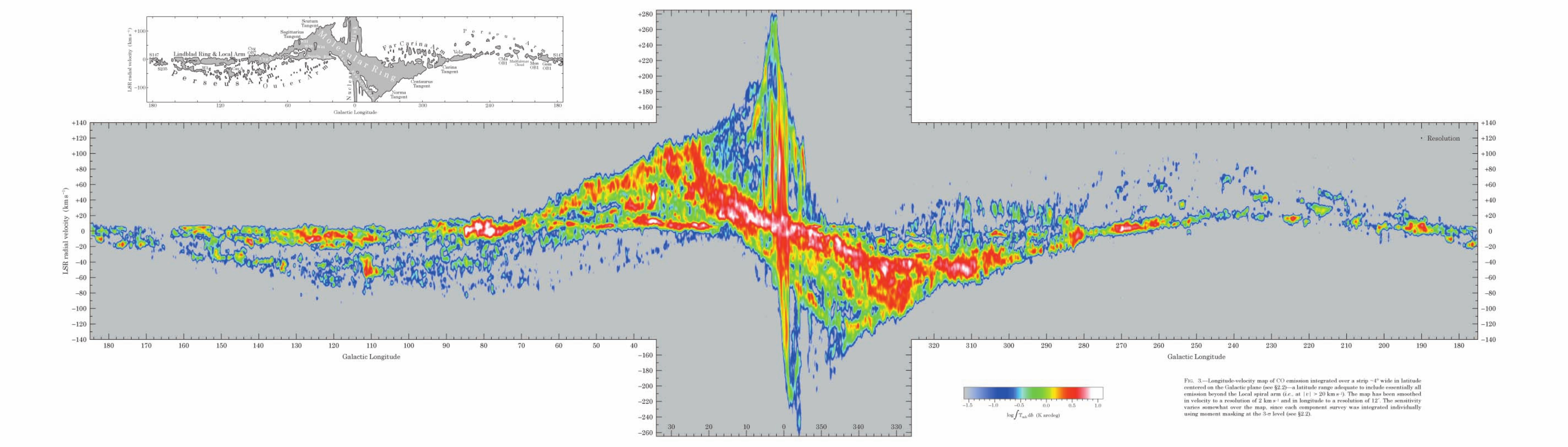}
	\end{subfigure}
	\begin{subfigure}[t]{0.99\textwidth}
	\includegraphics[width=0.97\hsize]{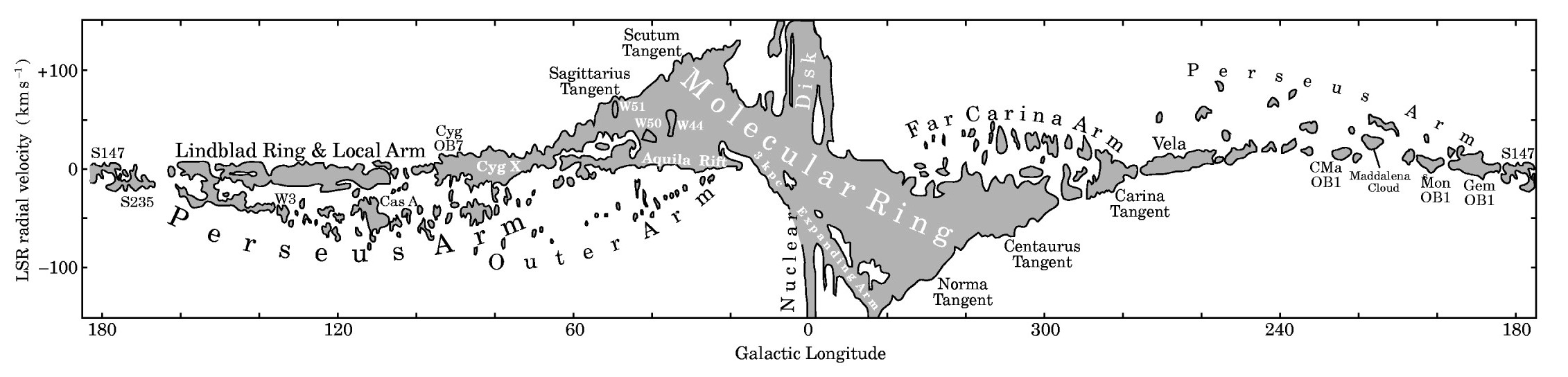}
	\end{subfigure}
	\end{minipage}
	\caption[Longitude-velocity map using CO(J=1-0) emission]{Longitude-velocity map of CO(J=1-0)) integrated for $|b| < 4$ and centered on the galactic plane. The map resolution is 2 $\mathrm{km.s^{-1}}$ in velocity and $12'$ in galactic longitude. The \textit{bottom} frame is a zoom on the annotated version of the  map. \textit{From }\citet{Dame_2001}.}
    \label{dame_figure}
\end{figure*}

Another suitable tracer of much denser ISM environments and that can be used in the same manner is the CO molecule. Its rotational transition line at 115 GHz is consider as easy to observe and CO is present in every molecular clouds, allowing for very complete detection. The study from \citet{Dame_2001} has been a reference in the identification of the galactic structures. From their observations they reconstructed a longitude velocity map of the Galactic Plane integrated over a $\pm 4^\circ$ latitude range, which is presented in Figure~\ref{dame_figure}. From this map they identified what could correspond to the spiral arm structures as illustrate in the bottom frame of the figure. With this approach it remains difficult to disentangle structures in the central region and behind. It also relies on the assumption that it is effectively possible to separate the arms in the velocity space, which might not always be the case especially if we consider the existence of bridges, gaps, and overall less continuous structures in the Milky Way.\\

	\begin{figure}[!t]
	\centering
	\includegraphics[width=0.71\hsize]{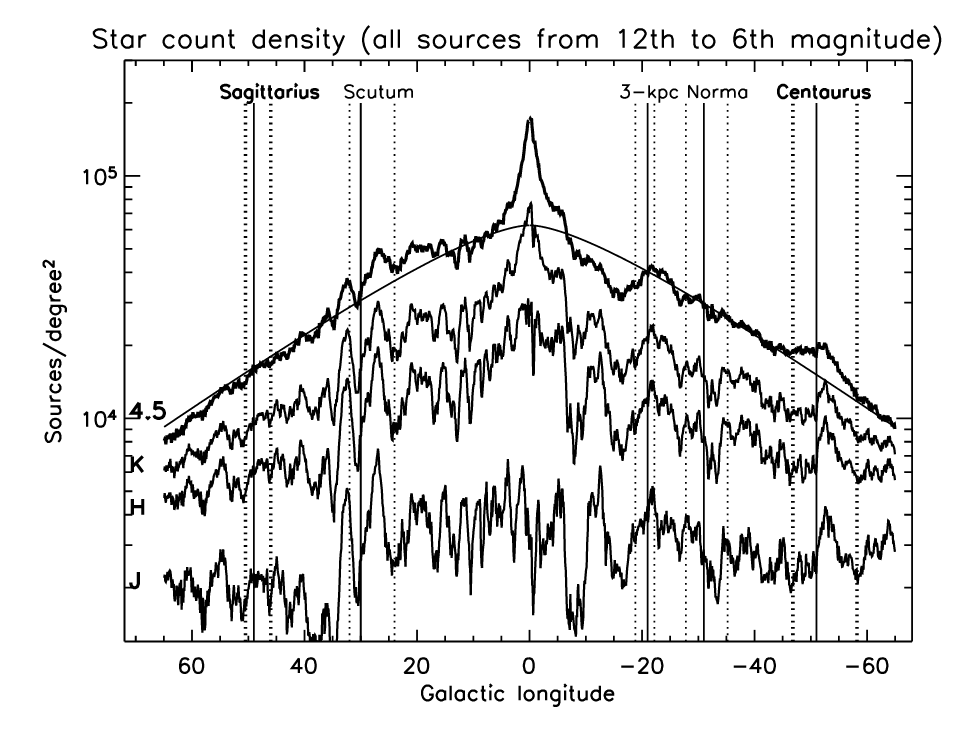}
	 \caption[Galactic plane source-count in the infrared from Spitzer]{Number of sources per $\mathrm{deg^2}$ as a function of galactic longitude using the 4.5 $\mathrm{\mu m}$ band of Spitzer, and the J, H, $\mathrm{K_s}$ bands from 2MASS. The vertical lines highlight the predicted position of some galactic arms. \textit{From }\citet{Churchwell_2009}.}
    \label{GLIMPSE_arms}
	\end{figure}

An other approach performed by \citep{Benjamin_2005, Churchwell_2009} using the Galactic Legacy Infrared Mid-Plane Survey Extraordinaire (GLIMPSE) performed with the Spitzer Space Observatory \citep{werner_spitzer_2004}, was to compare the observed star count in the galactic plane with the expected exponential disk population. Higher star count, especially from specific stellar population, can trace the tangent of the arms as illustrated in Figure~\ref{GLIMPSE_arms}. However, even if Spitzer uses infrared wavelength to observe through relatively dense environments, the star count remains affected by the extinction in a complex manner. Therefore, it requires an accurate extinction prescription or to choose carefully not too much extincted lines of sight.\\

\begin{figure*}[!t]
\hspace{-1.4cm}
\begin{minipage}{1.18\textwidth}
	\centering
	\begin{subfigure}[t]{0.49\textwidth}
	\includegraphics[width=1.0\hsize]{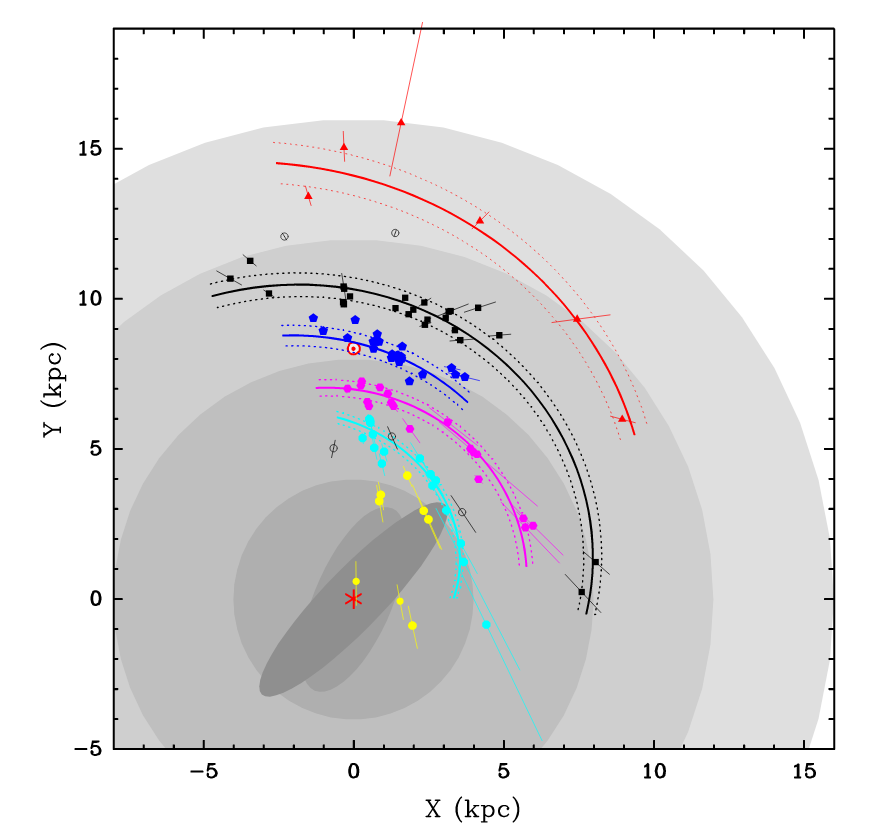}
	\end{subfigure}
	\begin{subfigure}[t]{0.48\textwidth}
	\includegraphics[width=0.95\hsize]{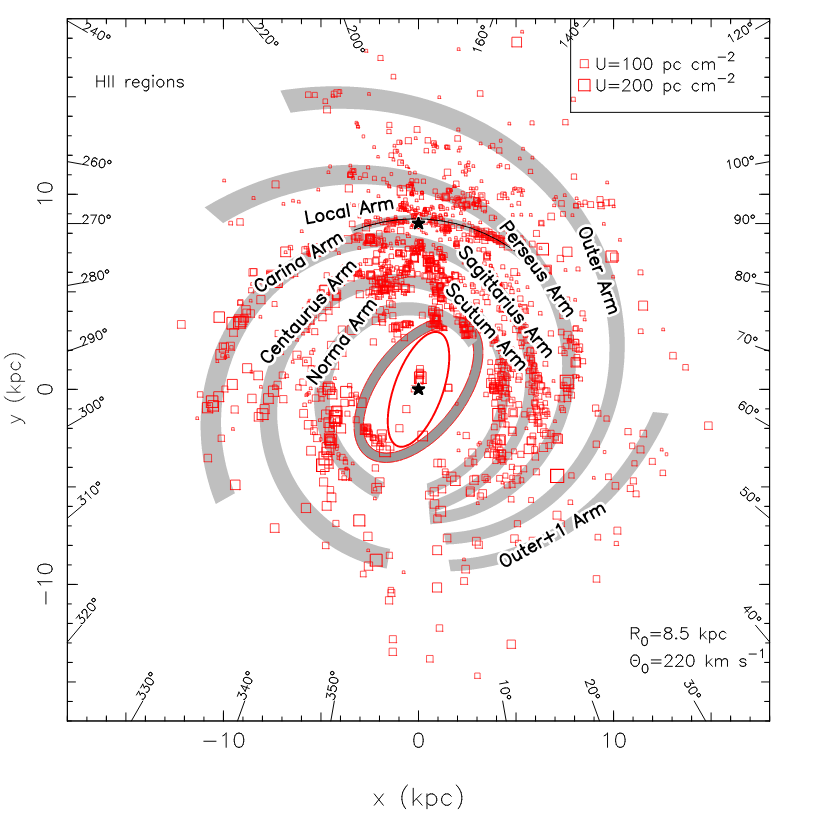}
	\end{subfigure}
	\end{minipage}
	\caption[Discrete catalogs for Galactic arm fitting]{Galactic structure from discrete object distribution using face-on views. On both images the Galactic Center is at the $X=0$,$Y=0$ coordinates and the Sun is around $X=0$,$Y=8$. {\it Left}: Maser with parallax observed using the VLBA, VERA and the EVN, from \citet{Reid_2014}. The color groups were made from velocity-longitude association, then the continuous lines are log-periodic spiral arms fitted using these groups. {\it Right}: HII regions collected by \citep{hou_and_han_2014}, their best 6-arm model fit on these regions is added in gray and associated to the usual arm names.}
    \label{punctual_dections_structure}
\end{figure*}

More discrete objects can also be used to reconstruct the galactic structures like HII regions, Giant Molecular Clouds (GMC) or masers. The latter are associated with young and massive stars that will therefore be present in active dense star forming regions, likely following the arms. They were used by \citet{Reid_2014} in combination with parallax measurements in order to add constraints of the portions of the arms that are relatively nearby as presented in the left frame of Figure~\ref{punctual_dections_structure}. From these results it is visible that for many arms the continuity is not straightforward. Therefore, it is only used to constrain an expected model and not to confirm its realism. In a similar fashion, a study from \citet{hou_and_han_2014} compiled more than 2500 HII regions and 1300 GMCs from the literature. They used the existing distance estimates when there was one and computed one from existing Milky Way rotation curves. This statistic allows them to try to fit various structures at a galactic scale, testing arm counts, logarithmic spiral arms, polynomial arms, influence of the rotation curves selection, etc. The right frame of Figure~\ref{punctual_dections_structure} shows their HII regions distribution along with their best fitting model re-associated to usual arm names. One drawback of this approach is that these types of regions are not always expected to follow the arms very tightly. Additionally the distance estimates for most regions still have an important uncertainty, and the one for which a rotation curve was used relies on its quality and could be biased.\\

\begin{figure}[!t]
	\centering
	\includegraphics[width=0.50\hsize]{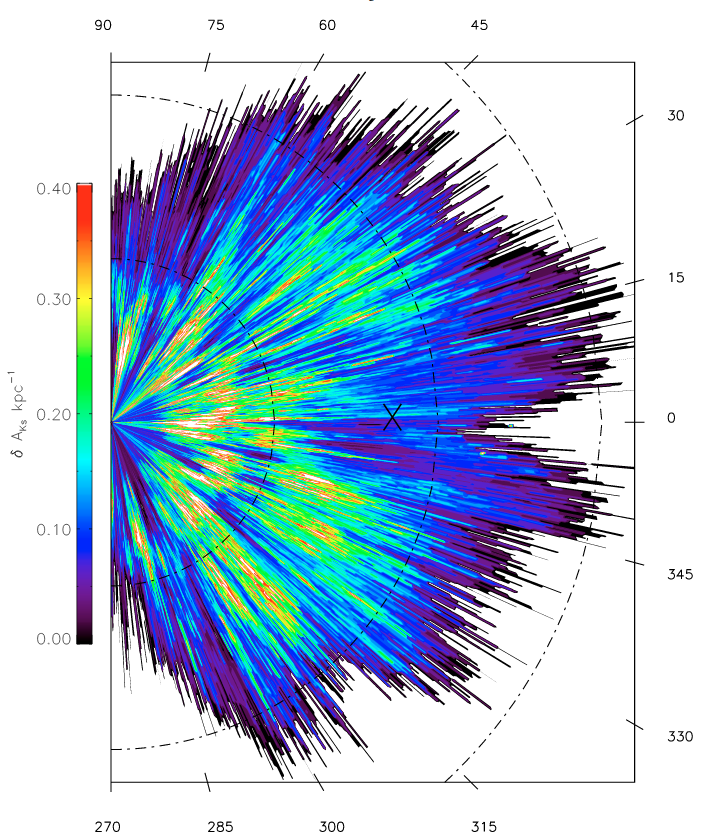}
	 \caption[Extinction map from \citet{Marshall_2006}]{Extinction distribution in the Milky Way using a face-on view of the Galactic Plane in longitude-distance coordinates. The Sun is at the center left and the Galactic Center is marked by the black cross. \textit{From }\citet{Marshall_2006}.}
    \label{marshall_early_2006}

	\end{figure}

Finally, another approach relies on the reconstruction of the extinction distribution in 3D in the Milky Way. This is the approach that will be explored in the Part III of the present study. It mainly relies on the fact that the extinction (see Sect.~\ref{intro_extinction}) directly depend on the dust density. Therefore reconstruction the extinction as a function of the distance is directly equivalent to map the dense structures of the ISM. Since it is one of our main application we delay the detailed discussion on present state-of-the-art extinction maps and the associated difficulties to Section~\ref{ext_properties_part3} as an introduction to our own approach. Still for illustration the Figure~\ref{marshall_early_2006} shows the widely used map from \citet{Marshall_2006}.\\

We note that all the presented observational constraints does not allow to firmly state that the Milky Way would correspond to the grand-design structure of galaxies. The present detection would be very representative of a more flocculent design with a lit of inter-arm structures and much less continuous large scale arms structures overall.

\clearpage
\section{The rise of AI in the current Big Data era}

In this section we describe some of the modern aspects about managing very large amounts of data. We start by highlighting some orders of magnitude that are becoming common for Big Tech companies. Then we draw a simple picture of the artificial intelligence usage and history in order to explain their recent and quick widespread adoption over the past few years. We will end by showing that the use of artificial intelligence has also grown in astronomy studies, and why they are becoming a must-have for recent and future paradigm-breaking large surveys.

\etocsettocstyle{\subsubsection*{\vspace{-1.2cm}}}{}
\localtableofcontents

\vspace{-0.3cm}
\subsection{Proliferation of data and meta-data}

	Data is raw information, usually in a numerical form, and that are uninterpreted. Data are acquired from an observation, or acquisition of some sort, or can sometimes be generated from other data. A simple example of a dataset would be a collection of words arranged in a certain way and stored. This data has intrinsic minimal information that is for example just a number corresponding to a letter, but are usually assembled to create more complex information. Using the same example, the order of the words in the dataset might form a sentence that has an associated meaning. In addition, there is also meta-data, that are considered to be data about other data. Again with the same example, a meta-data would be the time and date the sentence was written, or the time it took to write it. This allows to build context about the initial dataset.\\

	In the current all-numerical information exchange era, tremendous amounts of data and meta-data are generated or exchanged continuously. Every click, message, image, etc, is stored at some point, and more rarely deleted after its objective was achieved. This growing usage of numerical data is also sped up by the Internet Of Things (IOT) trend that consists in adding numerical elements in every objects, that acquire and share even more data than ever. In an attempt to provide orders of magnitude, the global IP traffic in 2017 was estimated at 122 exabytes ($10^{18}$ bytes) per month, and projections predict a value of almost 400 exabytes a month for 2022 (\href{https://www.cisco.com/c/dam/m/en_us/network-intelligence/service-provider/digital-transformation/knowledge-network-webinars/pdfs/1213-business-services-ckn.pdf}{Cisco Annual Internet Report, 2018–2023}). Also, as much of 60\% of the global internet traffic is related to Video On Demand (\href{https://www.sandvine.com/hubfs/Sandvine_Redesign_2019/Downloads/Internet%20Phenomena/Internet%20Phenomena%20Report%20Q32019%2020190910.pdf}{Sandvine, Global Internet Report 2019}). The data seem to follow a continuous increase in dimensionality following the increasing bandwith of all domestic and professional internet connection. More importantly, there are more and more statistics that are performed on the huge amount of data produced daily, and these statistics are in the form of new data as well. However, being able to interpret such a large amount of data is a very difficult task that is very challenging for classical statistical analysis algorithms. Consequently, more and more Big Tech companies like Google, Amazon or Microsoft are investing in Artificial Intelligence (AI) methods that are able to perform data mining in a very efficient way on large datasets and that scale well with their dimensionality.\\

\subsection{Artificial intelligence, a not-so-modern tool}

	\subsubsection{Beginnings of AI}
	\label{ai_begin}
	
	Due to the explosion of AI applications and demonstrated use cases in the past two decades, AI methods are often considered as modern methods that rely on very new technology. While we will provide a detailed definition of AI in Section~\ref{AI_definition} we will state for now that it is a category of methods that learn to solve a problem autonomously with no details given on the way to find the solution other than examples with the expected answer. For now, it can be seen as methods that learn from experience. The first research on what will progressively evolve to become the modern AI methods started in the years 1940. One of the first element was the paper from \citep{mcculloch_logical_1943} that described a mathematical model of what will be called an artificial neuron. For comparison, the first presentation of what will evolved as our modern computer paradigm was presented just a few years ago by \citep{Turing_1937}, also a precursor of modern AI concepts. From these basic elements, the research was mostly focused on artificial neural networks at the time, but other methods that are today considered as a part of the AI field was also designed around those years. The term AI was apparently adopted in 1956 from a conference on the topic of ``making machines behave intelligently'' \citep{McCorduck_2004}. One big step further was the publication from \citep{rosenblatt_perceptron:_1958} that described a model for connecting binary neurons in the form of a rudimentary network based on the weighted sum of input signal and that was named the Perceptron (see Sect.\ref{sect_perceptron}). Interestingly, this publication was made in the journal ``Psychological Review''. All the basic elements were in place for already very capable neural network predictions, and the following 20 years are known as the Golden Age of AI. During this time there was large money investment into the new field and a profusion of ideas about AI methods and techniques, but also dangerous claims about how these methods would reach near human performance in a matter of years. 
	
	\subsubsection{End of 20th century difficulties and successes}
	
	A few years latter the publication \citep{Minsky_1969} raised strong limitations in the present Perceptron formalism leading to a long rejection of all methods based on what was called ``connectionism''. Overall the difficulties mainly originate in the lack of computational power at the time in order to build large enough models to perform properly. During this time the AI research focused on other approaches. It is only in the early 1980 that the neural networks started to have a new support, mostly base on the publication from \citet{Hopfield_1982} that defines a new way of connecting and training a network architecture, that are now called Hopfield networks. Few years latter the publication from \citet{rumelhart_learning_1986} was a game changer since it summarized recent advances on neural networks and described the backpropagation algorithm for training neural network. This is still at this day one of the most used methods, even if it has been improved with some refinements (see Sects.~\ref{neuron_learn} and ~\ref{mlp_backprop}). Despite these important steps, funding agencies and companies lost their interest in AI as the field was not yet successful in providing industrial-scale applications that it had promised several years ago. The lack of computational power remained an issue, and the methods themselves were requiring an amount of training data that was not accessible at the time. Still, during the following years many adjustments were made behind the scene by some researchers who pushed the methods to the point where it could truly accomplish large-scale applications.

	\subsubsection{The new golden age of AI}
	
	In the late 1990's and early 2000's, it was possible to begin to have large datasets and the computational power of recent hardware started to reach a point that was compatible with AI techniques. An important mindset shift occurred in 1997 when the AI-dedicated system Deep Blue from IBM \citep[described in][]{Campbell_2002} defeated a world-class champion at chess. Compared to modern architectures, this machine was mainly performing a brute-force approach of decision tree comparison. Still, it was sufficient to put new lights on AI and on what the last two-decade improvements had led to. After that, large technology companies invested massively in AI, successfully applying it on problems that were predicted to be solvable with the methods decades ago like speech recognition, industrial robotics, data mining, computer vision, medical diagnosis, etc.\\
	
	At this point there was a very strong mutual interest from technology companies and AI researchers, that led to a very quick improvement of these method capacities supported by new dedicated hardware technologies (see Sects.~\ref{matrix_formal}). The "Deep Learning" field (see definition in Sect.~\ref{mlp_backprop} and~\ref{conv_layer_learn}) is the result of these recent advances by generalizing the approach from the past decades and overcoming many of the exposed difficulties, and also improving their numerical efficiency. We also note that it is easier than ever to use these methods from a completely external standpoint. A large variety of state-of-the-art user-friendly frameworks and pretrained models are freely accessible, even if it may occasionally imply some misuses (see Sect.~\ref{tool_boxes}). At these day the AI field is still moving very fast and the corresponding methods are progressively becoming the only suitable solution to work with ever growing datasets.\\

	\subsubsection{Astronomical uses of AI}
	\label{astro_ia_use}
	
	Artificial Intelligence is becoming a common tool for other research fields for which it is able to process large amount of data or learn complex correlations automatically from very high dimensionality spaces. For research other that in the AI field itself we rather use Machine Learning (ML) which is a subpart of the larger AI field that excludes many very specific tasks dedicated to reproducing realistic intelligence and cognition. ML is more focused on practical methods like regression, classification, clustering, etc, without aiming for high-level abstraction. This does not mean in any way that ML methods are less powerful, since they mostly rely on the same algorithms and architectures, but it is a switch in focus.\\
	
	As many other research fields, Astronomy has begun to use ML methods as an analysis tool a few years ago. Here we list a few works that had a significant impact in our community and that were relying on ML methods. The specific field of external-galaxy analysis and classification adopted ML methods earlier than others. The famous Galaxy Zoo study from \citep{Lintott2008} provided an unprecedentedly large catalog of galaxies with morphological classification performed by matching multiple human visual classification on SDSS images for each of them, providing very accurate labels. This dataset became a widely adopted playground for ML applications that attempted to automate the classification in order to create a high performance classifier that could be used on new galaxies. Various methods had been employed for this, including support vector machine, convolutional neural networks, Bayesian networks and more \citep[e.g][...]{Banerji_2010, Huertas-Company_2011, Huertas-Company2015, Dieleman_2015, Walmsley_2020}. On another topic, ML methods can sometimes be used to reproduce a very well defined problem for which there is an analytical solution but that is slow to compute. This way an efficient ML algorithm like a light neural network can be used to significantly speed up the prediction or can even be used as an accelerator for a larger computation \citep{Grassi_2011, Mijolla_2019}. By extending the previous approach it is possible to use ML methods to interpolate between predictions that are timely to make in order to provide a full parameter space predictor from a sample of examples \citep{Shimabukuro_2017}. A few other examples are, ISM structure classification with support vector machine \citep{Beaumont_2011}, molecular clouds clustering using the unsupervised Meanshift method \citep{Bron_2018}, or even differentiating ISM turbulence regimes using again neural networks \citep{Peek_2019}. This is a very incomplete view, since ML methods have become very common in astronomical studies the last few years.\\

\subsection{Astronomical Big Data scale surveys}
\label{astro_big_data}

	\subsubsection{Previous large surveys}
	
	Astronomical surveys are known to be very large datasets. Telescopes can produce very high resolution images, and point source catalog surveys are usually very large and present a high dimensionality. For example the Apache Point Observatory Galactic Evolution Experiment survey \citep[APOGEE][]{APOGEE_2017} contains $\sim 146000$ stellar spectra of resolution $R \sim 22,500$, which is a relatively small number of objects but with a very high dimensionality. At higher orders of magnitude, we can cite the Wide-field Infrared Survey Explorer \citep[WISE][]{wright_wide-field_2010} that contains $\sim 5.6 \times 10^8$ point source objects and a few parameters (4 bands, 4 uncertainties, sky position, other meta data ...) for each object. This size of dataset begins to be difficult to analyze on modest hardware infrastructures, and is clearly out of the scope of a domestic computer for a full dataset analysis. The Two Micron All Sky Survey \citep[2MASS][]{skrutskie_two_2006} is an other widely used survey that presents a similar size. At an even higher size scale there is the Spitzer space observatory  \citep{werner_spitzer_2004} point source catalog that contains almost $1.5\times 10^9$ objects with barely fewer parameters than the previous two surveys. In the same size category we can cite the U.S. Naval Observatory - B1 \citep[USNO-B1][]{Monet_2003} that contains also $~1 \times 10^9$ objects. \\
	
Such dataset sizes become very difficult to handle even for more advanced hardware using classical methods. Even for the smaller dataset we started with, Machine Learning methods could provide significant treatment time improvement or provide new insights due to new ways of exploring the parameter space. For the larger surveys, ML approaches really start to shine as they provide new analysis possibilities that are not possible using more classical tools. Additionally, one of the additional advantage of ML methods is the automation. Even if some task can be performed using carefully designed classical analysis tools, an ML method will either be able to find the most optimum approach granting even better results by itself or at least it could find a more efficient process to speed up the analysis. We discuss more deeply the advantages an ML approach can provide, even on "relatively small" dataset, in Section~\ref{yso_ml_motivation}.
	
	\newpage
	\subsubsection{A new order of magnitude with PanSTARRS and Gaia}
	\label{modern_large_scale_surveys}
	
	A recent very large scale survey is the Panoramic Survey Telescope and Rapid Response System \citep[Pan-STARRS][]{Chambers_2016} that is estimated to contains 1.6 PetaBytes of data in the form of very high resolution images. Dealing with such a dataset is a true challenge, and using methods that are able to work efficiently on image processing is absolutely necessary to perform analysis on the full dataset (See~\ref{image_process_section}). Another very popular survey right now is the second data release of the Gaia mission \citep{Gaia_Collaboration_2018} that contains more than $1.6 \times 10^9$ stars with astrometric parameters and a large variety of other acquired quantities like magnitudes, kinematics, or post processed quantities. This dataset is not only a challenge due to its size but also its raw data acquisition rate. Indeed, the spacecraft itself sends around 30 GB of raw data per day and the end of mission catalog is estimated to be of around 1 PetaByte. Additionally, Gaia data go though an advance processing of the data to reconstruct most of the parameters form observed star movement over successive scans. This process justified the creation of the Gaia Data Processing and Analysis Consortium (DPAC) that shares the data treatment in terms of compute resources and manpower into several European countries. The Gaia mission had justified the construction of large computer cluster totally dedicated to the mission data analysis and end product storage. An interesting fact is that the size of the intermediate results is so large and must be over so much storage nodes that the Gaia DPAC has adopted the Hadoop file system tool that is otherwise mostly used by Big tech companies for their country scale servers. Using Machine Learning to perform the analysis of such a large dataset then appears as a fantastic opportunity to extract unmatched problem complexity due the huge statistic in the dataset, and also as a necessity since many classical methods would scale too badly on such dataset sizes. We note that due to the distribution of the intermediate data over several storage clusters, using ML to perform intermediate computations requires very tricky distributed model learning as exposed in \citet{Hsieh_2017}.

	\subsubsection{The historical challenge of SKA and following surveys}
	
	Finally, we discuss succinctly the case of incoming very large scale astronomical instruments. The most illustrative example is what is expected to be produced by the Square Kilometer Array radio telescope interferometer \citep{Dewdney_2009}. This instrument will have such a wide angle of view and spectral bandwidth that it would output a total of 15 TB/s and is expected to produce 600 PetaByte per Year of at least partly pre-processed data. Currently, there is no detailed plan on how the data processing will be managed since there is no existing analysis method in the astronomical community that would have a sufficient computational performance. On the other hand, by the time the instrument will be finished, in a few years, computer technologies are expected to have significantly improved. Still, Machine Learning methods are at the middle of the attention of the SKA development teams since it appears that they have the most suitable design for this upcoming challenge.\\
	
	Presently several ML applications are being tested on SKA precursors to assess their capabilities to replace some pre-processing steps or as a posterior analysis tool. Indeed, even if the SKA data generation was tamed, methods must be spread to the astronomical community to enable them to analyze the data as well. Finally, the SKA instrument will certainly not be the only one of this scale to be built in the upcoming years, therefore these methods are likely to become a necessary part of an astronomer regular toolkit.

\setkeys{Gin}{draft=False}



\newpage
\clearpage
\thispagestyle{empty}
\null

\newpage
\hfill
\thispagestyle{empty}
\newpage
\hfill
\vspace{0.3\textheight}\\
\thispagestyle{empty}
\part{Young Stellar Objects classification}

\newpage
\null
\thispagestyle{empty}
\newpage
\etocsetnexttocdepth{2}
\etocsettocstyle{\section*{\vspace{-0.1cm}Part II: Young Stellar Objects classification}}{}
\localtableofcontents
\clearpage

\section{Young Stellar Objects as a probe of the interstellar medium}
\label{intro_yso}
\etocsettocstyle{\subsubsection*{\vspace{-1cm}}}{}
\localtableofcontents

\subsection{YSO definition and use}
\label{yso_def_and_use}

Young Stellar Objects (YSOs) refer to a relatively wide range of protostars. As described in Section~\ref{stellar_formation_dense_environment}, stars form in dense molecular clouds through gravitational collapse. In the presently accepted view \citep[for example, ][]{McKee_2007, Kennicutt_2012}, a typical low- to intermediate-mass YSO ($<8$ M$_\odot$) continuously accretes matter from its parent cloud through a disk that is quickly formed and partly maintained by its rotation. The disk progressively disappears due to its matter falling into the star because of friction, being photo-evaporate due to the star becoming energetic enough to radiate more, or being used to form planets. Simultaneously, the star accumulates a sufficient amount of matter to lead to the start of the nuclear fusion of hydrogen nuclei  in its core. A star for which the nuclear fusion is inactive or non-dominant in its energy production is a pre-main sequence star. YSOs correspond to all the steps from the gravitational collapse to the advanced stages of pre-main sequence stars. The formation sequence of massive stars is less well understood \citep{Motte_2018}, but it does not impact our study because the scarcity of those stars makes them less useful targets for the purpose of the present work.\\

Observing YSOs in stellar clusters and molecular clouds is a common strategy to characterize star forming regions. Their presence attests star formation activity, their spatial distribution within a molecular complex provides clues about its star formation history \citep{Gutermuth_2011}, and their surface density can be used as a measure of the local star formation rate \citep{heiderman_star_2010}. The youngest YSOs are the most interesting ones since they are more likely to be very close to their formation point, while more evolved ones had more time to drift from their original location. This was demonstrated, for example, by \citet{Hacar_2017} in the case of the low-mass star forming region NGC 1333 (see their figures 9 and 10) or by \citet{Buckner_2020} in the case of the massive star forming region NGC 2264. This is due to various ways of getting a different velocity than the one of the forming cloud, for example by interaction with other stars. For example, \citet{Stutz_2016} observed in Orion that the velocity of younger protostars is more coherent with their parent filament than that of more advanced protostars. Interestingly they concluded that protostars might be ejected by a slingshot-like mechanism from their oscillating original filament. This would also imply the interruption of the accretion mechanism, impacting the stellar IMF (Initial Mass Function). Recently, YSOs have also been combined with astrometric surveys like Gaia to recover the 3D structure and motion of star-forming clouds \citep{grossschedl_3d_2018}. This proves that YSOs may indeed be used to reconstruct more globally the Milky Way structure in 3D by the combination of large YSO catalogs and large astrometric surveys. There are two main difficulties to this approach, (i) the first being that YSOs are embedded in an envelope that progressively disappears during the protostar evolution making them usually very faint in the bands that are used to perform astrometry, (ii) the second being that, even using more suitable infrared observations as described bellow, it is still difficult to construct sufficiently large catalogs. Therefore, it is necessary to find efficient identification methods that are able to work on large catalogs that are sensitive enough to detect a large amount of them.

\subsection{YSO candidates identification}

\begin{figure}[!t]
	\centering
	\begin{subfigure}[t]{0.65\textwidth}
	\includegraphics[width=\textwidth]{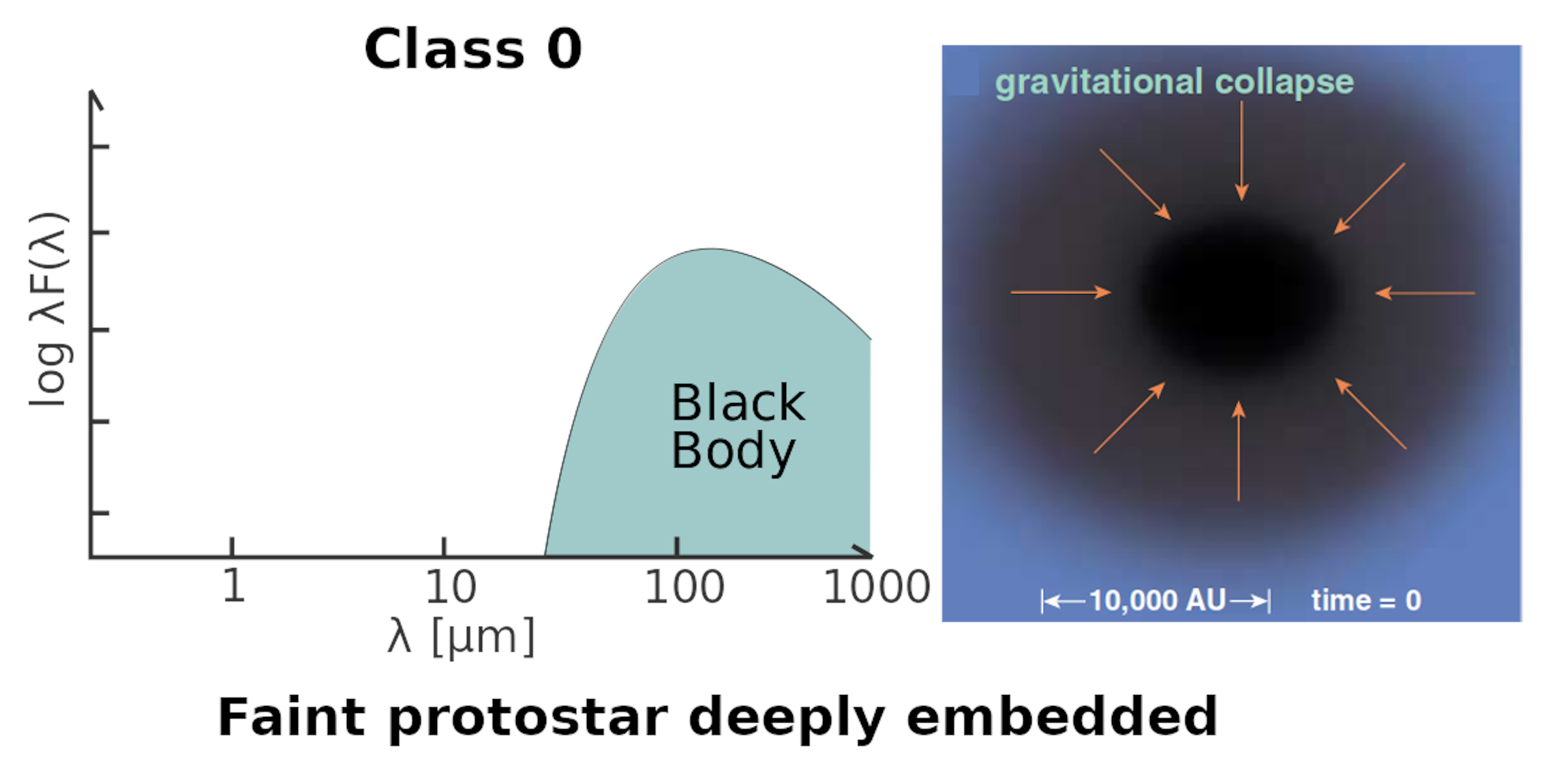}
	\end{subfigure}
	\vspace{0.2cm}
	\begin{subfigure}[t]{0.65\textwidth}
	\includegraphics[width=\textwidth]{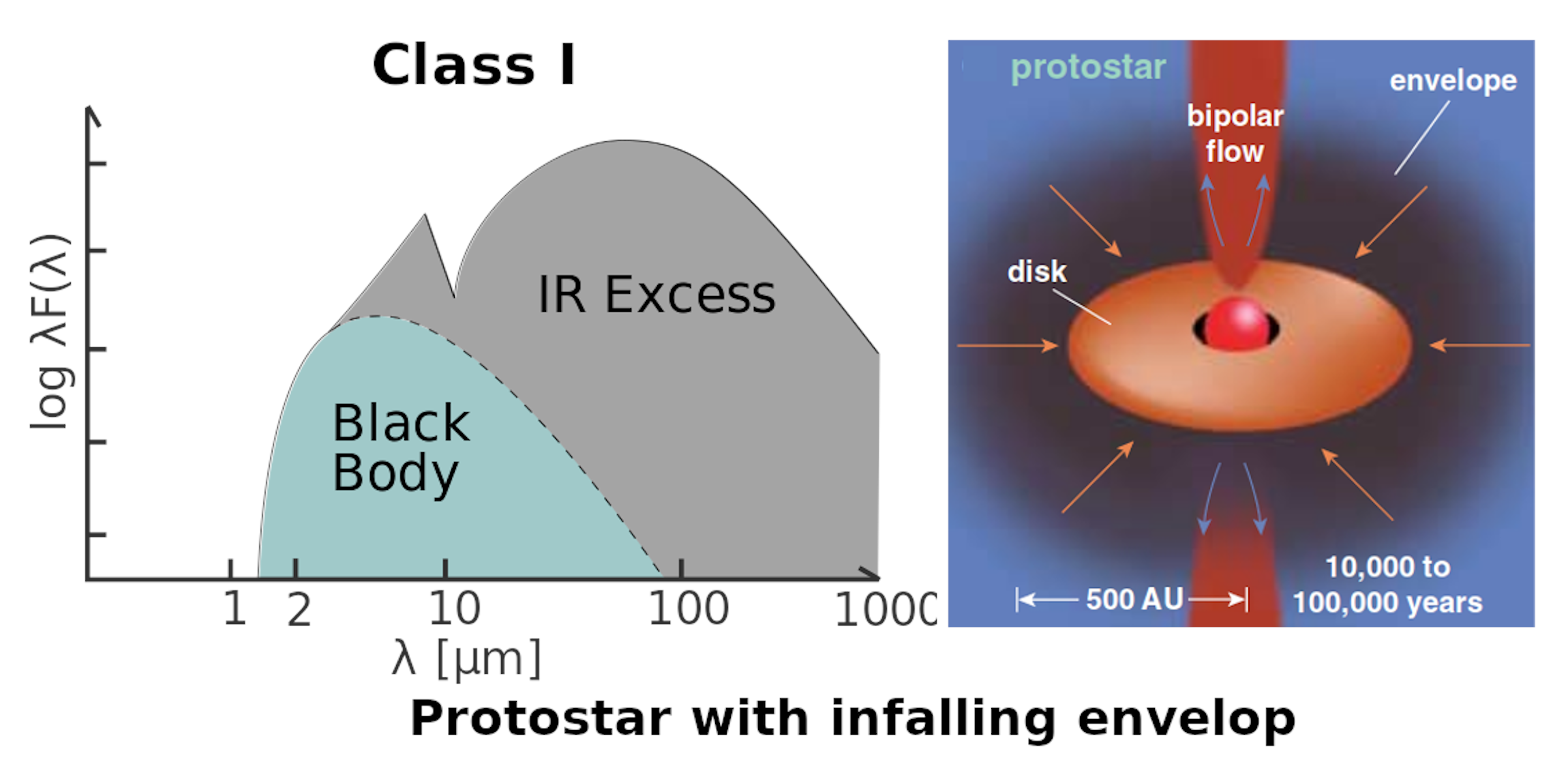}
	\end{subfigure}
	\vspace{0.2cm}
	\begin{subfigure}[t]{0.65\textwidth}
	\includegraphics[width=\textwidth]{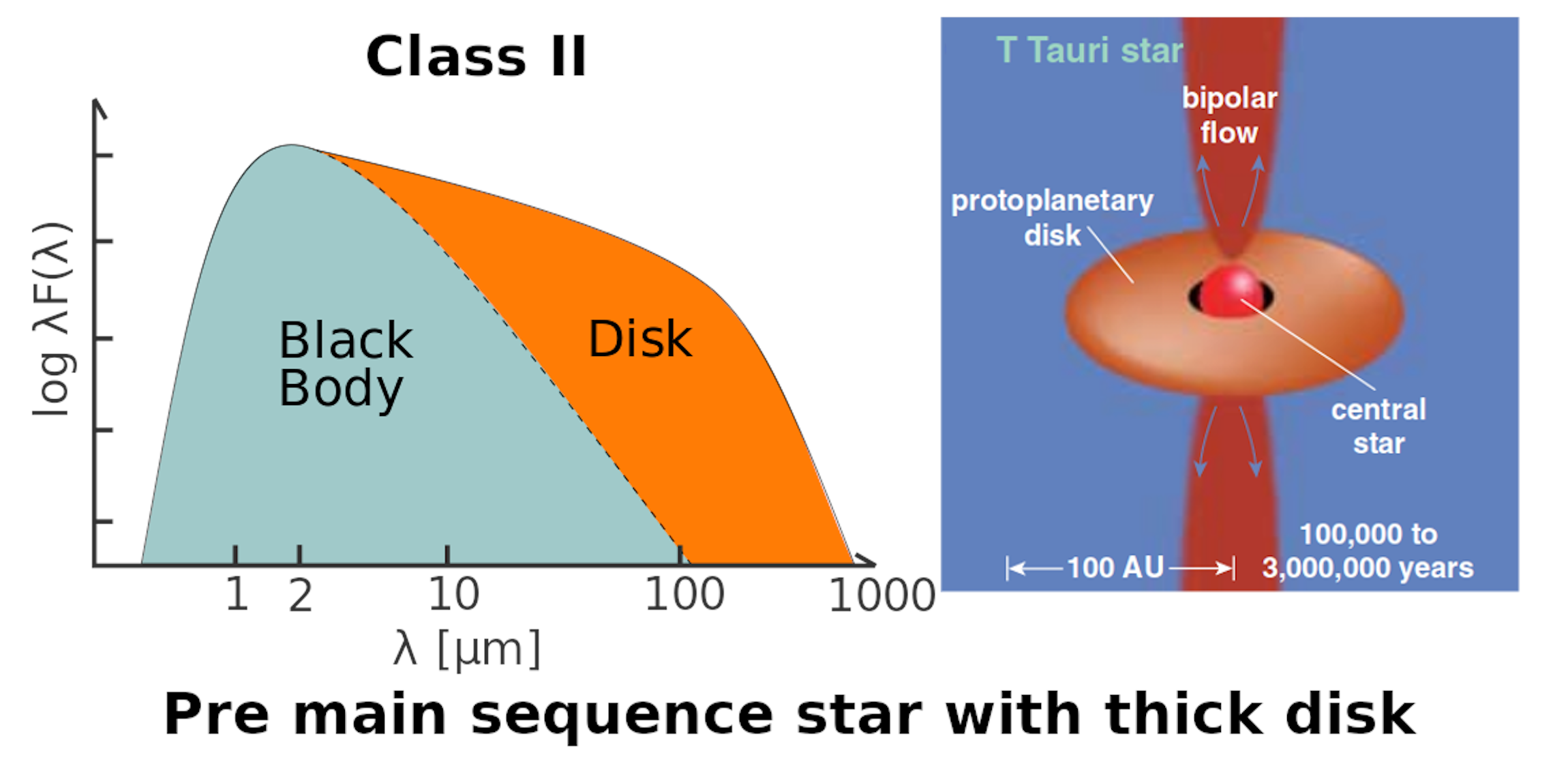}
	\end{subfigure}
	\vspace{0.2cm}
	\begin{subfigure}[t]{0.65\textwidth}
	\includegraphics[width=\textwidth]{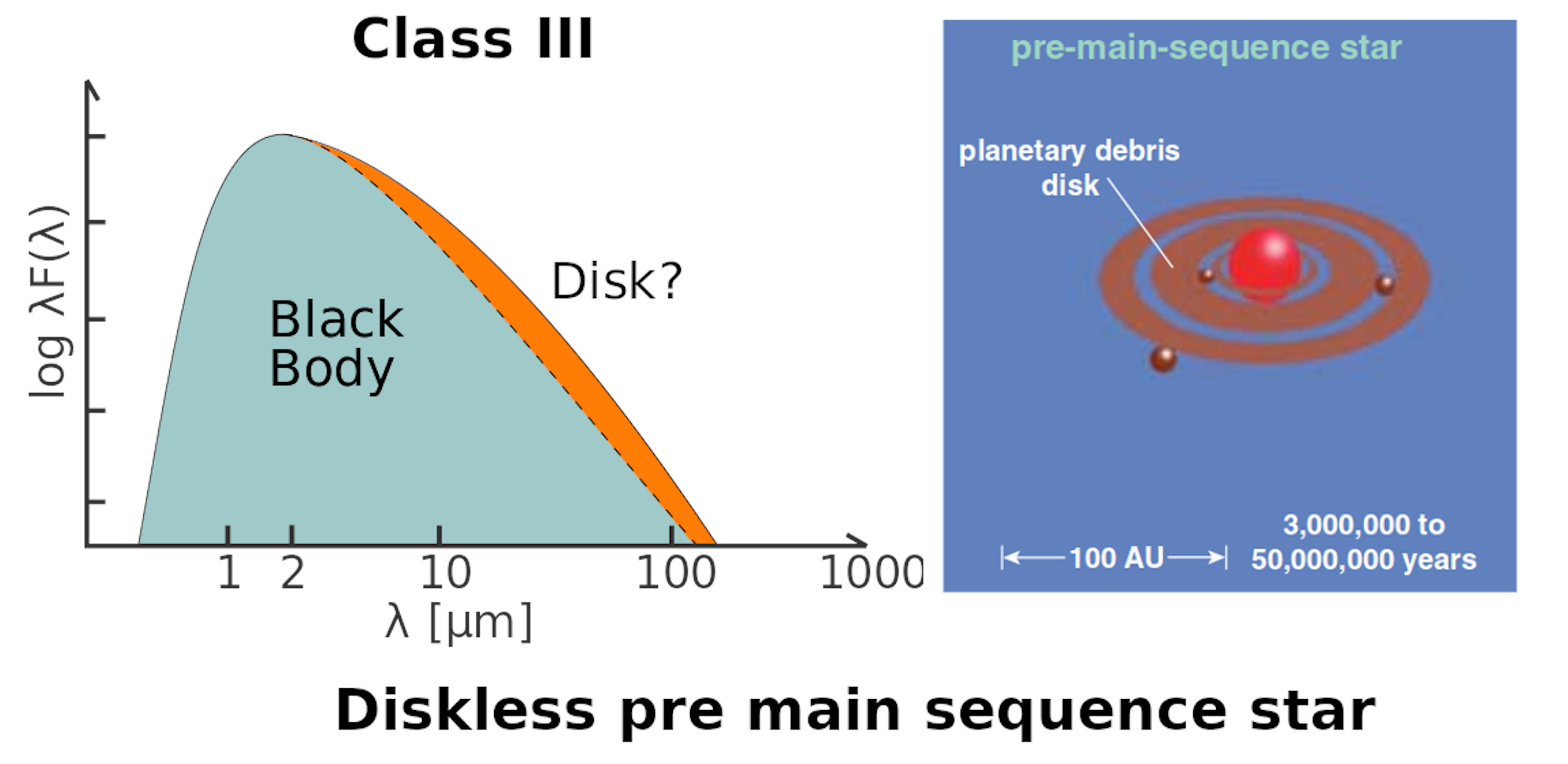}
	\end{subfigure}
	\caption[Simplified YSO SED for each class.]{Simplified YSO SED for each class. {\it Left}: The spectral energy as a function of the wavelength. The contribution of different elements of the system are shown colored. {\it Right}: Illustration of the star forming step associated with each class, which shows the different elements contributing to the SED. Adapted from \citet{Greene_2001} and \href{https://doi.org/10.6084/m9.figshare.1121574.v2}{Persson (2014)}.} 
	\label{yso_sed}
\end{figure}

YSO identification is often summarized as a classification problem. As they are cooler than more evolved stars and due to their dusty environment, YSOs are much simpler to detect using infrared wavelengths. Consequently, their classification relies mainly on their Spectral Energy Distribution (SED) in the IR, which allows one to distinguish evolutionary steps that range from the prestellar core phase to the main sequence. These steps were translated into classes ranging from 0 to III by \citet{Lada87} and \citet{allen_infrared_2004}, corresponding to the observed slope of the IR SED, characterized by its spectral index. Objects that present a black body spectrum in the far-IR, and that are quiet in the mid-IR are called Class 0 (C0) and correspond to dense cores or deeply embedded protostars. Objects that present a black body emission in the mid-IR and a strong excess in the far-IR are called Class I (CI) and correspond to protostars dominated by the emission of an infalling envelope. Objects that present a black body emission in the mid-IR but with a flattened emission in the far-IR are called Class II (CII) and correspond to pre-main sequence stars with an emissive thick disk. Objects that present a black body emission in the mid-IR and are devoid of far-IR emission are called Class III (CIII) and correspond to pre-main sequence stars without disks or too faint ones to be detected. Figure~\ref{yso_sed} shows simplified typical SEDs for each class, along with an illustration of the corresponding star formation step. We note that this classification is sometimes further refined to include other sub-classes like a flat-spectrum class between CI and CII, a Transition Disk class between CII and CIII \citep{gutermuth_spitzer_2009}, or even add a fainter class for deeply embedded CI YSOs \citep{megeath_spitzer_2012}. Overall, it is an efficient classification but it can still lead to misinterpretation for specific objects, like when a CII or CIII YSO is observed behind a thick cloud leading to a confusion with a CI. Similarly, when a CII YSO is observed edge-on, it is obscured by its disk and can be confused with a CI object. More subtle effects have also been identified that increase the confusion between pre-main sequence stars with a disk and flat spectrum objects \citep{Crapsi_2008}. Still, this classification is very efficient for statistical studies, like when studying the 3D ISM structure. This subclassification can then be used to provide additional information on the structure and evolution of star-forming regions since youngest (up to class~I) objects are more likely to be close to their formation position than more evolved YSOs.\\

\vspace{-0.25cm}
One of the most famous classification method that is based on the IR SED is the one described by \citep{gutermuth_spitzer_2009} based on data from the Spitzer space observatory  \citep{werner_spitzer_2004} and from the 2 Micron All Sky Survey \citep[2MASS, ][]{skrutskie_two_2006}, which is fully described in Section~\ref{data_prep}. The two papers that describe the first two versions of this classification \citep{Gutermuth_2008, gutermuth_spitzer_2009} have inspired other widely adopted methods on other surveys like the one by \citet{koenig_wide-field_2012}, for the use of data from the Wide-field Infrared Survey Explorer \citep[WISE, ][]{wright_wide-field_2010}. It is to be compared to other similar methods, for example \citet{allen_infrared_2004} or \citep{Robitaille_2008} that both use Spitzer.\\

\vspace{-0.25cm}
We note that despite the YSO classes being historically defined using infrared criteria, other identification methods can be used. For example centimeter or (sub)milimeter interferometers were used to assess the presence of the disk and its evolution state, like with the James Clerk Maxwell Telescope (JCMT) \citep{Brown_2000}, the SubMilimeter Array (SMA) \citep{Jorgensen_2009}, the Very Large Array (VLA) \citep{Segura-Cox_2018} or even more recently the Atacama Large Milimeter/submilimeter Array (ALMA) \citep{Yen_2014, Ohashi_2014}, etc. Many other similar studies are listed in \citet{tobin_2020}. YSOs are also known to be stronger X-ray emitter than more evolved stars. Such radiations are also capable of escaping dusty environment and can therefore be identified using a high-resolution X-ray telescope like the Chandra X-ray space Observatory \citep{Feigelson_2013}.

\subsection{Machine Learning motivation and previous attempts}
\label{yso_ml_motivation}
\vspace{0.4cm}
   As we already discussed in Section \ref{astro_big_data}, the astronomical datasets are becoming too large for traditional analysis methods, and more automated statistical approaches like ML are used. They are able to both work efficiently on large datasets using many dimensions, and take advantage of the increased statistics to often overcome limits of previously used methods. In this context, it is timely to try and design a classification method for YSOs, relying on current and future large surveys and taking advantage of ML tools. Such approaches have been attempted by \citet{marton_all-sky_2016}, \citet{marton_2019}, and \citet{miettinen_protostellar_2018}. The study by \citet{marton_all-sky_2016} used a supervised ML algorithm called Support Vector Machine (SVM) applied to the mid-IR ($3 - 22\ \mu$m) all-sky data of WISE \citep{wright_wide-field_2010}. The SVM used in this study offers great performance on linearly separable data. However, it is not able to separate more than two classes at the same time and has a less good scaling with the number of dimensions than other methods. Besides, the full-sky approach produced large YSO candidate catalogs, but suffers from the uncertainty and artifacts in star-forming regions of the WISE survey \citep{lang_unwise:_2014}. Additionally, the YSO objects used for training were identified using SIMBAD, resulting in a strong heterogeneity in the reliability of the training sample.\\
   
  \vspace{0.4cm}
In their subsequent study, \citet{marton_2019} added Gaia magnitudes and parallaxes to the study. Gaia is expected to add a large statistics and to complete the SED coverage \citep{Gaia_Collaboration_2018}, but the necessary cross match between Gaia and WISE excludes most of the youngest and embedded stars. The authors also compared the performances of several ML algorithms (SVM, Neural Networks, Random Forest, ...) and reported the random forest to be the most efficient with their training sample. This is a better solution as it overcomes the exposed limitations of the SVM. However, as in their previous study, the training sample compiles objects from different identification methods, including SIMBAD. This adds more heterogeneity and is likely to increase the lack of reliability of the training sample, despite the use of a larger training sample.\\

\citet{miettinen_protostellar_2018} adopted a different approach by compiling a large amount of ML methods applied to reliably identified YSOs using 10 photometric bands ranging from $3.6$ to $870\ \mathrm{\mu m}$. For this, he used the Herschel Orion Protostar Survey \citep{HOPS_2013}, resulting in just less than $300$ objects. Such a large number of input dimensions combined with a small learning sample is often highly problematic for most ML methods (Sect.~\ref{nb_neurons}). Moreover, this study focuses on the subclass distinction of YSOs and does not attempt to extract them from a larger catalog that contains other types of objects. In consequence, it cannot be generalized to currently available large surveys and relies on a prior YSO candidate selection. \\

\newpage
\subsection{Objective and organization}
\vspace{0.4cm}
	The aim of this first part of the manuscript (Part I)  is {\bf to propose a methodology to achieve YSO identification and classification, based on ML, and capable of taking advantage of present and future large surveys}. We describe some properties of the ML methods in general and explain our choice of method and our choice of building our own framework. We extensively detail the functioning of the selected ML method along with some basic application examples. We then describe how this method can be used to perform YSO classification using the Spitzer space observatory IR data, based on the widely used classification method developed by \citet{gutermuth_spitzer_2009}. We detail the data preparation phase and our choice of representations for the results along with their analysis, therefore exposing the encountered limitations. Finally, we discuss the caveats and potential improvements of our methodology, and propose a probabilistic characterization of our results. This Part ends with a reconstruction of the results from \citet{grossschedl_3d_2018} using our own YSO catalog to infer the 3D structure of the Orion molecular cloud. \\
	
 \vspace{0.4cm}
We emphasize that these results are to be published in A\&A \citet[][accepted]{cornu_montillaud_20} and that the present manuscript reproduces many sections of the published version, while also providing a large amount of additional material and a deeper analysis of the study. Also, this publication is associated with our catalog of YSO candidates that is described in the present manuscript in Section~\ref{proba_discussion} and that will be publicly available at the CDS.




\newpage
\section{Classical Artificial Neural Networks}
\label{global_ann_section}

In this section we describe the theoretical and technical aspects of Artificial Neural Networks (ANN) in detail. For this, we present ML in general along with the corresponding categories of algorithms. We will then describe a classical mathematical model to construct ANN and improve it until it is able to approximate any function. Finally, this section will describe a wide variety of in depth tuning of such network with common examples.

\etocsettocstyle{\subsubsection*{\vspace{-1cm}}}{}
\localtableofcontents

\newpage

\subsection{Attempt of ML definition}
\label{AI_definition}
	\subsubsection{"Animal" learning and "Machine" Learning}
	In order to define what ML means we need to find a definition of "Learning", which is often difficult as there are plenty of definitions depending on the field it is apply to (Psychology, Biology, Pedagogy, ...) or on the person you ask in a given field. The online Cambridge Dictionary holds the following definition:
\begin{displayquote}
\it ``The process of getting an understanding of something by experience''\footnote{\href{https://dictionary.cambridge.org/us/dictionary/english/learning?q=Learning}{https://dictionary.cambridge.org/}}
\end{displayquote}
while the \textit{Tresor de la langue Française informatisé} defines it with :
\begin{displayquote}
\it ``Acquérir la connaissance d'une chose par l'exercice de l'intelligence, de la mémoire, des mécanismes gestuels appropriés, etc.''\footnote{\href{http://stella.atilf.fr/Dendien/scripts/tlfiv5/advanced.exe?8;s=250699680;}{http://stella.atilf.fr}}
\end{displayquote}
that can be translated as :
\begin{displayquote}
\it ``Acquire the knowledge of something through the practice of intelligence, memory, appropriate gestures, etc. ''
\end{displayquote}
	The two definitions appear somewhat different, but both contain elements that are used to commonly define "Animal" learning. Both  definitions contain the idea of experience, or practice, meaning that in order to learn, an animal must face the appropriate situation, ideally several times. The second main element is the \textit{memory}, which is necessary in order to retain information about the experience. Then there is the understanding and correction, that is usually referenced to as the \textit{adaptability}. It means that the animal must be able to change its behavior in regard of the experience outcome. And finally the last point is the \textit{generalization} ability, that allows the animal to adapt its behavior based on non-identical but similar experience and dress a continuity of behaviors between them.\\

\vspace{-0.2cm}
Animal learning and intelligence are based on this elemental abilities but are obviously more complex and require many other complex faculties. We would need to define, for example, the reasoning or the logical deduction before starting to talk about intelligence. However, the previous basic capacities are enough to perform a lot of tasks, and this already justifies the attempt to reproduce them artificially. Here comes the ML (or Artificial Learning), for which there is, as well, no easy definition. Our personal definition, which merges many of other definitions, is the following :
\begin{displayquote}
\it ``Make a computer extract a statistical information and adapt to it through an iterative process.''
\end{displayquote}
This definition is vast in order to include all the common algorithms that are granted the ML label. It echoes the previously defined learning ability, as the machine will need :
\begin{itemize}[leftmargin=0.5cm]
\setlength\itemsep{0.2em}
\item \textit{Memory} to remember either the previous situations, or a reduced version of them.
\item A way to estimate if its behavior (output) is appropriate.
\item A way to \textit{adapt} its behavior during the learning process.
\item A way to \textit{generalize} its behavior to new outputs.
\end{itemize}
We acknowledge again that this is a very global view, and that it might not fit every algorithms that are considered as ML but it should correspond to most of them.
	
	\subsubsection{Types of artificial learning}
All ML methods do not work the same way. Therefore, it is necessary to identify what is the objective of the application in order to select a suitable method or algorithm. Usually ML methods are separated into two main families, {\it supervised} or {\it unsupervised}. It is however also common to add three other families, {\it semi-supervised}, {\it reinforced} and {\it evolutionary} learning.

\begin{itemize}[leftmargin=0.5cm]

\item The {\bf supervised} methods use a training dataset that contains the expected output of each example. The algorithm attempts to reproduce the target. It learns by comparing its current output with the target and correcting itself based on this comparison. After training, such algorithms are able to generalize to objects that are similar to those that were used for training. This is the most common type of algorithms because they are often simpler than the ones in other categories, and because it is easier to assess their prediction performance. They usually have a broad range of applications. 

\item The {\bf unsupervised} methods use a training dataset with no information on the expected output. The algorithm will then attempt to create categories in the dataset by itself based on some pre-defined proximity estimator. Most of these methods are clustering ones, that can either be used for classification or dimensionality reduction. Despite their reduced application range they are commonly used as well. Interestingly, many clustering methods that are widely used since decades have recently been rebranded as ML methods to follow the trend of this domain, which is legitimate considering the previous definition of ML.

\item The hybrid {\bf semi-supervised} methods are often a combination of an unsupervised algorithm that does the first part of the work, either to simplify the problem or to remove some bias in target definitions, and of a supervised part to benefit of the application range of these methods. They are fairly common as they often merge qualities of the two categories. Still, they are often more computationally heavy than purely supervised ones, and can be more difficult to constrain properly.

\item Instead of a labeled dataset, the {\bf reinforcement} methods use a reward function to measure how appropriate the output of the algorithm is. Usually, there is a distinction between an agent part that acts on an environment, and an interpreter part that provides the reward regarding the action that was performed. One key difference with other methods is that it often has a delayed reward since an action is often a series of interdependent actions, and its performance can only be assessed based on the final result. These methods are mostly used in robotics or to reproduce human tasks. It allows one to find a solution to a problem where one only knows some basics rules, and whether a specific output is better than another. For example, it was successfully used to make a robot learn how to walk without programming each motor, but just by encouraging the robot to test all possibilities to increase its velocity, in both simulation and real world applications \citep{Heess_2017,Haarnoja_2018}.

\item The {\bf evolutionary} methods are similar to the reinforced one. They work with a dataset, no target and only a global performance measurement. However, they explore the possibility space in a different way that mostly mimics biological evolution. The algorithm prediction is described as a population that starts with random properties. Then it learns by selecting the individuals that provide the best proximity with the solution and create a new population from them and repeat this process until convergence. It also often uses some kind of mutation to ensure exploration of new solutions. 

\end{itemize}

In addition to these families, the methods can be categorized either as discriminative or generative. This is a vast topic but it can be summarized as: discriminative models learn boundaries between cases, (i.e., they learn the probability distribution of the output given an input information), while generative models learn the actual distribution of the examples, (i.e., they learn the probability distribution of the input that corresponds to the output). It means that generative models can be used to create new mock objects or predict the output from an specific input, while discriminative models can only perform prediction tasks. However, generative algorithms are often more difficult to train, computationally more demanding, and are suitable to a relatively narrow range of problems.

	\subsubsection{Broad application range and profusion of algorithms}
	\label{ml_application_range}

ML methods are recognized for their vast application range. It is one of their strengths that is currently driving their global adoption in many computational and scientific fields. They can perform classical tasks like classification, regression and clustering. But they can also be applied to dimension reduction with unexpected results as an alternative to usual compression algorithms. There is a strong interest in there capacity in time series prediction, noticeably for economical markets \citep{Sezer_2019} or climate change predictions \citep[e.g][]{Feng_2016,Ise_2019}. The loudest application is obviously the image recognition with the strong appeal around autonomous vehicles \citep{Bojarski_2016} or facial recognition \citep{Wang_2018}, but with many other scientific applications in a lot of fields (See for Astronomy in Sect.~\ref{astro_ia_use}). There is also more and more high performance generative algorithms that allow to see deceased actors in new movies, to simulate aging of highly researched criminals, or to create realistic numerical instruments \citep[e.g][]{Zhu_2017,Sawant_2019,Engel_2017}. We can also cite ease-of-life applications like new spell checkers, real time vocal translators, real time media upscaling \citep[e.g][]{Bahdanau_2014,Ghosh_2017,Shi_2016}, or for scientists the IArxiv application that sorts papers by probable interest for the reader \citep{iarxiv}. Convinced or not by these applications, ML is becoming a part of the scientific landscape and understanding how its work is becoming an essential knowledge. We already highlighted some typical ML applications in astronomy in Section \ref{astro_ia_use}.\\

\begin{figure}[!t]
	\centering
	\includegraphics[width=0.92\textwidth]{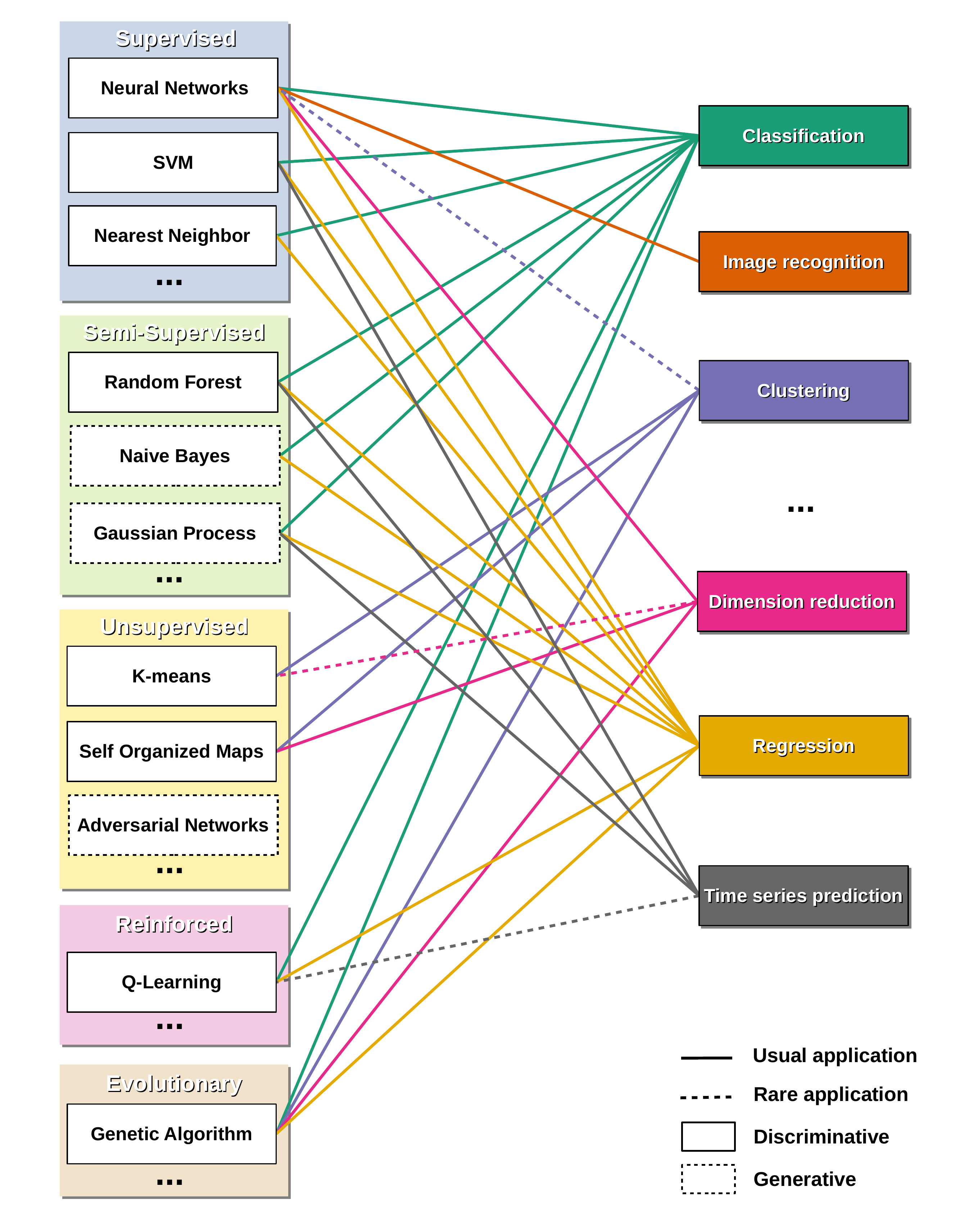}
	\caption[List of common ML algorithms packed by type]{List of common ML algorithms packed by type and linked to their usual application cases. We note that neither the list of methods, nor the list of applications, nor the links themselves are exhaustive.}
	\label{fig_ml_types}
	\end{figure}

As well as there are many application possibilities, there is also a profusion of algorithms. Among the famous ones are Artificial Neural networks, Random Forests, K-means, and many others. But there are also less known ones like Radial Basis Function Networks, Self Organizing Maps, Neural Gas, Deep Belief Networks, etc. Moreover, algorithms are more and more likely to be combined to achieve either better performance or new capabilities, like Reinforced Generative Adversarial Networks. Unfortunately, all those methods do not perform equally on each application, and are even unable to perform certain tasks at all. In Figure~\ref{fig_ml_types} we list some well-known algorithms arranged by family and linked to their usual application cases.

	\subsubsection{Toolboxes against home-made code}
	\label{tool_boxes}
	
	When searching how to use ML one faces an elemental dilemma that is the choice of the tool or library. There are many possible answers to this question, that we will address here, with some partiality. First of all one could choose to use none of the available frameworks and be tempted to program his/her own algorithm from scratch. This is not a common choice for many reasons including time efficiency, computational performance, ease of modifications, etc. Therefore we will first discuss the scenario of the use of a pre-existing framework.\\

\newpage
Nowadays, knowing a limited number of programming languages is not a problem to start using a ML framework since there are implementations in almost any common language. Moreover, most of the frameworks are very user friendly with a really small number of high-level functions to call in order to train a highly-efficient modern algorithm. Consequently, most people stick to the most common framework in their preferred language. For more experienced users, the question of the framework performance is more important. Even if most frameworks are equivalent in terms of application capability, some of them are more frequently upgraded with the latest innovations in the AI field. Additionally, they are not equivalent in terms of raw computational performance. Many ML algorithms can benefit a lot from Graphics Processing Unit (GPU) acceleration, which is not included in all the frameworks, as well as the capability to run in a scalable hardware environment like computer clusters. Among the most popular frameworks we can cite:
\begin{itemize}
\setlength\itemsep{0.3em}
\item \href{https://www.tensorflow.org/}{TensorFlow} \citep{tensorflow2015-whitepaper}: certainly the most popular framework, with native API in Python, C++, Java, etc. and many community supported APIs. Its popularity is mainly due to the fact that it is completely free and open source while being developed and maintained by Google. The updates are really frequent and the developers actively work with other software and hardware companies to include new capabilities concurrently to their official releases. See more capabilities in \citet{Abadi_2016}.
\item \href{https://keras.io/}{Keras}\citep{chollet2015keras}: A strong "addition" to TensorFlow, it is known for its very high level API that allows one to code complex ML algorithms with a minimal number of lines. It is also open source and developed by a large community. However, Keras being mainly included in TensorFlow as a higher level interface, many developers are employed by Google.
\item \href{https://pytorch.org/}{PyTorch} \citep{paszke2017automatic}: is an open source library with mainly Python and C++ APIs, but is more independent from Big-Tech companies than most of the other frameworks cited here, and has its own low-level computational algorithms. Like the previous one it remains computationally very efficient with most of the very modern hardware capabilities integrated.
\item \href{https://github.com/Microsoft/CNTK}{CNTK}: The Microsoft Cognitive Toolkit is also an open source framework that is supported by a big company. It is highly comparable to the previous ones with APIs in Python, C/C++ and Java. The choice of this framework is often motivated by very specific use case optimizations that are not included in other frameworks.
\item \href{https://www.r-project.org/}{R frameworks}: R is not as widespread as Python or C/C++ but it is widely used among statisticians. It contains some integrated functions that can be used to perform ML, but also contains various open source libraries that are mostly algorithm specifics. Additionally, as time is going it contains an increasing number of interfaces to more widely used APIs like Keras.
\item \href{https://developer.nvidia.com/cudnn}{cuDNN}: The Nvidia cuda Deep Neural Network library is somewhat different to the previous examples. It is neither open source nor applicable to many algorithms. It is also the only one to require a specific hardware with an Nvidia GPU and the specific programming language CUDA derived from C++ (also seen as an API but that requires a dedicated compiler), that is not very user friendly. However, it is worth noticing that, up to this day, it is the most computationally efficient way of building neural networks that make use of cutting edge modern hardware technologies. Many previously cited frameworks contain specific calls to this library in order to provide the best performance.

\end{itemize}

There are obviously plenty of other frameworks that we did not describe here like \href{https://caffe.berkeleyvision.org/}{Caffe}, \href{http://deeplearning.net/software/theano/#}{Theano}, \href{https://mxnet.apache.org/}{MXNET}, \href{https://scikit-learn.org/stable/}{scikit-learn} (very suitable for teaching), etc. \\

In contrast to the use of a pre-existing framework, one can develop a home made ML application using any programming language. One drawback of the use of a framework or library is the induced dependency to it. In software communities the choice of a library is sometimes compared to a wedding. One will invest time to understand practices that are framework specific, and develop specific applications. In the "relatively rare" case where the framework stops being developed, it can be very harmful for one's productivity as it will force one to completely re-develop many applications. For many of the previously cited frameworks, their open source characteristics strongly mitigates this issue. For the widespread frameworks that are supported by Big-Tech companies, this scenario seems very unlikely. Still, there are examples of widely adopted programmable solutions that stopped their activity. This is the case of the Adobe Flash platform that progressively stopped being supported, despite the fact that it had been a dominating development platform for web-based content between 2000 and 2010. Home-made codes, on the other hand, often rely only on the programming language or low-level libraries that can often be easily replaced if needed.\\

\vspace{-0.3cm}
Another general difference concerns the performance. It is commonly presumed that a framework developed by big-tech companies will be much more efficient than a home-made code, but it is not always the case. Particularly in the field of ML, the framework development is often more focused on ease-of-access and on widening the application range. It is true that many frameworks include automatic scalability or GPU acceleration which induces more computational power. However, a specific low-level implementation that makes use of the same hardware is often able to achieve better raw performance. One example would be the complex type conversion: when using a permissive framework one is able to plug almost any datatype. It eases the use of the framework, but will hamper the performance as the framework will have to assess datatypes and perform the appropriate conversion. This might add overhead at many steps of the computation. This example can be generalized to many small aspects of the algorithms. We emphasize here that such effects are stronger in highly iterative algorithms where each iteration can be dominated by such overhead, which is the case with most ML approaches.\\

\vspace{-0.3cm}
Finally, a more subjective argument, is that fully programming the algorithm is an efficient way to learn the underlying theory. The time investment is significant, but this is a much more transferable knowledge that can then be used with any framework that are mostly used as "black boxes". The final decision is mostly a matter of individual sensibility to all these aspects. In our case, {\bf we chose to develop our own home-made framework} due to our will of building a strong theoretical knowledge on ML applications and because we already had a solid High Performance Computing (HPC) experience. We still kept an interested eye toward the most used frameworks in order to compete in term of computational performance and capabilities. We implemented several algorithms, but we mainly focused on Artificial Neural Networks.\\

\vspace{-0.3cm}
Our framework, called CIANNA (Convolutional Interactive Artificial Neural Network for/by Astrophysicists), is a general purpose ANN framework that we started to develop independently to the present research but whose development was then driven by the present study. To date, our framework is capable of creating arbitrarily deep, arbitrarily thick, convolutional neural networks and is CUDA GPU accelerated, but it also support multi-threaded CPU computation through OpenBLAS and OpenMP. It is provided with Python and C high level APIs, and has been successfully applied to many applications: Classification, regression, clustering, computation acceleration, image recognition, detection in images, image generation, ... and is suitable for any ANN application. Its development is currently motivated by its ability to solve astrophysical problems. New elements are added as they are identified as useful for specific case studies. CIANNA is freely accessible on GitHub as open-source (ApacheV2) at \href{https://github.com/Deyht/CIANNA}{github.com/Deyht/CIANNA}. An in-depth presentation of the framework programming strategy and capabilities is given in the dedicated Appendix~\ref{cianna_app}. Additionally, some details of our implementation will also be discussed in various sections when it seems appropriate.

\subsection{Artificial Neuron}

\subsubsection{Context and generalities}

Artificial Neural Networks are one of the most famous Machine Learning algorithms. They belong to the supervised ML category and they have a really broad variety of applications. Their popularity might be explained by their intuitive definition and construction. They are also truly computationally efficient when applying an already trained network, which allows them to work on really light systems, or even on small embedded devices. Anyhow, they are perfectly suitable for really complex tasks with increasingly large networks. These days, despite other algorithms achieving better performance on specific tasks, there is a strong momentum toward the use of these methods. Many ML frameworks are even only dedicated to ANN. Their vast adoption by the machine learning community also makes them better documented and more optimized with many big-tech companies constantly breaking the limits of their performance.\\

As already described in Section \ref{ai_begin}, ANN are not new, despite the strong recent appeal about them, and about the AI field in general. While it is difficult to accurately date the appearance of ANN, one fundamental reference is \citet{mcculloch_logical_1943} where the authors attempt to summarize the behavior of biological neurons with a mathematical model. This reveals the obvious inspiration of the biological brain to try to reproduce intelligence artificially. The brain is a wonderful biological machine that performs really interesting tasks. It computes predictions based on a census of biological sensors and using many context information (social environment, time, previous topics and knowledge, ...). This means that the brain is able to compile highly dimensional data in an impressive small amount of time, which allows one to react within a quarter of second. It is also able to work with heavily noised data, like in a noisy bar where it is able to filter out irrelevant discussions to focus on a specific one. It also presents a high capacity of generalization, like when one is able to properly walk on previously unseen ground just by efficiently compiling previous walking knowledge. One final impressive capability is its resiliency to the loss of neurons with aging, with up to 10\% loss estimate between the age of 25 and 60 in certain specific brain areas \citep{Wickelgren96}. However, the brain maintains unchanged cognitive function in most cases, which demonstrates its ability to rearrange information in an efficient way. \\

More technically, the brain can be seen as a massively parallel computer with up to $10^{11}$ neurons, mostly binary compute units, and $10^{14}$ synaptic connections that retain information. With all these advantages and the apparent simplicity of the basic neuron behavior many attempts were made to reproduce it artificially. As stated in Section \ref{ai_begin}, despite the growing interest of ANN between 1940 and 1970, the lack of computational power refrained its adoption by a larger communities. It is between 1980 and 1990 that the interest started to rise again with new application outbreak due the increased computational power, while the real ANN boom started with the 21th century. In this section we describe the most common approach to construct ANN and subsequently increase its complexity up to modern architectures that are able to efficiently solve concrete problems. An extensive introduction can be found in \citet{Bishop:2006:PRM:1162264} or \citet{MarslandBook2} that relies on several reference papers including \citet{rosenblatt_perceptron:_1958,rumelhart_learning_1986,rumelhart_parallel_1986,widrow_30_1990}. 

\newpage
	\subsubsection{Mathematical model}
	\label{neuron_math_model}
	
\begin{figure}[!t]
\centering
\includegraphics[width=0.8\hsize]{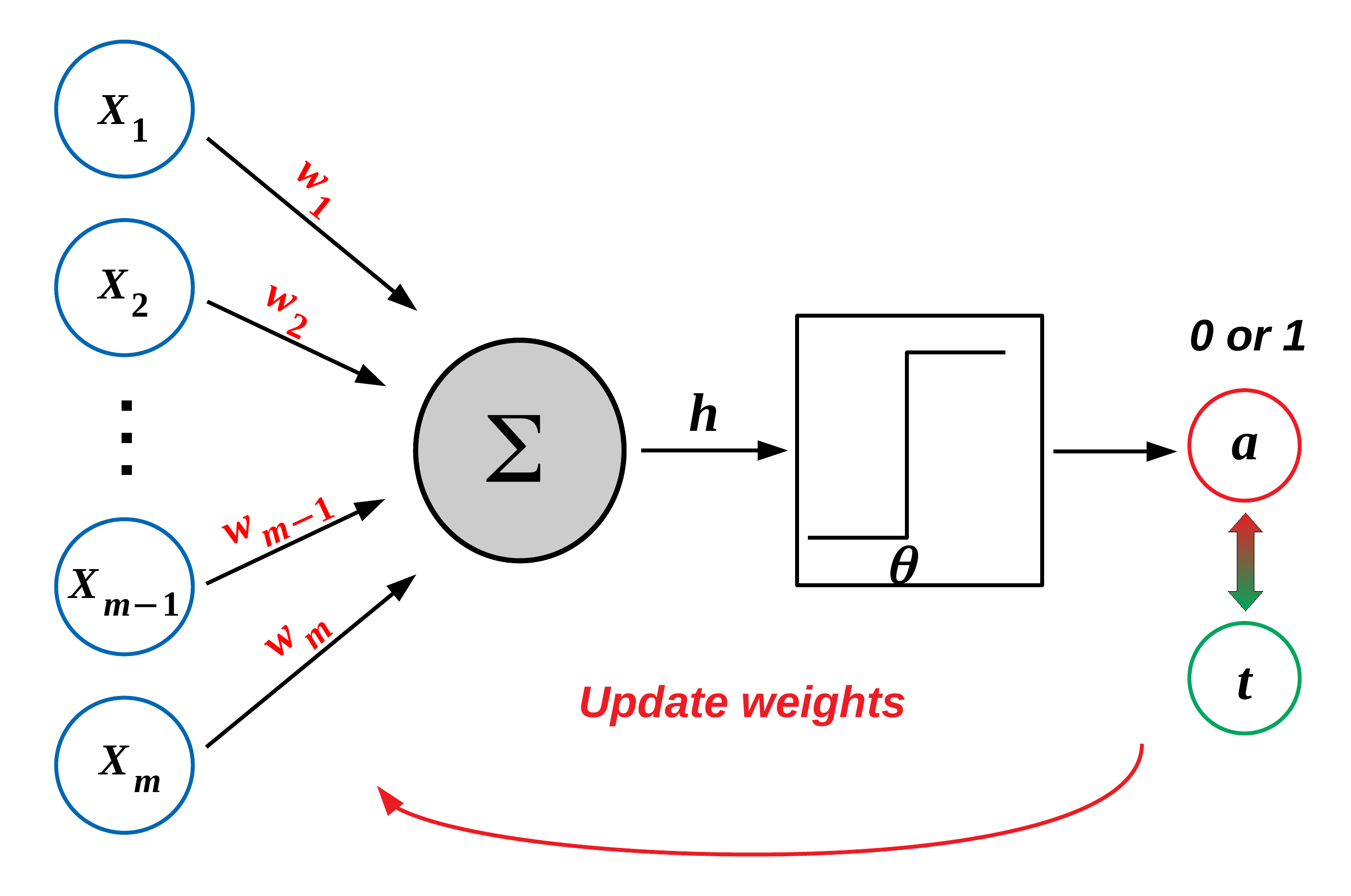}
\caption[Schematic view of a binary neuron]{Schematic view of a binary neuron. $X_ {[1,...,m]}$ are the input features for one object in the training dataset, $w_{[1,...,m]}$ are the weights associated with each feature. $\sum$ is the sum function and $h$ its result. $\theta$ is the threshold of the step function, and $a$ is the final neuron activation state that is compared to the target $t$ of the current object.}
\label{fig_neuron1}
\end{figure}
	
The following mathematical model is derived from \citet{mcculloch_logical_1943}. It consist of an input vector $X_i$ containing the $m$ dimensions of a specific object. In ML the input dimensions are called features, and the corresponding input dimension space is named the feature space. Each feature is associated with a weight $\omega_i$, to perform a weighted sum $h$:
	\begin{equation}
	\centering
		h = \sum_{i = 1}^{m}{X_i \omega_i}, \hspace{1.5cm}
	\label{eq_weighted_sum}
	\end{equation}
where the weights can take any positive or negative value. We note that there are many $X_i$ vectors while the $m$ weights are unique and shared for all the possible inputs. 
The neuron is associated to an activation function $g(h)$ which provides the activation $a=g(h)$ of the neuron, as a function of $h$. In very simple models a step function is often used:
	\begin{equation}
	\centering
		a = g(h) = \begin{cases} 1 & \text{if} \quad h > \theta, \\ 0 & \text{if} \quad h \leq \theta, \end{cases}
	\label{eq_activ_neuron}
	\end{equation}
where $\theta$ is the threshold value and is set by the user, generally to zero.
The weights quantify the correlations and anti-correlations of each input dimension with the current neuron. Figure~\ref{fig_neuron1} illustrates this model. This model mimics the behavior of a simple biological neuron which transforms an input signal into a binary response (often referred to as "firing" or "not-firing"). This action of computing the activation of a neuron, or more generally a network, for a given input vector is called a "forward" step.\\

Such simple neuron can already be used as a binary classifier with any number of input dimensions. It can be seen as a simple linear separator in the feature space with weights corresponding to the slope of the separation along each dimension.
	
	\subsubsection{Supervised learning of a neuron}
	\label{neuron_learn}
	
Training such a neuron consists in finding a suitable set of weight values that minimizes a given error function. Since it belongs to the supervised ML family it uses a training dataset of examples with pre-established solution. This is achieved in an iterative fashion: the neuron is activated for an input vector of the training dataset, and the result $a$ of the activation is compared to the expected result, the so-called target $t$. An error is computed by comparing the activation and the target, that is used to correct the weights, generally by a small amount. This step is called an update. The two steps, forward and update, are repeated for all the input vectors of the training dataset, making the weights converge toward suitable values. A learning phase performed on the complete dataset is called an "epoch". Many epochs are necessary to fully train such a neuron. In practice, the correction of the weights depends on the derivative of the chosen error function and is proportional to the relative input for each weight. For a binary neuron and a usual square error $ E = 0.5 \times (a - t)^2$ it can be computed using:
	\begin{equation}
	\centering
	\omega_i \leftarrow \omega_i - \eta \, \left(a - t\right) \, X_i
	\label{eq_update_neuron}
	\end{equation}
where $\eta$ is a learning rate that can be defined according to the problem to solve, or adjusted automatically (Sect. \ref{learning_rate}). Considering that learning with this neuron is an iterative process that searches for the optimal weight values one must define a starting state. Usually best performances are achieved when initializing the weights to small random values, which is discussed in Section \ref{weight_init}.

\subsection{The bias node}
\label{bias_node}

Equations~\ref{eq_weighted_sum} and~\ref{eq_update_neuron} show that the particular vector $X_i=0$ for all $i$ is a pathological point: its weighted sum $h$ is independent of the weights, and the weight correction is always null, regardless of the error function $a-t$. To circumvent this peculiarity, one approach consists in adding an $m+1$ value to the input vector, fixed to $X_{m+1}=-1$, and connected to the neuron by an additional weight $\omega_{m+1}$, which behaves as any other weight. This addition is equivalent to allow an adaptive value for the threshold $\theta$. Moreover, since the neuron acts as a linear separation, this additional weight on a fixed input allows a non-zero origin. We acknowledge that there are other possible implementations of this effect. For example, one can add a constant to the result of the weighted sum and share the responsibility for the additional activation over all inputs during the weight update process. This way the additional weight is shared by all input weights. Both approaches aim at minimizing the impact on the neuron formalism. Our choice for the first implementation can be justified by performance concerns, as exposed in Section \ref{matrix_formal}.\\

Because the input values $X_i$ are often called input nodes, this additional input dimension is generally referred to as "the bias node". The additional degree of freedom provided by the bias node enables the neuron to behave normally when $X_i=0$ for $1 \le i \le m$. \\

\begin{figure}[!t]
\centering
\includegraphics[width=0.75\hsize]{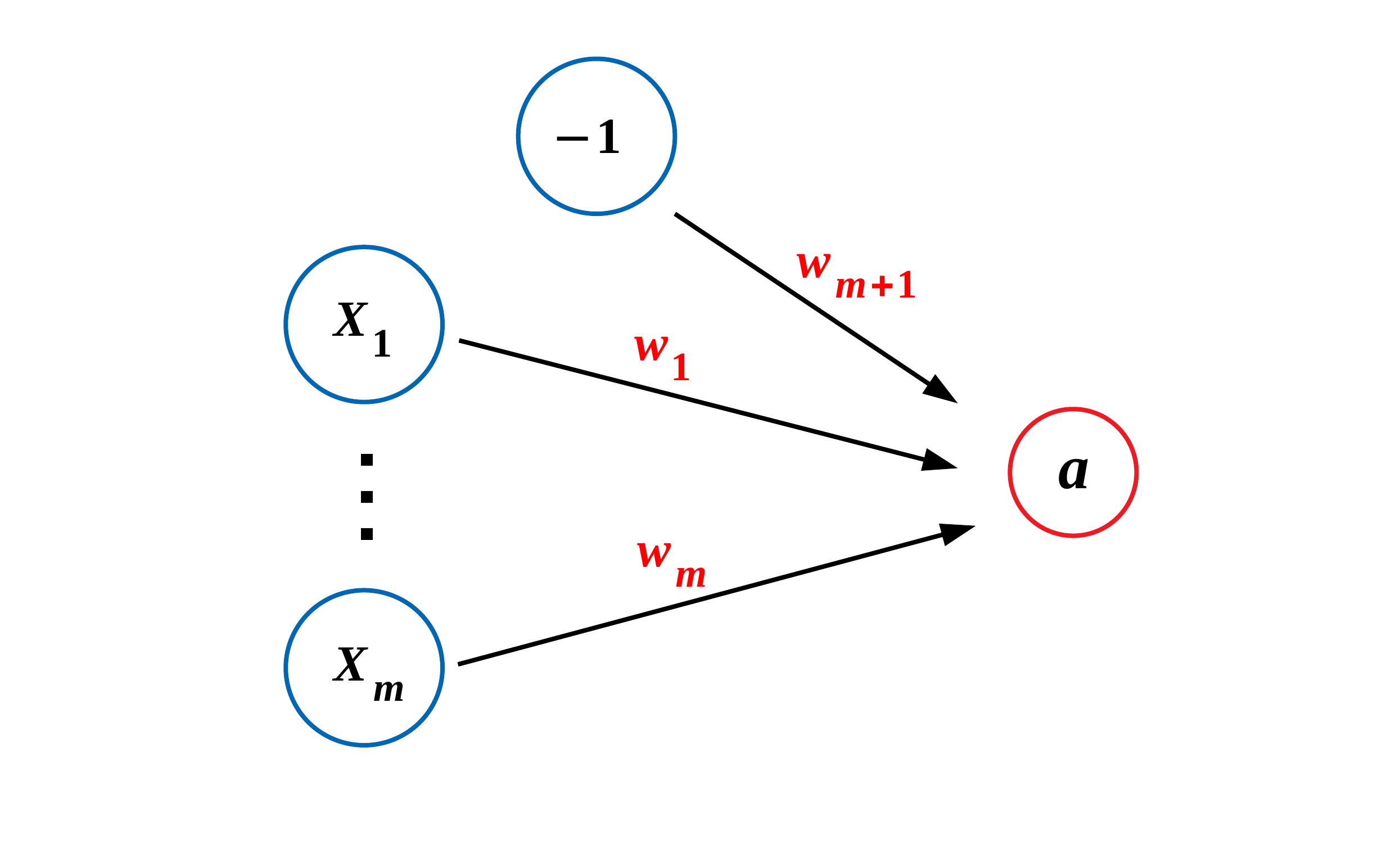}
\caption[Schematic view of a binary neuron with bias node]{Schematic view of a binary neuron with the addition of the bias node. $X_ {[1,...,m]}$ are the input features with the additional constant $-1$ bias input node, $w_{[1,...,m,m+1]}$ are the weights associated with each feature with the extra $w_{m+1}$ weight for the bias node. $a$ is the final neuron activation state. }
\label{bias_neuron}
\end{figure}

\begin{figure}[!t]
\vspace{2cm}
\hspace{-1.2cm}
\begin{minipage}{1.15\textwidth}
\centering
\includegraphics[width=0.45\hsize]{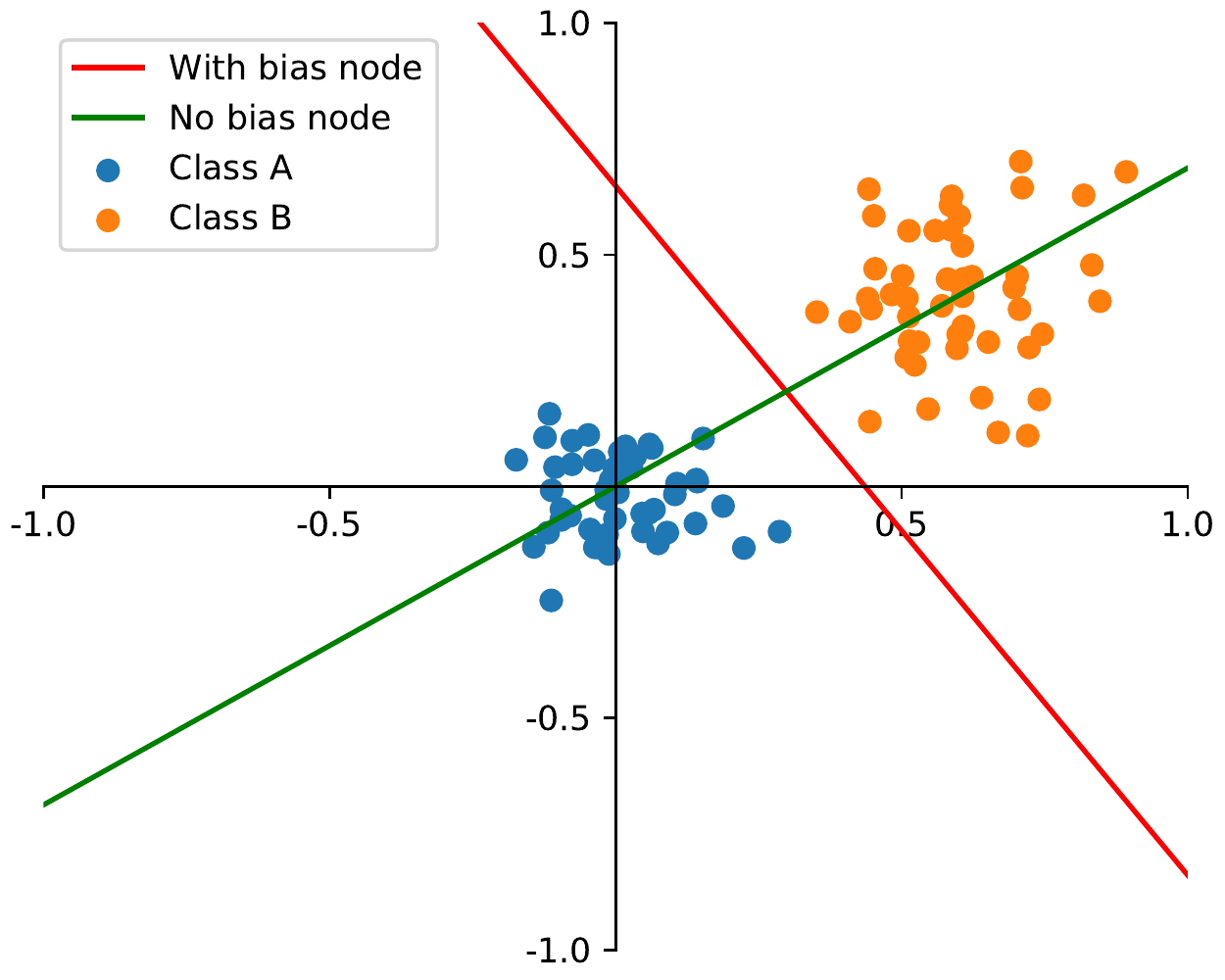}
\hspace{0.8cm}
\includegraphics[width=0.45\hsize]{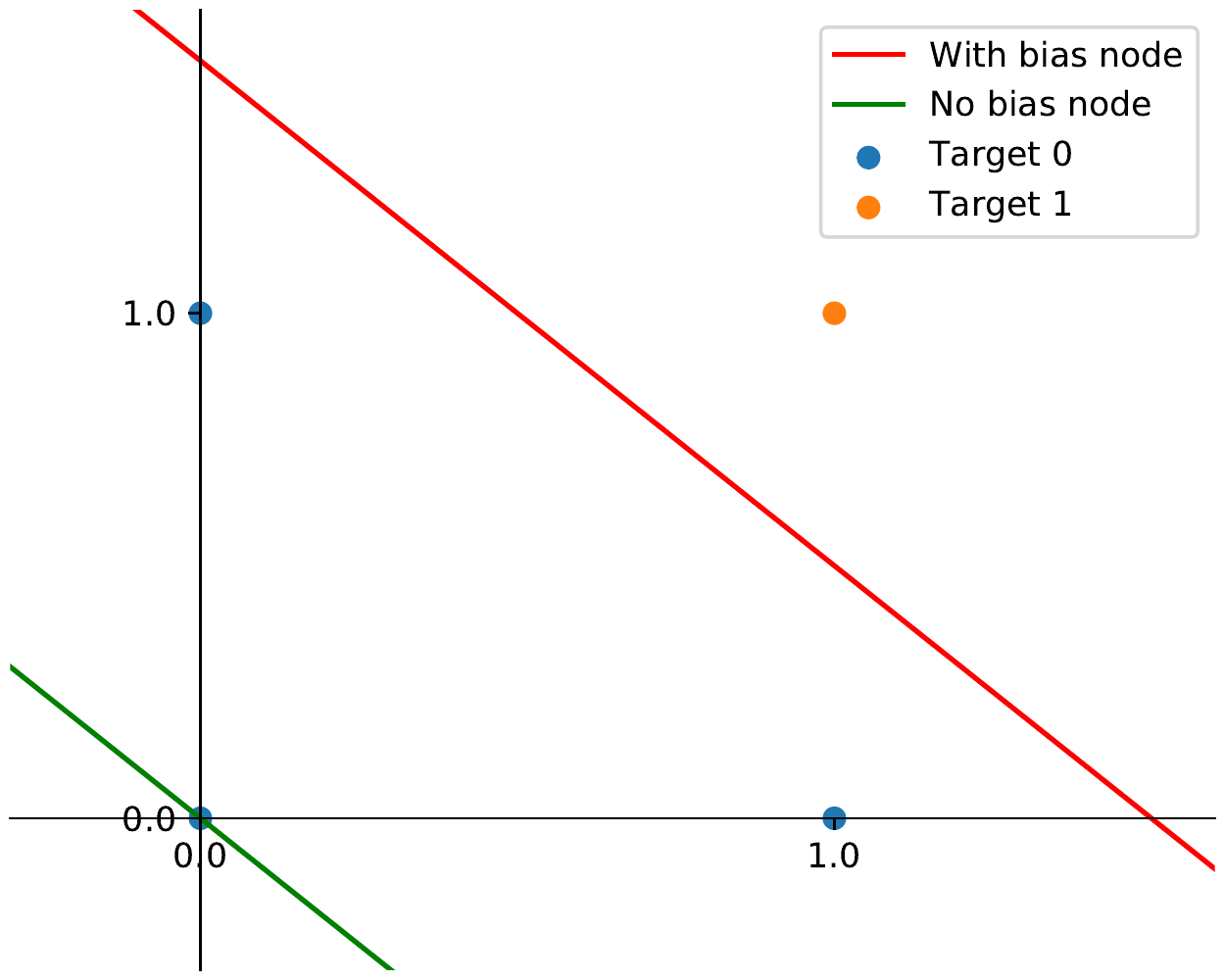}
\end{minipage}
\caption[Effect of a bias node on a binary separation]{Illustration of the effect of adding a bias input node to a single neuron binary separation. {\it Left}: two classes in blue and orange are separated by a trained binary neuron with or without a bias node in red and green, respectively. {\it Right}: reproduction of an AND logical gate using a trained binary neuron with or without a bias node in red and green, respectively.}
\label{bias_illus}
\end{figure}

Figure~\ref{bias_neuron} illustrates the addition of the bias node to the neuron schematic view and Figure~\ref{bias_illus} provides two examples where the use of a bias node is necessary. The first one is a simple binary classification, but with the best separation not being aligned with the origin of the frame axis, while the second attempts to reproduce the "AND" logical gate.\\

\newpage
\null
\newpage
\subsection{Perceptron algorithm: linear combination}
\label{sect_perceptron}
\vspace{0.2cm}
As we exposed (Sect.~\ref{neuron_math_model}), a single neuron is only able to perform a linear separation. To deal with complex problems more neurons are necessary. A simple way to combine neurons consists in adding them in the same layer. Each neuron is then fully connected to the input layer, but is not connected to the other neurons. They are thus fully independent and each of them has the exact same behavior as in the previous sections and are trained similarly as in the previous sections, except that the equations of the weighted sum and of the activation become matrix equations:
	\begin{equation} 
		h_j = \sum_{i = 1}^{m+1}{X_i \omega_{ij}} 
		\label{weighted_sum}
	\end{equation}
	and
	\begin{equation}
		a_j = g(h_j) = \begin{cases} 1 & \text{if} \quad h_j > \theta \\ 0 & \text{if} \quad h_j \leq \theta \end{cases}
	\label{eq_activ_perceptron}
	\end{equation}
where $j$ is the index of a neuron in the layer, and the sum runs from 1 to $m+1$ to account for the bias node. Similarly, the correction of the weights becomes:
	\begin{equation}
	\centering
	\omega_{ij} \leftarrow \omega_{ij} - \eta \left(a_j - t_j\right) \times X_i
	\label{eq_update_perceptron}
	\end{equation}
where $a_j$ and $t_j$ are the activation and target of neuron $j$, respectively.
This network and its training procedure altogether are called the Perceptron algorithm \citep{rosenblatt_perceptron:_1958}.\\

\begin{figure}[!t]
\centering
\includegraphics[width=0.8\hsize]{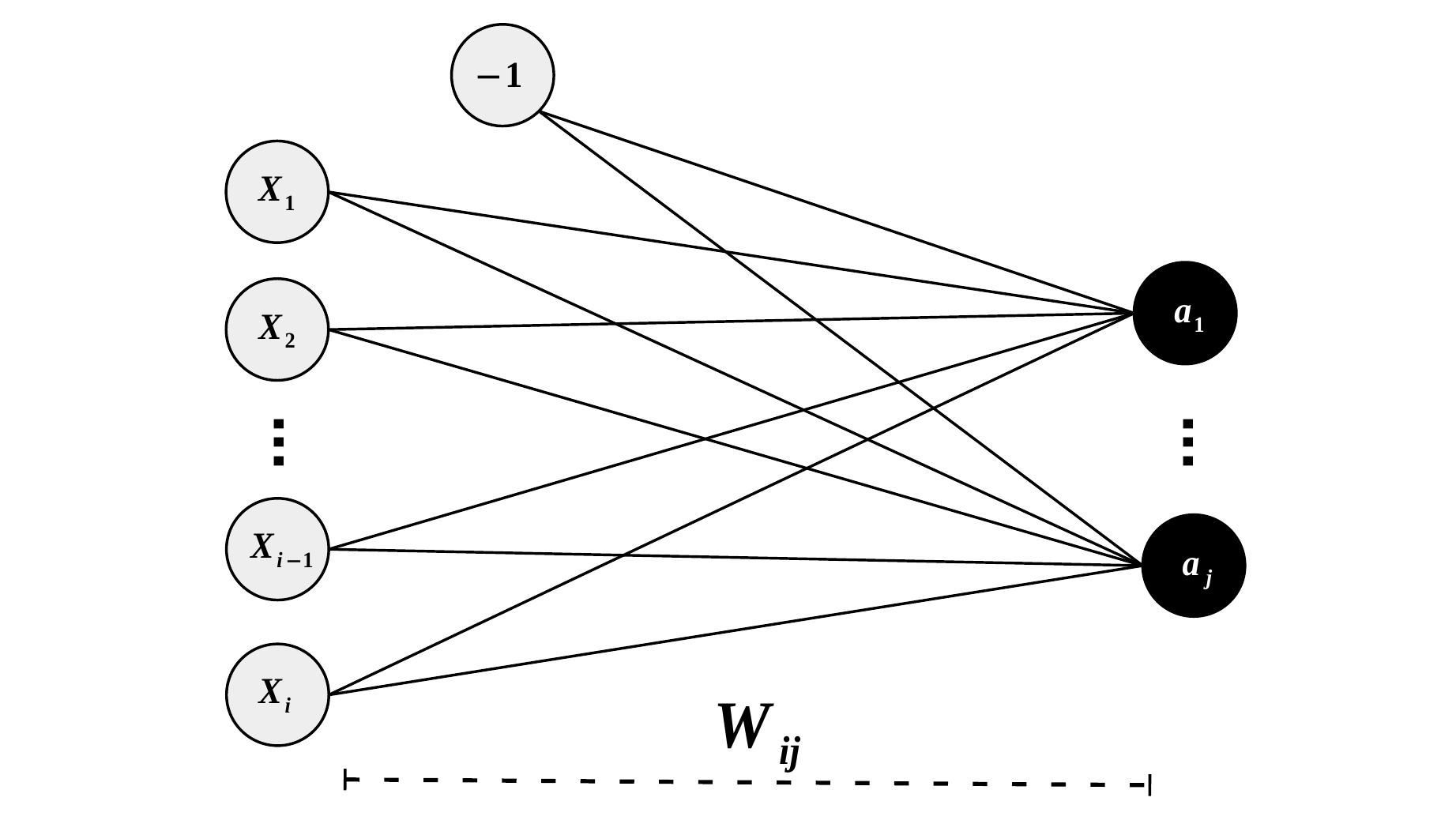}
\caption[Schematic view of a simple Perceptron neural network]{Schematic view of a simple Perceptron neural network. The light dots are input dimensions for one object of the training dataset. The black dots are neurons with the linking weights represented as continuous lines. Learning with this network relies on Eqs.~(\ref{weighted_sum}), (\ref{eq_activ_perceptron}) and (\ref{eq_update_perceptron}). $X_{[1,\dots,i]}$ are the dimensions for one input vector with an additional bias node, $a_{[1,\dots,j]}$ are the activations of the neurons, while $W_{ij}$ represents the weight matrices that connect the input vector to the neurons.}
\label{bias_perceptron}
\end{figure}

\begin{figure}[!t]
\centering
\includegraphics[width=0.75\hsize]{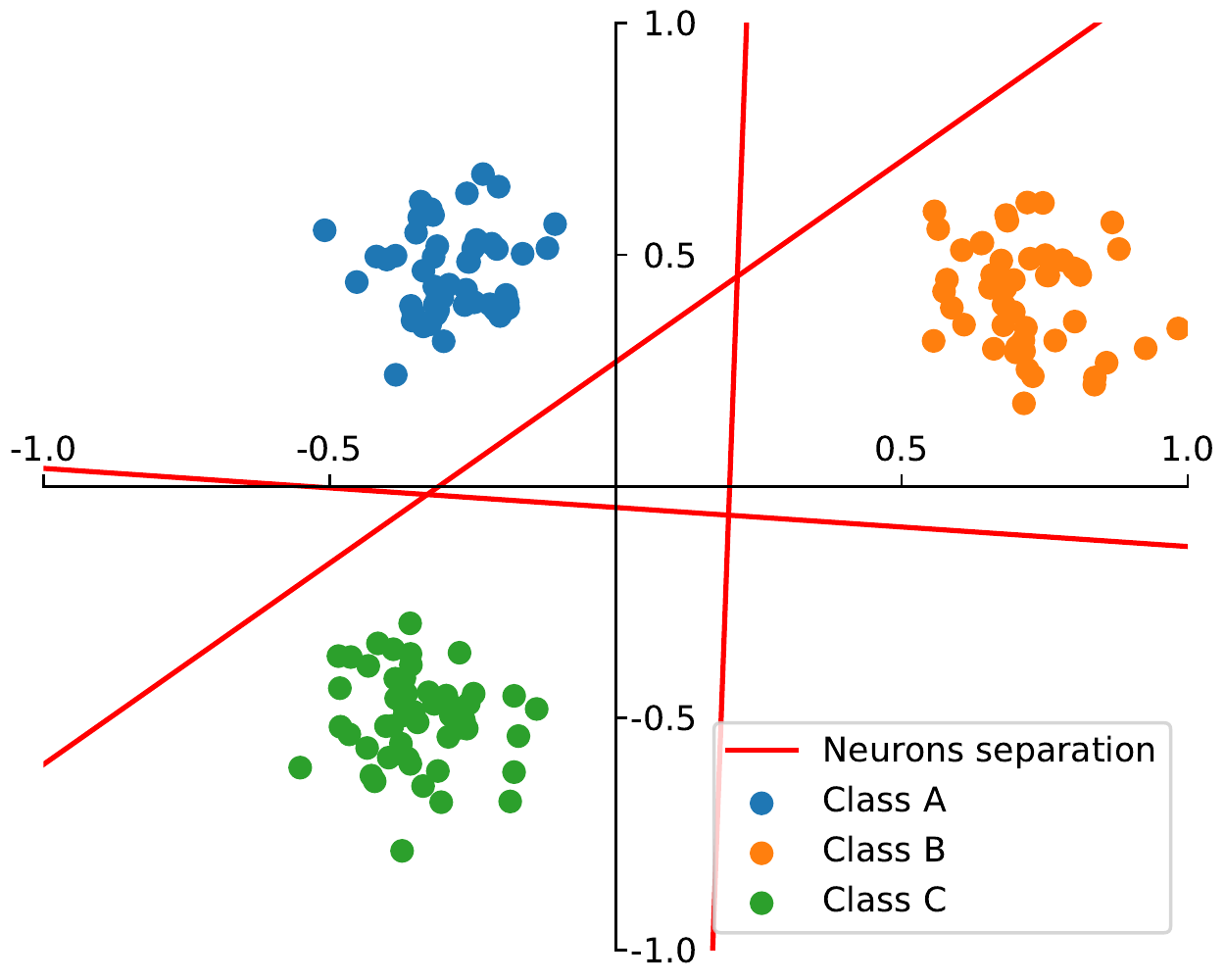}
\caption[Three class separation using a trained Perceptron]{Illustration of a three-class separation using a trained Perceptron with 3 neurons. Each red line corresponds to one binary neuron separation.}
\label{perceptron_illus}
\end{figure}

\vspace{0.4cm}
This architecture is illustrated in Figure~\ref{bias_perceptron}. Again, in this model, neurons are equivalent to hyperplanes in the weight space, each hyperplane performing a linear splitting between two classes. As before a slow training across multiple epochs is needed to let the weights of the network converge. In this structure, each neuron can learn a different part of the generalization \citep{rumelhart_parallel_1986}. One difficulty though, is to find a proper way to encode a global information into a set of binary neurons. When doing classification, one can set one neuron per output class and encode the target in the form of only one specific neuron being activated and the others set to 0. With a three-class example the possible outputs would be: A: (1-0-0), B: (0-1-0) and C: (0-0-1). This classification case is illustrated in Figure~\ref{perceptron_illus} that shows how the three neurons share the classification task. However, in order to work with the previous Perceptron algorithm, each class must be linearly separable from all the others. If it is not the case this encoding strategy must be refined in order to allow more than one neuron to represent each class. Another approach for a regression example would be to use binary value encoding, with for example 4 neurons as bits to encode a range of 16 values. One can also perform image processing by using one neuron per pixel.

\newpage
\subsection{Multi Layer Perceptron : universal approximation}
\label{mlp_sect}
	\subsubsection{Non linear activation function and neural layers stacking}		
	
\begin{figure}[!t]
\centering
\includegraphics[width=0.56\hsize]{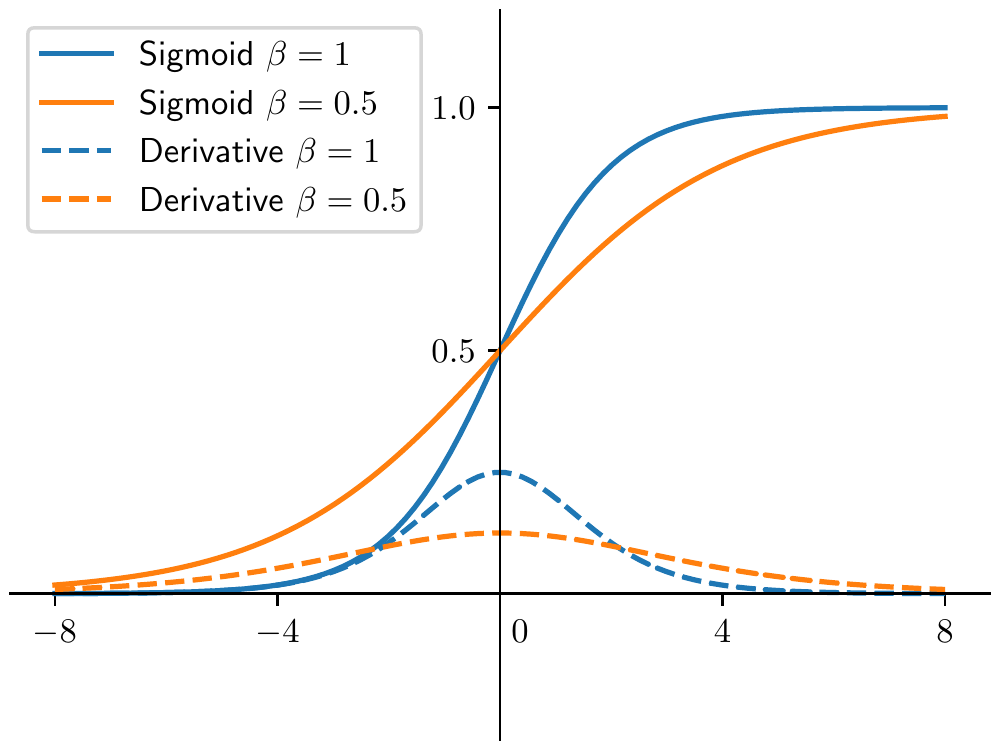}
\caption[Illustration of a sigmoid activation]{Illustration of sigmoid activations with $\beta = 1$ and $\beta = 0.5$ and their derivatives.}
\vspace{-0.1cm}
\label{sigmoid_fig}
\end{figure}

The Perceptron network remains too simple to learn complex problems. Firstly, it is restricted to linear combinations. Secondly, the neurons are limited to two states (0 and 1), and it can be a difficulty to encode physical values. Deep artificial neural networks can solve these two points. One major modification is to change the activation function to one which is continuous and differentiable, with the constraint that it must keep the global behavior of the neuron with two well distinct states. In a first step, we describe the sigmoid function \citep{rumelhart_parallel_1986}:
\begin{equation}
	\centering
	g(h) = \frac{1}{1+ \exp(-\beta h)}
	\label{eq_sigm}
\end{equation}
where $\beta$ is a positive hyperparameter that defines the steepness of the curve. This function has a $S$ shape with results between $0$ and $1$ and has a simple derivative form, which is illustrated in Figure~\ref{sigmoid_fig}. This addition noticeably allows easier regression with the Perceptron network, where each continuous variable can be represented with a single neuron.\\

\begin{figure*}[!t]
\centering
\includegraphics[width=0.9\hsize]{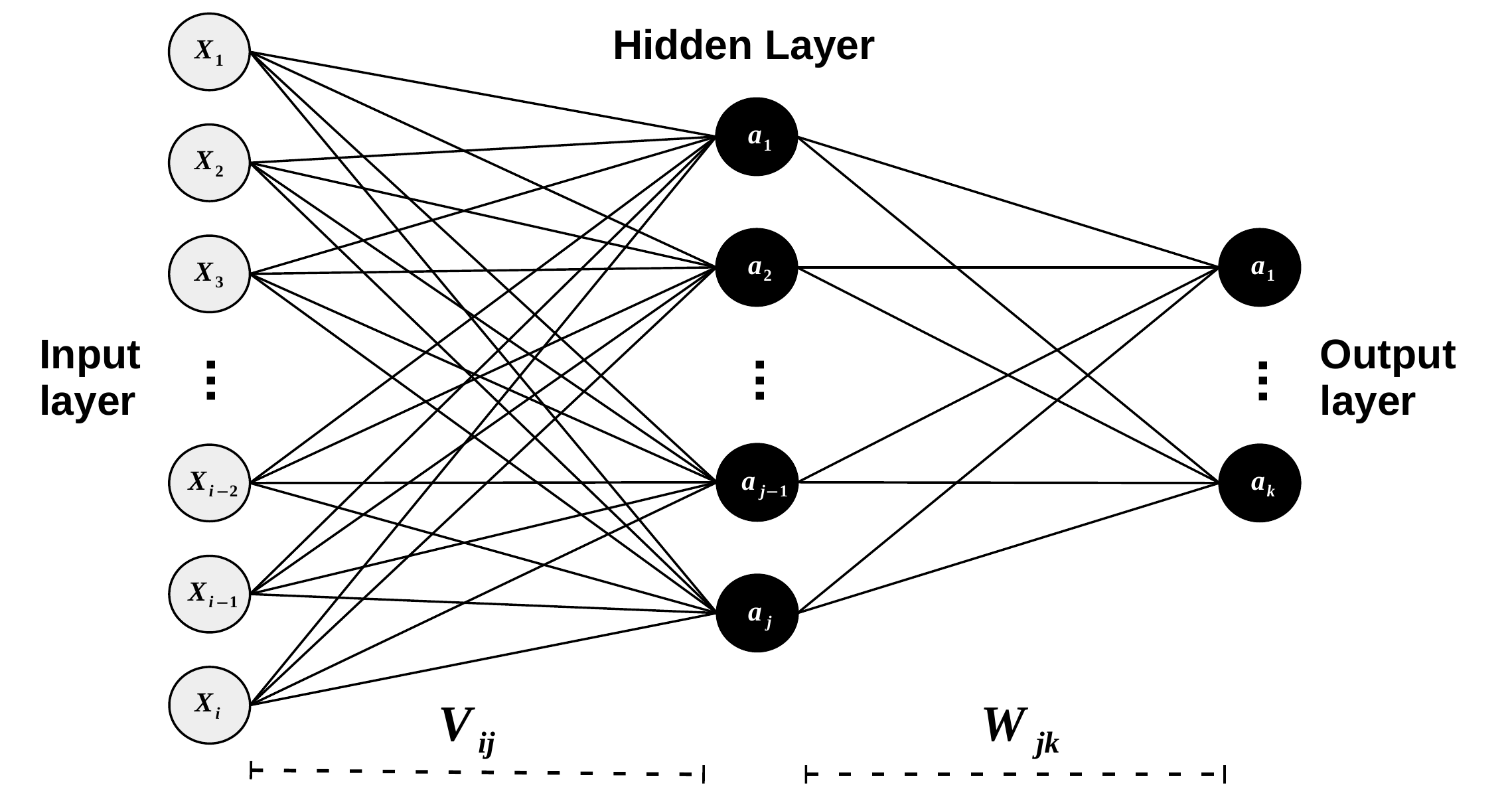}
\caption[Schematic view of a "deep" neural network with one hidden layer]{Schematic view of a simple "deep" neural network with only one hidden layer. The light dots are input dimensions. The black dots are neurons with the linking weights represented as continuous lines. Learning with this network relies on Eqs.~(\ref{eq_sigm}) to (\ref{eq_update_first}). $X_{[1,\dots,i]}$ are the dimensions for one input vector, $a_{[1,\dots,j]}$ are the activations of the hidden neurons, $a_{[1,\dots,k]}$ are the activations of the output neurons, while $V_{ij}$ and $W_{jk}$ represent the weight matrices between the input and hidden layers, and between the hidden and output layers, respectively.}
\label{fig_network}
\vspace{-0.2cm}
\end{figure*}

This addition is complemented by a second major modification: adding more layers. Neurons are added behind the previous layer, where they take as input the result of the activation of the neurons from the previous layer, as illustrated in Figure~\ref{fig_network}. Like in the Perceptron network, the neurons within one layer are independent one from the other, and their activation is computed following similar equations as Eqs.~(\ref{weighted_sum}) and (\ref{eq_sigm}). A bias node needs to be added to the previous layer to avoid any pathological behavior from the next one. This architecture is illustrated in Figure~\ref{fig_network}.\\

This procedure can be repeated to add multiple layers, constructing a "deep" network. The last layer is the output layer and the other neuron layers are the "hidden" layers. The input nodes are generally considered to form a first layer, dubbed input layer, although they are not neurons. While the input and output layers are mostly constrained by the problem to solve, the number and size of the hidden layers directly represent the computational strength of the network and must be adapted to the difficulty of the task. This kind of networks is called a Multi Layer Perceptron (MLP).\\

\begin{figure}[!t]
\centering
\includegraphics[width=1.0\hsize]{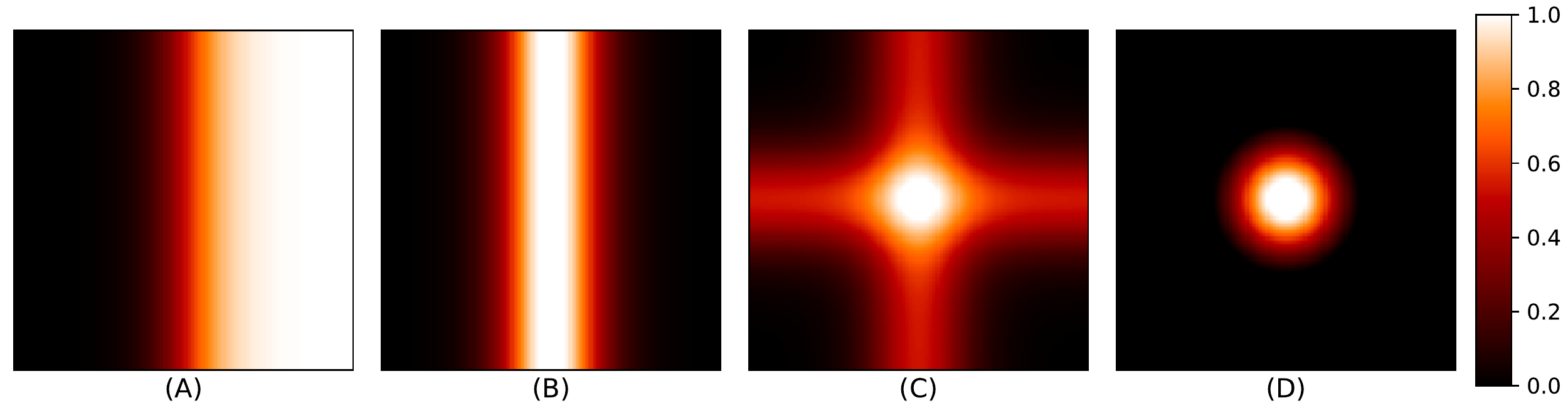}
\caption[Example of sigmoid combinations]{Example of sigmoid combinations. (A): one sigmoid alone, (B): two sigmoid combined into a hill shape, (C): two hill shape combined at $90\deg$ to form a "bump", (D): several rotated bump combined to obtain a localized pic function.} 
\label{sigmoid_comb}
\end{figure}

\begin{figure}[!t]
\centering
\includegraphics[width=0.68\hsize]{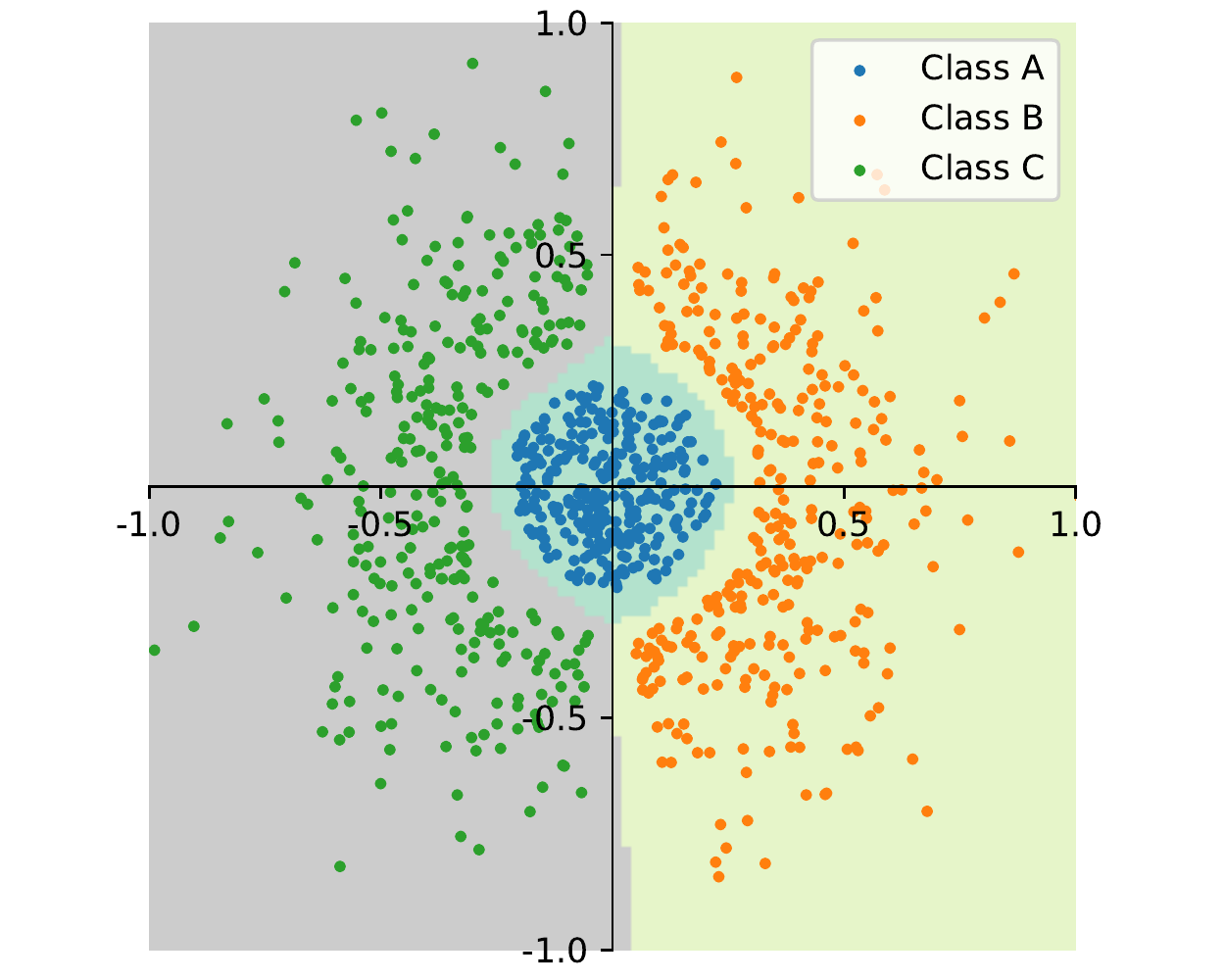}
\caption[Three class separation in a two dimensional space using a MLP]{Illustration of a three-class separation in a two dimensional feature space using a trained MLP with 3 output neurons and 8 hidden neurons, all with sigmoid activations. The light background colors indicate the regions of the feature space that the network has attributed to each class.}
\label{mlp_illus_class}
\end{figure}

This multilayer architecture allows the network to combine sigmoid functions in a non-linear way, each layer increasing the complexity of the achievable generalization. The combination of sigmoid functions can be used to represent any function, which means that this new network is a "Universal Function Approximator" as demonstrated by \citet{cybenko_approximation_1989}. The combination of sigmoid is illustrated in Figure~\ref{sigmoid_comb}. It shows that sigmoid can be combined into hill shapes, that can be combined to get bumps, and that these bumps can be combined to create any arbitrary point-like function. They also define the Universal Approximation Theorem. It demonstrates that only one hidden layer with enough neurons is able to approximate any function as accurately as an arbitrarily deep network. Therefore, it can be used to solve a very wide variety of problems as discussed in Section \ref{ml_application_range}.
We illustrate the capacity of such a network in Figure~\ref{mlp_illus_class}, where a network with two input dimensions learns non linear splittings between three classes. This network uses one output neuron per class similarly to Section \ref{sect_perceptron}, and only one hidden layer containing 6 sigmoid neurons.

	\subsubsection{Supervised network learning using backpropagation}
	\label{mlp_backprop}
	
Adding new layers introduces a difficulty to update the weights, since the targets are only available for the output layer. The "Backpropagation" algorithm \citep{rumelhart_parallel_1986} allows one to compute an error gradient descent, starting from the output layer, that can be propagated through the entire network. This gradient calculation depends on the error function, which is often the simple sum-of-squares error:
	\begin{equation}
	\centering
	E(a,t) = \frac{1}{2}\sum_{k=1}^{N}{(a_k -t_k)^2}
	\label{eq_error_fct}
	\end{equation}
where $k$ runs through the number $N$ of output neurons, $a_k$ is the activation of the $k$-th output neuron and $t_k$ the corresponding target. The weight corrections for a given layer $l$ are computed as follows:

	\begin{equation}
	\omega_{ij} \leftarrow \omega_{ij} - \eta \frac{\partial E}{\partial \omega_{ij}}
	\label{eq_grad_desc}
	\end{equation}
where the gradient $\frac{\partial E}{\partial \omega_{ij}}$ can be expanded as:

	\begin{equation}
	\frac{\partial E}{\partial \omega _{ij}} = \delta _l (j) \frac{\partial h_j}{\partial \omega_{ij}} 
	\quad \text{with} \quad
	\delta _l (j) \equiv \frac{\partial E}{\partial h_j} = \frac{\partial E}{\partial a_j}\frac{\partial a_j}{\partial h_j} = \frac{\partial a_j}{\partial h_j} \sum _j {\omega _{kj} \delta _{l+1}(k)}.
	\label{eq_update_full_network}
	\end{equation}

In these equations, the indices $i$ and $j$ run through the number of input dimensions of the current layer, and its number of neurons, respectively. These equations are the same for each layer. $\delta_l$ defines a local error term that can be defined for each layer of neurons, so that, for a hidden layer $l$, the error $E$ in eqs.~(\ref{eq_grad_desc}) and (\ref{eq_update_full_network}) is replaced by the multiplication of the following weight matrix and the next layer error $\delta_{l+1}$ with $k$ that runs through the number of neurons of the next layer. It also depends on the activation function $a=g(h)$ at each layer through the derivative $\frac{\partial a_j}{\partial h_j}$.
Thus, this kind of gradient can be evaluated for an arbitrary number of layers. These terms can be further simplified by considering the previous error, and a sigmoid activation for all neurons:
	\begin{equation}
	\frac{\partial h_j}{\partial \omega_{ij}} = a_i,
	\end{equation}
	the activation of the current layer,
	\begin{equation}
	\frac{\partial E}{\partial a_j} = (a_j - t_j),
	\end{equation}
	the derivative of the error to replace $\delta_l$ at the output layer,
	\begin{equation}
	\frac{\partial a_j}{\partial h_j} = \beta a_j (1 - a_j),
	\label{eq_sig_deriv}
	\end{equation}
	the derivative form of the sigmoid activation.
	Further details on the equations can be found in \citet{Bishop:2006:PRM:1162264} or \citet{MarslandBook2}.\\

While the {\bf definition of "deep learning"} is often fuzzy and varies from a user to another, some consider the MLP to be sufficient to fit in this category of ANN (see Sect~\ref{conv_layer_learn} for a more advanced definition). However, constructing a many-layer network is often challenging. The main difficulty is the so-called "vanishing gradient". When using the above equations to propagate the error through the network, each layer multiply the $\delta_l$ value received from the previous layer by the derivative of the activation function, in this case a sigmoid. The issue is that this factor is most of the time less than 1, and therefore the error information is smaller and smaller for the layers that are the closest to the inputs, up to the point where the corresponding weights are enable to be updated anymore. This issue, combined with the fact that one hidden layer is enough to approximate any function, induces that the most common MLP architecture only contains one hidden layer. However, such architectures are less employed these days, because most of the recent architectures overcome this limitation by adopting a different kind of activation function to remove the "vanishing gradient" issue, and therefore construct very deep networks. This point, along with a less disputed definition of "deep learning", will be covered later in the second part, Section \ref{conv_layer_learn}, of the present manuscript.\\

For the sake of simplicity, and due to the previous point, we detail here the upgrade procedure for a network with only one hidden layer like in Figure~\ref{fig_network}. The network is therefore composed of an input layer constrained by the input dimensions $m$ of the problem to solve, a hidden layer with a tunable number of neurons $n$, and an output layer with $o$ neurons. All the neurons use sigmoid activations in this example.
The gradient descent is computed from the backpropagation equations (Eqs.~(\ref{eq_error_fct}) to (\ref{eq_sig_deriv})) as follows. The local error $\delta_o(k)$ of the $k$-th output neuron is computed using:
	\begin{equation}
	\centering
	\delta_o(k) = \beta a_k (1 - a_k) (a_k - t_k).
	\label{eq_deltao}
	\end{equation}
The obtained values are combined with the weights between the hidden and output layers to derive the local error for neurons in the hidden layer, multiplied as before by the derivative of the sigmoid activation:
	\begin{equation}
	\centering
	\delta_h(j) =  \beta a_j(1-a_j) \sum_{k=1}^{o}{\delta_o(k)\omega_{jk}}
	\label{eq_deltah}
	\end{equation}
where $j$ is the index of a hidden neuron. Once the local errors are computed,
the weights of both layers are updated:
	\begin{eqnarray}
	\centering
	\omega_{jk} & \leftarrow & \omega_{jk} - \eta \delta_o(k)a_j
	\label{eq_update_second}\\
	v_{ij} & \leftarrow & v_{ij} - \eta \delta_h(j)x_i
	\label{eq_update_first}
	\end{eqnarray}
where $\omega_{jk}$ and $v_{ij}$ denote the weights between the hidden and output layers, and between the input and hidden layers, respectively, $a_j$ is the activation value of the $j$-th hidden neuron, and $x_i$ is the $i$-th input value.

\subsection{Limits of the model}

Before diving into some more advanced details, we discuss here some of the intrinsic limitations of the current model to reproduce the biological brain. There is a lot of unknowns that remain about the biological neurons and the brain, and it is not our subject here, but a first order comparison remains a good illustration of how simplified the ANN model is.\\

First of all, the mathematical neuron model itself is far less complex than a biological one. It is well established that a biological neuron does not fire a single impulsion, but rather a series of complex impulsions that can encode way more information. However, it conserves a "global" activated or non-activated state, but with an additional recovering time that adds an even more complex inhibition effect to its behavior. While in our model the activation function is fixed and only the weights can change, in the biological neuron the activation might change depending on environmental effects and even be function of time or of the number of activations. Another difference is that the model permits weights to change sign, while it is not possible for the biological neuron.\\

In a more global view, the biological neuron activations are also non-sequential (asynchronous), while our model is designed to activate each layer in a specific order. Moreover, the biological neurons are not nicely arranged in independent layers. Each neuron can be connected to many others in a complex architecture that creates many neural "paths", and connections can actively be remapped, which strongly contributes to the brain plasticity. Biological neurons can even reconnect to themselves (feedback). All of these aspects being only the tip of the iceberg, even regarding our partial understanding of the biological brain.\\

Yet, the simplicity of the model is not only a matter of difficulty to write algorithms that fulfill the brain capabilities. For example, feedback is a common feature of recent Recurrent Neural Networks (RNN) architectures \citep[already described by][]{rumelhart_parallel_1986}. We can also cite Neural Gas Networks \citep{Fritzke95} that help to construct complex non-layered architectures. The main objective of all the ANN models (and ML models in general) is to find an appropriate balance between computing performance and capabilities. While RNN has proven to be efficient for specific tasks like speech recognition \citep{Robinson1996, Waibel89}, it does not improve the generalization potential in all cases, although the computational cost excess of this method is substantial. It is, therefore, often advised to start by trying simple algorithms and properly assess if a more complex one would really improve the results. The exposed ANN model has proven its great efficiency in a vast variety of cases, which explains its popularity in many communities. We adopt this progressive strategy in the current manuscript with many advanced ANN capabilities being postponed to Part~2 where they are useful.

\subsection{Neural network parameters}

	\subsubsection{Network depth and dataset size}
	\label{nb_neurons}

All ML methods have in common that the quality of the prediction is directly linked to the number of examples provided. Most of the time it is useful to start by estimating the difficulty of the task to perform. For ANN, since individual neurons can be compared to linear separators, one can roughly estimate how many of such separations would be necessary to isolate all the groups of objects identified in the feature space. Most of the time it only provides a starting point, from which it is necessary to search the optimal number of neurons. Considering a network with a single hidden layer, only the number of neurons in this layer can be changed, therefore only this value has to be explored. Most commonly, the minimum error that can be achieved on the output layer is directly linked to the number of neurons. Increasing the number of neurons tends to improve the results (lower the minimum error), but the gain provided by a new neuron decreases with the total number of neurons in the layer. Beyond a certain amount of neurons, an error plateau is reached where adding more neurons only adds noise to the minimal error value. \\

However, increasing the number of neurons also increases the number of weights (or degrees of freedom) and, therefore, requires more data to be trained. A widely used empirical rule prescribes that the number of objects for each class in the training sample must be an order of magnitude larger than the number of weights in the network. As an example, for a network with $m$ features, $n$ hidden neurons in only one hidden layer, and $o$ output neurons, the number of weights in the network is $(m+1)\times n + (n+1)\times o$. It corresponds to the number of element of the weight matrices $V_{ij}$ and $W_{jk}$ of the example network in Figure~\ref{fig_neuron1}. We note that this estimate also depends on the type of neuron, i.e. their activation function. Binary or linear neurons being simpler than sigmoid ones they could be constrained with less examples. In order to reduce the number of weights needed, a deeper architecture with more but smaller hidden layer is often a successful strategy. However, such a network would be less stable even if capable of better absolute generalization capabilities and is more prone to the "vanishing gradient" issue. We also note that a network with several hidden layers will learn complex boundaries with less epochs than a single hidden layer one.\\

 \begin{figure}[!t]
\centering
\includegraphics[width=0.60\hsize]{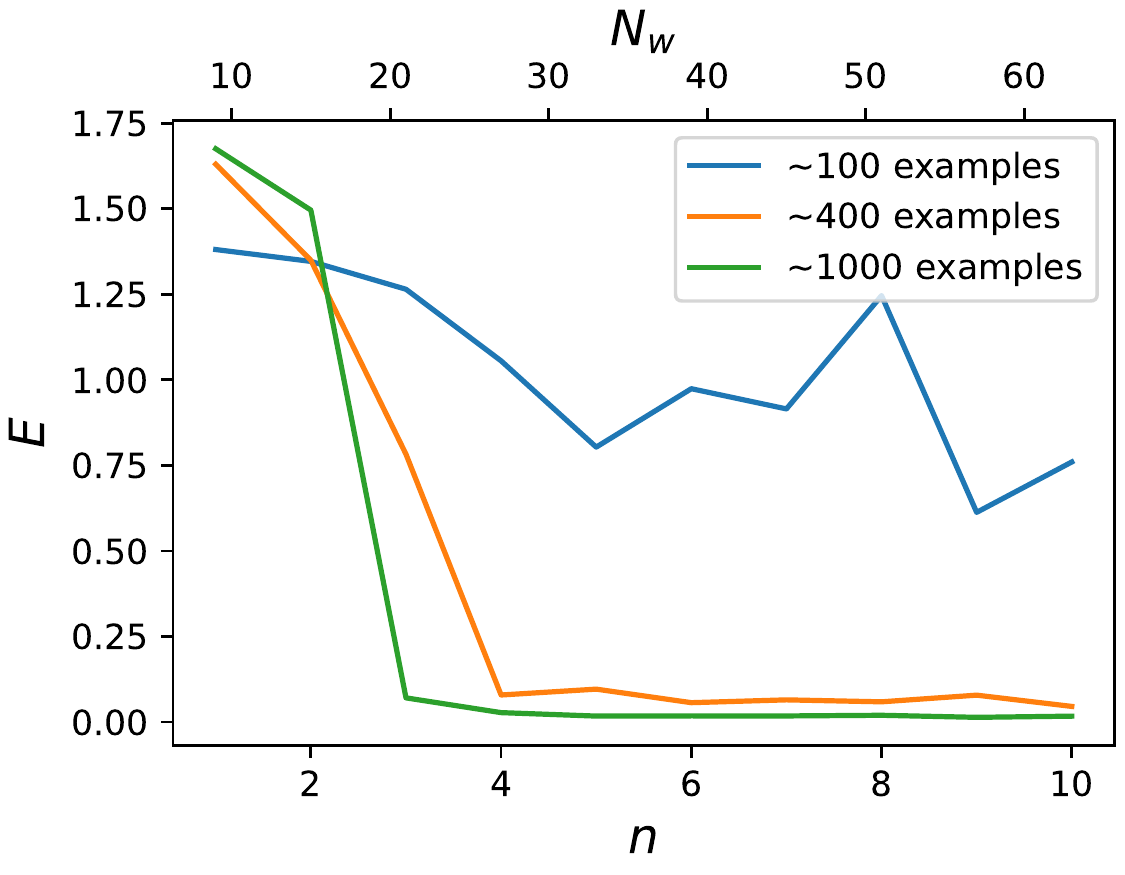}
\caption[Evolution of the error as a function of the number of neurons]{Evolution of the error as a function of the number of neurons in the network, for various numbers of examples. This figure uses the same network as for Figure~\ref{mlp_illus_class} with only one hidden layer. $n$ is the number of neurons in the hidden layer, $N_w$ is the total number of weights in the network depending on $n$, $E$ is the sum-of-square error of the output layer (equation~\ref{eq_error_fct}) averaged over all the given examples. Each error point is the average of 5 independent training with different weight initialization and training selection from the same distribution.}
\label{nb_neurons_opt}
\end{figure}

We used the same network as in the example in Section~\ref{mlp_sect} illustrated in Figure~\ref{mlp_illus_class}, where we separates three classes in a two dimensional feature space using one hidden layer. Figure~\ref{nb_neurons_opt} shows the search of the optimal number of neurons in the hidden layer by looking at the output-layer error-average for the corresponding dataset. Each point in this figure is the average of 5 independent training with the same network size in order to mitigate the effect of the random weight initialization and random example selection from the same distribution. We see that, when having few objects ($\sim 100$) in the training datasets, increasing the number of neurons fails to improve the results. This is due to the number of weights in the network being quickly of the same order as the number of examples. In this regime the network prediction is highly unstable, the weight updates being greatly underconstrained. The case with $\sim 400$ objects is more satisfying with a quick reduction of the error up to 4 neurons. It stabilizes itself to a constant value with small variations when adding more neurons. The last case is with $\sim 1000$ objects. With only three neurons it is already more efficient than the previous cases and is near its optimum value with five neurons. It is the case that also shows the less fluctuations when increasing the number of neurons, which indicates that they are properly constrained thanks to the larger training dataset.\\

We note here that these results are highly dependent on the other network parameters that are discussed on the following sections, such as the learning rate, activation function, number of layers, weight initialization, etc., and is very problem specific. Moreover, as we did in the simple example of Figure~\ref{nb_neurons_opt}, it is often necessary to perform several times the same training with the exact same parameters to see if the random selection of weights affects the result, and to compute an average, and ideally a dispersion. However, a strong limitation of this approach is that, as the number of hyper-parameters increases, the time required to train the network becomes very large, and it can become completely unrealistic to fully explore the impact of each hyperparameter on the results. This is why it is very common to find widespread architectures that are known to work nicely on many problems and that are reused blindly for other applications, with an exploration of parameters reduced to the learning rate and few other parameters. This also implies that the used network architectures are often over-sized for the considered problem which could lead to other difficulties as discussed in Section \ref{sect_overtraining}, and require special care to still work properly on too simple problems.

	\subsubsection{Learning rate}
	\label{learning_rate}

\begin{figure}[!t]
\centering
\includegraphics[width=0.495\hsize]{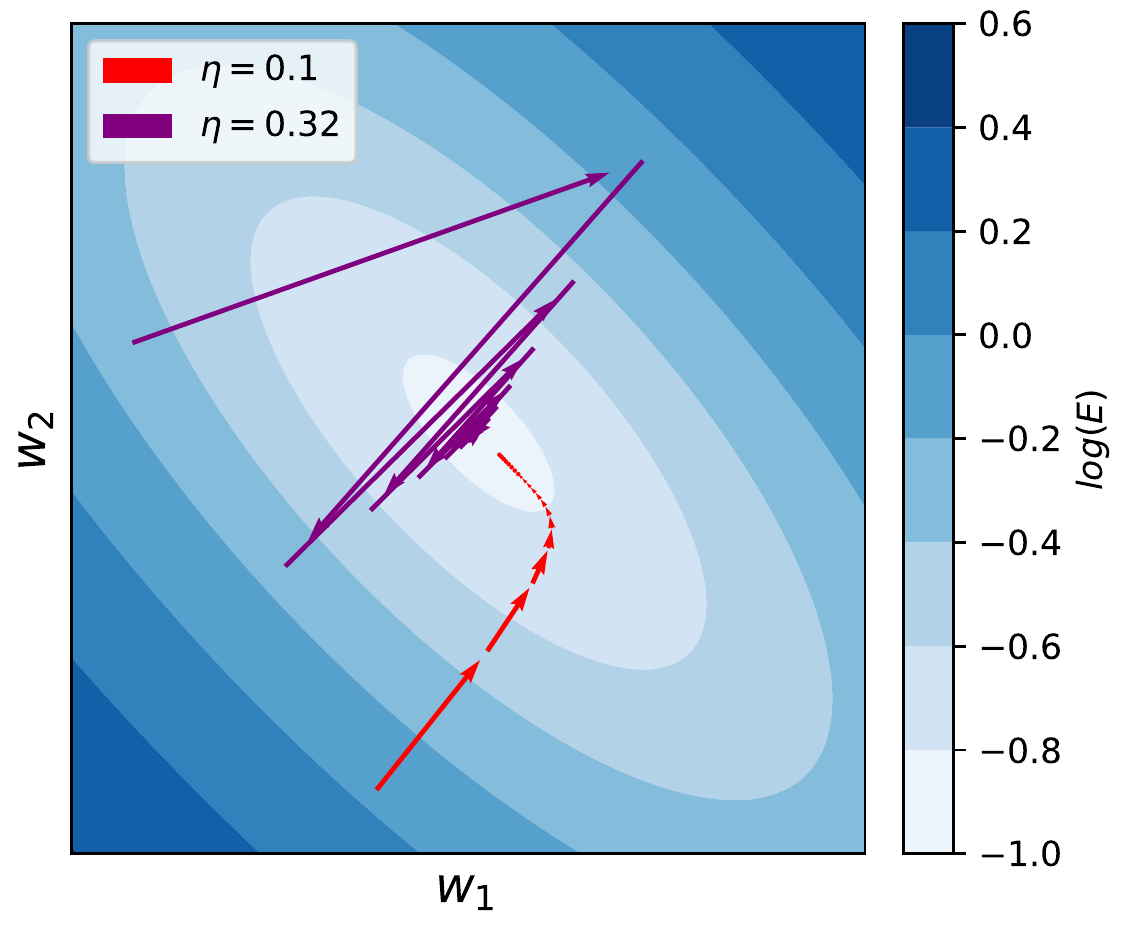}
\includegraphics[width=0.495\hsize]{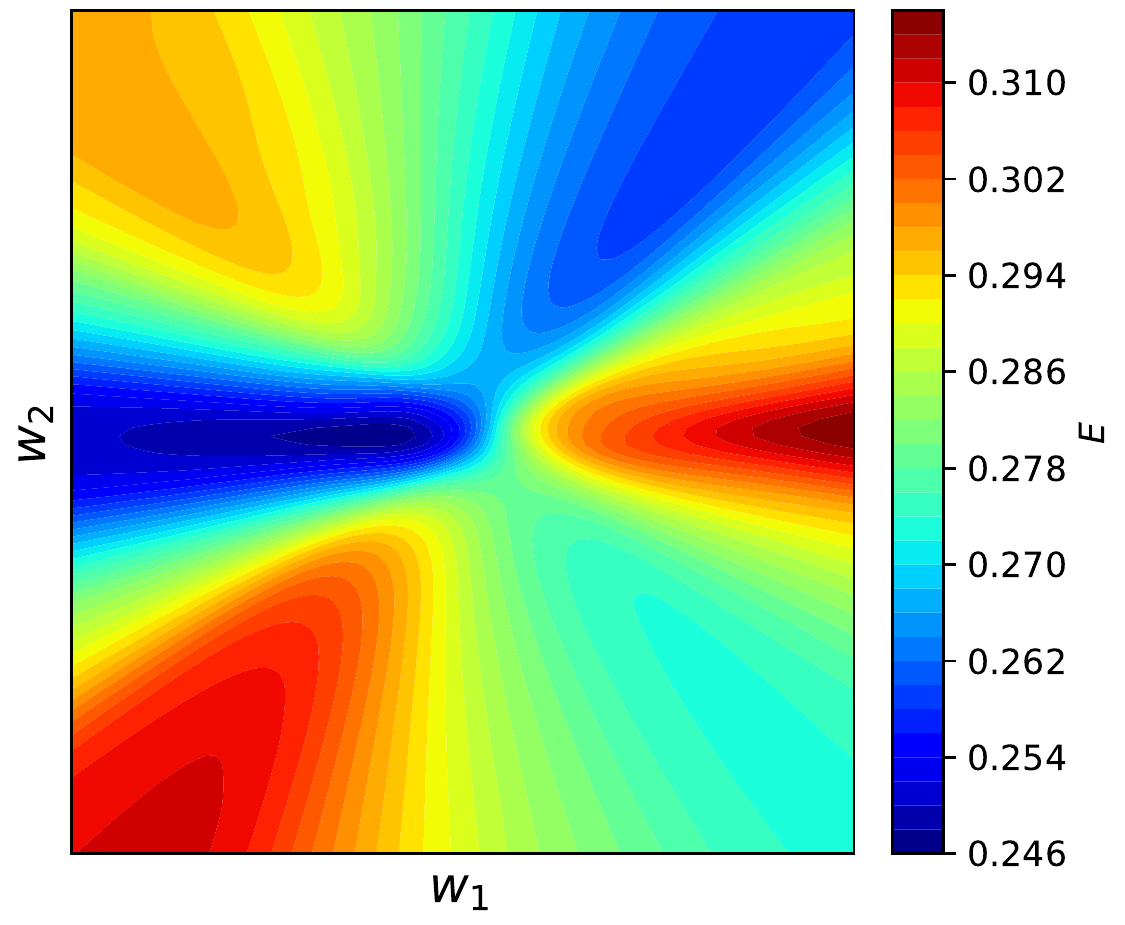}
\caption[Error value in the weight space]{Error value in the weight space. {\it Left}: Simple binary separation between two classes, from the example in Figure~\ref{bias_illus}. The background color represents the logarithm of the error of the neuron, averaged over all objects in the dataset, as a function of the 2D weight space of the neuron (excluding the bias node). Red and purple lines represent the first 20 weight values of the neuron during learning, with $\eta = 0.1$ and $\eta = 0.32$ respectively. {\it Right}: Example of the 2D weight space of a neuron in the hidden layer used for Figure~\ref{mlp_illus_class}, separating three classes in a two dimensional feature space.}
\label{weight_spaces}
\vspace{-0.2cm}
\end{figure}

The learning rate $\eta$ allows one to scale the weight updates in equations~(\ref{eq_update_perceptron}) and (\ref{eq_grad_desc}). This is useful to adapt the weight updates to the granularity of the feature space and of the various weight spaces. Lower values of $\eta$ increase the stability of the learning process, at the cost of a lower speed, and a higher chance for the system to get stuck in a local minimum. Conversely, larger values increase the speed of the learning process and its ability to roam from one minimum to another, but too large values might prevent it from converging to a good but narrow minimum. It should be noted that the correction also scales with the input value $X_i$, correcting more the weights of the inputs that are more responsible for the neuron activation. \\

The effect of the learning rate value is illustrated in the left frame of Figure~\ref{weight_spaces}, which shows the error of a neuron performing a linear separation between two classes in the two-dimensional weight space of that neuron. Two learning rates are compared. A small one at $\eta = 0.1$ leads to an efficient path in the weight space, illustrated by the red arrows, down to the minimum value. The second case with $\eta = 0.32$ leads to a zigzag path with very inefficient updates that overpass the optimal value. In the best case scenario, a too-large learning rate will only delay the convergence, but in most of the cases with a complex weight space it will not find the best minimum value. We note that values of $\eta$ larger than 0.32 most of the time induce divergence in this example, preventing the neuron to learn anything useful.\\

The right frame of Figure~\ref{weight_spaces} shows a more complex error distribution within a dimensional weight space that corresponds to one of the neuron in the hidden layer of the example given in Section \ref{mlp_sect} and Figure~\ref{mlp_illus_class}. In this case the hidden neuron activation is recombined with many other sigmoid neurons, which produces a complex error function. It is important to note that is a very partial visualization. In order to obtain this error in the weight space, we had to freeze all the network weights except those of the focused neuron. Indeed, this is a "snapshot" representation of the optimal error. During the full network training all the weights vary toward the optimal error, which produces a significantly different optimal error distribution in the weight space for the next iteration. It is then expected that the weights will converge and reach the optimum value. At this stage, the error distribution within the weight space is stable.\\

These examples show the importance of carefully choosing the learning rate value. However, a fixed learning rate is also a strong limitation. One easy addition consists of setting a large initial learning rate and a lower final learning rate with a progressive decrease between the two during training. This allows the network to go quickly to a part of the weight space that is near the optimum value, and then to have a smaller learning rate to properly resolve the optimal values. It is common to adopt an exponential learning rate decay \citep[for details, see][]{schedulers18}, which is the approach implemented in our framework CIANNA (see Sect.~\ref{cnn_hyperparameters}, Eq.\ref{eq_decay}). Following the same idea, advanced learning rate algorithms are getting more and more common. The aim is to dynamically change the learning rate to follow the evolution of weight update scale. Such a method aims at always providing the optimal learning rate regarding the current state of the network \citep{kingma2014}.\\

It should be noted here that there is no such thing as a perfect set of weight values. There are many sets of weights that could achieve the a similarly good prediction. It does not prevent any of these weight set solutions to be completely stable when the network has converged. This is a classic criticism against ANN methods, that are accused to be non-physical and impossible to interpret. However, there are many tools and techniques to link the input features to the predicted output by following the weights through the network. For example, assessing if there is a continuous path of large weight values between the inputs and outputs might help determining which inputs are useful to the network and which are not. Another common heuristic is to remove some input features to see if it impacts the prediction quality and infer the relative importance of each feature. Some applications also use an auto-encoder network to find the smallest number of dimensions necessary to represent all the input features, and draw insightful representations of the feature space into this reduced space \citep{Bengio_2012}.

	\subsubsection{Weight initialization}
	\label{weight_init}
	
We mentioned in Section \ref{neuron_learn} that the weights must be initialized to "small random values", but we did not explained why. Considering a network where all weights would be equal, for example to zero, then all the neuron activations would be identical, and so would be the weight updates and their propagation to the previous layers. This means that the corresponding network would only be able to solve problems that work with all the weights being equal. This is a symmetry issue. To break such a behavior the weights are randomly initialized. However, all initializations are not equal. It must guarantee that the weights are large enough to learn, and small enough to avoid divergence of the weights when the error of the neuron is large. Additionally, layers with many neurons will sum many individual errors. This could lead layers to learning at different pace, which increases the chances that network weights get stuck into a local minima, or at least slow down the network convergence. In the worst scenario, layers can even be stuck for many epochs in a saturated state where all neurons remain active, which is identical for all inputs, stopping the learning of all subsequent layers \citep{glorot_understanding_2010}. Therefore, it is often a good practice to scale the weight matrix between two layers accordingly to the size of the upward one. A common weight initialization is then a random uniform value in the range:
\begin{equation}
-1/\sqrt{N} < \omega < 1/\sqrt{N}
\end{equation}
where $N$ is the number of neurons of in the "input" layer. This way, the weight matrix keeps a zero mean with a dispersion that is properly scaled to the expected error on this layer.\\

On the other hand, the efficiency of a specific weight initialization is highly dependent on the chosen activation function. This is mainly due to the differences in mean value of the activation. For example a sigmoid activation has a 0.5 mean value, as well as a binary activation, while a linear activation will have a zero mean, as well as a hyperbolic-tangent activation. Additionally, it depends on the depth of the network, correlated with the activation function in a non trivial way. More informations about the effect of specific weight initialization regarding all those parameters can be found in \citet{glorot_understanding_2010} and \citet{he_delving_2015}, along the much more modern and widespread Xavier and He-et-al initialization methods. It is worth noting that the {\bf weight initialization is not just a small improvement} and can make the difference between a network that is able to learn something interesting and another that will not learn anything and diverge. A vast part of the improvement of the ANN methods over the last two decades came from changes in favor of newer combination of weight initialization and activation function.\\

\vspace{-0.6cm}
	\subsubsection{Input data normalization}
	\label{input_norm}
	
	The size of the weight updates is not only shaped by the weights themselves but also by the previous layer activation value and at some point by the value of the input features. This is visible in equations~(\ref{eq_update_neuron}) and (\ref{eq_update_first}) and is implicit in equation~(\ref{eq_update_full_network}). Therefore, it is important to scale the features to be of the same order as a typical layer activation. Otherwise, it could cause saturation or too large weight updates to converge smoothly, similarly to what can be observed with a too large learning rate (Sect.~\ref{learning_rate}). Additionally, all the features do not necessarily have the same range of values. For example, a feature that represents the age of a star will have a range of values between few millions and billion years, while a stellar apparent magnitude will be in the range of few tens. This will result in meaningless dominance of the numerically-larger features over the activation of the neurons and weight updates. \\

\vspace{-0.1cm}
A widespread solution is to normalize each feature individually, for example in an interval of $-1$ to $+1$ with a zero mean. This can be done by subtracting the mean value of a feature and then dividing by the new absolute maximum value. The network can then start with an equivalent importance of each feature, which we observed to be an efficient solution. Additionally this allows the weights to be of same order of magnitude for each feature, which tends to strongly stabilize the weight updates. It is usual to scale the inputs in a similar range than that achieved by the typical activation function used in the corresponding network, which leads to the same weight initialization for all the layers. We note that the normalization solutions we presented here are mostly suited for sigmoid activated neurons with $\beta=1$ and a $-1$ bias value as described in the previous sections.
	
	\vspace{-0.2cm}
	\subsubsection{Weight decay}
	\label{weight_decay}
	
	We already established that the weights should be non-equal to break symmetries, should have a zero mean to avoid initial direction bias, and should be small enough to prevent saturation. However, there is an additional benefit of having very small weights in some cases, because small weights are more likely to induce sigmoid neurons to be in their middle linear phase, since $g(0) = 0.5$. It is interesting to keep as much neurons in a linear state as possible to ease the error propagation as it is also where the sigmoid derivative reaches its maximum and where the "vanishing gradient" effects are the lowest. Ultimately, the network should use as few non-linear neurons as possible to represent the necessary non-linearity of the problem. \\

\vspace{-0.1cm}
To achieve such a behavior, it is common to add a weight decay \citep{Hanson_89} in the network algorithm. The most simple one consists in multiplying all the weights of the network by a decay factor $ 0 < d_c < 1$, generally very close to 1, after each epoch. This way the network will only keep large weights where it is absolutely necessary. Here, the natural dependency to the activation function is obvious, meaning that such improvement would not be necessary when using an activation function that behaves more like a linear activation. As before, very advanced weight decay methods can be found in some modern ANN architectures \citep{kingma2014}. It is worth noting that this technique can be used as a regularization (see Sect.~\ref{dropout_sect}) that helps to prevent overtraining on noisy data \citep{gupta_98}.

	\subsubsection{Monitor overtraining}
	\label{sect_overtraining}
	
As we began to expose with the previous section, tuning a network to make it learn can be tedious. But there can be difficulties even with a perfectly tuned network, a noticeable one being the overtraining of the network. In the most common scheme, the network weights start at a random inappropriate position and slowly converge to an optimal value. After this point, the learning process starts to over-fit the training data. It means that it is starting to learn the specificities of the dataset. In principle, it would not be an issue with a infinite dataset that perfectly suits the appropriate feature space. In a real case, the data distribution will always have gaps in the feature space that the network might try to fit, making it necessary to monitor the learning phase. A widely adopted solution consists in monitoring the overtraining using additional datasets. Most commonly the original dataset with labels is split in several parts:
\begin{itemize}
\setlength\itemsep{0.2em}
\item A {\bf training dataset} that contains the vast majority of the objects and that is used to effectively train the algorithm.
\item A {\bf validation set} that is used regularly during the training phase to compute an error, but without updating the weights, and enables one to monitor the training process.
\item A {\bf test dataset} that is used after the training phase to assess the quality of the generalization on data that were not seen during the training phase or the validation phase.
\end{itemize}

In practice, during the training, both the errors of the training and validation datasets are computed. Overall, the error on the training dataset tends to decrease down to an asymptotic value, around which it may fluctuate, regardless of whether the network is over-trained or not. In contrast, the error on the validation datasets generally varies in two phases. In a first phase, the error decreases, similarly to what is observed for the training set. In a second phase, when the network starts over-training, the error on the validation set starts to rise. Therefore, the training of the network must be stopped close to this point, which is often done by monitoring the error on the validation set on several consecutive steps.\\

\begin{figure*}[!t]
\centering
\begin{subfigure}[!t]{0.65\textwidth}
\centering
\includegraphics[width=\hsize]{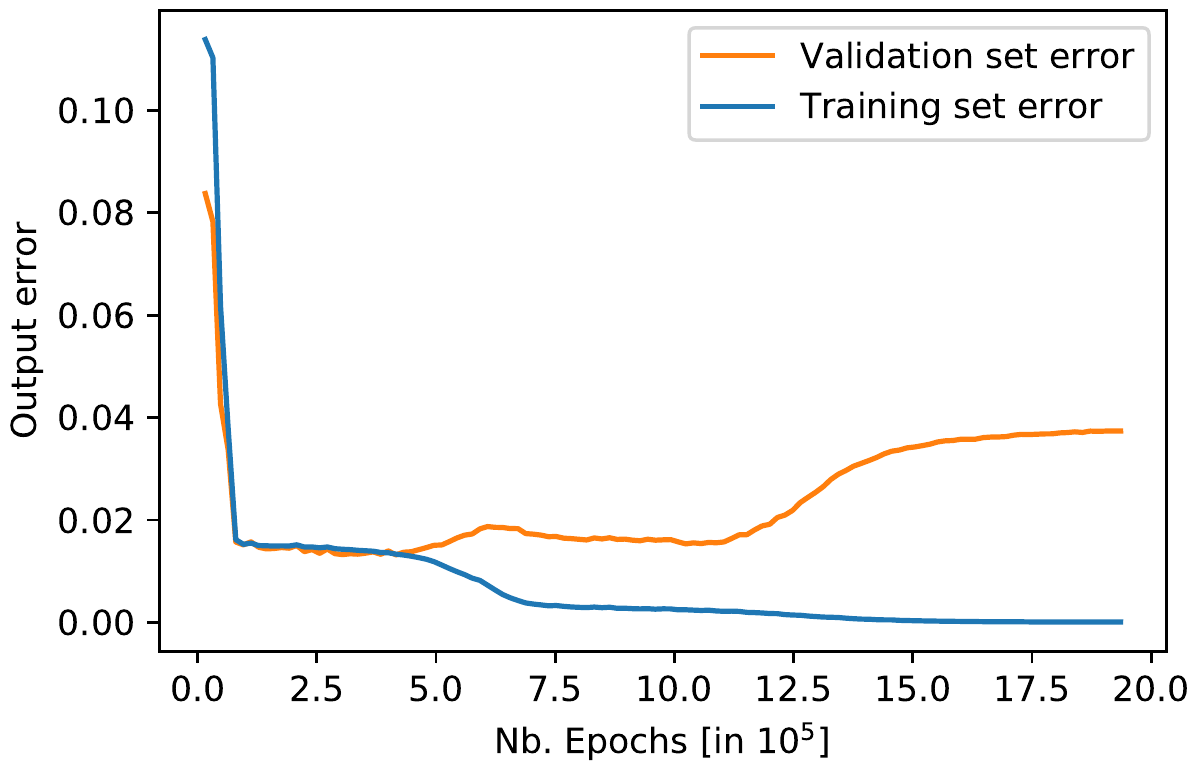}
\end{subfigure}
\begin{subfigure}[!t]{0.63\textwidth}
\centering
\includegraphics[width=\hsize]{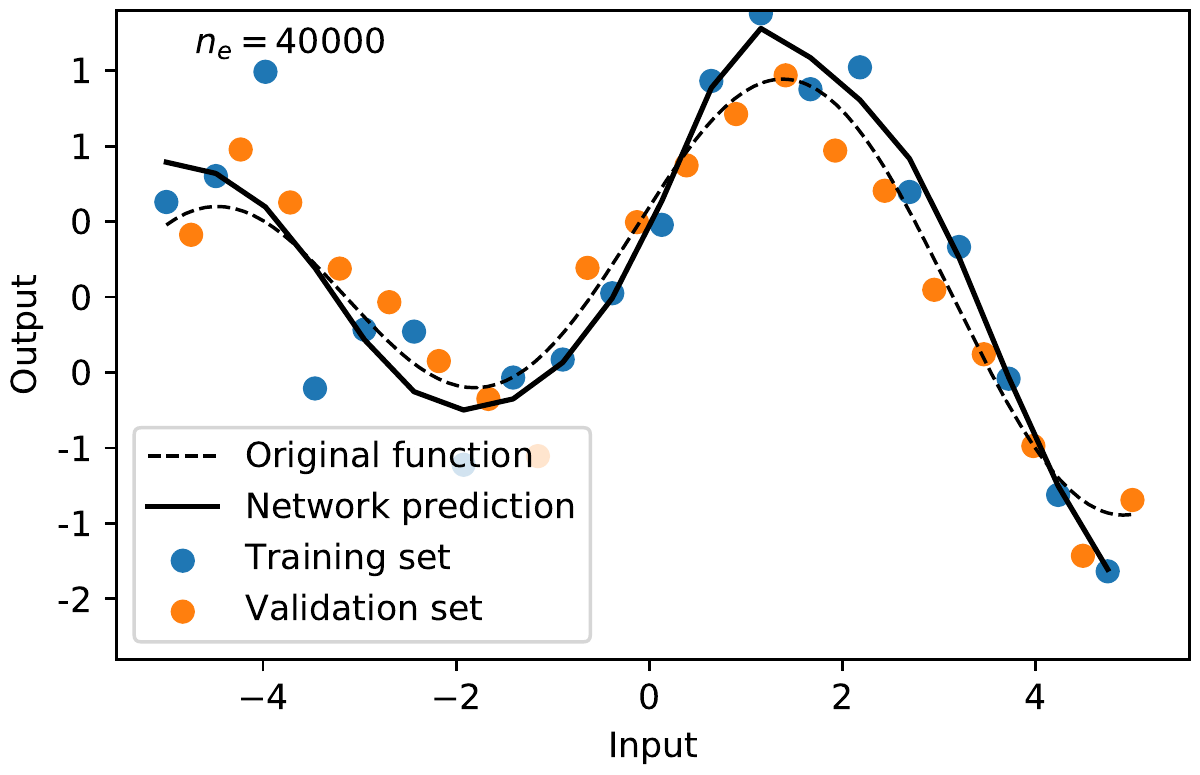}
\end{subfigure}
\begin{subfigure}[!t]{0.65\textwidth}
\centering
\includegraphics[width=\hsize]{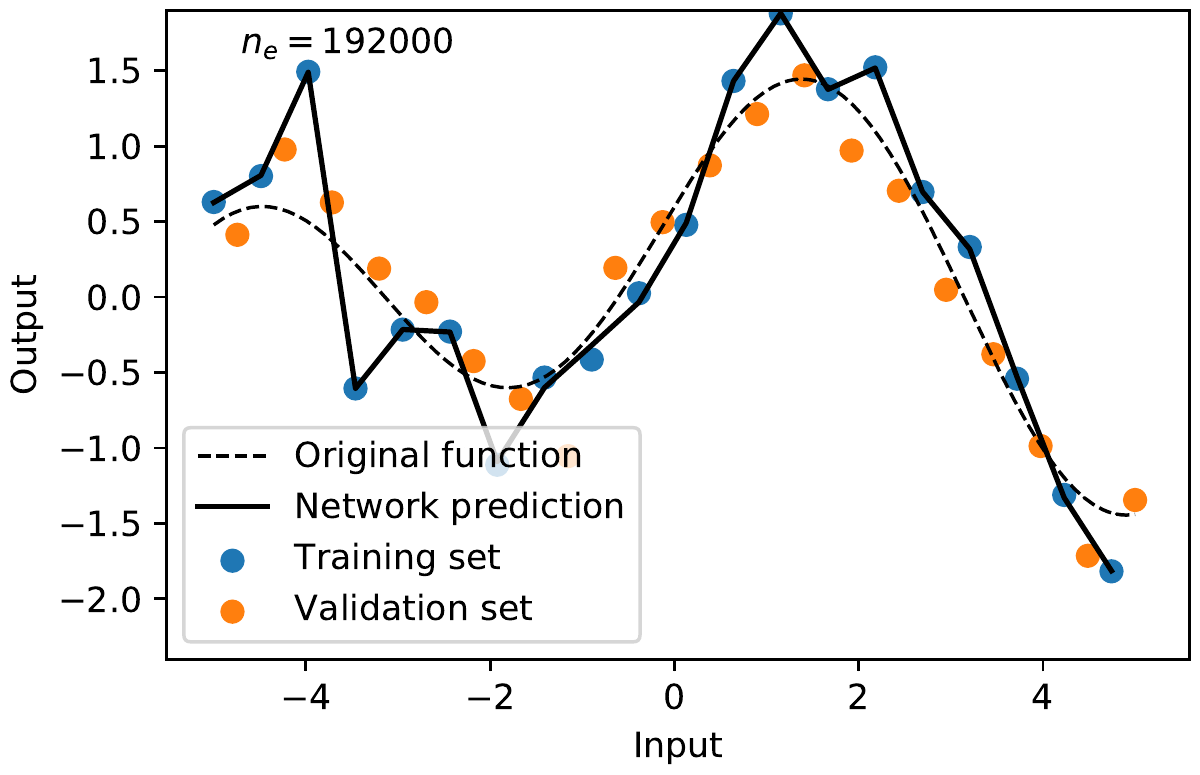}
\end{subfigure}
\caption[Effect of overtraining on a one dimensional regression]{Effect of overtraining on a one dimensional regression problem using an oversized network. {\it Top}: Evolution of the output error, averaged over the corresponding dataset, as a function of the number of epochs. {\it Middle}: Network prediction for the optimal epoch $n_e = 40000$. The points represents the training and validation datasets in blue and orange, respectively. The original function is the black dashed line, while the network prediction is the continuous black line. {\it Bottom}: Network prediction for the last epoch $n_e = 192000$.}
\label{overtrain_figs}
\end{figure*}

This behavior is illustrated in Figure~\ref{overtrain_figs} on a simple one dimensional regression example. In order to ease the reading, by exaggerating the effect, we deliberately used a very oversized network composed of two hidden layer of 32 sigmoid neurons each. Furthermore, we added a Gaussian noise to the original function to increase the disparity regarding the data selection. Half of the data are used as training set, and the other half as validation set. The bottom frame frame shows the the final network prediction at epoch $n_e=192000$. The strong over-training is visible since the prediction follows almost each point in the training dataset, loosing track of the general shape of the function. If the network is trained for a very large number of epochs, the prediction will end up being only linear links between the training data, using the sigmoid neurons only in their linear regime. The top frame illustrates the evolution of the errors of both the training and validation sets. It shows the specific point where the two errors start to diverge, and where the training should have been stopped, represented in the middle frame, despite the error of the training set being far from its minimum value. Several breaking steps in the error curve are due to the high granularity of this specific problem and to new sets of neurons reaching their linear state to fit an outlier point.\\

Usual distribution proportions between the subsets are 50:25:25, or 60:20:20. However, it strongly depends on the original size of the labeled dataset. For very large dataset, the training set is usually reduced to prevent the algorithms to use too much memory with little to no effect on the prediction quality. Large datasets can also be split into many subdatasets and training, validation and test are randomly picked into them. This allows one to repeat several trainings with different datasets and select the most efficient one which is the base of Bootstrap and Boosting methods \citep{efron_86, freund_97}, or even allows one to combine several networks to make a decision by committee \citep{xu_92}.\\

Since very complex problems require a lot of neurons and weights, and therefore a lot of training data, it is more common that the limitation of the generalization capacity is induced by having too few examples. In such conditions, it is preferable to have most of the data in the training set. Then, precautions must be taken, to avoid too small test and validation sets that are not representative of the feature space coverage of the problem or even suffer of small-number effects. In very stretched cases, test and validation sets can be merged into only one dataset that fulfill both roles \citep[e.g][]{lecun-98,Bishop:2006:PRM:1162264}. However, additional precautions should be taken to ensure that the network does not stop training at a point that is overly favorable to this dataset, for example by verifying that the predictions on the training and test sets are similar (see Sect.~\ref{training_test_datasets}).

	\subsubsection{Shuffle and gradient descent schemes}
	\label{descent_schemes}
	
Since the training set must be shown numerous times, the order of the objects and the frequency of the weight updates are two important parameters. In the previous section we assumed, following equations~(\ref{eq_update_perceptron}) and (\ref{eq_grad_desc}), that the network training process uses one object of the training sample at a time, computes the corresponding activation and corrects the weights accordingly, before switching to the next object. This is a usual approach called "on-line", however it can greatly suffer from ordering effects. Considering a simple one dimensional regression example, if data are given sorted, many of the first update steps will be biased toward the first part of the feature space area. Because the first steps are usually larger, it can trap the network weights in a local minimum for a long time, only fitting part of the problem. Such ordering effects can have non-desirable effects on a vast variety of applications. Therefore, it is advised to periodically shuffle the training dataset, up to after each epoch.\\

An additional difficulty raised by this effect is that each object pulls the weights toward its own {\it current} minimum. It might happen that objects push the network weights in opposite ways and slow the training process down. It is also very sensitive to other network aspects like the weight initialization and it often produces a noisy convergence toward the optimum value. For that reason, a widespread approach is to perform a so-called "Batch training". It consists in performing a forward step, therefore a prediction, for all the objects in the dataset, then summing or averaging the individual errors to perform a single weight update. It allows the network to better follow the general trend and often converges faster in the direction of the global error with much less noise in the prediction than the on-line approach. It is also less sensitive to initialization and normalization effects. It also removes all ordering effects, and can be written in a much more computationally-efficient way (Sect.~\ref{matrix_formal}). However, it was demonstrated that the on-line approach can be more safely used with larger learning rates, and that it better resolves the error surface, leading to a much faster global convergence \citep{wilson_general_2003}. This implies that the batch learning often takes much more epochs to learn, and is therefore inefficient in terms of the number of required computations to converge, despite being quicker to perform them. This is the reason why on-line training is still very common, especially when using lighter hardware.\\

An approach related to the on-line scheme is the Stochastic Gradient Descent (SGD) that randomly picks a training element in the dataset and uses it to perform one weight update, and then repeats the process until convergence. This is a drawing with replacement, which removes the necessity to shuffle and obsoletes the notion of batch. It has proven to be very efficient on large datasets and on light hardware. It usually minimizes the number of computation operations to be performed to reach convergence.\\

Considering this information, a recent and broadly adopted hybrid-scheme is the "mini-batch" scheme. It merges the best of both worlds by splitting the training dataset into several small groups that are used together to perform a weight update. It allows one to keep many small updates as in the on-line approach, and it reduces noise in the convergence by doing the average of a few errors. It can be implemented in two different ways: (i) either by selecting few random objects to construct the next mini-batch, in a SGD fashion or (ii) by splitting the whole dataset into groups that are used subsequently, then shuffling the complete dataset at each epoch to compose different groups. This scheme is implemented in most of the popular frameworks as it proved to have a good efficiency, both in terms of the number of required computations, and in terms of computational performance. In our framework CIANNA, the mini-batch is the default scheme and the size of the batches is a tunable parameter. The framework is also able to perform batch and SGD trainings if specified, using specific arguments. \\
	
	\subsubsection{Momentum conservation}
	\label{sect_momentum}
	
	Another optimization, still to avoid local minima, consists in adding a "momentum". This is a classic speeding up method for various iterative problems \citep{polyak_methods_1964}, that has been updated for modern problems \citep{qian_momentum_1999}. It consists in adding a fraction of the previous weight update to the next one, during the training phase. This memory of the previous steps helps keeping a global direction during the training especially in the first steps. This additional inertia also prevents the network from staying stuck in local minima without resorting to a higher learning rate (Section \ref{learning_rate}), and usually allows much faster convergence.\\
	
Moreover, it can be seen as another form of adaptive learning rate. Indeed, it usually increases the weight updates in the first training steps when the weights are the farthest away from their optimum values. Once closer to the global minimum error, the inertia will reduce, going back to smaller updates. Therefore, it allows a faster training even when taking a smaller learning rate. It also helps reducing the spread between repeated trainings. Again, there are many different implementations, but a simple one is to add an inertia term to the update equation~(\ref{eq_grad_desc}), which can then be expressed as:
\begin{equation}
\omega^{t}_{ij} \leftarrow \omega^{t-1}_{ij} - \Delta\omega^{t}_{ij},
\end{equation}S
where $\omega_{ij}$ is the weight matrix between two layers, $t$ marks the value for the current epoch update, and $\Delta\omega_{ij}$ is the computed weight update. The latter is expressed with the momentum term as:

\begin{equation}
\Delta \omega^{t}_{ij} = \eta \frac{\partial E}{\partial \omega^{t-1}_{ij}} + \alpha \Delta \omega^{t-1}_{ij},
\end{equation}
with $E$ the propagated error at the current layer, and where the hyperparameter $0 < \alpha < 1$ scales the momentum. Usual values are often between $0.6$ and $0.9$.

\subsection{Matrix formalism and GPU programming}
	\label{matrix_formal}

	\subsubsection{Hardware considerations for matrix operations}
	
We mentioned in Section~\ref{descent_schemes} that the batch, and consequently the mini-batch, gradient descent scheme is computationally more efficient. This is mainly due to the fact that batch methods can very efficiently be converted to matrix operations, and mostly to matrix multiplications. Indeed, matrix operations are extremely quick to compute, compared to scalar operations, regarding the raw number of computations to be performed. Indeed, matrix operations have been in the landscape of numerical computing for decades, and have been identified, especially by data and computer scientists, as computationally intensive while being necessary for a very wide range of workflows. Additionally, many computing operations, even if not explicitly written this way, can be and are computed as matrix operations. The very nature of computer hardware optimization (memory layout, batched operations, several operation per cycle, ...) works using the scheme of vector and matrix operations. Therefore, an enormous quantity of work and experience has been accumulated toward the optimization of matrix operations through the years. And at some point, it began to reversely influence hardware technology to make a better use of matrix operations and to express more applications into this formalism \citep{Du_2012, Cook_2013}.\\

The very expression of these optimizations is how matrix operations can be described as highly parallel operations, taking advantage of vector co-processors in Central Processing Units (CPU) and of the rising core count on the same processor chip (or multiple CPU chip on a single motherboard). Indeed matrix computations can be efficiently expressed as Single Instruction Multiple Data (SIMD) operations. It also takes a strong advantage of very quick low-level memory in CPUs as the same data are used several times subsequently. In summary, matrix operations influence and take advantage of many hardware novelty and optimizations. \\

The most extreme example of this situation is the Graphical Processing Units (GPU) that are almost dedicated to SIMD operations. As their name indicates, these chips are designed for graphical applications that rely on a pixel formalism. Therefore, dealing with images is intrinsically a matter of tables, or matrices, of pixels. For examples: dimming an image is equivalent to applying a factor to all pixel values, which is a pure SIMD operation; smoothing an image consists in the application of a filter to many zones of pixels in the image, which are made of many small matrix multiplications; the rotation of an image is also a SIMD operation to transform input coordinates into others; etc. For this purpose, GPUs are built in the form of a very large amount of very light compute units or cores with much more layered cache levels than CPUs. These cores are slow in terms of clock speed, which is around a GHz for high-end GPUs while it can be as high as above 5\,GHz for modern CPUs, and in terms of instructions per clock cycle and of general purpose capabilities. Noticeably, the GPU cores cannot perform or have very poor performance with double precision real numbers (excluding most advanced professional or science dedicated ones) and are limited to single precision float and integer operations. Such cores are simple and small enough to be stacked in large numbers in a single chip along with large amounts of very fast dedicated memory, allowing GPUs to reach accumulated performances that are orders of magnitude above regular CPUs. However, it makes such chips very application specific, while regular CPUs are able to handle much more diverse tasks, like handling an Operating System. This is why GPUs rely on a CPU to handle the general programming and are plugged as accelerators in more classical computer systems. For all these reasons, they are the most efficient way to speed up matrix operations to this date.
	
\newpage
	\subsubsection{Artificial Neural networks as matrix operations}
	\label{ANN_as_matrix}

\afterpage{%
    \clearpage
    \ifodd\value{page}
        \expandafter\afterpage
    \fi
    {
\begin{sidewaysfigure}
\centering
\includegraphics[width=0.80\hsize]{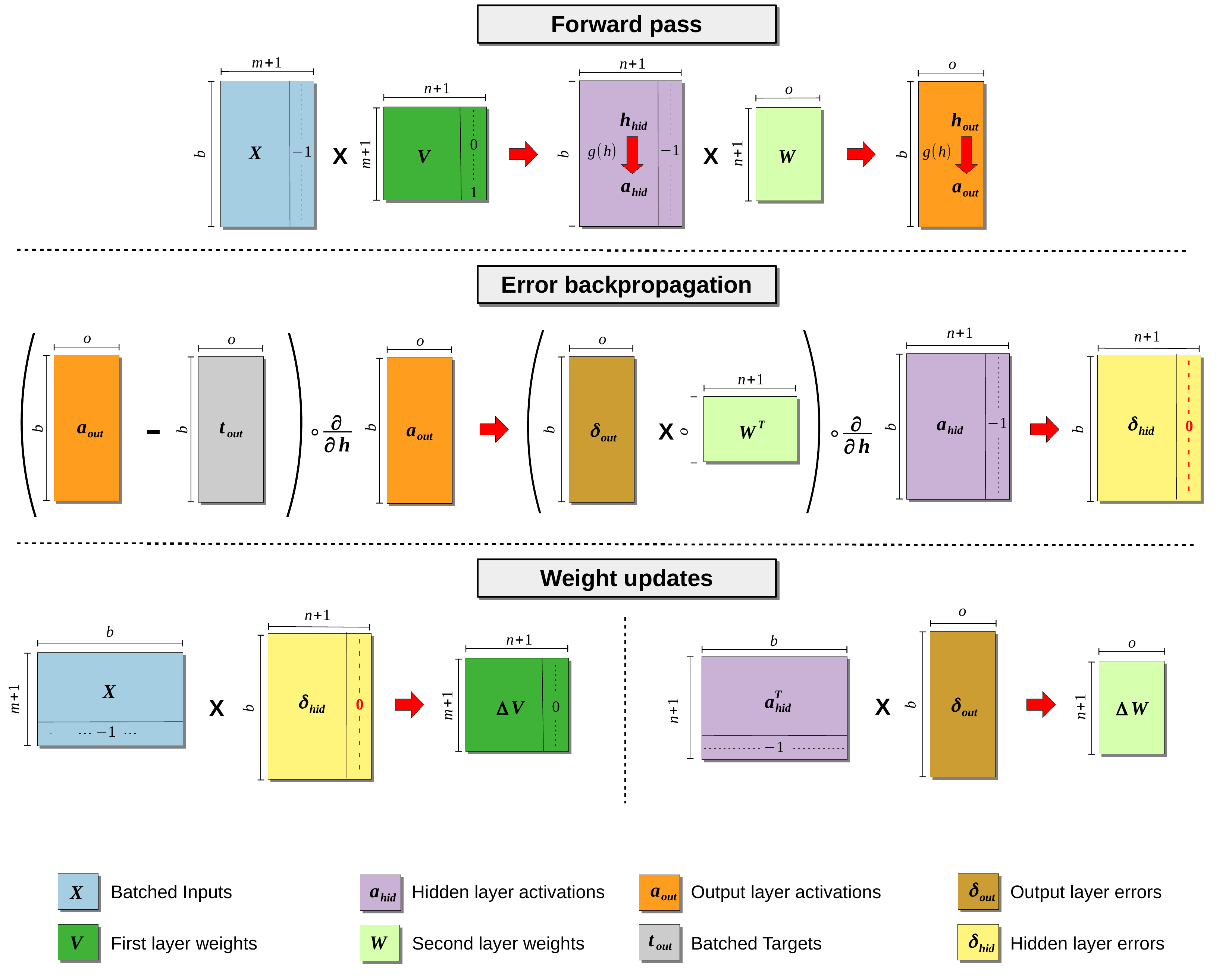}
\caption[Graphical representation of the matrix batch training]{Graphical representation of the matrix operations involved in the (mini-)batch training of a single hidden layer network. The large red arrows indicate the order in which the operations are performed. The result of a matrix multiplication is then used directly to perform the next operation. Red arrows inside matrix indicate activation function and are performed before the next operation. Large $\times$ symbols are matrix multiplications, while $\circ$ symbols stand for element wise multiplication. The matrix sizes are as follows: $b$ is the batch size, $m$ is number of input dimensions, $n$ is the number of neurons in the hidden layer, and $o$ is the number of output neurons.}
\label{matricial_form_fig}
\end{sidewaysfigure}
}
}

All the equations of Section \ref{global_ann_section} can be expressed as matrix operations relatively easily. Assuming a batch or mini-batch approach, the input vector is replaced by an input matrix $X_{bm}$ with $m$ input dimensions and $b$ training objects. Then, following equations~(\ref{weighted_sum}), it is multiplied with the weight matrix, which produces a matrix of all the sums $h_{bn}$ for $n$ neurons, where each element has to go through the activation function to produce all the activations of the first layer, corresponding to all the inputs in the current batch. This operation can then be repeated for each layer up to the end of the network where an output matrix $a_bn$ is produced. Following the same approach it can be compared to a target matrix using the selected error function, to produce an error matrix. Following the exact same approach, the back propagation can be done in a similar matrix formalism. However, it must be noted that during the back propagation either the current weight matrix or the produced $\delta_bn$ must be transposed to respect the ordering in equation~(\ref{eq_update_full_network}) when propagating the error. As well, either the current $\delta_bn$ or the current activation $a_bn$ must be transposed to respect the ordering when computing the weight updates.\\

One specific point in this formalism is the inclusion of the bias node. In many methods that use the matrix formalism it was decided to include it as a subsequent addition to the neuron weighted sum as evoked in Section~\ref{bias_node}. However, this implementation induces performance penalties when using GPUs, as explained in the following Section~\ref{gpus_prog}. Then, the difficulty using our approach of an extra node that acts as a supplementary constant feature is that, even if it works on the input layer by having a column of bias values, the multiplication of this extended input matrix by the weight matrix creates a hidden layer activation without bias node. One could solve this by appending a bias node column to the hidden activations, but this would result in even more performance penalties than the one we are trying to avoid. A possible solution, is then to also add an extra column in each weight matrix that is full of zeros except for on value defined to 1. Then the weighted sum of input automatically produces an extra column of bias with the same value as the input column.\\

The full procedure including the forward pass, the error back-propagation, and the weight update, is illustrated in Figure~\ref{matricial_form_fig} that shows the successive operations performed on the input data and the intermediate network results. We note that matrix operations are symbolized with the symbol $\times$ while the dot multiplication is denoted by $\circ$. This figure also illustrates the addition of the bias value using the previously described methodology in the input $X$ and weight matrix $V$, but not on the output layer for which it is unnecessary. In the back-propagation part, the red column in $\delta_{hid}$ is the echo of the extra bias column in $h_{hid}$ that comes from matrix size conservation of the process, but it does not contain meaningful information. This vector is then manually set to $0$ to preserve the extra column values in $\Delta V$ that enables the bias propagation. This solution to include the bias directly in the matrix could be seen as an unnecessary subsequent complexity and increase in memory usage. In practice, the difference in time needed for matrix multiplications, with or without the added bias, is completely negligible, and the memory usage increase is very marginal on a common application. But when considering the operation launch time overhead on a GPU, our solution is much more efficient since it requires significantly less operation launches than a subsequent bias addition, resulting in a very significant computation time reduction when using small batch sizes. Ultimately, this approach can be implemented using efficient Basic Linear Algebra Library (BLAS) which performs efficient multi-threaded matrix operations, like OpenBlas, or similar GPUs library like cuBLAS.

\newpage
	\subsubsection{GPUs variety}
	\label{gpus_variety}

It is important to note that historically the GPUs were not a programmable hardware. They were completely closed, with given APIs that only allows a limited number of tasks. The first uses of GPU for computation were then some sort of "hacks" where a problem was reformulated into operations on images so that the GPU could work on it. While ancestors of the GPUs originate in the 70s, it is only in the early 2000s that General-Purpose computing on Graphical Processing Units (GPGPU) began \citep{Du_2012}. They noticeably became more efficient than CPUs on some matrix operations in 2005 \citep{Galoppo_2015}.\\

At the moment, there are several manufacturers of GPUs, which leads to important variations in hardware architectures and software development tools. We list here the three major GPU manufacturers along with some of their specificities:\\

{\bf \large Intel}\vspace{-0.4cm}\\

Mostly known for their CPUs, they are in fact the company that sells the largest number of GPUs, in the form of integrated GPUs (iGPU) in their CPUs. However, despite this position they do not permit GPGPU! Indeed, the iGPU format is not the most adapted one for this approach as it shares resources with the CPU. However, Intel has recently made large investments in dedicated GPU solutions in order to deliver them to the market probably in the next years, and that would be suitable for GPGPU. We still note that the manufacturer has already made such attempts in the past that were not successful enough to be released. We also note that Intel has led some part of the research on "many cores" CPUs like their Xeon Phi lineup, but that are slowly being discarded in profit of GPUs.\vspace{+0.2cm}\\

{\bf \large AMD}\vspace{-0.4cm}\\

This manufacturer is mostly known in GPU technology due to its sub-brand Radeon. AMD is also a very well-known brand for its CPUs, however, unlike Intel, it produces both iGPUs and dedicated GPUs. Regarding GPGPU, the company is strongly invested in the development of OpenCL (or Open Computing Language), which is an open source framework mostly written in C that allows a very specific way of parallel computing. Indeed, the aim of OpenCL is to be usable on any device taking advantage of many various hardware: CPUs (even ARM CPUs), GPUs, Digital Signal Processors (DGS), Field-Programmable Gate Arrays (FPGA), ... It aims at unifying the programming to automatically use all these hardware. While this specificity along with the open source aspect of this solution is appealing, it suffers several caveats. Firstly, the OpenCL framework is known to be laborious to use. Secondly, this very general approach does not enable optimum performance for each hardware. For AMD GPUs, OpenCL is the only usable approach and still does not permit to use them at their maximum performance. Good performances are reachable, though, but at the cost of a very verbose OpenCL programming instead of the common approach. A last point is that AMD does not have much economical leverage as other companies, and therefore is often behind in terms of modern GPU techniques, despite being currently one of the most innovative brand regarding CPU.\vspace{+0.2cm}\\

\newpage
{\bf \large Nvidia}\vspace{-0.4cm}\\

This brand is the biggest one to be dedicated to GPU solutions (excluding some chips like the {\it Tegra} that are more close to full System on a Chip (SoC)). They are comparable to AMD in terms of volume of dedicated GPU sells, though usually slightly ahead. They also provide the most commonly adopted GPUs in professional environment and science with dedicated lineups. They are the main provider of GPU solutions for super computers, being included in 5 of the top 10 world supercomputers, including the first two positions. This very specialized expertise allows them to be on the edge of many GPU technologies. However, their GPGPU aspects is done by using a dedicated programming language slightly derived from C++ that is called CUDA (Compute Unified Device Architecture, that stands both for the hardware architecture and the associated language), which also contains a dedicated compiler based on gcc. However, it is only supported by Nvidia GPGPU, and therefore is much less general than OpenCL. Additionally this solution is not open source (with the induced dependency effect argued in Section~\ref{tool_boxes}), still, allowing for a very low level programming capacity. Many higher level functionalities are also proposed through not-open dedicated CUDA libraries even if they can be mostly reproduced using the low level CUDA language. It is worth noting that one can also use OpenCL for Nvidia GPUs, which often results in much lower performance considering the same time investment, but allows widespread OpenCL applications to work on such GPUs. Finally, Nvidia has a fairly aggressive approach on the professional market where some CUDA ``Pro'' accelerated applications can only be used with their professional GPU lineup that is often much more expensive.\\

\vspace{1cm}
We note that there are a few other software solutions for GPGPU on Nvidia or AMD hardware, namely the Microsoft Direct Compute that relies on their graphical library DirectX only available on the Windows OS; the latest versions of OpenMP that also support GPU acceleration; and the OpenACC solution that is very promising and progressively adopted by many recent super-computing infrastructures, and that rely on compiler directives to convert regular loops into GPU accelerated equivalent automatically with very efficient performance.\\

Regarding the specificity of each solution, we opted for Nvidia CUDA. First, we note that it was not motivated by the material at our disposal, since we had access to reasonable hardware from both AMD and Nvidia. The choice toward Nvidia was mostly motivated by the greater potential of Nvidia hardware for GPGPU. Additionally they are deeply involved in the field of AI, and more specifically in ANN, with many dedicated optimizations for these specific workflows. Moreover, they are currently adding more and more dedicated computing sub-units in their CUDA cores that are even more dedicated to small and light (lower bit-size numbers) matrix operations aiming at further improving neural network training speeds. These specific sub-units, namely the Tensor Cores, are present in the last two generations of Nvidia GPUs, and have been further improved in their upcoming Ampere architecture. Concurrently, the choice of the CUDA language has been made over OpenCL again for performance efficiency, and because CUDA is necessary to make use of all the dedicated AI materials we just presented. With most of the large computing clusters having modern Nvidia GPU, this appears as a strategic choice in anticipation of a potential future proposal to use large scale GPU facilities with dedicated access to AI research for example on the new \href{http://www.idris.fr/annonces/annonce-jean-zay-eng.html}{Jean Zay} supercomputer at GENCI (14 PFLOPS peak performance).\\

\newpage
	\subsubsection{Insights on GPU programming}
	\label{gpus_prog}

We describe here a few  aspects that are particularly important in our implementation in CIANNA (see Appendix~\ref{cianna_app}), but also for any GPU accelerated ANN. Our framework is able to handle CPU matrix operations though OpenBLAS, but its CUDA BLAS (cuBLAS) implementation drives most of its development. First, matrix operations are very nicely handled in cuBLAS, with a huge amount of small optimizations that make use of the accumulated knowledge on matrix computation and of the dedicated hardware. It is important to note that using these GPUs for matrix computation strongly drives the hardware development of Nvidia. The Tensor Cores are the best examples, but it works the same for cache optimization, choice of precision capabilities, memory bandwidth, memory-core layouts, etc. The general CUDA programming is used to construct extremely parallel kernels (i.e. individual functions that are executed simultaneously by many individual CUDA cores on the GPU corresponding to the SIMD formalism) for all the operations that are not matrix multiplications described in the previous section and illustrated in Figure~\ref{matricial_form_fig}. For example, we wrote kernels for activation functions, element wise derivative multiplication, etc. In practice cuBLAS operations also rely on underlying kernels but that are launched using the dedicated cuBLAS API, so we will most of the time refer as kernels for any function that executes on the GPU including matrix operations.\\

Overall, the programming scheme of a CUDA application consists in declaring all the useful variables on the CPU and on what is called the "host" RAM memory. Then all the useful data are moved on the GPU (also called the "device") RAM memory. Even if modern GPUs have very quick memory interface with the CPU, these data transfer operations still have an important time cost, therefore one must aim at minimizing them. For a neural network it is easy to create the weights and let them on the GPU memory. This is the same for many other network data like the layer activations and the layer errors. Even when computing the error, it is common to send back to the host as little data as necessary to monitor the error evolution. The input data is more tricky to manage because the size of the dataset can be large. With modern GPU that have a very large dedicated memory, the dataset can be fully loaded in the GPU and all the training epochs can be performed there. When the dataset is too large, chunck of data must be sent to the GPU regularly. There is a nice property of GPUs which is that memory transfer and computations can be launched concurrently without performance penalty. Therefore, one must aim at maximizing the overlap between the two, which can be easily done in this case. We note that this is considered as basic GPU programming knowledge and optimization by the \href{https://docs.nvidia.com/cuda/cuda-c-programming-guide/index.html}{CUDA programming guide}.\\

It is now the appropriate place to detail why the computation of the bias node can lead to a performance penalty (Sect.~\ref{ANN_as_matrix}). When using GPU to perform the matrix operations of a neural network, the computation themselves are very efficient, and this is the same for the kernel operation. However, due to the fact that the CPU leads the execution of the program and orders the GPU to execute the various kernels, there is a launch latency of the order of few tens of $\mathrm{\mu s}$. In most workflows that were used for years it was not much of an issue since GPUs were used with very large kernels or to perform very large matrix operations, for example in numerical simulations. However, regarding ANNs, an epoch cannot be expressed efficiently as a single very large kernel because the activations of a layer depend on the activations of the previous one. Therefore, ANN implementations are a succession of many smaller matrix or kernel operations, and it is even worsen by the fact that they often require many epochs to learn. Obviously, this effect is additionally scaled when using a small mini-batch size since an epoch is splitted into several small forward and back propagation passes on the network, vastly increasing the number of overheads. In this context it has become very common for ANN frameworks to be fully bottle-necked by the kernel launch latency rather than by the compute performance of the GPU. Despite that, it has been much more efficient to use GPUs than CPUs for neural networks for several years now. But it remains a good idea to try to minimize this effect to get closer to the full performance of the GPU. Nvidia has invested a lot of efforts recently in the last versions of CUDA (starting with 10.0) to reduce kernel launch latency and to provide solutions that allows the program to become independent of the CPU by having groups of kernels executed several times (CUDA Graphs). In this case the latency is reduced to the equivalent of only one kernel launch. However, reducing the number of kernels needed remains a good practice in order to obtain the best possible performance. This is what motivated our implementation of the bias node, which removes a kernel launch at each layer.\\

We also highlight, that we could have used the cuDNN toolkit or cuFFT, that were designed exactly for such cases. They include many optimizations of this kind for many modern ANN practices. But in contrast with cuBLAS, they usually impose constraints on the data format or the general approach. We therefore preferred to use cuBLAS to keep the freedom to try any unconventional method that could be appropriate for a future application of CIANNA.\\

Finally, we give here some details about the Tensor Cores that are present on the latest Nvidia GPUs architectures (Volta, Turing and Ampere). They are dedicated computing units that are able to perform a size-specific matrix multiplication per GPU clock cycle. The Tensor Cores even incorporate a subsequent addition to the result in the same clock cycle to account for the bias addition. These cores work on numbers with lower precision (FP16, INT8 or even INT4, and more recently TF32, BF16, ...) because they are dedicated to ANN workflows where the precision is far less important. We remind that many early models of ANNs used binary neurons! Therefore, most of the network parameters and results can be safely expressed using lower precision numbers. However, the matrix operation itself makes a lot of sums, exposing the user to the accumulation of rounding errors more dangerously than with higher precision variables. This is why Tensor Cores are "Mixed Precision" compute units, meaning that they include an internal sum accumulator that uses at least twice the number of bits than the input data. They can even store the results of the multiplication of two half-precision matrices into a full precision one. It is then possible to construct very efficient "mixed precision" networks by using these new hardware cores wisely, which is described nicely by Nvidia in \citet{micikevicius2018mixed}. Unfortunately, we did not have access to such modern GPU early enough for these capabilities to be present in CIANNA. However, we since had experience with them and it will be include in a near future.

\newpage

	\vspace{-0.3cm}
\subsection{The specificities of classification}
	
	In this section we discuss some specificities of a classification application of ANNs, along with some necessary precautions, that will be useful for the classification of Young Stellar Objects in Section~\ref{yso_datasets_tuning}. We finish this subsection with an illustration of a few common classification examples using ANNs.
	
	\vspace{-0.3cm}
	\subsubsection{Probabilistic class prediction}
	\label{proba_class_intro}
	
	While it is possible to do a classification that has a "binary" behavior using neural networks, it is much more common to represent the membership of a class using a continuous value. It has the nice property that it can measure the degree of certainty of the network in its prediction. As exposed in Section \ref{sect_perceptron}, a common approach is to define the output layer to have one sigmoid neuron per class and a target with only the neuron of the expected class having a value of one, the other being set to zero. Then, after the training of the network, a prediction of $0.9$ for one neuron can lead us to consider the corresponding object as a reliable member of the corresponding class. Indeed, the neuron is far off the linear regime of the sigmoid, which indicates that it has accumulated a significant signal toward its activation. In contrast, such a neuron with an activation value close to $0.6$ would not be considered as a very reliable classification result, even if it is the highest activation of all output neurons. We note that the 1.0 and 0.0 target are asymptotic values of the sigmoid meaning that these perfect predictions are very unlikely to be reached. To obtain a probability it additionally needs a normalization over all the outputs, so that the sum of the output values for one object is always 1 and each neuron provides a membership probability. This would give the network attributes of a Probabilistic Neural Network \citep[PNN; ][]{specht_probabilistic_1990, stinchcombe_universal_1989}. However, using sigmoid activation functions without normalization, there are no rules that prevent several output neurons from having simultaneously a value close to one, which might happen even if only one element of the target vector is set to one due to random weight initialization. Such prediction should naturally disappear during the training process but there is no strong condition that prevent such case to happen after the training for an underconstrained input.\\
	
	A more suitable activation function to perform a probabilistic classification is the so-called Softmax activation, also know as normalized exponential \citep{Bridle_90}. It is expressed as:
\begin{equation}
\centering
	a_k = g(h_k) = \frac{\exp(h_k)}{\sum_{k^\prime=1}^{o}{\exp(h_{k^\prime})}},
\label{Softmax_activation}
\end{equation}
where $k$ is the neuron index in the output layer. Thanks to the normalization over all the output neurons, the $k$-th output neuron provides a real value between zero and one, which acts as a proxy for the membership probability of the input object in the $k$-th class. The global behavior of this activation is also considerably less prone to saturation effects (Sect.~\ref{weight_init}), and therefore should be able to get excellent predictions with reasonably small weights. But this function is not without caveats. If used with an inappropriate error function, (as discussed in the next paragraph), it can push the weighted sum too high and therefore reach the overflow limit of the variables. It is specifically problematic for GPU implementation as it often uses low precision variables (e.g. FP32, FP16, or smaller, see Sect. \ref{gpus_prog}). A simple modification to the function is to subtract the maximum $h_k$ value to all the output values :
\begin{equation}
	a_k = g(h_k) = \frac{\exp(h_k - \max \{h_1, \dots, h_o\})}{\sum_{k^\prime=1}^{o}{\exp(h_{k^\prime} - \max \{h_1, \dots, h_o\})}}
\label{mod_Softmax_activation}
\end{equation}
This allows us to shift the number comparison in a less numerically problematic part of the exponential, and almost always prevents the overflow.\\

\begin{figure}[!t]
\centering
\includegraphics[width=0.6\hsize]{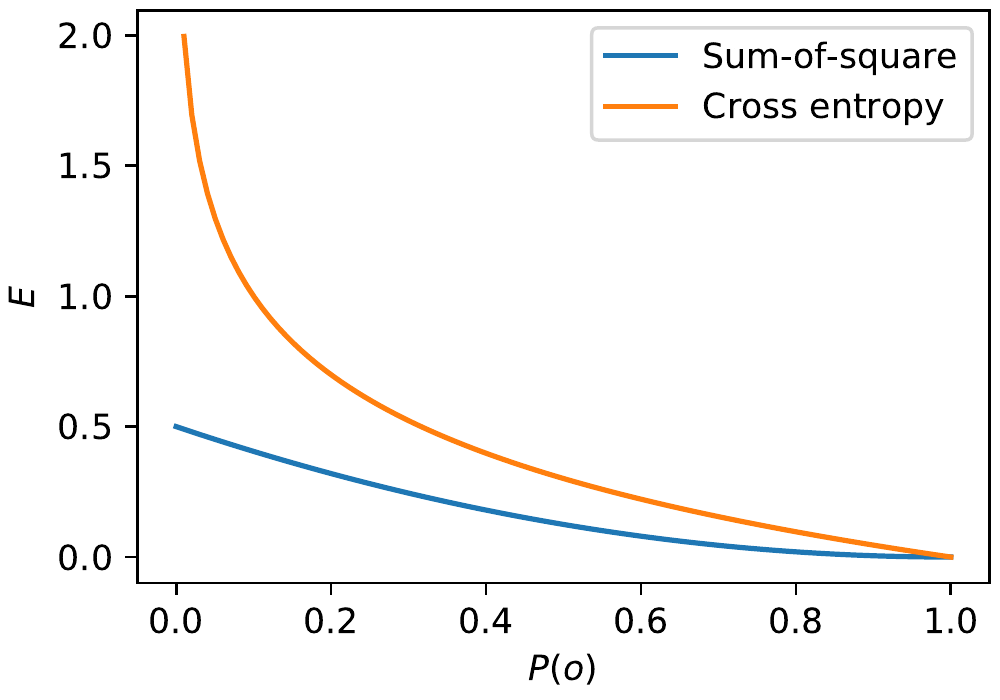}
\caption[Sum-of-square error against cross-entropy error]{Comparison between the behavior of the sum-of-square and cross-entropy error functions for a probabilistic output with target to 1.}
\label{errors_comparison}
\end{figure}

On the other hand, the sum-of-square error is less suitable for a probability output, especially with Softmax. It is most of the time replaced by the so-called Cross-Entropy error (or loss) that is much more efficient for probabilistic outputs and that produces a nice error propagation term when combined to the Softmax \citep{Bridle_90}. It is expressed as:
\begin{equation}
	E = - \sum^{o}_{k=1}t_k \log(a_k),
\label{cross_entropy_error}
\end{equation}
where $a_k$ are the activations of the output layer that count $o$ neurons and $t_k$ the corresponding targets, both of which are probability values. It can then be combined with the derivative of the Softmax activation function:
\begin{equation}
	\frac{\partial a_k}{ \partial h_{K} } = a_k(\delta_{k K } - a_K )
\label{soft_max_deriv}
\end{equation}
where $\delta_{k K }$ is the Kronecker symbol ($\delta_{k K } = 1$ if $k = K$ and 0 otherwise).
Following the equation~\ref{eq_update_full_network} the corresponding output error term is written as:
\begin{equation}
	\delta_o(k) = a_k - t_k.
\label{soft_max_output_error}
\end{equation}
Figure~\ref{errors_comparison} shows the difference between the cross-entropy and the sum-of-square errors on a probabilistic output for which the target is 1. It demonstrates that the cross-entropy quickly corrects the outputs that predict 0 when they should be 1. It also induces that the output error term $\delta_o$ is linear, which allows a better error propagation and balance between the various output during the weight updates. This avoids attributing to much importance to specific output neuron regimes.\\

Finally, a probability per class does not directly define what is the predicted output class. The easiest method consists in selecting the highest probability, but more advanced methods also exist. For example, one can require that the maximum value must be higher than the sum of the others, or that it is higher than the second maximum by a certain amount. The most selective one would be to require that the output probability is above a defined threshold. All these methods exclude objects that are too "confused" to be considered as reliable in order to improve the overall reliability of the selected objects.\\

However, it is important to emphasize again that this probability prediction only characterizes the network estimated probability regarding how it succeeded in fitting the presented distribution of objects in the feature space. Then, it must not be used as a genuine probability, and selecting all objects with a neuron activation probability above $0.9$ does not necessarily mean that these objects have a $90\%$ probability of being of the corresponding class. In Section~\ref{proba_discussion}, we will use these activation probabilities as a criteria to disentangle the confused and reliable objects, but the actual probabilities will be estimated using the tools presented in the next section.

	\newpage
	\subsubsection{The confusion matrix}
	
\begin{table}[t]
	
	\centering
	\caption{Confusion matrix example for a cats and dogs example.}
	\vspace{-0.1cm}
	\begin{tabularx}{0.55\hsize}{r l |*{2}{m}| r }
	\multicolumn{2}{c}{}& \multicolumn{2}{c}{\textbf{Predicted}}&\\
	\cmidrule[\heavyrulewidth](lr){2-5}
	\parbox[l]{0.2cm}{\multirow{5}{*}{\rotatebox[origin=c]{90}{\textbf{Actual}}}} & Class & Cat & Dog & Recall \\
	\cmidrule(lr){2-5}
	 &  Cat   & 93     & 7       & 93.0\% \\
	 &  dog   & 14      & 86     & 86.0\% \\
	\cmidrule(lr){2-5}
	 &  Precision & 86.9\% & 92.5\% & 89.5\%\\
	\cmidrule[\heavyrulewidth](lr){2-5}
	\end{tabularx}
	\vspace{-0.05cm}
	\label{balanced_confmat}
\end{table}
	
We define here the concepts necessary to present results statistically, and to characterize their quality. For this it is common to use the so-called "confusion" matrix. It is defined as a two dimensional table where the rows correspond to the targets, and the columns correspond to the classes predicted for the same objects by the classifier, in our context an ANN. A usual example is a classification between cats and dogs, as presented in Table~\ref{balanced_confmat} using made-up plausible results. It shows the corresponding $2 \times 2$ confusion matrix, where the number $14$ that appears in the second row, first column is the number of labeled dogs that were mistakenly classified as cats by the classifier. In this scheme, a $100\%$ "correct" classification gives a diagonal matrix, while the off-diagonal numbers count the misclassified objects. This representation directly provides a visual indication of the quality of the network classification. The confusion matrix allows us to define quality estimators for each class:

\begin{equation}
	\centering
	\text{Recall} = \frac{TP}{TP+FN} \qquad \text{Precision} = \frac{TP}{TP+FP}
	\label{eq_conf}
	\end{equation}
\vspace{-0.3cm}
	\begin{equation}
	\centering
	\text{Accuracy} = \frac{TP + TN}{TP+TN+FP+FN}
	\label{eq_accu}
	\end{equation}
where :
\begin{align*}
& TP \equiv \text{True Positive} & TN \equiv \text{True Negative}\\
& FP \equiv \text{False Positive} & FN \equiv \text{False Negative}\\
\end{align*}

The "recall" represents the proportion of objects from a given \textit{target} class that were correctly classified. The "precision" is a purity indicator of an output class. It represents the fraction of correctly classified objects in a \textit{predicted} class, as predicted by the network. And finally, the "accuracy" is a global quantity that gives the proportion of objects that are correctly classified with respect to the total number of objects. In our confusion matrix representation, we show the accuracy at the intersection of the recall column and the precision row. Limiting the result analysis to this latter quantity may be misleading, because it would hide the class-specific quality and would be strongly impacted by the possible imbalance between the output classes. The matrix format is particularly well-suited to reveal the weaknesses of a classification. It could, for example, reveal that the vast majority of a subclass is misidentified as a specific other subclass, which is also informative about some degeneracy between the two classes.\\

In our example with cats and dogs (Table~\ref{balanced_confmat}), the global accuracy is $89.5\%$ while the recall of cats is much better with $93\%$, and the recall of dogs is $86\%$, therefore there is a significant additional difficulty in fitting the distribution of dog examples in the feature space. This also has incidence on the precision values: the dogs being less well identified, the misclassified dogs induce a drop in the precision of cats, increasing the false positive rate of cat predictions. The differences between the two classes could even be exaggerated with the exact same global accuracy, which illustrates to what extent it is a bad indicator when doing classification. Interestingly, it also highlights a structural weakness in the way we defined our ANNs. Even if the error propagation is made by using each individual output error, one usually just monitors the average error over the output neurons, while it is possible that some are better fitted than others. It is possible that, during the training phase, some objects are overtraining a subset of the neurons, while others are still in the process of improving another subset of the neurons, with an average error that still decreases. This is why more elaborated error monitoring methods are sometimes used in some ANN implementations.\\

\subsubsection{Class balancing and observational proportions}
\label{class_balance}

Machine Learning methods are known to work better on balanced dataset \citep[e.g][]{Anand_93,yamin_2009}. In a classification case it means that it works better if all classes have the same number of examples in the training dataset. The main reason is that when more data of a specific class are used, they lead to proportionally more weight updates. The network therefore uses more weights and neurons to reconstruct this class in disadvantage of the others. However, this reasoning in based on the assumption that each class has the same intrinsic complexity in the feature space, which is a strong assumption for which we will a give counter example in Section~\ref{yso_results}. In cases where this assumption is valid, a common approach consists in rebalancing the training dataset to have equal class proportions despite the proportions of the original dataset. One should however be careful, because this can lead to several bad practices, the first being that the classification performance are also evaluated using a balanced dataset as validation and test datasets, while the true underlying proportions are strongly imbalanced. The study of the impact of imbalance in ML algorithm is a dedicated field of study called "imbalanced learning" which is known to be much more complicated than learning on classical balanced dataset and that is studied for more than two decades \citep{Anand_93, he_learning_2009}.\\

\newpage
To illustrate the effects of an imbalanced dataset, we selected a standard example of a specific virus immunity detection. Assuming that one wants to build a serological test for that purpose, it can be seen as a simple two-category classification with positive and negative test results. For the purpose of this example we assumed that only 1 out of 10 persons are truly infected. We did not make any assumption on how the classifier was built and just assumed a recall of $93\%$ for positive cases and $95.4\%$ for negative cases. Table~\ref{balanced_test_confmat} shows the confusion matrix using balanced proportions, that are likely to be used when tuning the method, for example when training the network. This gives apparently satisfying results with a $94\%$ global accuracy and a reasonable precision above $94.9\%$ to avoid false positive. However, as we said, only $10\%$ of people are truly positive in this example, therefore using the true proportions, which we will refer to as "observational proportions", it gives the confusion matrix in Table \ref{imbalanced_test_confmat}. This table shows that despite the $95\%$ recall of negative cases, it leads to a large contamination of the predicted positive cases whose precision drops to $66.9\%$, and therefore leads to a high false-positive rate. Such misleading results can have important consequences, like giving immunity passports to non-immune persons, or in similar medical studies, giving a high-risk treatment to healthy persons.\\

We also note a communication caveat that can happen even when using observational proportions. Some authors choose to normalize the confusion matrix and even to colorize it accordingly \citep[][...]{Richards2011, miettinen_protostellar_2018,Walmsley_2020}. It may help making the results more attractive and emphasizing some aspects of the results that are apparently easy to read. However, it implies to choose to normalize either on the lines or on the rows, respectively highlighting the recall or the precision value. This can subsequently hide the imbalance effect on either precision or recall. In the present work, we prefer to show the complete confusion matrix without color or arbitrary normalization to preserve their objectivity.\\

\begin{table}[t]
	\centering
	\caption{Imbalanced classification for a medical example using balanced proportions.}
	\vspace{-0.1cm}
	\begin{tabularx}{0.55\hsize}{r l |*{2}{m}| r }
	\multicolumn{2}{c}{}& \multicolumn{2}{c}{\textbf{Predicted}}&\\
	\cmidrule[\heavyrulewidth](lr){2-5}
	\parbox[l]{0.2cm}{\multirow{5}{*}{\rotatebox[origin=c]{90}{\textbf{Actual}}}} & Class & Positive & Negative & Recall \\
	\cmidrule(lr){2-5}
	 &  Positive    & 93     & 7       & 93.0\% \\
	 &  Negative   & 5      & 95     & 95.0\% \\
	\cmidrule(lr){2-5}
	 &  Precision & 94.9\% & 93.1\% & 94.0\%\\
	\cmidrule[\heavyrulewidth](lr){2-5}
	\end{tabularx}
	\vspace{-0.05cm}
	\label{balanced_test_confmat}
\end{table}

\begin{table}[t]
	\centering
	\caption{Imbalances classification for a medical example using imbalance proportion with similar recall for the two classes}
	\vspace{-0.1cm}
	\begin{tabularx}{0.55\hsize}{r l |*{2}{m}| r }
	\multicolumn{2}{c}{}& \multicolumn{2}{c}{\textbf{Predicted}}&\\
	\cmidrule[\heavyrulewidth](lr){2-5}
	\parbox[l]{0.2cm}{\multirow{5}{*}{\rotatebox[origin=c]{90}{\textbf{Actual}}}} & Class & Positive & Negative & Recall \\
	\cmidrule(lr){2-5}
	 &  Positive    & 93     & 7       & 93.0\% \\
	 &  Negative   & 46      & 954     & 95.4\% \\
	\cmidrule(lr){2-5}
	 &  Precision & 66.9\% & 99.27\% & 95.18\%\\
	\cmidrule[\heavyrulewidth](lr){2-5}
	\end{tabularx}
	\vspace{-0.05cm}
	\label{imbalanced_test_confmat}
\end{table}

\newpage
This example also highlights an interesting aspect of imbalanced classification, which is that to improve the overall quality of a rare class it can be necessary to further improve the recall of the dominant class. Using ANNs it implies that one might deliberately increase the proportion of negative cases in the training set in order to give the opportunity to the network to better constrain this case, increasing its recall. It does not have to be observational proportions because it depends on many factors, noticeably the expected results. We illustrate such a case with a classifier achieving a lesser recall of $87\%$ on the positive case and a much better recall of $99\%$ on the negative case. It results in the confusion matrix in Table \ref{imbalanced_test_confmat_improved} where the precision of the positive cases has increased to $89.7\%$ which is much closer to the positive recall of $87\%$.\\

This example raises an important point, which is the absence of a global quality estimator, since it depends on the end objective. As for any classification problem, one must choose the appropriate balance between reliability and completeness. Thus, in the last example we traded a recall drop by about $8\%$ for an improvement of $23\%$ in precision for the positive case. Therefore, acting on the training set proportions allows one to put the emphasis on certain quality indicators, while it remains necessary to test the results on observational proportions. We note that there are many tools that play on the balance of the datasets. For example, the so called "augmentation methods" can be used to artificially change the balance of the dataset by creating mock example that follows the same feature space distribution. Though, it would not improve the coverage of the class in the parameter space which is a strongly related issue. Also, there are often cases where all the classes do not have equivalent feature space coverage. In consequence some might be intrinsically more difficult to classify than others, which is completely overlooked when using balanced training proportions. These aspects are examined in detail in our main application of YSO classification in the dataset and result analyses in Sections~\ref{yso_datasets_tuning} to \ref{yso_discussion}.\\

\begin{table}[t]
	\centering
	\caption{Imbalanced classification for a medical example using imbalanced proportions with a better recall for the dominant class.}
	\vspace{-0.1cm}
	\begin{tabularx}{0.55\hsize}{r l |*{2}{m}| r }
	\multicolumn{2}{c}{}& \multicolumn{2}{c}{\textbf{Predicted}}&\\
	\cmidrule[\heavyrulewidth](lr){2-5}
	\parbox[l]{0.2cm}{\multirow{5}{*}{\rotatebox[origin=c]{90}{\textbf{Actual}}}} & Class & Positive & Negative & Recall \\
	\cmidrule(lr){2-5}
	 &  Positive   & 87     & 13       & 87.0\% \\
	 &  Negative   & 10     & 990     & 99.0\% \\
	\cmidrule(lr){2-5}
	 &  Precision & 89.7\% & 98.7\% & 97.9\%\\
	\cmidrule[\heavyrulewidth](lr){2-5}
	\end{tabularx}
	\vspace{-0.05cm}
	\label{imbalanced_test_confmat_improved}
\end{table}

	\newpage
\subsection{Simple examples}

In this section we present a few simple real examples. We pursue mainly two objectives: (i) provide additional insights in how a concrete problem can be expressed with the ANN formalism we described, using freely accessible data to ensure reproducibility, and (ii) to illustrate a few of the advanced effects we have described in the previous sections, like the use of some hyper-parameters. While numerous astrophysical examples could have been used, they usually employ large sets of parameters that need a careful preparation and would have required too much context introduction to be suitable as quick and simple examples.

\subsubsection{Regression}
\label{regression_expl}

We chose a regression that takes two inputs and also produces two outputs. The selected relation is fairly simple and expressed as:
\begin{equation}
o_1 = 0.7 \sin(x_1) + 0.6 \cos(0.6\,x_2) + (0.15\,x_1 + 0.1\,x_2)^2,
\end{equation}
\begin{equation}
o_2 = \sin(x_1) + \cos(0.8\,x_2) + 0.3(0.3\,x_1+ 0.6\,x_2),
\end{equation}
where $x_1$ and $x_2$ are the two input dimensions, and $o_1$ and $o_2$ the output target.\\

In this example, we constructed a grid of $x$ values ranging from $-5$ to $5$ with $60\times 60$ elements for each input. This is our full dataset. For the training set we selected only a grid of $20\times 20$ elements, which means that we have a training point for each 5 points of the original axis subdivision, for each axis. This translates into a training set size of $400$ couples of coordinates.\\

The corresponding network is inevitably constructed with 2 input nodes with the associated bias, one or several hidden layers, and 2 output neurons that are set to a linear activation, which is often much simpler for regression. We first decided to stick to a one hidden layer network with sigmoid-activated neurons. Regarding the fact that we have $400$ examples we first tried to be close to $40$ weights in the network, which is closely matched using $8$ neurons in the hidden layer since $3\times 8 + 9\times 2 = 42$ weights, accounting for the bias node in the input and hidden layers. In this example we used many of the previously described network optimizations. We chose to have a constant learning rate but with a momentum and the input dataset is normalized as we described in Section~\ref{input_norm}. The output is only divided by the maximum absolute value in the full dataset forcing it to be in the $-1$ to $1$ range, which works well for linearly activated output neurons. Finally the selected gradient descent scheme we used is the mini-batch one. Therefore, it gave us a list of hyperparameters to set. A quick manual exploration of these parameters made us choose $\eta = 0.004$ (considering that our updates are summed and not averaged in a minibatch), $\alpha = 0.8$ and a mini-batch size $b_s = 20$. We then trained the network for $4\times 10^5$ epochs while monitoring the error on the full $60\times 60$ grid.\\

\begin{figure}[!t]
\centering
\includegraphics[width=0.92\hsize]{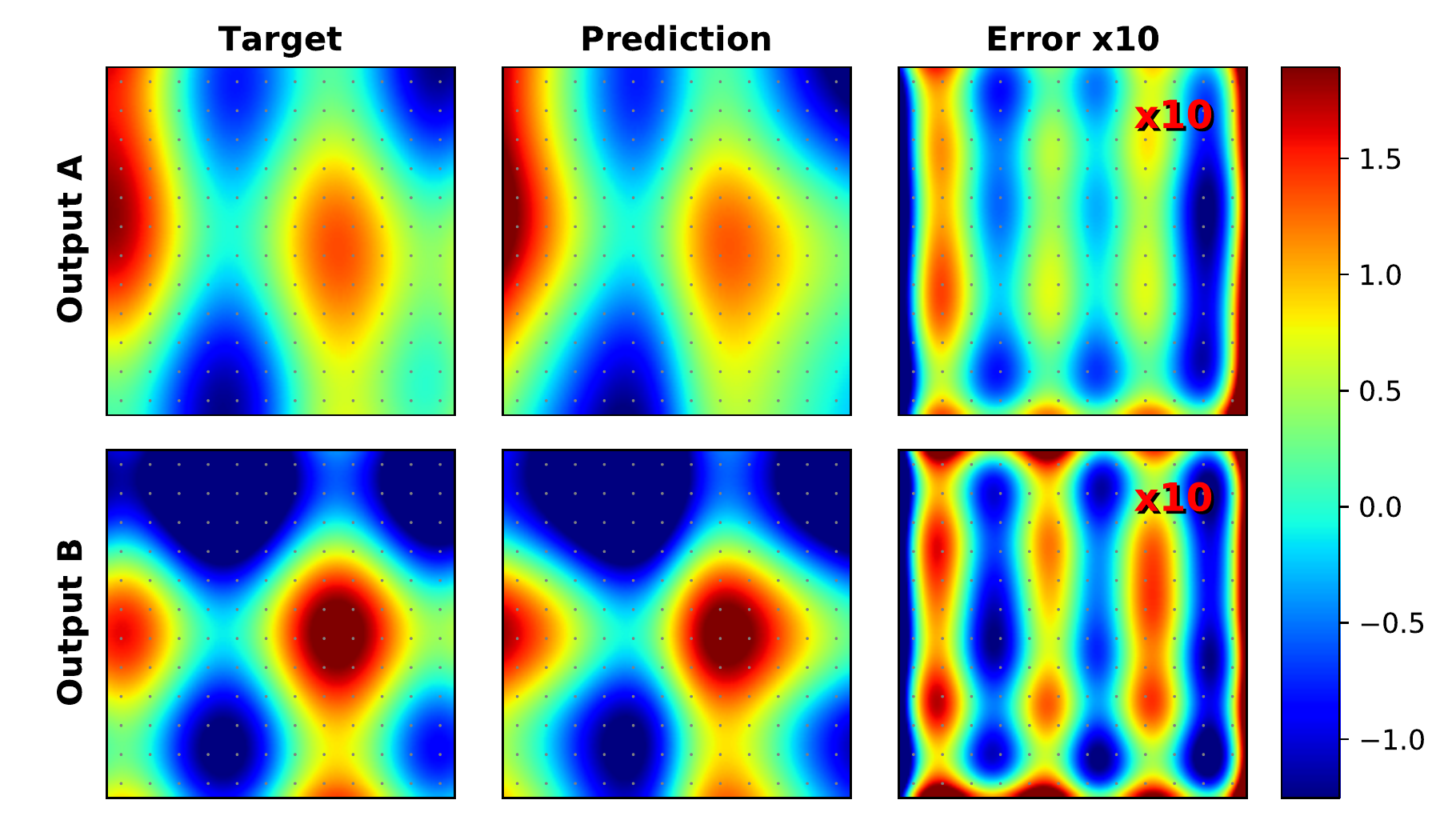}
\caption[Two dimensional regression example results.]{Two dimensional regression training results. Each frame represents the evolution of a quantity in the two dimension input space. {\it Rows} represent the two outputs of the network respectively. {\it Columns} are the target, prediction and the difference between the first two scaled by a factor of 10.}
\label{regre_expl2d}
\end{figure}

Figure~\ref{regre_expl2d} shows the results of this training. Because it is difficult to represent the evolution of the two outputs as a function of the two inputs in a single graph, we separated them into two rows. Three columns are used to represent the original function result, the prediction of the network and the last one shows 10 times the difference between the first two. The gray dots on the figures represent the training points. At first glance the prediction seems almost perfect, it requires a moment to spot the differences. The error frames strongly highlight that there is an edge effect, either with a stronger positive or negative difference that mostly depends on the surface slope at the edge. This is certainly due to the fact that it lacks constraints where there is less nearby learning points. Moreover, these error frames also show that the highest errors are in places where the surface curve dynamic is higher, and that it does not directly correlate with the places that have highest absolute values. Despite this, the error without the $\times 10$ factor looks really flat and the network reproduces nicely the shape of the function even with training points significantly distant from each other. \\

\begin{figure}[!t]
\centering
\includegraphics[width=0.58\hsize]{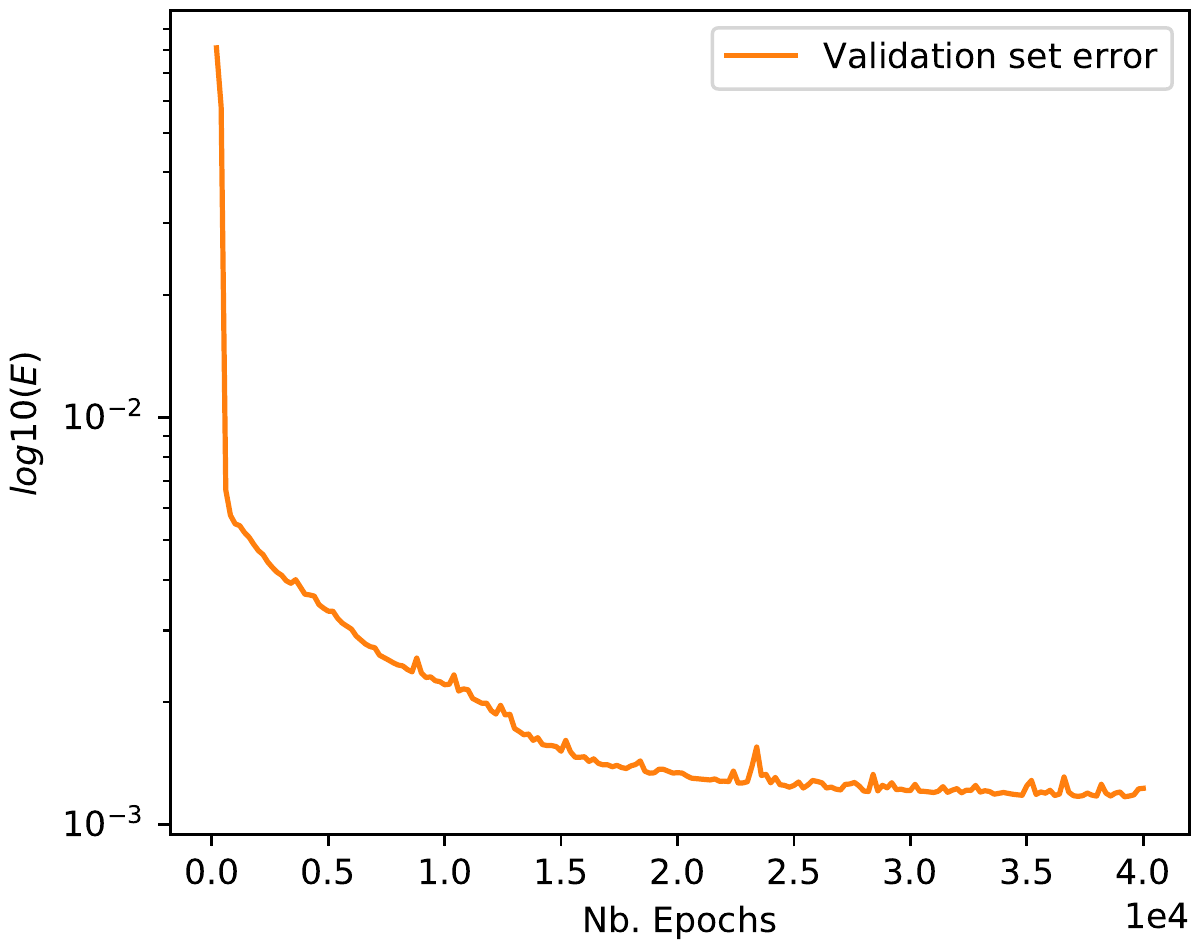}
\caption[Evolution of the error during training]{Evolution of the output layer error on the validation dataset during training as a function of the number of epochs.}
\label{regre_expl2d_error}
\end{figure}

\begin{figure}[!t]
\centering
\includegraphics[width=0.60\hsize]{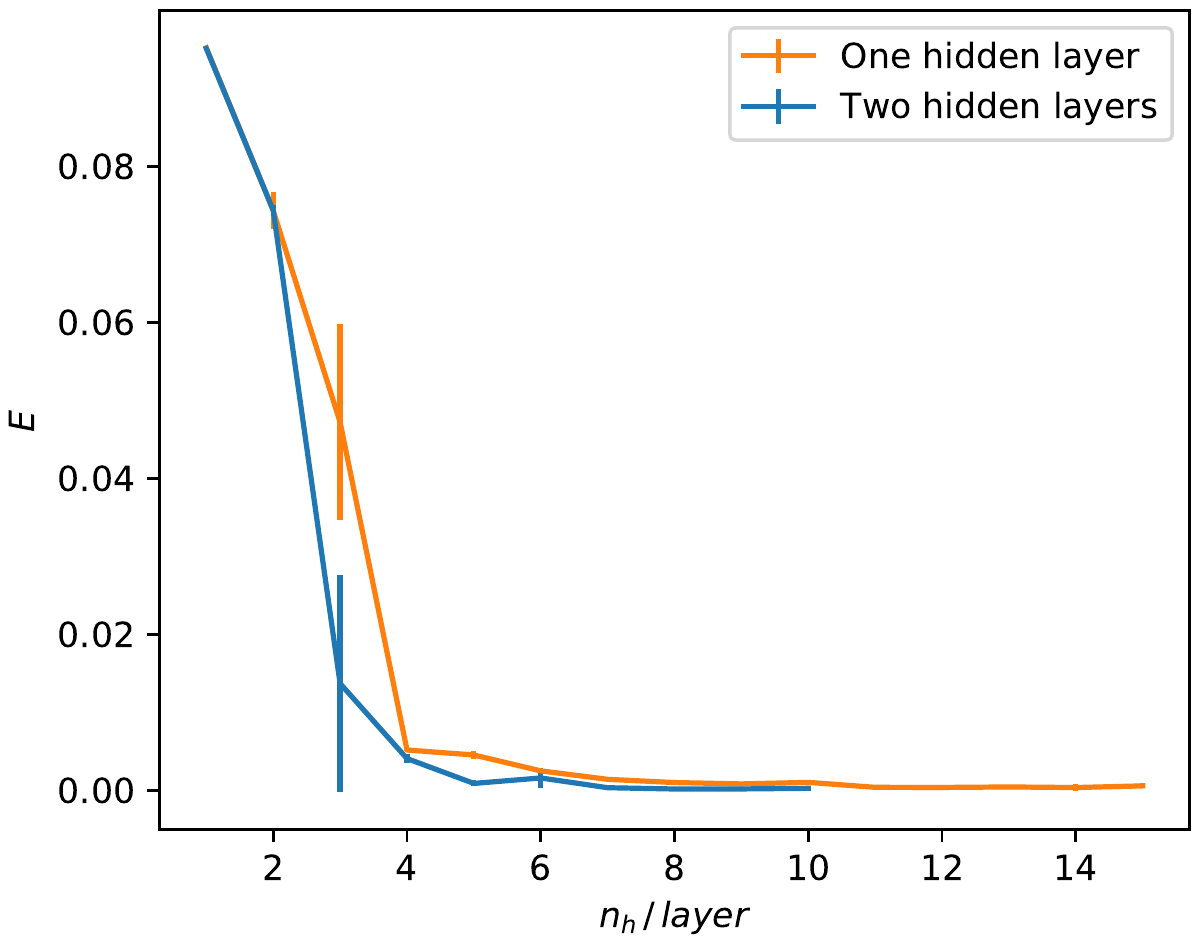}
\caption[Evolution of the error as a function of the number of neurons]{Evolution of the final output layer error on the validation dataset as a function of the number of neurons per layer. The orange and blue curves represent a one and two hidden layer network, respectively. The added uncertainty of each point is the standard deviation over few runs.}
\label{regr_expl_nb_neurons}
\end{figure}

\newpage
Figure~\ref{regre_expl2d_error} illustrates the output error value as a function of the number of epochs which shows that we are not overtraining. Nevertheless, we tested the effect of the number of neurons in this example as a complementary example of Section~\ref{nb_neurons}. Figure~\ref{regr_expl_nb_neurons} shows that, for one hidden layer, the optimal error is reached around $8$ neurons and additional neurons do not provide substantial improvement. We also tested a 2 hidden layer case, with this time the error as a function of the number of neurons per layer. It shows that, for this specific case, two hidden layers with 4 neurons are already very close to one hidden layer case with 8 neurons. We also note that, the two hidden layer solutions converged with less epochs. However, the comparison is very limited here since the limit of the number of examples is quickly reached.\\

This simple example illustrates how regression can be performed using the described model of ANNs. True useful problems can be more complicated, but they can usually be expressed in a very similar fashion as this example. One connected application is the modern use of ANNs as computation accelerators. If we imagine a heavy function that is used in a numerical application several times, there is a chance that this function could be reproduced up to a satisfying enough precision using ANN. Now, considering that modern ANNs use very simple activation functions and that they make use of powerful dedicated hardware, there is a strong chance that a pre-trained ANN could compute this approximated function much faster than it used to be done in the original application. This way ANNs can be used as computation accelerators even on well-known functions \citep{Baymani10, tompson16} or potential \citep[for example to replace a numerical solver for 3 body problems in][]{breen_2020}. This can be extended to training a network on a set of simulations that uses basic physics, from which it will extract higher level correlations, that can then be used to do new simulations quicker. The global application in Part III is a regression problem, still it requires more advance networks that are described in Section~\ref{cnn_global_section}.

\newpage
	\subsubsection{Classification}
	\label{classification_examples}
	
Prior to the YSO classification, we present simple examples to illustrate some effects that are important in the analysis of YSO classification. For this reason we present here two very simple common classification examples in order to focus on the technical aspects, both of which have their dataset freely accessible at the online \href{https://archive.ics.uci.edu/ml/index.php}{UCI Machine Learning repository}.\\

\textbf{A - Iris}\\

The first one is the Iris dataset, which is the most popular on the UCI repository. It consists of 3 types of Iris flowers, Setosa, Versicolour, and Virginica, based on 4 petal properties: sepal length, sepal width, petal length, petal width. The dataset contains 150 examples of these, perfectly balanced with 50 examples of each.\\

To perform the example in a proper way, we selected 35 examples of each class randomly to construct our training dataset (105), while keeping the rest, 15 examples of each class, to construct a merged validation-test dataset (45) due to the very small number of objects. The adopted network consists of 4 input nodes, one hidden layer with sigmoid activated neurons, and an output layer with 3 Softmax activated neurons. We identified that 6 hidden neurons were enough, corresponding to 51 weights, which is already above the standard recommendation. Less neurons provided less good results, while more neurons tended to strongly increase the instability of the network, and often led to a strong overtraining. The learning rate was set to $\eta = 5\times 10^{-4}$, the momentum to $\alpha = 0.8$, and we used mini-batches of size $b_s = 10$. This network was trained on 3200 epochs several times in order to characterize the result stability. Table~\ref{iris_confmat} shows the results of the network prediction for our test set. The very small number of objects in the test set makes it harder to interpret, because one misclassified object leads to a variation by several percents in the quality indicators. However, the network often succeeded in reaching a 100\% accuracy with this setting, and only misclassified three objects in two different classes in the worse case. Even if it is not a proper practice to look at the prediction for objects that were in the training sample, we still used the full dataset to ensure that the observed accuracy was not just a small number effect.\\

\begin{table}[t]
	\centering
	\caption{Iris classification, confusion matrix of the test set.}
	\vspace{-0.1cm}
	\begin{tabularx}{0.70\hsize}{r l |*{3}{m}| r }
	\multicolumn{2}{c}{}& \multicolumn{3}{c}{\textbf{Predicted}}&\\
	\cmidrule[\heavyrulewidth](lr){2-6}
	\parbox[l]{0.2cm}{\multirow{6}{*}{\rotatebox[origin=c]{90}{\textbf{Actual}}}} & Class & Setosa & Versicolour & Virginica & Recall \\
	\cmidrule(lr){2-6}
	 &  Setosa      & 15    & 0      & 0       & 100\% \\
	 &  Versicolor  & 0     & 14     & 1       & 93.3\% \\
	 &  Virgninica  & 0     & 1      & 14      & 93.3\% \\
	\cmidrule(lr){2-6}
	 &  Precision & 100\% & 93.3\% & 93.3\%& 95.6\% \\
	\cmidrule[\heavyrulewidth](lr){2-6}
	\end{tabularx}
	\vspace{-0.05cm}
	\label{iris_confmat}
\end{table}

It is noted by the authors of this dataset that one class is linearly separable from the two others which are not linearly separable from each other. This can be observed by using a very simple network with linear neurons as output and with no hidden layer. In this case, the degree of confusion between Versicolor and Virginica increased, while the Setosa remained well separated from the two others.\\

This example is also the occasion to illustrate the effect of probability membership. Looking at the maximum output probability distribution shows that almost all objects have a probability greater than $0.9$, and that there are only a few of them that are under this limit. As expected the objects that are properly classified (147) show a higher mean membership value, around $0.978 \pm 0.06$, than the mean value for misclassified objects (3), around $0.741 \pm 0.08$. However, one properly classified object has only a $0.52$ probability, while a misclassified one has a probability of $0.85$, which is also highlighted by the dispersion around the mean values. The underlying issue is that, even if an object is properly classified it can fall in a region of the feature space that is poorly constrained, and therefore less reliable. With applications that contain more objects, it is often a good choice to exclude less reliable ones by using a membership threshold, which often significantly improves the classification results (Sect.~\ref{proba_discussion}). \\

\textbf{B - Wines}\\

The second example is the Wine dataset, which is the third most popular from the UCI repository. It also consists in the separation of three classes that represent different unnamed vineyards. For this it provides a set of 13 features that correspond to many concentration measurements in the wine: Alcohol, Malic Acid, Ash, Magnesium, etc. The dataset contains 178 examples with classes being imbalanced with 59, 71, and 48 examples, respectively.\\

This example being imbalanced, we first extracted $\sim 28\%$ of objects in each class to construct our test/validation dataset. The purpose of such classifier being unclear, we assumed that it could be used to distinguish between the three vineyard on sold bottled and that the given proportions correspond to the true actual production proportion of each vineyard. Then, considering that they are in observational proportions, we tried to conserve them as much as possible in the test set. The adopted number of objects from each class in our test set are 16, 20, and 13, respectively, for a total of 49 objects. For the training dataset, we adopted the opposite approach by selecting the largest possible balanced dataset in the remaining data, which gave us 35 objects of each class for a training dataset with 105 objects. The adopted network is very close to the one of the previous example, with 13 input nodes, one hidden layer with sigmoid activated neurons, and an output layer with 3 Softmax activated neurons. We identified that 8 neurons are sufficient for this problem to reach its optimal prediction. Less neurons predicted less good results and more was not doing anything better since 8 neurons already led to an overtraining that must be monitored. As for the previous example, the learning rate was set to $\eta = 5 \times 10^{-4}$, the momentum to $\alpha = 0.8$, and we used mini-batches of size $b_s = 10$. This time, the network was trained several time on 4800 epochs. We observed that the network predictions were always above $93\%$ of global accuracy (3 objects misclassified) but that the results mostly depended on the random object selection for the training and test datasets. With the most appropriate selection the network reached $100\%$ prediction, which is maintained when forwarding the full dataset in the same network. Still, we present in Table~\ref{wines_confmat} an imperfect example for illustration purposes. Additionally, since the minimum error value depends on the data random selection in either the training or test dataset, we repeated 20 trainings and computed a mean minimum error averaged on the test dataset, of $0.086 \pm 0.062$. It illustrates the strong variability depending on the random selection of the training data over the dataset, and also that each individual neuron that contributes to this error is usually $\sim 0.03$ off the target value. This indicates that most objects will have a maximum prediction around $0.97$ which is guideline before the application of a threshold.\\

\begin{table}[t]
	\centering
	\caption{Wines classification, confusion matrix of the test set.}
	\vspace{-0.1cm}
	\begin{tabularx}{0.70\hsize}{r l |*{3}{m}| r }
	\multicolumn{2}{c}{}& \multicolumn{3}{c}{\textbf{Predicted}}&\\
	\cmidrule[\heavyrulewidth](lr){2-6}
	\parbox[l]{0.2cm}{\multirow{6}{*}{\rotatebox[origin=c]{90}{\textbf{Actual}}}} & Class & A & B & C & Recall \\
	\cmidrule(lr){2-6}
	 & A    & 16    & 0      & 0       & 100\% \\
	 & B    & 1     & 18     & 1       & 90.0\% \\
	 & C    & 0     & 1      & 12      & 92.3\% \\
	\cmidrule(lr){2-6}
	 &  Precision & 94.7\% & 94.7\% & 93.3\%& 93.9\% \\
	\cmidrule[\heavyrulewidth](lr){2-6}
	\end{tabularx}
	\vspace{-0.05cm}
	\label{wines_confmat}
\end{table}

Just like the previous example, it is very simple to reach $100\%$ each time with the exact same network when using the full dataset in just 1000 epochs. It only means that the classes are separable regardless of the complexity of their separation. However, as in the previous example, it means that the feature space is poorly sampled, which is expected using 13 dimensions with only $< 200$ objects. One approach to obtain better results out of less objects, would be to test if some input features are unnecessary to separate the classes. On the other hand, these imperfect results are another occasion to illustrate the membership probability. In this example, it is more visible that the misclassified objects are indeed predicted with a lesser probability. Taking an example with 3 misclassified objects in the full dataset, they have a mean membership value around $0.52 \pm 0.02$ while the properly classified ones have a mean of $0.97 \pm 0.06$. Using a threshold value of $0.9$ excludes 6 objects out of the 178 objects in the dataset, including all the misclassified ones. It is also interesting to note that some objects are more often misclassified than others, and also have a low probability when properly classified, making them probable outliers of the main distribution. The predicted output class of these objects, and therefore their membership probability, mostly depend on their random association to either the training or test dataset. Like in the previous example applying a membership threshold here would not significantly improve the results because they are already very good, but we efficiently used this approach on the YSO classification application with a clear success (Section~\ref{proba_discussion}).\\




\newpage
\section[Automatic identification of YSOs in infrared surveys]{Automatic identification of Young Stellar Objects in infrared surveys}
\label{yso_datasets_tuning}

In this section, we detail how we connected the general network presented in the previous sections with the YSO classification problem. We show how we arranged the data in a usable form for the network and describe the needed precautions for this process. We also explain how we defined the various datasets used to train our network.

\etocsettocstyle{\subsubsection*{\vspace{-1cm}}}{}
\localtableofcontents
\vspace{1cm}

\subsection{Problem description and class definition}
\label{data_prep}

As we exposed in Section~\ref{intro_yso}, our objective in this Part II is to design a methodology to perform YSO classification using ANN. A new classification scheme could be created using ML by mostly two approaches, (i) using a supervised ML algorithm that is trained on observed objects labeled without any ambiguity, for example YSOs for which the disk is directly observed, or by using simulated data (both discussed in Sect.~\ref{Method_discussion}), and (ii) using unsupervised learning to construct new, possibly more efficient, underlying classes (also in Sect.~\ref{Method_discussion}). These two approaches present major difficulties, due to the very small number of indisputably identified YSOs to train a ML algorithm for the first approach, and due to underlying feature space distribution properties of YSOs using infrared data for the second approach.\\


While these difficulties could be overcome, it would require a lot of work without guarantee of success. For this reason, as a first step, we set ourselves the objective to reproduce the classification by \citet{gutermuth_spitzer_2009} that makes use of Spitzer data, to evaluate the capacity of ML methods to perform such a task. It allowed us to assess the numerous elements reported here like the optimal number of data (Sects.~\ref{training_test_datasets}, \ref{NGC2264_results}), the minimal architecture requirement for an ANN method (Sect.~\ref{network_tuning}), the needed hardware resources (Sect.~\ref{network_tuning}), the degree of confusion between classes (Sects.~\ref{yso_results}, \ref{proba_discussion}), the effect of class imbalance (Sect.~\ref{training_test_datasets}), etc. But more importantly it can provide significant additions to the classification that is reproduced like identifying underconstrained feature space areas (Sects.~\ref{shocks_discussion}, \ref{proba_discussion}), generalizing the classification scheme to be used on very large catalogs efficiently (Sect.~\ref{yso_discussion}), or more importantly predicting a membership probability for the YSO candidates (Sect.~\ref{proba_discussion}).\\

Our choice of the \citet[][hereafter G09]{gutermuth_spitzer_2009} classification scheme to train from, is motivated by two main considerations: (i) the fact that we needed a method that works on large datasets in order to be used with a ML method, and (ii) our will to use a survey that is able to distinguish class I from class II YSOs.\\

\newpage
In the following section, we summarize the construction of the training sample using simplified version of this classification scheme. The G09 method combines data in the J, H, and K$_s$ bands from the Two Micron All Sky Survey \citep[2MASS, ][]{skrutskie_two_2006} and data between 3 and 24 $\mu$m from the Spitzer Space Telescope \citep{werner_spitzer_2004}. In our approach we did not use the 2MASS data and solely used Spitzer data in order to have a more homogeneous classification, which is discussed latter in the present section. We note that by using Spitzer data for YSO classification, rather than WISE data \citep[e.g][]{koenig_wide-field_2012}, we expect to cover only specific regions on the sky, but with a better sensitivity ($ \approx~1.6$ to $27\ \mathrm{\mu J}$ for the The Infrared Array Camera (IRAC) instrument) and spatial resolution ($1.2\arcsec$) than with WISE ($ \approx~80$ to $6000\ \mathrm{\mu J}$ and $ 6.1\arcsec$ to $12\arcsec$), the latter being used in \citet{marton_all-sky_2016} and \citet{marton_2019}. In the original G09 method, they performed the classification in several steps. In addition to Spitzer, they used 2MASS data as an additional step to refine the classification of some objects or to classify objects for which Spitzer bands are missing. Therefore, restricting our analysis to Spitzer data still allowed a reasonable classification. In our adapted method, we started with the four IRAC bands, at $3.6,\ 4.5,\ 5.8$ and $8\ \mathrm{\mu m}$, applying a preselection that kept only the sources with a detection in the four bands and with uncertainties $\sigma < 0.2$ mag, like in the original classification. Then this first classification is refined using the Multiband Imaging Photometer (MIPS) $24\ \mathrm{\mu m}$ band. With this classification, it is possible to identify candidate contaminants and to recover YSO candidates at the end. Similarly to G09, we used the YSO classes described in Section~\ref{intro_yso}.\\

\begin{figure*}[!t]
	\centering
	\begin{subfigure}[t]{0.43\textwidth}
	\includegraphics[width=\textwidth]{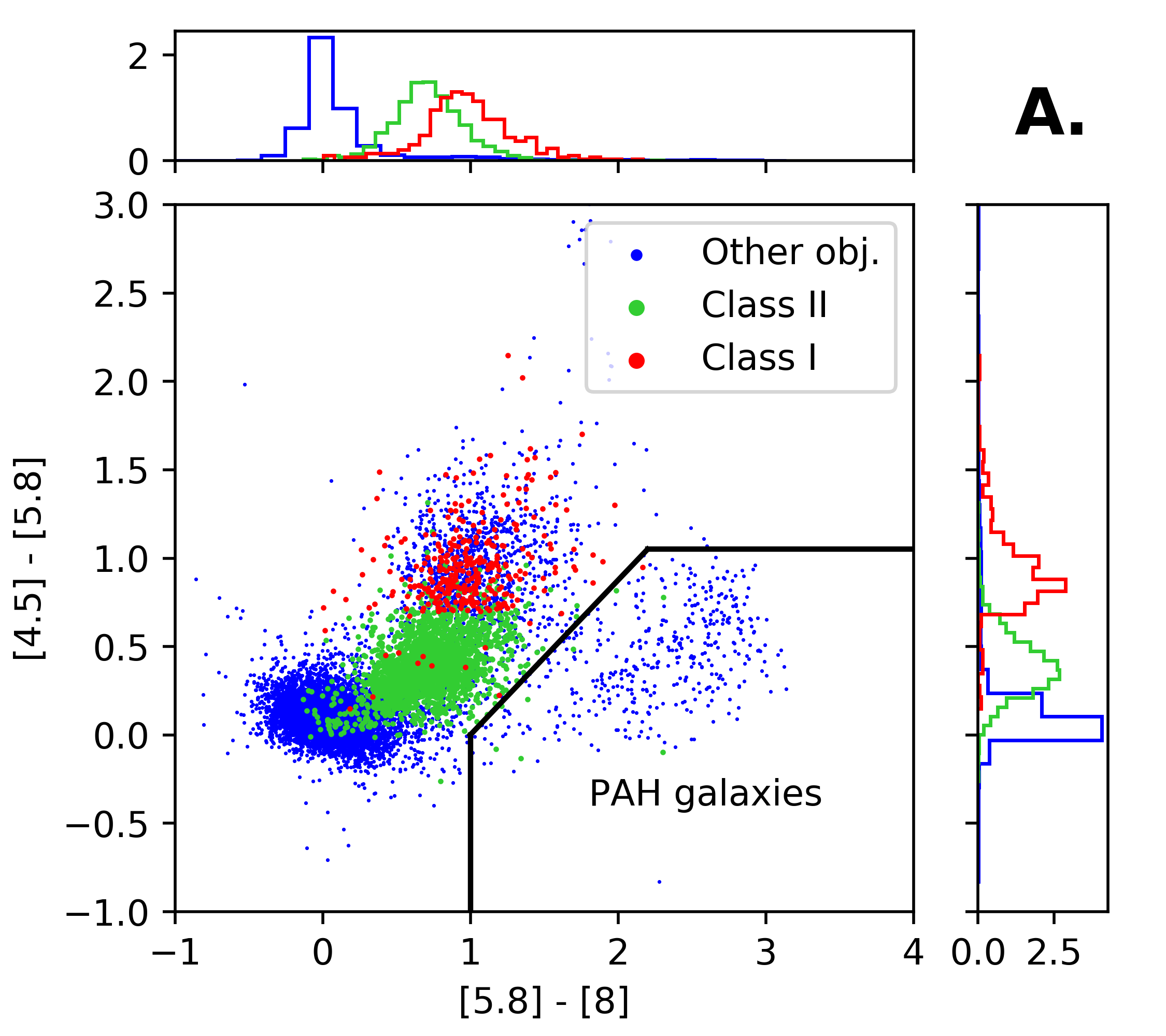}
	\end{subfigure}
	\begin{subfigure}[t]{0.43\textwidth}
	\includegraphics[width=\textwidth]{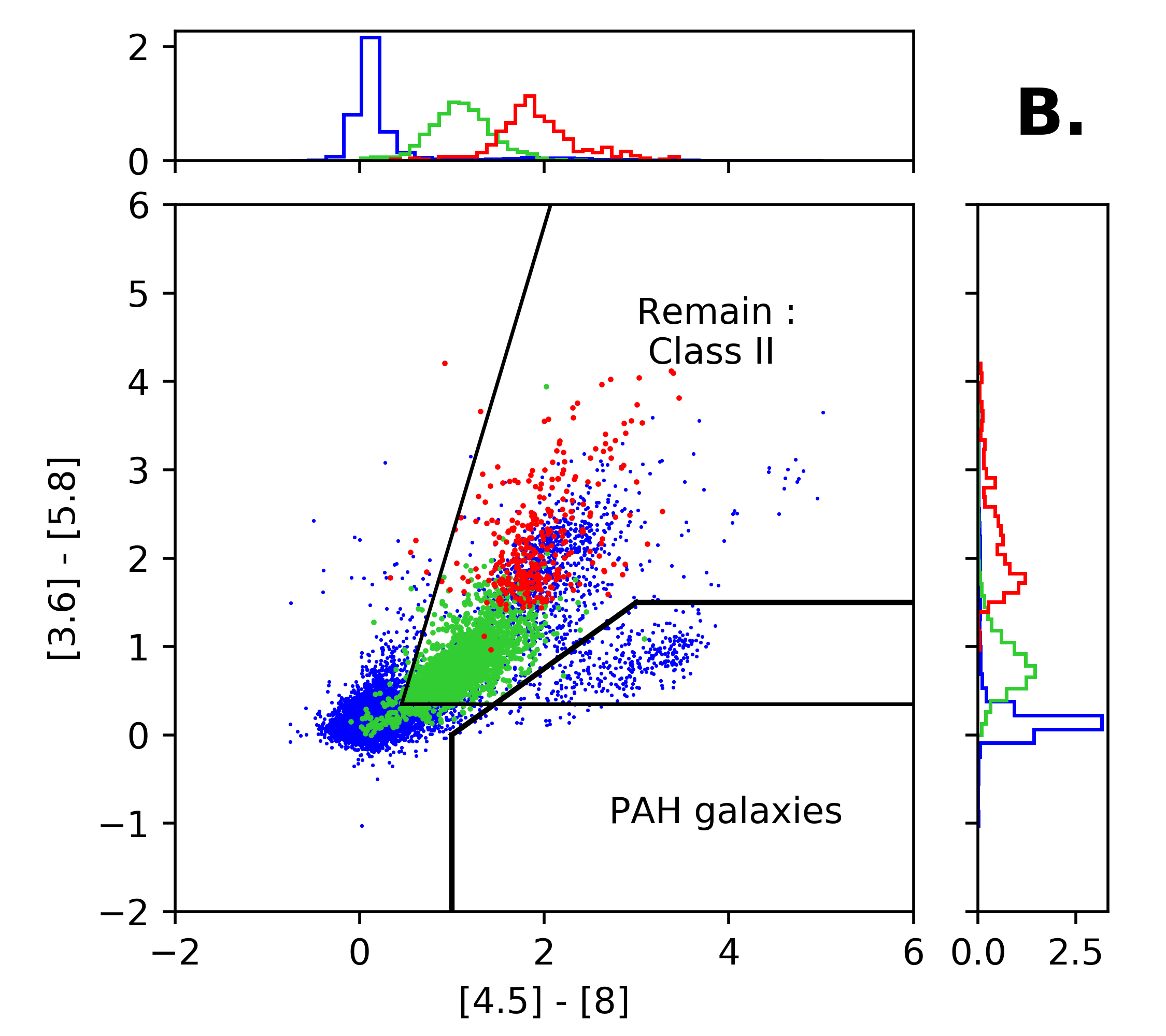}
	\end{subfigure}\\
	\begin{subfigure}[t]{0.43\textwidth}
	\includegraphics[width=\textwidth]{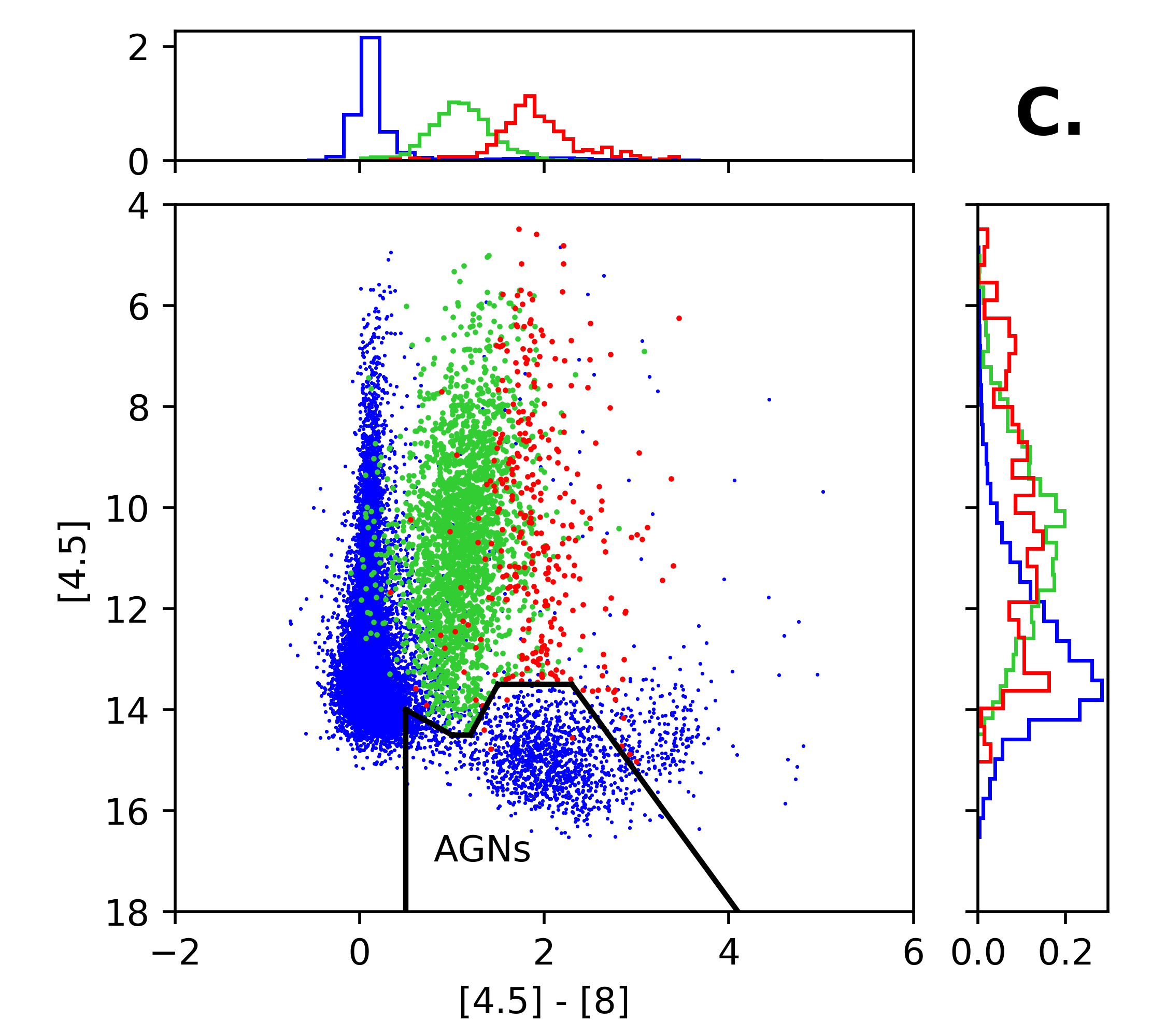}
	\end{subfigure}
	\begin{subfigure}[t]{0.43\textwidth}
	\includegraphics[width=\textwidth]{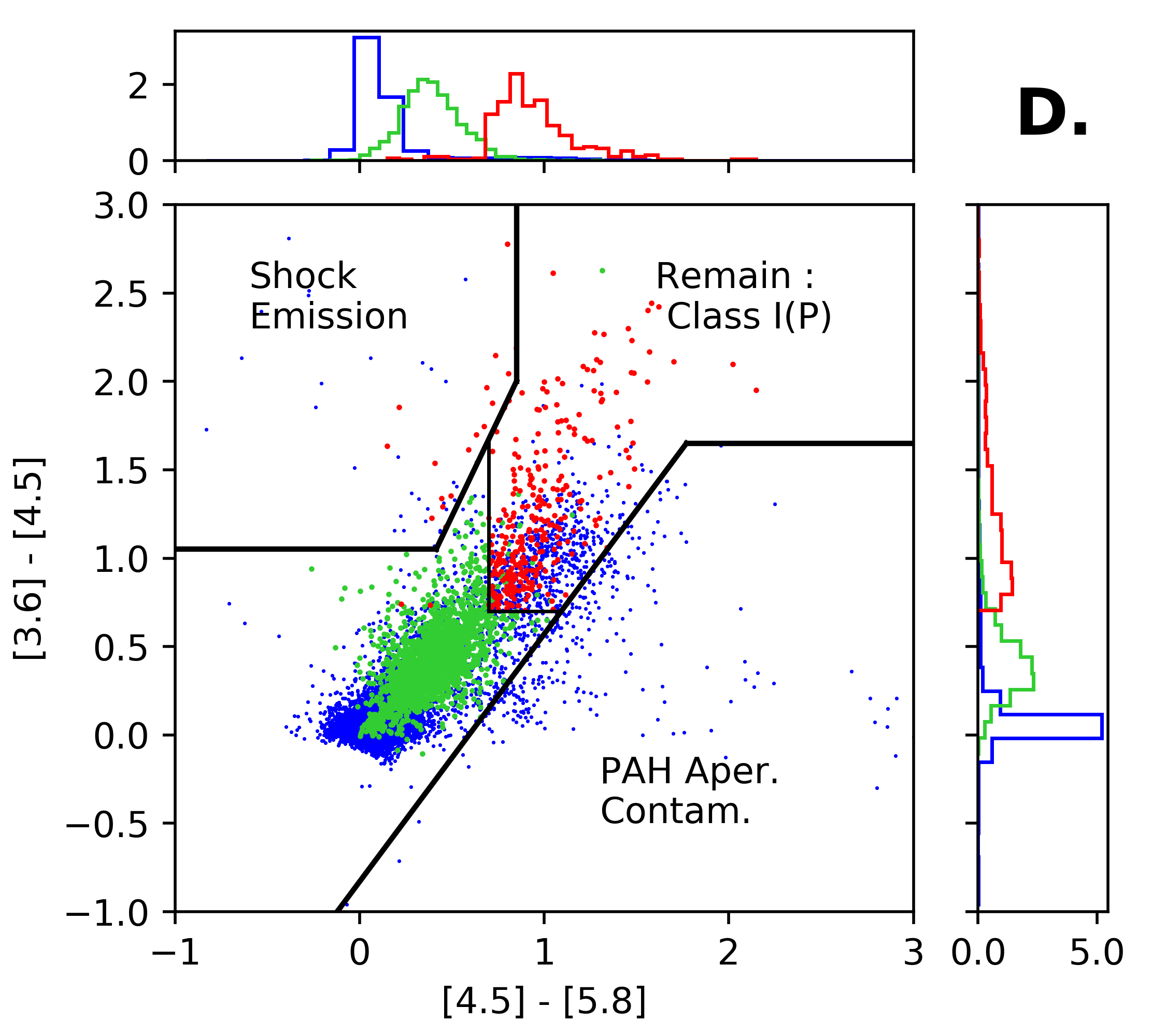}
	\end{subfigure}\\
	\begin{subfigure}[t]{0.43\textwidth}
	\includegraphics[width=\textwidth]{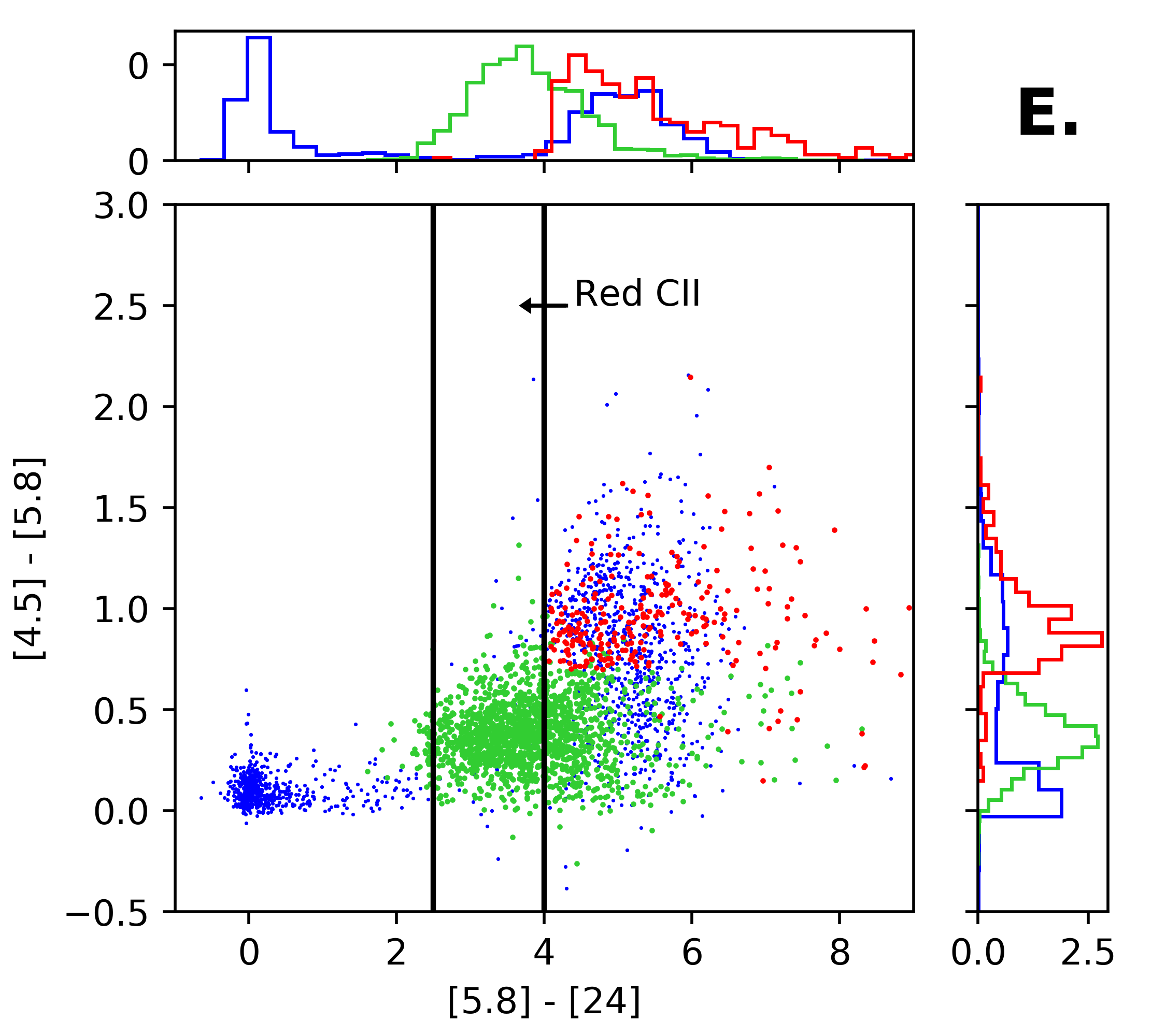}
	\end{subfigure}
	\caption[Selection of CMD diagrams from our simplified G09]{Selection of color-color and color-magnitude diagrams from our simplified multi-step G09 classification. The data used in this figure correspond to the Orion labeled dataset in Table~\ref{tab_selection}. The contaminants, CII YSOs, and CI YSOs are shown in blue, green, and red, respectively. They were plotted in that order and partly screen one another, as revealed by the histograms in the side frames. The area of each histogram is normalized to one. In frame A, some PAH galaxies are excluded. In frame B, leftover PAH galaxies are excluded based on another criteria. It also shows the criteria of class II extraction that is in a later step. In frame C, AGN are excluded. In frame D, Shocks and PAH aperture contaminants are excluded. It also show the last criteria of class I extraction. In frame E, one of the criteria from the MIPS $24\ \mathrm{\mu m}$ band is shown, that identifies reddened Class~II in the previously classified Class~I.}
	\label{fig_gut_method} 
\end{figure*}

Using mainly IRAC data prevented us from identifying class 0 objects, since they do not emit in the IRAC wavelength range (Fig.~\ref{yso_sed}). Similarly, because of Spitzer uncertainties, the class III objects are too similar to main sequence stars to be distinguished. For these reasons, we limited our objectives to the identification of CI and CII YSOs. We then proceeded to the so-called "phase~1" from G09 (their Appendix~A) to successively extract different contaminants using specified straight cuts into color-color and color-magnitude diagrams (CMDs) along with their respective uncertainties. This step enabled us to exclude star-forming galaxies, active galactic nuclei (AGN), shock emission, and resolved PAH emission. It ends by extracting the class I YSO candidates from the leftovers, and then again extracting the remaining class II YSO candidates, from more evolved stars. The cuts used on these steps are shown in Fig.~\ref{fig_gut_method}, with the final CI and CII YSO candidates from the Orion region (Sect.~\ref{data_setup}).\\

For the sake of simplicity, we adopted only 3 categories: CI YSOs, CII YSOs, and contaminants, which we also refer to as "others" in our tables. Doing so forced the network to focus on the separation of the contaminant class from the YSOs, rather than between different contamination classes. It can be seen as a simplification of the underlying classification if some subclasses of contaminants are close to each other in the feature space, reducing the required number of weights in the network. Therefore, we defined the output layer of our network with 3 neurons using a Softmax activation function, meaning that there was one neuron per class that returned a membership probability as described in Section~\ref{proba_class_intro}.\\

We chose not to use 2MASS, and therefore skipped the so-called "phase 2" of the G09 classification scheme. This is mainly motivated by the fact that it creates an artificial difference in degree of refinement between objects that have a 2MASS detection and objects that does not. Additionally, it creates several more classification paths that only contain very few objects, which is much more difficult to constrain (highlited at several places in Sects. ~\ref{yso_results} and ~\ref{yso_discussion}). However, G09 also proposed a "phase 3" that uses the MIPS $24\,\mu$m band, and which might be useful for our classification. In this last phase, some objects that were misclassified in the previous two phases are rebranded. Although this can raise difficulties, as discussed in Section \ref{sec:mips24}, we used it in our analysis because it relies only on Spitzer data.
Since MIPS $24\,\mu$m data are only used to refine the classification, we did not exclude objects without detection in this band. We only used it in phase 3 when it had an uncertainty $\sigma_{24} < 0.2$ mag. This additional phase ensured that the features identified in the SED with the four IRAC bands were consistent with longer wavelength data. It allowed: (i) to test the presence of a transition disk emission in objects classified as field stars, rebranding them as class II, (ii) to test the presence of a strong excess in this longer wavelength that is characteristic of deeply embedded class I objects, potentially misclassified as AGNs or Shocks,  (iii) to refine the distinction between class I and II by testing whether the SED still rises at wavelengths longer than $8\,\mu$m for class I, otherwise rebranding them as reddened class II. Those refinements explain the presence of objects beyond the boundaries in almost all frames in Fig.~\ref{fig_gut_method}. For example in frame B, some class II objects, shown in green, are located behind the boundary at the bottom-left part in a region dominated by more evolved field stars. In this figure, all the steps of this refinement are not shown, only the criterion on reddened class II identification is illustrated in frame E. Our adapted classification scheme was therefore composed of five bands (4 IRAC, 1 MIPS).\\
\vspace{-0.2cm}

Finally, we stress that the G09 classification uses all the band uncertainties ($[\sigma_{3.6}]$, $[\sigma_{4.5}]$, $[\sigma_{5.8}]$, $[\sigma_8]$, $[\sigma_{24}]$) to construct complementary conditions on the separation between stellar classes in all the phases. As an example, objects that satisfy the CI YSO condition but that have a high uncertainty in the bands used in this condition will be classified as CII YSOs. These uncertainties are therefore to be considered as direct input features of the classification. While such measurements could be used in a more complex way with modern ANN architectures, the present study focuses on using solely the G09 method as training object construction, forcing us to use band uncertainties as required input features. In summary, our labeled dataset was structured as a list of (input, target) pairs, one per point source, where the input was a vector with 10 values ($[3.6]$, $[\sigma_{3.6}]$, $[4.5]$, $[\sigma_{4.5}]$, $[5.8]$, $[\sigma_{5.8}]$, $[8]$, $[\sigma_8]$, $[24]$, $[\sigma_{24}]$), and the target was a vector of 3 values ($P(\text{CI}), P(\text{CII}), P(\text{Contaminant})$). The input features are normalized over the considered labeled dataset as described in Section~\ref{input_norm}. Here $P()$ denotes the membership probability normalized over the three neurons of the output layer. An alternative choice of input space is discussed in Sect.~\ref{sec:color_usage}.\vspace{-0.2cm}

\subsection{Labeled datasets in Orion, NGC 2264, 1\,kpc and combinations}
\label{data_setup}

We chose to use well-known and well constrained star forming regions, where YSO classification was already performed using Spitzer data. Although we employed an approach made of progressive steps by firstly drawing conclusions on one case before going to the next one, we summarize here all the datasets that were used. Therefore, in the following section we detail how we created the corresponding labeled dataset and what was the optimal proportions along with the network parameters. Additional information on how these parameters were found for each individual case are discussed in Section~\ref{yso_results}. This organization allows us to group information to ease the comparison and to better reflect the global approach while summarizing the dataset construction in one place.\\
\vspace{-0.2cm}

\begin{table*}[t]
\vspace{-0.5cm}
	\hspace{-1.2cm}
	\begin{minipage}{1.15\hsize}
	\footnotesize
	\centering
	\vspace{0.2cm}
	\caption{Results of our simplified G09 method for our various datasets.}
	\label{tab_selection}
\vspace{-0.2cm}
	\begin{tabularx}{1.0\hsize}{ l *{2}{x{0.06\hsize}} @{\hskip 0.065\hsize} *{5}{Y} @{\hskip 0.065\hsize} *{3}{x{0.075\hsize}}}
	\toprule
	\toprule
	\vspace{-0.3cm}\\
	\multirow{2}{*}{Dataset} & \multicolumn{2}{c}{\hspace{-0.6cm}Pre-selection} & \multicolumn{5}{c}{\hspace{-0.6cm}Detailed contaminants} & \multicolumn{3}{c}{\hspace{-0.6cm}Labeled classes}\\
	\cmidrule(l){2-11}
	& Total & Selected & Gal. & AGNs & Shocks & PAHs & Stars & CI YSOs & CII YSOs & Others\\
	\vspace{-0.3cm}\\
	\midrule
	\vspace{-0.15cm}\\
	Orion & 298405 & 19114 & 407 & 1141 & 28 & 87 & 14903 & 324 & 2224 & 16566\\
	\vspace{-0.15cm}\\
	NGC 2264 & 10454 & 7789 & 114 & 250 & 6 & 1 & 6893 & 90 & 435 & 7264\\
	\vspace{-0.15cm}\\
	Combined & 308859 & 26903 & 521 & 1391 & 34 & 88 & 21796 & 414 & 2659 & 23830\\
	\vspace{-0.15cm}\\
	1\,kpc* & 2548 & 2171 & 1 & 57 & 0 & 1 & 3 & 370 & 1735 & 67\\
	\vspace{-0.15cm}\\
	Full 1\,kpc & 311407 & 29074 & 522 & 1448 & 34 & 89 & 21799 & 784 & 4396 & 23897\\
	\vspace{-0.2cm}\\
	\bottomrule
	\bottomrule
	\end{tabularx}
	\end{minipage}
	\caption*{\vspace{-0.4cm}\\ {\bf Notes.} The third group of columns gives the labels used in the learning phase. The last column is the sum of the columns in the "Detailed contaminants" group.
	*The 1\,kpc sample contains only pre-identified YSOs candidates. Still, we classified some of them as contaminants because of the simplifications in our method.\vspace{-0.5cm}}
\end{table*}

\begin{figure}[!t]
\centering
\includegraphics[height=0.58\hsize]{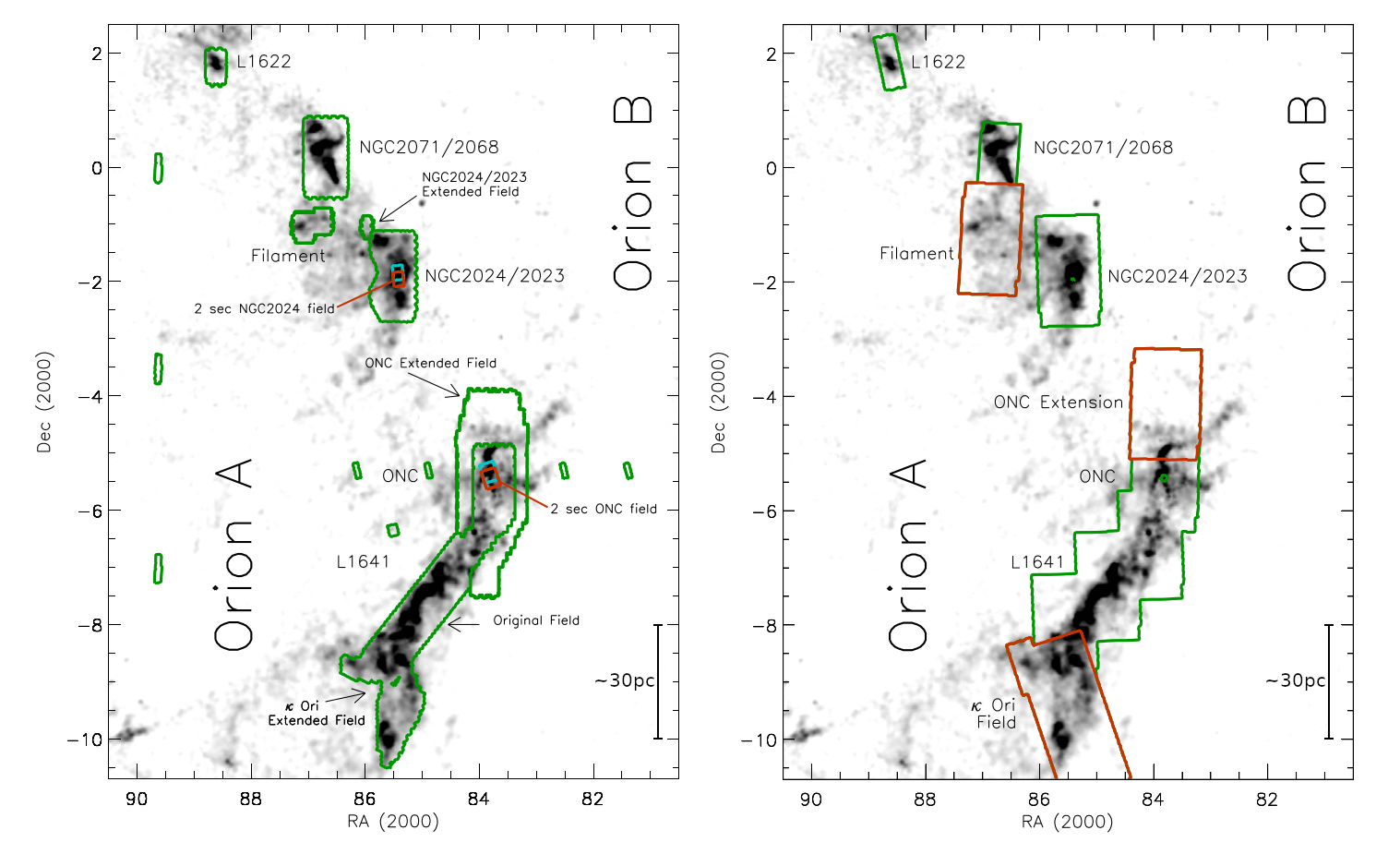}
\caption[Spitzer Coverage of the Orion cloud]{Spitzer coverage of the Orion cloud. The grayscale is an extinction map of the region obtained using the 2MASS point source catalog. {\it Left}: in green, the fields surveyed with IRAC. {\it Right}: in green and red, the fields surveyed with MIPS. {\it Adapted from} \citet{gutermuth_spitzer_2009}.}
\label{megeath_orion_cover}
\end{figure}

\begin{figure}[!t]
\centering
\includegraphics[height=0.58\hsize]{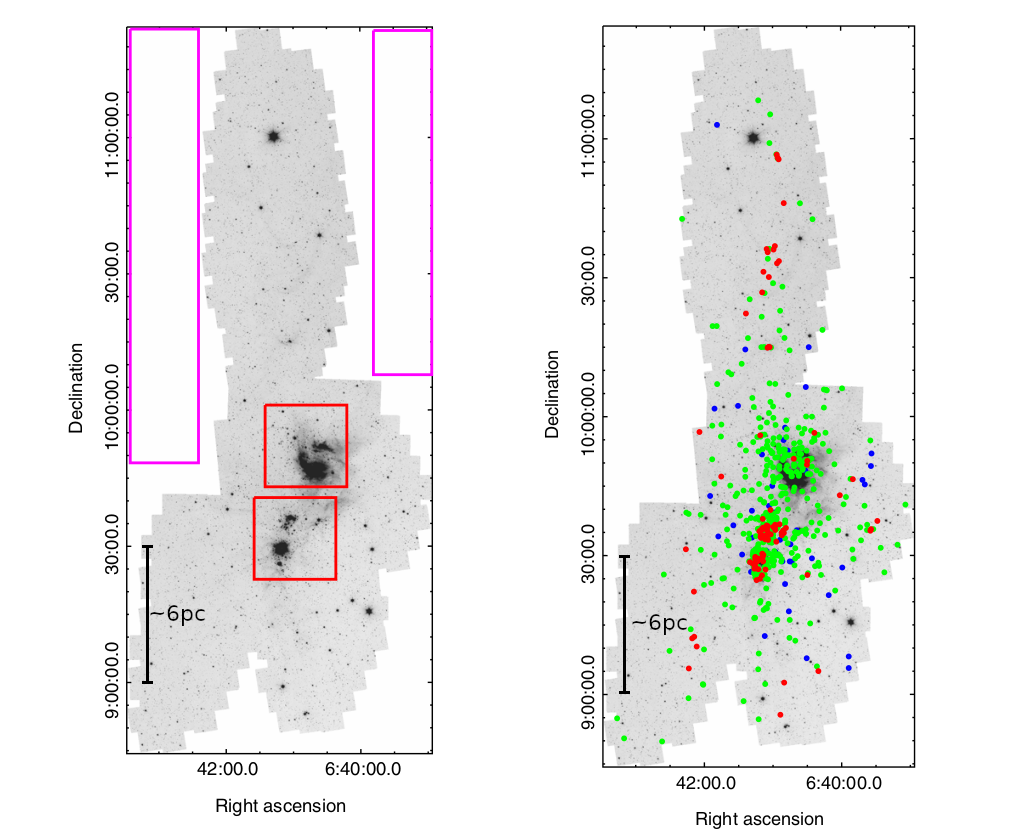}
\caption[Spitzer Coverage of Mon OB1. East]{Spitzer Coverage of Mon OB1 East. The grayscale is the inversion of the observed 4.5 $\mathrm{\mu m}$ band. {\it Left}: the red boxes define NGC 2264. The magenta boxes were used to estimate the field stars amount in the original study. {\it Right}: retrieved class 0/I YSOs are in red, class II YSOs are in green and "transition disks" are in blue, from the original study. {\it Adapted from} \citet{rapson_spitzer_2014}.}
\label{rapson_ngc2264_cover}
\end{figure}

\begin{figure*}[!t]
	\centering
	\begin{subfigure}[t]{0.49\textwidth}
	\caption*{\bf Orion}
	\includegraphics[width=\textwidth]{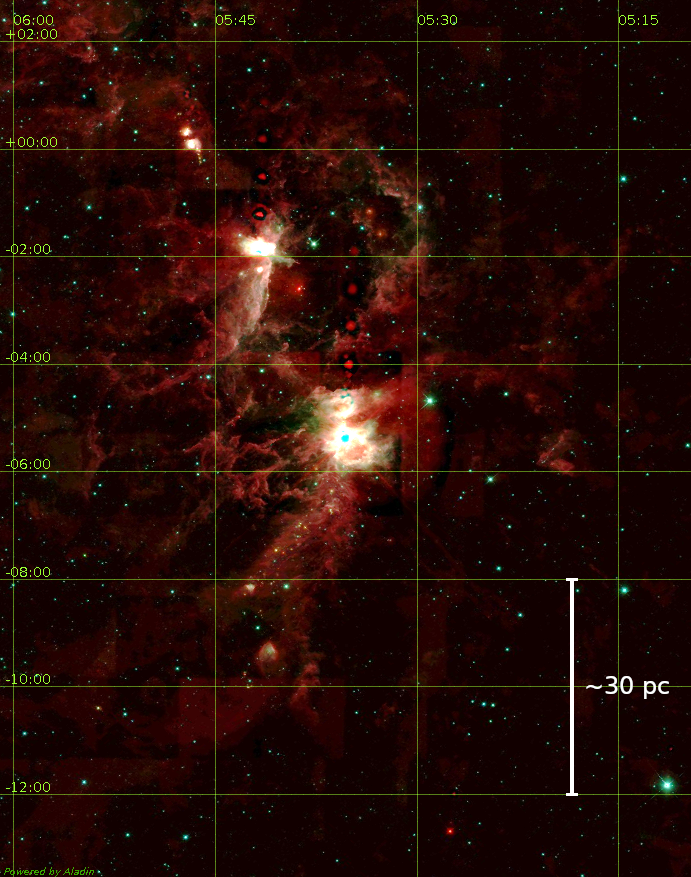}
	\end{subfigure}
	\begin{subfigure}[t]{0.49\textwidth}
	\caption*{\bf NGC 2264}
	\includegraphics[width=\textwidth]{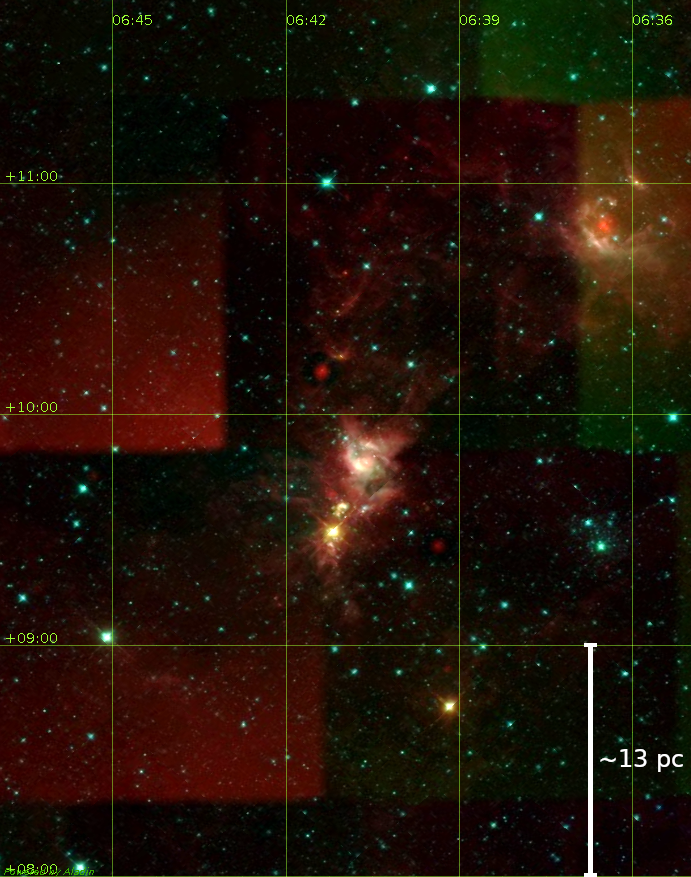}
	\end{subfigure}\\
	\caption[Orion and NGC 2264 as seen by AllWISE]{Orion and NGC 2264 as observed using the colored AllWISE data (Red, Green, Blue correspond to W4, W2, W1, respectively).}
	\label{orion_and_2264_wise} 
\end{figure*}

\begin{figure*}[!t]
\vspace{0.5cm}
\hspace{-0.7cm}
	\begin{minipage}{1.15\hsize}
	\centering
	\includegraphics[width=\textwidth]{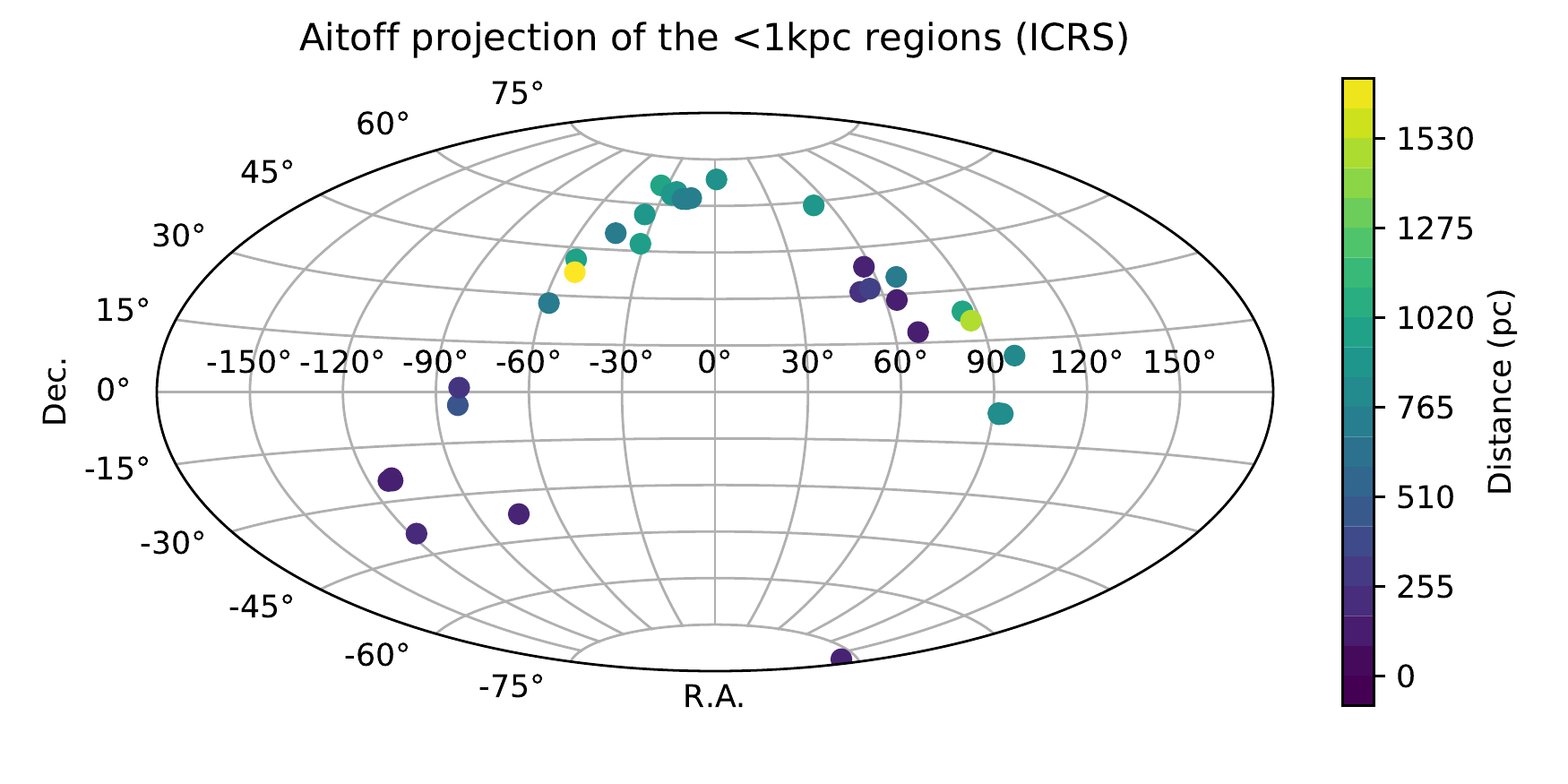}
	\end{minipage}
	\caption[Distribution of the < 1\,kpc regions]{Distribution of the < 1\,kpc regions from the corresponding catalog on the sky using an Aitoff projection with ICRS frame. The colors indicate the estimated distance of the regions, based on \citet{gutermuth_spitzer_2009}}
	\label{gutermuth_1kpc_dist} 
\end{figure*}

The main idea was to test the learning process on individual regions, and then compare it with various combinations of these regions. It is expected that, due to the increased diversity in the training set, the combination of regions should improve the generalization capacity of trained network and allows predictions on new regions. We selected regions analyzed in three studies, all using the original G09 method. However, some differences remain between the parameters adopted by the authors (e.g. the uncertainty cuts). Using our simplified G09 method, as presented in Section \ref{data_prep}, allowed us not only to base our study solely on Spitzer data, but also to build a homogeneous dataset with the exact same criteria for all regions. We present here the three selected regions and the corresponding catalogs:\vspace{-0.1cm}

\begin{itemize}[leftmargin=0.0cm]
\setlength\itemsep{0.2em}

\item[] {\hspace{0.3cm}\bf $\bullet \,$ Orion}: The first region we used was the Orion molecular cloud with the dataset from \citet{megeath_spitzer_2012} (Figs.~\ref{megeath_orion_cover} and \ref{orion_and_2264_wise}). This is the most studied star forming region, due to its relative proximity ($\sim 420$ pc), and because of its large mass (above $10^5 \mathrm{M_\odot}$) and size (more than $\sim 50 deg^2$ on the sky plane). The presence of several young stellar clusters, including the massive Trapezium cluster, makes it a bright target across the whole electromagnetic spectra and therefore an ideal target for most interstellar medium topics (for example, in massive star formation: \citet{Rivilla_2013}, in low-mass star formation: \citet{Nutter_2007}, in filament dynamics: \citet{Stutz_2016}, in photodissociation regions: \citet{Goicoechea_2016}, in astrochemistry: \citet{Crockett_2014}, etc.). It is composed of several parts in diverse evolutionary stages: Orion A is the most actively star forming part with a complex star formation history with various episodes from 12 Myr ago to this day \citep{Brown_94}; Orion B is in an earlier evolutionary stage, and is mostly quiescent in spite of a few well-known reflection nebulea \citep{Pety_2017}; and the $\Lambda$ Orionis shell is a spherical or toroidal structure shaped by the massive O-type star $\Lambda$ Orionis and a past supernova explosion \citep{Dolan_2001}, where the net effect of star formation feedback is debated \citep[][and reference therein]{Yi_2018}. The corresponding catalog covers only Orion A and B and contains all the elements we needed with the four IRAC bands and the MIPS $24\ \mu$m band and relies on the G09 method. The authors provide the full point source catalog they used to perform their YSO candidate extraction. This is one of the most important element in our study, since the network needs to see both the YSOs and the other types of objects to be able to learn the differences between them.

\item[] {\hspace{0.3cm}\bf $\bullet \,$ NGC 2264}: For the second dataset, we used the catalog by \citet{rapson_spitzer_2014} (Figs.~\ref{rapson_ngc2264_cover} and \ref{orion_and_2264_wise}), who analyzed Spitzer observations of NGC 2264 in the Mon OB1 region using the same classification scheme. This is also one of the largest star-forming region in the solar neighborhood \citep[$\sim 4 \times 10^4\,\mathrm{M_\odot}$, with an extent of $\sim 50$ pc, ][and references therein]{montillaud_2019_II} while being a bit more distant from us $\sim$ 723 pc \citep{Cantat-Gaudin_2018}. This region has sustained an active star formation for at least the last 5 Myr near the center of NGC 2264 \citep{Dahm_2005} that has occurred sequentially \citep{Buckner_2020}, and a secondary convergence center seems to be currently forming in the northern edge of the cluster \citep[around $\delta=10^\circ30'$][also visible in Fig.~\ref{rapson_ngc2264_cover} as a group of Class0/I YSOs]{montillaud_2019_II}. We note that, in contrast to the Orion dataset, this one does not provide the full point-source catalog, but a preprocessed object list compiled after performing band selection and magnitude uncertainty cuts. However, it should not affect the selection, since we used the exact same uncertainty cuts as them. 

\item[] {\hspace{0.3cm}\bf $\bullet \,$ Combined}: We then defined a dataset that is the combination of the previous two catalogs, which we call the "combined" dataset. We used it to test the impact of combining different star-forming regions in the training process, because distance, environment, and star formation history can impact the statistical distributions of YSOs in CMDs. 

\item[] {\hspace{0.3cm}\bf $\bullet \,$ Full 1\,kpc}: We pushed this idea further by defining an additional "1\,kpc" catalog, directly from \citet{gutermuth_spitzer_2009} (Fig.~\ref{gutermuth_1kpc_dist}). It contains a census of the brightest star forming regions closer than 1\,kpc, excluding both Orion and NGC 2264. However, this catalog only contains the extracted YSO candidates and not the original point source catalog with the corresponding contaminants. This is an important drawback, since it cannot be used to add diversity information in this category. Yet, it can be used to increase the number of class I and II and increase their respective diversity. We refer to the dataset that combines the three previous datasets Orion, NGC 2264 and 1\,kpc, as the "full 1\,kpc" dataset.
\end{itemize}

\vspace{-0.2cm}This first classification provided various labeled datasets, that were used as targets for the learning process. The detailed distribution of the resulting classes for all our datasets is presented in a common table (Table~\ref{tab_selection}), in order to ease their comparison. This table also shows the subclass distribution of the contaminants, as obtained with our modified G09 methods.\vspace{-0.2cm}\\

We examine here the discrepancies between our results and those provided in the respective publications. In the case of Orion, merging their various subclasses, \citet{megeath_spitzer_2012} found 488 class I and 2991 class II, no details being provided for the distribution of contaminant subclasses. This is consistent with our simplified G09 method, considering that the absence of the 2MASS phase prevented us from recovering objects that lacked detection in some IRAC bands, and that the authors also applied additional customized YSO identification steps. For the NGC 2264 region, \citet{rapson_spitzer_2014} report 308 sources that present an IR excess, merging class 0/I, II and transition disks. However, they used more conservative criteria than in the G09 method to further limit the contamination, which partly explains why our sample of YSOs is larger in this region. The authors do not provide all the intermediate numbers, but they state that they excluded 5952 contaminant stars from the Mon OB1 region, a number roughly consistent with our own estimate (6893). Finally, the 1\,kpc dataset only contains class I and II objects, which means that every object that we classified as contaminant is a direct discrepancy between the two classifications. This is again due to the absence of some refinement steps in our simplified G09 method. For the complete catalog containing 36 individual star-forming regions, \citet{gutermuth_spitzer_2009} report 472 class I and 2076 class II extracted, which is also consistent with our adapted method that extract 370 class I and 1735 Class II, taking into account the absence of the 2MASS phase.\vspace{-0.2cm}\\

From these results, the strong imbalance between the 3 labeled classes is striking. We detailed some of the difficulties that imbalance learning raises in Section~\ref{class_balance}. This aspect will be described and carefully handled in the following section, as it proved to have a very strong impact on our results (Sect.~\ref{orion_results}).

\clearpage
\newpage
\subsection{Construction of the test, valid and train dataset}
\label{training_test_datasets}

As described in Section~\ref{sect_overtraining}, the learning process requires not only a training dataset to update the weights, but also a test set and a validation set, which contain sources that were not shown to the network during the learning process. This provides a criteria to stop the learning process. The class proportions in those datasets can be kept as in the labeled dataset or can be rebalanced to have an even number of objects per class. However, our sample suffers from two limitations: its small size and its strong imbalance. To optimize the quality of our results, we needed to carefully define our training and test datasets. Since one of our class of interest (CI) is represented by a relatively small number of objects, the efficiency of the training strongly depends on how many of them we kept in the training sample. Therefore, we chose to adopt the strategy where the same dataset is used for both validation and test steps (Sect.~\ref{sect_overtraining}). It remains efficient to track overfitting, but it increases the risk to stop the training in a state that is abnormally favorable for the test set. As mentioned in Section~\ref{sect_overtraining}, discrepancies between results on the training and test datasets can be used to diagnose over-training and will be detailed for the present application in Section~\ref{orion_results}. Even with this strategy, it remains that the labeled dataset has only few objects to be shared between the training and the test set for some subclasses, considering the expected complexity of the classification.\\

\begin{figure}[!t]
\hspace{-1.2cm}
\begin{minipage}{1.1\hsize}
\centering
\includegraphics[width=\hsize]{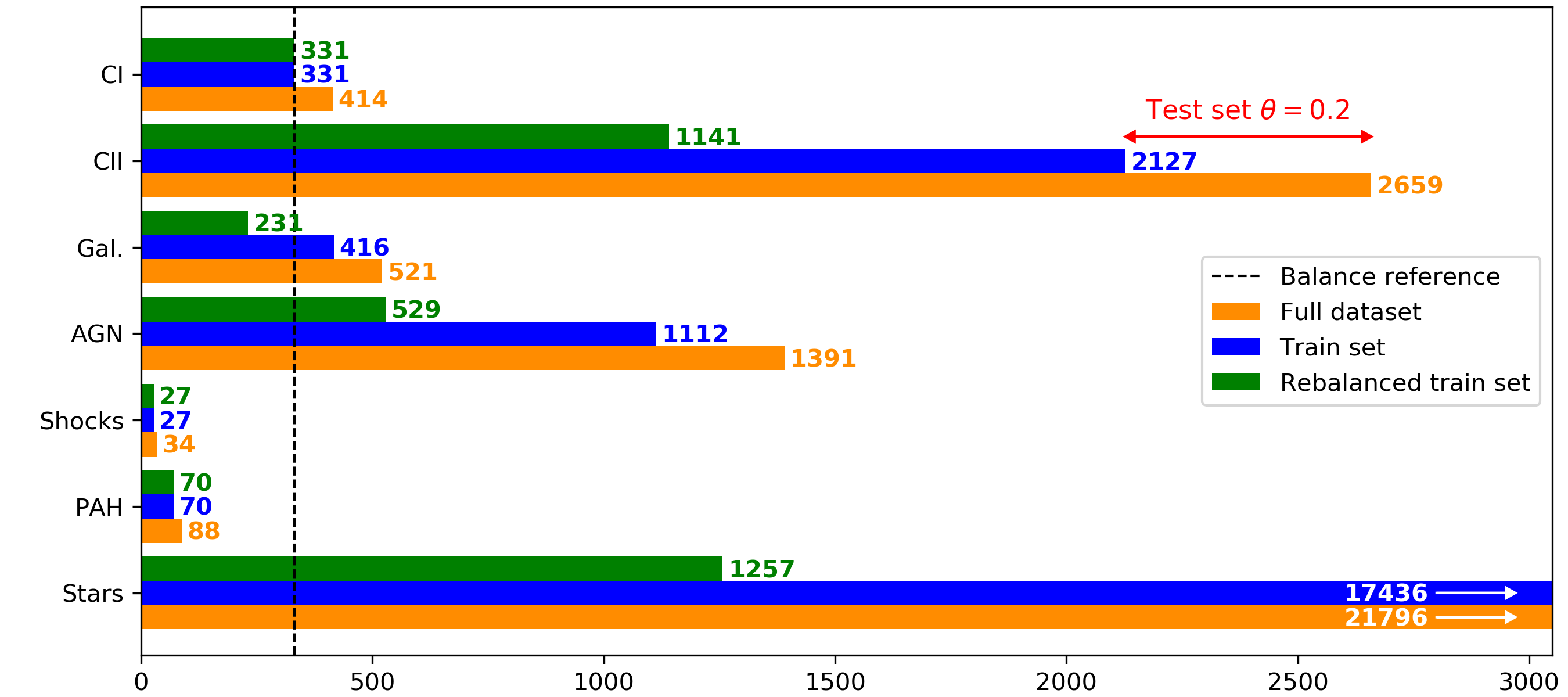}
\end{minipage}
\caption[Rebalancing process illustrated on the Combined dataset]{Rebalancing process illustrated on the Combined dataset. The orange color shows the number of objects in the complete dataset, the blue color represents the remaining objects after exclusion of the test set in observational proportion using $\theta = 0.2$, and the green color represents the training objects using the $\gamma_i$ corresponding to this case. The vertical dashed black line is the $1-\theta$ value corresponding to CI, which is the $\gamma_i = 1$ reference.}
\label{hist_train_rebalance}
\vspace{-0.2cm}
\end{figure}

In addition to the previous point, to evaluate the results quality, it was necessary for the test set to be representative of the actual problem. As before, this is difficult mainly because our case study is strongly imbalanced. Therefore, we needed to keep "observational proportions" for the test set as discussed in Section~\ref{class_balance}. We defined a fraction $\theta$ of objects from the labeled dataset that was taken to form the test dataset. This selection was made independently for each of the seven subclasses provided by the modified G09 classification. It ensured that the proportions were respected even for highly diluted classes of objects (e.g. for Shocks). The effect of taking such proportions is discussed for our various results in Section~\ref{yso_results}.\\

In contrast, the training set does not need to have observational proportions. It needs to have more numerous objects from the classes that have a greater intrinsic diversity and populate a larger volume in the input parameter space. It is also necessary to be more accurate on the most abundant classes, since even a small error on them induces a large contamination of the diluted classes (Sect.~\ref{class_balance}). We chose to scale the number of objects from each class to the number of CI YSOs. The choice of this class is motivated by the fact that the identification of CI YSOs is missing in several other YSO identification methods that do not use surveys with enough sensitivity to detect them. Therefore, we want this class to be predicted with the highest achievable quality. Additionally, they are rare in our labeled dataset, mainly due to the fact that they are much fainter. This induces that we wanted to have the maximum number of them in our training sample while also having a fine control over the degree of dilution of this class against the others. Consequently, the scaling is performed as follows. We shared all the CI objects between the training and the test samples as fixed by the fraction $\theta$, that is:
\begin{equation}
N^{\rm train}_{\rm CI} = (1-\theta) \times N^{\rm tot}_{\rm CI} \quad \text{and}
\end{equation}
\begin{equation}
N^{\rm test}_{\rm CI} = \theta \times N^{\rm tot}_{\rm CI},
\end{equation}
respectively, where $N^{\rm tot}_{\rm CI}$ is the total number of CI objects. Then, we defined a new hyperparameter, the factor $\gamma_i$, as the ratio between the number of selected objects from a given subclass $N^{\rm train}_i$ and the number $N^{\rm train}_{\rm CI}$ of class I YSOs in the same dataset:
\begin{equation}
\gamma_i = \frac{N^{\rm train}_i}{N^{\rm train}_{\rm CI}}.
\end{equation}

If $N^{\rm train}_i$ were computed directly from this formula, it may exceed $(1-\theta)\times N^{\rm tot}_i$ in some cases, a situation incompatible with keeping $N^{\rm test}_{\rm CI} = \theta \times N^{\rm tot}_{\rm CI}$ in the test set. Thus, we limited the values of $N^{\rm train}_i$ as follows:
\begin{equation}
N^{\rm train}_i = \min\Big( \big(\gamma_i\times (1-\theta) \times N_{\rm CI}^{\rm tot}\big), \big((1-\theta) \times N_i^{\rm tot}\big)\Big),
\label{eq:Ntrainmin}
\end{equation}
where the values of the $\gamma_i$ factors were determined manually by trying to optimize the results on each training set.
We note that for the most populated classes, this approach implies that only part of the sample was used to build the training and test sets. As discussed below, this was a motivation to repeat the training with various random selections of objects, and thus assess the impact of this random selection on the results.\\

\begin{table*}[t]
	\small
	\centering
	\vspace{0.2cm}
	\caption{Composition of the training and test datasets for each labeled dataset.}
	\label{sat_factors}
	\begin{tabularx}{1.0\hsize}{l l@{\hskip 0.05\hsize} x{0.085\hsize} x{0.105\hsize} *{5}{Y}  @{\hskip 0.07\hsize} c}
	\toprule
	\toprule
	\vspace{-0.3cm}\\
	& & CI & CII & Gal. & AGNs & Shocks & PAHs & Stars & Total\\
	\vspace{-0.3cm}\\
	\toprule
	\vspace{-0.3cm}\\
	\multicolumn{10}{c}{ Orion - $\theta = 0.3$}\\
	\cmidrule(lr){1-10}
	Test: & & 97 & 667 & 122 & 342 & 8 & 26 & 4470 & 5732\\
	\cmidrule(lr){2-10}
	\multirow{2}{*}{Train:} & $\gamma_i$ & 1.0 & 3.35 & 0.6 & 1.3 & 0.1 & 0.3 & 4.0 & \\
	& $N_i$ & 226 & 757 & 135 & 293 & 19 & 60 & 904 & 2394\\
	\vspace{-0.1cm}\\
	\multicolumn{10}{c}{ NGC 2264 - $\theta = 0.3$}\\
	\cmidrule(lr){1-10}
	Test: & & 27 & 130 & 34 & 75 & 1 & 0 & 2067 & 2334\\
	\cmidrule(lr){2-10}
	\multirow{2}{*}{Train:} & $\gamma_i$ & 1.0 & 2.5 & 0.3 & 0.6 & 0.1 & 0.3 & 3.5 & \\
	& $N_i$ & 62 & 155 & 18 & 37 & 4 & 0 & 217 & 493 \\
	\vspace{-0.1cm}\\
	\multicolumn{10}{c}{ Combined - $\theta = 0.2$}\\
	\cmidrule(lr){1-10}
	Test:& & 82 & 531 & 104 & 278 & 6 & 17 & 4359 & 5377\\
	\cmidrule(lr){2-10}
	\multirow{2}{*}{Train:} & $\gamma_i$ & 1.0 & 3.45 & 0.7 & 1.6 & 0.1 & 0.3 & 3.8 &\\
	& $N_i$ & 331 & 1141 & 231 & 529 & 27 & 70 & 1257 & 3586\\
	\vspace{-0.1cm}\\
	\multicolumn{10}{c}{ Full 1\,kpc - $\theta = 0.2$}\\
	\cmidrule(lr){1-10}
	Test**:& & 82 & 531 & 104 & 278 & 6 & 17 & 4359 & 5377\\
	\cmidrule(lr){2-10}
	\multirow{2}{*}{Train:} & $\gamma_i$ & 1.0/1.0* & 3.3/3.0* & 1.0 & 1.4 & 0.1 & 0.3 & 8.0 &\\
	& $N_i$ & 331/331* & 1092/993* & 331 & 463 & 27 & 70 & 2648 & 6286\\
	\vspace{-0.2cm}\\
	\bottomrule
	\bottomrule
	\end{tabularx}
	\caption*{\vspace{-0.3cm}\\ {\bf Notes.} *The first and second values of $\gamma_i$ are for YSOs from the Combined and 1\,kpc datasets, respectively. \newline **The 1\,kpc dataset does not add contaminants, therefore the Full 1\,kpc test set is the same as the Combined test dataset to keep realistic observational proportions.}
	\vspace{-0.5cm}
\end{table*}

The adopted values of $\theta$, $\gamma_i$, and the corresponding numbers of objects in the training sample are given in Table~\ref{sat_factors} for each dataset, while Figure~\ref{hist_train_rebalance} shows a graphical representation of the rebalancing process for the Combined dataset. The figure compares the sizes of the complete labeled dataset, the training set, and the test set. Like for the other parameters the choice of $\gamma_i$ values for each dataset was the result of an exploration and of the analysis of the results given for each case. The table shows the general trend that with larger labeled datasets, we can use smaller values of $\theta$ because it corresponds to a large enough number of objects in the associated test set. In addition, the number of objects in the training set of NGC 2264 is significantly smaller than in the other datasets. The sample also lacks some subclasses of contaminants, mainly due to the much smaller sky coverage of the region when compared to Orion, which impacted the results for the associated training. The fine tuning of the $\gamma_i$ values is discussed for each region in Section~\ref{yso_results} and aims at maximizing the precision for CI, while keeping a large enough value in recall (ideally $>90\%$ for both of them), and a good precision on CII as well. This choice strongly impacts the tuning of the $\gamma_i$ values, since they directly represent the emphasis given to a class against the others during the training phase, hence biasing the network toward the class that needs the most representative strength. This will slightly lower the quality of objects in other classes but always to a very acceptable extent. Moreover, it is still possible to isolate objects with the best classification reliability using the probability output to overcome this effect, as discussed in Section~\ref{classification_examples}. One last point is that the exploration of the optimal proportion of each class in the training dataset allows one to account for intrinsicly more complicated distributions in the input feature space for certain classes. Some subclasses might be very easy to isolate since they are linearly separable from the rest of the problem. Therefore, reducing the number of objects in this subclass will mostly conserve their quality but will free the space for other, more difficult, objects and will also reduce the dilution of other rare subclasses. This point was evoked in Section~\ref{class_balance} and will also be covered more deeply in the results (Sect.~\ref{yso_results})\\

\newpage
Finally, to ensure that our results are statistically robust, each training was repeated several times with different random selections of the testing and training objects based on the $\theta$ and $\gamma_i$ factors. This allowed us to estimate the variability of our results as discussed during the result analysis in Section~\ref{yso_results}. We checked these variability after each change in any of the hyperparameters. In the case of subclasses with many objects, some objects were included neither in the training nor in the test set. This ensured that the random selection could pick up various combinations of them at each training. In contrast, in the case of the rare subclasses, since they are entirely included in either the training or the test set, it is more difficult to ensure a large diversity in their selection to test the stability against selection. For each result presented in Section~\ref{yso_results}, we took care to also dissociate this effect from the one induced by the random initialization of weights by doing several trainings with the same data selection, which is an indication of the intrinsic stability of the network for a specific set of hyperparameters.

We acknowledge here that our approach to re-balancing might not be optimum. We have considered other approaches for this task like various data augmentation methods, setting a different error cost for each class, having some priors in the class distribution, etc. Our method based on $\gamma_i$ values still has the advantage of simplicity of implementation and it solely relies on observed data. However, it has the major flaw of not using a significant part of the labeled dataset.

\vspace{-0.3cm}
\subsection{Network architecture and parameters}
\label{network_tuning}
\begin{figure}[!t]
\centering
\includegraphics[width=0.95\hsize]{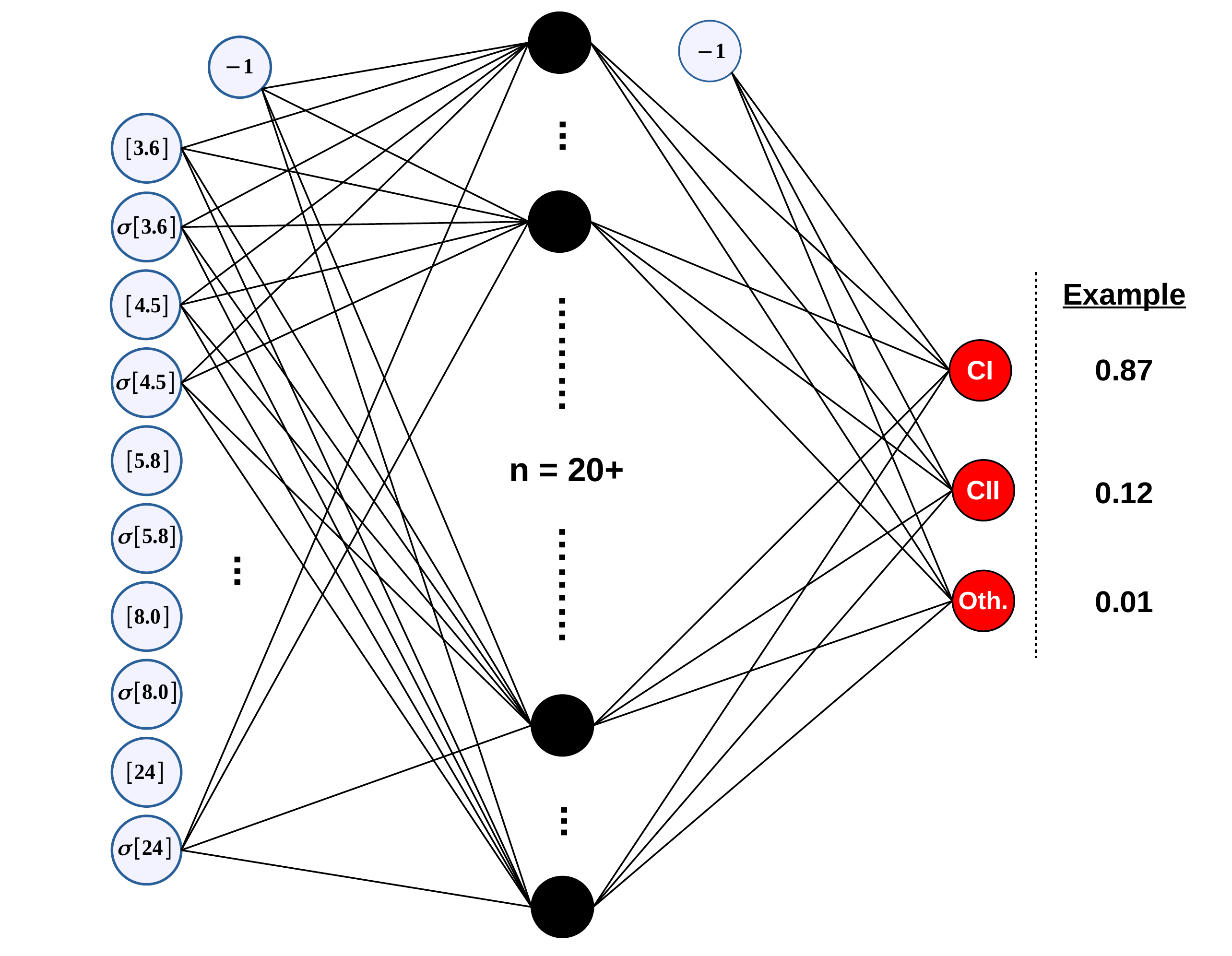}
\caption[Illustration of the ANN used for the YSO classification]{Illustration of the ANN actually used for the YSO classification. The light dots with blue border are input nodes representing each feature and the necessary bias nodes, the black dots are hidden neurons with sigmoid activation, and the red dots are the output neurons with a probabilistic Softmax activation. The black lines represent the linking weights. Only part of the hidden neurons and weights are represented to increase readability.}
\label{illustr_net_final}
\end{figure}

\vspace{-0.1cm}
We adjusted most of the network hyperparameters manually to find appropriate values for our problem. To ease the research of optimal values, we started with values from general recipes. To start, we defined the number of neurons in the hidden layer. The number of neurons can be roughly estimated with the idea that each neuron can be seen as a continuous linear separation in the input feature space (Sect.~\ref{nb_neurons}). Based on Figure~\ref{fig_gut_method} at least $n = 10$ neurons should be necessary, since this figure does not represent all the possible combinations of inputs. We then progressively raised the number of neurons and tested if the overall quality of the classification was improving as defined in the previous Section~\ref{training_test_datasets}, looking at CI and CII recall and precision. In most cases, it improved continuously and then fluctuated around a maximum value. The corresponding number of neurons and the maximum value can vary with the other network hyperparameters. The chosen number of neurons is then the result of a joint optimization of the different parameters. We observed that, depending on the other parameters, the average network reached its maximum value for $n \geq 15$ hidden neurons when trained on Orion. However, the network showed better stability with a slightly larger value. We adopted $n = 20$ hidden neurons for almost all the datasets, and increased it to $n = 30$ for the largest dataset, because it slightly improved the results in this case (Table~\ref{tab_hyperparam}). Increasing too much this number could lead to less stability and increases the computation time. The corresponding network architecture for this example is illustrated in Figure~\ref{illustr_net_final}. It shows all the input features described in Section~\ref{data_prep}, the "large" hidden layer with 20 to 30 sigmoid neurons and the 3 Softmax probability outputs.\\

\vspace{-0.1cm}
The optimum number of neurons and the maximum quality of the classification also depends on the number of objects in the training dataset. As discussed in Section~\ref{nb_neurons}, we checked if we satisfy the recommended rule of having 10 times more training objects that weights in the network. In our case, including the bias nodes, we would need $(m+1)\times n + (n+1)\times o \times 10$ objects in our training set, with the same notations as in Section~\ref{mlp_sect}. This gives us a minimum of $2830$ objects in the whole training set using our network structure with $n = 20$, assuming a balanced distribution among the output classes. As shown in Table~\ref{sat_factors}, some of our training samples are too small for the class I YSOs and critically small for various subclasses of contaminants. Still, each class does not get the same number of neurons from the network. As we already stated, we expect some classes to have a less complex distribution in the parameter space, meaning that they can be represented by a smaller number of weights, therefore with less training examples. The extra representative strength can then be used to better represent more complex classes that may be more abundant. Thus, it is a matter of balance between having a sufficient amount of neurons to properly describe our problem and the maximum number of available data.\\

\begin{table}
	\centering
	\caption{Non structural network hyperparameter values used in training for each dataset. }
	\vspace{-0.1cm}
	\begin{tabularx}{0.95\hsize}{l @{\hskip 0.1\hsize} *{4}{Y}}
	\toprule
	 & Orion & NGC 2264 & Combined & Full 1\,kpc \\
	\vspace{-0.4cm}\\
	\cmidrule(rr){2-5}
	\vspace{-0.3cm}\\
	Train size & 2394 & 493 & 3586 & 7476 \\
	\midrule
	$\eta$ & $3 \times 10^{-5}$ & $2 \times 10^{-5}$ & $4 \times 10^{-5}$ &  $8 \times 10^{-5}$ \\
	$\alpha$ & 0.7 & 0.6 & 0.6 & 0.8 \\
	$n$ & 20 & 20 & 20 & 30 \\ 
	$n_e$ & 5000 & 5000 & 5000 & 3000 \\
	\bottomrule
	\end{tabularx}
	\label{tab_hyperparam}
	\caption*{\vspace{-0.3cm}\\ {\bf Notes.} The size of the corresponding training set is put for comparison. $\eta$ is the learning rate, $\alpha$ the momentum, $n$ the number of neurons in the hidden layer, and $n_e$ the number of epochs between two control steps.}
	\vspace{-0.4cm}
\end{table}

Our datasets were individually normalized in an interval of $-1$ to $1$ as described in Section~\ref{input_norm}. Therefore, we set the steepness $\beta$ of the sigmoid activation of the hidden neurons to $\beta=1$, which worked well with the adopted normalization and our weight initialization that is the same as in Section~\ref{weight_init}. Regarding the gradient descent scheme (Sect.~\ref{descent_schemes}), all methods were compared at various steps of the study, but none has been outperforming significantly the others in terms of reached best prediction. Then, regarding the actual computation time necessary to converge we selected the full batch method.\\

Concerning the computational performance, it is worth noting that at the moment of the described application, all our computations were made using a much simpler precursor of CIANNA that also was GPU accelerated. At that time we used a now 7 years old NVIDA GTX 780 which is a non-professional GPU. Still, it was able to train our networks in about 10-15 minutes using the batch formalism, while our CPU parallel implementation using OpenBLAS and OpenMP on an intel 3770k (4C/8T, 3.5 Ghz) required approximately 1h. These results are for roughly $1.3 \times 10^6$ full epochs. This is consistent with a rough estimate of 537 GFLOPS for our overclocked 3770k against 4.93 TFLOPS for our version of the GTX 780 (both for FP32). This is a nice example of the aspects discusses in Section~\ref{matrix_formal}, showing that GPUs are very efficient to perform such tasks. As a matter of fact, this GPU draws up to 250 W of power while the 3770k CPU is rated for 77W, which is certainly vastly underestimated considering our 4.2GHz all core overclock. Therefore, while having a 3.24 times higher power usage, the GPU has a raw compute capability 9.18 times higher. To end the comparison, we note that the used framework at that time was much more naive than our current version of CIANNA and was using an old CUDA version (<10.0) that did not take advantage of some kernel latency improvements of the subsequent versions, while it would have allow for mini-batch to efficiently converge with less epochs, i.e less raw computations. Also, while modern GPU now have a compute power above 15 TFLOPS in FP32, their power consumption is very similar. In contrast, the modern high-end CPUs power consumption tends to increase (up to 250 W) following their higher core count. Also we did not account for any modern hardware ANN specific use in this comparison (Sect.~\ref{gpus_prog}), which would have further reinforce the advantage of GPUs.\\

More practically, the learning rates we finally adopted are in the range $\eta = 3 - 8 \times 10^{-5}$ and used momentum ranging from $\alpha = 0.6$ to $ \alpha = 0.8$ depending on the dataset, as shown in Table~\ref{tab_hyperparam}. We note that during training and in contrast with what is described in Section~\ref{descent_schemes}, we summed the weight update contributions from each object in the training set as in \citet{rumelhart_parallel_1986} over a batch, instead of averaging them as, for example, in \citet{glorot_understanding_2010}. This implies that our learning rate must be accordingly small when using large datasets. We observed that, for this specific study, the learning rate could instead be progressively increased when the training dataset was larger. This indicates that, in small training sets, the learning process is dominated by the lack of constraints, causing a less stable value of convergence. This translates into a convergence region in the weight space that contains numerous narrow minima due to the relatively larger granularity of the objects in a smaller dataset. The network can only properly resolve it with a smaller learning rate and will be less capable of generalization. This is an expected issue, because we intentionally included small datasets in the analysis to assess the limits of the method with few objects.\\

Finally, one less important hyperparameter is the number of epochs between two monitoring steps which was set from $n_e = 3000$ to $ n_e = 5000$. It defines at which frequency the network state is saved and checked, leaving the opportunity to decrease the learning rate $\eta$ if necessary.

\subsection{Convergence criteria}

Since training the network is an iterative process, a convergence criteria must be adopted.
In principle, this criteria should enable one to identify an iteration where the training has sufficiently converged for the network to capture the information in the training sample, but is not yet subject to over-training, as stated in Section~\ref{sect_overtraining}. However, in our case this global error is affected by the proportions in the training sample and does not necessarily reflect the underlying convergence of each subclass. Our approach to this issue was to let the network learn for an obviously unnecessary amount of steps and regularly save the state of the network. This allowed us to better monitor the long term behavior of the error, and to compare the confusion matrices at regular steps. In most cases, the error of the training and test sets both converged to a stable value and stayed there for many steps before the second one started to rise. During this apparently stable moment, the prediction quality of the classes oscillated, switching the classes that get the most representativity strength from the network. Because we want to put the emphasis on CI YSOs, we then manually selected a step that was near the maximum value for CI YSO precision, with a special attention to avoid the ones that would be too unfavorable to CII YSOs.\\

As one would expected, we observed that the convergence step changed significantly with the network weight random initialization, even with the exact same dataset and network, ranging from 100 to more than 1000, where each step corresponds to several thousands epochs (Table~\ref{tab_hyperparam}). Most of the time, the error plateau lasted around 100 steps. We emphasize that the number of steps needed to converge has no consequences on the quality of the results; it only reflects the length of the particular trajectory followed by the network during the training phase.

\newpage
\section{Subsequent application to multiple star-forming regions}
\label{yso_results}

This section presents the results of our YSO classification using ANN, obtained for the various labeled datasets described in Section~\ref{yso_datasets_tuning}. To ease the reading of this section, we summarize all the cases in Table~\ref{results_cases}. This section also includes some analysis of each case and some comparison of the results that allows to explain the motivations behind our choice of parameters.\\

\etocsettocstyle{\subsubsection*{\vspace{-1.5cm}}}{}
\localtableofcontents

\begin{table*}[!ht]
	\centering
	\caption{List of case studies regarding the dataset used to train the network and the dataset to which it was applied to provide predictions.}
	\vspace{-0.1cm}
	\def\arraystretch{1.3}
	\begin{tabularx}{0.8\hsize}{r l *{4}{m}}
	\multicolumn{2}{c}{} & \multicolumn{4}{c}{\textbf{\large Forward dataset}}\\
	\cmidrule[\heavyrulewidth](lr){2-6}
	\parbox[l]{0.2cm}{\multirow{6}{*}{\rotatebox[origin=c]{90}{\textbf{\large Training dataset}}}} & & Orion & NGC 2264 & Combined & Full 1\,kpc \\
	\cmidrule(lr){2-6}
	 & Orion & O-O & \hspace{0.19cm}O-N* & & \\
	 & NGC 2264 & \hspace{0.1cm} N-O* & N-N & & \\ 
	 & Combined & & & C-C & \\
	 & Full 1\,kpc & \hspace{0.12cm}F-O* & \hspace{0.13cm}F-N* & F-C & \\
	\cmidrule[\heavyrulewidth](lr){2-6}
	\end{tabularx}
	\label{results_cases}
	\caption*{\vspace{-0.1cm}\\ {\bf Notes.} *These cases were only forwarded on the full corresponding dataset with no need for a test set. There was no forward on the full 1\,kpc dataset since, as a combination of a complete catalog and a YSO-only catalog, it is not in observational proportions.}
\end{table*}

\newpage
	\subsection{First training on one specific region: the Orion molecular cloud}
	\label{orion_results}

In this section, we consider the case where both the training and forward datasets were built from the Orion labeled dataset, hereafter denoted the "O-O" case. This first application is expected to {\bf draw a baseline of results on the simplest case possible}, since Orion is the star-forming region of our sample that contains the most YSOs. \\

\vspace{-0.8cm}
\subsubsection{Hyper-parameter and training proportion evaluation}
The network hyperparameters used for Orion are described in the previous Section~\ref{yso_datasets_tuning} and Table~\ref{tab_hyperparam}. The resulting confusion matrix is shown in Table~\ref{tab:OO} using the test set in observational proportions from Table~\ref{sat_factors}.
The optimal $\gamma_i$ factors found for Orion show a stronger importance of the CII YSOs ($\gamma_{\text{CII}} = 3.35$) and of the Stars subclass ($\gamma_{\text{Stars}} = 4.0$) than for any other subclass ($\gamma_i \lesssim 1$). In contrast, the optimal values for Shocks and PAHs are saturated in the sense that in Eq.~(\ref{eq:Ntrainmin}), $N_i^{\rm train} = (1-\theta) \times N_i^{\rm tot}$, but they appeared to have a negligible impact on the classification quality in this case. Galaxies and PAHs appeared to be easily classified with a rather small number of them in the training sample. This is convenient since adding too many objects of any class hampers the capacity of the network to represent CI objects, i.e. the most diluted class of interest, in the network, degrading the reliability of their identification. Therefore, Stars and CII objects could be well represented with a large fraction of them in the training sample, still limiting their number to avoid an excessive dilution of CI YSOs (Sect.~\ref{class_balance}).\\

We note that we have explored different values for the $\theta$ parameter. It revealed that the network predictions improve continuously when increasing the number of objects in the training sample. However, to keep enough objects in the test dataset, we had to limit $\theta$ to 0.3 (Table~\ref{sat_factors}). The only classes for which the number of objects in the training sample is limited by the $\theta$ value rather than their respective $\gamma_i$ values are CI YSO, Shocks and PAHs. Since Shocks and PAHs are rare in the observational proportions, they are unlikely to have a dominant impact on the prediction quality; we discuss more deeply the results for these rare subclasses in Section~\ref{shocks_discussion}. This leads to the outcome that the results on Orion are currently mainly limited by the number of CI YSOs, because their dilution prevents the use of more objects of the other subclasses for which we have numerous examples.\\

\subsubsection{Main result}

\begin{table}[!t]
	\small
	\centering
	\caption{Confusion matrix for the O-O case for a typical run.}
	\vspace{-0.1cm}
	\begin{tabularx}{0.65\hsize}{r l |*{3}{m}| r }
	\multicolumn{2}{c}{}& \multicolumn{3}{c}{\textbf{Predicted}}&\\
	\cmidrule[\heavyrulewidth](lr){2-6}
	\parbox[l]{0.2cm}{\multirow{6}{*}{\rotatebox[origin=c]{90}{\textbf{Actual}}}} & Class & CI YSOs & CII YSOs & Others & Recall \\
	\cmidrule(lr){2-6}
	 &  CI YSOs    & 88     & 4       & 5       & 90.7\% \\
	 &  CII YSOs   & 7      & 651     & 9       & 97.6\% \\
	 &  Others     & 11     & 58      & 4899    & 98.6\% \\
	\cmidrule(lr){2-6}
	 &  Precision & 83.0\% & 91.3\% & 99.7\% & 98.4\% \\
	\cmidrule[\heavyrulewidth](lr){2-6}
	\end{tabularx}
	\vspace{+0.1cm}
	\label{tab:OO}
\end{table}

\begin{table}[!t]
	\small
	\centering
	\caption{Subclass distribution for the O-O case.}
	\vspace{-0.1cm}
	\label{rep_hist_orion}
	\begin{tabularx}{0.75\hsize}{r l *{6}{Y} l}   
	\multicolumn{2}{c}{}& \multicolumn{7}{c}{\textbf{Actual}}\\      
	\cmidrule[\heavyrulewidth](lr){2-9}
	\parbox[l]{0.2cm}{\multirow{5}{*}{\rotatebox[origin=c]{90}{\textbf{Predicted}}}} & & CI & CII & Gal & AGNs & Shocks & PAHs & Stars\\
	\cmidrule(lr){3-9}
	 &  CI YSOs & 88 & 7 & 1 & 2 & 3 & 3 & 2 \\
	 &  CII YSOs & 4 & 651 & 5 & 0 & 2 & 4 & 47 \\
	 &  Others & 5 & 9 & 116 & 340 & 3 & 19 & 4421 \\
	\cmidrule[\heavyrulewidth](lr){2-9}
	\end{tabularx}
	\vspace{-0.8cm}
	\label{tab:OO-sub}
\end{table}

The global accuracy of this case is 98.4\%, but the confusion matrix (Table~\ref{tab:OO}) shows that this apparently good accuracy is unevenly distributed between the three classes. The best represented class is the contaminant class, with an excellent precision of 99.7\% and a very good recall of 98.6\%. The results are slightly less satisfying for the two classes of interest, with recalls of 90.7\% and 97.6\%, and precisions of 83.0\% and 91.3\% for the CI and CII YSOs, respectively. In spite of their very good recall, due to their widely dominant number, objects from the Other class are the major contaminants of both CI and CII YSOs, with 11 out of 18, and 58 out of 62 contaminants, respectively. Therefore, improving the relatively low precision of CI and CII objects mainly requires to better classify the Other objects. In addition, less abundant classes are more vulnerable to contamination. This is well illustrated by the fact that the 7 CII YSOs misclassified as CI YSOs account for a loss of 7\% in precision for CI objects, while the 9 CII YSOs misclassified as Others account for a loss of only 0.2\% in the Others precision. Those properties are typical of classification problems with a diluted class of interest, where it is essential to compute the confusion matrix using observational proportions. Computing it from a balanced forward sample would have led to apparently excellent results, which would greatly overestimate the quality genuinely obtained in a real use case. Moreover, it illustrates the necessity of a high $\gamma_i$ value for dominant classes regardless of their interest (e.g Stars), as we need to maximize the recall of these classes to enhance the precision of the diluted ones.\\

\subsubsection{Test on a balanced dataset}

To illustrate the interest of selecting our training proportions with the $\theta$ and $\gamma_i$ factors, we made a test with a balanced training set, where all three classes were represented by an equal number of objects. The result of this test for a typical training run is provided in Table \ref{tab:OO_balanced}. The best we could achieve this way was not more than $\sim 55\%$ precision on CI YSOs, showed in red in our table, and $\sim 87\%$ for CII YSOs, which is considerably less than the results with the rebalanced dataset. This was mostly due to the small size of the training sample (681 objects with $\theta = 0.3$), which was constrained by the less abundant class, and to the poor sampling of the Other class compared to its great diversity. Especially with only $227$ contaminants it is impossible to represent all the subclasses. We could also have attempted a balanced training with more output classes in our network definition, to account for this diversity, but it results in other issues, CI recall getting way too low regarding our expectations. In contrast, when using our more complex sample definition, despite the reduced proportion of YSOs in the training sample, the precision and recall quantities for both CI and CII remained above $80\%$ and $90\%$, respectively. This means that we found an appropriate balance between the representativity of each class and their dilution in the training sample.\\

\begin{table}[!t]
	\small
	\centering
	\caption{Confusion matrix for a balanced training on the O-O case for a typical run.}
	\vspace{-0.1cm}
	\begin{tabularx}{0.65\hsize}{r l |*{3}{m}| r }
	\multicolumn{2}{c}{}& \multicolumn{3}{c}{\textbf{Predicted}}&\\
	\cmidrule[\heavyrulewidth](lr){2-6}
	\parbox[l]{0.2cm}{\multirow{6}{*}{\rotatebox[origin=c]{90}{\textbf{Actual}}}} & Class & CI YSOs & CII YSOs & Others & Recall \\
	\cmidrule(lr){2-6}
	 &  CI YSOs    & 94     & 3       & 0       & 96.9\% \\
	 &  CII YSOs   & 28     & 613     & 26      & 91.9\% \\
	 &  Others     & 50     & 93      & 4826    & 97.1\% \\
	\cmidrule(lr){2-6}
	 &  Precision & \textcolor{red}{\bf 54.7\%} & 86.5\% & 99.5\% & 96.5\% \\
	\cmidrule[\heavyrulewidth](lr){2-6}
	\end{tabularx}
	\vspace{+0.1cm}
	\label{tab:OO_balanced}
\end{table}

\subsubsection{Prediction stability}

As discussed in Section \ref{training_test_datasets}, we tested the stability of those results regarding (i) the initial weight values using the exact same training dataset, and (ii) the random selection of objects in the training and test set. For point (i), we found that in Orion, the weight initialization has a weak impact with approximately $\pm 0.5\%$ dispersion in almost all the quality estimators. For point (ii), we found the dispersion to average around $\pm 1\%$ for the recall of YSO classes. Contaminants were found to be more stable with a recall dispersion under $\pm 0.5\%$. Regarding the precision value, there is more instability for the CI YSOs, because they are weakly represented in the test set and one misclassified object changes the precision value by typically $1\%$. Overall, we observed values ranging from $77\%$ to $83\%$ for the CI YSOs precision. For the better represented classes, we obtained much more stable values with dispersions of $\pm 0.5$ to $\pm 1\%$ on class II, and less than $\pm 0.5\%$ on Other objects. This relative stability is strongly related to the proper balance between classes, controlled by the $\gamma_i$ parameters, since strong variations between runs imply that selection effects are important, and that there are not enough objects to represent the input parameter space properly.\\

\subsubsection{Detailed sub-classes distribution}

We also looked at the detailed distribution of classified objects regarding their subclasses from the labeled dataset. These results are shown for Orion in Table~\ref{rep_hist_orion}. It is particularly useful to detail the distribution of contaminants across the three network output classes. For CI YSOs, the contamination appears to originate evenly from various subclasses, while for CII there is a strong contamination from non-YSO stars, though this represents only a small fraction ($\sim 1\%$) of the Stars population. The distribution of Other objects among the subclasses is very similar to the original one (Table~\ref{tab_selection}). Interestingly, the Shocks subclass is evenly scattered across the three output classes, which we interpret as a failure by the network to find a proper generalization for these objects. More generally, Table~\ref{rep_hist_orion} shows that the classes that are sufficiently represented in the training set like AGNs or Stars are well classified, while the Galaxies, Shocks and PAHs are less well predicted. This is directly related to the fact that the training dataset does not fully covers their respective volume in the input parameter space or to the fact that they are too diluted in the dataset. Additionally, Stars and Galaxies mainly contaminate the CII class. This is a direct consequence of the proximity of theses classes in the input parameter space, as can be seen in Figure~\ref{fig_gut_method}.

\subsubsection{Full dataset result}

To circumvent the limitations due to the small size of our test set, we also applied our network to the complete Orion dataset. The corresponding confusion matrix is in Table~\ref{tab:OO_all} and the associated subclass distribution is in Table~\ref{tab:OO_all-sub}. It may be considered to be a risky practice, because it includes objects from the training set that could be over-fitted, so it should not be used alone to analyze the results. Here, we used it jointly with the results on the test set as an additional over-fitting test. If the classes are well constrained, then the confusion matrix should be stable when switching from the test to the complete dataset. For Orion, we see a strong consistency between Tables~\ref{tab:OO} and \ref{tab:OO_all} for the Other and CII classes, both in terms of recall and precision. For CI YSOs, the recall has increased by 3.4\%, and the precision has decreased by 1.2\%. These variations are of the same order as the variability observed when changing the training set random selection, indicating that over-fitting is unlikely here. If there is over-fitting it should be weak and restricted to CI YSOs. Therefore, the results obtained from the complete Orion dataset appear to be reliable enough to take advantage of their greater statistics. Table~\ref{tab:OO_all} gives slightly more information than Table~\ref{tab:OO}, and mostly confirms the previous conclusions on the contamination between classes. Table~\ref{tab:OO_all-sub} provides further insight. AGNs, which seemed to be almost perfectly classified, are revealed to be misclassified as YSOs in 1.8\% of cases. It also shows that the missed AGNs are equally distributed across the CI and CII YSO classes. Shocks are still evenly spread across the three output classes. Regarding PAHs, Table~\ref{tab:OO_all-sub} reveals that there is more confusion with the CII YSOs than with the CI YSOs.

\begin{table}[!t]
	\small
	\centering
	\caption{Confusion matrix for the O-O case forwarded on the full dataset.}
	\begin{tabularx}{0.65\hsize}{r l |*{3}{m}| r }
	\multicolumn{2}{c}{}& \multicolumn{3}{c}{\textbf{Predicted}}&\\
	\cmidrule[\heavyrulewidth](lr){2-6}
	\parbox[l]{0.2cm}{\multirow{6}{*}{\rotatebox[origin=c]{90}{\textbf{Actual}}}} & Class & CI YSOs & CII YSOs & Others & Recall \\
	\cmidrule(lr){2-6}
	 &  CI YSOs    & 305     & 11      & 8        & 94.1\% \\
	 &  CII YSOs   & 34      & 2157    & 33       & 97.0\% \\
	 &  Others     & 34      & 201     & 16331    & 98.6\% \\
	\cmidrule(lr){2-6}
	 &  Precision & 81.8\% & 91.1\% & 99.7\% & 98.3\% \\
	\cmidrule[\heavyrulewidth](lr){2-6}
	\end{tabularx}
	\vspace{+0.1cm}
	\label{tab:OO_all}
\end{table}

\begin{table}[!t]
	\small
	\centering
	\caption{Subclass distribution for the O-O case forwarded on the full dataset.}
	\vspace{-0.1cm}
	\begin{tabularx}{0.75\hsize}{r l *{6}{Y} l}   
	\multicolumn{2}{c}{}& \multicolumn{7}{c}{\textbf{Actual}}\\      
	\cmidrule[\heavyrulewidth](lr){2-9}
	\parbox[l]{0.2cm}{\multirow{5}{*}{\rotatebox[origin=c]{90}{\textbf{Predicted}}}} & & CI & CII & Gal & AGNs & Shocks & PAHs & Stars\\
	\cmidrule(lr){3-9}
	 &  CI YSOs & 305 & 34 & 2 & 11 & 11 & 7 & 3 \\
	 &  CII YSOs & 11 & 2157 & 10 & 9 & 9 & 18 & 155 \\
	 &  Others & 8 & 33 & 395 & 1121 & 8 & 62 & 14745 \\
	\cmidrule[\heavyrulewidth](lr){2-9}
	\end{tabularx}
	\vspace{-0.1cm}
	\label{tab:OO_all-sub}
\end{table}

\newpage
	\subsection{Effect of the selected region: training using NGC 2264}
\label{NGC2264_results}
\vspace{0.3cm}
After having established base results using Orion, we wanted to {\bf test if the learning process could be performed on another region}. As Orion is the largest star-forming region of our sample, it implies selecting a region with less YSOs, which was expected to have a strong impact on the results. Training on an individual region is very limited as they are very few to have a sufficient amount of stars. However, we wanted to have two distinct one-region cases in order to ease the comparison of the two, and to assess the presence of region specific features.\\

\subsubsection{Main result}
For this, we used the training and forward datasets for NGC 2264 as described in Table~\ref{sat_factors} with the corresponding hyperparameters (Table~\ref{tab_hyperparam}). The results for this region alone, obtained by a forward on the test set, are shown in Table~\ref{tab:NN}, with the subclass distribution in Table~\ref{tab:NN-sub}. We refer to this case as the N-N case. The major differences with Orion are expected to come from the differences in input parameter space coverage and from the different proportions of each subclass. This N-N case is also useful to see how difficult it is to train our network with a small dataset. We notice that the recall and precision of CI YSOs are greater ($96.3\%$ and $89.7\%$, respectively) than in Orion, but the corresponding number of objects is too small to draw firm conclusions. For CII YSOs, the recall and precision are lesser than in Orion by approximately $4\%$ and $10\%$, respectively. The Other class shows similar values as in Orion.\\

\begin{table}[!t]
	\small
	\centering
	\caption{Confusion matrix for the N-N case for a typical run.}
	\vspace{-0.1cm}
	\begin{tabularx}{0.65\hsize}{r l |*{3}{m}| r }
	\multicolumn{2}{c}{}& \multicolumn{3}{c}{\textbf{Predicted}}&\\
	\cmidrule[\heavyrulewidth](lr){2-6}
	\parbox[l]{0.2cm}{\multirow{6}{*}{\rotatebox[origin=c]{90}{\textbf{Actual}}}} & Class & CI YSOs & CII YSOs & Others & Recall \\
	\cmidrule(lr){2-6}
	 &  CI YSOs    & 26      & 1       & 0      & 96.3\% \\
	 &  CII YSOs   & 1       & 121     & 8      & 93.1\% \\
	 &  Others     & 2       & 31      & 2144   & 98.5\% \\
	\cmidrule(lr){2-6}
	 &  Precision & 89.7\% & 79.1\% & 99.6\% & 98.2\% \\
	\cmidrule[\heavyrulewidth](lr){2-6}
	\end{tabularx}
	\vspace{+0.1cm}
	\label{tab:NN}
\end{table}

\begin{table}[!t]
	\small
	\centering
	\caption{Subclass distribution for the N-N case.}
	\vspace{-0.1cm}
	\begin{tabularx}{0.75\hsize}{r l *{6}{Y} l}   
	\multicolumn{2}{c}{}& \multicolumn{7}{c}{\textbf{Actual}}\\      
	\cmidrule[\heavyrulewidth](lr){2-9}
	\parbox[l]{0.2cm}{\multirow{5}{*}{\rotatebox[origin=c]{90}{\textbf{Predicted}}}} & & CI & CII & Gal & AGNs & Shocks & PAHs & Stars\\
	\cmidrule(lr){3-9}
	 &  CI YSOs & 26 & 1 & 0 & 2 & 0 & 0 & 0 \\
	 &  CII YSOs & 1 & 121 & 4 & 5 & 1 & 0 & 21 \\
	 &  Others & 0 & 8 & 30 & 68 & 0 & 0 & 2046 \\
	\cmidrule[\heavyrulewidth](lr){2-9}
	\end{tabularx}
	\vspace{+0.2cm}
	\label{tab:NN-sub}
\end{table}

\newpage
\subsubsection{Small dataset issues}
We highlight here how having a small learning sample is problematic for this classification. First of all, the training set contains only 62 CI YSOs, which is far from enough in regard of the size of the network (Sect. \ref{nb_neurons}). This difficulty is far worse than for Orion, because, to avoid dilution, we had to limit the number of objects in the two other classes, leading to the small size of the training sample (493 objects), and consequently to worse results for all classes. To mitigate these difficulties and because the dilution effect occurs quickly, we adopted lower $\gamma_i$ values for CII YSOs and Stars, thus reducing their relative strength. This results in too small training set sizes for all the subclasses compared to the number of weights in the network. However, we observed that a decrease in the number of neurons still reduced the quality of the results. Although a lower number of hidden neurons tended to increase stability, we chose (i) to keep them at $n = 20$ to get the best results, and (ii) to reduce the learning rate to achieve better stability. We note that, due to the use of batch training with the sum of contributions, the smaller size of the dataset than for the O-O case is somewhat equivalent to an additional lowering of the learning rate (Sect.~\ref{descent_schemes}). For this dataset, slight changes on the $\gamma_i$ values happened to lead to great differences in terms of results and stability, which indicates that the classification lacks constraints.\\

\subsubsection{Prediction stability}
Even for a given good $\gamma_i$ set, there is a large scatter in the results when changing the training and test set random selection. It leads to a dispersion of about $\pm 4\%$ in both recall and precision for the CI YSOs. This can be due to a lack of representativity of this class in our sample, but it can also come from small-number effects in the test set that are stronger than in Orion. These two points show that the quality estimators for CI YSOs are not trustworthy with such a small sample size. The results shown in Tables~\ref{tab:NN} and \ref{tab:NN-sub} correspond to one of the best training on NGC 2264, that achieves nearly the best values for CI quality estimators. The CII precision dispersion is about $\pm 2\%$, and its average value is around $80\%$, which is higher than in the specific result given in Table~\ref{tab:NN}, but still significantly lower than for Orion. In contrast, the CII recall is fairly stable with less than $\pm 1\%$ dispersion. Contaminants seem as stable as for Orion using these specific $\gamma_i$ values. However, it could come from the artificial simplification of the problem due to the quasi-absence of some subclasses (Shocks and PAH, see Table~\ref{sat_factors}) in the test set. We note that the network would not be able to classify objects from these classes if this training were applied to any other region that contained such objects.\\

As in the previous section, we studied the effect of the random initialization of the weights. We found that both precision and recall of YSO classes are less stable than for the O-O case with a dispersion of $\pm 1.5\%$ to $\pm 2.5\%$. The Other class shows a similar stability than for Orion, with up to $\pm 0.5\%$ dispersion on precision and recall, which could again be biased by the fact that the absence of some subclasses simplifies the classification. These results indicate as before that our network is not sufficiently constrained using this dataset alone with respect to the architecture complexity that is needed for YSO classification. \\

\newpage
\subsubsection{Full dataset result}

The forward on the complete NGC 2264 dataset is crucial in this case, since it may overcome small-number effects for many subclasses. The corresponding results are shown in Tables~\ref{tab:NN_all} and \ref{tab:NN-sub_all}. It is more difficult in this case than in the O-O one to be sure that there is no over-training, even with a careful monitoring of the error convergence on the test set during the training, because the small-number effects are important. As a precaution, in all the results for the N-N case, we chose to stop the training slightly earlier in the convergence phase in comparison to Orion, for which we found over-training to be negligible or absent (Sect.~\ref{orion_results}). We expect this strategy to reduce over-training, at the cost of a higher noise.\\

With this assumption, the results show more similarities to the Orion case than those obtained with the test set only (comparing Tables~\ref{tab:OO_all} and \ref{tab:NN_all}). Because NGC 2264 contains less CI and CII YSOs than Orion, their boundaries with the contaminants in the parameter space are less constrained. This results in a lower precision for YSO classes, which is mainly visible for the CII YSOs with a drop in precision down to $83.7\%$. For NGC 2264, we have smaller optimal $\gamma_i$ values for the contaminants (especially the Stars) than in Orion. Since it implicitly forces the network to put the emphasis on CI and CII, it should result in better, or at least equivalent, values for recall on these classes than on Orion. It appears to be the case for CI ($\approx 98\%$). It is less clear for CII ($93.3\%$), possibly because of their lesser $\gamma_i$ value than for the Orion case. For the sub-contaminant distributions, the statistics are more robust than in Table~\ref{tab:NN-sub}, and the Galaxies and AGNs are properly represented. Still, it appears that the AGN classification quality is not sufficient and has a stronger impact on the CI precision than in the case of Orion. The other behaviors are similar to those identified in Orion.\\

\begin{table}[!t]
	\small
	\centering
	\caption{Confusion matrix for the N-N case forwarded on the full dataset.}
	\vspace{-0.1cm}
	\begin{tabularx}{0.65\hsize}{r l |*{3}{m}| r }
	\multicolumn{2}{c}{}& \multicolumn{3}{c}{\textbf{Predicted}}&\\
	\cmidrule[\heavyrulewidth](lr){2-6}
	\parbox[l]{0.2cm}{\multirow{6}{*}{\rotatebox[origin=c]{90}{\textbf{Actual}}}} & Class & CI YSOs & CII YSOs & Others & Recall \\
	\cmidrule(lr){2-6}
	 &  CI YSOs    & 88      & 2       & 0        & 97.8\% \\
	 &  CII YSOs   & 7       & 406     & 22       & 93.3\% \\
	 &  Others     & 12      & 77      & 7175     & 98.8\% \\
	\cmidrule(lr){2-6}
	 &  Precision & 82.2\% & 83.7\% & 99.7\% & 98.4\%\\
	\cmidrule[\heavyrulewidth](lr){2-6}
	\end{tabularx}
	\vspace{+0.1cm}
	\label{tab:NN_all}
\end{table}

\begin{table}[!t]
	\small
	\centering
	\caption{Subclass distribution for the N-N case forwarded on the full dataset.}
	\label{tab:NN-sub_all}
	\vspace{0.1cm}
	\begin{tabularx}{0.75\hsize}{r l *{6}{Y} l}   
	\multicolumn{2}{c}{}& \multicolumn{7}{c}{\textbf{Actual}}\\      
	\cmidrule[\heavyrulewidth](lr){2-9}
	\parbox[l]{0.2cm}{\multirow{5}{*}{\rotatebox[origin=c]{90}{\textbf{Predicted}}}} & & CI & CII & Gal & AGNs & Shocks & PAHs & Stars\\
	\cmidrule(lr){3-9}
	 &  CI YSOs & 88 & 7 & 0 & 8 & 3 & 0 & 0 \\
	 &  CII YSOs & 2 & 406 & 8 & 10 & 1 & 0 & 58 \\
	 &  Others & 0 & 22 & 106 & 232 & 2 & 0 & 6835 \\
	\cmidrule[\heavyrulewidth](lr){2-9}
	\end{tabularx}
	\vspace{+0.1cm}
	\label{tab:NN_all-sub}
\end{table}

\newpage
	\subsection{Generalization capacity: crossed application}
\label{cross_forward}

In this section, we {\bf tested the generalization capacity of the trained networks} by using the network trained on one region to classify the sources of the other one. This test is important because this is a typical use case: training the network on well-known regions, and use it on a new one. This is also a way to highlight more discrepancies between the datasets.

\subsubsection{Cross forward considerations}

For this, we used the obtained trained networks from the O-O and N-N cases described in Sects. \ref{orion_results} and \ref{NGC2264_results}. Since they are both built from the same original classification scheme (Sect. \ref{data_prep}), we applied directly one training to the other labeled dataset, which resulted in the two new cases O-N and N-O (see Table~\ref{results_cases}). However, the forwarded dataset must be normalized in the same way as the training set (Sect.~\ref{network_tuning}). Omitting this step would lead to deviations and distortions of our network class boundaries in the input parameter space, with a strong impact on the network prediction. One difficulty is that some objects end up with parameters outside the $[-1;1]$ range, corresponding to areas of the feature space where the network is not constrained. One could partly hide this effect by excluding those out-of-boundary objects. However, they give an additional information about which kind of objects are missing in the respective training datasets and about the corresponding input feature space areas. Therefore, we preferred to keep them in the forward samples. It is legitimate here to use the full dataset directly to test the networks, because none of its objects were used during the corresponding training. It also means that we forwarded datasets with different proportions than the ones they were trained with, but this is the expected end use of such networks. Moreover, both datasets are the results of observations, which means that our tests measured the effective performance of the trained network on a genuine observational use case with the corresponding proportions of classes.\\

\begin{table}[!t]
	\small
	\centering
	\caption{Confusion matrix for the O-N case forwarded on the full NGC 2264 dataset.}
	\vspace{-0.1cm}
	\begin{tabularx}{0.65\hsize}{r l |*{3}{m}| r }
	\multicolumn{2}{c}{}& \multicolumn{3}{c}{\textbf{Predicted}}&\\
	\cmidrule[\heavyrulewidth](lr){2-6}
	\parbox[l]{0.2cm}{\multirow{6}{*}{\rotatebox[origin=c]{90}{\textbf{Actual}}}} & Class & CI YSOs & CII YSOs & Others & Recall \\
	\cmidrule(lr){2-6}
	 &  CI YSOs    & 74      & 2       & 14      & 82.2\% \\
	 &  CII YSOs   & 6       & 402     & 27      & 92.4\% \\
	 &  Others     & 9       & 52      & 7203    & 99.2\% \\
	\cmidrule(lr){2-6} 
	 &  Precision & 83.1\% & 88.2\% & 99.4\% & 98.6\% \\
	\cmidrule[\heavyrulewidth](lr){2-6}
	\end{tabularx}
	\vspace{+0.1cm}
	\label{tab:ON_all} 
\end{table}

\begin{table}[!t]
	\small
	\centering
	\caption{Subclass distribution for the O-N case forwarded on the NGC 2264 dataset.}
	\vspace{-0.1cm}
	\begin{tabularx}{0.75\hsize}{r l *{6}{Y} l}   
	\multicolumn{2}{c}{}& \multicolumn{7}{c}{\textbf{Actual}}\\      
	\cmidrule[\heavyrulewidth](lr){2-9}
	\parbox[l]{0.2cm}{\multirow{5}{*}{\rotatebox[origin=c]{90}{\textbf{Predicted}}}} & & CI & CII & Gal & AGNs & Shocks & PAHs & Stars\\
	\cmidrule(lr){3-9}
	 &  CI YSOs & 74 & 6 & 0 & 3 & 5 & 0 & 1 \\
	 &  CII YSOs & 2 & 402 & 6 & 2 & 0 & 0 & 44 \\
	 &  Others & 14 & 27 & 108 & 245 & 0 & 1 & 6848 \\
	\cmidrule[\heavyrulewidth](lr){2-9}
	\end{tabularx}
	\vspace{-0.2cm}
	\label{tab:ON_all-sub}
\end{table}

One must note that, in order to properly compare the results, we needed to keep the exact same networks that produced the results in Tables~\ref{tab:OO}, \ref{tab:OO_all}, \ref{tab:NN}, \ref{tab:NN_all}. Therefore, we did not estimate the dispersion of the prediction regarding the weight initialization, and the training set random selection on the O-N and N-O cases. \\

\subsubsection{O-N main result}

\begin{figure*}[t]
	\centering
	\begin{subfigure}[t]{0.44\textwidth}
	\caption*{\textbf{Orion}}
	\includegraphics[width=\textwidth]{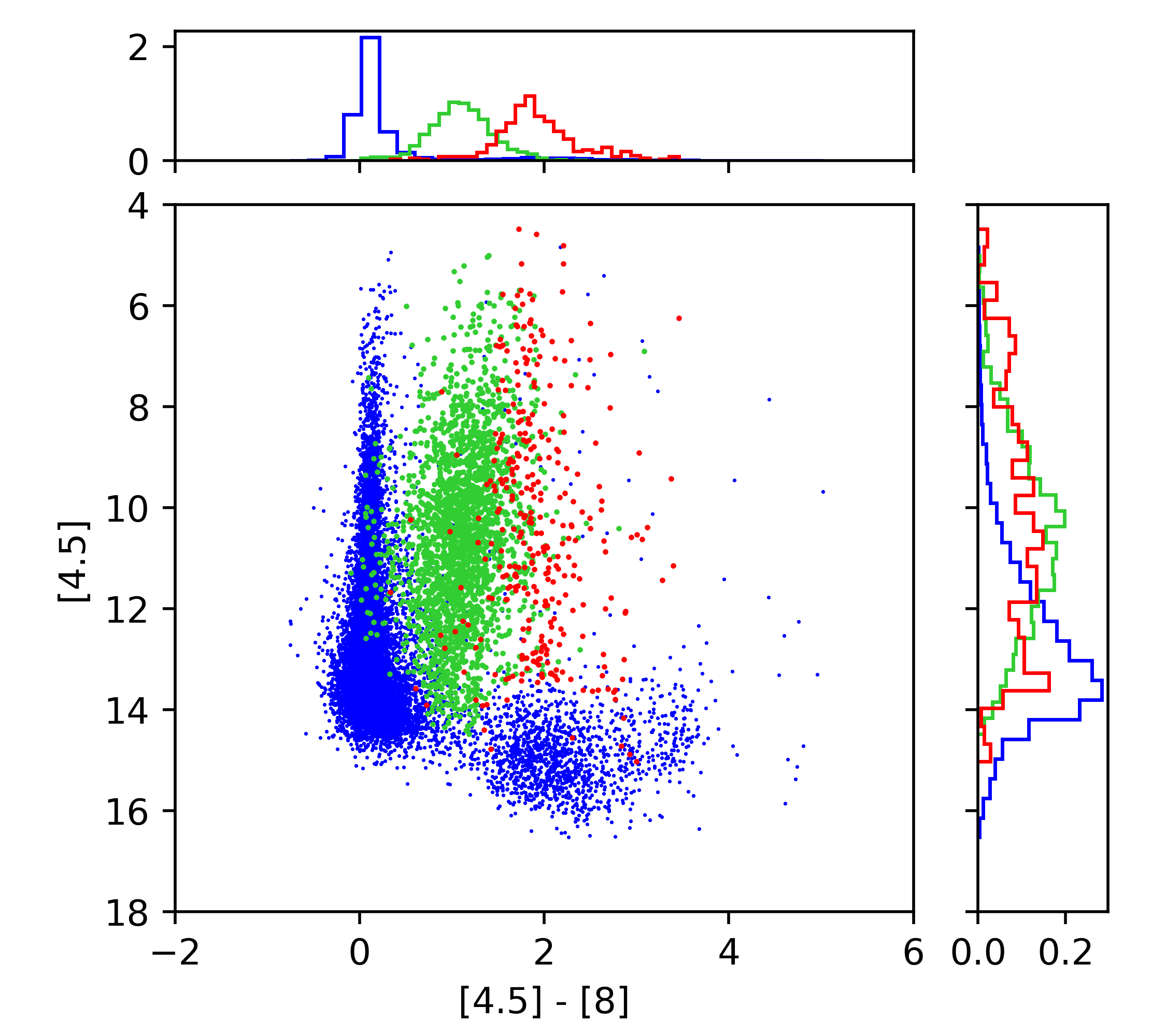}
	\end{subfigure}
	\begin{subfigure}[t]{0.44\textwidth}
	\caption*{\textbf{NGC 2264}}
	\includegraphics[width=\textwidth]{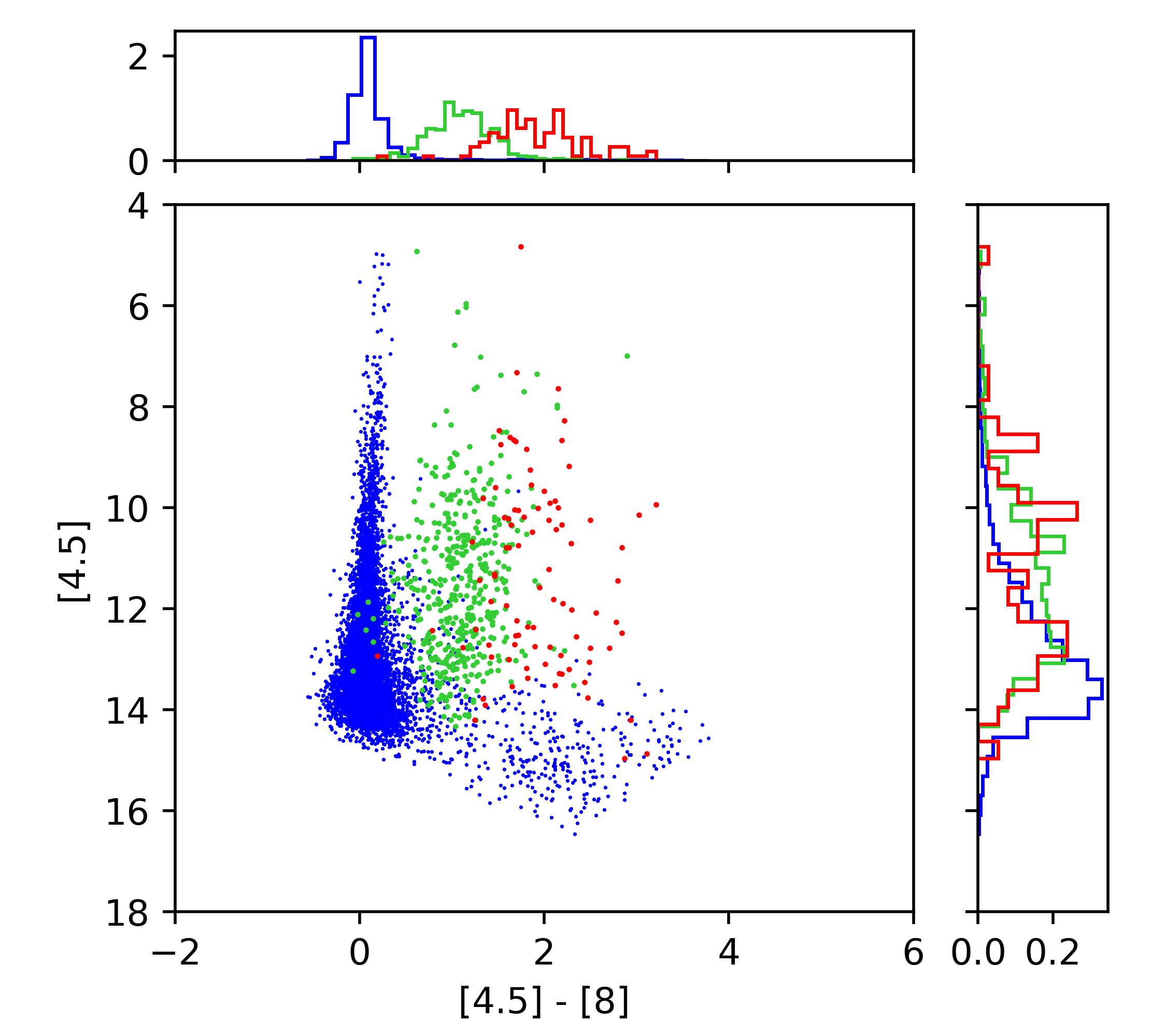}
	\end{subfigure}
	\\\vspace{0.3cm}
	\begin{subfigure}[t]{0.44\textwidth}
	\caption*{\textbf{Combined}}
	\includegraphics[width=\textwidth]{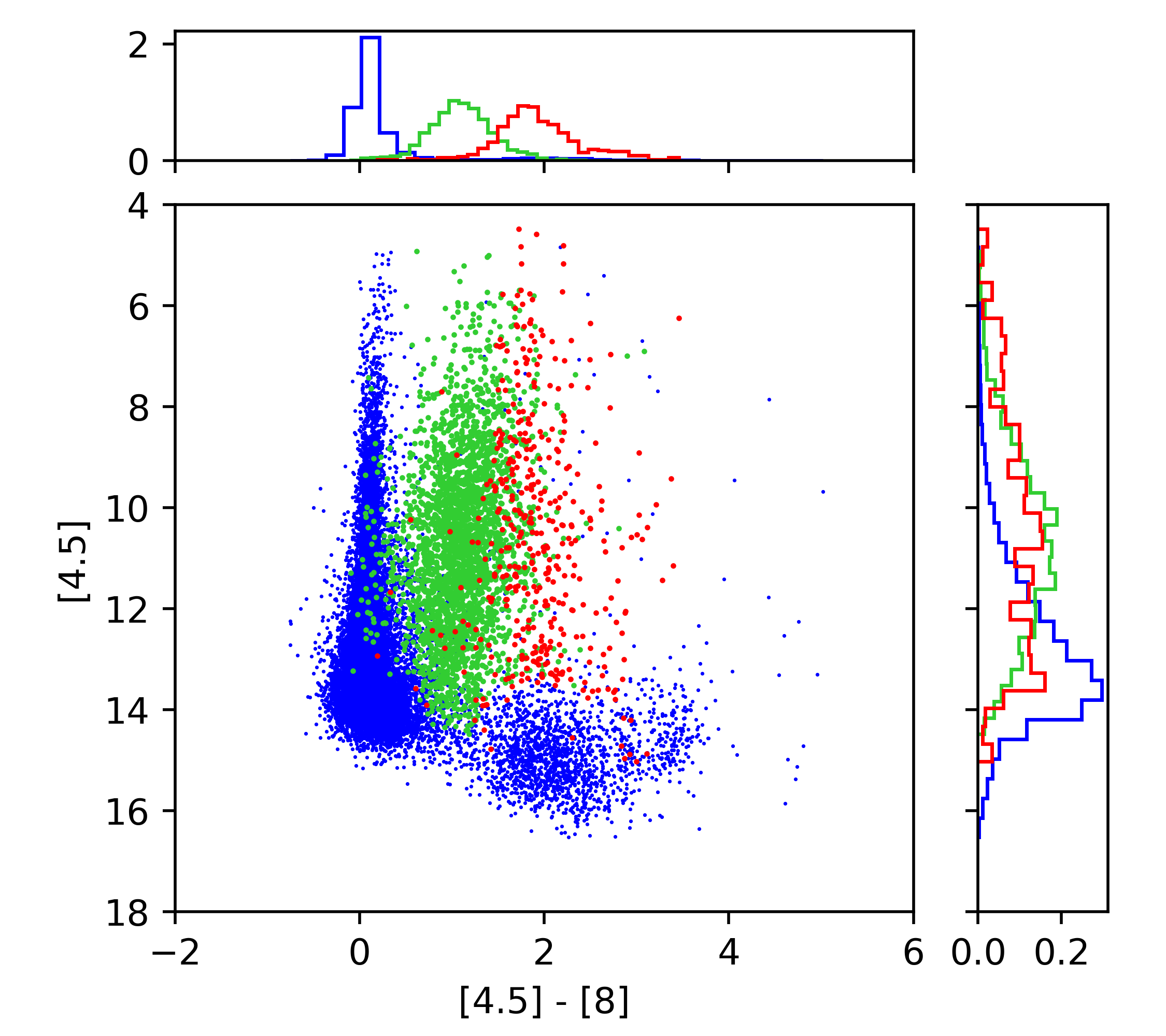}
	\end{subfigure}
	\begin{subfigure}[t]{0.44\textwidth}
	\caption*{\textbf{Combined + 1\,kpc}}
	\includegraphics[width=\textwidth]{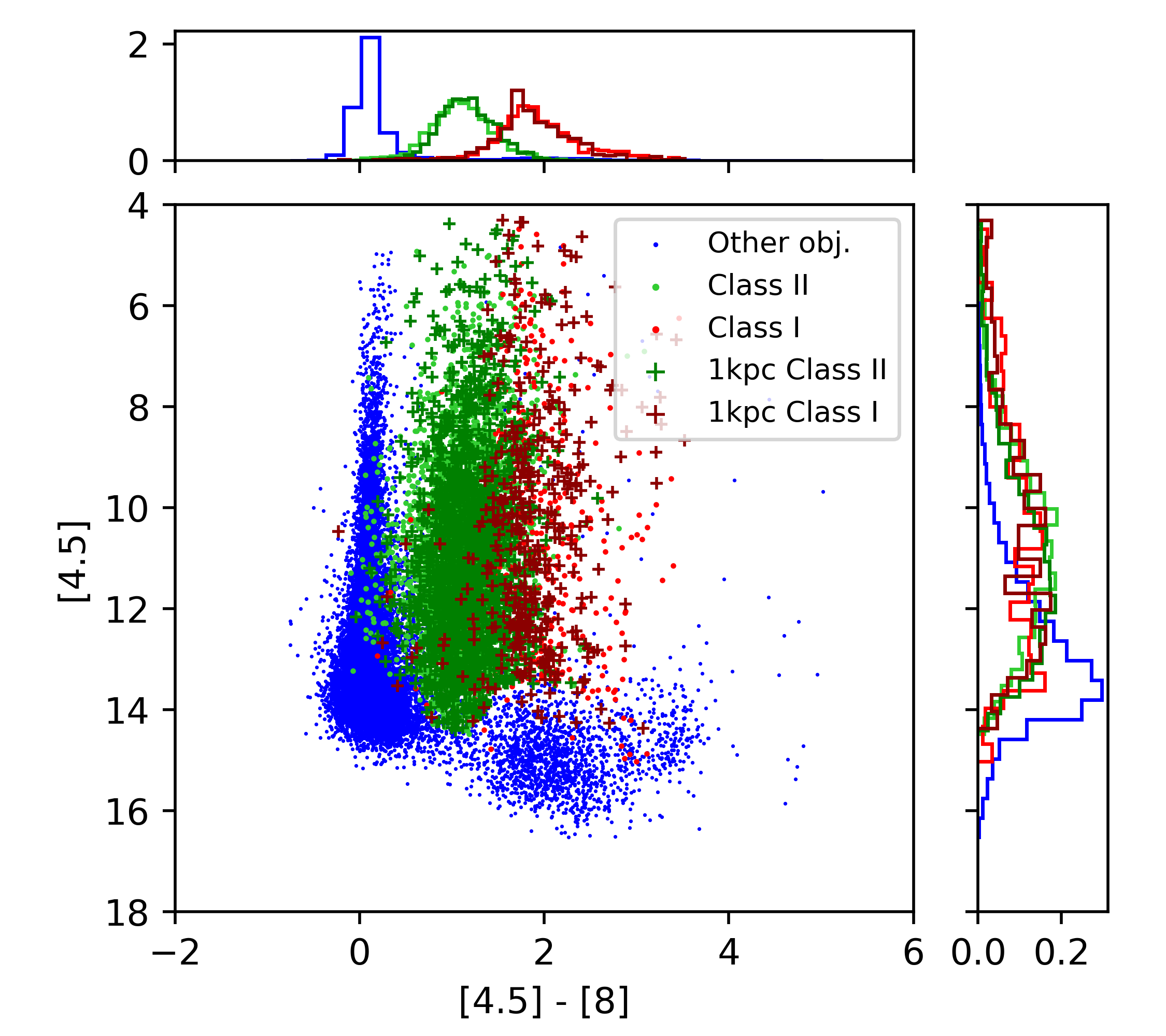}
	\end{subfigure}
	\caption[Differences in feature space coverage for our datasets]{Illustration of the differences in feature space coverage for our datasets. The CI YSOs, CII YSOs, and contaminants are shown in red, green, and blue, respectively, according to the simplified G09 classification scheme. The crosses in the last frame show the YSOs from the 1\,kpc sample. In the side frames, the area of each histogram is normalized to one.}
\vspace{-1cm}
\label{datasets_space_coverage}
\end{figure*}

\begin{figure*}[!t]
	\centering
	\begin{subfigure}[t]{0.49\textwidth}
	\caption*{\textbf{Missed}}
	\includegraphics[width=\textwidth]{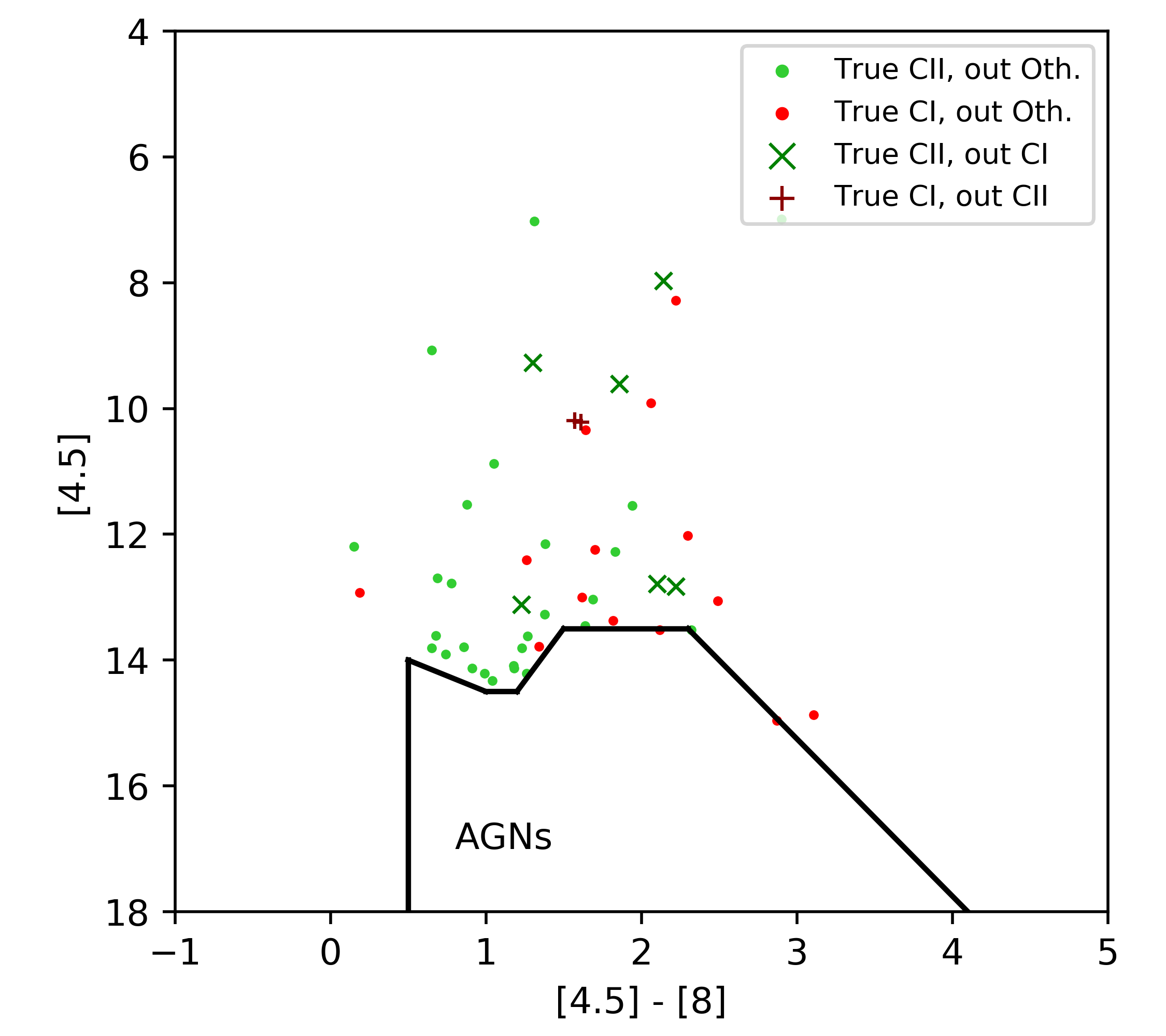}
	\end{subfigure}
	\begin{subfigure}[t]{0.49\textwidth}
	\caption*{\textbf{Wrong}}
	\includegraphics[width=\textwidth]{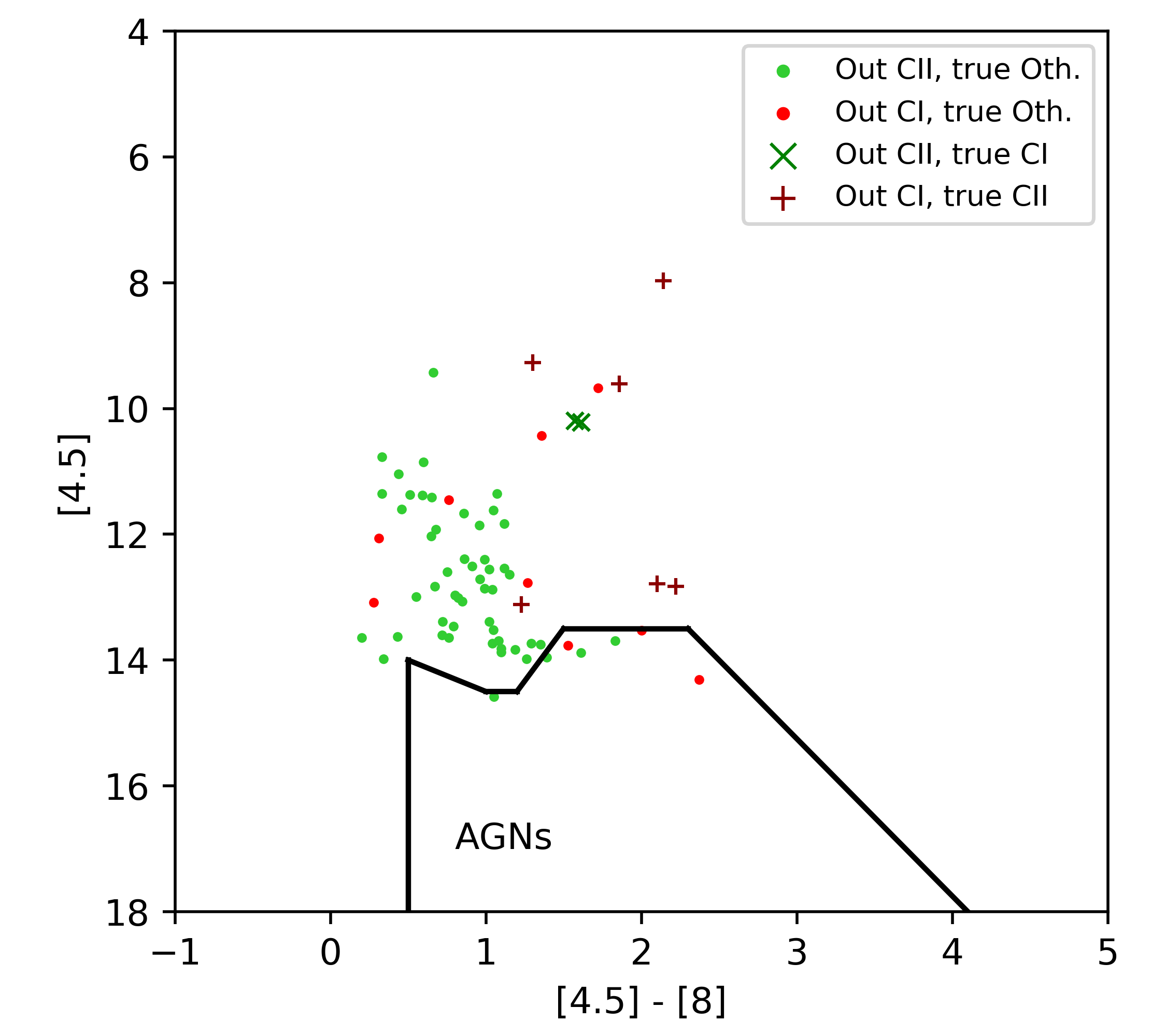}
	\end{subfigure}
	\caption[Space coverage of misclassified objects in the O-N case]{Space coverage of misclassified objects in the O-N case. \textit{Left}: Genuine CI and CII according to the labeled dataset that were misclassified by the network. Green is for CII YSOs, red for CI YSOs. The points and crosses indicate the network output as indicated in the legend. \textit{Right}: Predictions of the network that are known to be incorrect based on the labeled dataset. Green is for predicted CII YSOs, red for predicted CI YSOs. The points and crosses indicate the genuine class as indicated in the legend.}
	\label{orion_fwd_2264_space_coverage}
\end{figure*}

Regarding the results from O-N in Tables~\ref{tab:ON_all} and \ref{tab:ON_all-sub}, we see that the recall for CI YSOs is lower by approximately $8\%$ than the one on the O-O case (Table~\ref{tab:OO}) and lower by approximately $12\%$ when compared to the Orion full dataset results (Table~\ref{tab:OO_all}). Similarly, CII YSOs have a recall lower by approximately $5\%$. This difference being much greater than the dispersion of our results on the O-O case, indicates that the Orion data lack some specific information that is contained in NGC 2264 for these classes. This should correspond to differences in feature space coverage, but these differences might be subtle in the limited set of CMDs considered in the G09 method, whereas the network works directly in the 10-dimension space composed of the 5 bands and 5 errors. For example, as shown in Fig.~\ref{datasets_space_coverage}, it is striking that both YSO classes cover less the upper part of the diagram ([4.5] < 9) in the NGC 2264 case than for Orion. The slopes of the normalized histograms in this figure also illustrate that the density distributions are different between Orion and NGC 2264, especially for CI YSOs. For this population, Orion presents a virtually symmetrical peaked distribution of [4.5]-[8] centered near [4.5]-[8] = 1.9 mag, while NGC 2264 shows a flatter and more skewed distribution. Although subtle, this specificity of the parameter space coverage is in line with the drop in the CI YSO recall in the O-N case, since, in Orion, the area located at [4.5]-[8] $>2$ is less constrained than for [4.5]-[8] $\approx 1.9$, while, in NGC 2264, the area at [4.5]-[8] $>2$ contains a larger fraction of CI YSOs. This interpretation is also consistent with the fact that in the O-N case, CI YSOs are mostly confused with objects from the Other class, in contrast with the O-O and N-N cases, suggesting a lack of constraint for the boundary between the CI and Other classes in the lower-right area of the CI distribution in Fig.~\ref{datasets_space_coverage}, although the differences in class proportions may also contribute. From the perspective of the network, it is likely that the weight values were more influenced by the more abundant updates from objects near the CI peak at [4.5]-[8] = 1.9 mag.\\

\subsubsection{Detailed feature space analysis for O-N}
\label{detailed_feature_space_analysis_ON}

To confirm the previous analysis, Figure~\ref{orion_fwd_2264_space_coverage} shows the distribution of misclassified objects for this O-N case using the same ([4.5]-[8], [4.5]) CMD. To ease the comparison, the misclassified objects are separated in two categories, with "Missed" objects standing for misclassified genuine YSOs, and "Wrong" objects standing for any class being wrongly predicted as a YSO. This representation is equivalent to take either the YSO rows or the YSO columns in the confusion matrix, respectively. As a consequence, CI YSOs misclassified as CII, and CII YSOs misclassified as CI both appear in the two representations. Despite the very small number of CI YSOs of the NGC 2264 dataset, and therefore the few CI YSOs that are misclassified, some trends can still be observed. Regarding missed CI YSOs, more that half of them are misclassified as Other in the bottom part ([4.5]< 12) of the CMD. This indicates that this region is less constrained when training on Orion than when training on NGC 2264, or at least that the learned boundary is not favorable to the NGC 2264 dataset. While the latter region has much less CI training examples in this part of the CMD, the more homogeneous CI YSOs distribution over the feature space of NGC 2264 allows more weights to be dedicated to this specific part. We also noted that the number of CI misclassified as CII is almost the same in N-N and O-N with 3 and 2 objects, respectively (Tables~\ref{tab:NN_all} and \ref{tab:ON_all}). Therefore the drop in CI recall is dominated by these misclassified CI as contaminants. Interestingly, the O-N subclass distribution shows that the misclassified CI are mainly predicted as CII and Shocks, the confusion with AGN being of the same order than the N-N case. Regarding the CII YSOs, the O-N case is more compelling since their recall only slightly dropped, and their precision increased. Therefore, the CII distribution in Orion appears suitable to constrain the CII YSO boundaries which was not the case of NGC~2264.\\

Physically, the observed differences in this CMD are likely to come from the different star formation histories and from the difference in distance between the two regions, respectively between $\sim$ 420 pc for Orion \citep{megeath_spitzer_2012}, and $\sim$ 760 pc for NGC 2264 \citep{rapson_spitzer_2014}. In contrast, the Other class appears to be well represented, suggesting that the Orion training set contains enough objects to represent properly the inherent distribution of this class also in NGC 2264.\\

The changes in precision are less significant than those in recall, due to the differences in class proportions between the two datasets. For example, there is a  $1.58$ factor in the CI over Other ratio between Orion and NGC 2264. The number of misclassified Other as CI is then expected to rise, with a consequent impact on CI precision. However, for this case the improved Other recall between the O-O and O-N case of 0.6\% seems to overcome this effect partly. In contrast, the CII YSOs, for which the proportions are lowered by a $2.24$ factor, indeed suffer a $\sim 8\%$ drop in precision. This strong interplay between proportions and changes in recall for each class makes the differences in precision less prone to analysis.

\subsubsection{N-O main result}
Concerning the results from N-O in Tables~\ref{tab:NO_all} and \ref{tab:NO_all-sub}, the precision of CI YSOs dropped to $65.2\%$, in spite of the number of objects, sufficient not to be affected by small-number effects. This is the worst quality estimator value we observed in the whole study. The precision drop in CII YSOs is less important and only $2\%$ lesser than the NGC 2264 full dataset results. The impact of the differences in feature space coverage is even stronger than for the O-N case, since there are almost no YSOs brighter than [4.5]=9 mag in NGC 2264, therefore a large part of the feature space where many Orion objects lie is left unconstrained. Moreover, the NGC 2264 dataset lacks shocks and PAHs that are present in non negligible proportions in the Orion dataset. Therefore, the NGC 2264 trained network did not constrain them, as confirmed in Table~\ref{tab:NO_all-sub}, where PAHs are evenly scattered in all output classes, and where shocks are completely misclassified as YSOs. In addition to these flaws, the number of objects in the training set is too small to properly constrain the overall network architecture that suits this problem (Sect. \ref{network_tuning}).

\newpage
\subsubsection{Detailed feature space analysis for N-O}

\begin{table}[!t]
	\small
	\centering
	\caption{Confusion matrix for N-O case forwarded on the full Orion dataset.}
	\vspace{-0.3cm}
	\begin{tabularx}{0.65\hsize}{r l |*{3}{m}| r }
	\multicolumn{2}{c}{}& \multicolumn{3}{c}{\textbf{Predicted}}&\\
	\cmidrule[\heavyrulewidth](lr){2-6}
	\parbox[l]{0.2cm}{\multirow{6}{*}{\rotatebox[origin=c]{90}{\textbf{Actual}}}} & Class & CI YSOs & CII YSOs & Others & Recall \\
	\cmidrule(lr){2-6}
	 &  CI YSOs    & 285     & 33      & 6       & 88.0\% \\
	 &  CII YSOs   & 54      & 1967    & 203     & 88.4\% \\
	 &  Others     & 98      & 293     & 16175   & 97.6\% \\
	\cmidrule(lr){2-6} 
	 &  Precision & 65.2\% & 85.8\% & 98.7\% & 96.4\%\\
	\cmidrule[\heavyrulewidth](lr){2-6}
	\end{tabularx}
	\vspace{-0.1cm}
	\label{tab:NO_all} 
\end{table}
	
\begin{table}[!t]
	\small
	\centering
	\vspace{-0.1cm}
	\caption{Subclass distribution for the N-O case forwarded on the full Orion dataset.}
	\begin{tabularx}{0.75\hsize}{r l *{6}{Y} l}   
	\multicolumn{2}{c}{}& \multicolumn{7}{c}{\textbf{Actual}}\\      
	\cmidrule[\heavyrulewidth](lr){2-9}
	\parbox[l]{0.2cm}{\multirow{5}{*}{\rotatebox[origin=c]{90}{\textbf{Predicted}}}} & & CI & CII & Gal & AGNs & Shocks & PAHs & Stars\\
	\cmidrule(lr){3-9}
	 &  CI YSOs & 285 & 54 & 8 & 37 & 12 & 39 & 2 \\
	 &  CII YSOs & 33 & 1967 & 18 & 34 & 15 & 27 & 199 \\
	 &  Others & 6 & 203 & 381 & 1070 & 1 & 21 & 14702 \\
	\cmidrule[\heavyrulewidth](lr){2-9}
	\end{tabularx}
	\vspace{-0.3cm}
	\label{tab:NO_all-sub}
\end{table}

\begin{figure*}[!t]
	\centering
	\begin{subfigure}[t]{0.49\textwidth}
	\caption*{\textbf{Missed}}
	\includegraphics[width=\textwidth]{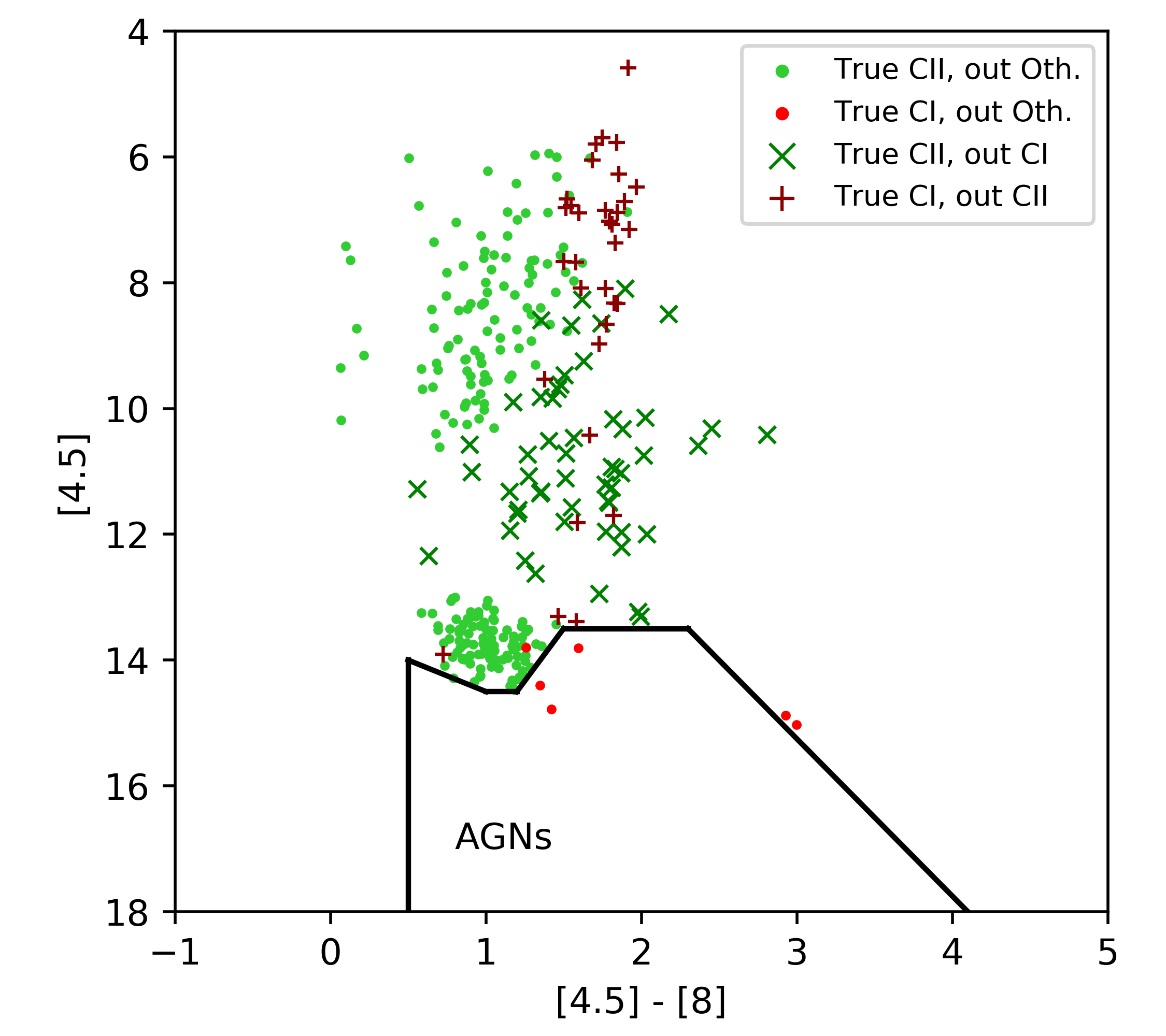}
	\end{subfigure}
	\begin{subfigure}[t]{0.49\textwidth}
	\caption*{\textbf{Wrong}}
	\includegraphics[width=\textwidth]{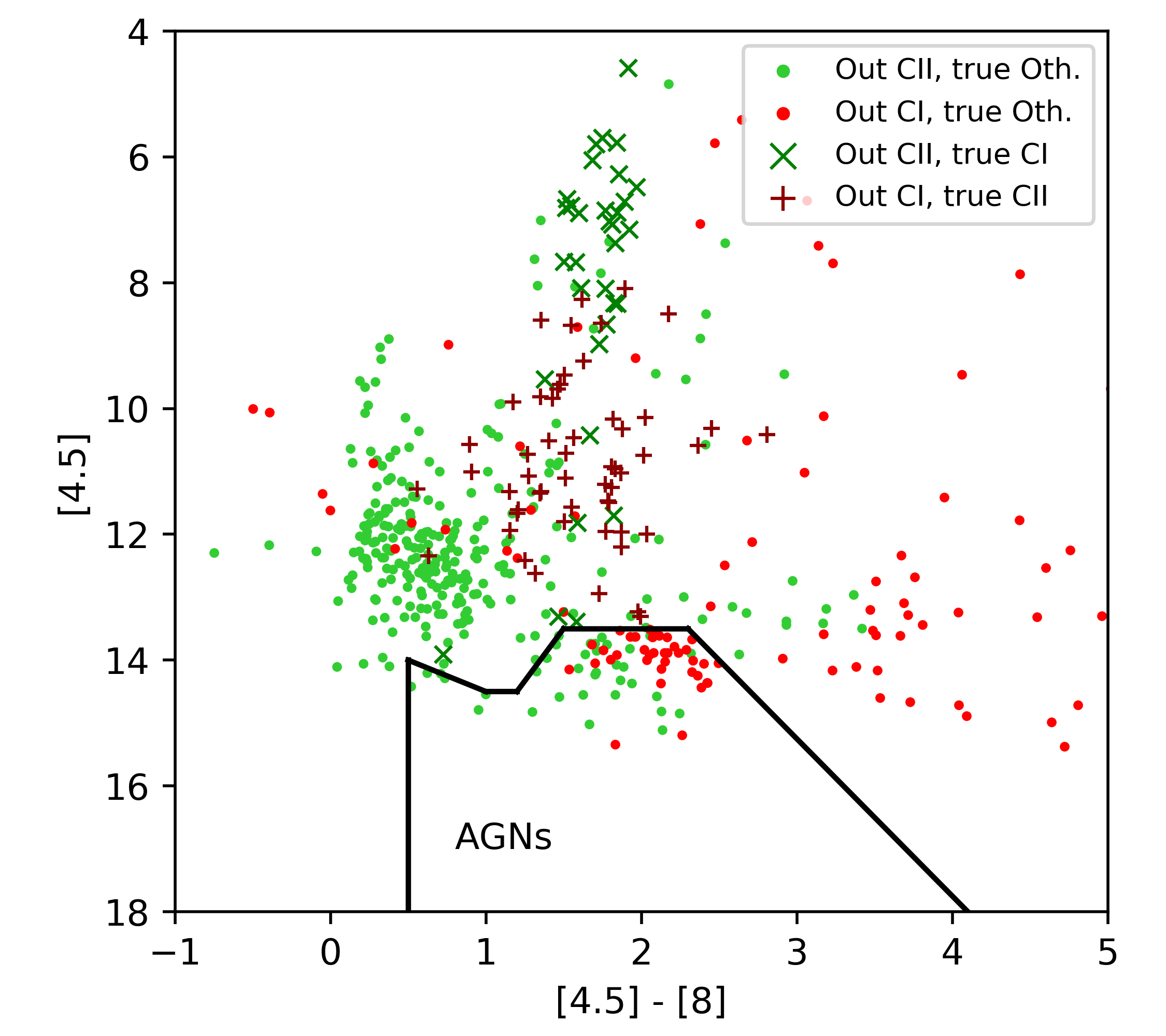}
	\end{subfigure}
	\caption[Space coverage of misclassified objects in the N-O case]{Space coverage of misclassified objects in the N-O case. \textit{Left}: Genuine CI and CII YSOs according to the labeled dataset that were misclassified by the network. Green is for CII YSOs, red for CI YSOs. The points and crosses indicate the network output as indicated in the legend. \textit{Right}: Predictions of the network that are known to be incorrect based on the labeled dataset. Green is for predicted CII YSOs, red for predicted CI YSOs. The points and crosses indicate the genuine class as indicated in the legend.}
	\label{2264_fwd_orion_space_coverage}
\end{figure*}

\vspace{-0.2cm}
Similar to the previous O-N case, Figure~\ref{2264_fwd_orion_space_coverage} shows the distribution of misclassified objects for this N-O case. Thanks to the much larger Orion dataset size, it is much easier to extract the trends regarding the class distribution within the feature space. On the main vertical separation between CI and CII YSOs, around [4.5]-[8] = 1.8, there is a noticeable change in behavior at [4.5] = 9. Bellow this limit ([4.5] > 9), there is a large amount of missed CII classified as CI, while above this limit ([4.5] < 9) it is reversed with more CI misclassified as CII. It perfectly illustrates the fact that in the N-N training it was acceptable to consider every object above this limit and with [4.5]-[8] > 1.8 as CII. In the same way, the highest density of CI and CII near the vertical splitting between CI and CII YSOs that is present in Orion (top left frame in Figure~\ref{datasets_space_coverage}), is smoothed in NGC 2264 (top right frame in Figure~\ref{datasets_space_coverage}). Therefore, this boundary is less constrained, and the class proportions gave the advantage to CI YSOs due to the much lower $\gamma_{CII}$ value in NGC 2264 that allowed to reach good CI recall in the N-N case, but is unsuitable for a generalization to Orion. This behavior strongly decreases the predicted CI YSO precision but is not sufficient to explain the drop to $~65\%$. Most of the contamination comes from contaminants misclassified as CI. In the figure there are two main regions for these objects, below the AGNs limit and on the far right side of the CMD, [4.5]-[8] > 3. This is confirmed by the subclass distribution (Table~\ref{tab:NO_all-sub}), where there is a lot of misclassified AGNs and PAHs as CI, which correspond to these regions. This is directly due to the complete absence of identified PAHs in the NGC 2264 dataset and to the very few number of AGNs that must not be enough to provide a complete coverage of their feature space. Finally, the CII YSOs are contaminated by genuine CI YSOs in the upper part of the diagram, but the main sources of contaminants are the Stars. As before the two frames of Figure~\ref{2264_fwd_orion_space_coverage} show how a lot of misclassified objects fall at the boundary between the two classes. Interestingly, the region [4.5] > 13 that contains genuine CII classified as contaminants (group of green dots in the left frame), is continuous with the region [4.5] < 13 where Contaminants (mainly stars) are misclassified as CII (group of green dots in the right frame), illustrating the misplacement of the network boundary. The upper part of the CMD at [4.5] < 9, contains as before a lot of CII that were missed as contaminants, most likely bright stars. The remaining contamination is visible in the "wrong" frame showing many CII predictions in the AGNs and PAH region, again due to this regions not being constrained properly by the NGC 2264 training.

	\subsection{Improving diversity: combined training}
	\label{cross_train}

The two major limitations identified in the cases of Orion and NGC 2264 are (i) the lack of CI YSOs in the training datasets to be properly constrained by the network, with the associated reduction of other types of objects to avoid dilution, and (ii) the differences in feature space coverage for the two different regions, which induce a lack of generalization capacity toward new star-forming regions. A simple solution to overcome those limitations is to {\bf perform a combined training with the two clouds} (Fig.~\ref{datasets_space_coverage}). We refer to this case, where we merged the labeled samples from Orion and NGC 2264, and used it both to train the network and perform the forward step, as the C-C case. Since the two labeled datasets were obtained with our modified G09 classification, they formed a homogeneous dataset and it was straightforward to combine them. We normalized this new combined dataset as explained in Section~\ref{network_tuning}. The detailed subclass distribution of the target sample for this dataset is presented in Table~\ref{tab_selection}. Thanks to the larger number of CI YSOs in the labeled dataset, we were able to adopt a lower value of $\theta$ ($\theta = 0.2$) to build the test set, which proved to be large enough to mitigate the small-number effects for our output classes. It conserved most data in the training set, where they were needed to improve the classification quality. We note that merging the datasets led to slightly different observational proportions, still realistic enough.

\newpage
\subsubsection{Hyper-parameter and training proportion changes}
\vspace{-0.2cm}
Table~\ref{sat_factors} shows the optimal $\gamma_i$ values obtained with the combined dataset. The $\gamma_i$ values are very similar to those of Orion, as a result of Orion providing two to five times more objects than NGC 2264 to the combined dataset. The dataset is globally larger, so that the optimal number of neurons could have been raised to represent the expected more complex boundaries in the parameter space. However, increasing the number of hidden neurons did not show any improvement of the end results. Thus, we kept 20 hidden neurons for this C-C case. Nevertheless, the larger size of the training set tended to stabilize the convergence of the network during the training, which allowed us to increase the learning rate to $\eta = 4\times 10^{-5}$. As exposed in Section~\ref{network_tuning}, this is counter-intuitive. Indeed, since the weight updates are computed as a sum over the objects in the training sample, they should be greater here than in previous cases, increasing the probability that the network misses potentially good but narrow minima, usually forcing to lower the learning rate. On the other hand, the larger statistics improves the weight space resolution, mitigating meaningless local minima that originate in the limited number of objects. It appears that the latter effect was dominant with smaller dataset forcing a smaller learning rate to properly explore all these minima and find the best one. Using larger dataset then allowed us to raise the learning rate even more. We kept the momentum value at $\alpha = 0.6$ (Sect.~\ref{sect_momentum}) because a greater value happened to make the network diverge in the first steps of training, when the weight corrections were too large.\\

\vspace{-0.9cm}
\subsubsection{Main result}

\begin{table}[!t]
	\small
	\centering
	\caption{Confusion matrix for the C-C case for a typical run.}
	\vspace{-0.2cm}
	\begin{tabularx}{0.65\hsize}{r l |*{3}{m}| r }
	\multicolumn{2}{c}{}& \multicolumn{3}{c}{\textbf{Predicted}}&\\
	\cmidrule[\heavyrulewidth](lr){2-6}
	\parbox[l]{0.2cm}{\multirow{6}{*}{\rotatebox[origin=c]{90}{\textbf{Actual}}}} & Class & CI YSOs & CII YSOs & Others & Recall \\
	\cmidrule(lr){2-6}
	 &  CI YSOs    & 77     & 2       & 3       & 93.9\% \\
	 &  CII YSOs   & 9      & 514     & 8       & 96.8\% \\
	 &  Others     & 9      & 49      & 4706    & 98.8\% \\
	\cmidrule(lr){2-6} 
	 &  Precision & 81.1\% & 91.0\% & 99.8\% & 98.5\%\\
	\cmidrule[\heavyrulewidth](lr){2-6}
	\end{tabularx}
	\vspace{-0.2cm}
	\label{tab:CC}
	\end{table}
	
\begin{table}[!t]
	\small
	\centering
	\caption{Subclass distribution of the C-C case.}
	\vspace{-0.2cm}
	\begin{tabularx}{0.75\hsize}{r l *{6}{Y} l}   
	\multicolumn{2}{c}{}& \multicolumn{7}{c}{\textbf{Actual}}\\      
	\cmidrule[\heavyrulewidth](lr){2-9}
	\parbox[l]{0.2cm}{\multirow{5}{*}{\rotatebox[origin=c]{90}{\textbf{Predicted}}}} & & CI & CII & Gal & AGNs & Shocks & PAHs & Stars\\
	\cmidrule(lr){3-9}
	 &  CI YSOs & 77 & 9 & 1 & 3 & 3 & 2 & 0 \\
	 &  CII YSOs & 2 & 514 & 0 & 3 & 3 & 4 & 39 \\
	 &  Others & 3 & 8 & 103 & 272 & 0 & 11 & 4320 \\
	\cmidrule[\heavyrulewidth](lr){2-9}
	\end{tabularx}
	\vspace{-0.4cm}
	\label{tab:CC-sub}
\end{table}

\vspace{-0.2cm}
The results of this C-C case, presented in Table~\ref{tab:CC} and \ref{tab:CC-sub}, are very close to those on the O-O case with the full Orion dataset. The largest difference is by $0.7\%$ for the precision of CI YSOs. The other differences are $\leq 0.2\%$. The stability of the results regarding both the weight initialization and the random selection of the test and training sets is also very similar to that of the O-O case (Sect. \ref{orion_results}), with recall and precision values scattered by typically $\pm 0.5\%$, except for CI precision which scattered by about $\pm 1\%$. These fluctuations exceed the differences between the O-O and C-C case, as observed from their confusion matrices, when considering the full-dataset forward. This stability was not guaranteed, since, on the one hand, the combined training set is more general than previous training sets, and, on the other hand, the combined training is a more complex problem than a single cloud training, due to the expected more complex distribution of objects in the input parameter space, especially for YSOs.

\newpage
\subsubsection{Generalization capacity evaluation}
If the latter effect dominates, the results could be expected to be poorer than both the O-O and N-N results individually, or any linear combination of them. We illustrate this idea with the following conservative reasoning. If, when using the combined training dataset, the network had only learned from Orion objects, as might be argued due to their dominance in the combined sample, then the state of the network should be very similar to that obtained in the O-O case. The C-C confusion matrix should then be a linear combination of those of the O-O (Table \ref{tab:OO_all}) and O-N cases (Table \ref{tab:ON_all}), weighted by the respective abundances of Orion and NGC 2264 in the forward sampling. The recall of CI YSOs in the O-O and O-N were 94.1\% and 82.2\%, respectively. Since, in the Combined dataset, 78.3\% of CI YSOs come from Orion, the expected recall from an Orion dominated network would be 91.5\%. This results can also be seen as the cell-wise sum of the two matrices, from which the recall and precision are re-computed using the new proportions. Considering the obtained value of 93.9\% in the C-C case(Table \ref{tab:CC}), with a $\pm 1\%$ dispersion, the network has indisputably learned information from the NGC 2264 objects, and the increased complexity of the problem was more than balanced by the increased generality of the sample. In other words, the fact that the results of the C-C test are as good as those of the O-O test in spite of the increased complexity implies that the network managed to take advantage of the greater generality of the combined sample to find a better generalization.\\

The analysis of the other two classes does not contradict those conclusions, although the improvement for CII objects is only marginal, since the same reasoning applied to CII YSOs leads to a recall of 96.2\%, to be compared to the C-C value of 96.8\%, with $\pm 0.5\%$ dispersion. This is in line with the fact that the CII YSO coverage in Orion was already close to that of NGC 2264, as highlighted by the less than $1\%$ difference between the CII YSO recall in O-N and N-N. Finally, contaminants are dominated by subclasses that were already nicely constrained in the O-O and N-N cases.\\

\begin{table}[!t]
	\small
	\centering
	\caption{Confusion matrix for the C-C case forwarded on the full dataset.}
	\vspace{-0.1cm}
	\begin{tabularx}{0.65\hsize}{r l |*{3}{m}| r }
	\multicolumn{2}{c}{}& \multicolumn{3}{c}{\textbf{Predicted}}&\\
	\cmidrule[\heavyrulewidth](lr){2-6}
	\parbox[l]{0.2cm}{\multirow{6}{*}{\rotatebox[origin=c]{90}{\textbf{Actual}}}} & Class & CI YSOs & CII YSOs & Others & Recall \\
	\cmidrule(lr){2-6}
	 &  CI YSOs    & 389    & 14      & 11      & 94.0\% \\
	 &  CII YSOs   & 53     & 2570    & 36      & 96.7\% \\
	 &  Others     & 50     & 254     & 23526   & 98.7\% \\
	\cmidrule(lr){2-6}  
	 &  Precision & 79.1\% & 90.6\% & 99.8\% & 98.4\% \\
	\cmidrule[\heavyrulewidth](lr){2-6}
	\end{tabularx}
	\label{tab:CC_all} 
	\vspace{-0.1cm}
\end{table}

\begin{table}[!t]
	\small
	\centering
	\caption{Subclass distribution for the C-C case forwarded on the full dataset.}
	\vspace{-0.1cm}
	\begin{tabularx}{0.75\hsize}{r l *{6}{Y} l}   
	\multicolumn{2}{c}{}& \multicolumn{7}{c}{\textbf{Actual}}\\      
	\cmidrule[\heavyrulewidth](lr){2-9}
	\parbox[l]{0.2cm}{\multirow{5}{*}{\rotatebox[origin=c]{90}{\textbf{Predicted}}}} & & CI & CII & Gal & AGNs & Shocks & PAHs & Stars\\
	\cmidrule(lr){3-9}
	 &  CI YSOs & 389 & 53 & 2 & 10 & 22 & 11 & 5 \\
	 &  CII YSOs & 14 & 2570 & 4 & 16 & 11 & 15 & 208 \\
	 &  Others & 11 & 36 & 515 & 1365 & 1 & 62 & 21583 \\
	\cmidrule[\heavyrulewidth](lr){2-9}
	\end{tabularx}
	\vspace{-0.1cm}
	\label{tab:CC_all-sub}
\end{table}

\newpage
The fact that the network results for the C-C case are as good as, or better than, for the Orion case despite the added complexity confirms that the number of objects was a strong limitation in the O-O and N-N cases. It also confirms that the O-O training might have provided better results with more observed objects in the same region, which was already established from the improvement of results with lower $\theta$ values in Section~\ref{orion_results}. Moreover, the absence of positive effect when raising the number of neurons demonstrates that the network efficiently combined their respective input parameter space coverage and that $n = 20$ is not limiting these results. The change in observational proportions that occurred by merging the two datasets seems to have a negligible impact as they are still close to the Orion ones, but adding more regions with less YSOs is expected to decrease the precision values for YSOs by increasing their dilution by the Other class.

\vspace{-0.2cm}
\subsubsection{Full dataset result and analysis of rare sub-classes prediction}
The results for the complete combined dataset are presented in Tables~\ref{tab:CC_all} and \ref{tab:CC_all-sub}. As before, the results appear to be free of over-training, since there is no noticeable increase in recall for any of our classes. These results are very similar to the previous ones, with differences in quality estimators of the same order as the dispersion observed with random weight initialization. The slight decrease in precision of CI YSOs is also of the same order as the dispersion obtained from the random selection of our training and test samples. The contaminants that are not sufficiently constrained, like Shocks, could also be affected by selection effects between the two sets, which could lead to such a dispersion in precision for CI YSOs. This seems to be confirmed by the fact that two thirds of the shocks were misclassified as CI YSOs. \\
\label{shocks_discussion}

\vspace{-0.1cm}
Interestingly, this suggests a change in the network behavior, compared to the O-O case, where shocks were almost evenly distributed among the three output classes. We interpret the difference in shock distributions as a consequence of the difference in the relative abundance of this subclass compared to the rest of the training set, and to its strong dependancy to the MIPS rebranding step. Indeed, the special location of Shocks in the feature space, close to CII YSOs and mixed with the MIPS-identified CI YSOs (Fig.~\ref{fig_gut_method} D), makes this subclass identification sensitive to its small relative abundance during the learning process. Thus, in the O-O case, the number of shocks in the sample enabled the network to place the boundary in the vicinity of the Shocks region, but in an inaccurate way, hence the even distribution. Conversely, the lower fraction of shocks in the C-C sample probably made the network find an optimum where most of its representative strength was used for other parts of the feature space. In this situation, the majority of shocks are likely to be included in one specific output class, which can vary according to the random training set selection, but is more likely to be a YSO class, and even more likely to be CI due to the MIPS rebranding step.\\

\vspace{-0.1cm}
To summarize the results of this combined training, we exposed that {\bf combining two star-forming clouds has improved the underlying diversity of our prediction}, and therefore the generalization capability of our network over possible new regions. The added complexity was largely overcome by the increased statistics on our classes of interest, CI and CII YSOs, which allowed to conserve very good accuracy and precision for them. However, some rare contaminant subclasses suffered from their increased dilution. Also, it is worth mentioning that, despite the very good recall we obtained on this C-C result that could convincingly be used to predict other regions, it is still composed of just two star-forming regions that are much more massive than any other at our disposal. This was our motivation to include more regions in the next Section despite the noted limitations of the corresponding 1\,kpc dataset (Sect. \ref{data_setup}).

\newpage
	\subsection{Further increase in diversity and dataset size: nearby regions (< 1kpc)}
	\label{1kpc_train}

In this section, we present the advantages of the 1\,kpc dataset to {\bf further improve the network generalization capacity by increasing the underlying diversity} of the object sample. As discussed in Section~\ref{data_setup}, this dataset only contains YSOs. This is not a major issue, because most of our contaminant subclasses are already well constrained, while we have shown that it is not the case for YSOs, since adding more of them led to a better generalization. Moreover, as the dataset contains several regions, it should ensure an even better diversity and input parameter space coverage for YSOs than the previous C-C case, but it might also increase again the underlying distribution complexity (Fig.~\ref{datasets_space_coverage}). In this section, we study the F-C case, that is a training on the full 1\,kpc dataset (combined + 1\,kpc YSOs) and a forward on the combined dataset, to keep a realistic test dataset with almost observational proportions. As before, the full 1\,kpc dataset is normalized as described in Section~\ref{network_tuning}.\\

\subsubsection{Hyper-parameter and training proportion changes}
The detailed $\gamma_i$ selection for this more complicated dataset is presented in Table~\ref{sat_factors}. As we added YSOs, we had to increase the number of contaminants to preserve their dominant representation in the training sample. However, some subclasses of contaminants were already too few in the C-C case and already included in the training set as much as possible. Therefore, we did not add all the CI YSOs at our disposal to avoid a too strong dilution of these subclasses of Contaminants. For objects from the combined dataset, we kept $\theta = 0.2$, giving the now usual $(1-\theta)$ CI YSOs in the training sample, which are doubled using the 1kpc dataset. For CII YSOs, results were better when taking a slightly fewer proportion of them from the 1\,kpc dataset. In the same manner as for the other datasets, we tried various numbers of neurons in the hidden layer, with for the first time a higher optimum value around $n=30$. This means that we certainly have sufficiently raised the number of objects to break previously existing limitations regarding the size of the network. We also took advantage of the larger dataset and adopted greater values for $\eta = 8 \times 10^{-5}$ and $\alpha = 0.8$, which proved to stabilize more the network than smaller value, following the trend already described in Section~\ref{cross_train}.\\

\subsubsection{Main result}
The results for this F-C case are presented in Tables~\ref{tab:FC} and \ref{tab:FC-sub}. The precision of CI YSOs has dropped by $2.5\%$, but all the other precisions have slightly improved. Compared to the C-C case (Table~\ref{tab:CC}), the precision of CI YSOs raised by $2.8\%$, but the recall is significantly lowered with a drop by nearly $5\%$. In contrast, the precision of CII YSOs dropped by $1.2\%$, and the recall improved by $0.8\%$. Overall, these results are similar to the previous C-C case, despite the increase in complexity coming from the addition of YSOs from new star forming regions. Similarly to the combination of Orion and NGC 2264, we could have observed a stronger drop in quality estimators, because the problem becomes more general and therefore more difficult to constrain. It is worth noting that the stability of the network somewhat decreased in comparison to the O-O and C-C cases. We observed a dispersion of recall regarding the weight random initialization of about $\pm 1\%$ for CI and $\pm 0.7\%$ for CII YSOs. This dispersion affects less the Other class with a value around $\pm 0.15\%$. The precision is less reliable with a dispersion of nearly $\pm 1.5\%$ for CI YSOs. The precision dispersion for CII YSOs is around $\pm 0.5\%$ and is less than $\pm 0.1\%$ for the Other class.\\

\vspace{-0.4cm}
\subsubsection{More detailed analysis}
More generally, the sources of contamination of the YSO classes have not changed, their overall effect has just risen. The fact that raising the number of neurons from $20$ to $30$ in the network leads to better results is certainly an indication of the increased complexity of this problem. This means that the network uses more refined splittings in the input parameter space. However, there might not be enough objects in our dataset to perfectly constrain this larger network, despite the added YSOs. This naturally leads to a stronger sensitivity to the weight initialization. In contrast, the dispersion over the training set random selection is similar to the one observed on the C-C case and is of the same order as the weight initialization dispersion. As in the previous cases, the results show that the main source of contamination of CI YSOs are the CII YSOs, while the latter are mostly contaminated by the Other class. This is, again, an indication of the respective proximity of the three classes in the input parameter space.\\

\begin{table}[!t]
	\small
	\centering
	\caption{Confusion matrix for the F-C case for a typical run.}
 \vspace{-0.1cm}
	\begin{tabularx}{0.65\hsize}{r l |*{3}{m}| r }
	\multicolumn{2}{c}{}& \multicolumn{3}{c}{\textbf{Predicted}}&\\
	\cmidrule[\heavyrulewidth](lr){2-6}
	\parbox[l]{0.2cm}{\multirow{6}{*}{\rotatebox[origin=c]{90}{\textbf{Actual}}}} & Class & CI YSOs & CII YSOs & Others & Recall \\
	\cmidrule(lr){2-6}
	 &  CI YSOs    & 73      & 4       & 5       & 89.0\% \\
	 &  CII YSOs   & 9       & 518     & 4       & 97.6\% \\
	 &  Others     & 5       & 55      & 4704    & 98.7\% \\
	\cmidrule(lr){2-6}
	 &  Precision & 83.9\% & 89.8\% & 99.8\% & 98.5\% \\
	\cmidrule[\heavyrulewidth](lr){2-6}
	\end{tabularx}
	\vspace{-0.1cm}
	\label{tab:FC} 
\end{table}

\begin{table}[!t]
	\small
	\centering
	\caption{Subclass distribution for the F-C case.}
	\vspace{-0.1cm}
	\begin{tabularx}{0.75\hsize}{r l *{6}{Y} l}   
	\multicolumn{2}{c}{}& \multicolumn{7}{c}{\textbf{Actual}}\\      
	\cmidrule[\heavyrulewidth](lr){2-9}
	\parbox[l]{0.2cm}{\multirow{5}{*}{\rotatebox[origin=c]{90}{\textbf{Predicted}}}} & & CI & CII & Gal & AGNs & Shocks & PAHs & Stars\\
	\cmidrule(lr){3-9}
	 &  CI YSOs & 73 & 9 & 0 & 0 & 1 & 2 & 2 \\
	 &  CII YSOs & 4 & 518 & 1 & 6 & 5 & 6 & 37 \\
	 &  Others & 5 & 4 & 102 & 272 & 0 & 9 & 4321 \\
	\cmidrule[\heavyrulewidth](lr){2-9}
	\end{tabularx}
	\vspace{-0.6cm}
	\label{tab:FC-sub}
\end{table}

The increased number of objects allowed us to see more details on the subclass distribution across the output classes. Similarly to the C-C case, the Shocks behave as completely unconstrained, since they end up in mostly one class, which changes randomly when the training is repeated. Compared to the C-C case, this effect is stronger, most likely because we did not add any Shocks in the training sample, therefore increasing their dilution. For almost any of the other subclasses, the variations are quite within the dispersion, with a slight trend for contaminant subclasses (Galaxies, AGNs, Shocks, PAH) to be less well classified, and CII YSOs and Stars to be better classified. One may expect these results, because we increased the number of YSOs and Stars in the training sample. On the other hand, we also have increased the YSO distribution complexity, which could lead to worse overall results. Possibly, this induced the slight drop in CI YSO recall observed from C-C to F-C, whereas CII YSOs and Others kept their quality indicators stable, either due to the increased statistics, or because their input feature space was already properly constrained by the Combined dataset (C-C case).\\

\begin{table}[!t]
	\small
	\centering
	\caption{Confusion matrix for the F-C case forwarded on the full combined dataset.}
	\vspace{-0.2cm}
	\begin{tabularx}{0.65\hsize}{r l |*{3}{m}| r }
	\multicolumn{2}{c}{}& \multicolumn{3}{c}{\textbf{Predicted}}&\\
	\cmidrule[\heavyrulewidth](lr){2-6}
	\parbox[l]{0.2cm}{\multirow{6}{*}{\rotatebox[origin=c]{90}{\textbf{Actual}}}} & Class & CI YSOs & CII YSOs & Others & Recall \\
	\cmidrule(lr){2-6}
	 &  CI YSOs    & 378     & 22      & 14      & 91.3\% \\
	 &  CII YSOs   & 45      & 2584    & 30      & 97.2\% \\
	 &  Others     & 43      & 244     & 23543   & 98.8\% \\
	\cmidrule(lr){2-6}
	 &  Precision & 81.1\% & 90.7\% & 99.8\% & 98.5\% \\
	\cmidrule[\heavyrulewidth](lr){2-6}
	\end{tabularx}
	\vspace{-0.1cm}
	\label{tab:FC_all}
\end{table}

\begin{table}[!t]
	\small
	\centering
	\caption{Subclass distribution for the F-C case forwarded on the full combined dataset.}
	\vspace{-0.2cm}
	\begin{tabularx}{0.75\hsize}{r l *{6}{Y} l}   
	\multicolumn{2}{c}{}& \multicolumn{7}{c}{\textbf{Actual}}\\      
	\cmidrule[\heavyrulewidth](lr){2-9}
	\parbox[l]{0.2cm}{\multirow{5}{*}{\rotatebox[origin=c]{90}{\textbf{Predicted}}}} & & CI & CII & Gal & AGNs & Shocks & PAHs & Stars\\
	\cmidrule(lr){3-9}
	 &  CI YSOs & 378 & 45 & 0 & 15 & 8 & 15 & 5 \\
	 &  CII YSOs & 22 & 2584 & 6 & 22 & 25 & 14 & 177 \\
	 &  Others & 14 & 30 & 515 & 1354 & 1 & 59 & 21614 \\
	\cmidrule[\heavyrulewidth](lr){2-9}
	\end{tabularx}
	\vspace{-0.3cm}
	\label{tab:FC_all-sub}
\end{table}

\newpage
\vspace{-0.3cm}
\subsubsection{Full dataset result}
\vspace{-0.1cm}
The results of a forward of the complete combined dataset using this network are shown in Table~\ref{tab:FC_all}, with the subclass distributions in Table~\ref{tab:FC_all-sub}. These results show a $2.3\%$ increase in the CI YSO recall compared to Table~\ref{tab:FC} and a $2.8\%$ drop in precision for the same class. As for all the previous cases, the Other class remained almost identical. For CII YSOs and Other, the variations in precision and recall are within the weight initialization dispersion. The case of CI YSOs is less clear as their recall increase is greater than their dispersion, which could mean that there is a slight over-training. However, when searching for the optimum set of $\gamma_i$ values, we observed that the sets leading to a lesser over-training of CI YSOs also degraded the overall quality of the results. Still, it suggests that the genuine CI YSO recall is between the values of Table~\ref{tab:FC} and Table~\ref{tab:FC_all}.\\

\vspace{-0.8cm}
\subsubsection{Misclassified objects distribution}
\vspace{-0.1cm}
\begin{figure*}[!t]
	\centering
	\begin{subfigure}[t]{0.48\textwidth}
	\caption*{\textbf{Missed}}
	\includegraphics[width=\textwidth]{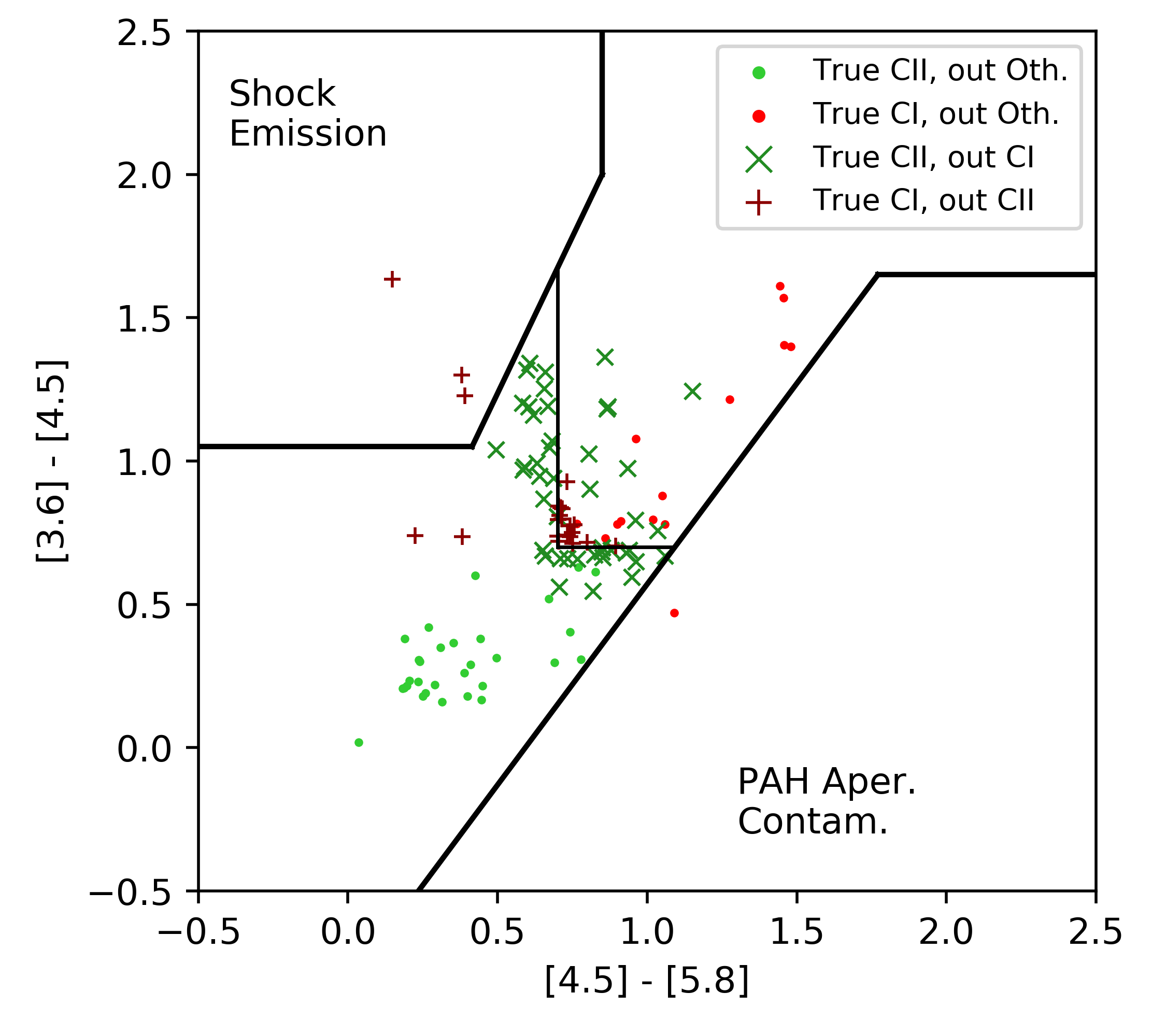}
	\end{subfigure}
	\begin{subfigure}[t]{0.48\textwidth}
	\caption*{\textbf{Wrong}}
	\includegraphics[width=\textwidth]{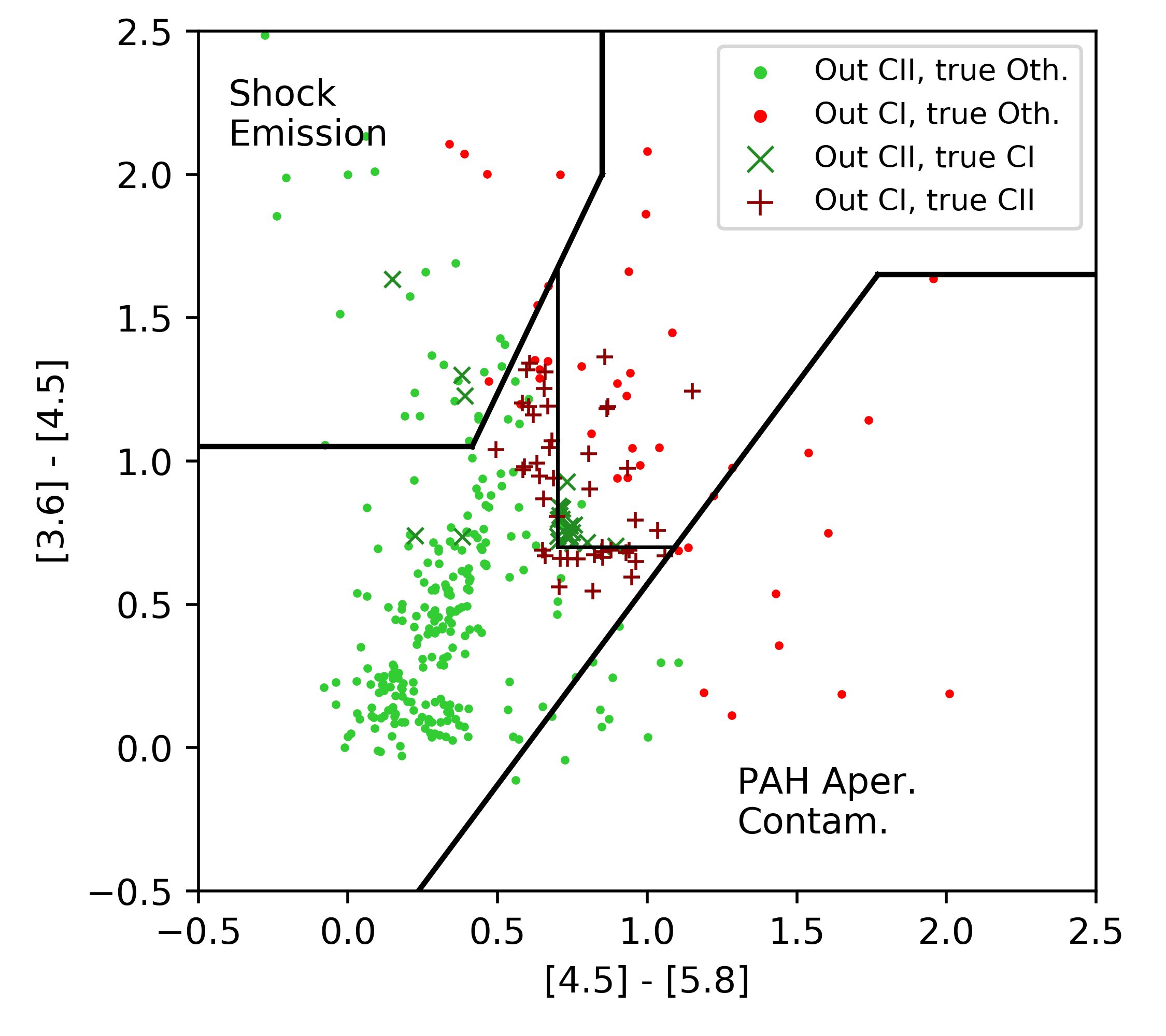}
	\end{subfigure}
	\caption[Zoom on misclassified objects in the F-C case]{Zoom on the $[4.5] - [5.8]\, \text{vs.}\, [3.6] - [4.5]$ graph, for misclassified objects in the F-C case. \textit{Left:} Genuine CI and CII YSOs according to the labeled dataset that were misclassified by the network. Green is for CII YSOs, red for CI YSOs. The points and crosses indicate the network output as indicated in the legend. \textit{Right} Predictions of the network that are known to be incorrect based on the labeled dataset. Green is for predicted CII YSOs, red for predicted CI YSOs. The points and crosses indicate the genuine class as indicated in the legend.}
	\label{missed_wrong_zoom}
\end{figure*}

Since {\bf the present F-C case is our most complete result}, and in order to provide additional verification of the limitations we exposed, we checked the distribution of the misclassified objects in the same fashion as in Section~\ref{cross_train}. For that, Figure~\ref{missed_wrong_space} at the end of the section shows the same five usual CMDs as in Section~\ref{data_prep} for the labeled distribution from our G09 method, the predicted distribution by the F-C network, the "missed" and the "wrong" CI and CII YSOs, all for the full Combined dataset. Figure~\ref{missed_wrong_zoom} presents a zoom on the fourth row to ease the comparison. As for the previous result we did not forward on the 1\,kpc sample since we do not have contaminant estimates and therefore we cannot provide quality estimates. They are just used to enhance the YSO diversity through their feature space coverage in the training process. From this figure, we observe that there are two main misclassification zones. The first one between CI and CII YSOs, which can mostly be seen in the fourth row (Fig.~\ref{missed_wrong_zoom}). And the second one between CII YSOs and more evolved Stars, which can mostly be seen in the second row. This is strongly consistent with our previous interpretation of the object distribution in the various confusion matrix of the F-C case. Some other contamination area can also be seen like in the AGNs exclusion region close to the corresponding cut, or like in the Shocks exclusion region. The fact that the misclassified objects mostly stacks along the cuts is a strong indication of where the network is less precise,this is also a first indication that the membership probability could be used to improve the results.

\subsubsection{Forward of the trained network on Orion and NGC 2264}

We also looked at the F-C trained network prediction over Orion and NGC 2264 individually, hereafter the F-O and F-N cases. This allowed us to verify that they are both properly represented by the F-C network and that one is not responsible for the majority of the misclassified objects. We present in Tables \ref{tab:FO} and \ref{tab:FN} the confusion matrices that represent the predictions on the F-O and F-C cases, respectively. For this, we used the full labeled datasets of the each region, meaning that they must be compared to the full combined dataset prediction from Table \ref{tab:FC_all}. Regarding the F-O result, we can see that the changes in recall and precision are very small and within the dispersion for each class. For the F-N result, it is also very stable but with an average of 2\% change with respect to Table \ref{tab:FC_all} in all the YSO quality estimators. However, the two regions are expected not to have the same prediction due to the fact that they can sometimes constrain different parts of the feature space. Therefore, the full combined dataset result in Table~\ref{tab:FC_all} is a sum of the two individual predictions from Tables~\ref{tab:FO} and \ref{tab:FN}, which implies that the quality estimator for both of them may individually get out of the dispersion range estimated around the mean value of the quality estimator of the full dataset. Overall, both the individual predictions remain satisfying.\\

Interestingly, these F-O and F-N predictions can be compared to the individual training on Orion and NGC 2264, with the O-O case in Table~\ref{tab:OO_all} and the N-N case in Table~\ref{tab:NN_all}, respectively.
For Orion, the F-O results are very similar to the O-O results, with a slightly better recall on the CII YSOs with a 0.6\% increase. The CI YSOs, however, are slightly less well represented with a drop by 3.4\% in recall, which seems mainly caused by an increased confusion with CII YSOs. Still, we suspected a slight overtraining of the CI YSOs in this O-O case that might explain the difference and could be confirmed by the fact that the CI recall between F-O and the test dataset only O-O results are almost identical, the dispersion of the F-C training being lower (Table~\ref{tab:OO}). However, the O-O case did not show this stronger confusion between CI and CII, which tends to indicate that the numerical similitude between F-O and O-O does not correspond to the same underlying classification properties. It means that the two results cannot be directly compared, and prevents a strong conclusion on the overtraning of the O-O. Still, this result remains very strong since we achieved almost identical prediction, and better CII prediction, with our much more generalist ANN training. Indeed, as we stated previously we could have expect to obtain a classifier that works well enough on all regions but that would be significantly poorer than any individual training.\\

\begin{table}[!t]
	\small
	\centering
	\caption{Confusion matrix for the F-O case.}
	\vspace{-0.1cm}
	\begin{tabularx}{0.65\hsize}{r l |*{3}{m}| r }
	\multicolumn{2}{c}{}& \multicolumn{3}{c}{\textbf{Predicted}}&\\
	\cmidrule[\heavyrulewidth](lr){2-6}
	\parbox[l]{0.2cm}{\multirow{6}{*}{\rotatebox[origin=c]{90}{\textbf{Actual}}}} & Class & CI YSOs & CII YSOs & Others & Recall \\
	\cmidrule(lr){2-6}
	 &  CI YSOs    & 294     & 20      & 10      & 90.7\% \\
	 &  CII YSOs   & 35      & 2170    & 19      & 97.6\% \\
	 &  Others     & 36      & 191     & 16339   & 98.6\% \\
	\cmidrule(lr){2-6}
	 &  Precision & 80.5\% & 91.1\% & 99.8\% & 98.4\% \\
	\cmidrule[\heavyrulewidth](lr){2-6}
	\end{tabularx}
	\vspace{-0.1cm}
	\label{tab:FO}
\end{table}

\begin{table}[!t]
	\small
	\centering
	\caption{Confusion matrix for the F-N case.}
	\vspace{-0.1cm}
	\begin{tabularx}{0.65\hsize}{r l |*{3}{m}| r }
	\multicolumn{2}{c}{}& \multicolumn{3}{c}{\textbf{Predicted}}&\\
	\cmidrule[\heavyrulewidth](lr){2-6}
	\parbox[l]{0.2cm}{\multirow{6}{*}{\rotatebox[origin=c]{90}{\textbf{Actual}}}} & Class & CI YSOs & CII YSOs & Others & Recall \\
	\cmidrule(lr){2-6}
	 &  CI YSOs    & 84      & 2       & 4       & 93.3\% \\
	 &  CII YSOs   & 10      & 414     & 11      & 95.2\% \\
	 &  Others     & 7       & 53      & 7204    & 99.2\% \\
	\cmidrule(lr){2-6}
	 &  Precision & 83.2\% & 88.3\% & 99.8\% & 98.9\% \\
	\cmidrule[\heavyrulewidth](lr){2-6}
	\end{tabularx}
	\vspace{-0.1cm}
	\label{tab:FN}
\end{table}

The results are much more striking for NGC 2264, with the F-N prediction being much better than the N-N prediction for CII YSOs (+1.9\%) and contaminants (+0.4\%), which allows an increase of CI precision of 1\% despite their 4.5\% drop in recall. This is not very surprising that the CI YSOs from NGC 2264 are less well represented since their distribution is very peculiar and that they are not numerous enough in the full 1\,kpc training dataset to be better constrained than the N-N case that focus only on this specific distribution. Still, the confusion between CI and CII YSOs is not very impacted and the confusion between CII and contaminants is significantly improved, which is as before a very strong results considering that it is the result of the more generalist network. This F-N case can also be compared to the O-N case (Table~\ref{tab:ON_all}), where it is striking from the 11.1\% CI recall increase and the 2.8 \% CII recall increase, that this full 1\,kpc training is much more suitable than Orion alone to generalize over other regions. \\

\begin{sidewaysfigure}
	\centering
	\begin{subfigure}[t]{0.24\textwidth}
	\caption*{\textbf{\hspace{0.3cm}Actual}}
	\includegraphics[width=\textwidth]{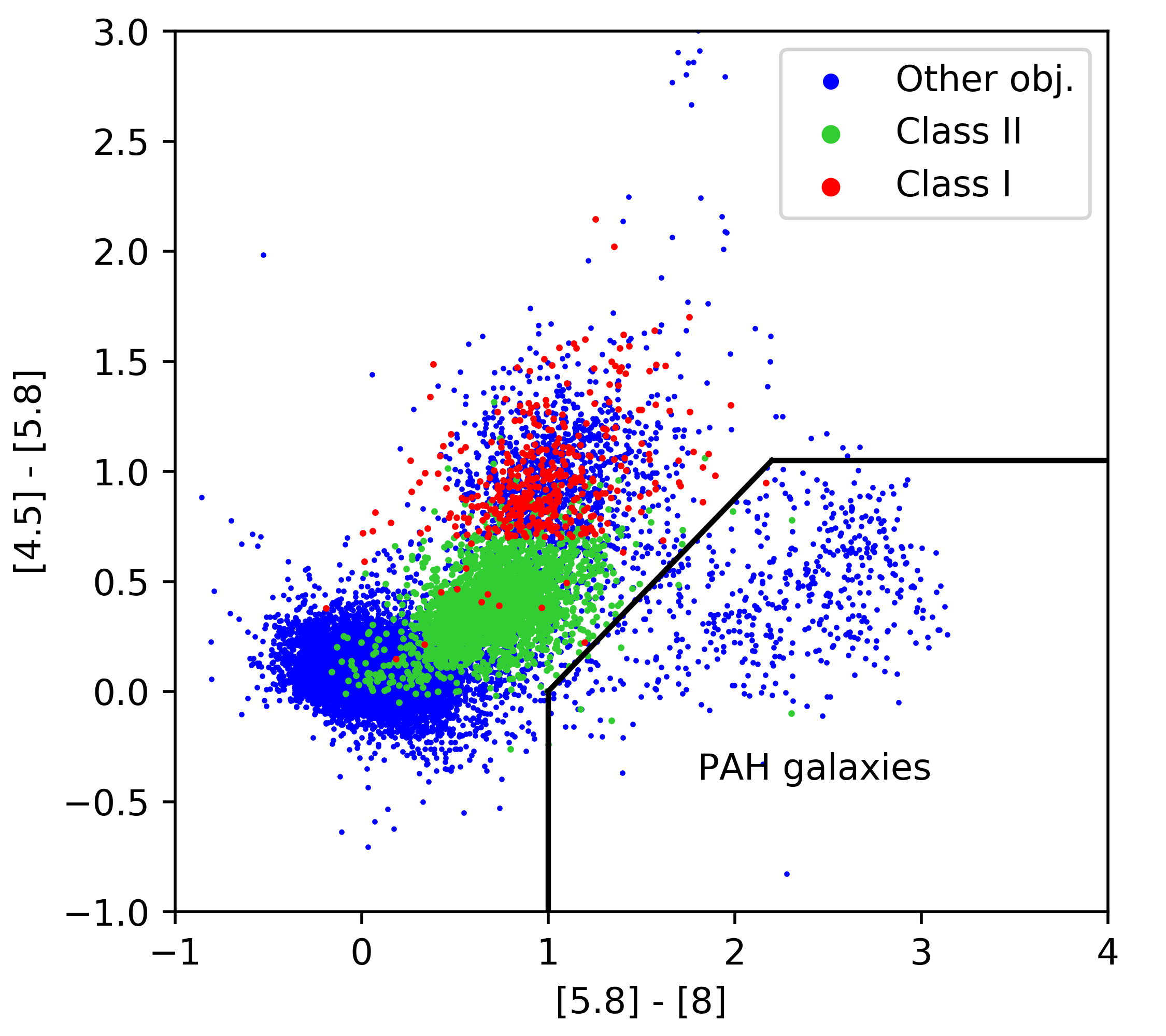}
	\end{subfigure}
	\begin{subfigure}[t]{0.24\textwidth}
	\caption*{\textbf{\hspace{0.3cm}Predicted}}
	\includegraphics[width=\textwidth]{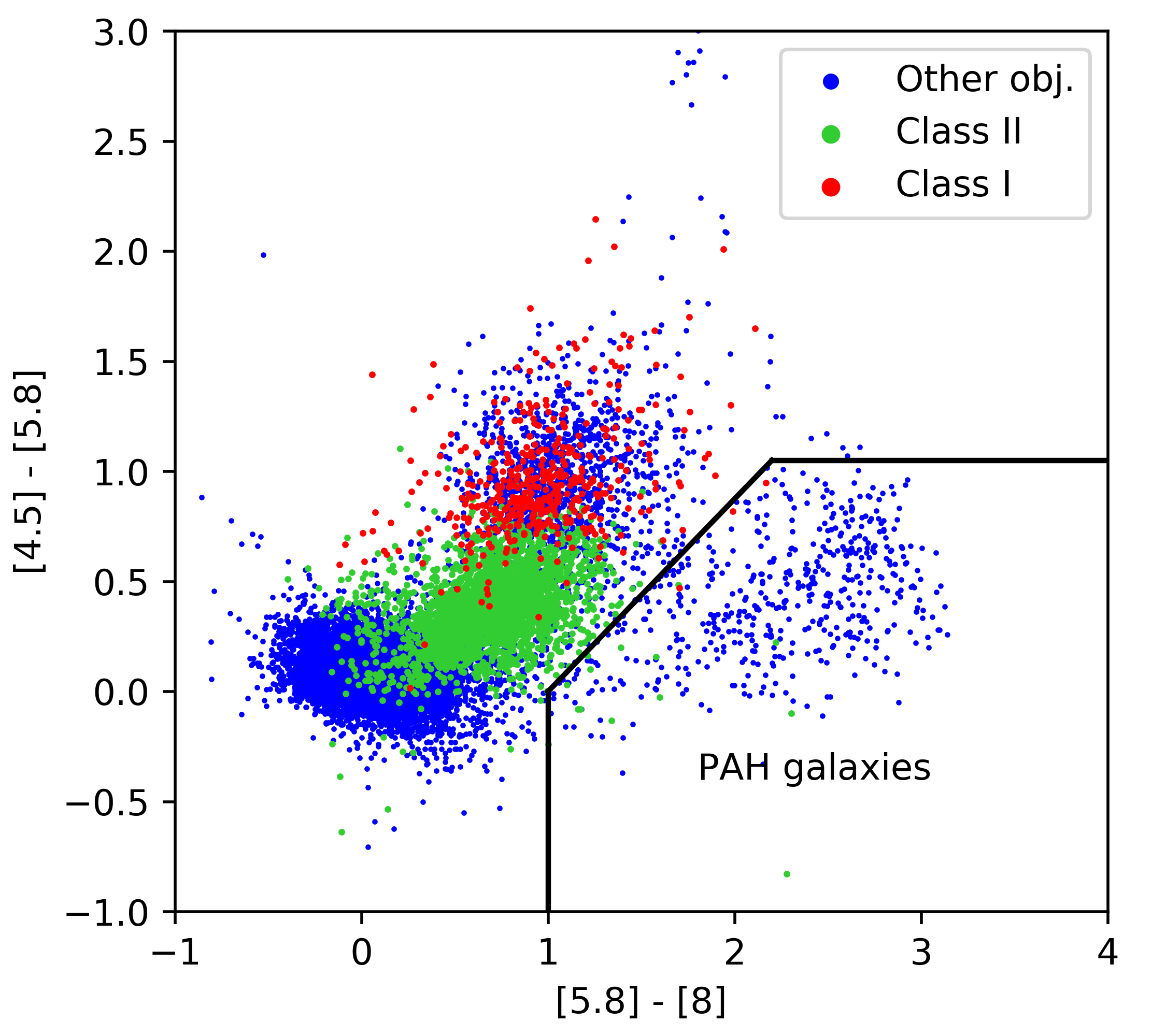}
	\end{subfigure}
	\begin{subfigure}[t]{0.24\textwidth}
	\caption*{\textbf{\hspace{0.3cm}Missed}}
	\includegraphics[width=\textwidth]{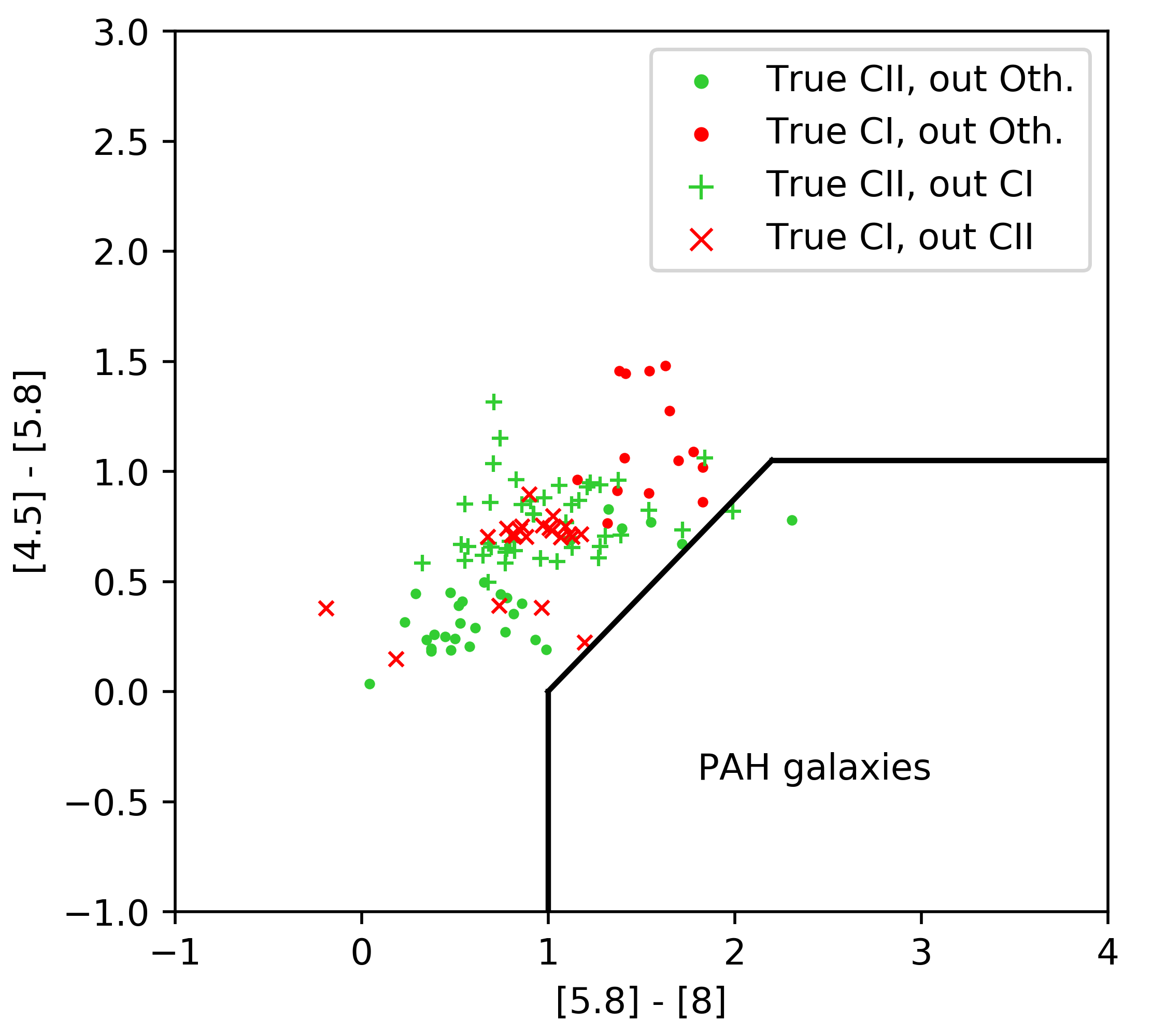}
	\end{subfigure}
	\begin{subfigure}[t]{0.24\textwidth}
	\caption*{\textbf{\hspace{0.3cm}Wrong}}
	\includegraphics[width=\textwidth]{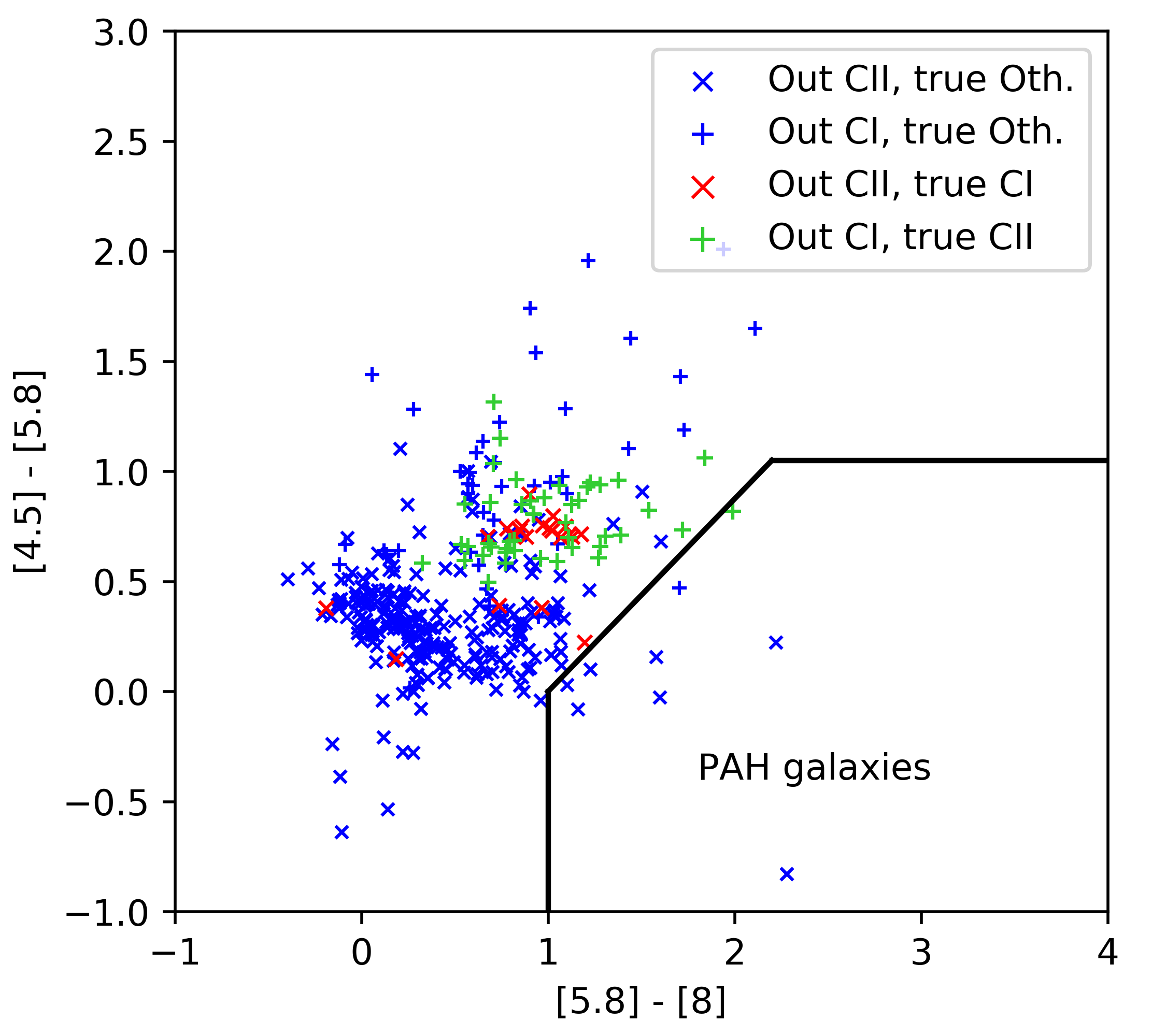}
	\end{subfigure}\\
	\begin{subfigure}[t]{0.24\textwidth}
	\includegraphics[width=\textwidth]{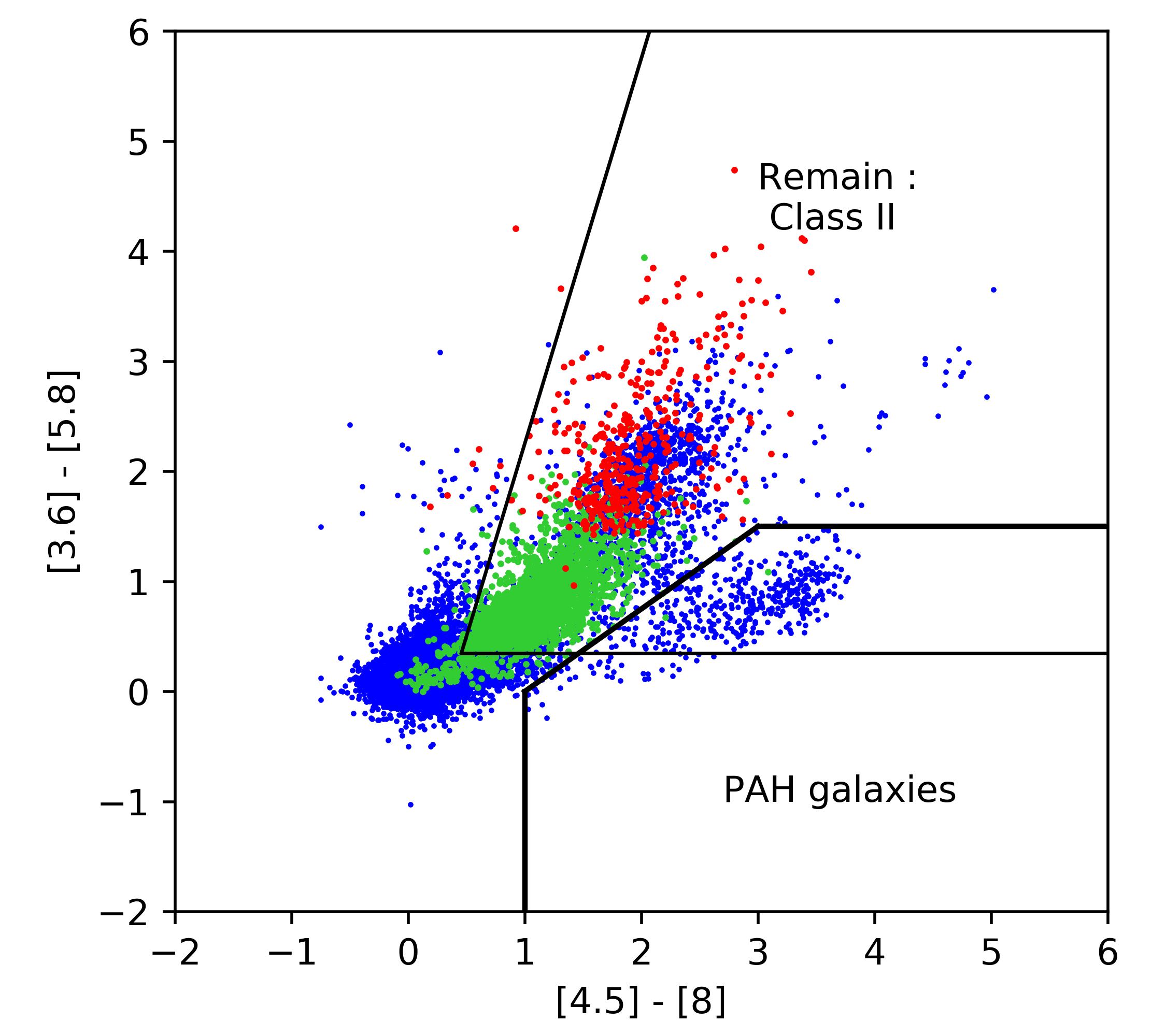}
	\end{subfigure}
	\begin{subfigure}[t]{0.24\textwidth}
	\includegraphics[width=\textwidth]{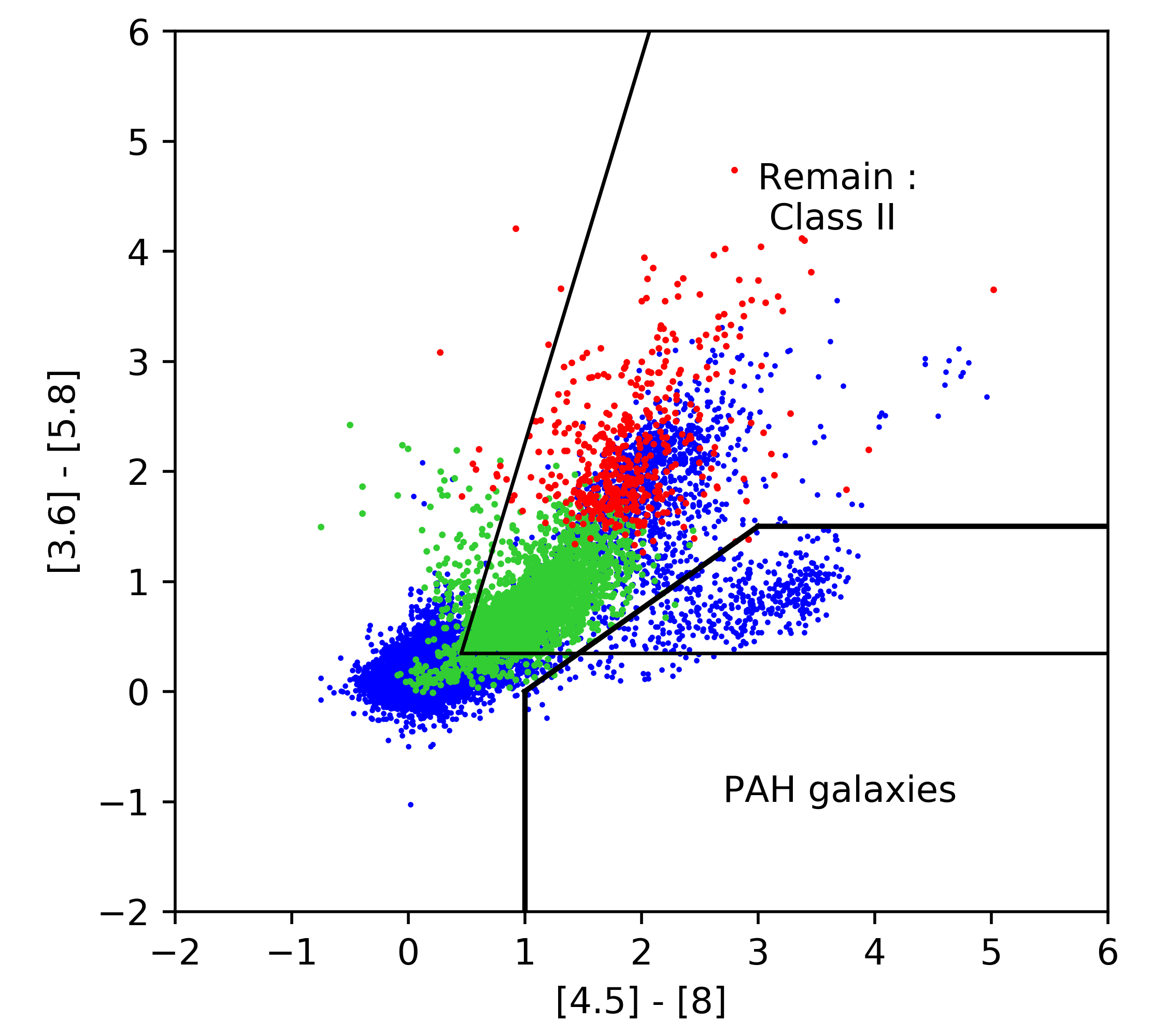}
	\end{subfigure}
	\begin{subfigure}[t]{0.24\textwidth}
	\includegraphics[width=\textwidth]{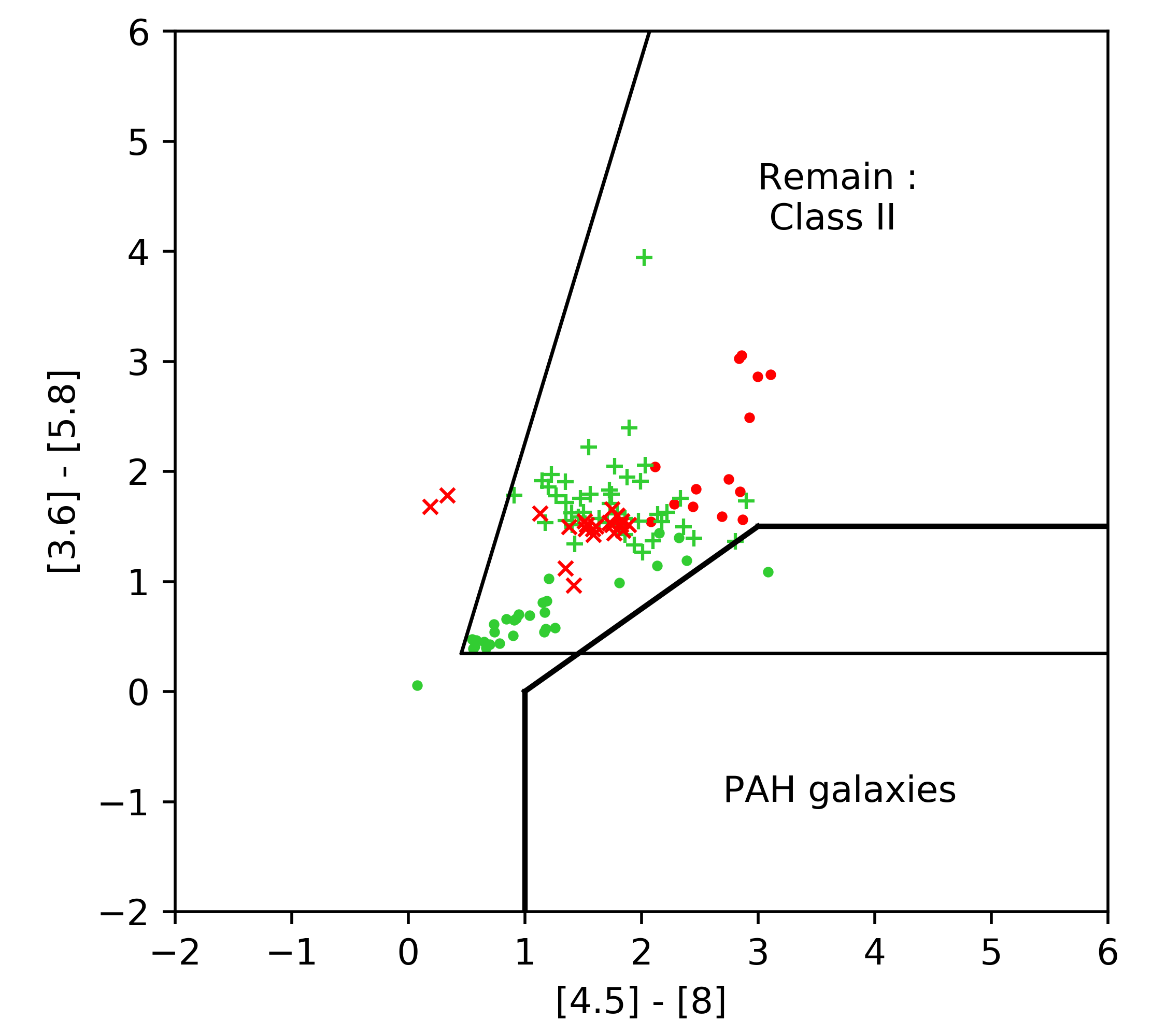}
	\end{subfigure}
	\begin{subfigure}[t]{0.24\textwidth}
	\includegraphics[width=\textwidth]{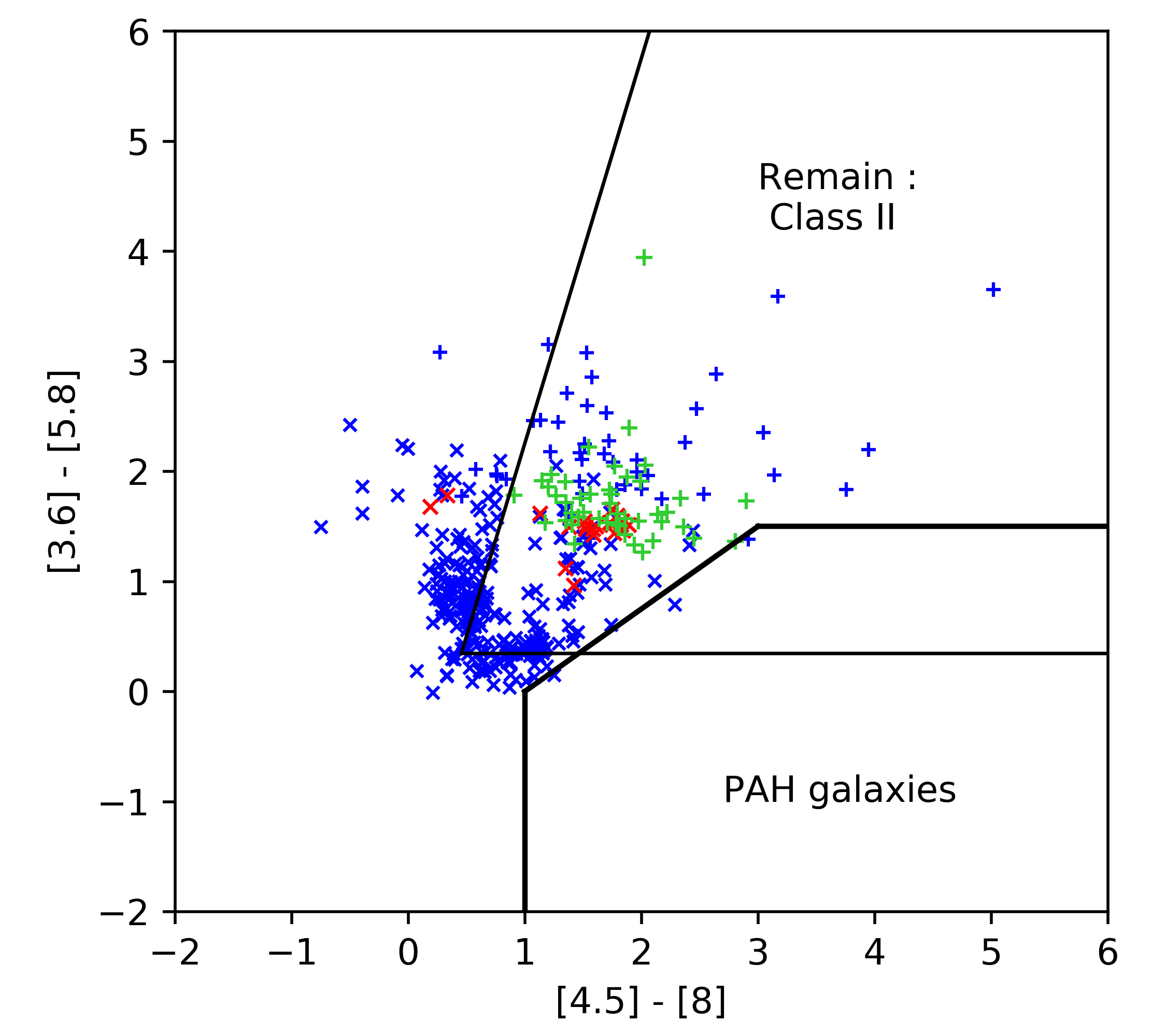}
	\end{subfigure}\\
	\begin{subfigure}[t]{0.24\textwidth}
	\includegraphics[width=\textwidth]{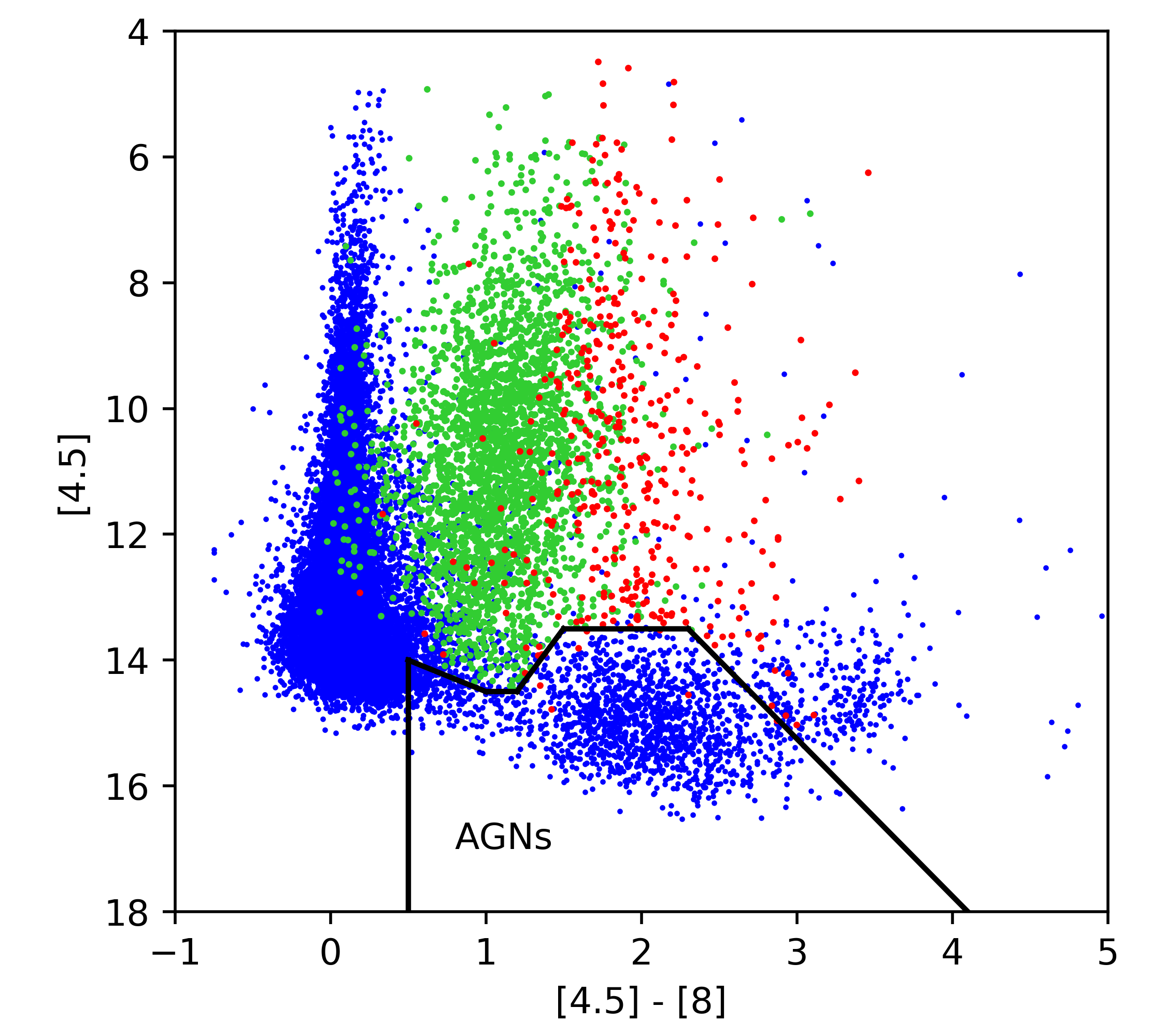}
	\end{subfigure}
	\begin{subfigure}[t]{0.24\textwidth}
	\includegraphics[width=\textwidth]{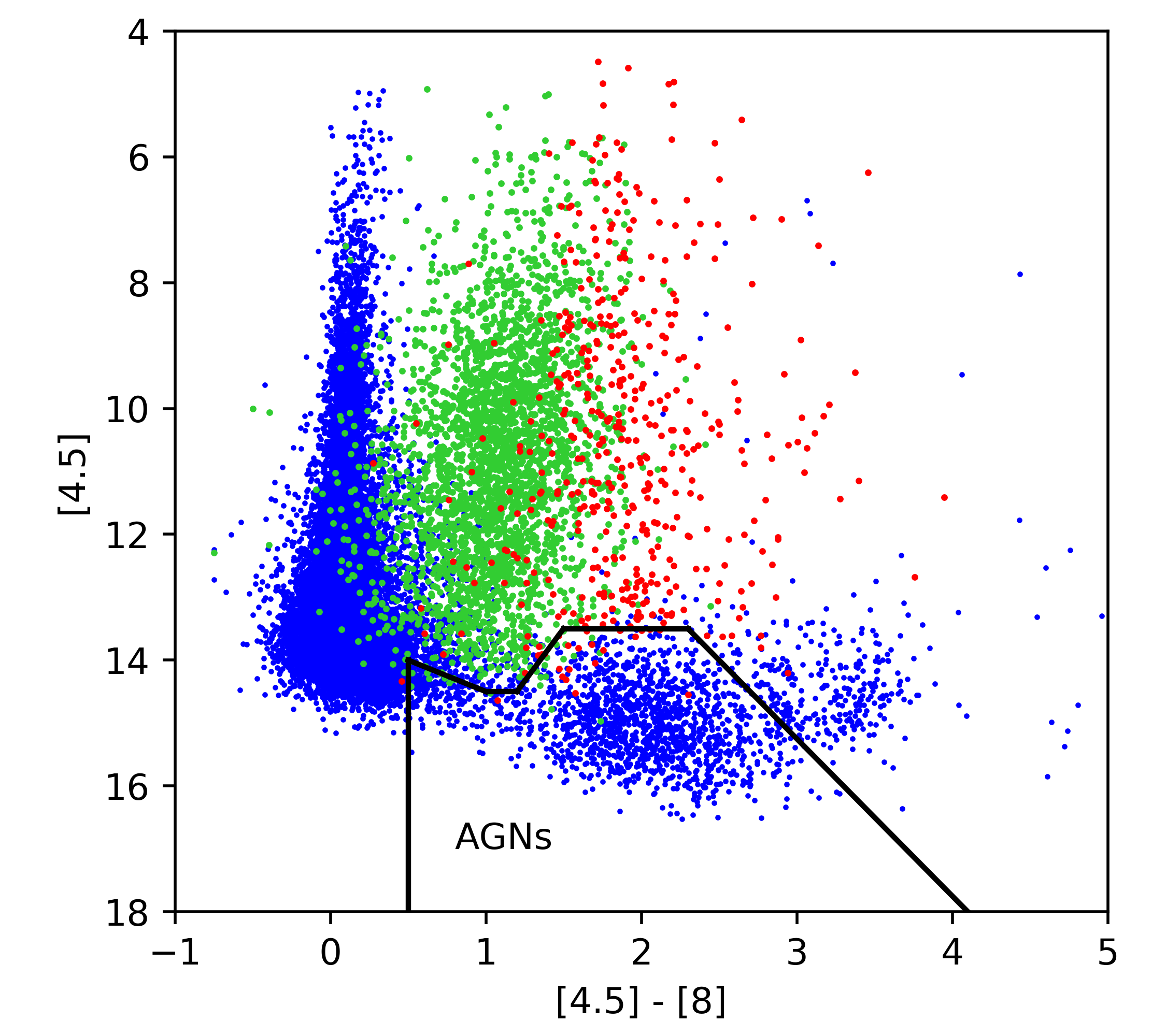}
	\end{subfigure}
	\begin{subfigure}[t]{0.24\textwidth}
	\includegraphics[width=\textwidth]{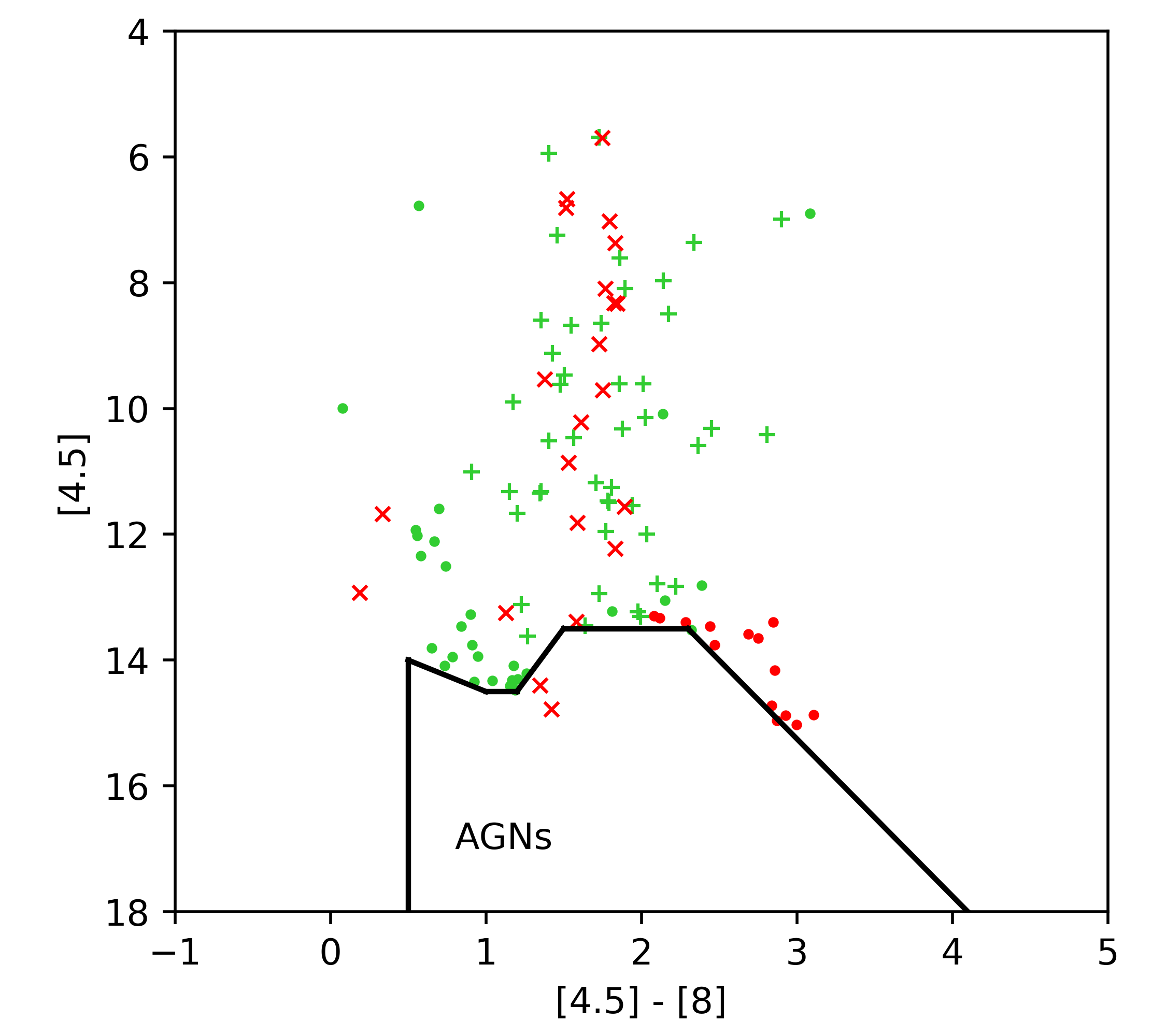}
	\end{subfigure}
	\begin{subfigure}[t]{0.24\textwidth}
	\includegraphics[width=\textwidth]{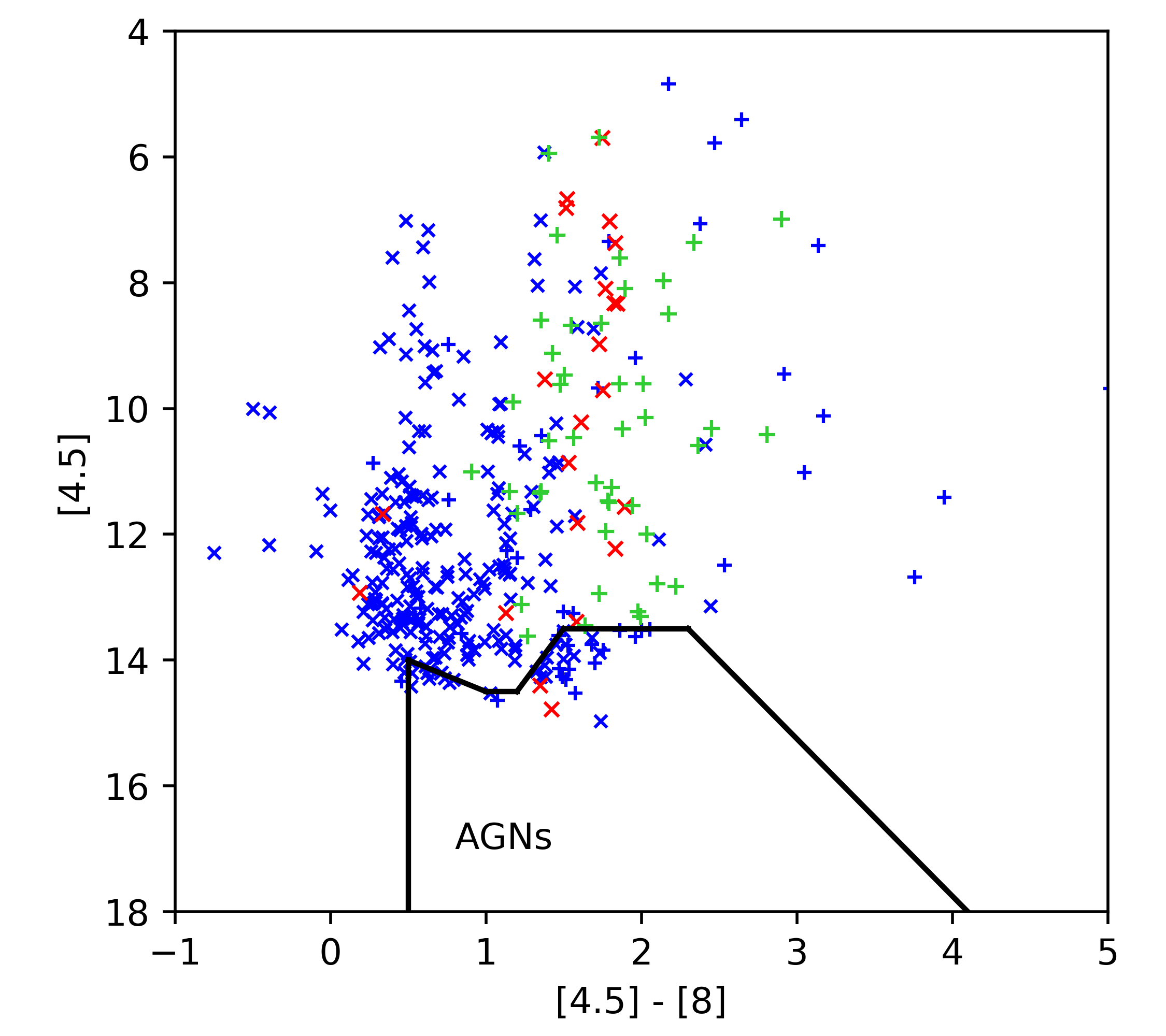}
	\end{subfigure}
	\caption*{}
\end{sidewaysfigure}

\addtocounter{figure}{-1}

\begin{sidewaysfigure}
	\centering
	\begin{subfigure}[t]{0.24\textwidth}
	\caption*{\textbf{\hspace{0.3cm}Actual}}
	\includegraphics[width=\textwidth]{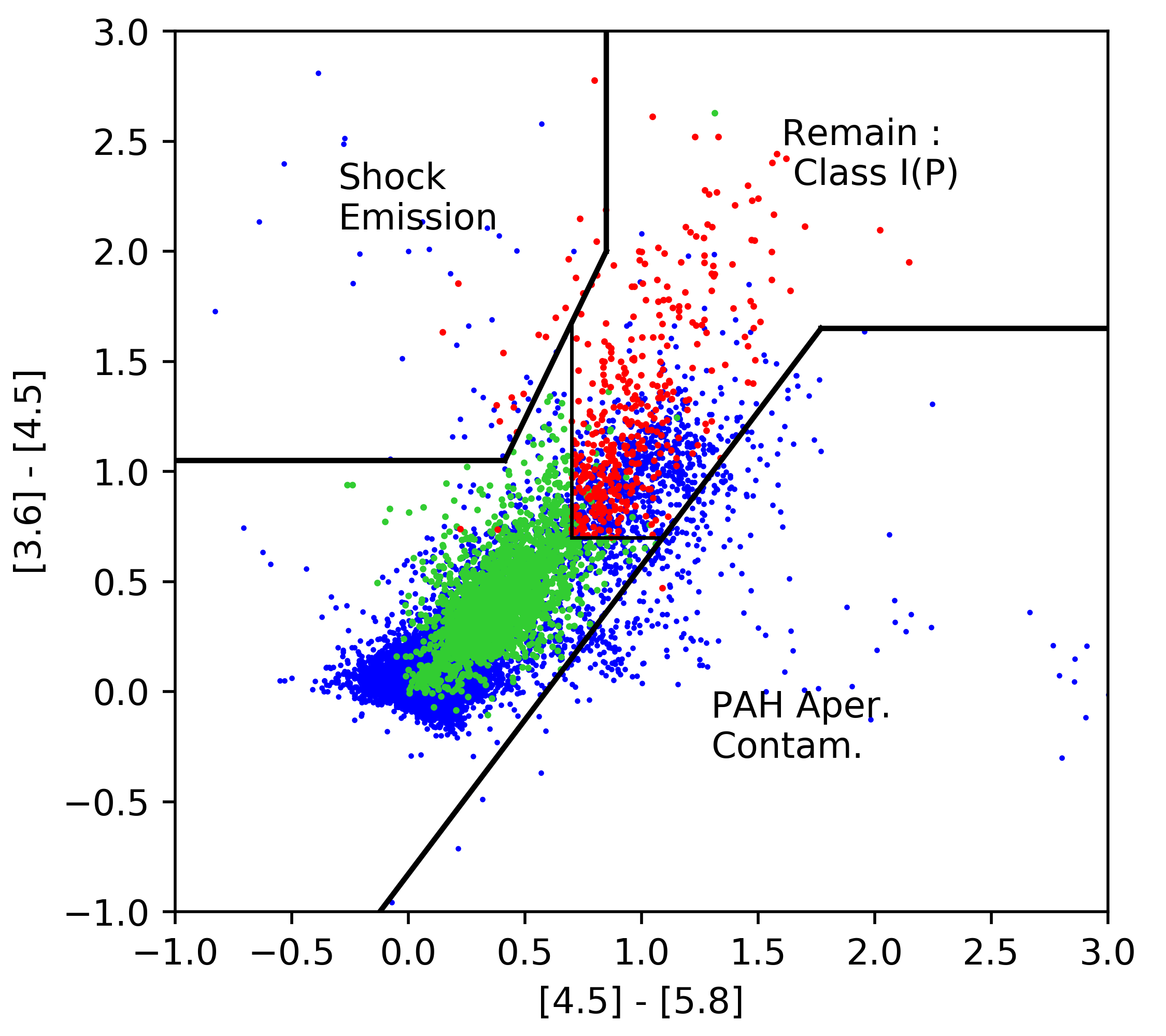}
	\end{subfigure}
	\begin{subfigure}[t]{0.24\textwidth}
	\caption*{\textbf{\hspace{0.3cm}Predicted}}
	\includegraphics[width=\textwidth]{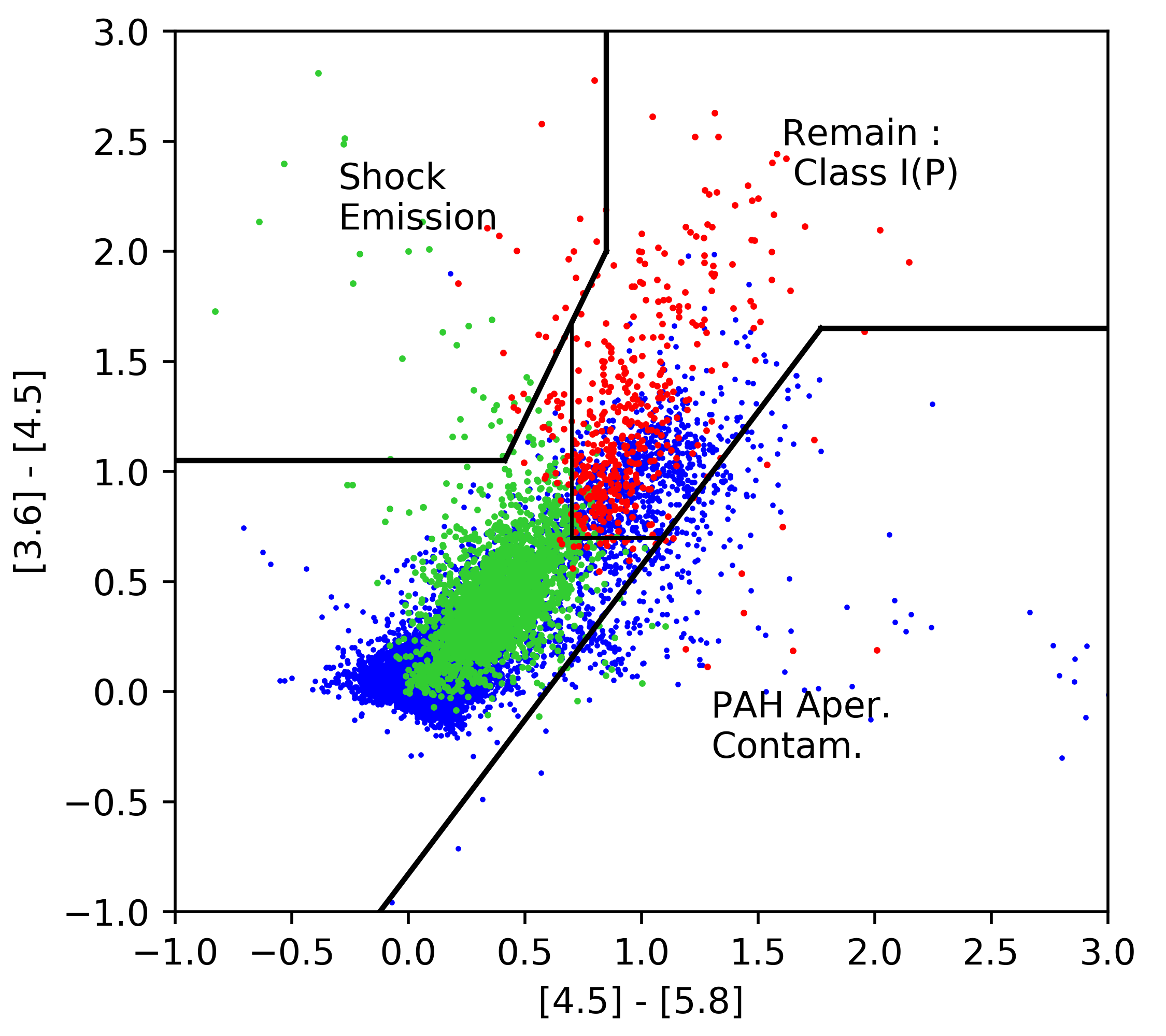}
	\end{subfigure}
	\begin{subfigure}[t]{0.24\textwidth}
	\caption*{\textbf{\hspace{0.3cm}Missed}}
	\includegraphics[width=\textwidth]{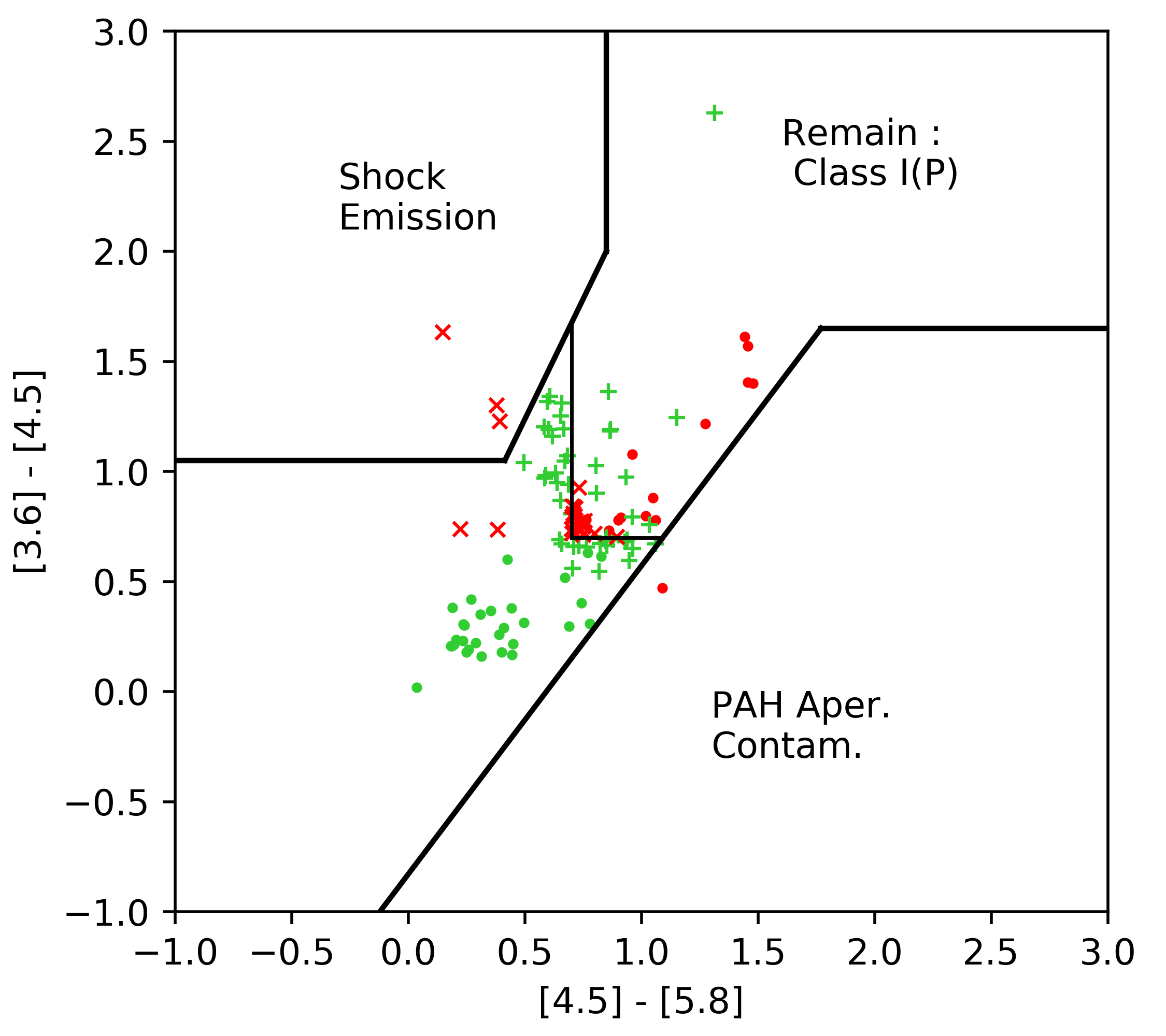}
	\end{subfigure}
	\begin{subfigure}[t]{0.24\textwidth}
	\caption*{\textbf{\hspace{0.3cm}Wrong}}
	\includegraphics[width=\textwidth]{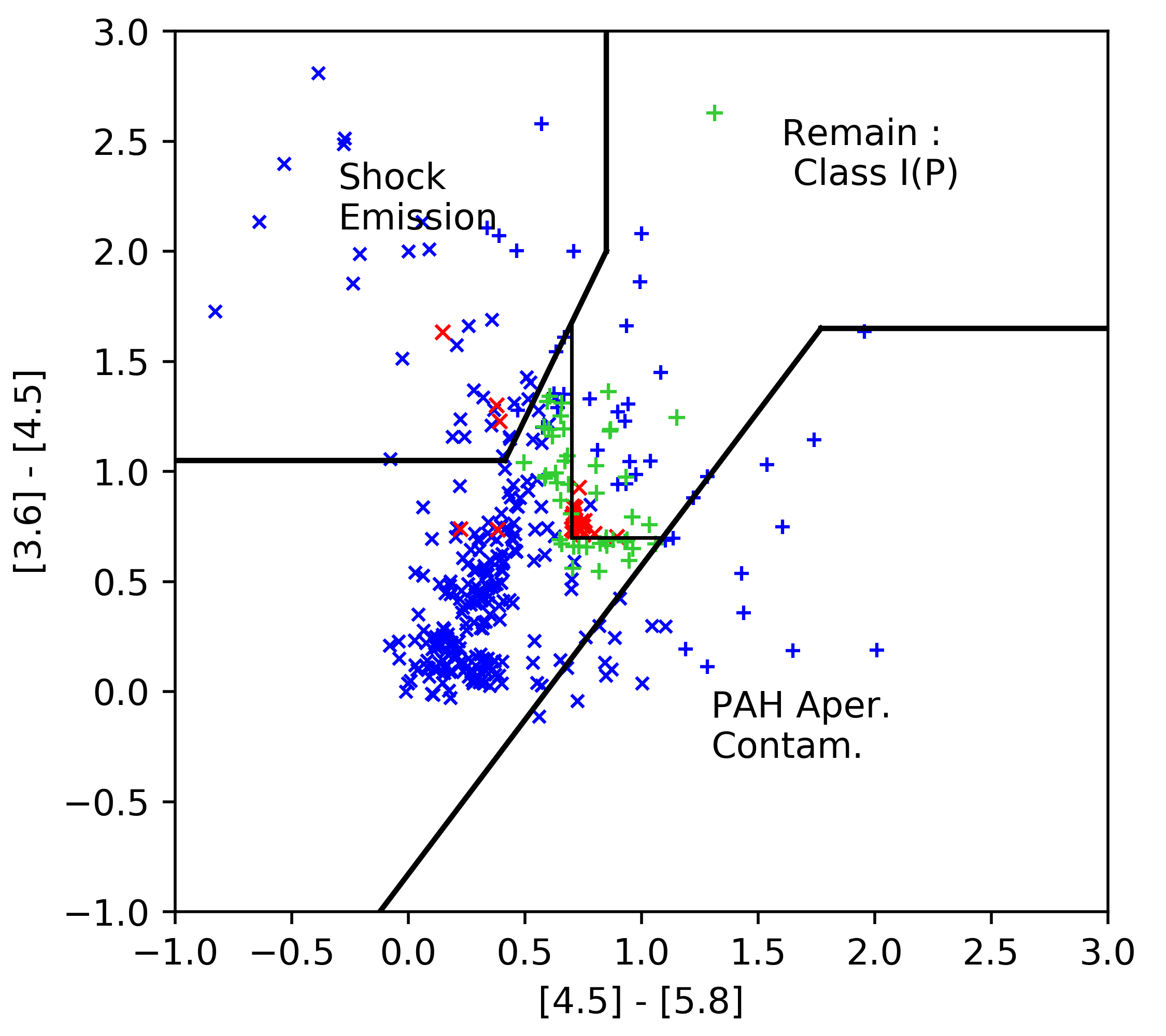}
	\end{subfigure}\\
	\begin{subfigure}[t]{0.24\textwidth}
	\includegraphics[width=\textwidth]{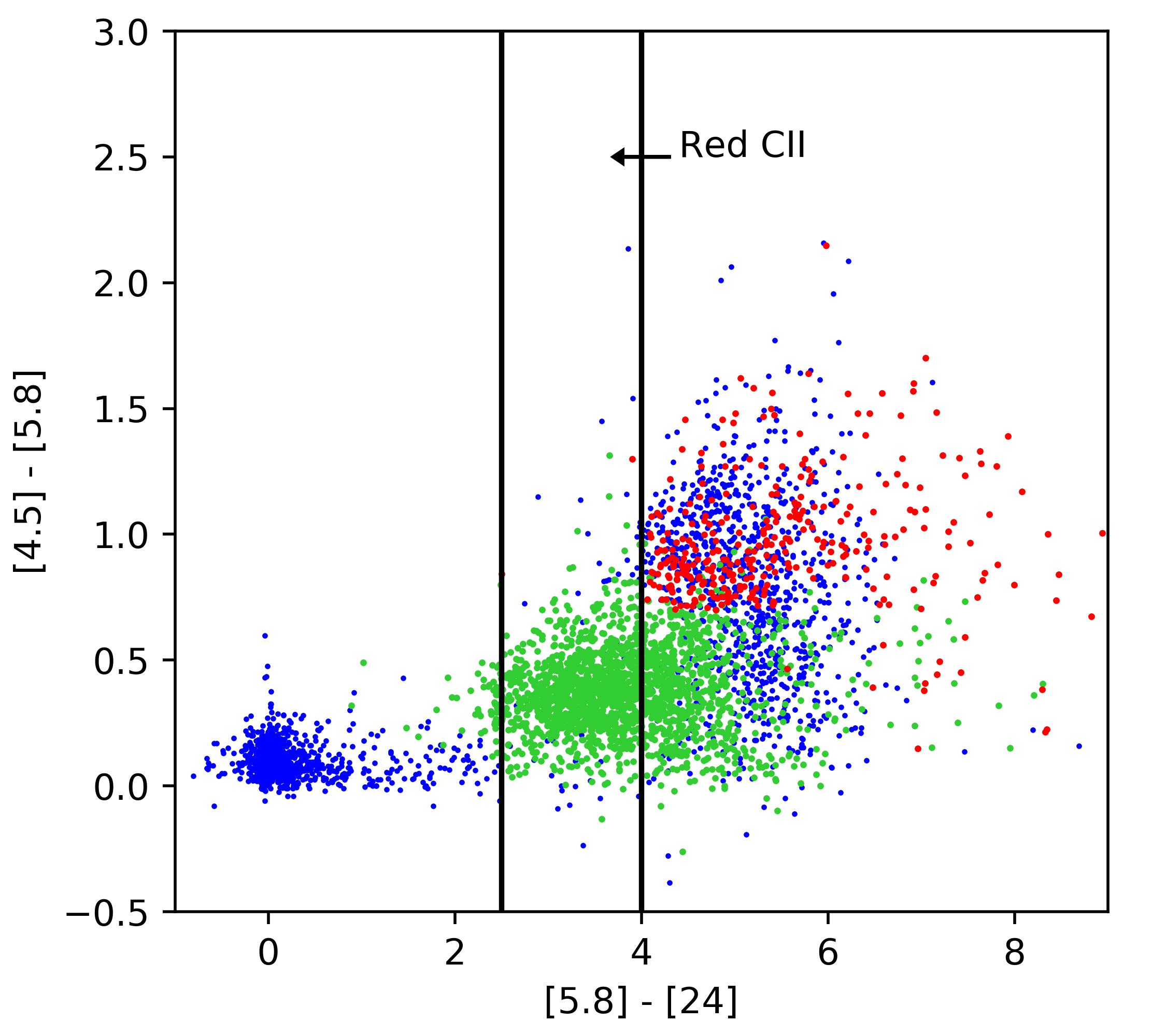}
	\end{subfigure}
	\begin{subfigure}[t]{0.24\textwidth}
	\includegraphics[width=\textwidth]{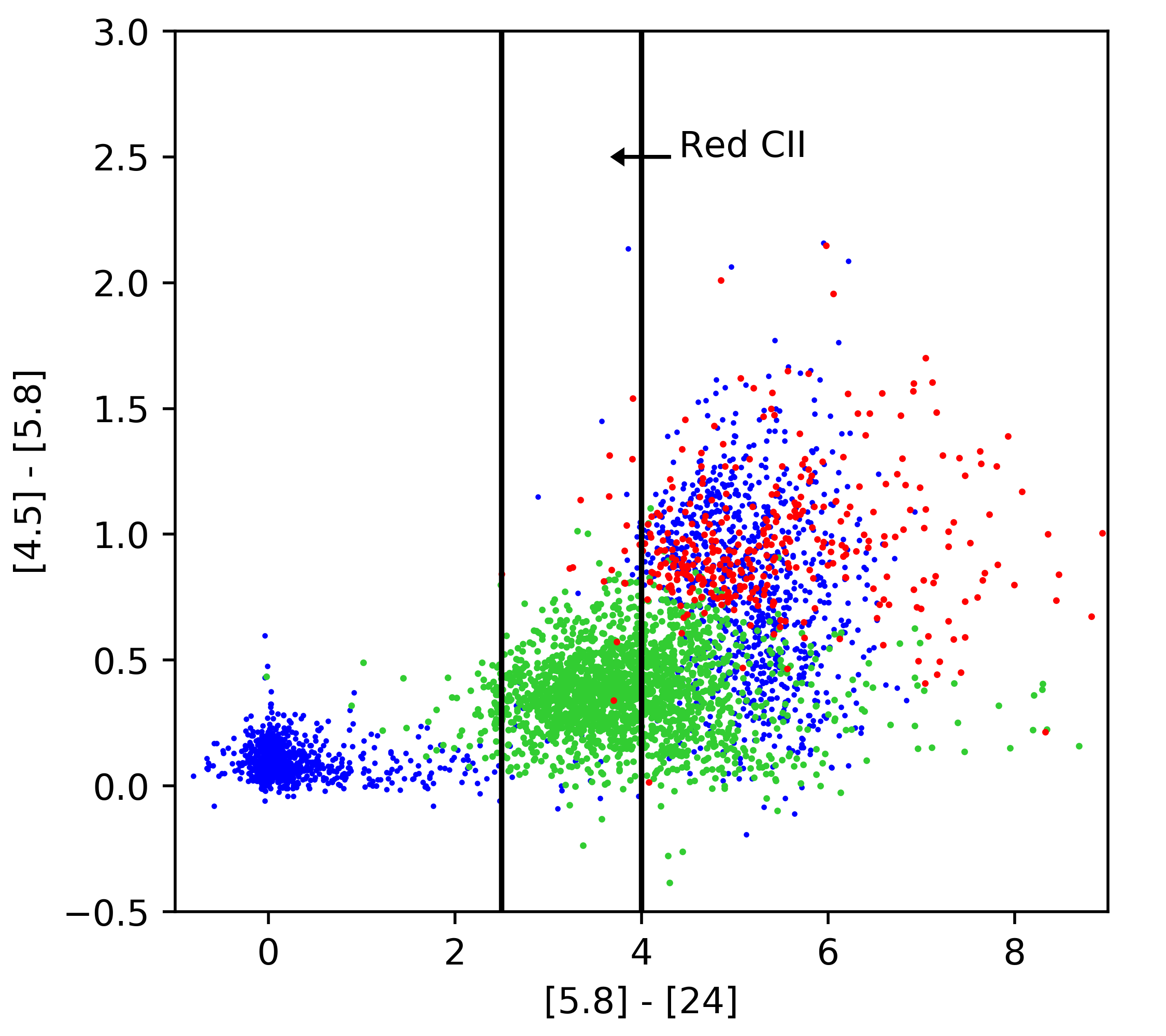}
	\end{subfigure}
	\begin{subfigure}[t]{0.24\textwidth}
	\includegraphics[width=\textwidth]{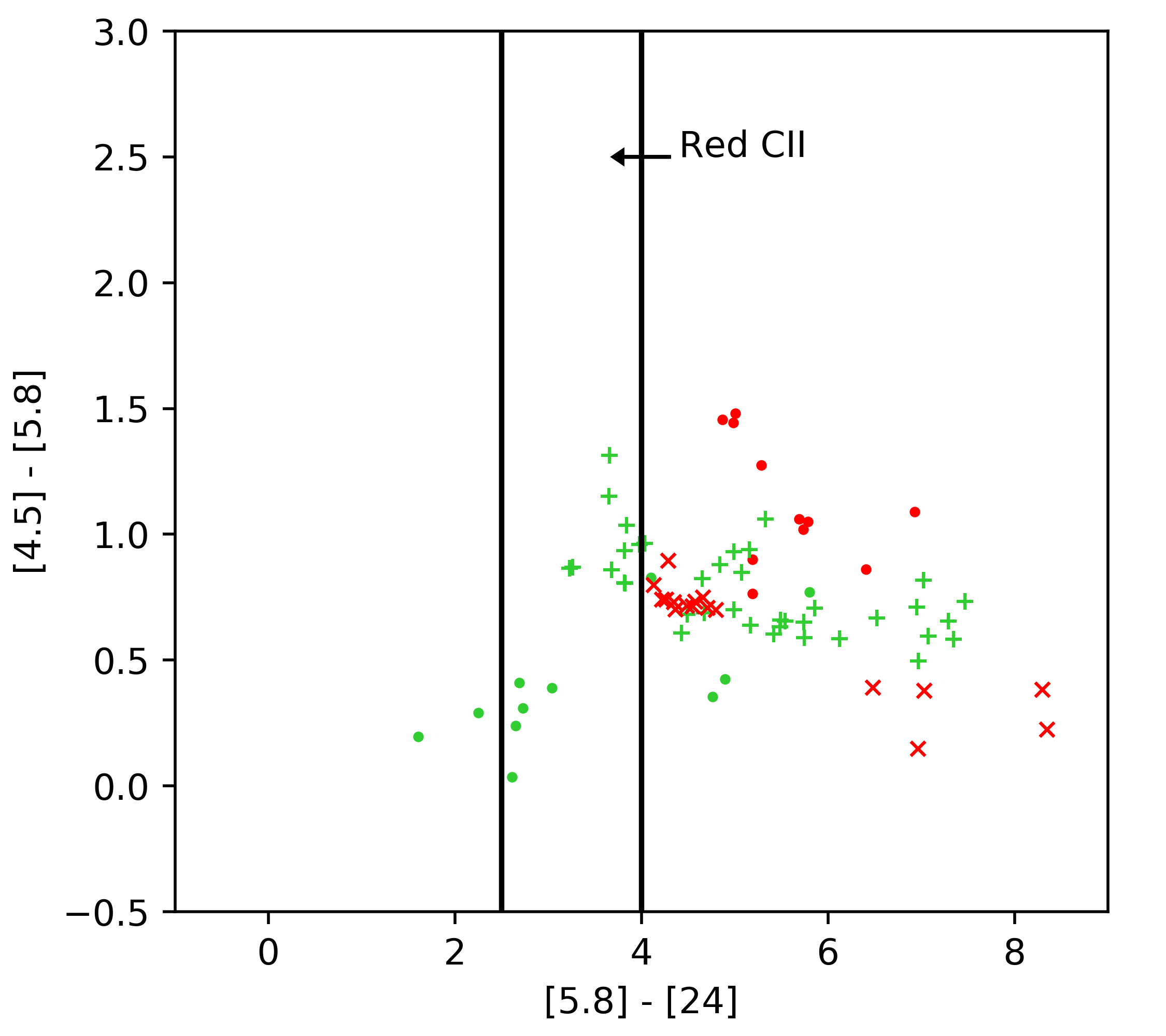}
	\end{subfigure}
	\begin{subfigure}[t]{0.24\textwidth}
	\includegraphics[width=\textwidth]{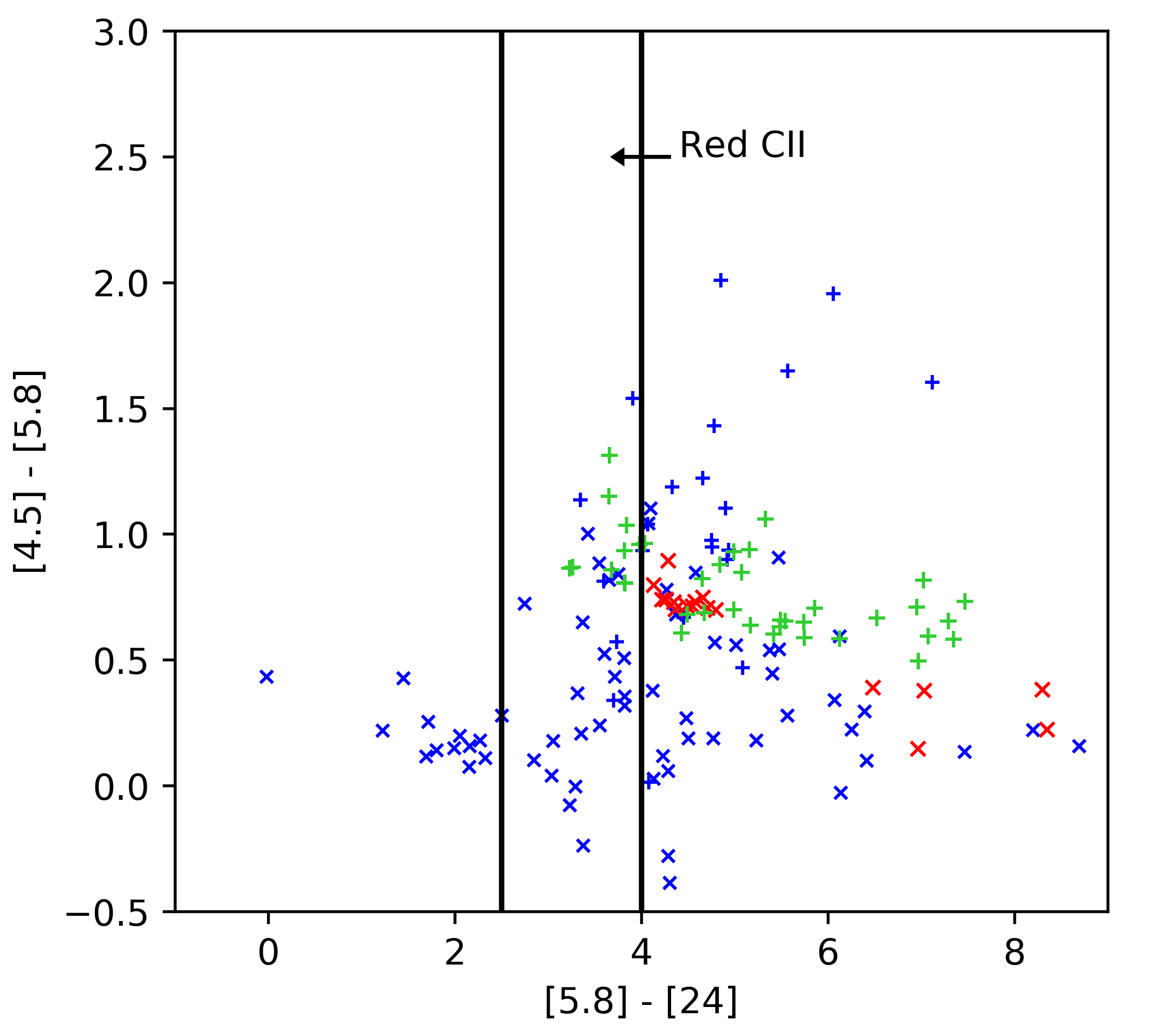}
	\end{subfigure}
	\caption[Input parameter space coverage in the F-C case]{Input parameter space coverage in the CMDs used for the G09 method in the F-C case on the full dataset regarding different populations. \textit{Actual:} distribution of genuine classes. CI YSOs are in red, CII YSOs are in green and Others are in blue. \textit{Predicted:} prediction given by the network with the same color-code as for the \textit{actual} frames. { \textit{Missed:} Genuine CI and CII according to the labeled dataset that were misclassified by the network. Green is for genuine CII YSOs, red for genuine CI YSOs. The points and crosses indicate the network output as specified in the legend. \textit{Wrong:} YSO predictions of the network that are known to be incorrect based on the labeled dataset. Green is for genuine CII YSOs, red for genuine CI YSOs and blue is for genuine contaminants. The two types of crosses indicate the predicted YSO class as specified in the legend.}}
\label{missed_wrong_space} 
\end{sidewaysfigure}

\clearpage

	\subsection{Orion and NGC 2264 YSO candidates distribution maps}
	\label{yso_candidates_maps}

With the prediction from the full 1\,kpc training over Orion and NGC 2264 we were able to look at the distribution of CI and CII YSOs in the corresponding regions. To represent the density of the regions we chose to use data from the Herschel space observatory \citep{herschel_2010}, especially the Spectral and Photometric Imaging REceiver (SPIRE) 500 $\mathrm{\mu m}$ band that is a reasonable proxy of the total gas column density.\\

\begin{figure*}[!t]
\begin{minipage}{1.0\textwidth}
	\centering
	\begin{subfigure}[t]{1.0\textwidth}
	\includegraphics[width=\hsize]{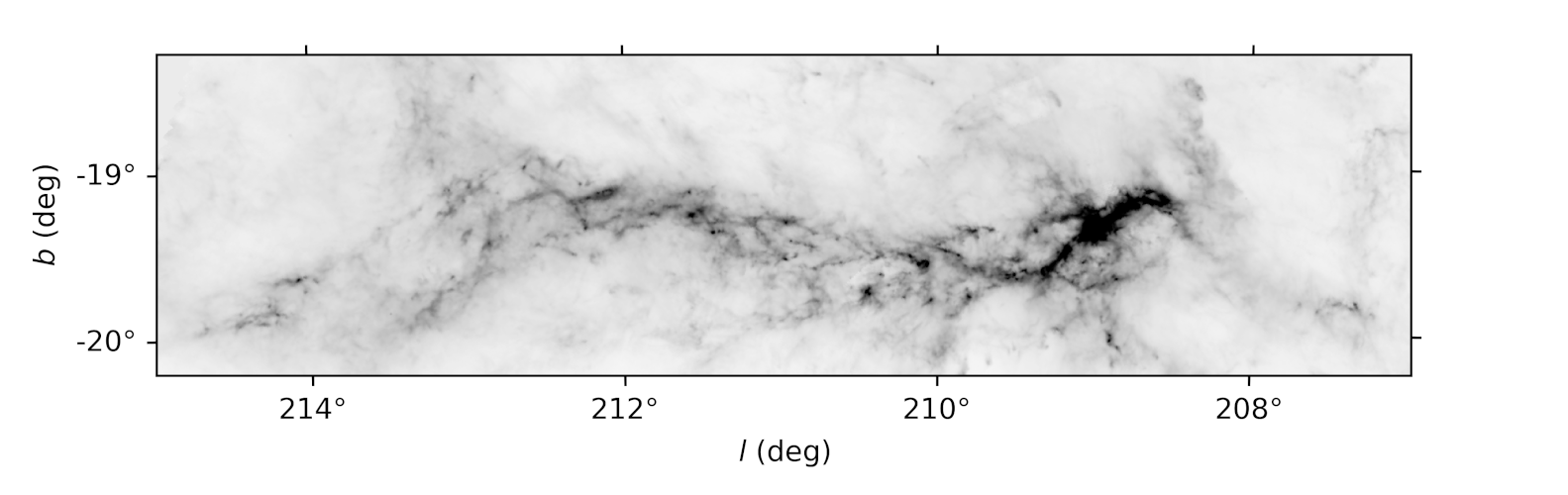}
	\end{subfigure}\\
	\vspace{-0.7cm}
	\begin{subfigure}[t]{1.0\textwidth}
	\includegraphics[width=\hsize]{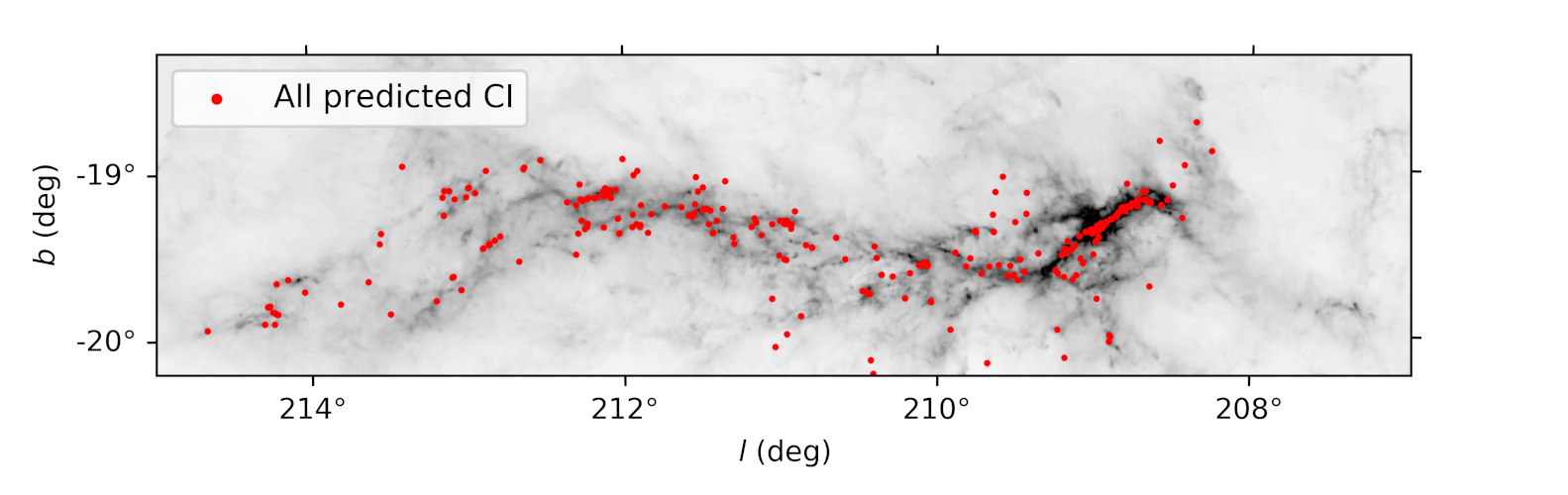}
	\end{subfigure}\\
	\vspace{-0.7cm}
	\begin{subfigure}[t]{1.0\textwidth}
	\includegraphics[width=\hsize]{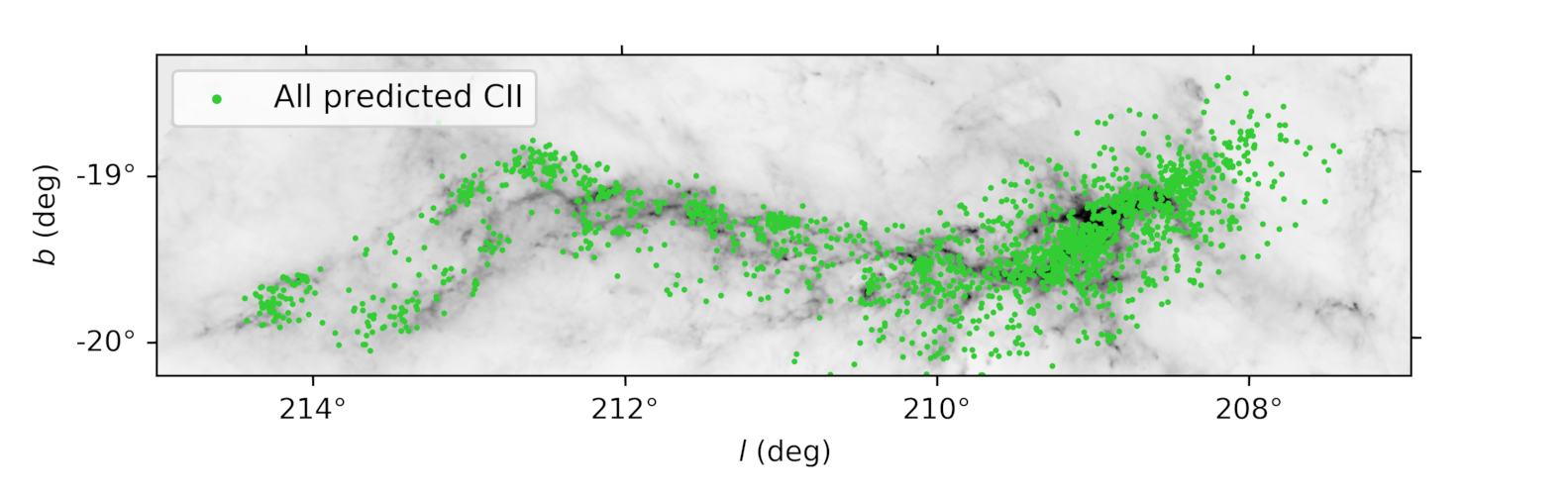}
	\end{subfigure}
	\end{minipage}
	\caption[Orion A YSO candidate distribution]{Distribution of YSO candidates in the Orion A part of the molecular complex. The grey scale shows the Herschel SPIRE 500 $\mathrm{\mu m}$ map. CI and CII YSOs are shown in the middle and lower frames in red and green, respectively.}
\label{orion_A_yso_dist} 
\end{figure*}

Figure~\ref{orion_A_yso_dist} shows the distributions for the Orion A region that contains the Orion nebula (Messier 42). It shows that our CI YSOs follow the main dense filament very closely, especially in the so-called "integral" shaped filament (between $l=208$ and 210 degrees). As expected, the CII YSOs, which are more evolved, spread more widely on the observed region. Still, the high CII density nicely maps the densest part of the Orion molecular cloud and traces the parts that are expected to form stars more actively.\\ 

\begin{figure*}[!t]
\hspace{-0.5cm}
	\begin{minipage}{1.05\hsize}
	\centering
	\begin{subfigure}[t]{0.32\textwidth}
	\includegraphics[width=1.0\hsize]{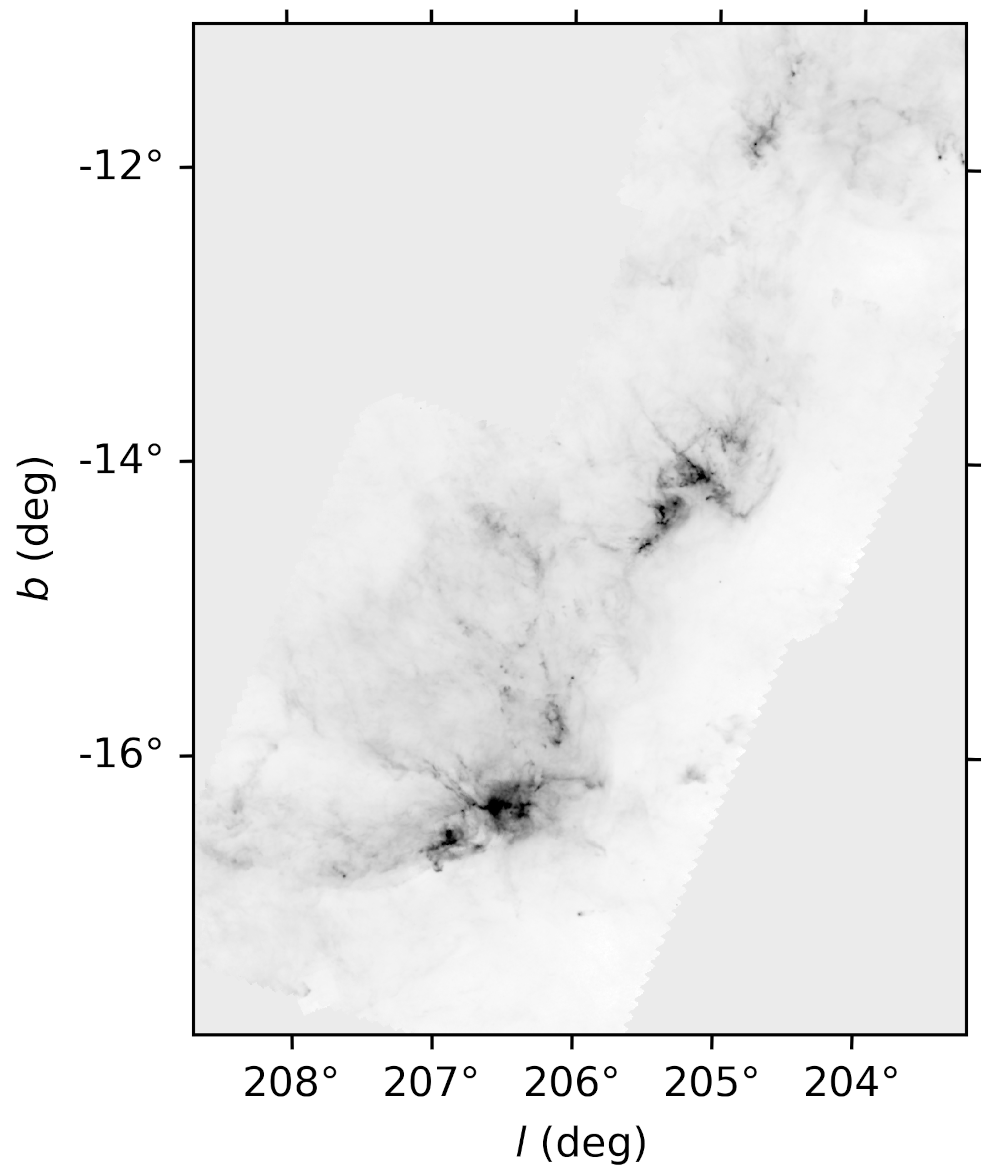}
	\end{subfigure}
	\begin{subfigure}[t]{0.32\textwidth}
	\includegraphics[width=1.0\hsize]{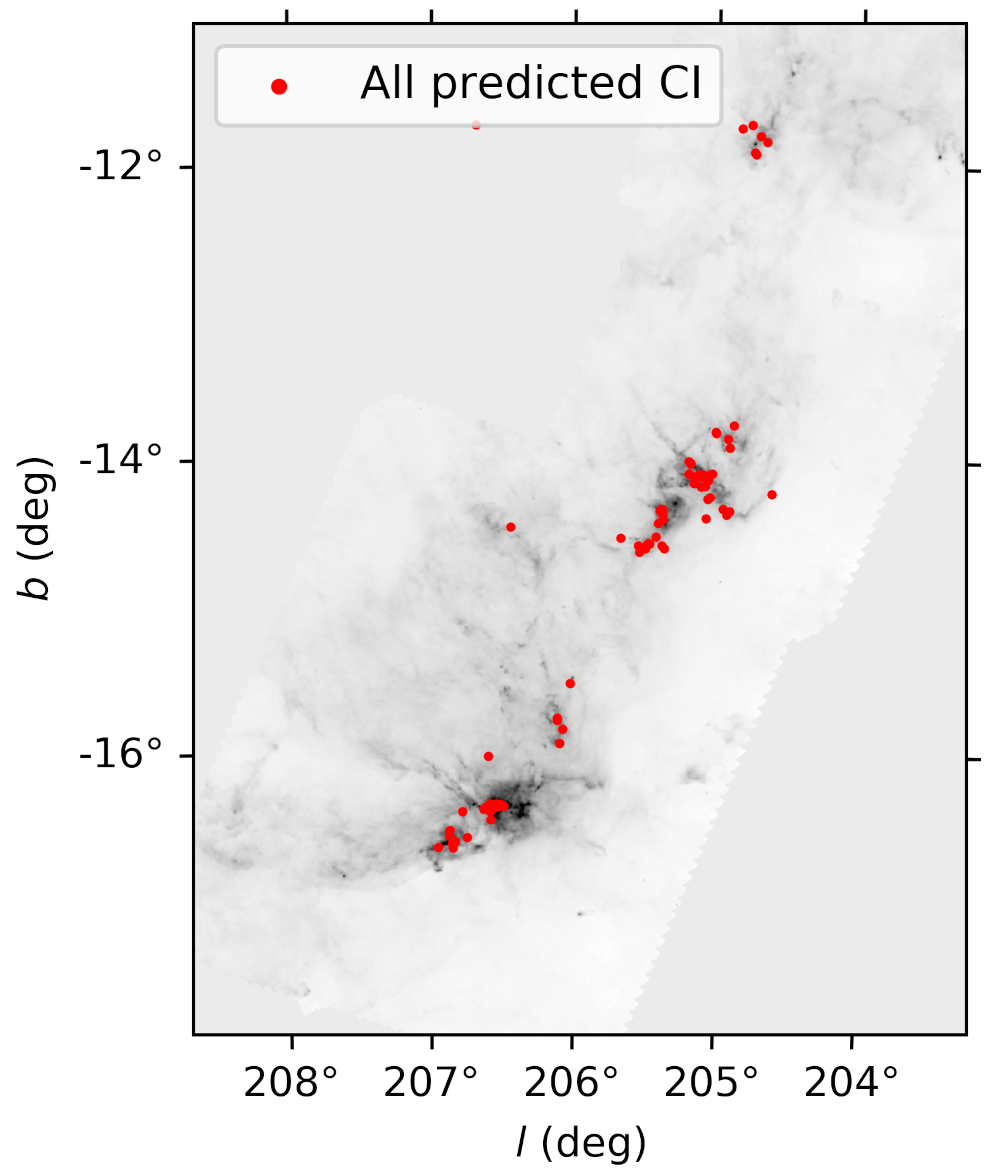}
	\end{subfigure}
	\begin{subfigure}[t]{0.32\textwidth}
	\includegraphics[width=1.0\hsize]{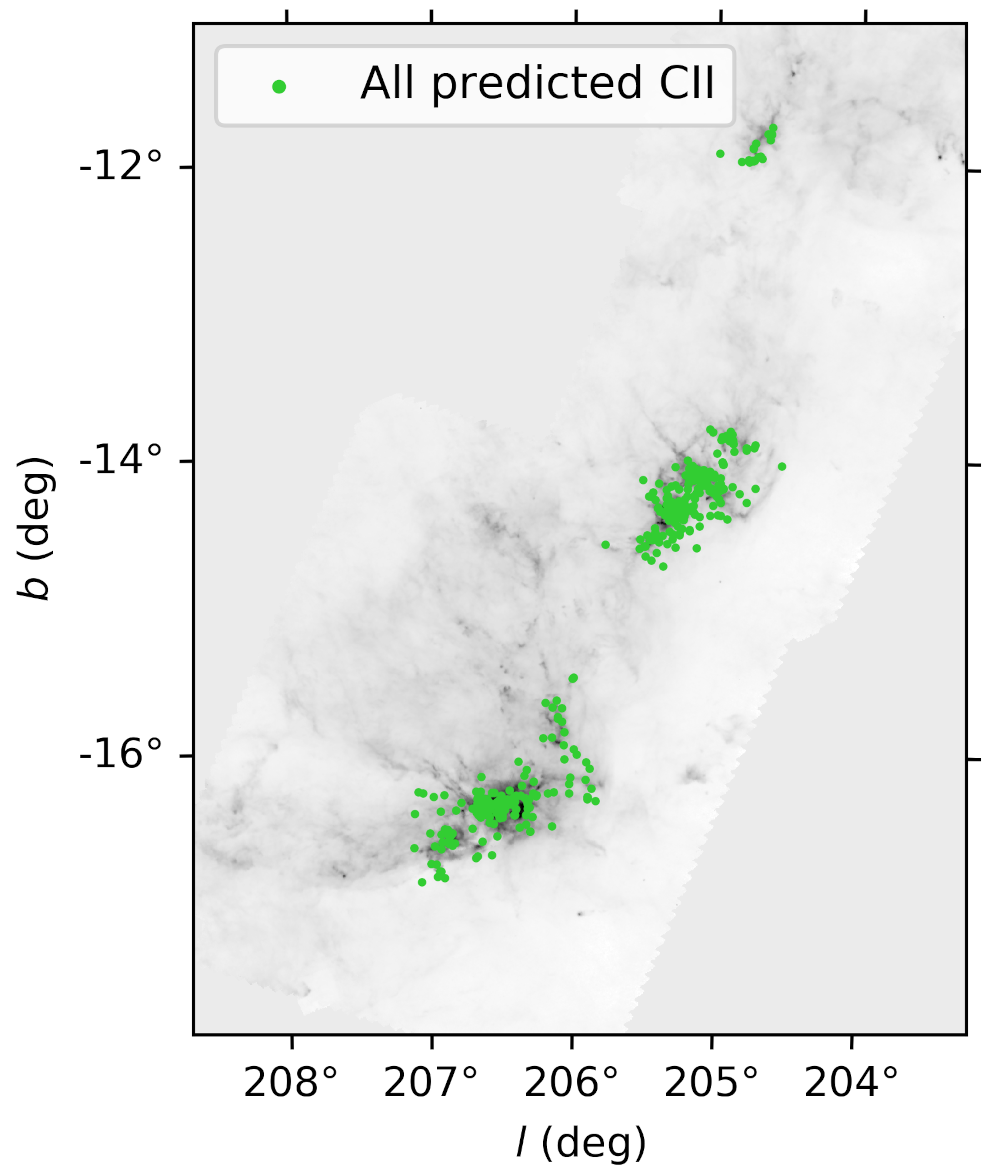}
	\end{subfigure}
	\end{minipage}
	\caption[Orion B YSO candidates distribution]{Distribution of YSO candidates in the Orion B part of the molecular complex. The grey scale shows the Herschel SPIRE 500 $\mathrm{\mu m}$ map. CI and CII YSOs are shown in the middle and right frames in red and green, respectively.}
\label{orion_B_yso_dist}
\end{figure*}

The Orion B part of the molecular complex is presented in Figure~\ref{orion_B_yso_dist}. The number of YSOs in this part of the cloud is lesser than in Orion A. This is line with Orion B being in an earlier evolutionary stage, but the comparison is hampered by the fact that in this region the Spitzer observations were not continuous (see Fig.~\ref{megeath_orion_cover}). Yet, as for Orion A the CI YSOs tightly follow the densest parts of the star-forming region, while CII show a greater dispersion around the density peaks. We note that the number of YSOs found in each part is large enough to hope for some of them having a Gaia counterpart that would allow us to estimate their distance (Sect.~\ref{3d_yso_gaia}). \\

\begin{figure*}[!t]
\hspace{-2.0cm}
\begin{minipage}{1.25\textwidth}
	\centering
	\begin{subfigure}[t]{0.49\textwidth}
	\includegraphics[width=1.0\hsize]{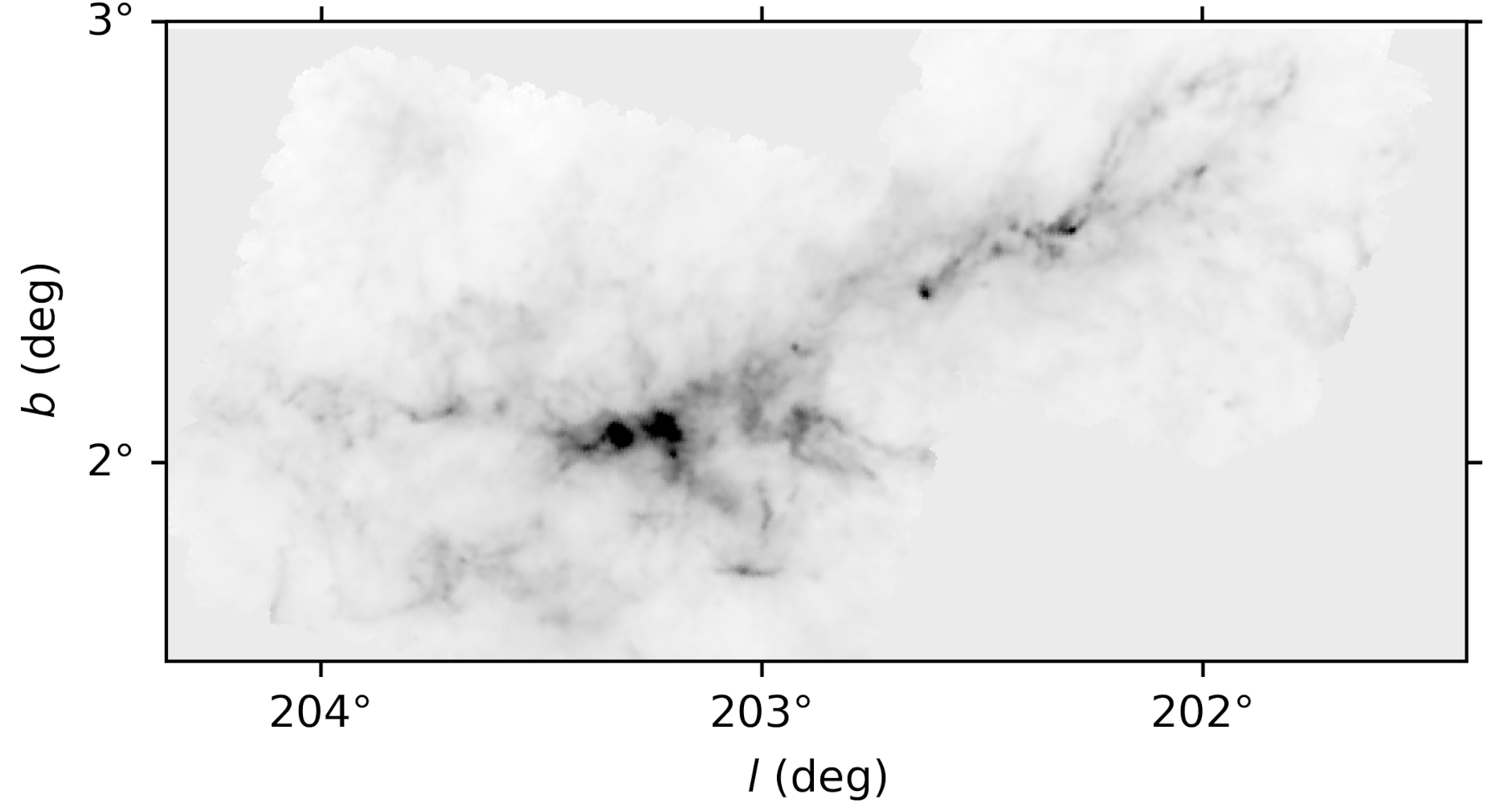}
	\end{subfigure}\\
	\begin{subfigure}[t]{0.49\textwidth}
	\includegraphics[width=1.0\hsize]{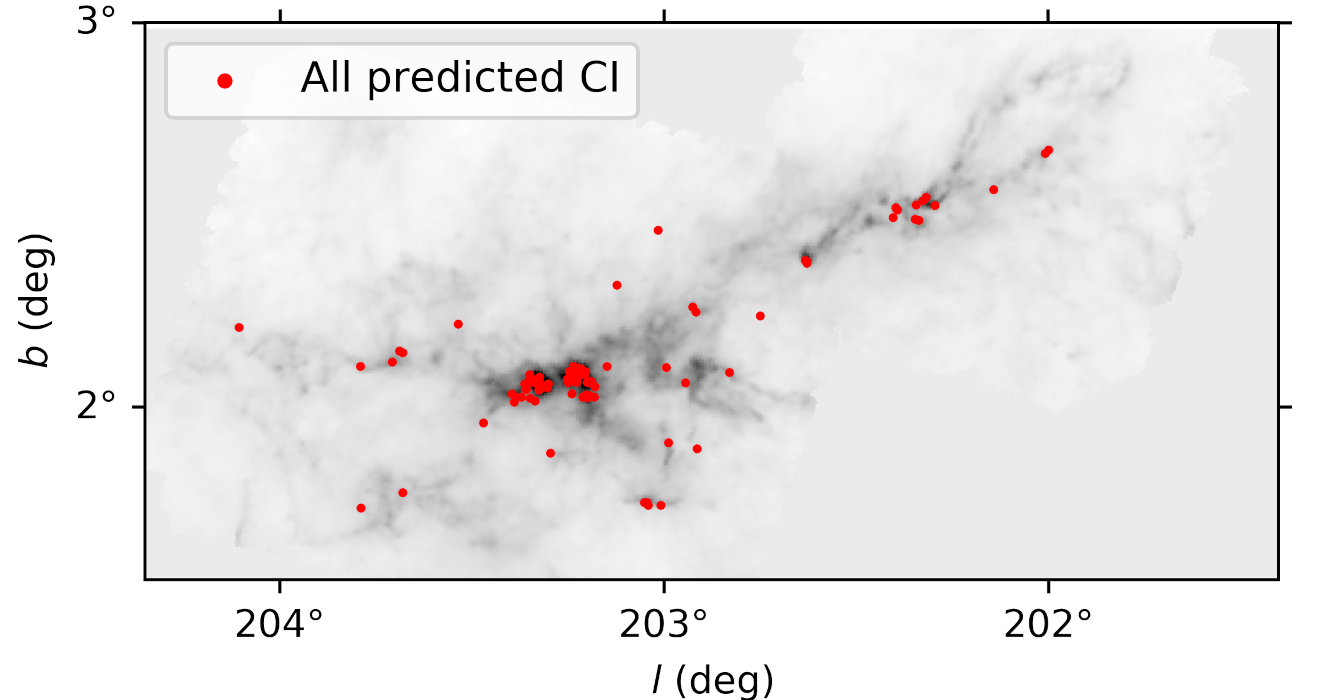}
	\end{subfigure}
	\begin{subfigure}[t]{0.49\textwidth}
	\includegraphics[width=1.0\hsize]{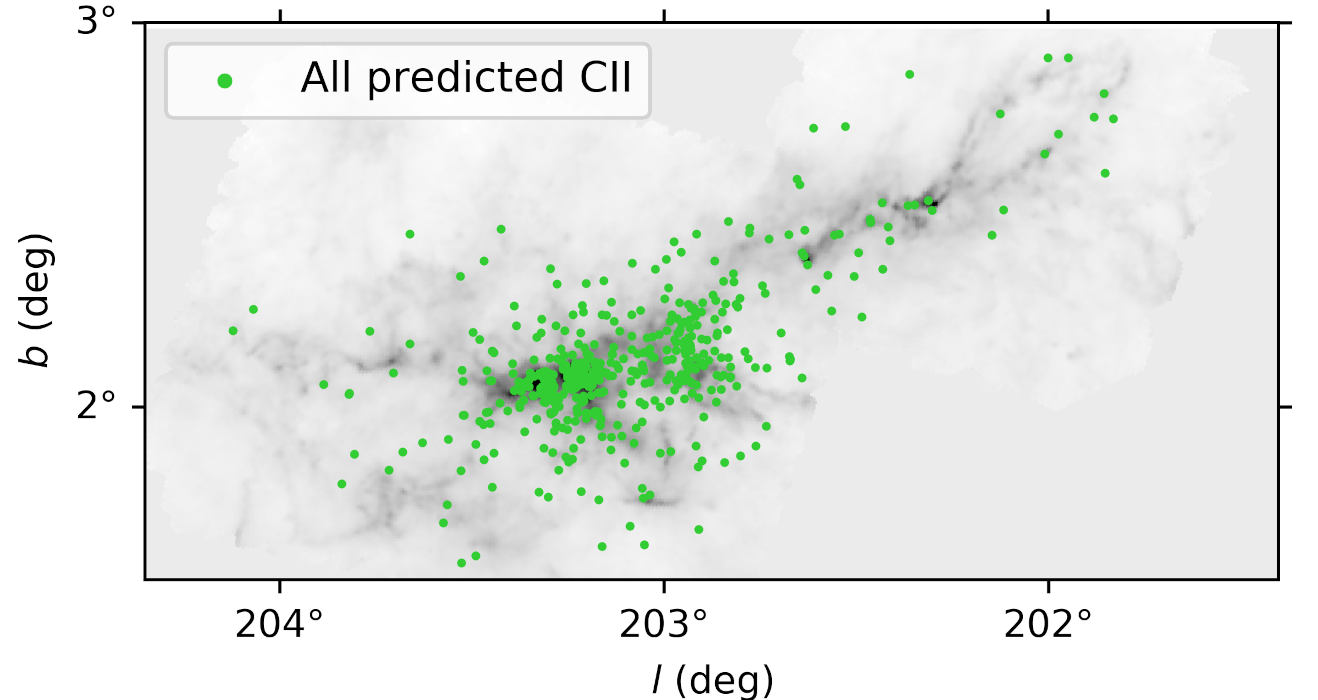}
	\end{subfigure}
	\end{minipage}
	\caption[NGC 2264 YSO candidates distribution]{Distribution of YSO candidates in the NGC 2264 region. The grey scale shows the Herschel SPIRE 500 $\mathrm{\mu m}$ map. CI and CII YSOs are shown in the middle and lower frames in red and green, respectively.}
\label{ngc2264_yso_dist}
\end{figure*}

The NGC 2264 region is presented in Figure~\ref{ngc2264_yso_dist}. As for the two others, the main star-forming region is well traced by our CI YSOs, with CII being more dispersed \citep[as in]{Buckner_2020}. As for the rest of the study the smaller number of YSOs in the region makes it slightly more difficult to analyze.\\

\begin{figure}[!t]
	\includegraphics[width=0.9\hsize]{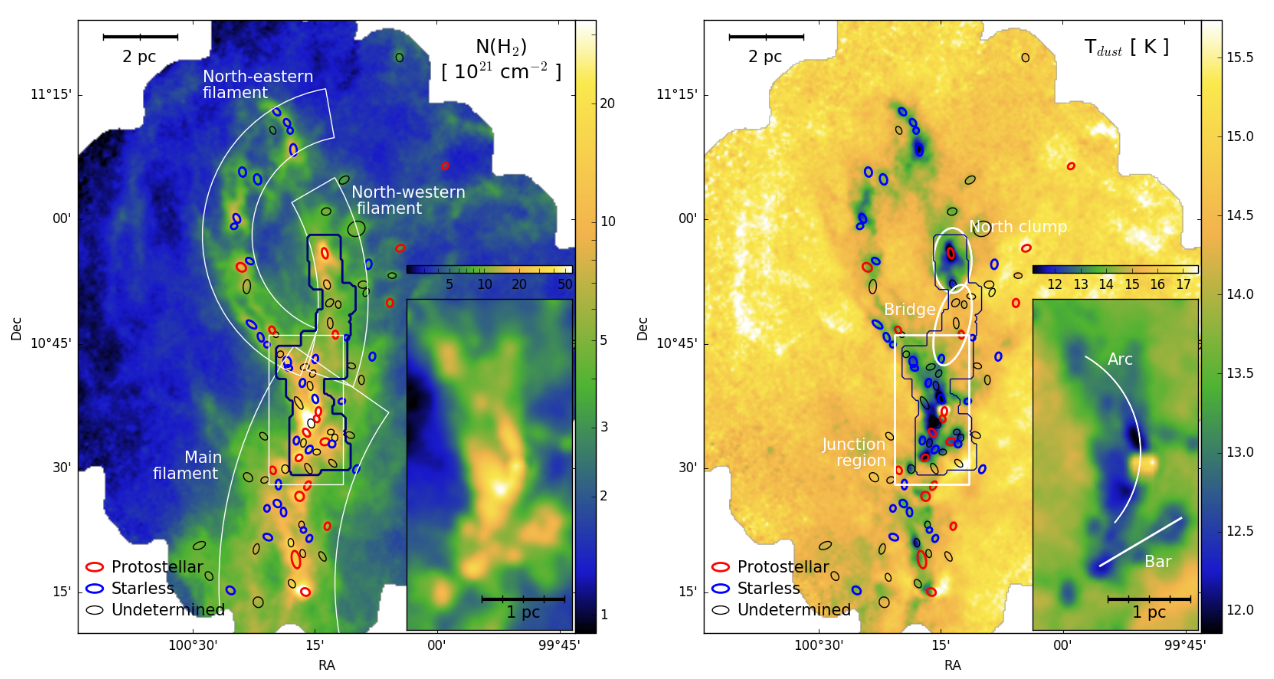}
	\caption[Herschel view of the G202.3+2.5 region]{Herschel view of the G202.3+2.5 region, about 1 deg north to the open cluster NGC 2264. {\it Left}: column density of molecular hydrogen derived from the SED fit of SPIRE bands. {\it Right}: dust temperature from the same SED fit. In both frames, the ellipses show the submillimeter compact sources extracted by Montillaud et al. (2015). The junction region, along with other important structures of the cloud are indicated with white shapes. In both frames, the inset shows a zoom to the junction region. {\it From} \citet{montillaud_2019_II}.}
	\label{montillaud_2019b_fig2}
\end{figure}

Interestingly, there is a small concentration of CI YSOs around $l = 202.3$, $b = 2.5$ which corresponds to the $G202.3+02.50$ molecular cloud region where we showed that two filaments of the cloud are colliding \citep[][and Fig.~\ref{montillaud_2019b_fig2}]{montillaud_2019_I, montillaud_2019_II}. It is remarkable that this small CI cluster coincides tightly with the junction region between the two merging filaments, whereas the CII distribution seems mostly independent. Based on the typical time scale of the CI protostellar phase, and assuming that the formation of the small CI cluster was triggered by the filament collision, we conclude that the collision would have started typically $\lesssim 5 \times 10^5$ years ago \citep{Evans_2009}. This age is compatible with the age estimate of $\sim 10^5$ yr obtained from N$_2$H$^+$ observations of the junction region (Montillaud+19b).\\

Finally, we do not provide any prediction from the 1\,kpc dataset, because it is impossible for us to construct a confusion matrix and to provide quality estimators from them, due to the absence of contaminants. However, we have made some attempts to use our trained network over other star-forming regions using Spitzer data, which is partly discussed in Section~\ref{conclusion_perspectives}.\\

To conclude this section, we are confident that our Full 1\,kpc trained network contains a sufficient diversity of subclasses to be efficiently applied to most nearby ($\lesssim 1$ kpc) star forming regions. Our results show that one can expect nearly $90\%$ of the CI YSOs to be properly recovered with a precision above $80\%$, while near $97\%$ of CII YSOs are expected to be recovered with a $90\%$ precision.

\clearpage
\section{Probabilistic prediction contribution to the analysis}
\label{proba_discussion}

\etocsettocstyle{\subsubsection*{\vspace{-1cm}}}{}
\localtableofcontents

\vspace{0.3cm}

\begin{figure*}[!t]
\vspace{-0.2cm}
	\centering
	\begin{subfigure}[t]{0.35\textwidth}
	\caption*{\textbf{Output}}
	\includegraphics[width=\textwidth]{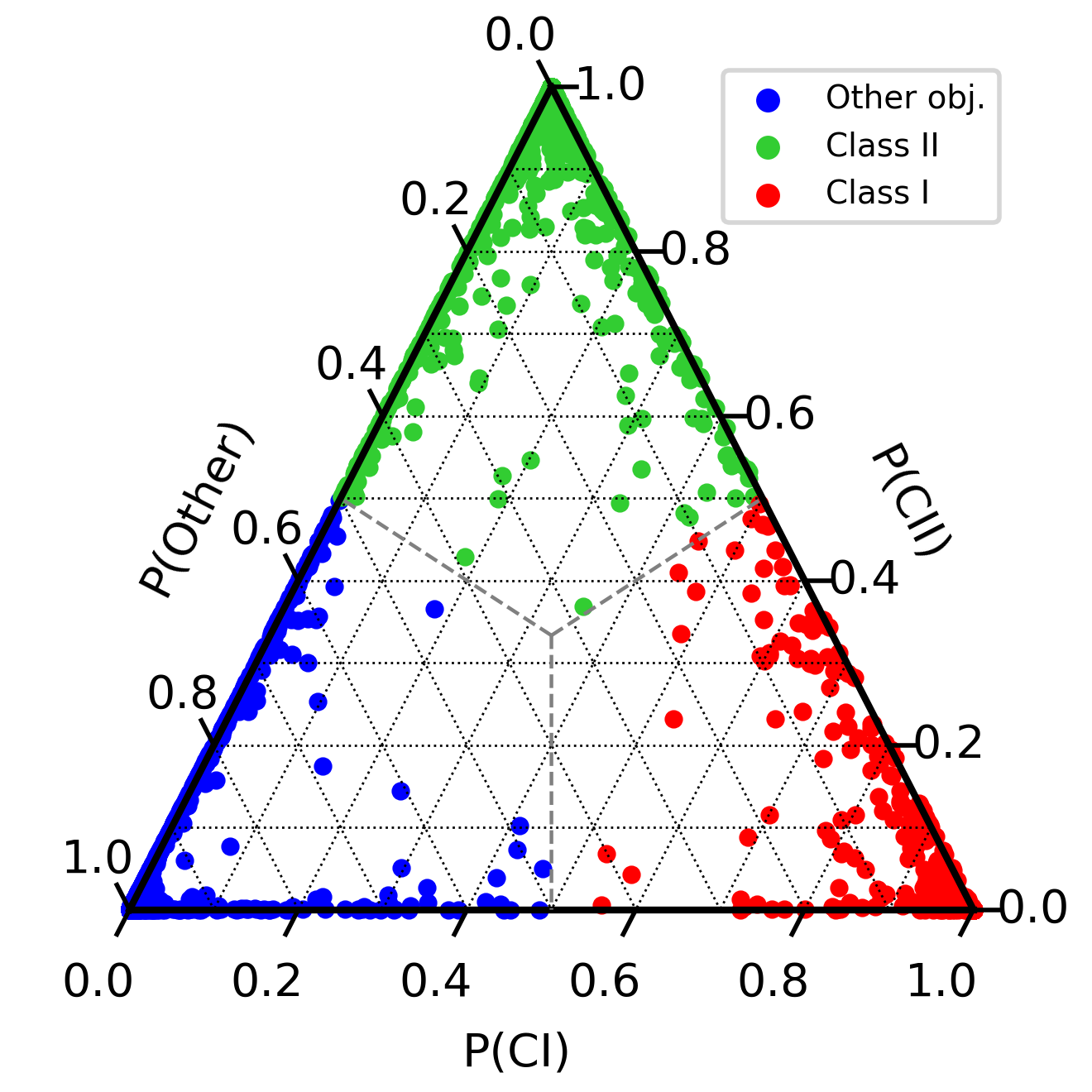}
	\end{subfigure}
	\hspace{0.2cm}
	\begin{subfigure}[t]{0.35\textwidth}
	\caption*{\textbf{Correct}}
	\includegraphics[width=\textwidth]{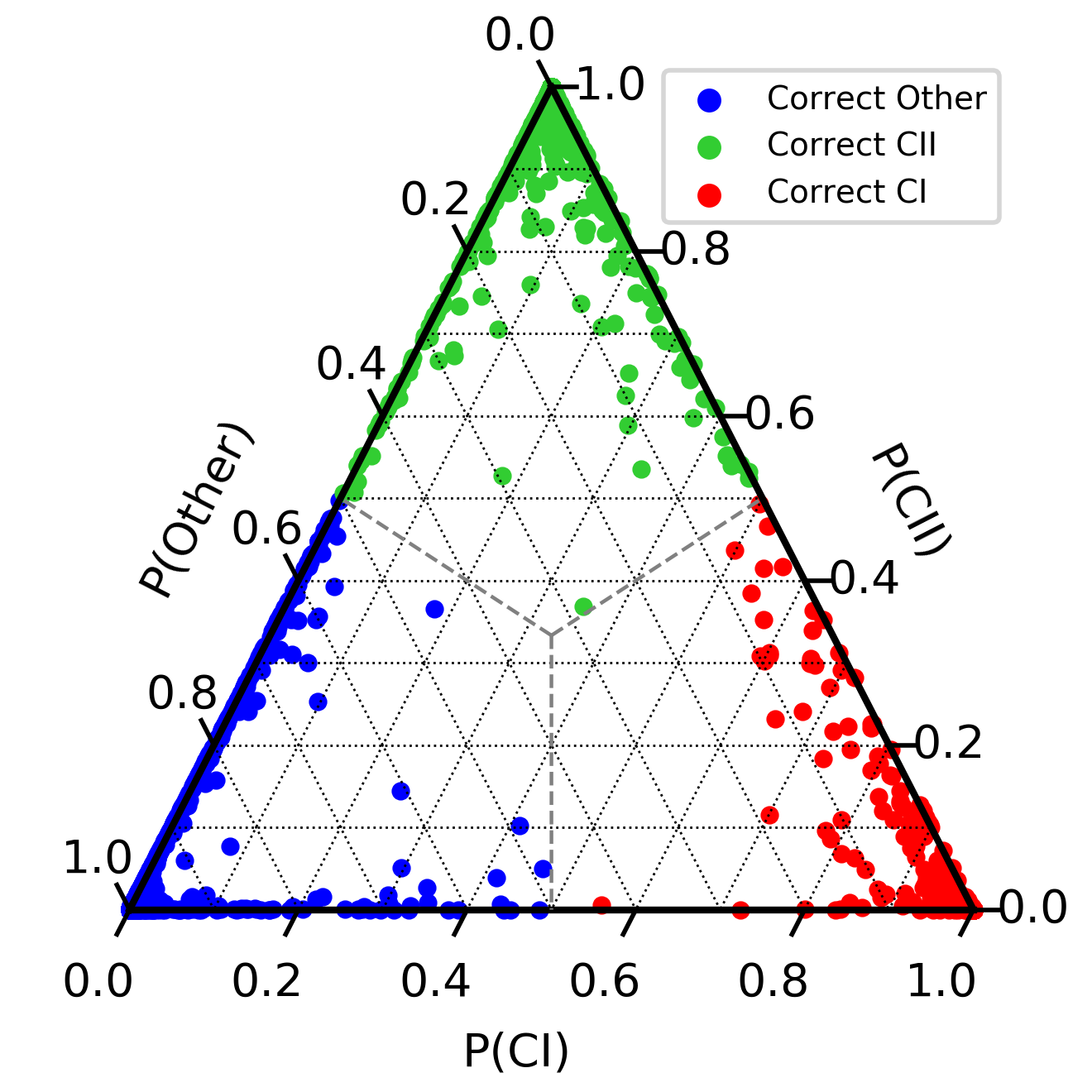}
	\end{subfigure}\\
	\vspace{0.1cm}
	\begin{subfigure}[t]{0.35\textwidth}
	\caption*{\textbf{Missed}}
	\includegraphics[width=\textwidth]{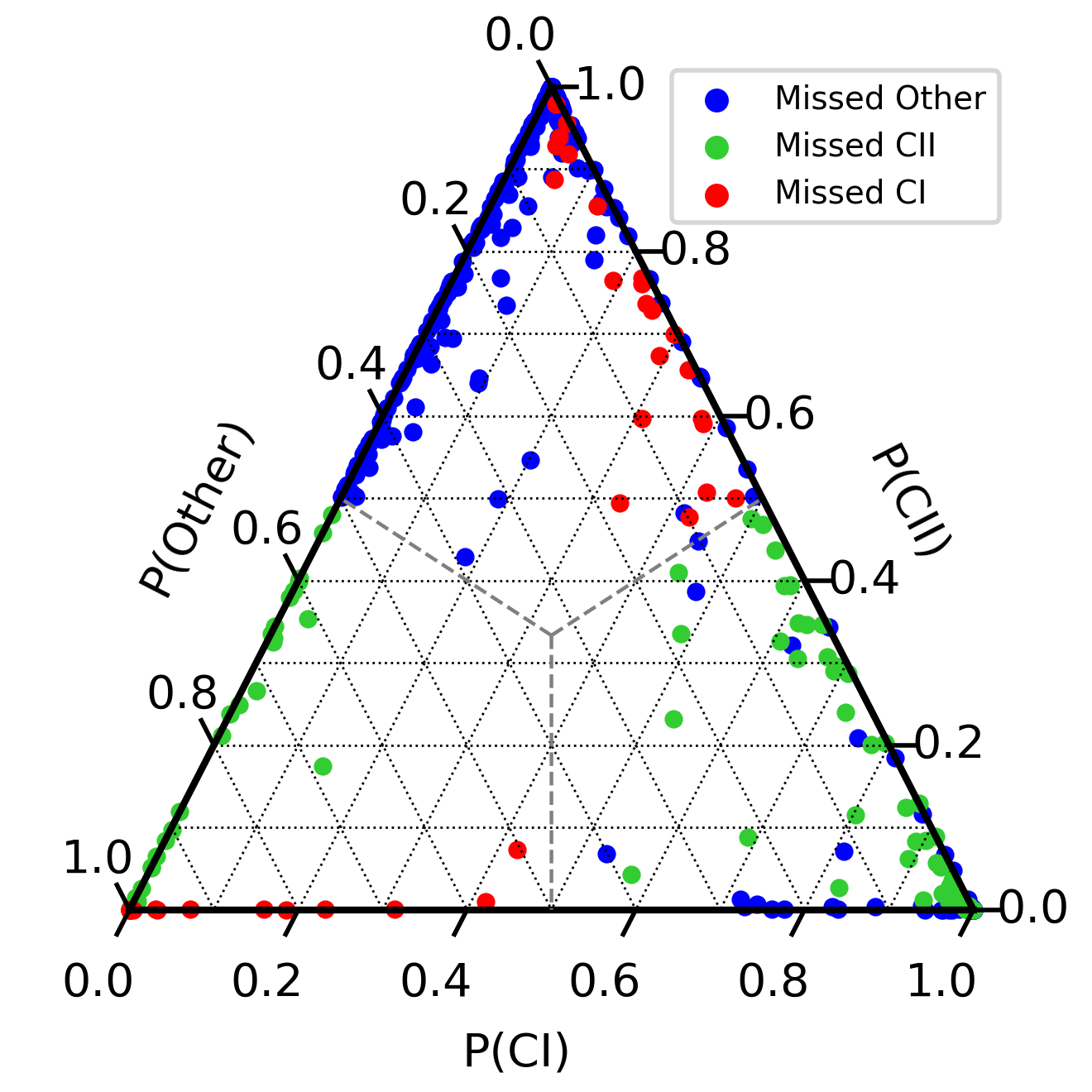}
	\end{subfigure}
	\hspace{0.2cm}
	\begin{subfigure}[t]{0.35\textwidth}
	\caption*{\textbf{Wrong}}
	\includegraphics[width=\textwidth]{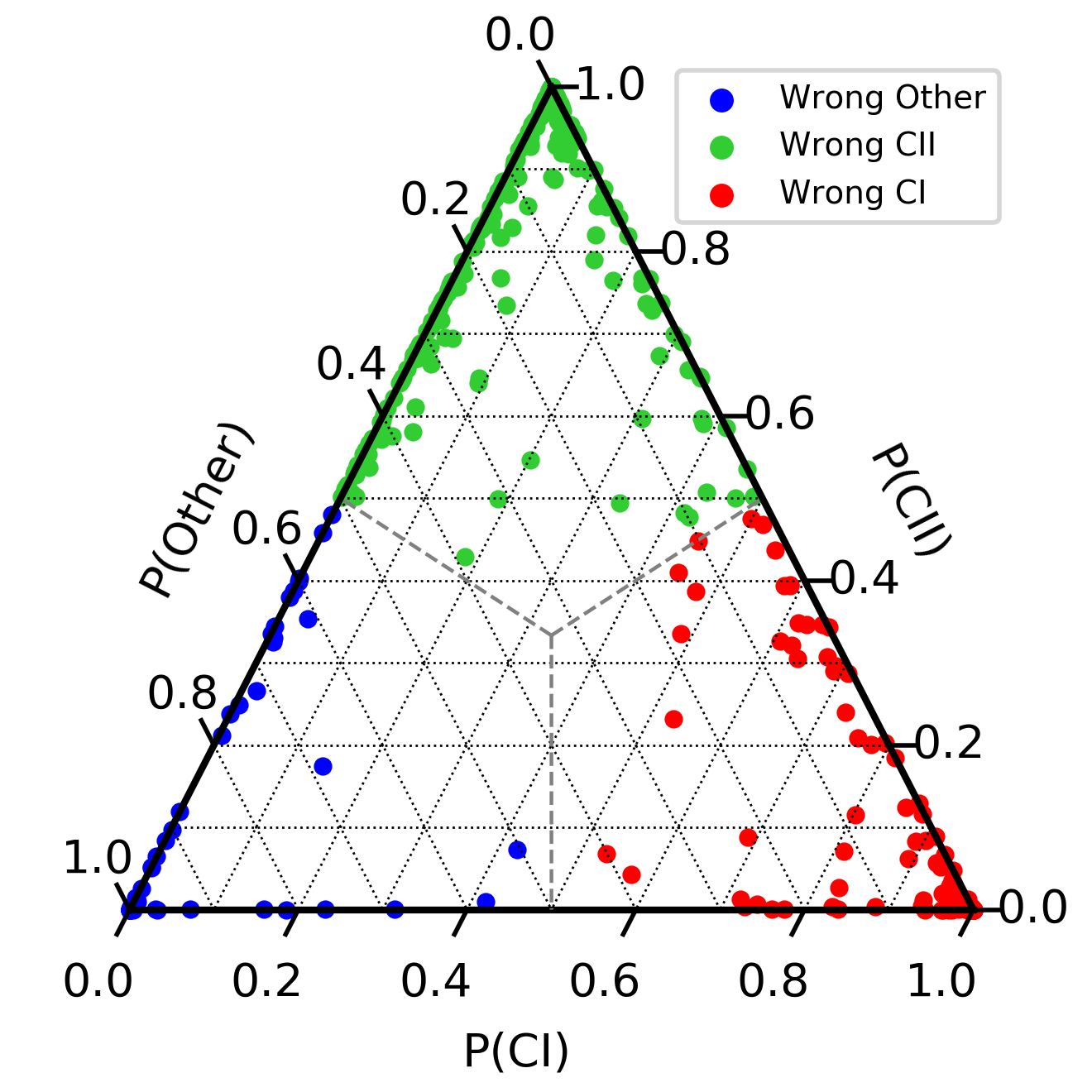}
	\end{subfigure}
	\caption[Ternary plots of output membership probability in the F-C case]%
	{\href{https://doi.org/10.5281/zenodo.2628066}{Ternary plots} of output membership probability for each class in the F-C case forwarded on the full dataset. \textit{Output:} all objects. \textit{Correct:} genuine and predicted classes are identical. \textit{Missed:} misclassified objects colored regarding their genuine class. \textit{Wrong:} misclassified object colored regarding their predicted class.}
	\label{ternary_plots}
\end{figure*}

In this section, we discuss the inclusion of a membership probability prediction into our network. If we assumed that the original classification were absolutely correct, the discrepancies would only correspond to errors. However, as illustrated by the effect of the MIPS band, the original classification has its own limitations. Therefore, the objects misclassified by our network might highlight that they were already less reliable in the original classification or may even have been misclassified. A membership probability allows one to refine this idea by quantifying the level of confidence of the network on each prediction, directly based on the observed distribution of the objects in the input parameter space. In practice, as already illustrated in Figure~\ref{missed_wrong_zoom}, where misclassified objects stack around the inter-class boundaries, the classification reliability of individual objects is mostly a function of their distance to these boundaries. One strength of the probabilistic output presented in Section~\ref{proba_class_intro} is that the probability values provided by the network take advantage of the network ability to combine the boundaries directly in the ten dimensions of the feature space.\\

\subsection{Interpretation of the membership probability}

We used the probabilistic predictions to measure the degree of confusion of an object between the output classes. This is illustrated by the ternary plots in Figure~\ref{ternary_plots}, where the location of the objects corresponds to their predicted probability to belong to each class. On these plots, an object with a high confidence level lies near the peaks. Objects that are in the inner part of the graph are the most confused between the three classes, while objects on the edges illustrate a confusion between only two classes. The sample size obviously plays a role in this representation, but each class clearly shows a level of confusion that is higher with one specific other class. The graph for all outputs shows that the confusion between CI and Other is the lowest, followed by the confusion between CI and CII YSOs, with the highest confusion level being between the CII and Other classes. Those observations are strongly consistent with our previous analysis based only on the confusion matrix.\\

\begin{figure*}[!t]
\hspace{-1.2cm}
\begin{minipage}{1.17\textwidth}
	\centering
	\begin{subfigure}[t]{0.48\textwidth}
	\caption*{\textbf{Output}}
	\includegraphics[width=\textwidth]{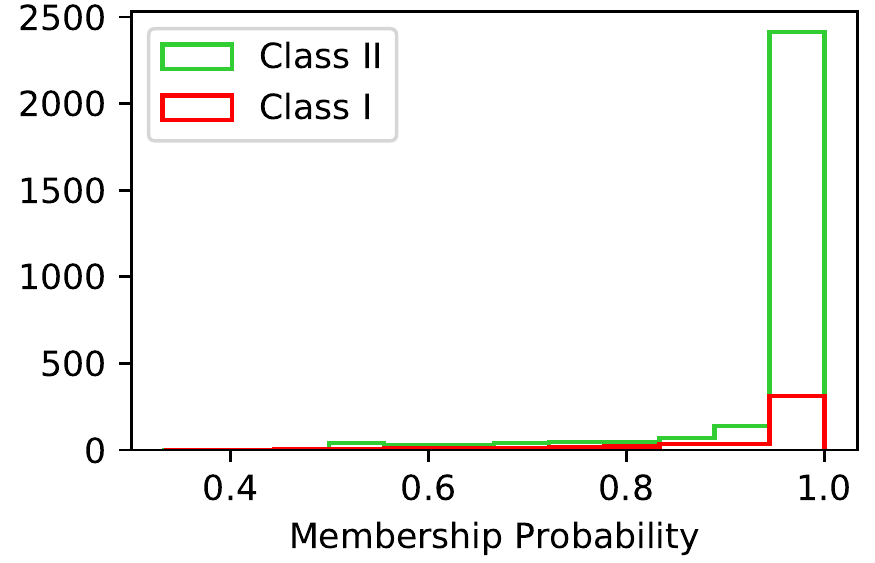}
	\end{subfigure}
	\begin{subfigure}[t]{0.48\textwidth}
	\caption*{\textbf{Correct}}
	\includegraphics[width=\textwidth]{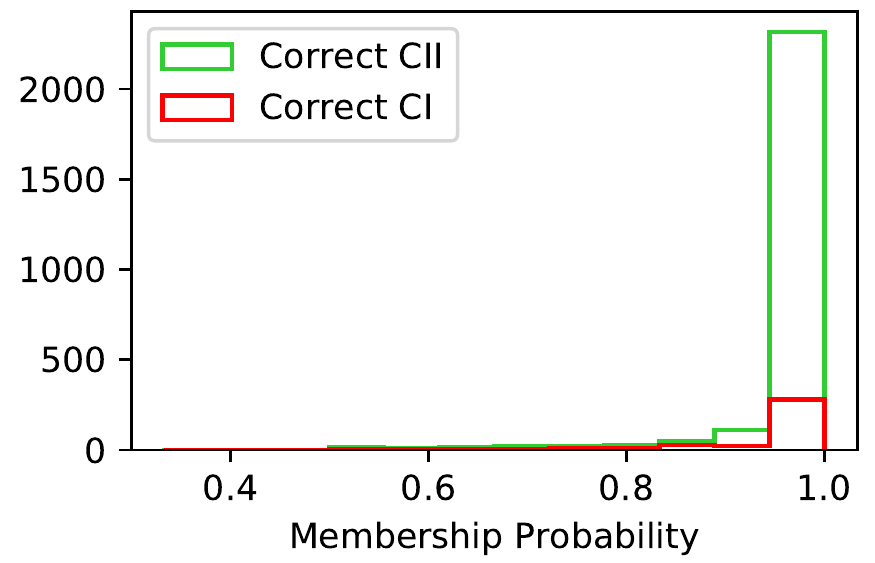}
	\end{subfigure}\\
	\vspace{0.2cm}
	\begin{subfigure}[t]{0.48\textwidth}
	\caption*{\textbf{Missed}}
	\includegraphics[width=\textwidth]{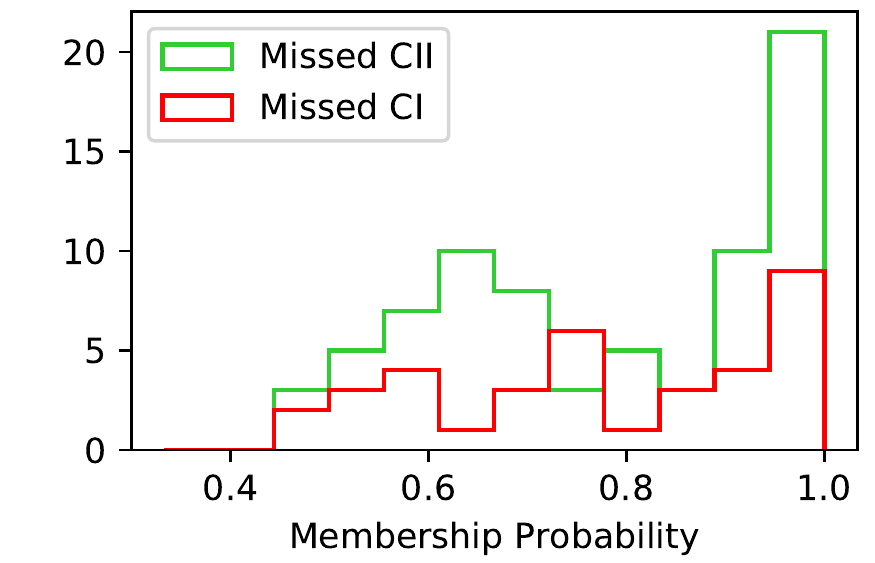}
	\end{subfigure}
	\begin{subfigure}[t]{0.48\textwidth}
	\caption*{\textbf{Wrong}}
	\includegraphics[width=\textwidth]{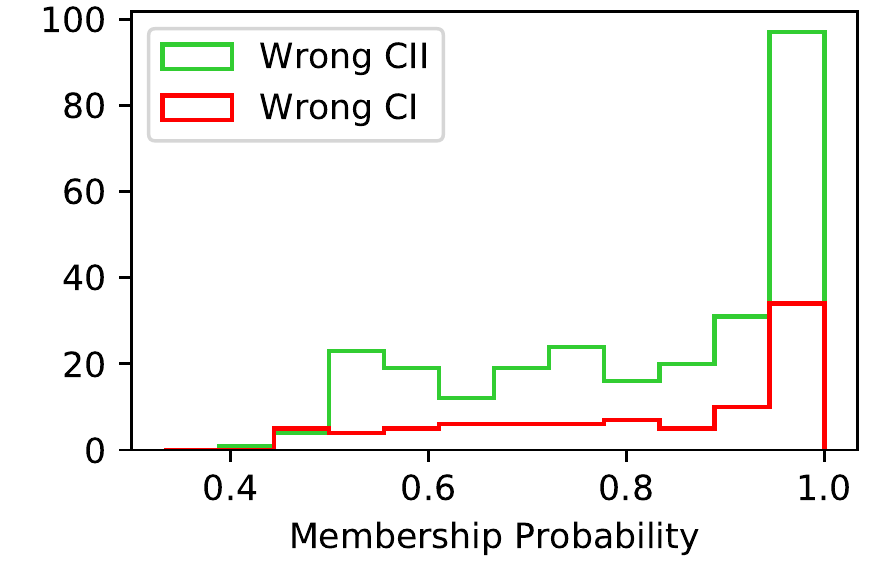}
	\end{subfigure}
	\end{minipage}
	\caption[Histograms of membership probability in the F-C case]{Histograms of membership probability for YSO classes regarding different populations in the F-C case forwarded on the full dataset. \textit{Output:} all objects. \textit{Correct:} genuine and predicted classes are identical. \textit{Missed:} misclassified objects colored regarding their genuine class. \textit{Wrong:} misclassified object colored regarding their predicted class.}
	\label{hist_proba}
\end{figure*}

\begin{table}[!t]
	\small
	\centering
	\caption{F-C case forwarded on the full dataset with membership probability $> 0.9$. }
	\vspace{-0.1cm}
	\begin{tabularx}{0.65\hsize}{r l |*{3}{m}| r }
	\multicolumn{2}{c}{}& \multicolumn{3}{c}{\textbf{Predicted}}&\\
	\cmidrule[\heavyrulewidth](lr){2-6}
	\parbox[l]{0.2cm}{\multirow{6}{*}{\rotatebox[origin=c]{90}{\textbf{Actual}}}} & Class & CI YSO & CII YSO & Other & Recall \\
	\cmidrule(lr){2-6}
	 &  CI YSO    & 297     & 5       & 8       & 95.8\% \\
	 &  CII YSO   & 16      & 2412    & 13      & 98.8\% \\
	 &  Other     & 26      & 118     & 23247   & 99.4\% \\
	\cmidrule(lr){2-6}
	 &  Precision & 87.6\% & 95.1\% & 99.9\% & 99.3\% \\
	\cmidrule[\heavyrulewidth](lr){2-6}
	\end{tabularx}
	\caption*{\vspace{-0.3cm}\\ {\bf Notes.} The selection removed 104 CI ($-25.1\%$), 218 CII ($-8.2\%$), and 439 Other ($-1.8\%$).}
	\vspace{-0.1cm}
	\label{conf_proba_09} 
\end{table}

\begin{table}[!t]
	\small
	\centering
	\caption{F-C case forwarded on the full dataset with membership probability $> 0.95$. }
	\vspace{-0.1cm}
	\begin{tabularx}{0.65\hsize}{r l |*{3}{m}| r }
	\multicolumn{2}{c}{}& \multicolumn{3}{c}{\textbf{Predicted}}&\\
	\cmidrule[\heavyrulewidth](lr){2-6}
	\parbox[l]{0.2cm}{\multirow{6}{*}{\rotatebox[origin=c]{90}{\textbf{Actual}}}} & Class & CI YSO & CII YSO & Other & Recall \\
	\cmidrule(lr){2-6}
	 &  CI YSO    & 274     & 2       & 7       & 96.8\% \\
	 &  CII YSO   & 11      & 2302    & 8       & 99.2\% \\
	 &  Other     & 23      & 92      & 23136   & 99.5\% \\
	\cmidrule(lr){2-6}
	 &  Precision & 89.0\% & 96.1\% & 99.9\% & 99.4\% \\
	\cmidrule[\heavyrulewidth](lr){2-6}
	\end{tabularx}
	\caption*{\vspace{-0.3cm}\\ {\bf Notes.} The selection removed 131 CI ($-31.6\%$), 338 CII ($-12.7\%$), and 579 Other ($-2.4\%$).}
	\vspace{-0.1cm}
	\label{conf_proba_095} 
\end{table}

\begin{table}[!t]
	\small
	\centering
	\caption{F-C case forwarded on the full dataset with membership probability $> 0.99$.}
	\vspace{-0.1cm}
	\begin{tabularx}{0.65\hsize}{r l |*{3}{m}| r }
	\multicolumn{2}{c}{}& \multicolumn{3}{c}{\textbf{Predicted}}&\\
	\cmidrule[\heavyrulewidth](lr){2-6}
	\parbox[l]{0.2cm}{\multirow{6}{*}{\rotatebox[origin=c]{90}{\textbf{Actual}}}} & Class & CI YSO & CII YSO & Other & Recall \\
	\cmidrule(lr){2-6}
	 &  CI YSO    & 203     & 0       & 5       & 97.6\% \\
	 &  CII YSO   & 4       & 1970    & 4       & 99.6\% \\
	 &  Other     & 14      & 51      & 22747   & 99.7\% \\
	\cmidrule(lr){2-6}
	 &  Precision & 91.9\% & 97.5\% & 99.9\% & 99.7\% \\
	\cmidrule[\heavyrulewidth](lr){2-6}
	\end{tabularx}
	\caption*{\vspace{-0.3cm}\\ {\bf Notes.} The selection removed 206 CI ($-49.8\%$), 681 CII ($-25.6\%$), and 1018 Other ($-4.3\%$).}
	\vspace{-0.1cm}
	\label{conf_proba_099} 
\end{table}

Additionally, as discussed in Section~\ref{proba_class_intro}, the probabilistic predictions can be used to remove objects that are not reliable enough. The misclassified objects show a higher degree of confusion, and therefore a maximum value of membership probability lower than the objects properly classified. This characteristic is illustrated by Figures~\ref{ternary_plots} and \ref{hist_proba}. The latter compares histograms of the highest output probability for properly and wrongly classified objects. This figure reveals that the great majority of correctly classified YSOs have a membership probability greater than 0.95, whereas most missed or wrong YSOs have a probability membership below that threshold. In this context, applying a threshold on the membership probability will proportionally remove more misclassified objects than properly classified ones, therefore improving the recall and precision of our network. The threshold value is arbitrary, depending on the application. \\

\newpage
We illustrate this selection effect on the F-C case in Tables~\ref{conf_proba_09}, \ref{conf_proba_095}, and \ref{conf_proba_099}. These tables represent the confusion matrix of the complete combined dataset after selecting objects with membership probability above 0.9, 0.95 and 0.99 respectively. In the 0.9 case (Table~\ref{conf_proba_09}), 25\% (104) of the CI YSOs were removed, while their recall increased by 4.5\%. In the same way, 8.2\% (218) of the CII YSOs were removed leading to a 1.2\% increase in their recall. Contaminants were less affected with only 1.8\% of objects removed, which still increased the recall by 0.6\%. This is an additional demonstration of the CI YSOs being less constrained than the other output classes. In the 0.95 case (Table~\ref{conf_proba_095}), the output classes have lost 31.6\% (131), 12.7 (338), and 2.4\% (579) of objects, respectively. This still improved the recall of the two YSO classes with a 1\% increase for CI and an 0.4\% increase for CII, when compared to the 0.9 case. This result is also the first one to be close to have all quality estimators above 90\%, since the CI YSO precision is 89\%, while losing an acceptable fraction of them. The 0.99 case (Table~\ref{conf_proba_099}) is more extreme, since almost 50\% (206) of CI YSOs were removed, but the recall of the remaining one reached 97.6\%, that is a 6.3\% improvement over the regular F-C full dataset case. However, the CII YSOs are also strongly affected, with 25.6\% (681) of them removed, and only yielding a 0.4\% improvement in comparison to the 0.95 case. Another illustration that this strategy effectively excludes objects that are near the cuts is presented by Figure~\ref{membership_threshold_comparison} where the objects above or below a $0.9$ membership threshold are plotted separately for a usual set of CMDs. This effect is particularly visible in the ([4.5]-[8],[3.6]-[5.8]) (second frame) and the ([4.5]-[5.8],[3.6]-[4.5]) (forth frame) diagrams. This figure illustrates that membership probability less than 0.9 can be considered unreliable. \\

\begin{figure*}
	\centering
	\begin{subfigure}[t]{0.49\textwidth}
	\caption*{{\bf \Large Above 0.9}}
	\includegraphics[width=\textwidth]{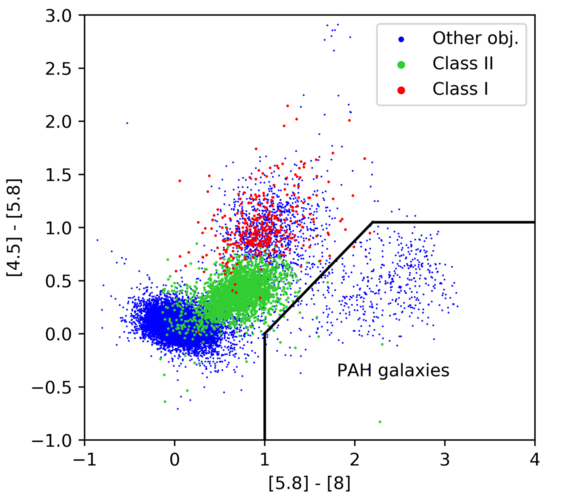}
	\end{subfigure}
	\begin{subfigure}[t]{0.49\textwidth}
	\caption*{{\bf \Large Bellow 0.9}}
	\includegraphics[width=\textwidth]{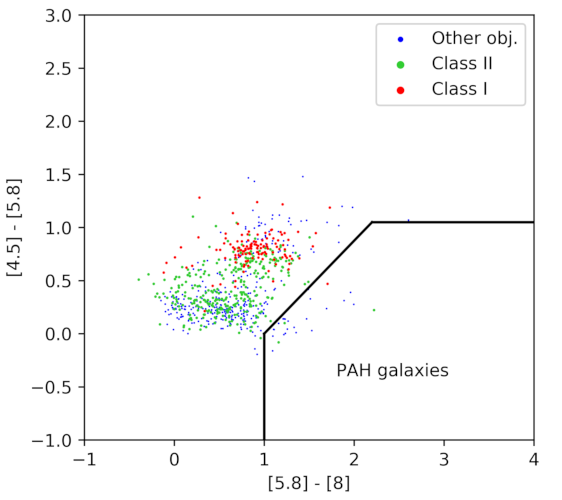}
	\end{subfigure}\\
	\begin{subfigure}[t]{0.49\textwidth}
	\includegraphics[width=\textwidth]{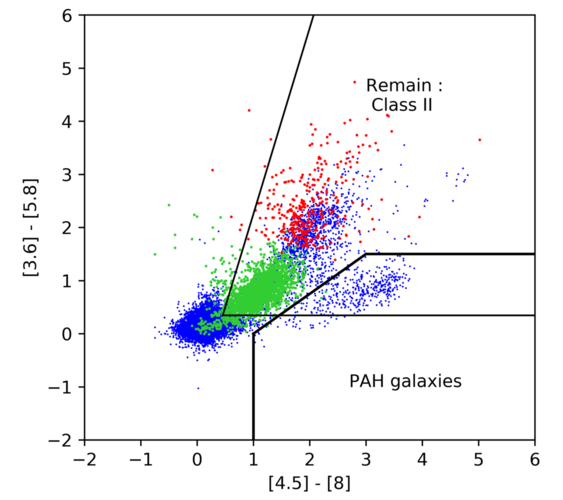}
	\end{subfigure}
	\begin{subfigure}[t]{0.49\textwidth}
	\includegraphics[width=\textwidth]{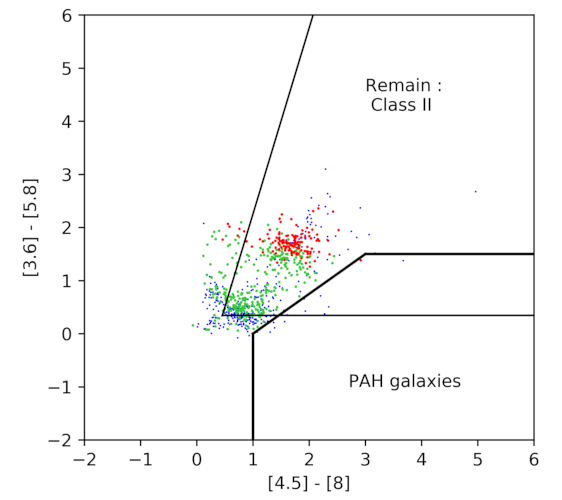}
	\end{subfigure}\\
	\begin{subfigure}[t]{0.49\textwidth}
	\includegraphics[width=\textwidth]{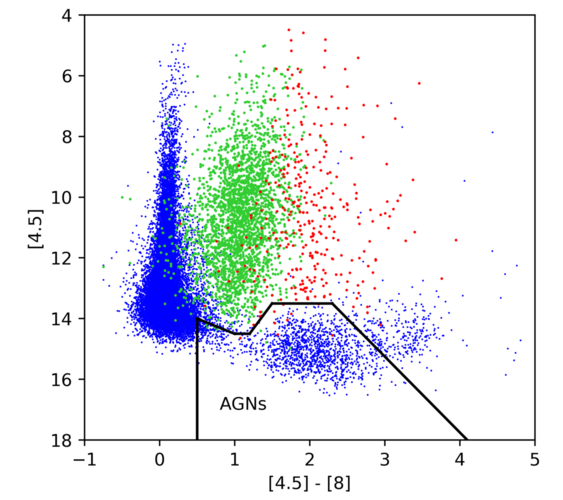}
	\end{subfigure}
	\begin{subfigure}[t]{0.49\textwidth}
	\includegraphics[width=\textwidth]{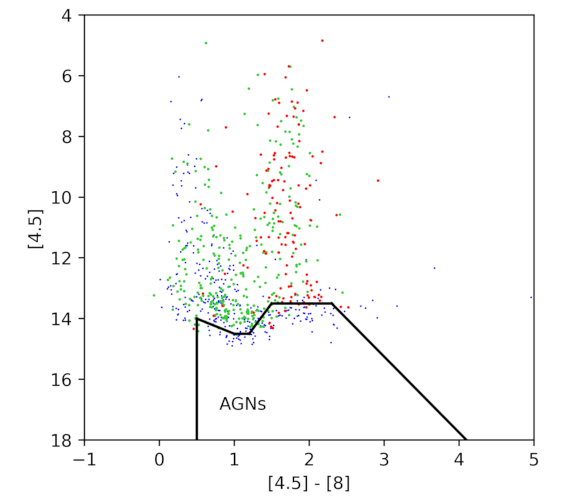}
	\end{subfigure}
	\caption{See caption on second half of the figure next page.}
\end{figure*}

\addtocounter{figure}{-1}

\begin{figure*}
	\centering
	\begin{subfigure}[t]{0.49\textwidth}
	\includegraphics[width=\textwidth]{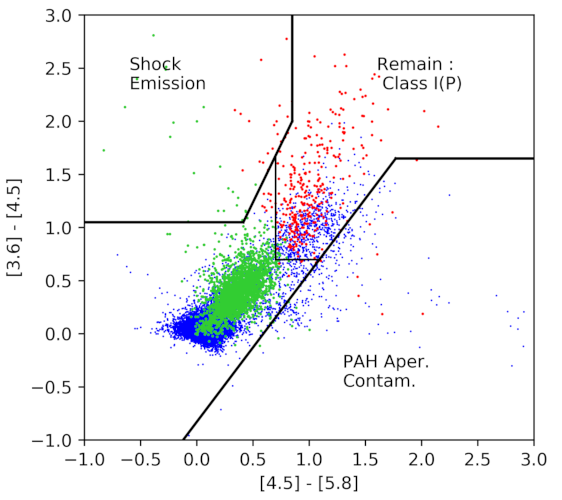}
	\end{subfigure}
	\begin{subfigure}[t]{0.49\textwidth}
	\includegraphics[width=\textwidth]{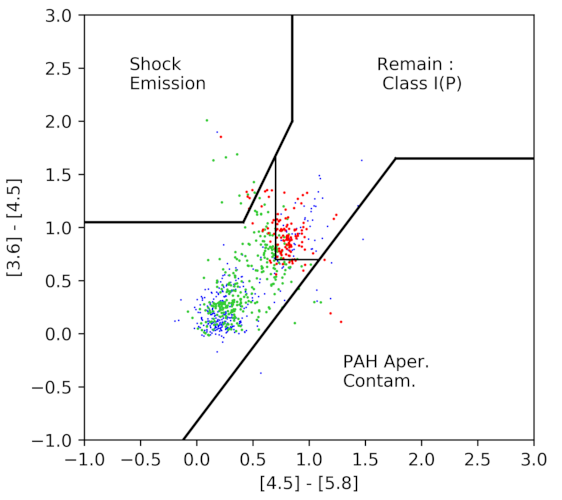}
	\end{subfigure}\\
	\begin{subfigure}[t]{0.49\textwidth}
	\includegraphics[width=\textwidth]{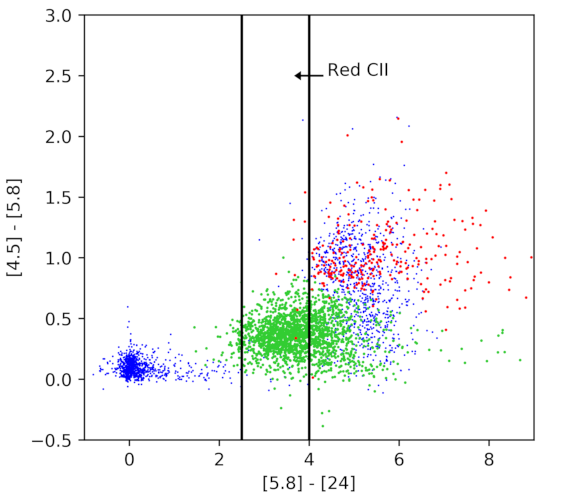}
	\end{subfigure}
	\begin{subfigure}[t]{0.49\textwidth}
	\includegraphics[width=\textwidth]{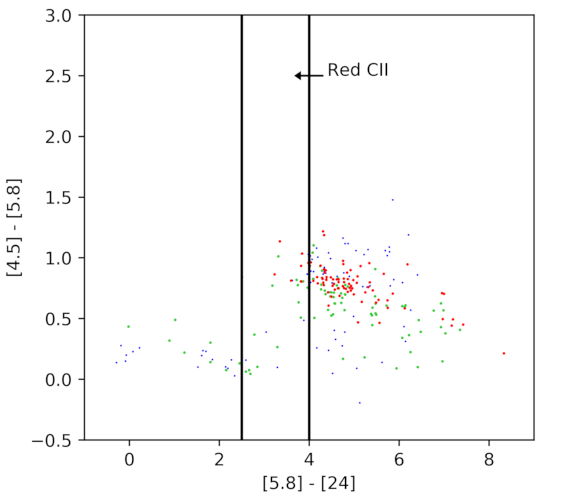}
	\end{subfigure}
	\caption[Probability threshold effect on feature space coverage]{Input parameter space coverage using the usual G09 diagrams in the F-C case on the full dataset regarding their predicted membership probability. CI YSOs are in red, CII YSOs are in green while Other are in blue. \textit{Left:} objects with membership probability greater than $0.9$. \textit{Right:} objects with membership probability less than $0.9$.}
	\label{membership_threshold_comparison} 
\end{figure*}

\vspace{0.3cm}
It is important to emphasize again, as stated in Section~\ref{proba_class_intro}, that the membership probability output is not a direct physical probability. It is a probability regarding the network knowledge of the problem, which can be biased or incomplete or both. Therefore, selecting a $0.9$ membership probability does not necessarily correspond to a $90\%$ certainty prediction level. The only usable probability is the one given by the confusion matrix. Consequently, according to Table~\ref{conf_proba_09}, when applying a $0.9$ membership limit, the probability that a predicted class I YSO is correct is estimated to be $87.6\%$, while, with the same limit, class II YSOs are correct in $96.1\%$ of the cases. These two values are not equivalent and one must not use the network output membership probability as a true estimate of the reliability of an object. It can only be used to compare objects from the same network training, and must be converted as a true quality estimator using the confusion matrix. \\

\newpage
\subsection{Graphical analysis of the membership probability}

\begin{figure}[!t]
\hspace{-1.8cm}
	\begin{minipage}{1.20\hsize}
	\centering
	\begin{subfigure}[t]{0.49\textwidth}
	\includegraphics[width=\hsize]{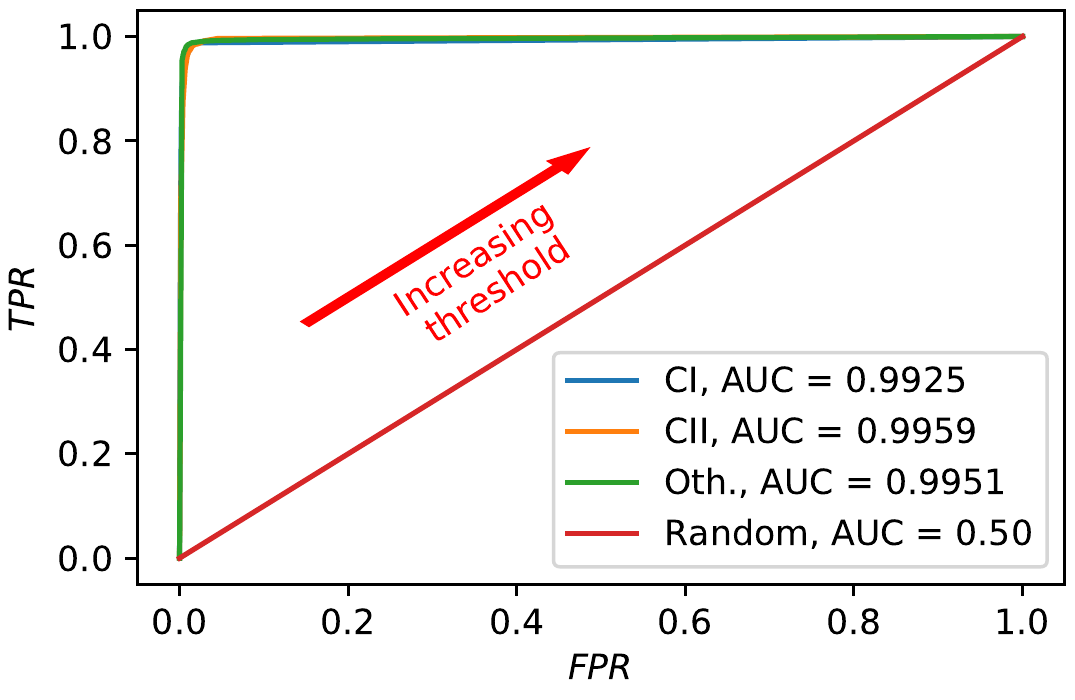}
	\end{subfigure}
	\begin{subfigure}[t]{0.49\textwidth}
	\includegraphics[width=\hsize]{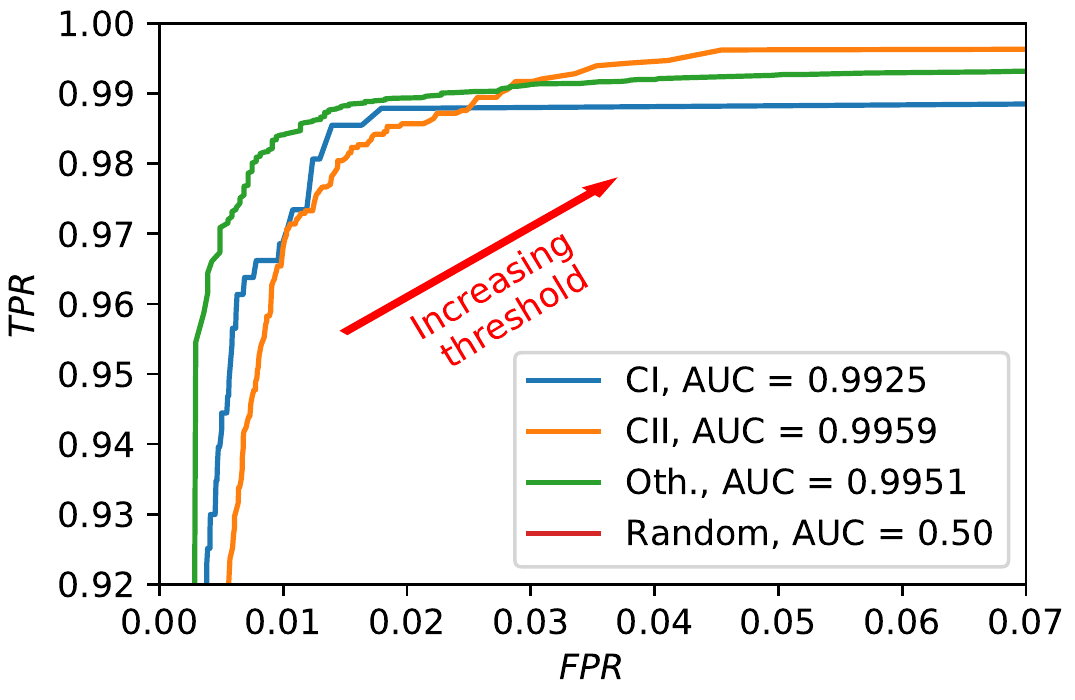}
	\end{subfigure}
	\end{minipage}
	\caption[ROC curves for output classes in the F-C case]{ROC curves for each of our output classes, CI in blue, CII in orange and other in green. The red curve illustrates a random classifier. Each point of the curve is obtained from a given threshold limit. The {\it right} frame is a zoom of the upper left part of the {\it left} frame.}
	\label{roc_curve}
	\vspace{1.5cm}
\end{figure}

Printing the confusion matrix for each threshold value of the membership probability is the optimal way to get the direct performance information at a given threshold value. Despite this, there are some common representations that can be made using the threshold value. The most common one is the Receiver Operating Characteristic (ROC) curve, which is a standard tool to assess the prediction quality of a binary probabilistic classifier. In our case it is possible to plot the corresponding ROC curve for each output class by considering it as binary output against the other two classes. The ROC curve is usually defined as the False Positive Rate (FPR) or 1 - specificity, against the True Positive Rate (TPR) or sensitivity, which is the equivalent of our previously defined recall (Sect.~\ref{class_balance}). To produce this curve, the threshold value is sampled and the previous two values are computed for each point for a specific class. We stress that, during this process the predicted class depends solely on the fact that the corresponding neuron has a value higher than the threshold and does not mean that it is the maximum value of the output neurons. It produced the Figure~\ref{roc_curve} that contains the corresponding ROC for each class. In this figure, a random classifier would produce a linear response, while a perfect classifier would have only one point at the top left edge of the graph, meaning that it has both a perfect sensitivity and a perfect specificity. This ROC curve allows to compute the Area Under the Curve or AUC, that is an estimate of the global binary classifier performance. The regular ROC plot being generally used for less efficient classifier, we made a zoom on the interesting part for our case in the bottom frame of Figure~\ref{roc_curve}. It is striking by looking at the AUC and the curves that our CI YSO class is less well represented. Interestingly, the CII and Other classes seem more or less equivalent using this quality estimator, which was not the case when looking at our confusion matrix.\\

\begin{figure}[!t]
	\centering
	\begin{subfigure}[t]{0.70\textwidth}
	\includegraphics[width=\hsize]{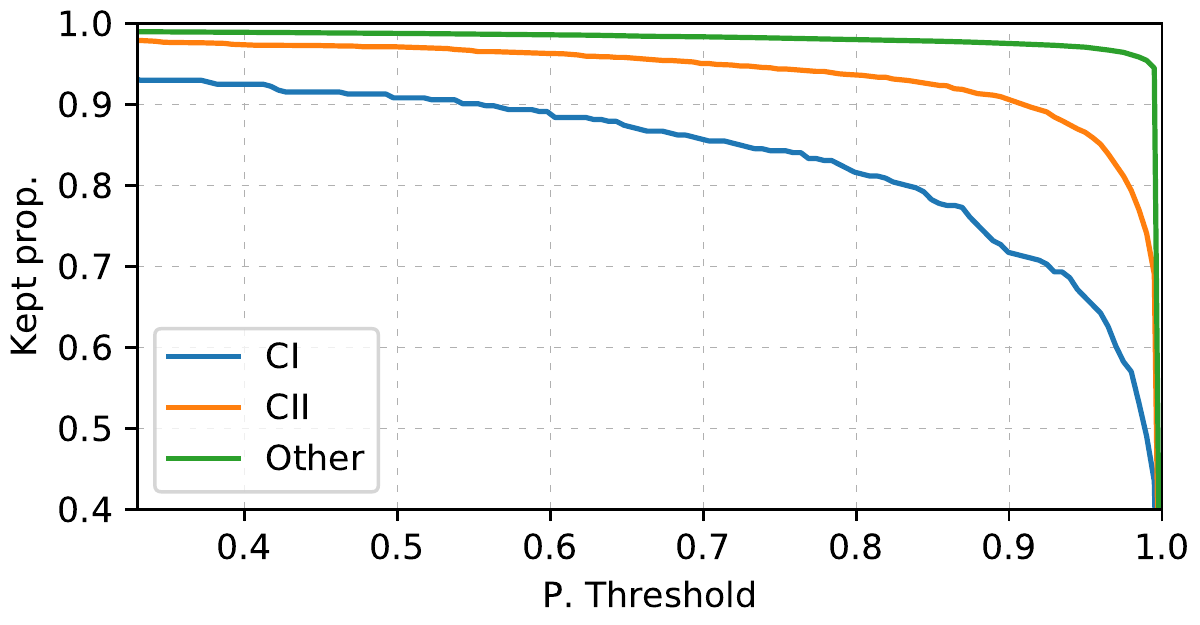}
	\end{subfigure}\\
	\vspace{0.4cm}
	\begin{subfigure}[t]{0.70\textwidth}
	\includegraphics[width=\hsize]{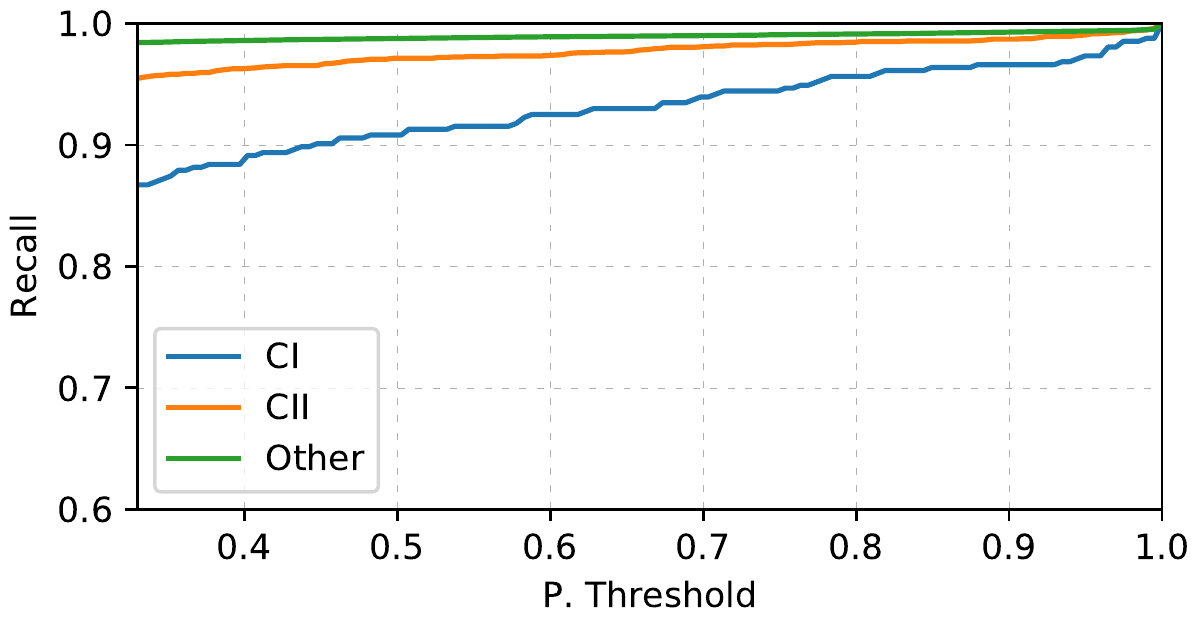}
	\end{subfigure}\\
	\vspace{0.4cm}
	\begin{subfigure}[t]{0.70\textwidth}
	\includegraphics[width=\hsize]{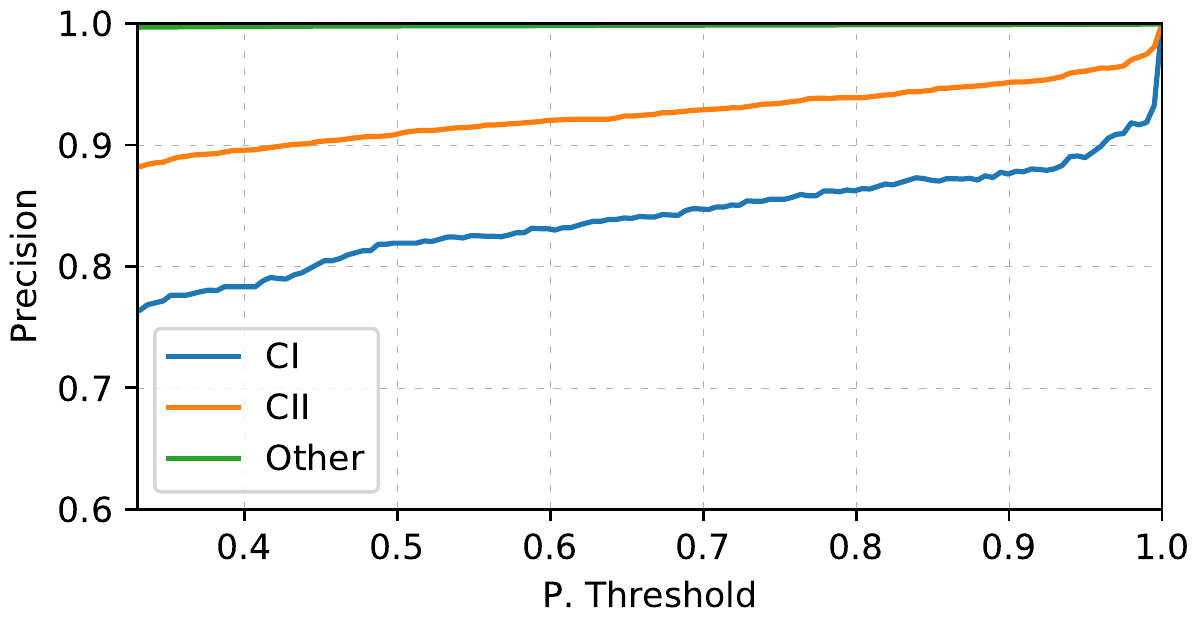}
	\end{subfigure}
	\caption[Fraction of exclusion, recall and precision as a function of probability treshold]{Evolution of quality estimators for each class regarding the membership probability threshold. {\it Top}, {\it Middle} and {\it Bottom} frame shows the evolution of the kept proportion, recall, and precision, respectively.}
	\label{threshold_curves}
\end{figure}

\clearpage
The ROC is an interesting indicator when comparing different classifiers, but it is less useful in our case with just one final classifier. However, it motivated the estimate of other quantities as a function of the probability threshold. There are mainly three quantities that are interesting in our case, which are the proportion of objects that are excluded, the recall and the precision. Figure~\ref{threshold_curves} shows all these quantities for our three output classes. We note that the curve does not go bellow a probability value of $1/3$ since, for a normalized output with three classes, when a class has a probability lesser than $1/3$, another necessarily has a probability greater than this value. These curves perfectly illustrate the lesser network confidence level on our CI prediction against the two other classes. Ultimately, they can be used to predict the threshold to apply in order to get a given recall or precision on a given class, and what would be the number of objects that are lost and the impact on the other classes.\\

However, we note that this curve is unable to reproduce our F-C full dataset result since the class association that was used for it consisted in taking the maximum of the three probability output. With this threshold approach, considering an object as a CI as soon as its membership probability is above 0.4 does not prevent another class to be at 0.6 and therefore do a misclassification. However, doing so allows to select objects that are at least close to the CI category. In order to avoid misclassification, the threshold value must be above 0.5, since no other class alone can be higher. Still, it will miss some objects that the maximum probability association would have found, as for example a (0.4,0.3,0.3) probability output. These examples highlight the threshold approach limits, especially for lower thresholds values.\\

\begin{figure*}[!t]
\hspace{-1.5cm}
	\begin{minipage}{1.25\hsize}
	\centering
	\begin{subfigure}[t]{\textwidth}
	\includegraphics[width=\hsize]{images/orion_herschel_all_CI.png}
	\end{subfigure}\\
	\vspace{-0.2cm}
	\begin{subfigure}[t]{\textwidth}
	\includegraphics[width=\hsize]{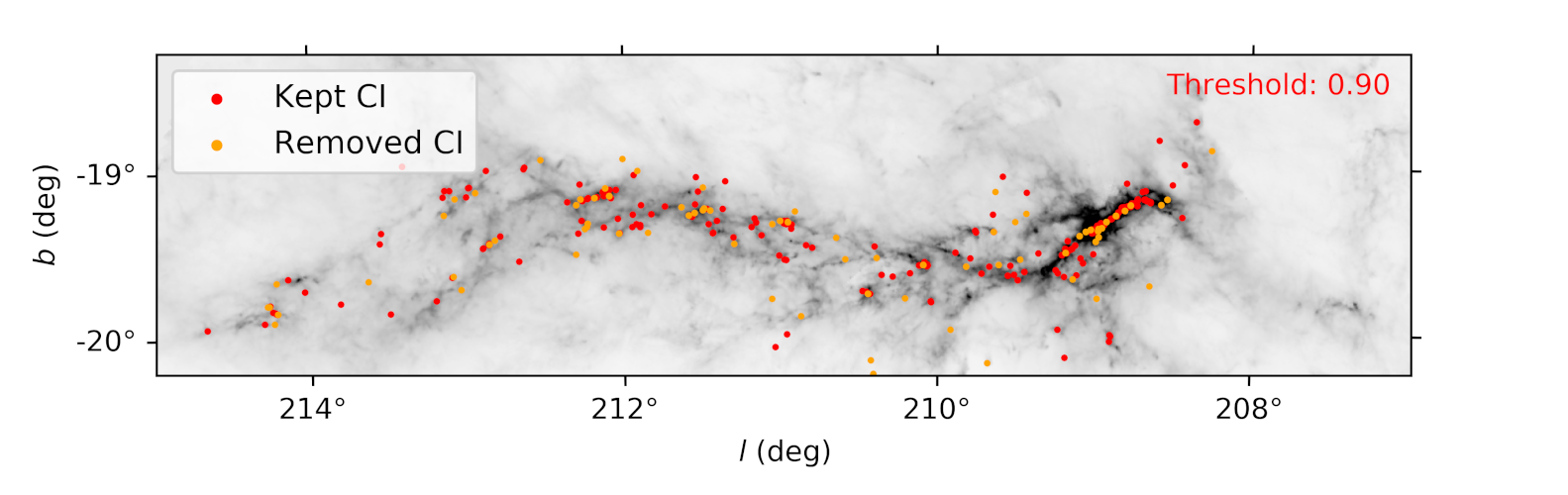}
	\end{subfigure}\\
	\vspace{-0.2cm}
	\begin{subfigure}[t]{\textwidth}
	\includegraphics[width=\hsize]{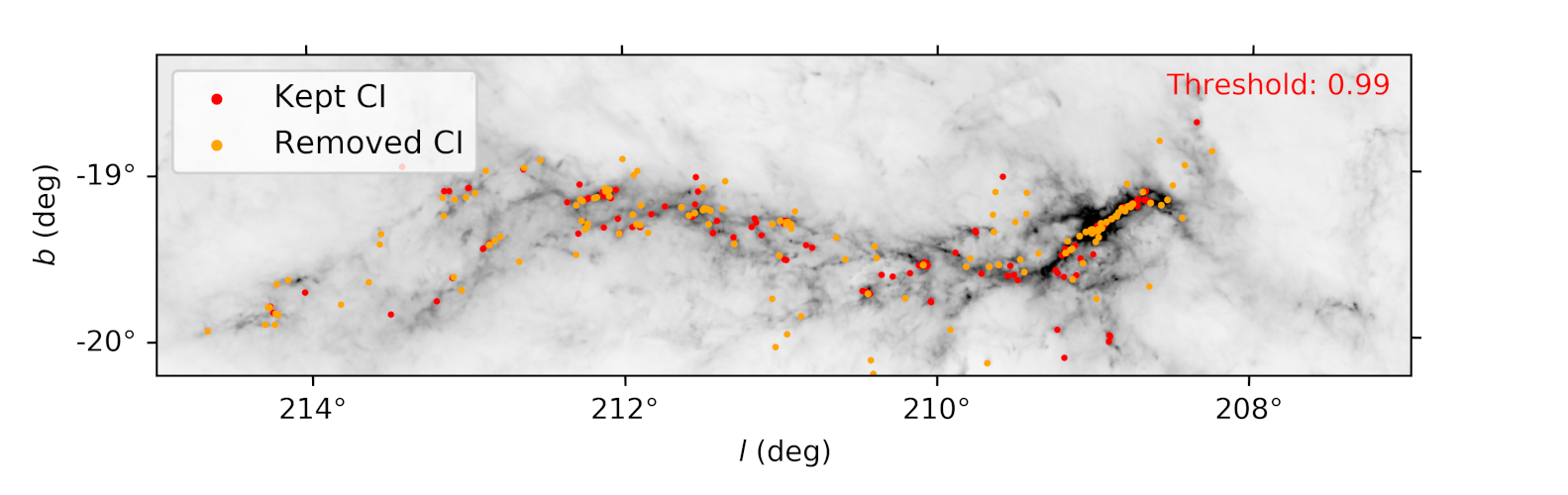}
	\end{subfigure}
	\end{minipage}
	\caption[Probability filter on Orion CI YSO candidate distribution]{Distribution of CI YSO candidates for Orion A after a membership probability threshold. The background grayscale is the Herschel SPIRE 500 $\mathrm{\mu m}$ map. {\it Top}, {\it Middle} and {\it Bottom} frames are for 0.9, 0.95 and 0.99 probability threshold, respectively.}
\label{yso_ci_dist_proba_orion}
\end{figure*}

\begin{figure*}[!t]
\hspace{-1.5cm}
	\begin{minipage}{1.25\hsize}
	\centering
	\begin{subfigure}[t]{\textwidth}
	\includegraphics[width=\hsize]{images/orion_herschel_all_CII.png}
	\end{subfigure}\\
	\vspace{-0.2cm}
	\begin{subfigure}[t]{\textwidth}
	\includegraphics[width=\hsize]{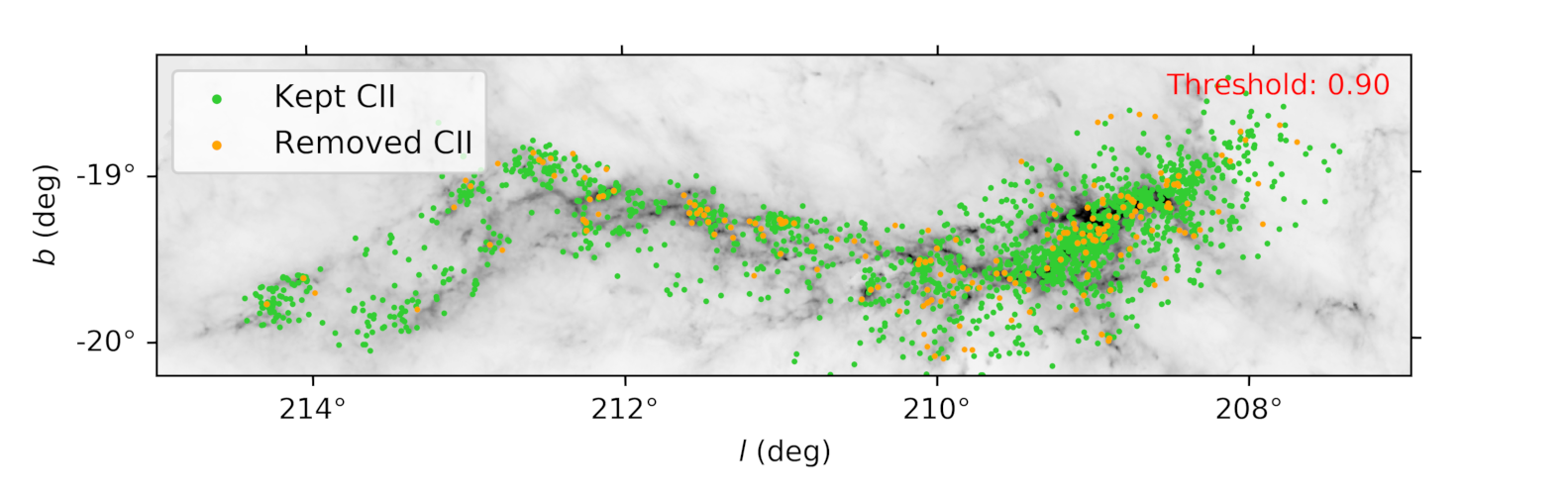}
	\end{subfigure}\\
	\vspace{-0.2cm}
	\begin{subfigure}[t]{\textwidth}
	\includegraphics[width=\hsize]{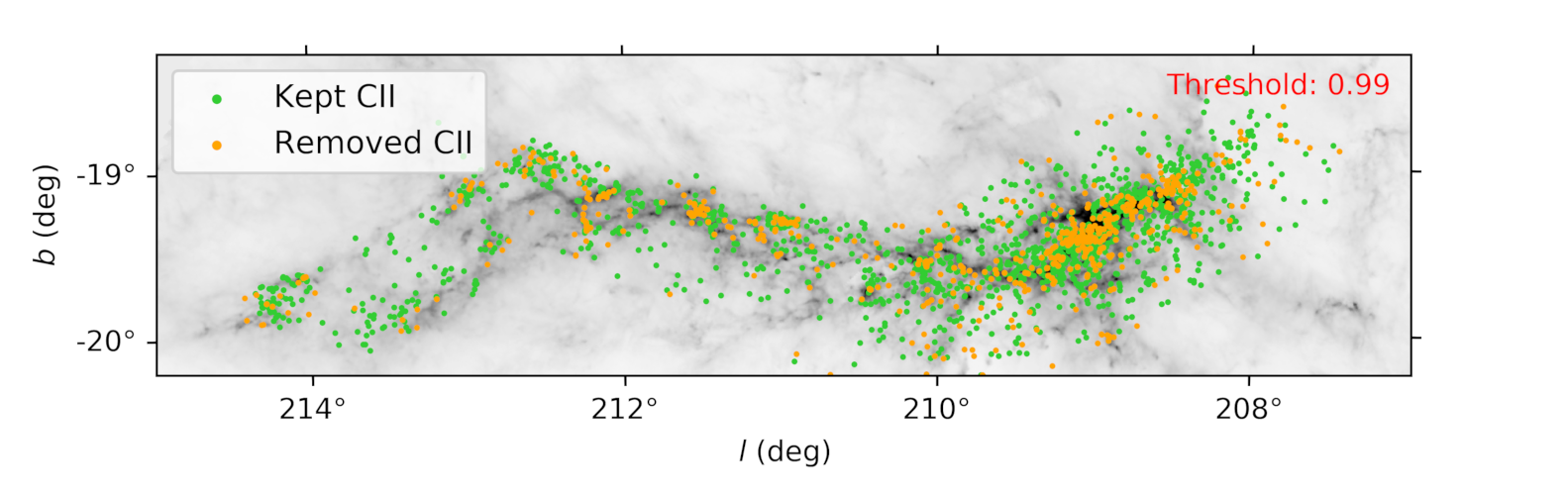}
	\end{subfigure}
	\end{minipage}
	\caption[Probability filter on Orion CII YSO candidate distribution]{Distribution of CII YSO candidates for Orion A after a membership probability threshold. The background grayscale is the Herschel SPIRE 500 $\mathrm{\mu m}$ map. {\it Top}, {\it Middle} and {\it Bottom} frames are for 0.9, 0.95 and 0.99 probability threshold, respectively.}
\label{yso_cii_dist_proba_orion}
\end{figure*}

\begin{figure*}[!t]
	\hspace{-1.7cm}
	\begin{minipage}{1.2\hsize}
	\centering
	\begin{subfigure}[t]{0.48\textwidth}
	\includegraphics[width=1.0\hsize]{images/ngc2264_herschel_all_CI.png}
	\end{subfigure}
	\begin{subfigure}[t]{0.48\textwidth}
	\includegraphics[width=1.0\hsize]{images/ngc2264_herschel_all_CII.png}
	\end{subfigure}\\
	\vspace{+0.4cm}
	\begin{subfigure}[t]{0.48\textwidth}
	\includegraphics[width=1.0\hsize]{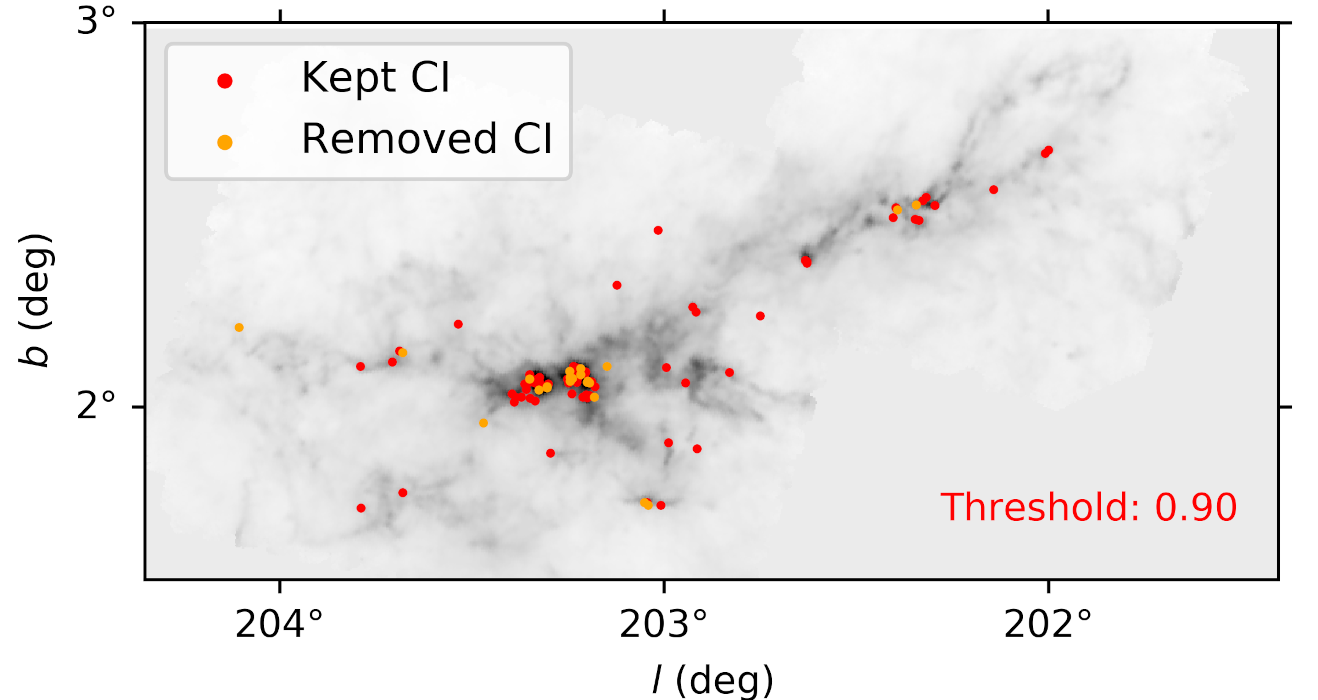}
	\end{subfigure}
	\begin{subfigure}[t]{0.48\textwidth}
	\includegraphics[width=1.0\hsize]{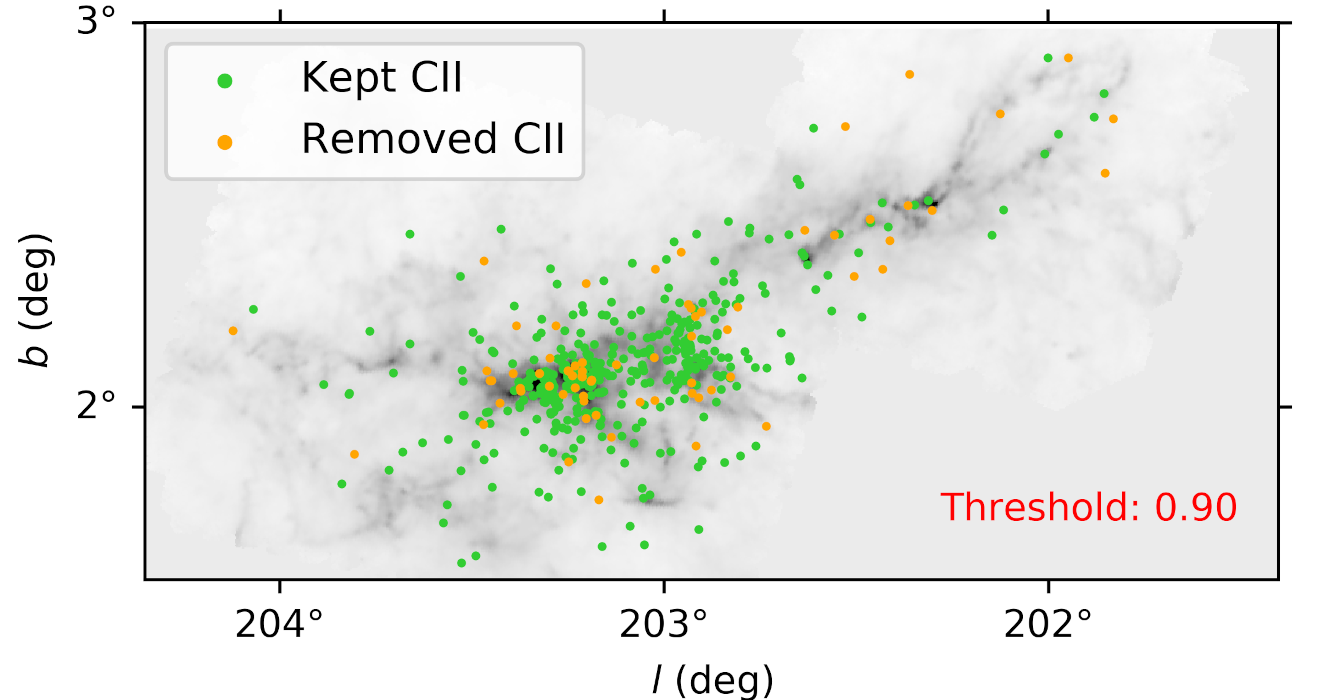}
	\end{subfigure}\\
	\vspace{+0.4cm}
	\begin{subfigure}[t]{0.48\textwidth}
	\includegraphics[width=1.0\hsize]{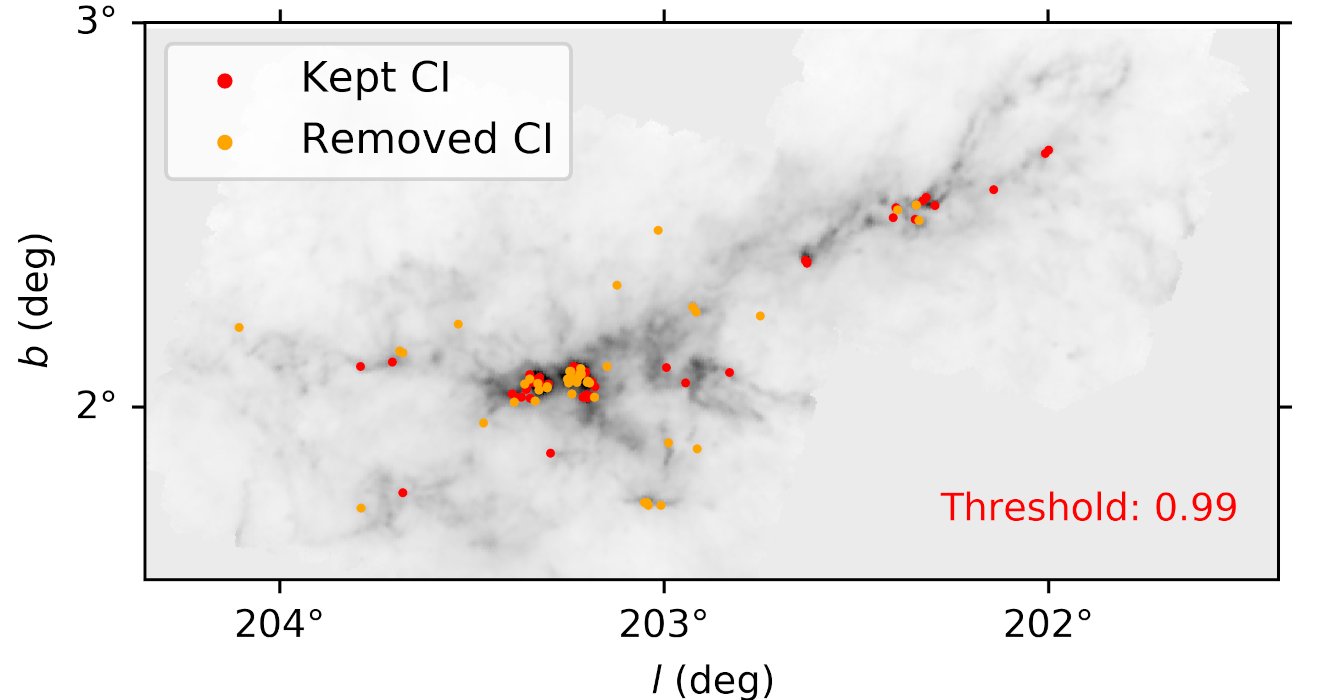}
	\end{subfigure}
	\begin{subfigure}[t]{0.48\textwidth}
	\includegraphics[width=1.0\hsize]{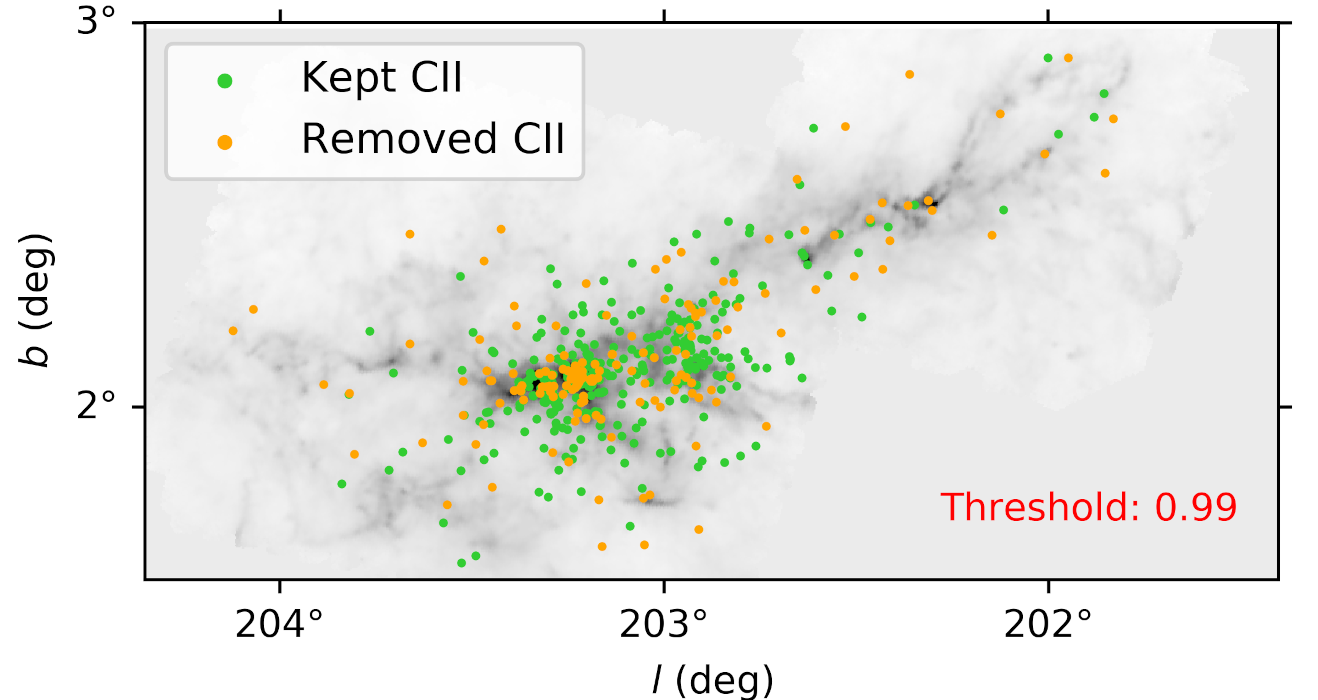}
	\end{subfigure}
	\end{minipage}
	\caption[Probability filter on NGC 2264 YSO candidate distribution]{Distribution YSO candidates for NGC 2264 after a membership probability threshold. The background grayscale is the Herschel SPIRE 500 $\mathrm{\mu m}$ map. {\it Left}: distribution of the CI YSOs. {\it Right}: distribution of the CII YSOs. {\it Top}, {\it Middle} and {\it Bottom} frames are for 0.9, 0.95 and 0.99 probability threshold, respectively.}
\label{yso_dist_proba_ngc2264}
\end{figure*}

Finally we looked at the effect of the membership probability threshold on the distribution of the remaining YSOs in Orion and NGC 2264. Figures~\ref{yso_ci_dist_proba_orion} and \ref{yso_cii_dist_proba_orion} show the distribution of CI and CII YSOs, respectively. Despite the fact that the objects classified as CI but removed by a probability threshold of 0.9 have a greater chance not to be genuine CI YSOs, it does not translate into evident correlation in the distribution of these objects in the sky plane. Indeed, the removed excluded objects seem to mostly follow the global distribution of the same class. Still, due to the higher concentration of objects in the densest part of the cloud, applying a threshold will result in an apparently narrower distribution on the filament. Figure~\ref{yso_dist_proba_ngc2264} shows similar results for NGC 2264. We note that such a cut is only interesting when looking at the statistical properties of the clouds. For more local, or per star study, it could be useful to keep all candidates and proceed to further individual inspection.\\

To conclude, with the inclusion of this probability in our results, we provide a substantial addition to the original G09 classification, for which it might be more difficult to identify the reliable objects. The results of the F-C case will be published in the form of a public catalog available at CDS and associated with our paper \citep{cornu_montillaud_20}, which contains the class prediction along with the membership probability for each object in the Combined dataset. It includes all objects from the catalogs by \citet{megeath_spitzer_2012} and \citet{rapson_spitzer_2014}, as described in Section~\ref{data_setup}, and Table~\ref{yso_catalog} shows an excerpt from our catalog.\\

\clearpage

\begin{sidewaystable*}
	\scriptsize
	\centering
	\caption{First 20 and last 20 elements of the catalog of network prediction in the F-C case using the full dataset.}
	\vspace{-0.1cm}
	\begin{tabularx}{0.93\hsize}{*{19}{c}}
	\toprule
	\toprule
	RA & DEC & Catalog & Orig. Class & 3.6   & e3.6  & 4.5   & e4.5  & 5.8   & e5.8  & 8.0   & e8.0  & 24    & e24   & Targ. & Pred. & P(CI) & P(CII) & P(Oth.)\\
	(deg)   & (deg)   &         &             & (mag) & (mag) & (mag) & (mag) & (mag) & (mag) & (mag) & (mag) & (mag) & (mag) &             &             &       &        & \\
	\vspace{-0.3cm}\\
	\toprule
100.792999 & +8.7531472   & 0 & III/F & 10.32 & 0.003 & 10.17 & 0.003  & 10.07 & 0.005  & 10.03 & 0.008 & \dots & \dots & 6 & 2 & 0.0         & 5.7e-5    & 0.9999\\
100.677625 & +8.7556250   & 0 & III/F & 11.62 & 0.003 & 11.60 & 0.004  & 11.55 & 0.016  & 11.56 & 0.035 & \dots & \dots & 6 & 2 & 0.0         & 0.0       & 1.0   \\
100.760958 & +8.7566528   & 0 & III/F & 13.38 & 0.007 & 13.28 & 0.011  & 13.27 & 0.05   & 13.78 & 0.155 & \dots & \dots & 6 & 2 & 0.0         & 0.0       & 1.0   \\
100.757875 & +8.7589389   & 0 & III/F & 12.52 & 0.005 & 12.52 & 0.006  & 12.4  & 0.03   & 12.41 & 0.053 & \dots & \dots & 6 & 2 & 0.0         & 0.0       & 1.0   \\
100.724500 & +8.7606944   & 0 & III/F & 13.71 & 0.009 & 13.66 & 0.013  & 13.6  & 0.069  & 13.67 & 0.148 & \dots & \dots & 6 & 2 & 0.0         & 0.0       & 1.0   \\
100.728917 & +8.7609722   & 0 & III/F & 13.23 & 0.007 & 13.17 & 0.008  & 12.99 & 0.042  & 13.16 & 0.081 & \dots & \dots & 6 & 2 & 0.0         & 0.0       & 1.0   \\
100.744958 & +8.7630750   & 0 & III/F & 11.28 & 0.003 & 11.37 & 0.004  & 11.32 & 0.011  & 11.22 & 0.027 & \dots & \dots & 6 & 2 & 0.0         & 0.0       & 1.0   \\
100.667167 & +8.7653722   & 0 & III/F & 13.5  & 0.015 & 13.36 & 0.029  & 13.39 & 0.075  & 13.43 & 0.11  & \dots & \dots & 6 & 2 & 2.0e-6      & 1.4e-4    & 0.9998\\
100.670250 & +8.7691222   & 0 & III/F & 8.64  & 0.002 &  8.54 & 0.002  & 8.4   & 0.002  & 8.36  & 0.002 & 8.29  & 0.043 & 6 & 2 & 0.0         & 5.9e-4    & 0.9994\\
100.792083 & +8.7692694   & 0 & III/F & 13.39 & 0.007 & 13.40 & 0.010  & 13.41 & 0.052  & 13.01 & 0.09  & \dots & \dots & 6 & 2 & 0.0         & 4.6e-5    & 0.9999\\
100.769292 & +8.7704556   & 0 & III/F & 12.93 & 0.006 & 12.87 & 0.008  & 12.89 & 0.038  & 12.7  & 0.058 & \dots & \dots & 6 & 2 & 0.0         & 3.0e-6    & 0.9999\\
100.757708 & +8.7710500   & 0 & III/F & 10.58 & 0.002 & 10.65 & 0.003  & 10.59 & 0.007  & 10.56 & 0.012 & \dots & \dots & 6 & 2 & 0.0         & 0.0       & 1.0   \\
100.811250 & +8.7714556   & 0 & III/F & 7.75  & 0.002 &  7.83 & 0.002  & 7.63  & 0.002  & 7.63  & 0.002 & \dots & \dots & 6 & 2 & 0.0         & 8.2e-3    & 0.9917\\
100.768208 & +8.7728194   & 0 & III/F & 13.83 & 0.009 & 13.80 & 0.014  & 13.99 & 0.088  & 13.56 & 0.117 & \dots & \dots & 6 & 2 & 0.0         & 0.0       & 1.0   \\
100.773667 & +8.7744222   & 0 & III/F & 11.68 & 0.004 & 11.87 & 0.004  & 11.61 & 0.012  & 11.61 & 0.032 & \dots & \dots & 6 & 2 & 0.0         & 0.0       & 1.0   \\
100.672208 & +8.7765889   & 0 & III/F & 13.37 & 0.007 & 13.31 & 0.010  & 13.32 & 0.055  & 13.21 & 0.089 & \dots & \dots & 6 & 2 & 0.0         & 1.0e-6    & 0.9999\\
100.768375 & +8.7775694   & 0 & III/F & 12.49 & 0.005 & 12.52 & 0.006  & 12.44 & 0.026  & 12.41 & 0.053 & \dots & \dots & 6 & 2 & 0.0         & 0.0       & 1.0   \\
100.697292 & +8.7783972   & 0 & III/F & 10.78 & 0.003 & 10.79 & 0.003  & 10.74 & 0.007  & 10.57 & 0.014 & \dots & \dots & 6 & 2 & 0.0         & 3.0e-6    & 0.9999\\
100.684208 & +8.7784639   & 0 & III/F & 12.87 & 0.005 & 12.85 & 0.007  & 12.81 & 0.033  & 12.74 & 0.065 & \dots & \dots & 6 & 2 & 0.0         & 0.0       & 1.0   \\
100.792542 & +8.7796389   & 0 & AGN   & 16.18 & 0.047 & 5.12  & 0.035  & 14.37 & 0.141  & 13.00 & 0.079 & \dots & \dots & 3 & 2 & 4.9e-5      & 2.0e-6    & 0.9999\\
\dots & \dots & \dots & \dots & \dots & \dots & \dots & \dots & \dots & \dots & \dots & \dots & \dots & \dots & \dots & \dots & \dots & \dots & \dots   \\
86.8015397 & -0.7217830   & 1 & Other & 13.79 & 0.011 & 13.73 & 0.017  & 13.66 & 0.106  & 13.54 & 0.153 & \dots & \dots & 6 & 2 & 0.0         & 1.0e-6    & 0.9999\\
86.7227924 & -0.7204420   & 1 & Other & 11.88 & 0.005 & 11.85 & 0.006  & 11.87 & 0.026  & 11.73 & 0.036 & \dots & \dots & 6 & 2 & 0.0         & 0.0       & 1.0   \\
86.7296191 & -0.7189594   & 1 & Other & 14.54 & 0.019 & 14.14 & 0.026  & 13.84 & 0.147  & 10.98 & 0.022 & 8.44  & 0.110 & 2 & 2 & 0.0         & 0.0       & 1.0   \\
86.6185832 & -0.7163786   & 1 & Other &  9.60 & 0.002 &  9.60 & 0.003  &  9.57 & 0.006  &  9.53 & 0.007 & \dots & \dots & 6 & 2 & 0.0         & 2.1e-5    & 0.9999\\
86.8822281 & -0.7111607   & 1 & Other & 13.16 & 0.009 & 13.13 & 0.012  & 13.02 & 0.056  & 13.00 & 0.110 & \dots & \dots & 6 & 2 & 0.0         & 2.0e-6    & 0.9999\\
86.8187251 & -0.7086041   & 1 & Other & 13.40 & 0.010 & 13.32 & 0.011  & 13.24 & 0.072  & 13.34 & 0.123 & \dots & \dots & 6 & 2 & 0.0         & 0.0       & 1.0   \\
86.8938200 & -0.7075397   & 1 & Other & 11.36 & 0.004 & 11.38 & 0.005  & 11.27 & 0.017  & 11.29 & 0.027 & \dots & \dots & 6 & 2 & 0.0         & 0.0       & 1.0   \\
86.7451751 & -0.7074037   & 1 & Other & 11.61 & 0.004 & 11.54 & 0.004  & 11.51 & 0.024  & 11.53 & 0.024 & \dots & \dots & 6 & 2 & 0.0         & 0.0       & 1.0   \\
86.6627309 & -0.7060398   & 1 & Other & 12.99 & 0.008 & 12.93 & 0.008  & 12.86 & 0.056  & 12.95 & 0.064 & \dots & \dots & 6 & 2 & 0.0         & 0.0       & 1.0   \\
86.6652294 & -0.7036116   & 1 & Other & 11.52 & 0.004 & 11.47 & 0.004  & 10.59 & 0.013  & 11.45 & 0.022 & \dots & \dots & 5 & 2 & 0.0         & 0.0       & 1.0   \\
86.6478710 & -0.7028939   & 1 & Other & 11.40 & 0.004 & 11.44 & 0.005  & 11.35 & 0.018  & 11.28 & 0.027 & \dots & \dots & 6 & 2 & 0.0         & 0.0       & 1.0   \\
86.6734974 & -0.7025317   & 1 & Other & 12.68 & 0.006 & 12.64 & 0.009  & 12.60 & 0.045  & 12.48 & 0.064 & \dots & \dots & 6 & 2 & 0.0         & 1.0e-6    & 0.9999\\
86.6593266 & -0.6985667   & 1 & Other & 13.54 & 0.010 & 13.50 & 0.015  & 13.55 & 0.095  & 13.36 & 0.147 & \dots & \dots & 6 & 2 & 0.0         & 0.0       & 1.0   \\
86.8586910 & -0.6948155   & 1 & Other & 12.77 & 0.007 & 12.75 & 0.010  & 12.63 & 0.043  & 12.72 & 0.077 & \dots & \dots & 6 & 2 & 0.0         & 0.0       & 1.0   \\
86.6522543 & -0.6913875   & 1 & Other &  8.95 & 0.007 &  8.95 & 0.002  &  8.87 & 0.004  &  8.83 & 0.004 & 8.94  & 0.168 & 6 & 2 & 1.0e-6      & 3.1e-3    & 0.9969\\
86.7718531 & -0.6903623   & 1 & Other & 13.83 & 0.014 & 13.78 & 0.020  & 13.63 & 0.104  & 13.65 & 0.184 & \dots & \dots & 6 & 2 & 0.0         & 1.0e-6    & 0.9999\\
86.8164438 & -0.6901289   & 1 & Other & 13.39 & 0.010 & 13.29 & 0.011  & 13.12 & 0.066  & 13.08 & 0.084 & \dots & \dots & 6 & 2 & 1.0e-6      & 2.5e-5    & 0.9999\\
86.7218342 & -0.6863691   & 1 & Other & 14.15 & 0.015 & 13.90 & 0.022  & 13.08 & 0.076  & 10.32 & 0.014 & 8.32  & 0.095 & 2 & 2 & 0.0         & 0.0       & 1.0   \\
86.6604287 & -0.6855791   & 1 & Other & 10.51 & 0.003 & 10.46 & 0.002  & 10.40 & 0.009  & 10.42 & 0.010 & \dots & \dots & 6 & 2 & 0.0         & 0.0       & 1.0   \\
86.8976492 & -0.6839529   & 1 & Other &  8.97 & 0.006 &  8.97 & 0.002  &  8.85 & 0.004  &  8.81 & 0.004 & 8.82  & 0.152 & 6 & 2 & 1.0e-6      & 3.0e-3    & 0.9970\\
	\bottomrule
	\end{tabularx}
	\caption*{\vspace{-0.4cm}\\ {\bf Notes.} The full catalog is publicly available at CDS. The columns are: 
	(1-2) The source coordinates (J2000); 
	(3) the original catalog (0: \citet{megeath_spitzer_2012}, 1: \citet{rapson_spitzer_2014});
	(4) the original classification;
	(5-14) IRAC and MIPS magnitudes and corresponding uncertainties;
	(15) the target classification obtained with our simplified G09 scheme (0: CI YSOs, 1: CII YSOs, 2: Galaxies, 3: AGNs, 4: Shocks, 5: PAHs, 6: Stars);
	(16) the classification predicted by the ANN in the F-C case (0: CI YSOs, 1: CII YSOs, 2: contaminants);
	(17-19) the corresponding membership probabilities.
	}
	\vspace{-0.3cm}
	\label{yso_catalog}
\end{sidewaystable*}

\clearpage
\section{3D cloud reconstruction using cross-match with Gaia}
\label{3d_yso_gaia}

\etocsettocstyle{\subsubsection*{\vspace{-0.5cm}}}{}
\localtableofcontents

\vspace{0.5cm}

\subsection{Orion A distance and 3D information}
\label{sect:orion_A_dist}

\vspace{0.2cm}
We aim to demonstrate that our catalog of YSO candidates can be used to retrieve distance information about molecular clouds, and even reveal more subtle 3D structural characteristics. For this we mainly followed the approach exposed by \citet{grossschedl_3d_2018} (hereafter GR18) who performed a 3D reconstruction of Orion A based on the \citet{megeath_spitzer_2012} catalog (and some refinements from \citet{Megeath_2016}) and an additional 200 YSOs sample from the ESO-VISTA near-infrared survey \citep{Meingast_2016}, all cross-matched with Gaia DR2 in order to obtain parallax measurements. Therefore, using our own YSO candidate catalog, we tried to reproduce the results on Orion A, and subsequently applied a similar approach to Orion B and NGC 2264. Our approach is described in the following paragraphs.\\

We performed a cross-match with Gaia DR2 \citep{Gaia_Collaboration_2018} that associate objects based on a $1\arcsec$ sky distance as in GR18. Since YSOs are embedded in an environment that is often optically thick for the optical Gaia G band, we only recovered a fraction of them, and lost almost all the CI YSOs for all datasets. Still following GR18 we applied cuts in parallax quality with the $\sigma_{\varpi}/\varpi < 0.1$ condition. We also stress that we conserved their assumption that the direct inverse of parallax is a sufficiently good distance approximation in the case of Orion A, with less than 1\% difference with the \citet{bailer-jones_2018} Bayesian inference distance estimate. Then, instead of applying subsequent Gaia astrometry filters and an additional color excess exclusion like in GR18, we preferred to test if improving the YSO selection quality leads to similar results. For this we applied a 0.95 membership threshold that has shown to conserve a large fraction of the objects while having a $\sim$90\% precision on our CI YSOs (Table~\ref{conf_proba_095}), which is important for such application. This threshold further reduces the number of objects recovered after the cross match with Gaia. The rest of the analysis is then specific to each region.\\

\begin{table}
	\centering
	\caption{Orion A sample size for different selection criteria.}
	\vspace{-0.1cm}
	\begin{tabularx}{0.9\hsize}{l @{\hskip 0.1\hsize} *{2}{Y}}
	\toprule
	 & CI YSOs & CII YSOs\\
	\midrule
	(0): Raw catalog & 275 & 1957\\
	(1): Raw X-match & 49 & 1612\\
	(2): 1 with $\varpi$ & 36 & 1457\\
	(3): 2 with $\sigma_{\varpi}/\varpi < 0.1$ & 12 & 1006 \\ 
	(4): 2 with $P(X) > 0.99$ & 2 & 1038 \\
	(5): 2 with $P(X) > 0.95$ and $\sigma_{\varpi}/\varpi < 0.1$ & 5 & 932\\
	
	\bottomrule
	\end{tabularx}
	\label{orion_A_select_crit}
	\vspace{1.5cm}
\end{table}

\newpage

\begin{figure}[!t]
	\hspace{-1.7cm}
	\begin{minipage}{1.2\hsize}
	\centering
	\includegraphics[width=1.0\hsize]{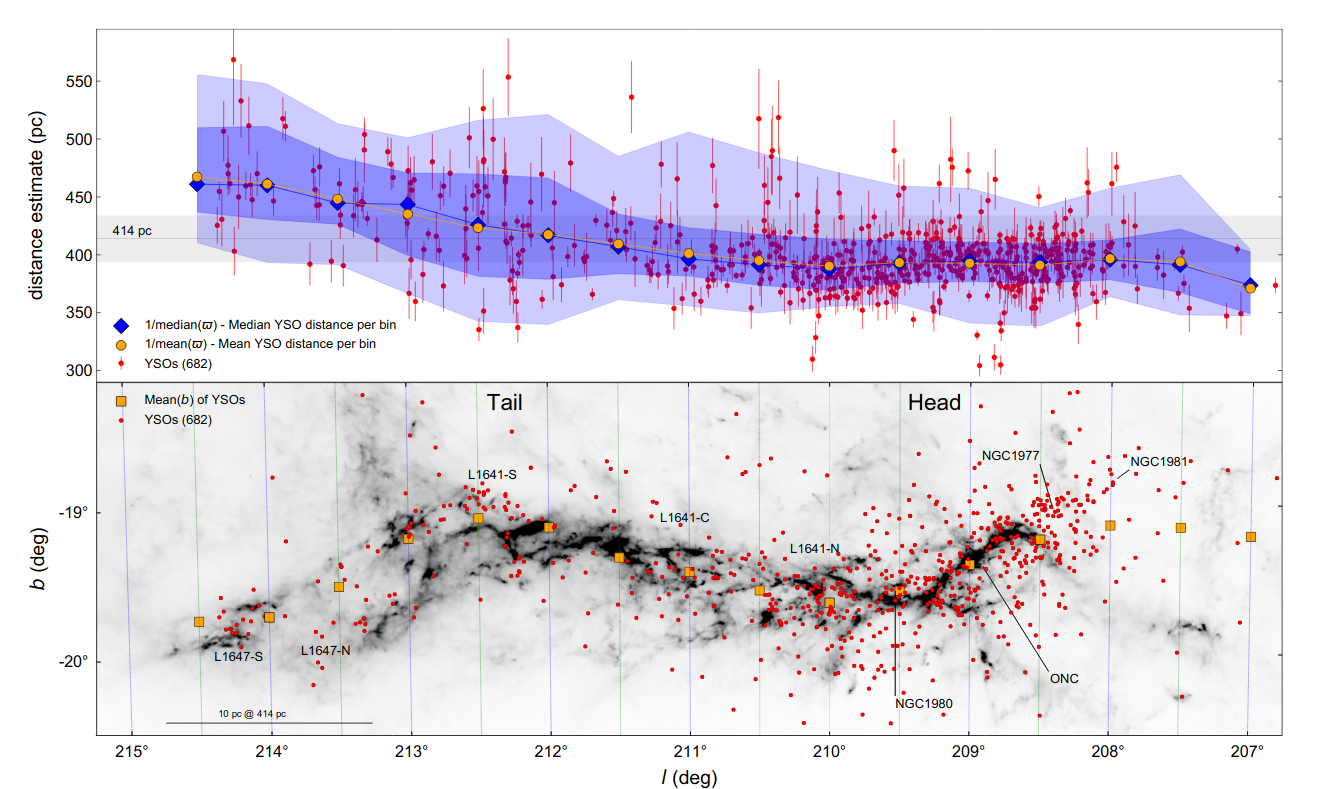}
	\end{minipage}
	\caption[Orion A distance distribution from \citet{grossschedl_3d_2018}]{Orion A distance distribution from GR18, with their full selection. Error bars correspond to $(\sigma_{\varpi}/\varpi^2)$. The orange and blue markers are the mean and median distance for each $\Delta l = 1^{\circ}$ distance bin, respectively. The light and dark blue areas represents the $2\sigma$ and $1\sigma$ percentiles for each bin, respectively. The background grayscale is an Herschel map. {\it From} \citet{grossschedl_3d_2018}}
	\label{orion_A_dist_GR18}
\end{figure}

\begin{figure*}[!t]
	\centering
	\begin{subfigure}[t]{0.70\textwidth}
	\caption*{\bf Selection criteria (5)}
	\includegraphics[width=1.0\hsize]{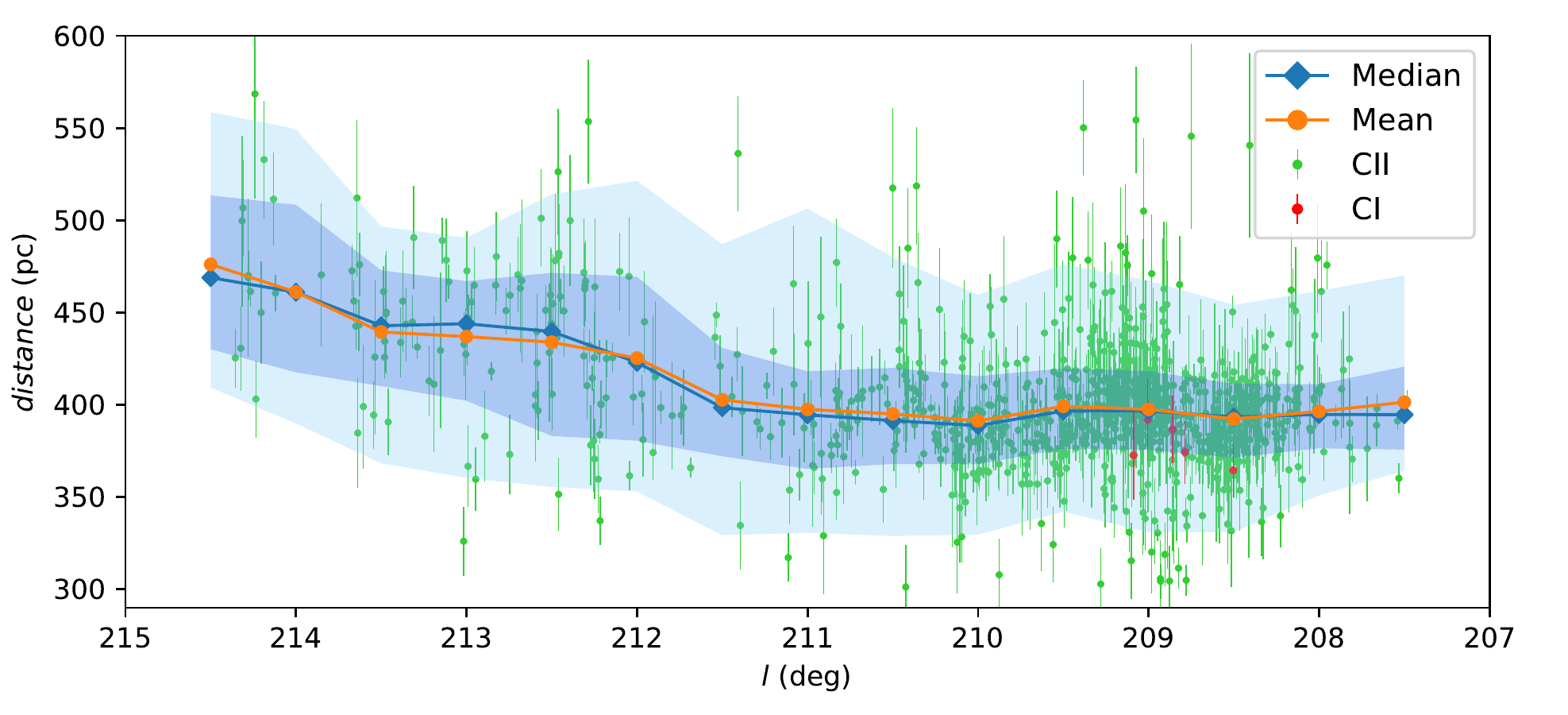}
	\end{subfigure}\\
	\vspace{-0.3cm}
	\begin{subfigure}[t]{0.70\textwidth}
	\includegraphics[width=1.0\hsize]{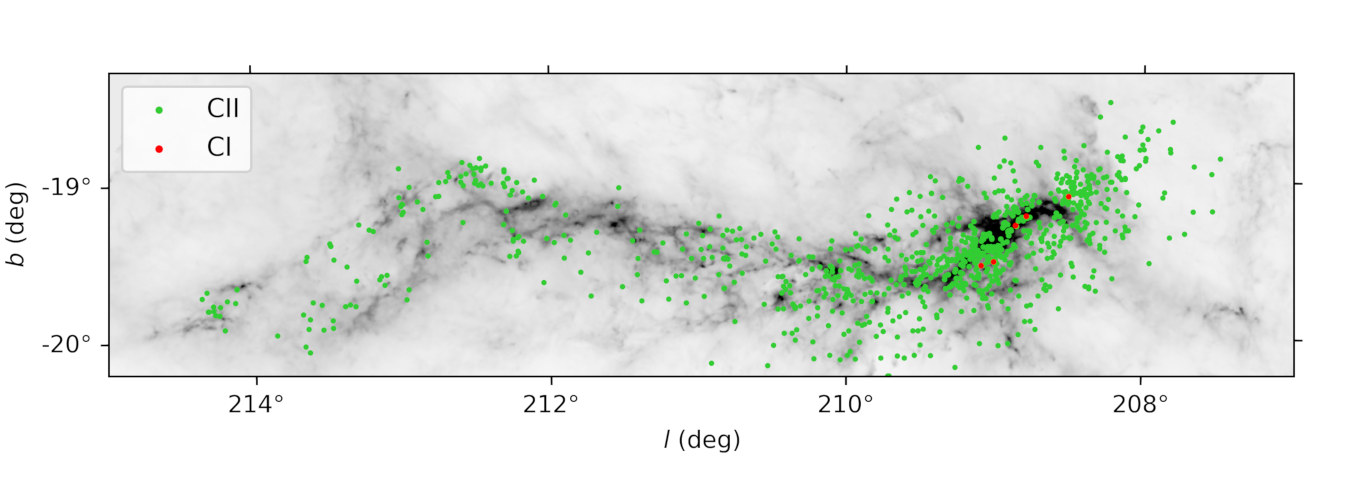}
	\end{subfigure}\\
	\vspace{0.2cm}
	\begin{subfigure}[t]{0.70\textwidth}
	\caption*{\bf Selection criteria (4)}
	\includegraphics[width=1.0\hsize]{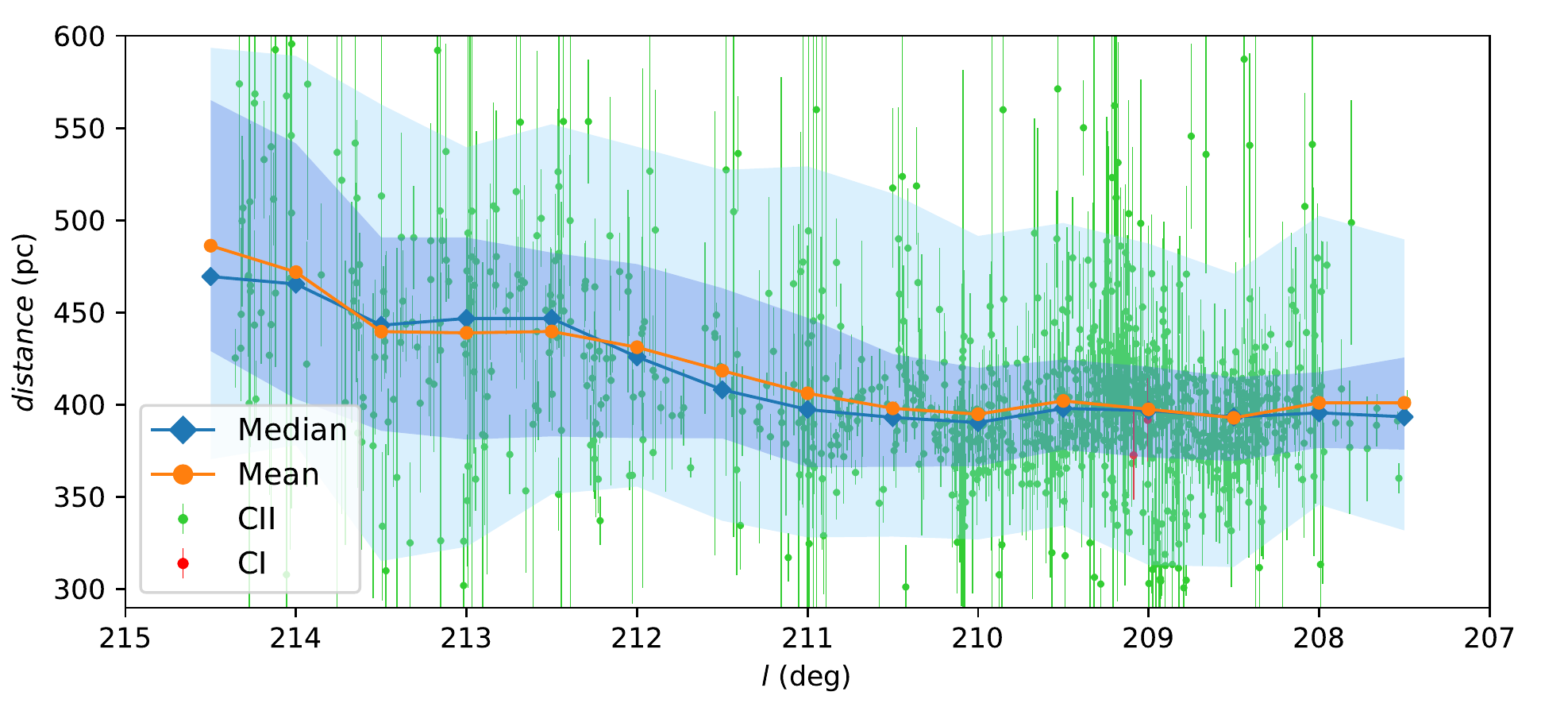}
	\end{subfigure}\\
	\vspace{-0.3cm}
	\begin{subfigure}[t]{0.70\textwidth}
	\includegraphics[width=1.0\hsize]{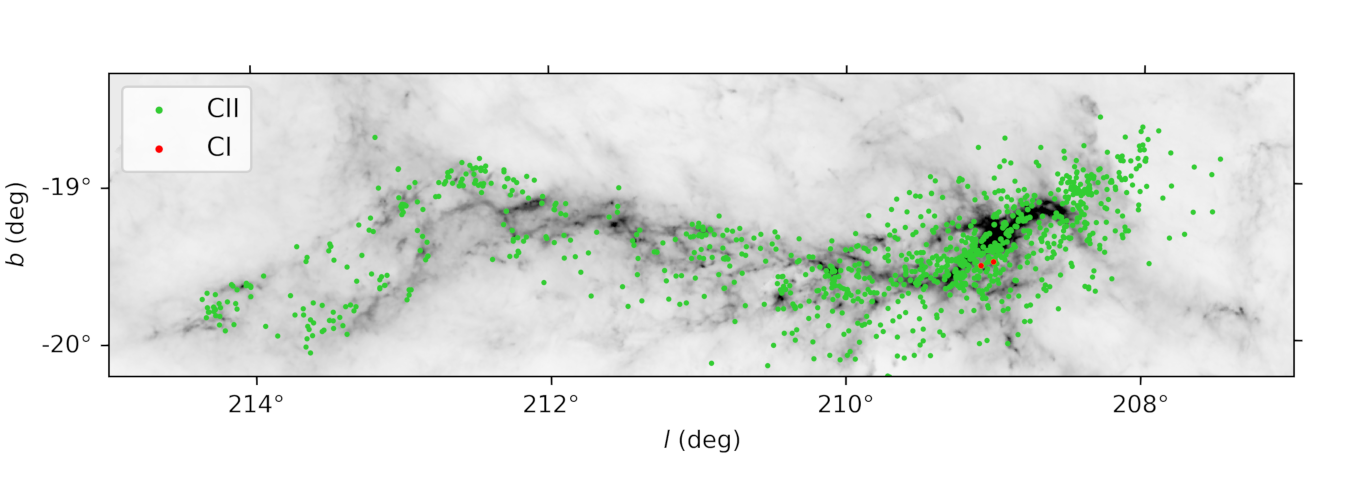}
	\end{subfigure}
	\caption[Orion A distance distribution from the F-C YSO candidates]{Orion A distance distribution from the F-C YSO candidates, along with the sky distribution over the region. Error bars correspond to $(\sigma_{\varpi}/\varpi^2)$. CI and CII selected YSOs are in red and green, respectively. The orange and blue markers are the mean and median distance for each $\Delta l = 1^{\circ}$ distance bin, respectively. The light and dark blue areas represents the 2 and 1 $\sigma$ percentiles for each bin, respectively. The background grayscale is the Herschel SPIRE 500 $\mathrm{\mu m}$ map. {\it Top duo}: YSO selection and distribution in case (5). {\it Bottom duo}: YSO selection and distribution in case (4).}
\label{orion_A_dist}
\end{figure*}

\newpage
For Orion A, we used galactic coordinates to properly compare with GR18, and because it is a wise choice since the filament structure is mostly aligned with the galactic longitude axis. A few YSO selection criteria have been tested and are summarized in Table~\ref{orion_A_select_crit}. From 275 CI and 1957 CII YSO candidates corresponding to Orion A in our catalog (case (0)), the Gaia cross match only preserves 49 CI and 1612 CII YSOs (case (1)), with only 36 and 1457 of them, respectively, with a parallax measurement (case (2)). Afterward, the use of our "good YSO" condition that include the membership probability threshold at 0.95, combined with an exclusion of objects with parallaxes clearly unrelated with Orion ($\varpi > 3.333$ or $\varpi < 1.666$) like in GR18), left only 5 CI and 932 YSOs (case (5)). We observed that the threshold is only responsible for a subsequent removal of 7 CI and 74 CII, while all the other removals are induced by the parallax condition (case (3)). Nevertheless, selecting YSOs with a membership probability $>0.95$ with no cut in parallax quality still led to a sample that is smaller that the raw cross match, with 23 CI and 1441 CII YSOs. A 0.99 membership cut (case(4)) resulted in a sample with 2 CI and 1038 CII YSOs, on which the addition of the parallax condition left only 2 CI and 846 CII YSOs. This tends to indicate that there is an overlap between the two conditions, meaning that selecting very firmly established YSOs could be a sufficient enough criteria in the case of Orion A.\\

\newpage

\begin{sidewaystable*}
	\footnotesize
	\centering
	\caption{Orion A distance estimates and dispersion for each galactic longitude bin.}
	\begin{tabularx}{1.0\hsize}{l @{\hskip 0.01\hsize} | Y l *{6}{Y} @{\hskip 0.05\hsize} | Y l *{6}{Y}}
	\toprule
	\multicolumn{17}{c}{}\vspace{-0.2cm}\\
	\multicolumn{1}{c}{} & \multicolumn{8}{c}{{\bf Case (5): X-match with $\bm{ P(X) > 0.95}$ and $\bm{\sigma_{\varpi}/\varpi < 0.1}$}} & \multicolumn{8}{c}{{\bf Case (4): X-match with $\bm{P(X) > 0.99}$}}\\
	\multicolumn{1}{c}{}\vspace{-0.2cm}\\
	\midrule
	$\Delta l$ center & $N_{YSO}$ & Mean & Median & StD & $p(-2\sigma)$ & $p(-1\sigma)$ & $p(+1\sigma)$ & $p(+2\sigma)$ & $N_{YSO}$ & Mean & Median & StD  & $p(-2\sigma)$ & $p(-1\sigma)$ & $p(+1\sigma)$ & $p(+2\sigma)$ \\
	 (deg) &  & (pc) & (pc) & (pc) & (pc) & (pc) & (pc) & (pc) &  & (pc) & (pc) & (pc)  & (pc) & (pc) & (pc) & (pc) \\
	\midrule
	
	207.5 & 2   & $401.4\pm 19.4$ & 394.7 & 29.1 & 364.3 & 375.7 & 420.7 & 470.1   & 18 & $401.0\pm 38.4$ & 393.5 & 41.3 & 331.9 & 375.7 & 425.6 & 489.6 \\
	208.0 & 155 & $396.4\pm 17.2$ & 394.7 & 25.5 & 350.9 & 376.4 & 411.2 & 461.7  & 155 & $401.1\pm 42.9$ & 395.7 & 35.7 & 345.9 & 376.7 & 417.6 & 502.5 \\
	208.5 & 321 & $392.3\pm 17.9$ & 393.4 & 29.0 & 331.1 & 371.4 & 411.4 & 454.1  & 314 & $392.9\pm 35.4$ & 393.4 & 36.7 & 312.1 & 369.5 & 414.9 & 471.0 \\
	209.0 & 416 & $397.5\pm 18.5$ & 396.7 & 32.0 & 330.8 & 375.5 & 418.3 & 467.3  & 410 & $397.6\pm 29.5$ & 396.9 & 38.2 & 313.2 & 371.4 & 420.7 & 487.4 \\
	209.5 & 337 & $399.2\pm 18.5$ & 396.6 & 30.4 & 342.1 & 376.1 & 419.8 & 476.1  & 344 & $402.1\pm 35.3$ & 398.0 & 37.4 & 334.6 & 375.6 & 424.5 & 498.5 \\
	210.0 & 180 & $391.3\pm 18.4$ & 388.7 & 29.8 & 329.6 & 367.8 & 415.5 & 459.6  & 195 & $394.9\pm 38.9$ & 390.4 & 37.5 & 327.0 & 366.7 & 420.0 & 491.5 \\
	210.5 & 115 & $395.1\pm 19.5$ & 391.4 & 33.8 & 328.9 & 367.9 & 420.0 & 479.9  & 136 & $398.1\pm 60.7$ & 393.1 & 40.4 & 328.6 & 366.3 & 427.5 & 514.4 \\
	211.0 & 56  & $397.5\pm 20.4$ & 394.6 & 38.8 & 330.6 & 365.3 & 418.2 & 506.4   & 78 & $406.2\pm 84.8$ & 397.5 & 47.7 & 328.1 & 366.3 & 447.1 & 529.2 \\
	211.5 & 31  & $402.7\pm 20.2$ & 398.4 & 39.8 & 329.3 & 372.1 & 430.8 & 487.0   & 48 & $418.5\pm 66.6$ & 408.2 & 47.0 & 337.1 & 381.8 & 463.2 & 527.3 \\
	212.0 & 52  & $425.3\pm 22.0$ & 423.0 & 44.5 & 353.0 & 380.5 & 469.4 & 521.4   & 66 & $431.2\pm 51.7$ & 426.0 & 48.0 & 355.7 & 381.9 & 476.2 & 539.8 \\
	212.5 & 65  & $433.9\pm 23.4$ & 439.7 & 43.8 & 355.5 & 383.1 & 471.5 & 514.0   & 89 & $439.7\pm 61.4$ & 446.8 & 50.6 & 351.6 & 382.9 & 482.5 & 552.0 \\
	213.0 & 48  & $436.9\pm 23.9$ & 443.9 & 35.8 & 360.3 & 402.2 & 467.1 & 490.5   & 81 & $439.0\pm 66.5$ & 446.8 & 56.8 & 323.1 & 381.2 & 490.7 & 539.5 \\
	213.5 & 32  & $439.5\pm 24.5$ & 442.9 & 36.0 & 368.3 & 410.1 & 472.9 & 496.6   & 59 & $439.7\pm 69.8$ & 443.1 & 62.1 & 315.3 & 385.9 & 490.7 & 562.7 \\
	214.0 & 24  & $461.1\pm 29.1$ & 461.0 & 44.6 & 389.9 & 417.6 & 508.4 & 549.5   & 52 & $471.9\pm 95.8$ & 465.4 & 63.4 & 378.1 & 403.2 & 541.7 & 589.1 \\
	214.5 & 13  & $476.1\pm 30.2$ & 468.9 & 44.5 & 409.3 & 430.1 & 513.5 & 558.6   & 30 & $486.2\pm 104$ & 469.5 & 64.9 & 370.6 & 429.1 & 565.1 & 593.5 \\
	
	\bottomrule
	\end{tabularx}
	\label{tab_orion_a_dist}
\end{sidewaystable*}

\clearpage

To help the comparison of our results with GR18, Figure~\ref{orion_A_dist_GR18} from \citet{grossschedl_3d_2018} depicts the distribution of their YSO selection on Orion A using their full selection, resulting in a 682 CII YSOs sample. The top frame shows the distance distribution following the galactic longitude with each point being a YSO with its uncertainty $(\sigma_{\varpi}/\varpi^2)$. To extract a continuous 3D information, they chose to make bins of longitude with a $\Delta l = 1^\circ$ width and evaluated the mean and median distance for such interval every $0.5^\circ$. This conducted to their main result with a clear depth evolution toward the molecular complex with an estimated angle of $\sim 70^\circ$ with the plane of the sky. This also allowed them to estimate that the physical length of Orion A is $\sim 90$ pc. These results strongly underline the limits of the commonly adopted 414 distance estimate for the whole molecular complex \citep{Menten2007}.\\

Using our catalog we were able to observe a very similar distance distribution. Figure~\ref{orion_A_dist} shows our result with the combined selection based on probability and parallax uncertainty, corresponding to case (5), and for our membership selection alone, i.e. case (4). The first one, despite being more noisy that the GR18 result, follows the same global trend, with an $1\sigma$ dispersion that is very similar. We used the same longitude binning approach and represented the same mean, median and 1-2 $\sigma$ percentiles. While the distribution over the plane of the sky is slightly more crowded, it seems that we have less objects that are far away from the main filament. It is interesting to note that, in this example we have $~37\%$ more CII than GR18, which could result in a better overall statistic. Also, we remind that, in this case, we only applied one criterion on astrometry quality, while there are many other filters applied in GR18. Still, we reached the same conclusion as GR18, with at least around 80 pc distance difference between the closest ($\sim$395) and the farthest ($\sim$475) points and very similar bin averages and standard deviations. \\

The lower part of Figure \ref{orion_A_dist}, shows our attempt to use the case (4), with no parallax quality criteria at all. As expected the uncertainties are much larger, while the overall trend is mostly conserved. The fact that there is no parallax quality filter at all makes the mean uncertainty estimation very large for some regions for which very uncertain objects are used. We observed that, in the longitude bins that contains many YSOs, the $1\sigma$ percentile of the distance estimate is very similar to case (5), while it significantly increases for areas that are less dense in YSOs, at greater longitude. Still, the $2\sigma$ percentile is always larger, which is expected with the increased dispersion. Nevertheless, the distribution over the plane of the sky remains convincing with very few objects far off the main filament. Additionally, we have noticed that other selections did not lead to significant changes. A combination of the two cases with both the 0.99 threshold and the astrometry criteria only reduced the dispersion by a small amount, with no other significant changes. For in depth comparison between our two cases and the GR18 case, we provide all measurements for each of our galactic longitude bins in Table~\ref{tab_orion_a_dist}, containing YSO counts, mean and median values, and the standard deviation along with the four percentiles values.\\

\newpage
\subsection{Distances to Orion B sub-regions}

\begin{figure}[!t]
	\centering
	\includegraphics[width=0.68\hsize]{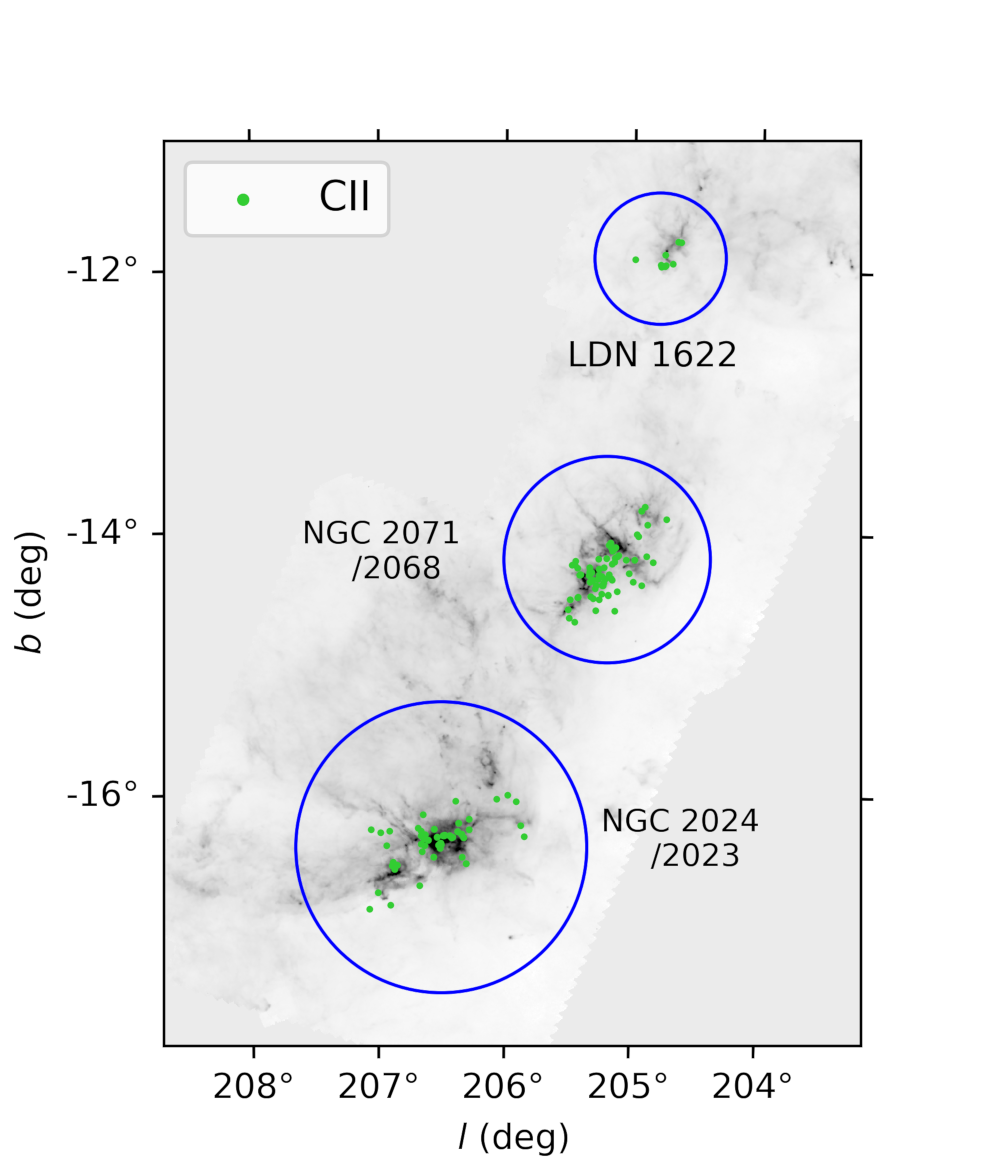}
	\caption[Orion B distribution from the F-C YSO candidates]{Distribution of the YSO candidates that are kept by the selection criteria to compute the distance of each region. The blue circles show the three regions for which a distance is estimated.}
	\label{orion_B_selected_dist}
\end{figure}

\vspace{-0.2cm}
We applied the same methodology to the Orion B part of the molecular complex. As before we used the 0.95 membership threshold along with the $\sigma_{\varpi}/\varpi < 0.1$ condition. We also removed the maximum and minimum parallax limits since there was much less objects that were evident distance outliers. We considered the following sub-regions: NGC 2024/2022, NGC 2071/2068 and LDN 1622, as in \citet{megeath_spitzer_2012}. We applied our selection to each of these regions and produced a distance estimate for the three of them. The results are summarized in Table~\ref{orion_B_select_crit_and_dist}. First, we observed that, despite the fact that no CI passed through the full selection criteria, NGC 2024/2022 and NGC 2071/2068 still contain a reasonable number of CII YSOs to estimate a distance with 51 and 67 objects, which is larger than in several bins we used for Orion A. The LDN 1622 region, despite being actively forming stars on a very localized spot (see CI in Fig.~\ref{orion_B_yso_dist}), is only characterized with 9 CII YSOs after the selection which is similar to our less represented Orion A bin. We show in Figure~\ref{orion_B_selected_dist} the selected YSOs within Orion B.\\

\vspace{-0.1cm}
The NGC 2024/2022 region, that is the closest one to Orion A in the plane of the sky, shows a mean $382 \pm 20$ pc and median 397 pc estimated distance that is very close to the one of the Orion A nebula. We emphasize that all the errors that are given along the mean prediction correspond to the propagated parallax errors. In this case, the standard deviation of $47$ pc is similar to the largest values observed in Orion A (Table~\ref{tab_orion_a_dist}) and could indicate either that the selection of YSO candidates in this region in not as good as expected, or that this cloud is particularly extended along the line of sight. Using the mean distance, it is possible to estimate the physical width of the region considering a $2^\circ$ width on the plane of the sky (as the circle in Fig.~\ref{orion_B_selected_dist}), which provides $14\,$pc. Considering this results it is unlikely that our dispersion represents the cloud depth. Also, a recent estimation of the distance of this region was made using the VLA and led to an average distance of $423 \pm 15$ pc using the stellar parallax estimation of 5 very well identified YSOs \citep{Kounkel2017}. Despite the relatively small given uncertainty, they have values ranging from 356 to 536 pc associated to this region with a very small sample size. Interestingly, they refined their distance estimate in \citet{Kounkel2018}, where they used clustering in a 6D space that merges Gaia and Apogee data with several YSO catalogs including \citet{megeath_spitzer_2012}. In this study they found a $403\pm 4$ pc estimate that is slightly more compatible with our own result. \\

\vspace{-0.1cm}
For the NGC 2071/2068 region, we found a mean distance of $431\pm 26$ pc, with an even larger standard deviation of 53 pc. This result suggests that the region is further away by around 30  to 40 pc than the Orion nebula region, which is more than what their alignment in the plane of the sky suggested. Comparing to \citet{Kounkel2017}, they estimated a distance of $388 \pm 10$, which is the opposite of the trend we observed. Still, this estimate was only based on 3 YSOs with again strong differences between their sources estimated at 383, 392 and 455 pc, respectively. Their distance refinement from \citet{Kounkel2018} is more compatible with our result with a $417\pm 5$ pc estimation. \\

\vspace{-0.1cm}
The LDN 1622 region appears much closer than expected with a mean distance of $343\pm 13$ pc and a much smaller standard deviation of $\pm 17$ pc. This result clearly separates the region from the rest of the molecular complex. The same conclusion is obtained by \citet{Kounkel2018}, where they found a distance of $345 \pm 6$ pc that is almost identical to our own estimate. These results highlight that the widely adopted 400 to 420 pc distance estimate might often be incorrect since the star-forming regions in the Orion molecular complex can spread over a 80 pc distance range, and some of them that were though to be linked to the complex are actually up to 60 pc closer to us than the Orion nebula.

\begin{table}
	\centering
	\caption{Orion B distance estimates and dispersion for each identified region.}
	\vspace{-0.1cm}
	\begin{tabularx}{0.80\hsize}{l @{\hskip 0.05\hsize} @{\hskip 0.08\hsize} *{3}{Y}}
	\toprule
	 & NGC 2024 & NGC 2071  & LDN 1622\\
	 & /2022 & /2068  &  \\
	\midrule
	Raw catalog & 31/182 & 52/223 & 7/18\\
	Raw X-match & 5/131 & 4/151 & 1/15\\
	Full selection  & 0/51 & 0/67 & 0/9\\
	Mean (pc)     & 381.8 & 431.5 & 343.4\\
	Median (pc)   & 397.0 & 424.9 & 337.8\\
	StD (pc)      & 46.9  & 52.9  & 17.4 \\
	$p(-2\sigma)$ & 261.9 & 346.5 & 323.3\\
	$p(-1\sigma)$ & 340.7 & 399.1 & 327.3\\
	$p(+1\sigma)$ & 421.7 & 464.2 & 357.8\\
	$p(+2\sigma)$ & 443.3 & 509.1 & 375.7 \\
	\bottomrule
	\end{tabularx}
	\caption*{\vspace{-0.4cm}\\ {\bf Notes.} The object numbers are given for CI/CII YSOs. The full selection applied here are the same than the case (5) of Table~\ref{orion_A_select_crit}.}
	\label{orion_B_select_crit_and_dist}
\end{table}

\newpage
\subsection{NGC 2264 distance and 3D information}

The last region for which we were able to provide a distance estimate is NGC 2264. As for Orion A, it is possible to make slices using the galactic longitude that mostly follow the sub-structures of the region. We used the same selection as for Orion B with a 0.95 membership threshold and the same $\sigma_{\varpi}/\varpi < 0.1$ criteria. We summarize in Table~\ref{ngc2264_select_crit} the effect of the basic selections. We note that there was no CI remaining after our final selection criteria. First, we made an average distance estimate over the whole region that gave a mean distance of $738 \pm 43$ pc and a median distance of 742 pc, and standard deviation of 48 pc and the following 1 and 2 $\sigma$ percentiles 630, 699, 781, 827 pc. This global estimate is consistent with the \citet{rapson_spitzer_2014} estimates, but is even  closer to the estimated distance of the open cluster with Gaia of $723 \pm 50$ that has a similar dispersion to our result \citep{Cantat-Gaudin_2018}. We note that this last estimate was not made using YSOs and therefore our estimated distance should better trace the star-forming region inside NGC 2264.\\

\begin{figure*}[!t]
	\hspace{-0.7cm}
	\begin{minipage}{1.1\hsize}
	\centering
	\begin{subfigure}[t]{1.0\textwidth}
	\includegraphics[width=1.0\hsize]{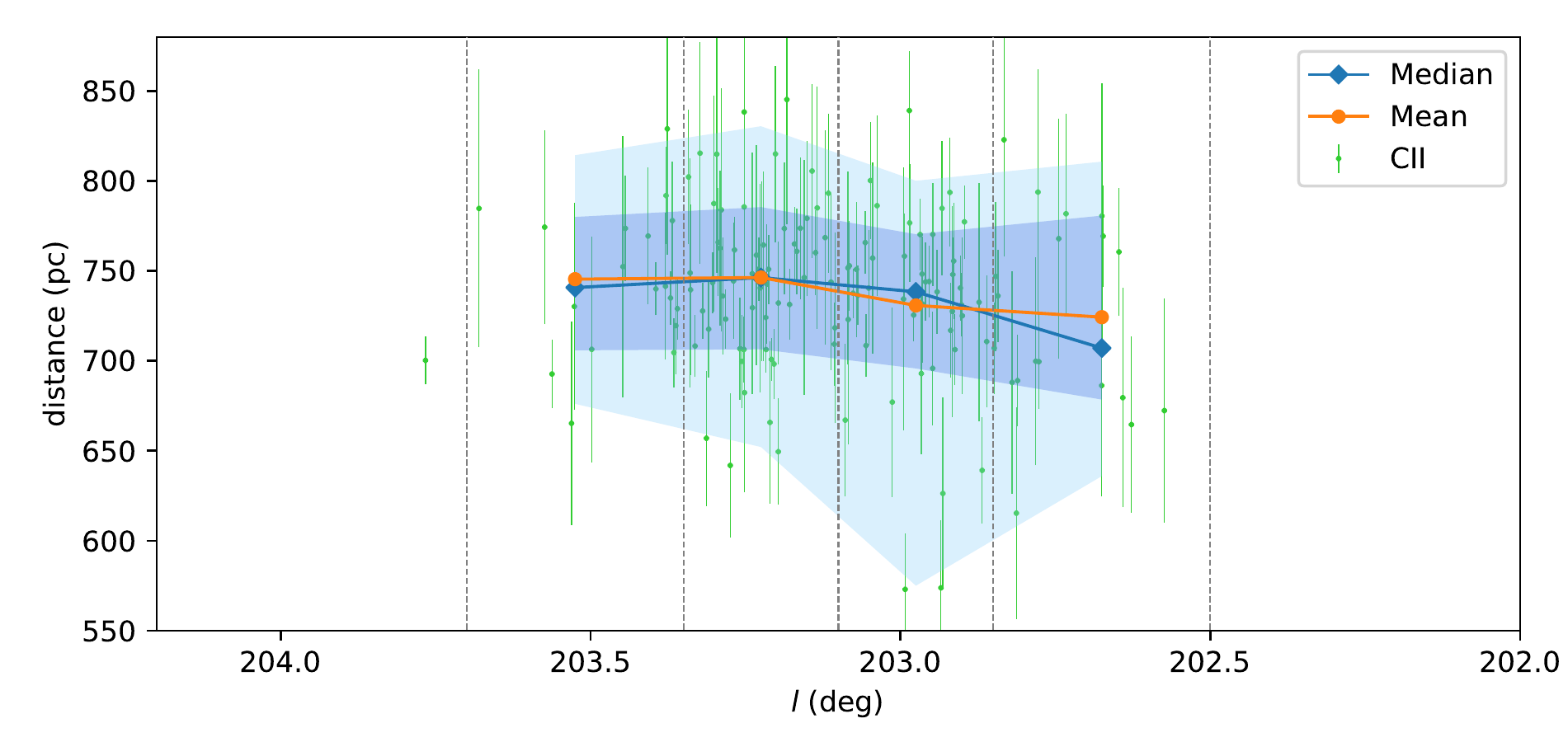}
	\end{subfigure}\\
	\vspace{-0.0cm}
	\begin{subfigure}[t]{1.0\textwidth}
	\includegraphics[width=1.0\hsize]{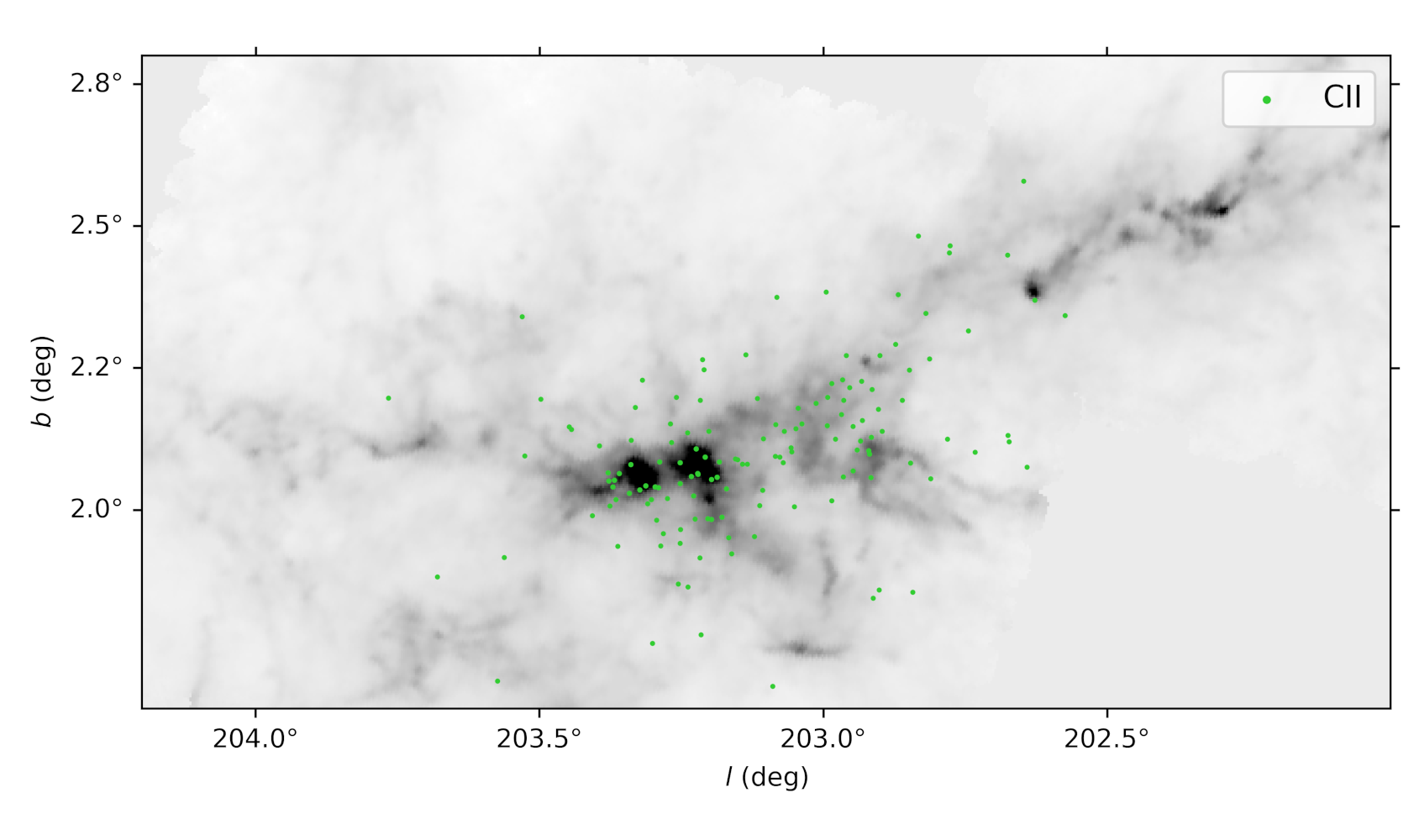}
	\end{subfigure}
	\end{minipage}
	\caption[NGC2264 distance distribution from the F-C YSO candidates]{NGC 2264 distribution of the F-C YSO candidates, along with the sky distribution over the region. Error bars correspond to $(\sigma_{\varpi}/\varpi^2)$. The selected YSOs (all CII) are in green. The orange and blue markers are the mean and median distance for each longitude bin, respectively. The light and dark blue areas represent the 2 and 1 $\sigma$ percentiles for each bin, respectively. The background grayscale is the Herschel SPIRE 500 $\mathrm{\mu m}$ map.}
\label{ngc2264_dist}
\end{figure*}

Since the number of selected CII is low we limited ourselves to four longitude bins that do not have the same size in order to get more stars in the less crowded regions. The selected bins are $l = [202.50,202.85], [202.85,203.10], [203.10,203.35], [203.35, 203.70]$. Figure~\ref{ngc2264_dist} shows the distance distribution of our YSO selection using the same visualization as for Orion A in Section~\ref{sect:orion_A_dist}. We observed that the distance estimate is much more constant in longitude with only a variation of around 20 pc in the mean values. We summarize in Table \ref{tab_ngc2264_dist} the results for each bin including the YSO count, the mean and median values, and the standard deviation along with the 4 percentiles values. This detailed values indicate a small variation in distance along the star-forming cloud, however our dispersion are significantly larger than for Orion making this trend difficult to confirm. We also note that our selection criteria removed most of the stars in the region of the filament junction region G202.3+2.5 \citep{montillaud_2019_II}.

\begin{table}
	\centering
	\caption{NGC 2264 sample size for different selection criteria.}
	\vspace{-0.1cm}
	\begin{tabularx}{0.8\hsize}{l @{\hskip 0.1\hsize} *{2}{Y}}
	\toprule
	 & CI YSOs & CII YSOs\\
	\midrule
	Raw catalog & 101 & 469\\
	Raw X-match & 8 & 390\\
	with $\varpi$ & 6 & 355\\
	with $P(X) > 0.95$ and $\sigma_{\varpi}/\varpi < 0.1$ & 0 & 142\\

	\bottomrule
	\end{tabularx}
	\label{ngc2264_select_crit}
\end{table}

\begin{table}
	\small
	\caption{NGC 2264 distance estimates and dispersion for each galactic longitude bin.}
	\hspace{-0.9cm}
	\begin{tabularx}{1.1\hsize}{l @{\hskip 0.04\hsize} l l *{6}{Y} }
	\toprule
	$l$ interval & $N_{YSO}$ & Mean & Median & StD & $p(-2\sigma)$ & $p(-1\sigma)$ & $p(+1\sigma)$ & $p(+2\sigma)$ \\
	 (deg) &  & (pc) & (pc) & (pc) & (pc) & (pc) & (pc) & (pc)\\
	\midrule
	$[202.50,202.85]$ & 19 & $724.3\pm 45.6$ & 707.1 & 52.6 & 636.0 & 678.5 & 780.8 & 810.8 \\
	$[202.85,203.10]$ & 45 & $730.8\pm 42.5$ & 738.4 & 51.7 & 575.0 & 695.8 & 770.4 & 800.1 \\
	$[203.10,203.35]$ & 59 & $746.3\pm 36.9$ & 743.3 & 43.1 & 652.1 & 706.4 & 785.5 & 830.4 \\
	$[203.35,203.70]$ & 18 & $745.4\pm 52.7$ & 740.7 & 39.2 & 676.1 & 705.9 & 780.0 & 814.3 \\
	\bottomrule
	\end{tabularx}
	\label{tab_ngc2264_dist}
\end{table}

\clearpage
\section{Additional discussion and further improvements}
\label{yso_discussion}

\etocsettocstyle{\subsubsection*{\vspace{-1cm}}}{}
\localtableofcontents

	\subsection{Identified limitations to our results}

With the dataset selected for this study, the quality of our results is mostly dependent on the proper choice of the $\gamma_i$ factors, that is to say that the main limitation comes from the construction of our labeled dataset. It is indeed expected to be the most critical part of any ML application, because the network only provides results as good as the input data. One of our major issues is that some subclasses of rare contaminants remain poorly constrained, like Shocks or PAHs, which leads to a significant contamination of the YSO classes. The non-homogeneity between the 1\,kpc small cloud dataset and the other datasets worsen this effect by increasing the dilution of these rare subclasses (Sect. \ref{cross_train}). They are almost evenly distributed across output classes in the O-O case, revealing that the network was not able to identify enough constraints on those objects. In contrast, for the C-C and F-C cases, they are randomly assigned to an output class. This means that they are completely unconstrained by the network, which failed to disentangle them from the noise of another class. This effect appeared in those specific cases due to the increased dilution of those subclasses of contaminants.\\

On the other hand, the main source of contamination for CII YSOs is the Star subclass. Adding more of them has proven to improve their classification quality (Sects.~\ref{orion_results} and \ref{cross_train}), but at the cost of even more dilution of all the other subclasses, which has a stronger negative impact on the global result. Similarly, YSO classes themselves should be more present to further improve their recall, but again at the cost of an increased dilution of the contaminant subclasses. The confusion between CI and CII YSOs is illustrated by Figure~\ref{missed_wrong_zoom}, where the misclassified YSOs of both CI and CII accumulate at the boundary between them in the input parameter space. This figure also illustrates the CII contamination from Other with the same kind of stacking, where the two classes are close to each other. A similar representation for all the CMDs is provided in Figure~\ref{missed_wrong_space}.\\

Overall, we lack data to get better results. Large Spitzer point source catalogs are available, but the original classification from \citet{gutermuth_spitzer_2009} was tailored for relatively nearby star-forming regions, where YSOs are expected to be observed. Therefore, using a non-specific large Spitzer catalog would mostly add non star-forming regions, which would create a significant number of false positive YSOs. In practice, these false positive YSOs would overwhelmingly contaminate the results, and the network performance would drop to the point where more than $50\%$ of CI YSOs are false positive. However, since one of our main limitation is the number of contaminants, a large Spitzer catalog could be used to increase the number of rare contaminants in the training sample by selecting areas that are known to be clear of YSOs. Unfortunately, this approach would mostly provide us with more Stars, Galaxies and AGNs, which are already well constrained, while the two most critical contaminant subclasses, Shocks and PAHs, originate mostly from star forming regions. \\
	
\newpage
\subsection{MIPS 24 micron band effect on the results}
	
\label{sec:mips24}
We investigate here the impact of the MIPS $24\ \mathrm{\mu m}$ band on the original classification, and therefore on the results of the network. As stated in Section~\ref{data_prep}, this band is used as a refinement step of the G09 method. Considering the classification performed using the four IRAC bands, it ensures that it is consistent with the $24\ \mathrm{\mu m}$ emission where available, for example by testing whether the SED still rises at long wavelength to better distinguish between different YSO classes. However, it adds heterogeneity in the classification scheme, since objects that do not present a MIPS emission cannot be refined. It makes the results harder to interpret and gives more work to the network as it has to learn an equivalent of this additional step. Moreover, the effect of this band on the end classification strongly affects some subclasses that are very rare in the dataset. For example, almost half of the objects initially classified as Shocks are reclassified as CI YSOs after this refinement step. Therefore, as it corresponds to a significant increase in complexity on very few objects, it is difficult to get the network to constrain them, considering the other limitations. It results in a strong contamination of the CI YSOs, as highlighted multiple times in our results. \\

On the other hand, most of the Spitzer large surveys miss a $24\ \mathrm{\mu m}$ MIPS band measurement, preventing us from generalizing our network to those datasets. Nevertheless, we chose to keep this band in our study to have the most complete view of its effect on our network. To quantify this effect, we have trained networks that did include neither the MIPS refinement step, nor the $24\ \mathrm{\mu m}$ in input. These networks have shown a small increase in performance, especially for CI YSOs with $2\%$ to $3\%$ improvement in recall and precision in the F-C case. This can mainly be explained by the simplification of the problem, but also by the greater number of objects in rare subclasses like Shocks. Moreover, such results could be generalized over larger datasets. In this case, a MIPS refinement step could still be performed \textit{a posteriori} on the network output, for objects where this band is available. Interestingly, although the absence of the MIPS refinement step could be expected to degrade the absolute reliability of the classification, the potentially large increase in the number of rare subclasses may improve the overall network performance sufficiently for the net effect on the absolute accuracy of the classification to be positive. \\

We emphasize here that the inherent difficulties that come from the use of the MIPS band can be generalized to the addition of any other band from a different survey. As we exposed in Section~\ref{yso_ml_motivation} with the study from \citet{miettinen_protostellar_2018}, cross-matching several surveys to have more bands comes at the cost of much less objects or divergent classification paths that are very difficult to constrain using ML methods. Still, very well identified YSOs with several bands that better reconstruct the SED could be used as a training dataset to construct a single large scale infrared survey classification. In this case, selection effects should be looked into with care, and it would still require a significant amount of examples.
	
\subsection{Usage of Spitzer colors instead of bands}
\label{sec:color_usage}

We stated in Section~\ref{data_prep} that we chose to use IRAC and MIPS bands, along with their respective uncertainties, as direct input features. The obvious alternative would have been to use colors, that present the advantage to be robust to many environmental properties of the star-forming regions of interest like the distance. While this approach could be efficient in principle, the G09 classification was constructed using a few direct band criteria like in the C frame of Figure~\ref{fig_gut_method}. This allows the G09 method to more robustly exclude some extra-Galactic contaminants, but it also excludes the faintest YSOs and consequently limits the method to close star forming regions. Since the present study only sticks to the G09 scheme for training, we were constrained to use these magnitudes in a way or another. The next section discusses possible alternative methods to construct the training dataset, in which case the use of colors as input features could be much more relevant.\\

From the network standpoint, using bands or colors as input features is somehow identical since it is able to reconstruct one from the other. Still, it would have an effect on the normalization of the features since colors are, in principle, less prone to variations in feature space between various star-forming regions. Another argument for using colors is that it should already be a more appropriate space for the problem we want to solve. However, all our attempts to use various color combinations or a mixture of bands and colors, plus uncertainty combinations, never outperformed our training where we used solely the bands and uncertainties directly. In any case, the prediction results were very similar but the network tends to train slightly faster when using colors but at the cost of a small increase in result dispersion. For this last reason, we chose to keep the bands as input features for the present study.

	\subsection{Method discussion}
	\label{Method_discussion}

Our approach has several caveats, the main one being that we built our labeled dataset from the preexisting G09 classification that has its own limitations including the placement of the cuts, the fact that it was constrained only on few star-forming regions, the use of magnitude cuts limiting the distance range, etc. As a consequence, our prediction is likely to inherit several of these limitations. The membership probability discussed in the previous section provides a first but limited view of the uncertainties of the original classification scheme. One approach to completely release our methodology from its dependence on the G09 scheme would consist in building our training set from a more conclusive type of observations, like visible spectroscopy to detect the $H_\alpha$ line that is usually attributed to gas accretion by the protostar \citep{Kun_2009}, or (sub-)millimeter interferometry to detect the disks \citep[e.g.][]{ruiz-rodriguez_2018, alma_disk_yso, tobin_2020}. Alternatively, a large set of photometric bands could be gathered to reconstruct the SED across a wider spectral range, as in \citet{miettinen_protostellar_2018}. Unfortunately, for now, too few objects have been observed that extensively to build a labeled sample large enough to efficiently train most of the ML algorithms.\\

Another approach would be to use simulations of star-forming regions \citep[e.g.][]{padoan_2017, vazquez-semadini_2019} and of star-forming cores \citep[e.g.][]{robitaille_interpreting_2006} to provide a mock census of YSOs and emulate their observational properties. This option would enable us to generate large training catalogs, and would provide additional control on the YSO classes, but at the cost of other kinds of biases coming from the simulation assumptions. An additional difficulty of this approach would be to find a way to generate the required large variety of contaminant objects, each of which would require a dedicated treatment.\\

A different strategy could consist of improving the method itself. With feedforward neural networks like in this study, there may still be improvement possibilities by using deeper networks with, for example, a different activation function, weight initialization, or a more complex error propagation. We explore this aspect later in the present manuscript (see Sect.~\ref{cnn_hyperparameters}) as it needs subsequent introduction to more complex networks (Sect.~\ref{cnn_global_section}. By choosing a completely different, unsupervised method, one could work independently of any prior classification. However, there is a risk that the classes identified by the method do not match the classical ones. In particular, the continuous distribution from CI to CII YSOs, and then to main sequence stars, is likely to be identified as a single class by such algorithms. A middle-ground could be the semi-supervised learning algorithms such as Deep Belief Networks \citep{Hinton504}. Such algorithms were designed to find a dimensional reduction of the given input feature space that is more suitable to the problem, therefore making its own classes based on the proximity of objects in the feature space. It could then be connected with a regular supervised feedforward neural network layer, that would combine the found classes into more usual ones. This approach would reduce the impact of the original classification on the training process, and therefore its impact on the final results.\vspace{-0.8cm}\\

\subsection{Conclusion and perspectives}
\label{conclusion_perspectives}

\vspace{-0.2cm}
We have presented a detailed methodology to use Neural Networks to extract and classify YSO candidates from several star-forming regions using Spitzer infrared data, based on the method described by \citep{gutermuth_spitzer_2009}. This study led to the following conclusions.\vspace{-0.25cm}\\

Neural Networks are a suitable solution to perform an efficient YSO classification using the Spitzer four IRAC bands and the MIPS 24 $\mu$m band. When trained on one cloud only, the prediction performance mostly depends on the size of the sample. Fairly simple networks can be used for this task with just one hidden layer that only consists of 15 to 25 neurons with a classical sigmoid activation function.\vspace{-0.25cm}\\

The prediction capability of the network on a new region strongly depends on the properties of the region used for training. Therefore, the study revealed the necessity to train the network on a census of star forming regions to improve the diversity of the training sample. A network trained on a more diverse dataset has been able to maintain a high quality prediction, which is promising for its ability to be applied to new star-forming regions.\vspace{-0.25cm}\\

The dataset imbalance has a strong effect on the results, not only on the classes of interest, but also for the hidden subclasses considered as contaminants. Carefully rebalancing each subclass in the training dataset, according to its respective feature space coverage complexity and to its proximity with other classes of interest, has shown to be of critical importance. The use of observational proportions to measure the quality of the prediction has been exposed to be of major importance to properly assess the quality of the prediction.\vspace{-0.25cm}\\

This study showed that the network membership probability prediction complements the original G09 classification with an estimate of the prediction reliability. It allows one to select objects based on their proximity to the whole set of classification cuts in a multi-dimensional space, using a single quantity. In addition, the identification of objects with a higher degree of confusion highlights parts of the parameter space that might lack constraints and that would benefit from a refinement of the original classification. The corresponding catalog of YSO candidates in Orion and NGC 2264 predicted by our final ANN, along with the class membership probability for each object, is publicly available at CDS.\vspace{-0.25cm}\\

We showed that our prediction can efficiently be used in combination to a survey like Gaia to recover distance information on the star forming regions. In the most favorable cases, it allowed to reconstruct a continuous distance information, while in the other cases it provided competitive global distance estimates of the star forming clouds. We also exposed that the more interesting younger CI YSO candidates do not have an optical emission that is strong enough or that the recovered parallaxes has too large uncertainty on these objects. It would be interesting to be able to recover these objects with a good parallax measurement as the showed to better trace the star forming regions than more evolved CII YSOs.\vspace{-0.25cm}\\

\vspace{-0.5cm}
The current study contains various limitations, mainly the lack of additional near star-forming region catalogs, that contain the sub-contaminant distinction to construct complete training samples. Moreover, some sub-classes, namely Shocks and PAHs, remain strongly unconstrained due to their scarcity. Identifying additional shocks and resolved PAH emission in Spitzer archive data could significantly improve their classification by our networks, and consequently improve the YSO classification. The attention has also been drawn toward the use of simulations to compile large training datasets, that might be used in ensuing studies.\vspace{-0.25cm}\\

Finally, our method could be improved by adopting more advanced networks which would probably overcome some difficulties, for example by avoiding local minima more efficiently, and would improve the raw computational performance of the method. Semi-supervised or fully unsupervised methods may also be promising tracks to predict YSO candidates which may overthrow the supervised methods in terms of prediction quality. On the other hand, we have highlighted that most of the difficulties come from the training set construction, which is mostly independent of the chosen method. Therefore, future improvements in YSO identification and classification from ML applied to mid-IR surveys will require compilation of larger and more reliable training catalogs, either by taking advantage of current and future surveys from various facilities, like the Massive Young Star-Forming Complex Study in Infrared and X-ray \citep[MYStIX,][]{Feigelson_2013} and the VLA/ALMA Nascent Disk and Multiplicity survey \citep[VANDAM,][]{tobin_2020}, or by synthesizing such catalogs from simulations.\vspace{-0.25cm}\\

Most of the {\bf methods and results discussed in this first part (Part I) are published} in the section "Numerical methods and codes" of the journal Astronomy and Astrophysics \citep[][accepted]{cornu_montillaud_20}.

\setkeys{Gin}{draft=False}




\clearpage
\null
\thispagestyle{empty}
\newpage
\thispagestyle{empty}
\hfill
\vspace{0.3\textheight}
\part[Reconstruction of the 3D interstellar extinction of the MW]{Reconstruction of the 3D interstellar extinction of the Milky Way}

\newpage
\null
\thispagestyle{empty}
\newpage
\etocsetnexttocdepth{2}
\etocsettocstyle{\subsection*{Part III: Reconstruction of the 3D interstellar extinction of the Milky Way}}{}
\localtableofcontents{}

\clearpage

\section{Using interstellar extinction to infer the 3D Milky Way structure}
\label{ext_map_first_section}

\etocsettocstyle{\subsubsection*{\vspace{-0.6cm}}}{}
\localtableofcontents

\subsection{Current state of 3D extinction maps}
\label{ext_properties_part3}

We introduced in Section~\ref{intro_extinction} that the extinction is a physical process which reduces the apparent luminosity of an astronomical object. The latter is also reddened according to an extinction law that correlates the effect of the extinction with the wavelength according to a given dust grain size and composition, which is considered constant in the diffuse ISM of the Milky Way. For large-scale Galactic studies or stellar population studies, having a good measurement of the extinction is a necessity, since all the light observed has traveled through at least a small piece of ISM. It can also be important for extra-galactic and cosmological studies to remove the Milky Way foreground extinction in order to measure absolute magnitudes, for example to estimate the distance to standard candles like type Ia supernovae or Cepheids \citep{Nataf_2016}.\\

One major issue with extinction is that it is an integrated quantity, cumulated along the whole light path from the source to the observer. Still, it requires to know what is the true emitted spectra of the source in order to measure its reddening properly. Since the extinction value is directly linked to the dust density, being able to estimate the differential extinction as a function of the distance is a way to reconstruct the Milky Way dust structure. For these reasons, the reconstruction of 3D extinction maps of the Milky Way has been an active topic for several years involving many research groups.\\

\afterpage{%
\begin{figure*}[!t]
	\hspace{-1.1cm}
	\begin{minipage}{1.15\textwidth}
	\centering
	\begin{subfigure}[!t]{0.48\textwidth}
	\hspace{+0.2cm}
	\includegraphics[width=\textwidth]{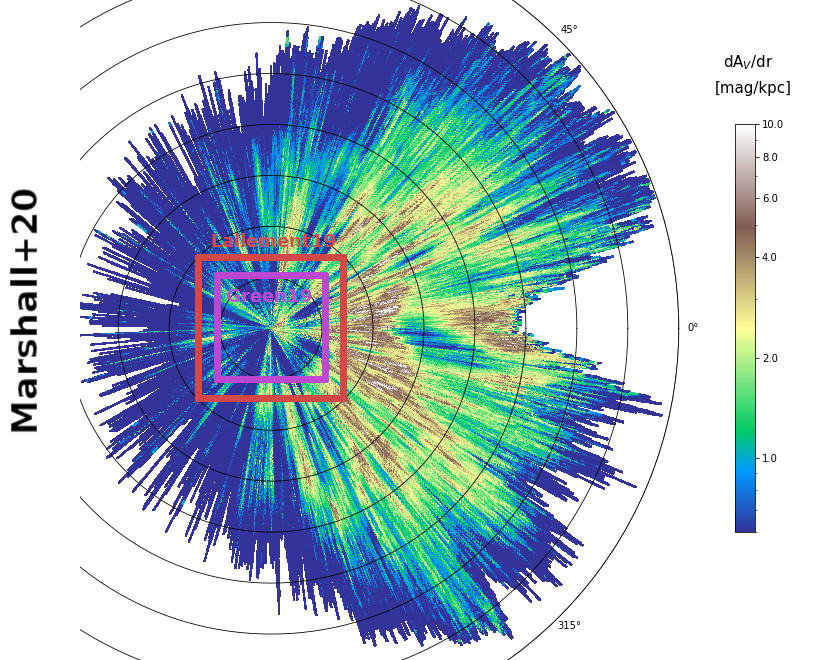}
	\end{subfigure}
	\vspace{0.2cm}
	\begin{subfigure}[!t]{0.48\textwidth}
	\includegraphics[width=\textwidth]{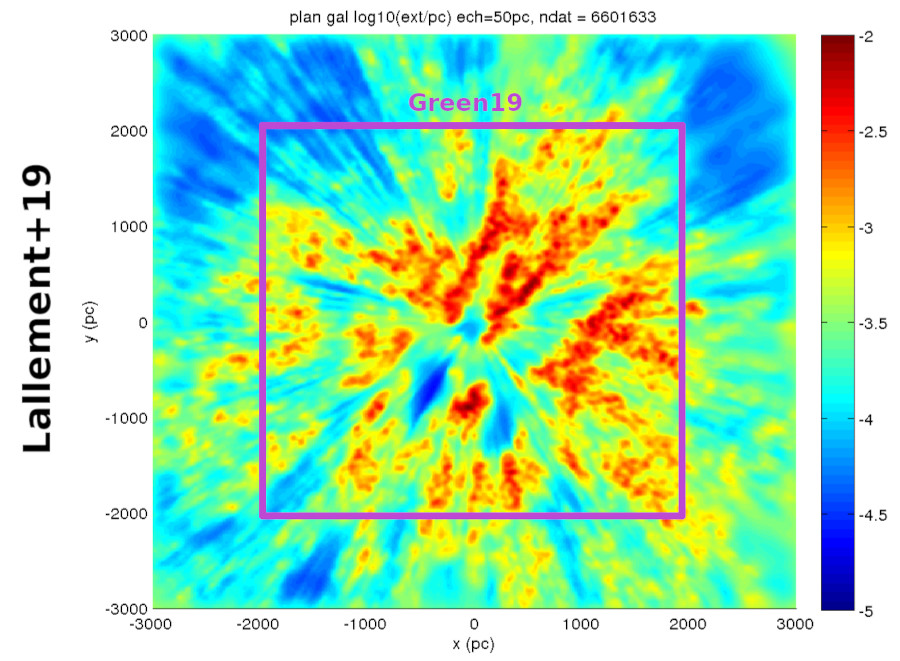}
	\end{subfigure}\\
	\vspace{0.2cm}
	\begin{subfigure}[!t]{0.48\textwidth}
	\includegraphics[width=\textwidth]{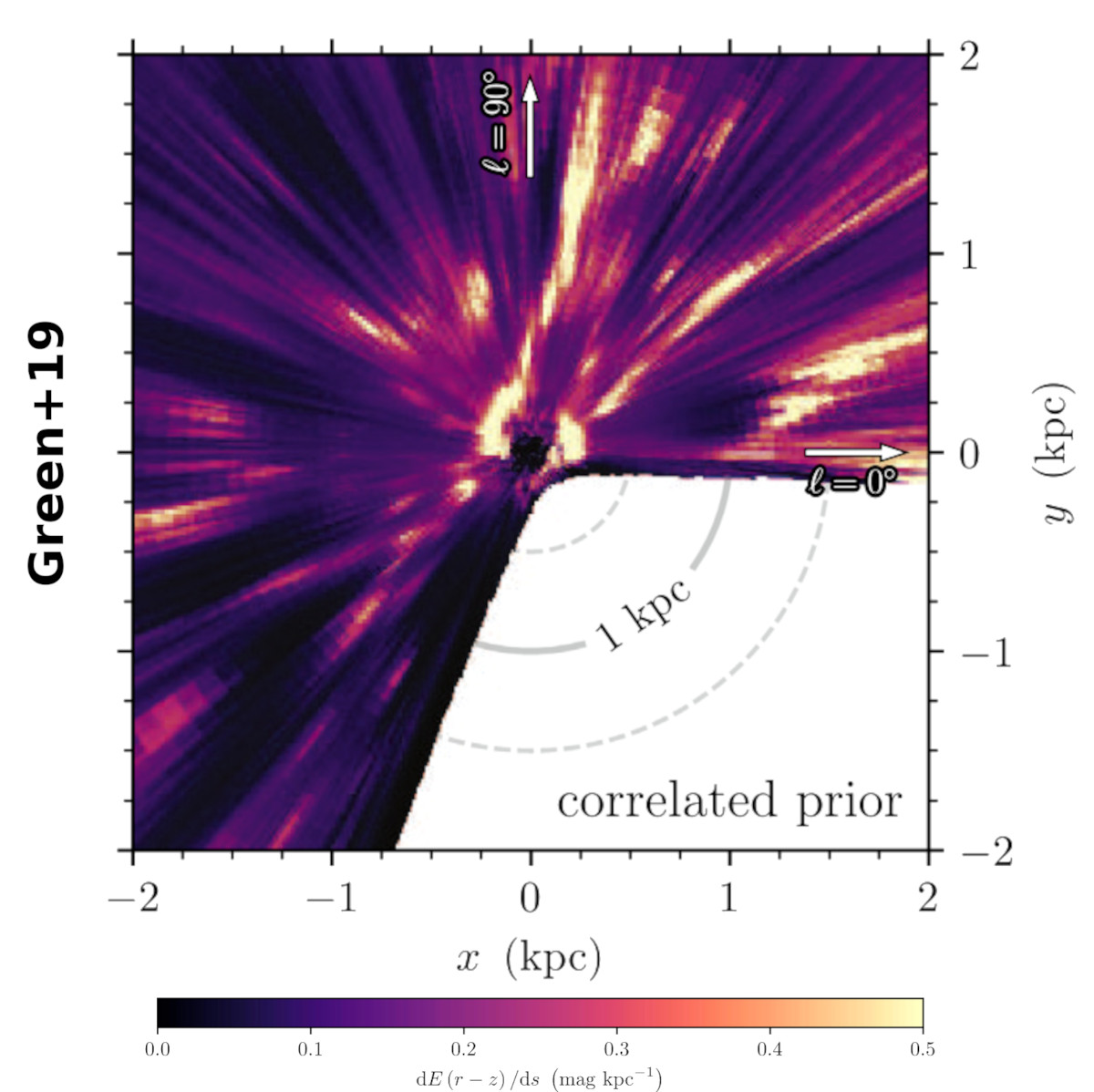}
	\end{subfigure}
	\hspace{0.5cm}
	\begin{subfigure}[!t]{0.48\textwidth}
	\vspace{-1.0cm}
	\includegraphics[width=\textwidth]{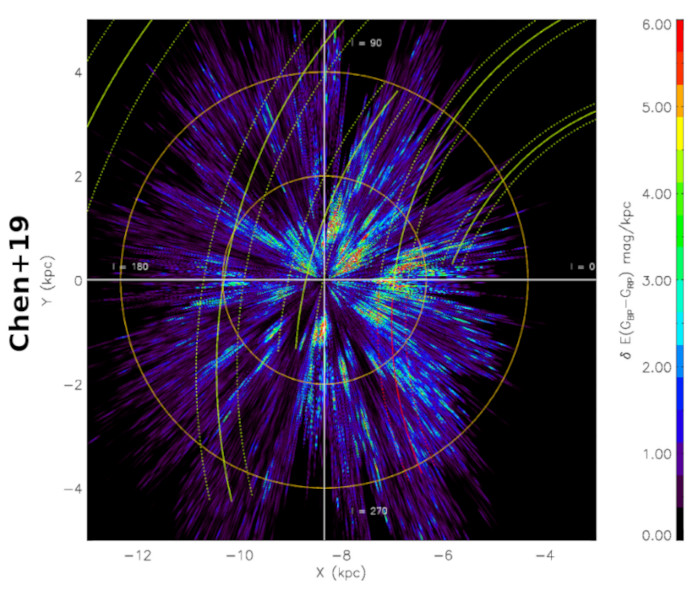}
	\end{subfigure}
	\vspace{0.2cm}
	\end{minipage}
		\caption[Examples of recent 3D extinction maps]{Four recent extinction maps based on different methods and data. These maps all represent a face-on view of differential extinction in the Milky Way Disk integrated over a given galactic height or latitude. In all maps the Sun is in the middle and the galactic center is to the right. {\it Top-left}: Map from \citet[][in prep.]{Marshall_2020} integrated for $|b| < 1$ deg using solely 2MASS. Each circle is a 2 kpc radius, the purple and red squares represent the range of two of the other maps as indicated.  {\it Top-right}: Map from \citet{Lallement_2019} integrated for $|z|< 300$ pc using 2MASS and Gaia DR2 cross matched data, the purple square represents the range of the bottom left map. {\it Bottom-left}: Map from \citet{Green_2019} integrated for $|z|< 300$ pc using Pan-STARRS, 2MASS and Gaia DR2 cross matched data. {\it Bottom-right}: Map from \citet{Chen_2019} integrated for $|b|< 0.1$ deg using WISE, 2MASS and Gaia DR2 cross matched data.}
	\label{extinction_maps}
	
\end{figure*}
\clearpage
}

The usual approach consists in estimating the extinction for each star, or group of stars, along a line of sight (LOS). The extinction is usually measured in the infrared since it is less affected by the extinction than optical wavelength, which provides a deeper view in dense environments. The approaches that rely on star parallaxes to get the distance of each star and reconstruct the extinction distribution from it, are usually limited to short distance estimates and are not able to provide constraints on galactic-scale structures other than the closest arms (Local, Perseus, Sagittarius). Still, such approaches usually provide a better resolution at short distances. In contrast, methods that rely solely on infrared data are usually able to make predictions at much greater distances, but suffer from a lower resolution and a quickly increasing distance uncertainty, which produces elongated artifacts that are known as "fingers of God".\\

Here we discuss some of the most known extinction maps in order to identify what are their present limitations. Among the most known extinction maps we can cite \citet{Marshall_2006} with its last refinement on which we participated \citet[][in prep.]{Marshall_2020}. This map is done using a per line of sight approach and works by comparing statistical predictions of the Besançon Galaxy Model (see Sect.~\ref{BGM_sect}) with the equivalent observed quantities. The map is made using solely 2MASS data allowing a greater distance range (up to 14 kpc) due to the lesser optical depth of interstellar clouds in the infrared than in the visible. The last iteration of this map is visible on the top-left frame of Figure~\ref{extinction_maps} using a face-on view of the Galactic disk. It has been successfully used to identify some large-scale structures that are coherent with several of the expected Galactic arms. Still, they can be difficult to distinguish from one another due to the relatively low distance resolution in some places of the map (e.g. the Scutum-Crux and Norma arms at a distance of $\sim$4 kpc in the direction of the Galactic Center). It also detects a first part of the Galactic bar centered around 8 kpc. The main limitations of this map are that the anti-center region is not strongly constrained due to a lesser star count, and that the fingers of god artifacts remain significant.\\

The map from \citet{Lallement_2019} is not based on a line of sight approach, but on a global inversion from a given star list. The aim is to find the extinction value for each star and to reconstruct a 3D spatially coherent distribution from it. In practice the last iteration of the map is based on a cross-match between Gaia and 2MASS and uses the magnitudes from both surveys. The combination of: (i) a meticulously-constructed hierarchical inversion method that is based on Bayesian processes, and (ii) the very large set of individual Gaia distances from stars that have been carefully selected, leads to an unmatched map resolution. The corresponding map is shown in the top-right frame of Figure~\ref{extinction_maps} also using a face-on view. This map efficiently disentangles several extinction regions that are aligned on the same line of sight, which is difficult to achieve with other methods. The structures present little to no stretching in distance. This map as been observed to reconstruct well-know structures at close range and highlight a global curved and continuous structure associated to the local arm. Some sub-structures also tend to match the expected position of some other arms, namely Perseus, and a Sagittarius-Carina foreground. Still, the limits of this map are the possible biases in the star selection and the low distance range of the prediction, up to 3 kpc only, which is mainly due to the cross-match with Gaia that is much more affected by extinction than 2MASS. The authors also highlight that the distance estimates from the parallax inversion might be underestimated, and that there is a lower limit in structure size induced by the method that remains unrealistic, which is more problematic with larger distances.\\

Another common map we can cite is \citet{Green_2019} that uses individual lines of sight but with a prior on the correlation between adjacent ones. The method is made of several steps that consist in finding the extinction and distance modulus for each star using the observed parallax and photometric magnitudes using a set of priors. Then another step reconstructs the line of sight extinction distribution using the star list sampled into distance bins. Finally a Gaussian process is added to correlate adjacent lines of sight. In practice they used a cross match between Pan-STARRS 1 \citep{Chambers_2016}, 2MASS and Gaia DR2 (parallax only), inducing and even lower maximum distance estimate than the map from \citet{Lallement_2019} with only a 2kpc range of confident prediction. The bottom-left frame of Figure~\ref{extinction_maps} also shows a face-on view of this map. The authors observed a convincing match between their prediction and well-known star-forming regions that are associated to different arms that appear roughly aligned in the map. The local arm structure is the most convincing with Perseus, Sagittarius and Scutum being either not strongly predicted by the map or less well defined by the reference star-forming regions that present a high distance uncertainty. The limitations of this map are that the survey selection prevents any prediction in more than a quarter of the MW, the short distance prediction, and finally several individual priors that could accumulate biases. We also note that, despite the added correlation between lines of sight, there is still significant variations between adjacent ones, especially for distances greater than 1\,kpc.\\

Lastly, we mention the map by \citet{Chen_2019} that uses a machine learning method based on the very efficient random forests algorithm that is trained using well constrained example stars. They use this method to predict various color excesses for individual stars using the 2MASS and Gaia magnitudes. Their full star list is then decomposed into lines of sight for which a color excess vs. distance profile is fitted using Gaia distances from \citep{bailer-jones_2018}. The bottom-right frame of Figure~\ref{extinction_maps} shows a face-on view of their prediction. The overploted arms are from \citep{Reid_2014} and mostly match what would be the local arm an the close part of Sagittarius. The match with the Perseus arm that the authors claim to observe in their map seems more uncertain, because many similar structures as those used for this assessment are observed outside any arm structure. Still, the map is mostly in good agreement with \citep{Lallement_2019} when looking at comparable distance range, which is expected since they used very similar input data and dimensions. The limits of this map are mainly the relatively small distance range as well, and strong fluctuations between adjacent lines of sight due to the absence of correlation between them.\\

We do note perform an exhaustive extinction map census here, and we will stop with these four more detailed maps, but there are still a few works that we find worth mentioning. The work by \citet{Drimmel_2001} that was a precursor for the present extinction maps, the map from \citet{Sale_2014} and all its refinements that also relies on a hierarchical Bayesian approach but using other types of surveys like in $H_\alpha$, or the recent new work from \citet{Rezaei_2017, Rezaei_2018} that uses Gaussian processes to overcome several difficulties of the previous maps and that is slightly more discussed in Section~\ref{other_ext_map_ML}.\\

For the present study the aim is to construct a method that is able to be efficient at an intermediate scale between large distance from \citet[][in prep.]{Marshall_2006,Marshall_2020} and the closer range ones like \citet{Lallement_2019}. More details on our objectives are given in Sect.~\ref{cnn_maps_objective}.

\subsection{Per line of sight approach}
\begin{figure}[!t]
	\centering
	\includegraphics[width=1.0\hsize]{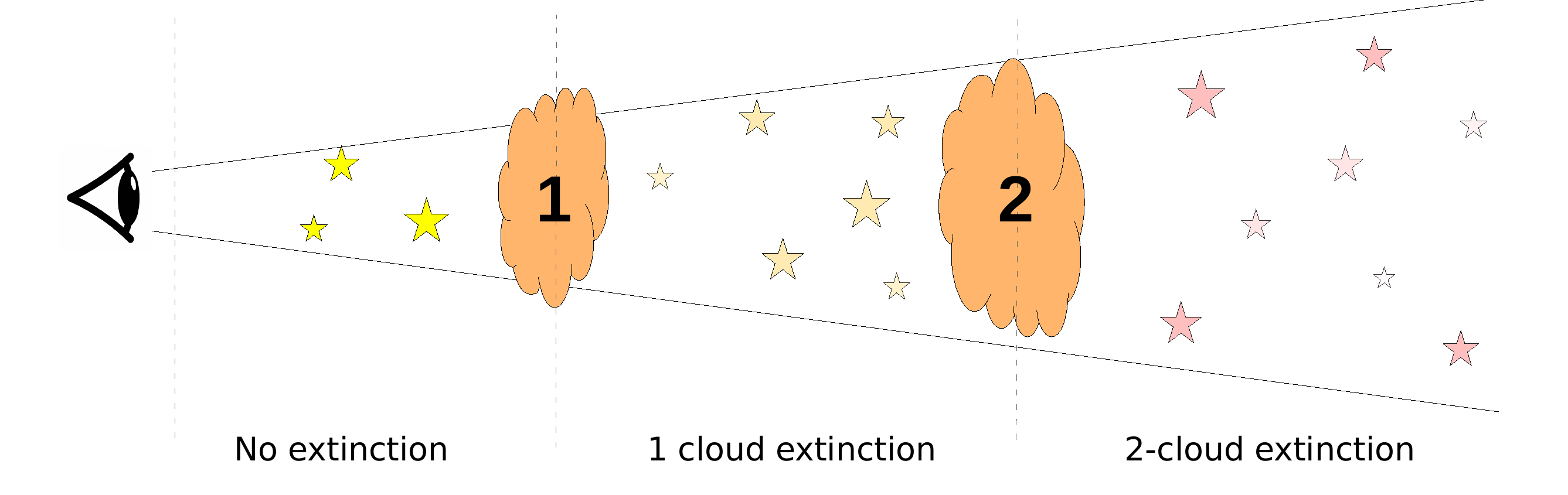}
	\caption[Illustration of a simple LOS with two clouds]{Simple line of sight (cone view) example that contains two clouds, with the observer to the left. The star colors are reddened and faded according the observed cumulative extinction effect on them from the observer's point of view.}
	\label{los_illustration}
	\end{figure}

We describe here the approach that consists in selecting a small cone observation in the sky that contains several stars. This cone is designated as a Line Of Sight (LOS), that is defined by its center position on the sky and by its radius or solid angle. We illustrate in Figure~\ref{los_illustration} a simple case where all the extinction is packed in two individual clouds. In this simple case it is visible that the stars before the first cloud do not suffer of extinction, the stars between the two clouds are extincted only by the first one, while the stars that are behind the second cloud are extincted by both clouds.\\

\begin{figure}[!t]
	\centering
	\includegraphics[width=0.75\hsize]{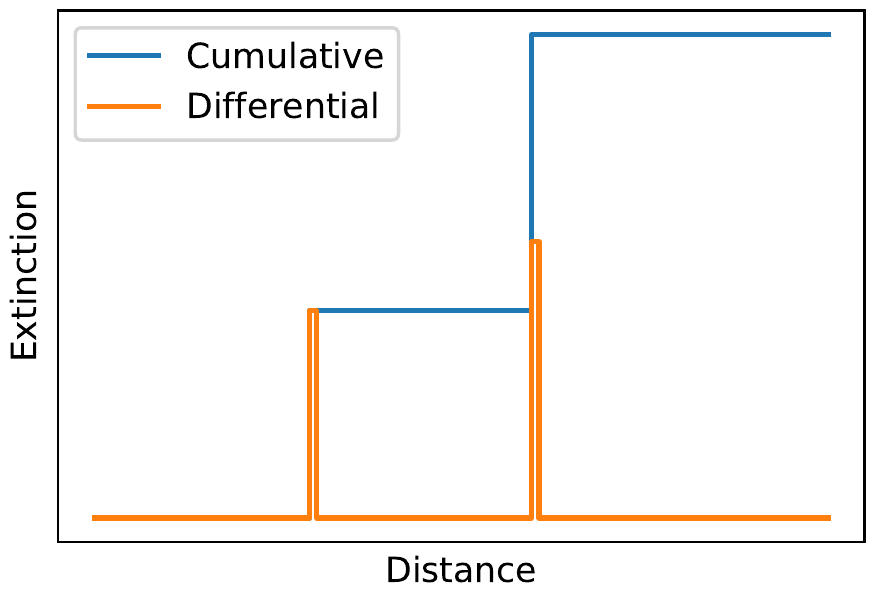}
	\caption[Extinction profile example]{Simple extinction profile example corresponding to the LOS with two clouds of Figure~\ref{los_illustration}. The profile is represented using both the cumulative and differential extinction from the same underlying quantity.}
	\label{simple_profile_expl}
\end{figure}

In this context, we want to reconstruct the distribution of the extinction as a function of the distance for this LOS. If the majority of the extinction is concentrated in dense clouds then it is equivalent to find the position of the clouds along the LOS. The corresponding cumulative or differential extinction profile, assuming that the clouds have a negligible extent along the LOS, is illustrated in Figure~\ref{simple_profile_expl}. To construct a large 3D extinction map it is then possible to decompose the plane of the sky into several small individual lines of sight.\\

\newpage
Inferring the extinction along the LOS can be done using different methods and data. They all have in common to work from observed stars but using different quantities. One approach consists in estimating the distance and the intrinsic extinction to the stars and then infer the extinction profile along the LOS. This approach was adopted by \citet{Green_2018} who used Bayesian inference with a Markov Chain Monte Carlo (MCMC) technique to estimate the distance-extinction pairs of a sample of stars built by cross-matching Pan-STARRS-1, Gaia DR2, and 2MASS stars. Another MCMC step is then used to infer an extinction profile compatible with the distance-extinction pairs in each conical LOS. In \citet{Lallement_2019} a catalog made of stars from the cross-match of 2MASS and Gaia DR2 was compiled, where the distances were derived from Gaia parallaxes with uncertainties better than 20\%, and the extinction from a fit of the intrinsic colors in Gaia-2MASS colors by adjusting the extinction parametrization. The differential extinction distribution is then inferred by a hierarchical, multi-scale Bayesian inversion in 3D where a 3D Gaussian kernel, whose size depends on the current scale, is used to ensure the spatial coherence.  \\

In contrast, \citep[][in prep.]{Marshall_2006, Marshall_2009, Marshall_2020} forgo determining the distance and extinction to individual stars. Instead, they rely on a stellar population model of the Milky Way (see next section) which provides the statistical distributions of the intrinsic stellar observational properties for each LOS. The extinction profile of each LOS is inferred from the statistical comparison between the intrinsic and observed distributions of stars. Different methods were attempted, including genetic algorithm \citep{Marshall_2009} and MCMC \citep[][in prep.]{Marshall_2020}.\\

Figure~\ref{extinction_maps} compares some of the 3D extinction maps obtained by the authors mentioned above. They reveal some of the major limitations of these approaches: (i) it is difficult to recover spatially coherent structures between line of sights without adding an ad-hoc correlation, (ii) once a first front of extinction has been localized it is much more difficult to reliably detect additional extinction beyond this front, and (iii) the large difference between the uncertainties parallel and perpendicular to the LOS leads to an elongated radial artifact often called "finger of gods", with a strong variation of the prediction between adjacent LOS. \\

In the present work, we elaborate on the approach by Marshall et al., comparing 2MASS and Gaia data to a stellar population model of the Milky Way: the Besançon Galaxy model.

\newpage

\clearpage
\subsection{The Besançon Galaxy Model}
\label{BGM_sect}

\begin{figure*}[!t]
	\centering
	\begin{subfigure}[t]{0.48\textwidth}
	\includegraphics[width=\textwidth]{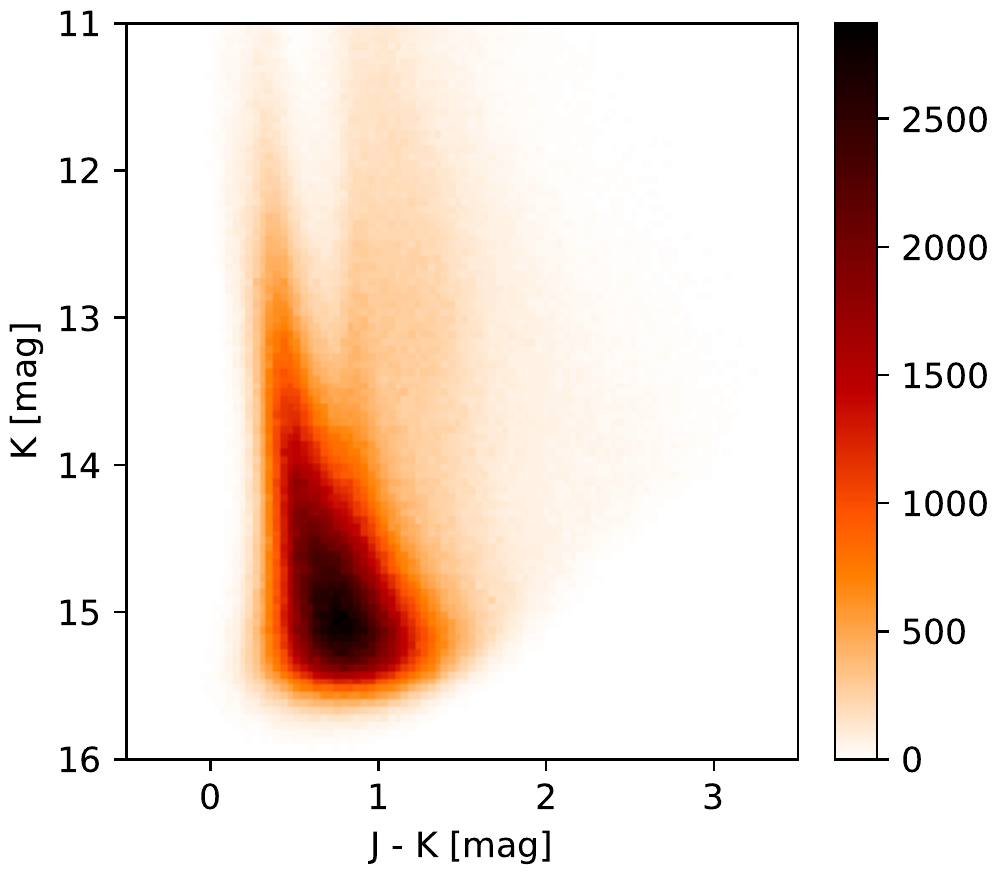}
	\end{subfigure}
	\begin{subfigure}[t]{0.48\textwidth}
	\includegraphics[width=\textwidth]{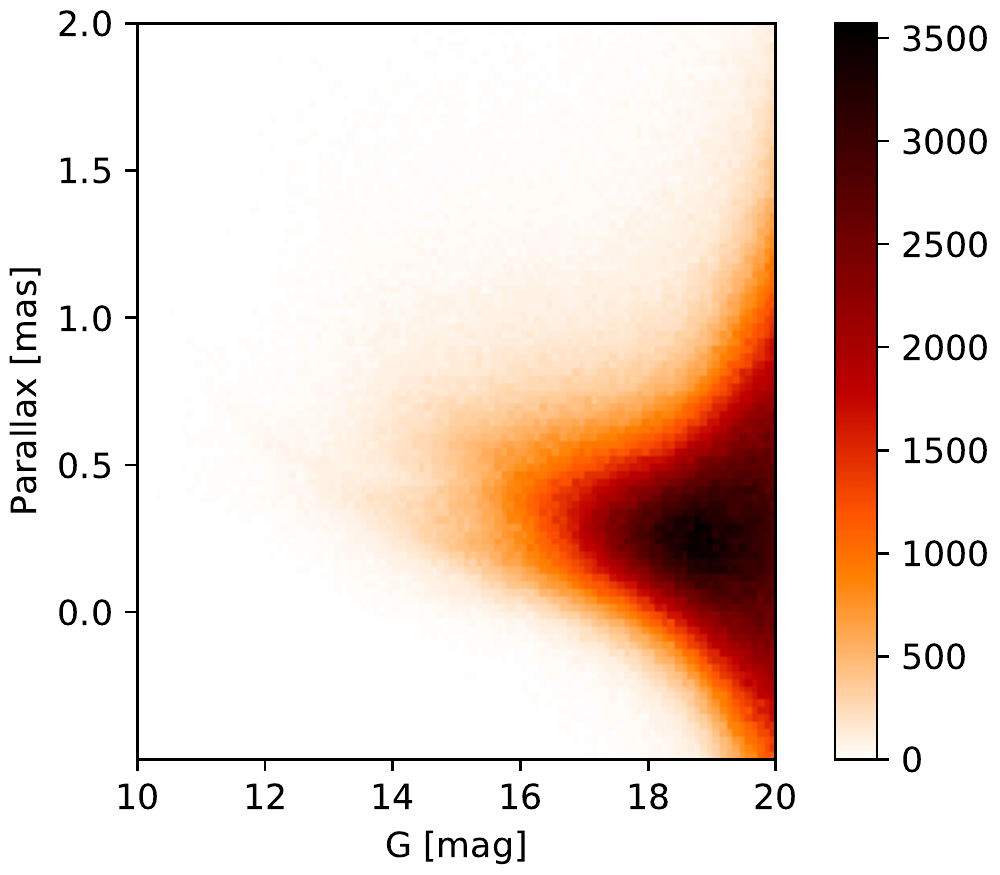}
	\end{subfigure}
	\caption[Observed diagrams for comparison with the BGM]{Illustration of observed diagrams that can be reproduced using the Besançon Galaxy Model. The two diagrams are obtained from a $4^\circ$ degree radius centered at galactic coordinates $l=280$ deg, $b=0$ deg. {\it Left:} 2MASS observed [J-K]-[K] CMD. {\it Right:} Gaia DR2 observed [Gmag]-[Parallax] diagram.}
	\label{expl_observed_cmds}
	\vspace{-0.2cm}
\end{figure*}

\vspace{-0.1cm}
We are lucky to have a favored access to the Besançon Galaxy Model (BGM), a world-class stellar population synthesis model \citep{Robin_2003,Robin_2012}. It was noticeably adopted to anticipate Gaia results \citep{Robin_2012b} and is still used as a validation tool for Gaia catalogs \citep{Arenou_2018}. This model is able to generate 3D stellar distributions that are statistically representative of many observables. It is based on four distinct stellar populations: a thin disk, a thick disk, a bulge and a halo. The model is constrained by both observations and theoretical recipes that account for stellar evolution \citep{Lagarde_2012}, dynamics \citep{Bienayme_1987}, initial mass function \citep{Haywood_1997a, Haywood_1997b}, etc. A BGM computed realization takes the form of a star list that contains various physical quantities for each modeled star, like mass, velocity, age, magnitudes, stellar count, distance, etc. Regarding the star emission, the BGM uses color tables based on stellar atmosphere models to accurately reproduce stellar colors \citep[based on and refined in][]{Lejeune_1997, Westera_2002}.\\

\vspace{-0.1cm}
It is interesting to note that, in order to convert absolute quantities to observable ones appropriately, the BGM must use an extinction map. Depending on the model version, it uses different extinction maps, and it is even possible to select the most appropriate map depending on the region of the Milky Way. Overall it relies mainly on those of \citet{Marshall_2006} and more recently \citet{Lallement_2019}. This is another example of the importance of producing good quality extinction maps.\\

\vspace{-0.1cm}
In order to be representative of the real Milky Way, the BGM prediction must be used only with statistical representations and with a large enough number of stars. A suitable representation is a Color-Magnitude Diagram, that is similar to 2D histogram of the star list, or any similar representation involving a stellar observable (e.g. parallaxes).  Figure~\ref{expl_observed_cmds} shows two examples for observed data, with a 2MASS [J-K]-[K] CMD and a Gaia [Gmag]-[Parallax] diagram using a $4^\circ$ radius line of sight centered on the galactic coordinates $l=280$ deg, $b=0$ deg. We detail in Section~\ref{cmds_construction_section} and \ref{gaia_diag_constuction} how the BGM can be used to reconstruct theses diagrams in a realistic fashion.

\subsection{Mesuring extinction using the BGM}
\label{extinction_with_bgm_intro}

\begin{figure}[!t]
	\centering
	\vspace{-0.3cm}
	\includegraphics[width=0.93\textwidth]{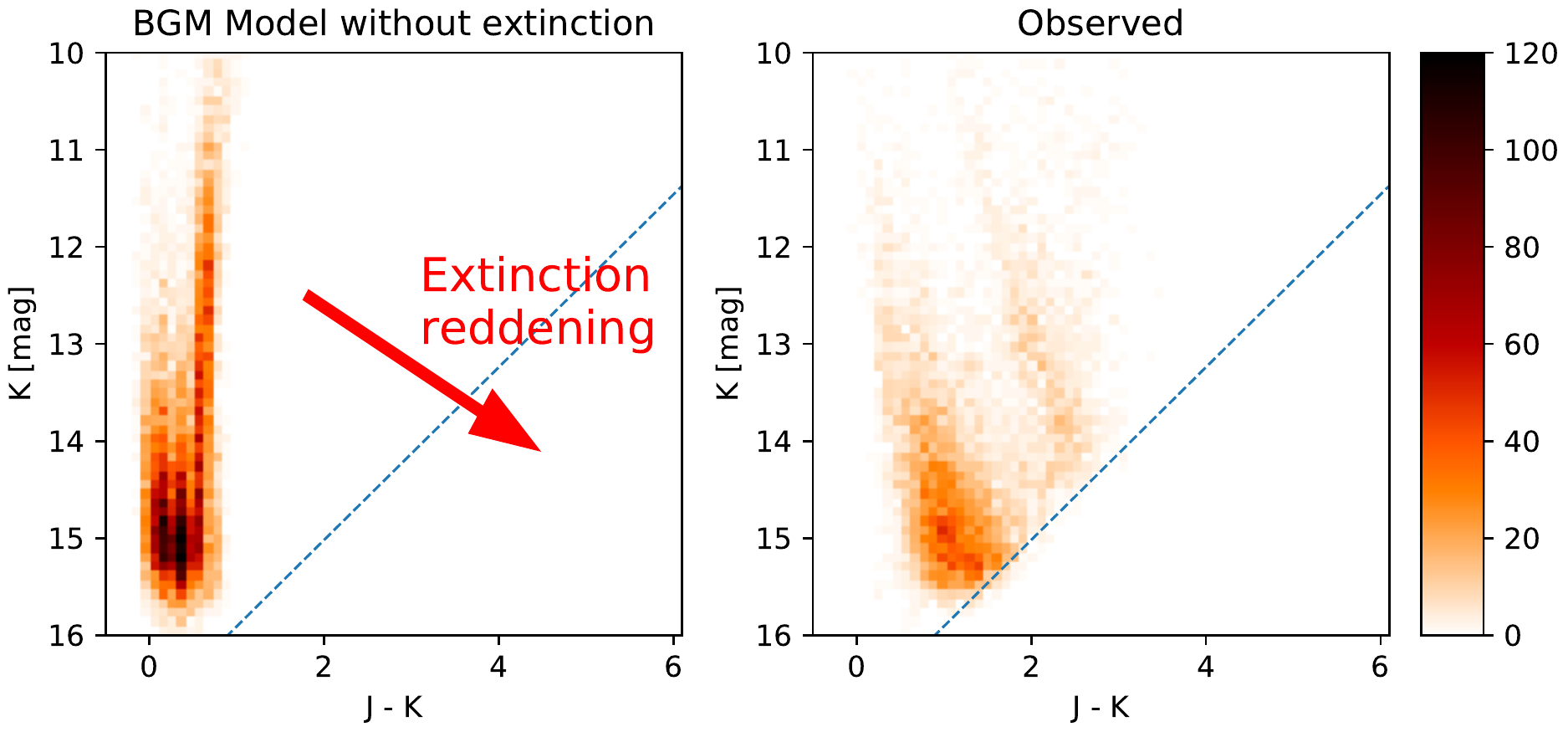}
	\caption[Model without extinction and observed extinction for the same LOS]{2MASS [J-K]-[K] CMD comparison between model prediction without extinction and observed extinction for the same LOS. The two diagrams are obtained from a $0.25^\circ$ degree radius centered at galactic coordinates $l=280$ deg, $b=0$ deg. The red arrow illustrates the extinction translation direction for individual stars. The dashed blue line corresponds to the observation limit cut as described in Sect.~\ref{ext_profile_and_cmd_realism}. {\it Left:} 2MASS model without extinction. {\it Right:} 2MASS observed [J-K]-[K] CMD.}
	\label{obs_model_ext_comparison}
\end{figure}

The fact that the BGM first produces stellar quantities without the extinction contribution is very useful feature in our case. On the left frame of Figure~\ref{obs_model_ext_comparison} we show the same [J-K]-[J] 2MASS CMD predicted by the model as in Figure~\ref{expl_observed_cmds}, but for a smaller radius of $0.25^\circ$ and without the extinction. Following the physical properties of the extinction exposed in Sections~\ref{intro_extinction} and~\ref{ext_properties_part3}, the stars will both be fainted and reddened by the extinction creating a translation toward lower [K] and larger [J-K], due to the fact that the wavelength of J ($1.235 \mu m$) is lower than the one of $\mathrm{K_s}$ ($2.159 \mu m$) inducing a stronger extinction in the J band. The right frame of Figure~\ref{obs_model_ext_comparison} illustrates this effect by showing the equivalent observed 2MASS CMD. \\

From these results it is possible to formulate the following hypothesis: {\bf assuming a perfectly representative model, the difference between the BGM and the observed CMD is solely due to interstellar extinction}. Considering this, it should be possible to design a method that infers the extinction profile from the differences between synthetic and observed CMDs. This model-observation comparison is already at the heart of the \citet{Marshall_2020} method which also uses the BGM as a reference. We emphasize that, even if the effect of extinction on a single star in this diagram is a simple translation, there is a complex distribution of these stars along the LOS entangled with the extinction distribution. It means that all stars will move following their own local cumulative extinction, inducing a much more complex transformation in this diagram including translation, stretching and overlap. It also has to account for a cut in magnitude that corresponds to the limits of the 2MASS observations (see Sect.\ref{ext_profile_and_cmd_realism}).\\

\newpage
\subsection{Using Machine Learning for this task}
\label{other_ext_map_ML}

There are many various methods that were used to perform a similar comparison in order to reconstruct the extinction distribution (see references in Sect.~\ref{ext_properties_part3}), some recent ones including machine learning as well. Still, we observed that few attempts were made using classical algorithms and that the solutions proposed are usually too computationally intensive for large maps reconstructions. For example, in \citet{Marshall_2009} they used a Genetic Algorithm (GA) method (see Sect.~\ref{ml_application_range}, and Fig.~\ref{fig_ml_types}) to reconstruct individual dark clouds distance from extinction. Still, they used GA by fitting each LOS individually in a very similar fashion as the recent work by \citet{Marshall_2020}, and to not construct a high resolution extinction map. Another recent example comes from \citet{Rezaei_2017} and \citet{Rezaei_2018} who use Gaussian Process (GP) (Sect.~\ref{ml_application_range}, and Fig.~\ref{fig_ml_types}) to reconstruct the 3D extinction. In this approach there is no LOS consideration and the large regions are reconstructed at once. It noticeably reconstructs 3D spatially coherent structures in a very smooth way. The main difficulty is that, because it relies on a one time large inversion that scales non linearly with the number of data points and the resolution, it is very computationally heavy. To overcome this, the authors mostly used restricted datasets, which negatively impacted the statistical representation of the problem. Another approach using Random Forest is described by \citet{Chen_2019} where they used it to fit the Gaia color excess of individual stars that are then positioned in distance using Gaia parallaxes. This method is certainly computationally efficient but the stars that have the largest reddening will have very uncertain distance estimate, which results in important finger of gods effects.\\

We highlight that, to our present knowledge, there was no published application that uses any kind of Artificial Neural Network to reconstruct extinction profiles, or that uses the extinction distribution to infer distances, and even less a large 3D structural reconstruction of or from the extinction. The only links between ANN and extinction we found are in applications that assess the cumulative extinction as a single quantity for given stellar clusters \citep{Bialopetravivius_2020} or for galaxy observations \citep{Almeida_2010}. We believe that this method may have been considered too computationally heavy under the intuitive approach where each LOS would be fit individually by an ANN, following a similar approach to \citet{Marshall_2009} with GA method. The difficulty is that training an ANN for each LOS with a sufficient resolution on the plane of the sky to build a map is an unrealistic solution due to the huge cumulative training dataset size that it would require and similarly the massive cumulative training time. The dataset size and computation time for a good single LOS training that are exposed later, in Section~\ref{2mass_single_los}, perfectly illustrates this point. However, {\bf we will demonstrate in the present study that it is possible to design an ANN formalism that can be trained one single time on various lines of sight simultaneously and that can still predict individual LOS extinction profiles} (Sect.\ref{los_combination}). Additionally, this type of method being capable of combining different quantity at the same time and find the correlation automatically, it should permit a combination of photometric and astrometric surveys without the necessity of a cross match (Sect.~\ref{gaia_2mass_ext_section}). We also note that, in opposition to a widespread belief, simple ANN architectures can be tweaked to provide result uncertainties in the form of a posterior probability distribution just like a Gaussian process method (Sect.~\ref{dropout_error}).

\newpage
\subsection{Objective and organization}
\label{cnn_maps_objective}

The aim of this second part of the manuscript (Part II)  is {\bf to propose an ANN architecture that is capable of sharing information from various lines of sight in a single training and that can be used to predict large extinction maps}. For this we extensively describe a more advanced ANN formalism that is based on the redundancy of information when using images as input, namely Convolutional Neural Networks. We then describe how it can be used to reconstruct extinction profiles using CMDs from the Besançon Galaxy Model and from observational surveys. We also detail the construction of the training dataset, which requires several precautions that has considerable impact on the prediction capability. We analyze the effect of several properties of the network on the prediction quality. We finally use this formalism to predict extinction maps for a portion of $45^\circ$ of the Milky Way disk using both a 2MASS only dataset, and a {\bf 2MASS plus Gaia DR2 dataset without cross match}.\\

Like for the previous session we emphasize that the results of the following part will soon be published in \citet{cornu_montillaud_20b} in the form of a short letter to the Astronomy and Astrophysics journal. The present manuscript provides a large amount of additional material and analysis.

\clearpage
\section{Convolutional Neural Networks}
\label{cnn_global_section}

In this section we describe several additions to the ANN formalism introduced in Section~\ref{global_ann_section} which can be used to construct much more complex network architectures and that can efficiently process images. We will first describe how classical neural networks can be used on images and what are the corresponding limitations. Then, we will define the convolution operation and the associated convolutional layers and explain their training procedure. Finally, we will discuss the construction of deep ANN architectures along with descriptions of various necessary parameters.

\etocsettocstyle{\subsubsection*{\vspace{-1cm}}}{}
\localtableofcontents

\subsection{The image processing impulse}
\label{image_process_section}
	
	Machine Learning is historically tightly related to the field of signal processing. Similar methods are used in both fields and they regularly influence each other. Image processing is certainly the strongest bond between the two, with a long history in signal processing and with most modern corresponding applications being made using ML. In this section we focus on describing how to use images as input for ANN.
	
	\subsubsection{Spatially coherent information}
	\label{spatial_coherence}

\begin{figure}[!t]
	\centering
	\includegraphics[width=0.7\hsize]{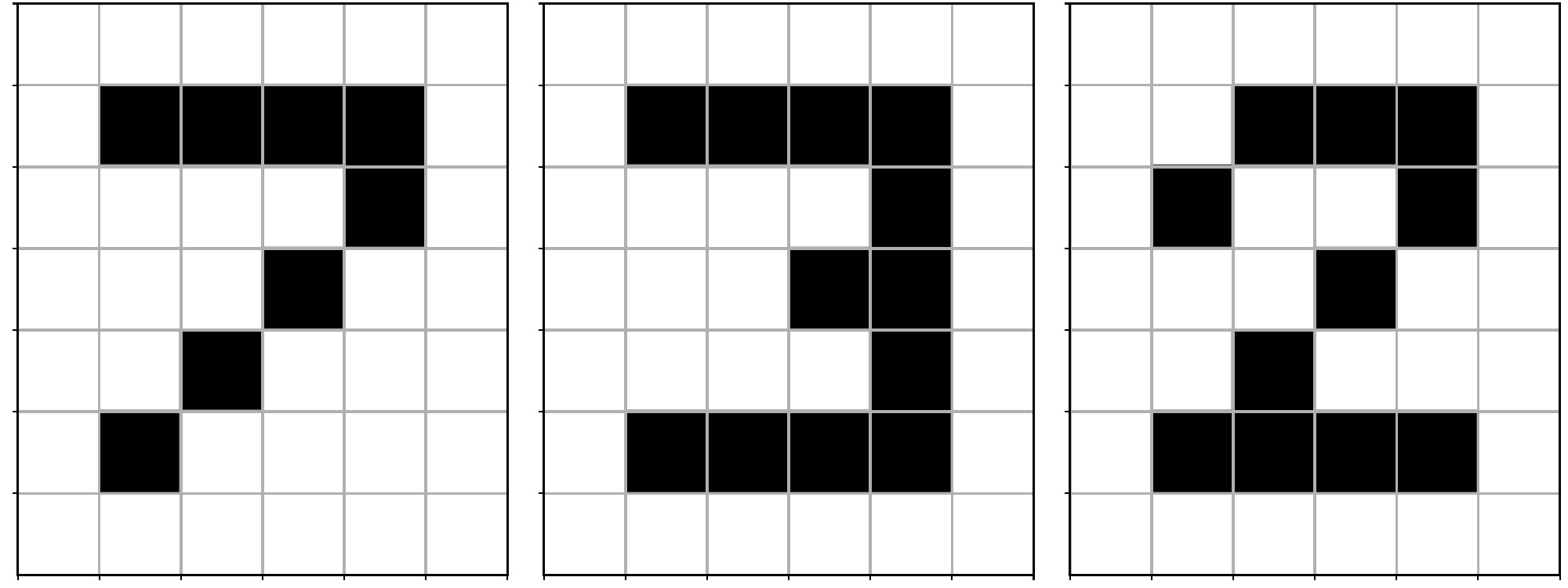}
	\caption[Simple digit image $6\times 7$]{Simple digit representation as a $6\times 7$ binary pixel image.}
	\label{im_proc_numbers}
	\end{figure}

	As we discussed in the GPU Section~\ref{matrix_formal}, images can be considered as two dimensional arrays of pixels. Each pixel contains at least one value that is often an integer encoded using a 8 bit format (0-255) and more rarely up to 16 or 32 bit. To obtain colored images, at least 3 of these 8 bit arrays must be superimposed, with each layer corresponding to a specific color intensity in each pixel often following the Red-Green-Blue (RGB) encoding. Fundamentally, an image is a decomposition of an information, for example a physical object, captured into flat ``2D'' discrete representation. Image recognition performs the opposite operation, that is to find coherent information in an array of pixel and associate it to a more abstract object that the image represents. For example, Figure~\ref{im_proc_numbers} represent $6\times 7$ binary pixel representations of digits, from which one may want to identify the digit that is represented. This is a typical image classification example that will be discussed for a more concrete example in Section~\ref{mnist_example}.\\

	\begin{figure}[!t]
	\centering
	\includegraphics[width=0.55\hsize]{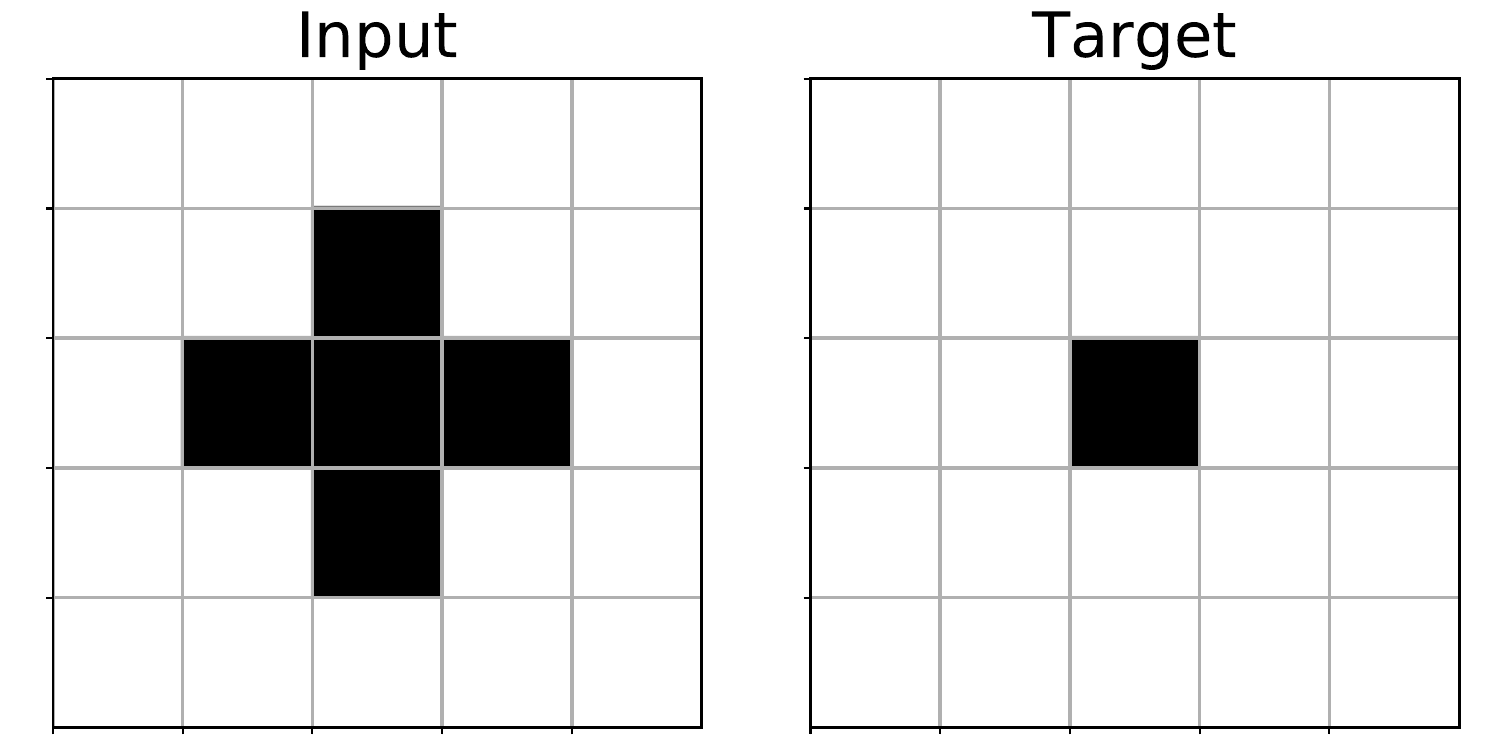}
	\caption[Cross pattern localization example]{Representation of a simple cross pattern on a $5\times 5$ image as input, and the corresponding localization prediction on an equivalent size output image.}
	\label{im_proc_loc}
	\end{figure}
	
	\begin{figure}[!t]
	\vspace{0.3cm}
	\centering
	\includegraphics[width=0.97\hsize]{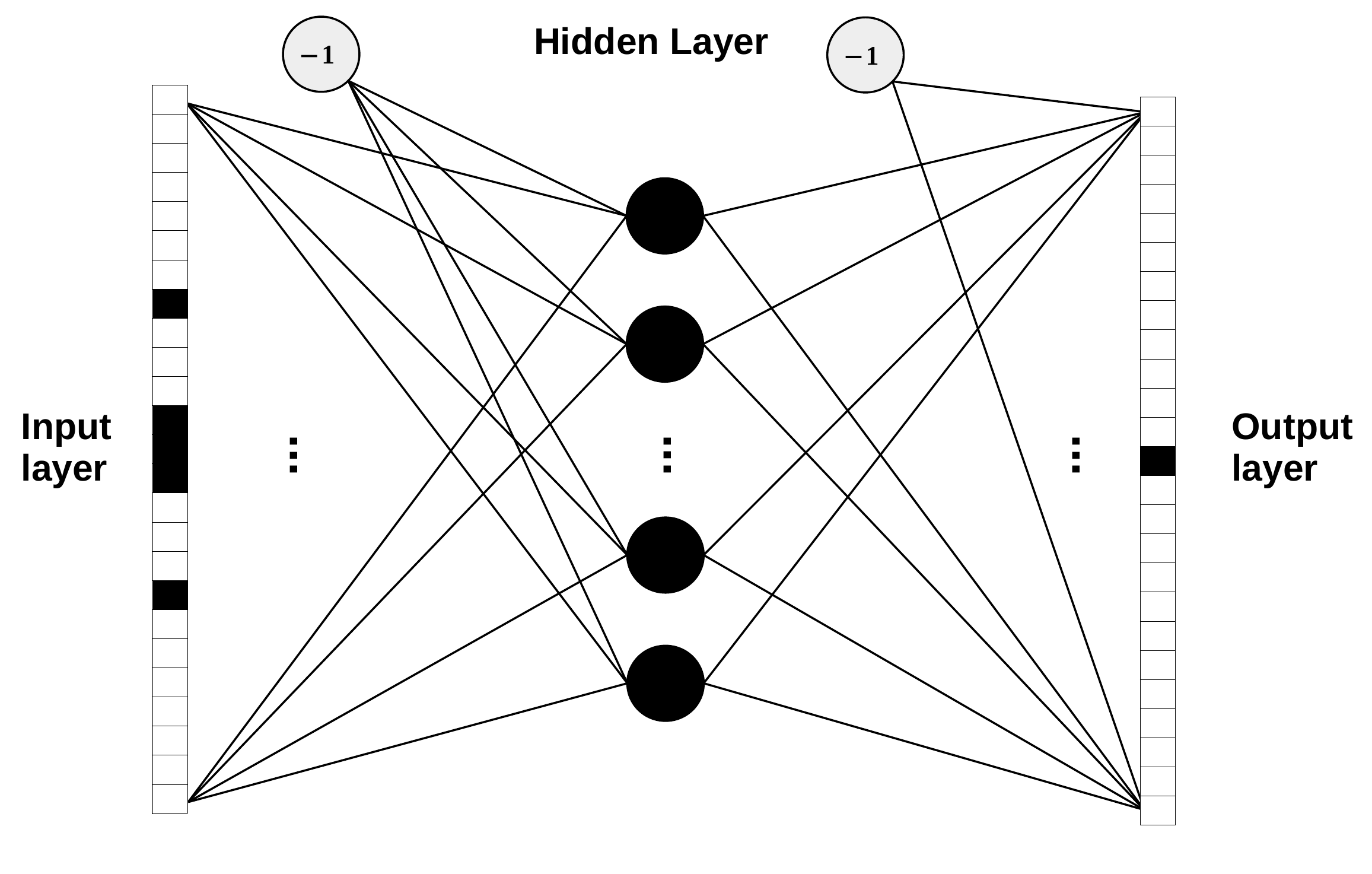}
	\caption[FC network for the cross pattern localization example]{Fully connected network corresponding to the cross pattern identification and localization problem. The images are flattened and each pixel is connected to each neuron of the hidden layer acting as a feature. The pixels of the output layer are considered neurons.}
	\label{im_proc_illustr_net}
	\end{figure}

\begin{figure}[!t]
	\centering
	\includegraphics[width=0.9\hsize]{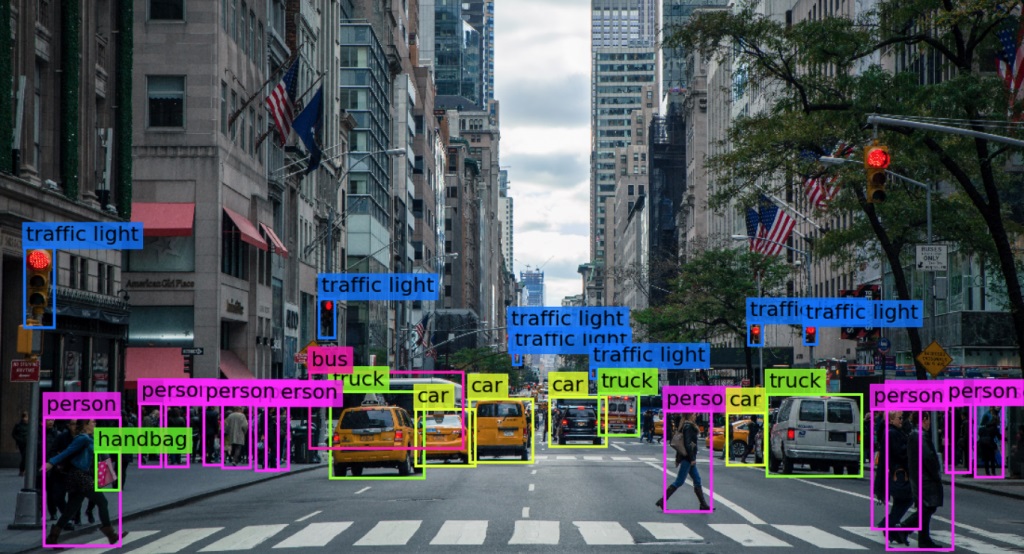}
	\caption[Computer vision example for autonomous vehicle]{Image of the prediction of a computer vision deep learning algorithm for autonomous vehicle. Objects are classified based on a list of useful elements for driving and are localized into a boundary box. {\it Image credits \href{https://www.lebigdata.fr/computer-vision-definition}{www.lebigdata.fr}.}}
	\label{computer-vision}
	\end{figure}

One key element to the extraction of image information is that it is spatially coherent. To illustrate why spatial coherence can be difficult, we first show how the classical ANN algorithm described in Section~\ref{global_ann_section} can be used to compute information using images as input. We consider a network for pattern recognition and localization. An image, that may or may not contain a specific pattern, is presented to this network. The algorithm task is then to predict the position of the pattern in the image when it is present. This application is illustrated by the Figure~\ref{im_proc_loc} where the pattern to identify is a $3\times 3$ binary pixel cross and the expected output is an identically sized image that contains a positive value at the pixel that represents the center of the cross. Although we chose a single and very simple pattern, this application illustrates many modern ML uses in computer vision, where the objective is to find if an image contains a specific object from a given set (classification) and to predict its location or a boundary box around it. This is the typical example that is shown for self driving vehicles in Figure~\ref{computer-vision}.\\

Using this representation, the easiest approach to connect an input image to an ANN is to consider that each pixel is an individual feature. An input vector can then be constructed, with the same size as the image. In this specific example the output vector has the same dimension. Figure~\ref{im_proc_illustr_net} shows a corresponding single hidden layer network that takes as input and output the corresponding flatten images. This network is actually suitable to perform this task. In order to ease the comparison with other network architectures, we will now refer to such a network arrangement as a "fully connected" or "dense" network, and we will also use these formulation as adjectives for a "fully connected layer" or "dense-layer" to depict a MLP like weight connection.\\

The first striking limitation with this representation is that the number of weights is very large. It has a cost in terms of computing performance and requires a lot of examples to be properly constrained. The second limitation is that, in order to train the weights of each pixel, examples of a cross on each pixel must be contained in the training set. It means that such a network on this specific case will need to be provided all the possible positions as example, and is therefore inefficient to generalize the problem.
	
	\subsubsection{Information redundancy: pattern recognition}

	\begin{figure}[!t]
	\hspace{-1.1cm}
	\begin{minipage}{1.15\textwidth}
	\centering
	\includegraphics[width=1.0\hsize]{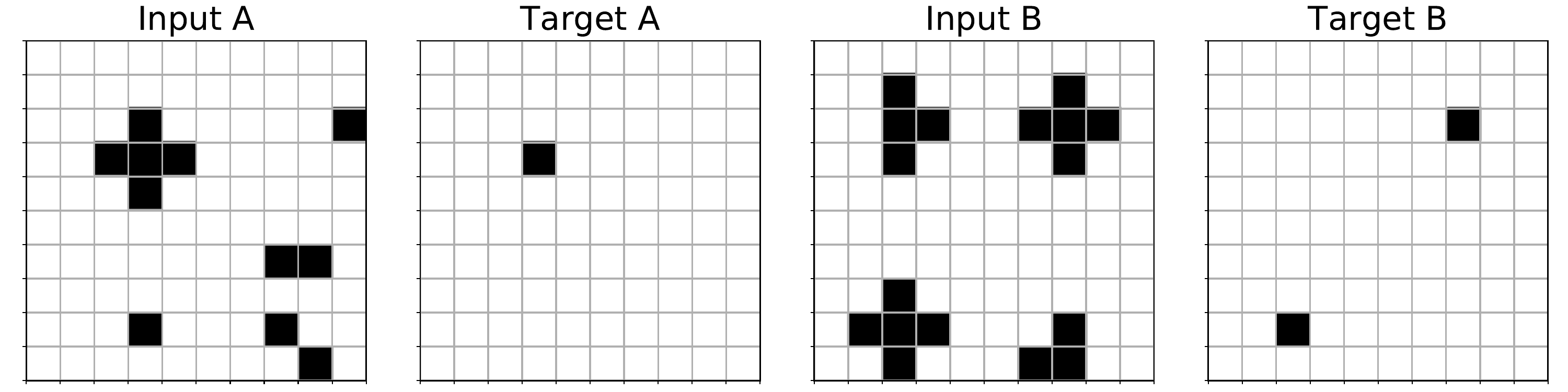}
	\end{minipage}
	\caption[Multiple cross pattern detection in a large image with noise]{Two examples, A and B, of input and target images for cross pattern localization. Multiple objects in the same $10\times 10$ image are permitted and the complexity is raised by the addition of noise or other patterns.}
	\label{im_proc_loc_mult}
	\end{figure}
	
	We now consider an example where the image is larger and that it can contain multiple times the looked-up pattern at the same time. This case is illustrated by Figure~\ref{im_proc_loc_mult} where we represented two different $10\times 10$ cases in which we have added some irrelevant input pixels that could be noise or non looked-for patterns to make the example more realistic. The images in this case can be plugged as before to a fully connected network. This time, considering that there are irrelevant patterns increasing the complexity of the problem, the network should be able to generalize information. Indeed, even if it is mandatory to provide at least one example of a searched pattern at each point, it is not necessary to add examples that correspond to all the possible irrelevant information combination on all other pixels. A fully connected ANN trained on this problem might be more time efficient to predict the solution than a naive algorithm that searches for the presence of a cross at each pixel position in the image. A similar illustration on differentiating T from C letters representations independently of translation and rotation using a MLP is presented in \citet{rumelhart_learning_1986}. \\
	

However, an instinctive reaction to this problem is to notice that there is only one pattern to look for, and that learning this pattern once and for all, and then search for it at different places, would be much more efficient than learning how to react to this pattern at every possible positions in the image. This is because one, as a human being, is sensitive to redundancy of the information. Then it should be possible to construct a network architecture that is able to perform the same task. The objective is to build an operation that is capable of detecting a pattern in a way that is invariant by translation in the image. This can be done by creating a unique artificial receptive field that can be applied at several places on the image, strongly reducing the number of parameters that are needed by sharing them over the full image. In practice this is done using an operation that is called a convolution and that relies on a filter (also called a kernel).

\newpage
\subsubsection{Convolution filter}
\label{conv_filter}

\begin{figure}[!t]
	\hspace{-0.9cm}
	\begin{minipage}{1.0\textwidth}
	\centering
	\includegraphics[width=1.0\hsize]{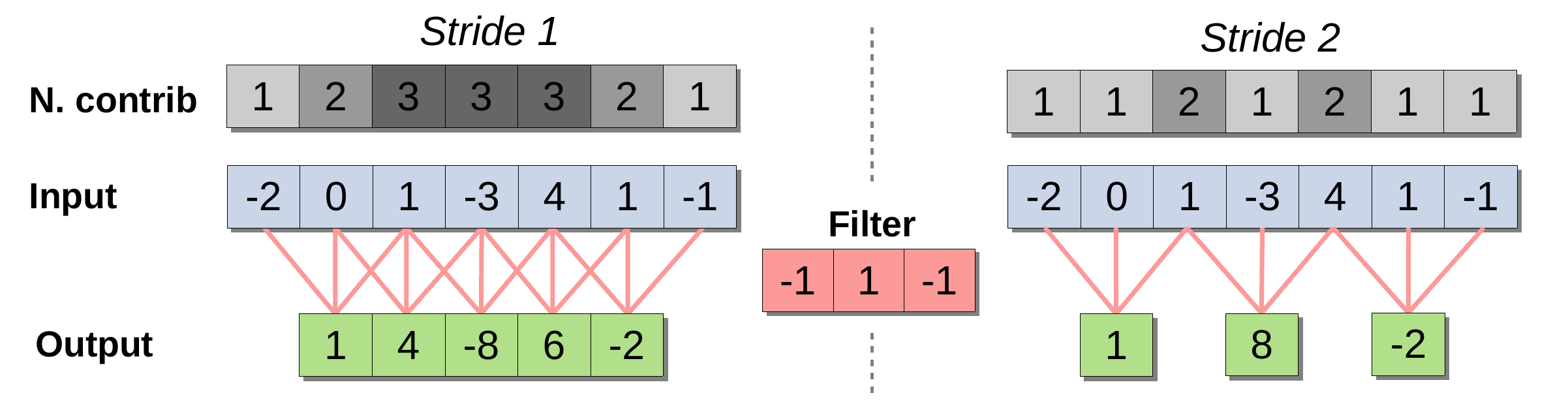}
	\end{minipage}
	\caption[1D convolution example]{Illustration of a 1D convolution operation on a 7-element input vector (in blue) using a 3-element filter (in red). The output result is in green and the grayscale table represents the number of times each input was used by the operation. Two examples are given for $S=1$ (left) and $S=2$ (right) on the same input vector and using the same filter.}
	\label{im_proc_1d_conv}
	\end{figure}

A convolution operation consists in the application of a filter to an image through a decomposition in sub-regions. For this, the filter is considered as a set of numerical values that has the size of the wanted receptive field. The values of this filter are then multiplied element-wise to a subset of pixels and the results are summed to obtain a single value. This operation can be performed at several places in the image to produce an output vector that contains all the corresponding results. Usually, the filter is applied at regular intervals on the input vector with a shift in pixels between each application that is called the stride $S$. We illustrate this operation in a one dimensional example in Figure~\ref{im_proc_1d_conv} with a 7-pixel input vector (here with positive or negative integer values) that is convolved using a 3-pixel filter. This figure shows two examples with strides $S=1$ and $S=2$ leading to output vectors of 5 and 3 elements, respectively. We stress that, in the $S=1$ case, each input value is used between 1 and 3 times depending on the filter overlap and with a continuous contribution pattern, while in the $S=2$ case, each input is only used 1 or 2 times with a periodic overlap pattern. While such an overlap pattern from each input pixel is expected in this operation, it could lead to concerns in specific cases. To better visualize this property of the convolution operation, we included a representation of the contribution number of each pixel in Figure~\ref{im_proc_1d_conv} and in subsequent figures.\\
	
For images, the information is spatially coherent in two dimensions, therefore the convolution operation is performed using a 2D filter that "slides" over the image along one axis (say, along a line) with a shift between each application that is defined by the stride. When the end of the line is reached, the operation is repeated for another line according to the stride. This way the filter is applied regularly in both dimensions. One side effect of the convolution is to reduce the size of the image. Although this can be useful to reduce the dimensionality of the problem, in some cases it is preferable to conserve the image dimension by adding a zero-padding (often just refered to as padding) around the input image. It results in the following relation between input and output dimensions:
\begin{align}
	& w_{out} = \frac{w_{in} - f_s + 2P}{S} + 1
	& h_{out} = \frac{h_{in} - f_s + 2P}{S} + 1
	\label{width_height_relation}
\end{align}
where $w$ and $h$ denote the input and output widths and heights, respectively, $f_s$ is the filter size, $P$ is the padding, and $S$ is the stride, considering that the last three quantities have the same values for both axes. \\

\begin{figure}[!t]
	\hspace{-1.5cm}
	\begin{minipage}{1.20\textwidth}
	\centering
	\includegraphics[width=\hsize]{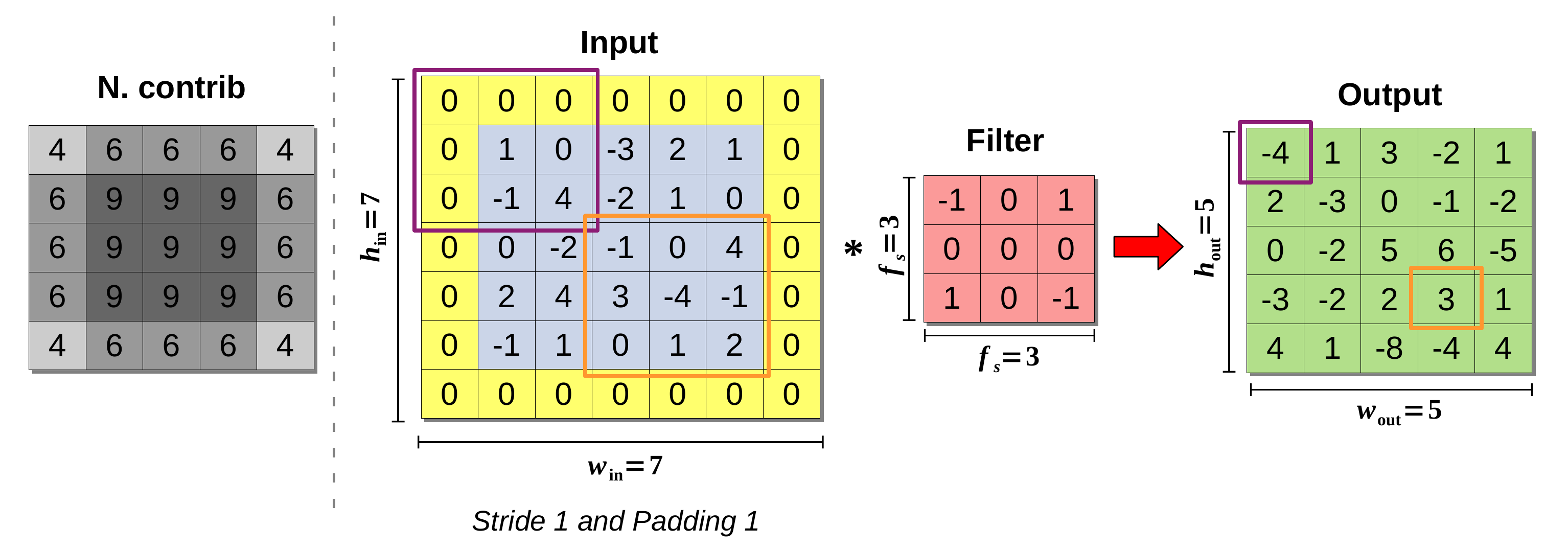}
	\end{minipage}
	\caption[2D convolution example]{Illustration of a 2D convolution operation on a $5 \times 5$ input table (in blue) with an added zero padding $P=1$ represented in yellow, using a $f_s=3$ filter (in red). The resulting $5\times 5$ table is in green and the grayscale table is the number of contributions from each input element. Purple and orange squares highlight two specific sub-region products and their corresponding output pixels.}
	\label{im_proc_2d_conv}
	\end{figure}

We illustrate this 2D convolution operation in Figure~\ref{im_proc_2d_conv} using a $7\times 7$ input 2D table and a filter with $f_s=3$, $S= 1$ and $P=1$, where the $\bm *$ symbol represents the convolution operation. To ease the understanding of the operation, the figure also presents two colored squares that are used to highlight two specific sub-regions that are individually multiplied by the filter and result in the corresponding colored elements in the output table \footnote{A very nice animation of the convolution operation on a randomly generated input table can be found in the excellent ANN online course of the Stanford university at \href{https://cs231n.github.io/convolutional-networks/}{cs231n.github.io}}.\\

\afterpage{
	\begin{sidewaysfigure}
	\centering
	\begin{minipage}[t]{0.18\textwidth}
		\centering
		{\bf \large No filter}
	\end{minipage}
	\begin{minipage}[t]{0.18\textwidth}
		\centering
		{\bf \large Sharpen}
		\begin{equation*}
		\begin{bmatrix*}[r]
		0 & -1 & 0 \\
		-1 & 5 & -1 \\
		0 & -1 & 0
		\end{bmatrix*}
		\end{equation*}
		\vspace{-0.1cm}
	\end{minipage}
	\begin{minipage}[t]{0.18\textwidth}
		\centering
		{\bf \large Gaussian blur}
		\begin{equation*}
		\frac{1}{16}\begin{bmatrix*}[r]
		1 & 2 & 1 \\
		2 & 4 & 2 \\
		1 & 2 & 1
		\end{bmatrix*}
		\end{equation*}
		\vspace{-0.1cm}
	\end{minipage}
	\begin{minipage}[t]{0.18\textwidth}
		\centering
		{\bf \large Edge detector}
		\begin{equation*}
		\begin{bmatrix*}[r]
		-1 & -1 & -1 \\
		-1 &  8 & -1 \\
		-1 & -1 & -1
		\end{bmatrix*}
		\end{equation*}
		\vspace{-0.1cm}
	\end{minipage}
	\begin{minipage}[t]{0.18\textwidth}
		\centering
		{\bf \large Axis elevation}
		\begin{equation*}
		\begin{bmatrix*}[r]
		-1 & -1 & -1 \\
		 0 &  0 &  0 \\
		 1 &  1 &  1
		\end{bmatrix*}
		\end{equation*}
		\vspace{-0.1cm}
	\end{minipage}\\
	\begin{subfigure}[t]{0.18\textwidth}
	\includegraphics[width=\textwidth]{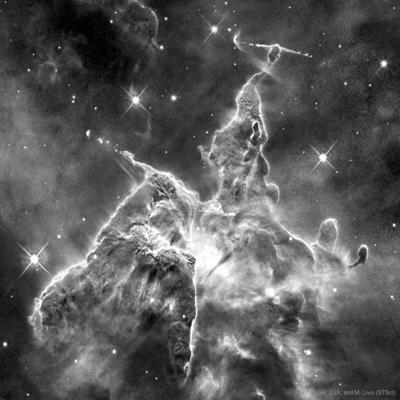}
	\end{subfigure}
	\begin{subfigure}[t]{0.18\textwidth}
	\includegraphics[width=\textwidth]{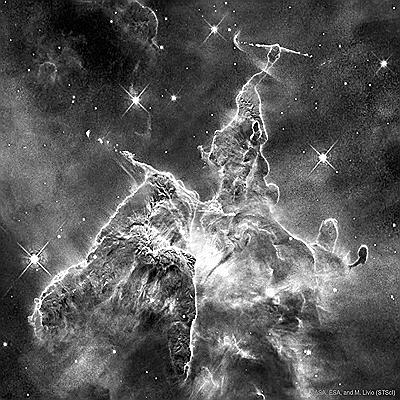}
	\end{subfigure}
	\begin{subfigure}[t]{0.18\textwidth}
	\includegraphics[width=\textwidth]{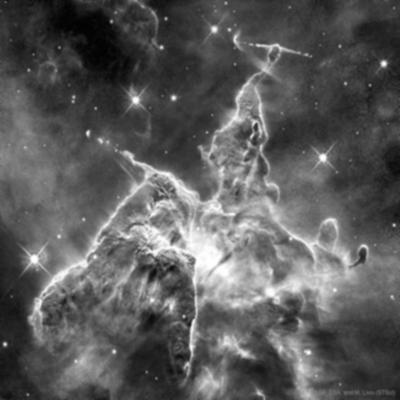}
	\end{subfigure}
	\begin{subfigure}[t]{0.18\textwidth}
	\includegraphics[width=\textwidth]{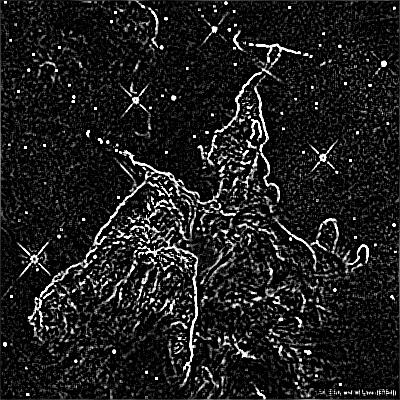}
	\end{subfigure}
	\begin{subfigure}[t]{0.18\textwidth}
	\includegraphics[width=\textwidth]{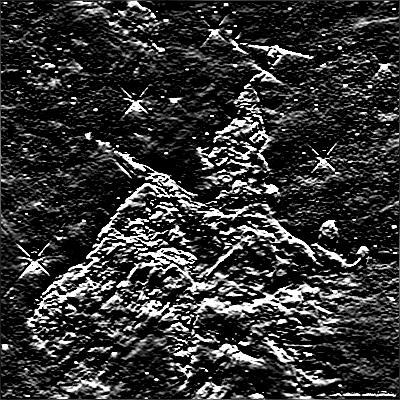}
	\end{subfigure}\\
	\begin{subfigure}[t]{0.18\textwidth}
	\includegraphics[width=\textwidth]{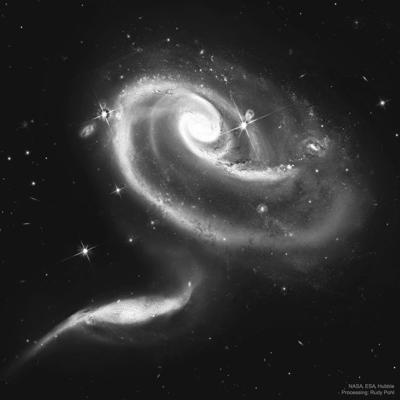}
	\end{subfigure}
	\begin{subfigure}[t]{0.18\textwidth}
	\includegraphics[width=\textwidth]{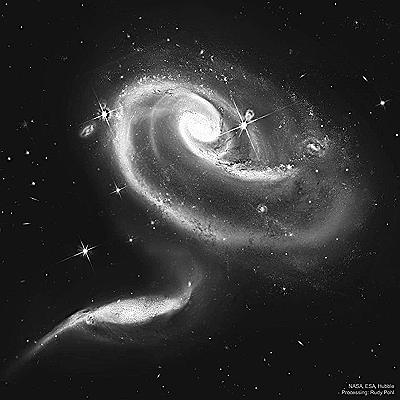}
	\end{subfigure}
	\begin{subfigure}[t]{0.18\textwidth}
	\includegraphics[width=\textwidth]{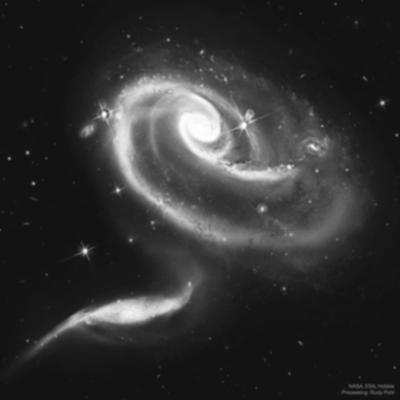}
	\end{subfigure}
	\begin{subfigure}[t]{0.18\textwidth}
	\includegraphics[width=\textwidth]{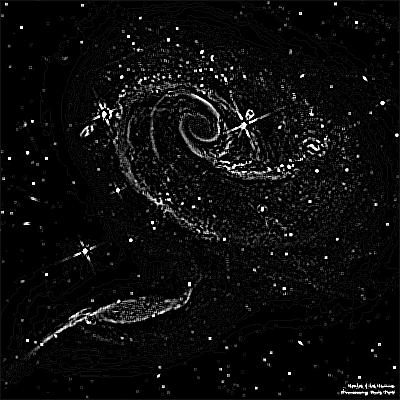}
	\end{subfigure}
	\begin{subfigure}[t]{0.18\textwidth}
	\includegraphics[width=\textwidth]{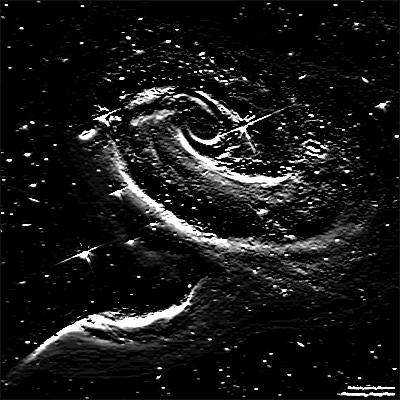}
	\end{subfigure}\\
	\begin{subfigure}[t]{0.18\textwidth}
	\includegraphics[width=\textwidth]{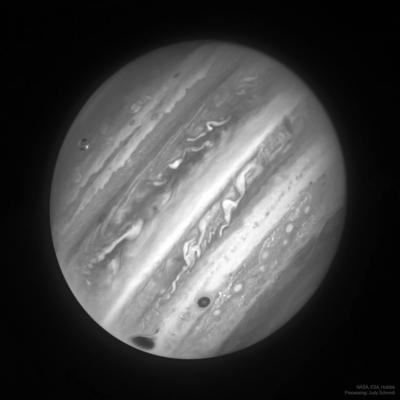}
	\end{subfigure}
	\begin{subfigure}[t]{0.18\textwidth}
	\includegraphics[width=\textwidth]{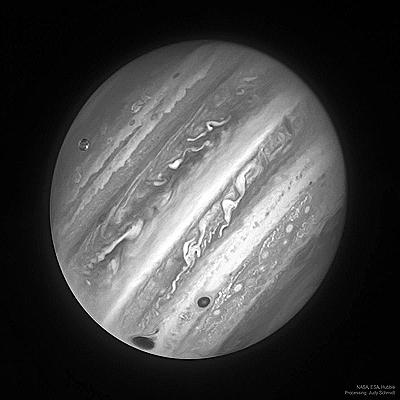}
	\end{subfigure}
	\begin{subfigure}[t]{0.18\textwidth}
	\includegraphics[width=\textwidth]{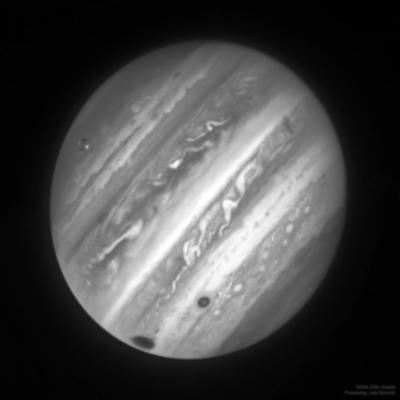}
	\end{subfigure}
	\begin{subfigure}[t]{0.18\textwidth}
	\includegraphics[width=\textwidth]{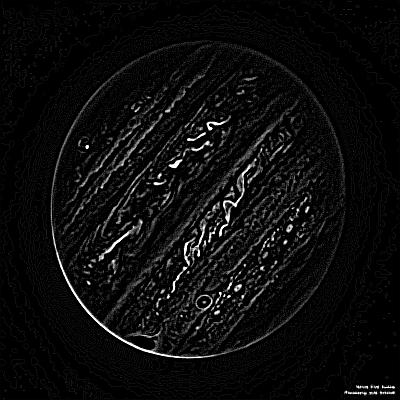}
	\end{subfigure}
	\begin{subfigure}[t]{0.18\textwidth}
	\includegraphics[width=\textwidth]{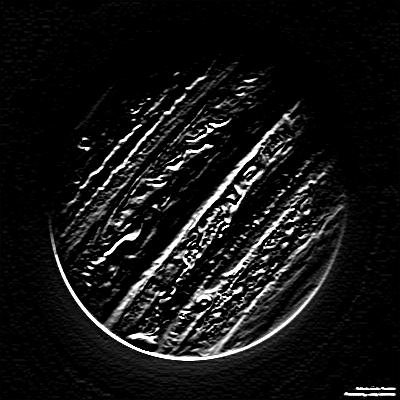}
	\end{subfigure}
	\caption[Common filters applied to a selection of astronomical images]{Convolution results using common filters applied to a selection of three astronomical images from the Hubble Space Telescope (HST) representing objects of very different physical size. The base images on the left are the red channels from the original images. {\it Top}: HST image of the \href{https://apod.nasa.gov/apod/ap170702.html}{Carina Nebula}. {\it Middle}: HST image of \href{https://apod.nasa.gov/apod/ap191120.html}{UGC 1810} along with the smallest UGC 1813 galaxies. {\it Bottom}: HST image of \href{https://apod.nasa.gov/apod/ap181016.html}{Jupiter}.}
	\label{conv_filter_examples}
\end{sidewaysfigure}
}
	
Consequently, the convolution operation can be used to detect a specific pattern at any place in an image by using a filter that is a replica of the pattern. This property is very convenient for the example of the previous section, which is automatically solved using only a convolution with the appropriate padding to conserve the image size in the output. However, for more complex images the convolution can be seen as a specific image processing. The most common convolution operations are the ones that blur an image or apply an interpolation in order to resize the image. We illustrate in Figure~\ref{conv_filter_examples} the effect of standard convolution filters on three different astronomical images that represent very different physical scales. While the sharpen and blur operations are common in every day life, the edge detector and the axis elevation ones demonstrate how convolution filters can be used to extract patterns in an image. Still, one convolution would often be insufficient to solve a complexe problem directly, leading to a necessary combination of several convolution operations.\\
	
	\newpage
	\subsubsection{Convolutional layer}
	
	\begin{figure*}[!t]
	\centering
	\begin{subfigure}[!t]{0.80\textwidth}
	\centering
	\caption*{\large {\bf Input images}}
	\includegraphics[width=\hsize]{images/im_proc_numbers.pdf}
	\end{subfigure}
	\vspace{0.6cm}\\
	\begin{subfigure}[!t]{0.6\textwidth}
	\centering
	\caption*{\large {\bf Filters}}
	\includegraphics[width=\hsize]{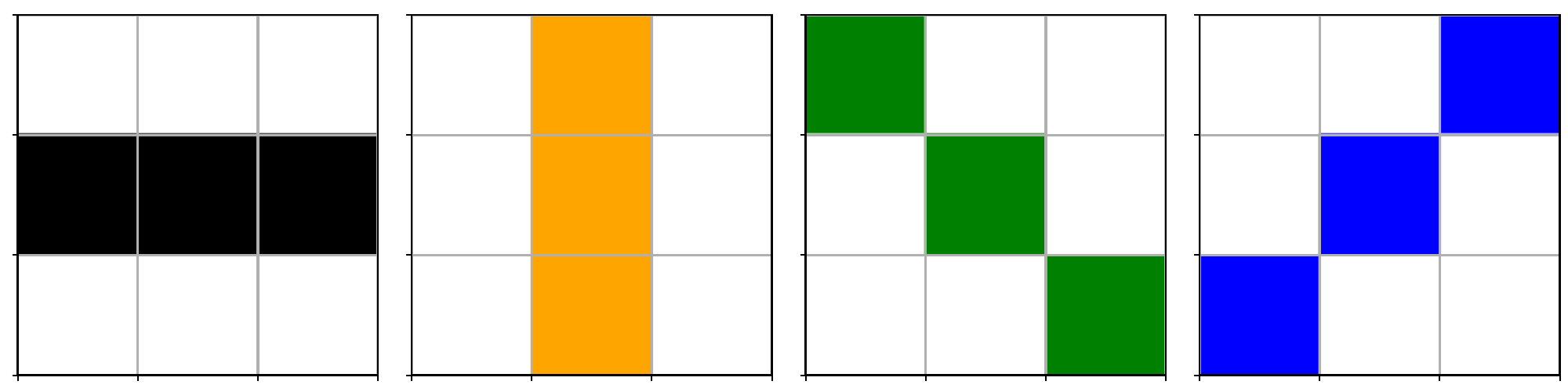}
	\end{subfigure}
	\vspace{0.6cm}\\
	\begin{subfigure}[!t]{\textwidth}
	\centering
	\caption*{\large {\bf Superimposed output images}}
	\includegraphics[width=0.70\hsize]{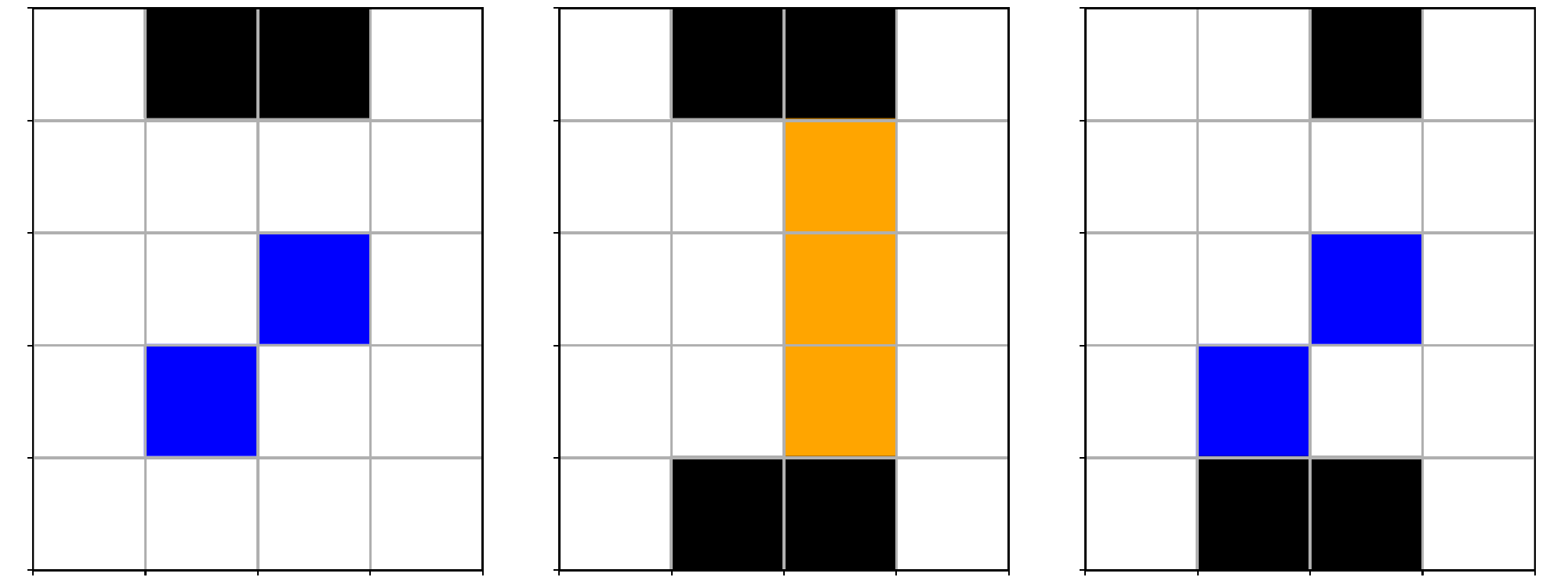}
	\end{subfigure}
	\caption[Multiple filters for simple digit recognition]{Multiple simple $3 \times 3$ filters applied to digit recognition on three $6\times 7$ images. Each filter can extract patterns at several point in each image and can be used for different digits. A filter predict a positive output if there is a full match between its activated pixels and the activated pixels of each input sub-region. {\it Top}: the three input images. {\it Middle}: the colored filters. {\it Bottom}: output from all filters are stacked and color-coded in one $4\times 5$ output image for each input example.}
	\label{mult_filt}
	\end{figure*}
	
	There are two main cases that justify the use of several convolution filters. The first is when one filter alone, that directly represents the looked up pattern, is too large compared to the size of the image. The second is when multiple patterns are looked for. In such cases it is more efficient to use several small convolution filters that represent sub-parts of the pattern and that can be shared in the case of multiple-object identification. If we consider the simple digit-identification example from Section~\ref{spatial_coherence}, it is manifest that they can be decomposed into much easier redundant parts. We illustrate this case in Figure~\ref{mult_filt}, where we proposed 4 simple filters that are just different continuous and aligned pixel layouts. When used in a convolution operation, each of these filters will produce an individual output image. In this case we consider that there must be a perfect match between the filters colored pixel and the pixel of the input image at given sub-region for the corresponding output to be activated. If we consider that blank pixels are 0 and colored ones are 1, it means that there is a threshold of 3 to be reached for the output to be set to 1 as well. In this figure we used different colors for each filter and stacked up the color-coded output images in order to ease the representation. We note that there is no overlap in this specific case, but that nothing prevents multiple filters to produce the same output at a given pixel location. The figure illustrates that, by using adequate filters, it is possible to construct a smaller and simpler space that represents all digits differently.\\

Using the defined convolution operation, it is now possible to link it to our ANN formalism. There is a direct analogy between the dot product and sum of the result that occurs in the convolution operation and the weighted sum of input in a neuron operation that we described in Section~\ref{neuron_math_model}. Indeed, the sub-region of the image, when flattened, can be considered as our input vector $X_i$, the filter considered as the weights $W_i$ also considering it flattened, and the result is summed into a $h$ quantity. The only missing part for a convolution output pixel to act as a neuron is the activation function along with a bias value that are very simple to add to the operation. The choice of an appropriate activation function that works well with the convolution operation on the weights is discussed in Section \ref{sect_relu}. The convolution operation is then a suitable approach to share a small group of weights over the all image, strongly reducing the number of parameters that need to be constrained. This formalism of neural receptive field is long known and was already introduced in a very similar fashion by \citet{rumelhart_parallel_1986}.\\

\begin{figure}[!t]
\vspace{+0.6cm}
	\hspace{-2.1cm}
	\begin{minipage}{1.25\textwidth}
	\centering
	\includegraphics[width=\hsize]{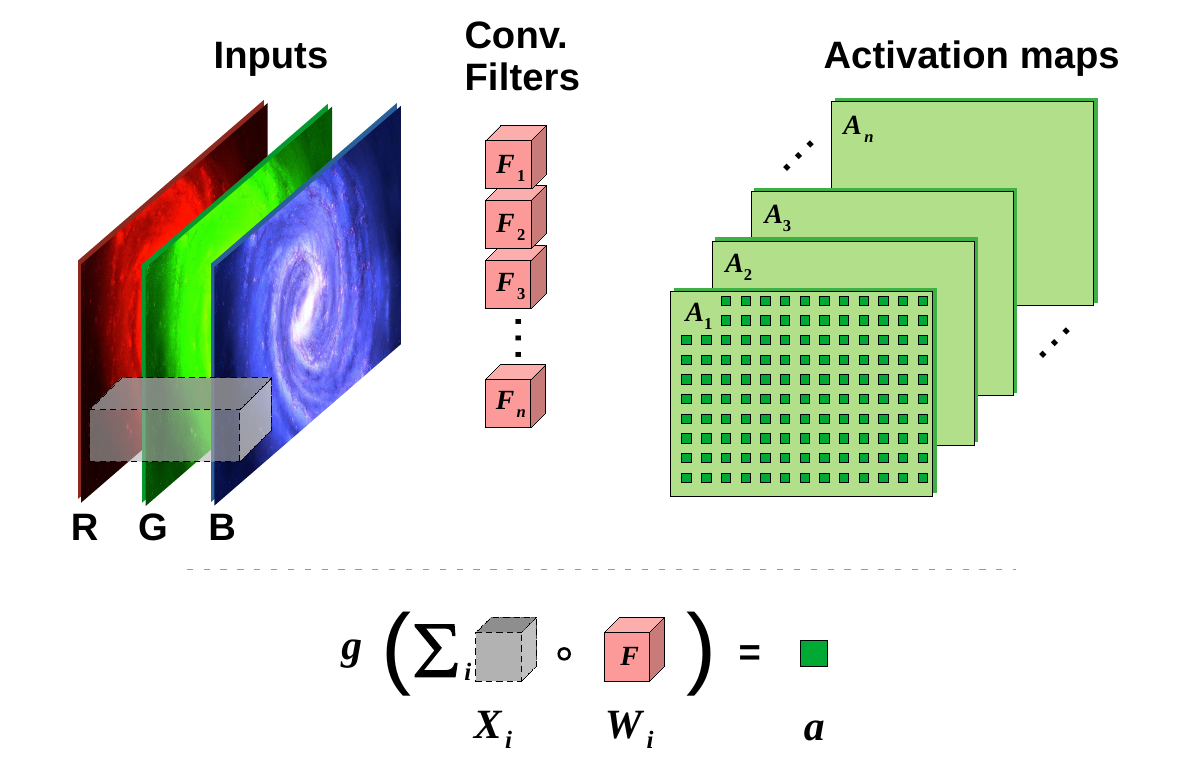}
	\end{minipage}
	\caption[Detailed convolutional layer]{Detailed representation of a convolutional layer. The input image is made of $d=3$ layers (RGB), the filters (in light red) are cubes of dimension $f_s\times f_s \times d$. The transparent gray box denotes the region of the image that is multiplied by the filter to obtain a specific neuron activation. The activation maps from each filter are in green, with the first one showing each activated neuron pixel as a dark green square.}
	\label{convolutional_layer_im}
	\vspace{0.7cm}
	\end{figure}

To be usable for real image processing, this model of weight filters still has to cover the case of multiple color channels in an image, which can be considered as an added depth $d$ to the input, resulting in an input volume of $w_{in} \times h_{in} \times d$. This can be done by considering that the weight filter is a 3D table that has a width and height equivalent to the filter size $f_s$ and an additional depth $d$ that corresponds to the number of input depth channels. This way the weight filter combines spatial information that comes from all the input depth channels at the same place in the images and still only moves in a 2D space following the stride parameter. We note that this 3D filter still sums its $ f_s \times f_s \times d$ contributions into only one result for each area of the images, and therefore produces a 2D output. To stick to the network representation, each pixel of this output image goes through an activation function, which results in a so-called activation map. The input images can be convolved by a certain number $n_f$ of these 3D weight filters, which results in a number of activation maps equal to the number of filters, with a size that follows the relations of Equation~\ref{width_height_relation}. The output volume is then of $w_{out} \times h_{out} \times n_f$. This complete operation constitutes what is called a \textbf{convolutional layer} and is illustrated in Figure~\ref{convolutional_layer_im}.\\

\vspace{0.5cm}
The resulting activation maps that constitute a re-arrangement and often a dimensional reduction of the input images can then be fully connected to an MLP architecture with each pixel of each activation map considered as a feature. If we consider an example with input images sized as $w_{in} = 100$, $h_{in} = 100$ et $d = 3$, a full connection to an $n = 20$ hidden MLP layer would require $6\times 10^5$ weights. Using the same input images and $n_f = 6$ filters of size $f_s=4$ that are used with a stride of $S=3$, it is only $48$ weights per filter or $288$ weights for all the filters of the convolutional layer. This convolution produces a $32\times 32\times 6$ activation map volume that requires "only" $1.23 \times 10^5$ weights to be fully connected to the same $n = 20$ MLP layer.\\

\newpage
This architecture already presents several advantages: (i) the exclusion of irrelevant patterns, (ii) the dimension reduction that occurs with a large stride and filter size, and (iii) a huge reduction of the number of weights compared to a fully connected layer thanks to the possibility to share the small filters that are applied over the whole image . While these additions are already interesting enough to justify such an architecture, the true interest is the possibility to easily stack such layers to extract more complex representations of the input, which is covered in detail in Section~\ref{convolution_stacking}.
	
	\newpage
	\subsubsection{A simpler activation function : the rectified linear unit}
	\label{sect_relu}
	
	\vspace{-0.1cm}
	While a convolution layer can technically be built using any activation function, some of them proved to be more efficient than others. The main issue is that adding a convolutional layer over a fully connected network, and moreover stacking several convolutional layers as we will discuss in the next section, add depth to the network which can result in a vanishing gradient issue (Sect.~\ref{mlp_backprop}). Additionally, since each output pixel is considered as a neuron, and therefore is activated, the computational efficiency of the activation function becomes more important when working with large images.\\
	
\begin{figure}[!t]
	\begin{subfigure}[!t]{0.49\textwidth}
	\centering
	\includegraphics[width=1.0\hsize]{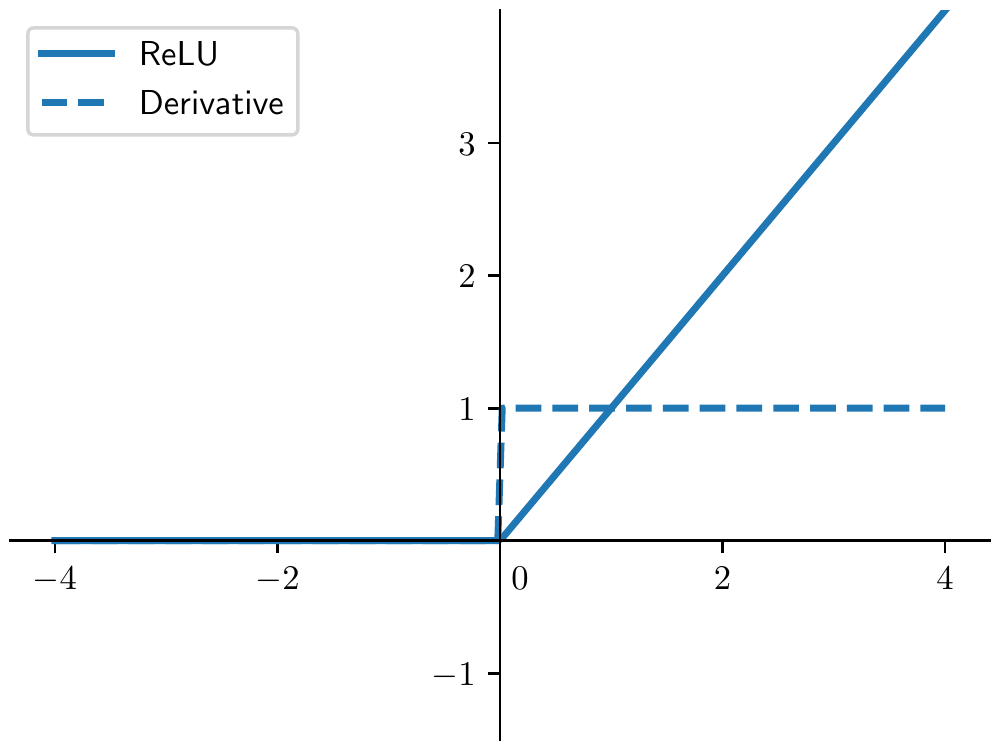}
	\end{subfigure}
	\begin{subfigure}[!t]{0.49\textwidth}
	\centering
	\includegraphics[width=1.0\hsize]{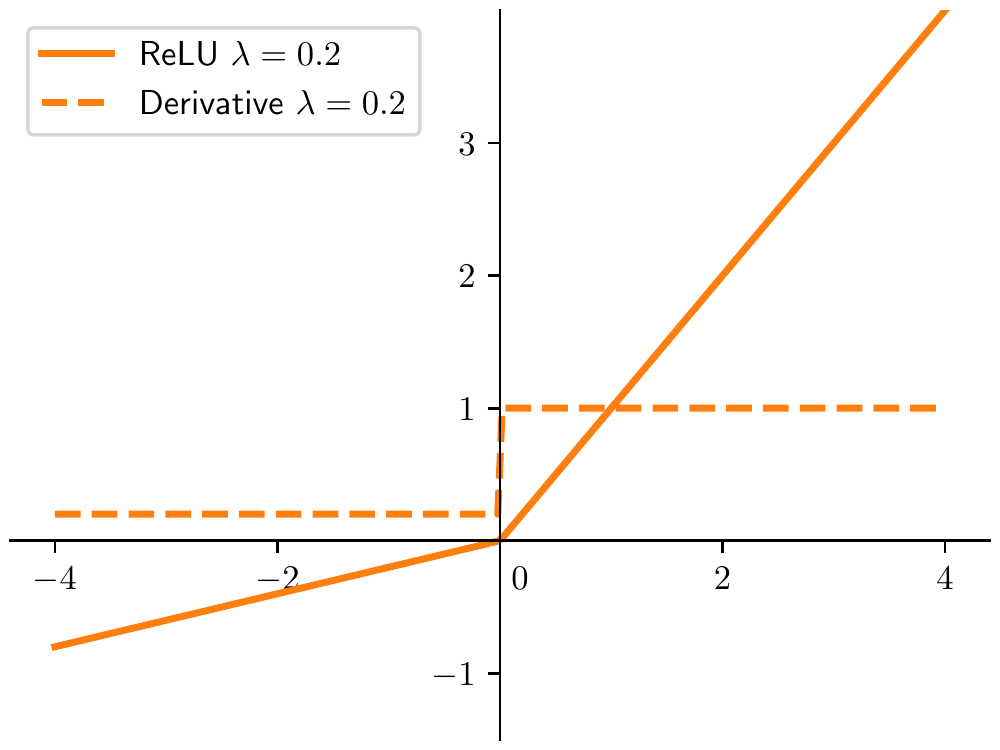}
	\end{subfigure}
	\caption[Illustration of a ReLU activation]{Illustration of the ReLU activation function as plain line and its derivative as dashed line. {\it Left}: classical ReLU activation. {\it Right}: leaky ReLU activation with $\lambda = 0.2$.}
	\label{relu_fig}
	\end{figure}

Because it is able to solve this problem, the most widely used activation function in deep network with convolution is the Rectified Linear Unit activation, or ReLU \citep{nair_2010}. This function is simply linear for any input value above zero and is equal to zero in the negative input part, which is summarized as:
\begin{align}
	& a_j = g(h_j) = \begin{cases} h_j & \text{if} \quad h_j \geq 0 \\ 0 & \text{if} \quad h_j < 0 \end{cases}
	& \text{or} \hspace{1.5cm}
	& a_j = g(h_j) = \max(0,h_j)
\label{relu_activ}
\end{align}
following the same notations as Equation~\ref{eq_activ_perceptron}.
Despite its simplicity this function presents several advantages, (i) it conserves the global non linearity with two states, (ii) it is scale invariant, because it does not saturate, (iii) it is easy to compute with only a comparison and a memory affectation in the negative input case, (iv) it has a constant derivative equal to one, meaning that there is no loss of gradient information when the neuron is in its activated state, which also speeds up the learning process. We illustrate this function along with its derivative in the left frame of Figure~\ref{relu_fig}.\\

There is in fact a family of ReLU activation functions, starting with the leaky ReLU \citep{Maas_2013} that defines a leaking factor $\lambda$ for the negative input part of the function making it linear as well. While in the original paper they suggested a leaking factor of $\lambda = 0.01$ recent applications have demonstrated that other leaking factor could produce better results. This activation is summarized as:
\begin{align}
	&a_j = g(h_j) = \begin{cases} h_j & \text{if} \quad h_j \geq 0 \\ \lambda h_j & \text{if} \quad h_j < 0 \end{cases}
	& \text{or} \quad \quad
	& a_j = g(h_j) = \max(0,h_j) + \min(0, \lambda h_j).
\label{leaky_relu_activ}
\end{align} 
The leaky ReLU activation is illustrated in the right frame of Figure~\ref{relu_fig}. The idea behind this addition is that neurons that are not activated with the classical ReLU are not updated at all, which could lead to neurons that remain stuck in this state depending on the input features and weight values. Using a leaky parameter, neurons that where not responsible of a given activation can slowly get involved if they are not currently used to constrain another part of the feature space, or if they would be more useful to the present example than to the ones they where constraining before. A small leaking factor value ensures that activated ReLU remains the main responsible for the output, preserving the global propagation scheme of the network. We note that, this small propagation to non-activated neurons is sensitive to the vanishing gradient since the derivative is equal to $\lambda$. Therefore the propagation remains driven by the activated neurons in the previous layer, which is mostly a good thing as we want the global propagation to mostly follow the activated neurons path. Still, it can be an issue in the very rare case of a continuous path of never activated neurons. \\

For now, the leaky ReLU activation is our preferred solution with $\lambda$ as a tunable hyperparameter. Other ReLU activations are for example the parametric ReLU (or PReLU) that is equivalent to the leaky case but with the optimal leaking factor being learned during the training process \citep{he_delving_2015}, the randomized leaky ReLU that randomly selects a $\lambda$ value for each neuron, or the exponential linear unit (ELU) that smooth the negative input part with an exponential. Empirical performance comparison for networks with convolutional layers did not shown any significant superiority of these variants of ReLU, while they all perform decisively better than the basic ReLU \citep{xu_2015}.\\

While the ReLU activation was introduced to help convolutional networks it can efficiently be used on fully connected MLP networks. As we discussed in Section \ref{weight_decay}, the sigmoid neurons are expected to work mainly in their close-to-linear regime to help error propagation and network stability, while the non-linear part of the function should only be used when necessary. These constraints are released by the ReLU activation with a very nice error propagation even for deep networks. The trade-off is that much more ReLU neurons must be used when the boundaries to find are truly non linear. However, ReLU neurons are much simpler to constrain and we observed in our applications that less data are usually necessary to properly constrain the network, in spite of the greater number of weights than in the case of sigmoid neurons (Sect. \ref{nb_neurons}). We show an example of a fully connected layer that uses ReLU in Section~\ref{dropout_error}.

	\newpage
	\subsubsection{Stacking Convolutional layers}
	\label{convolution_stacking}
	
	Even if one convolutional layer can already identify a lot of pattern if provided with many filters, it remains limited to patterns of the size of the filter and is often enable to learn efficiently complex non-linear patterns \citep{Lecun-95}. Therefore, convolutional layers are usually repeated in an MLP-like fashion, each layer considering the outputs of the previous layer as its own input images. This is a very efficient way to identify more and more complex patterns as the network gets deeper. Usually the first layers act as basic detectors for edges, lines, colors, luminosity, ... that correspond to low-level representations; then the mid network layers detect more advanced features of the image like textures, repetitive patterns, and very basic shapes that act as mid-level representations; the final network layers act as sub-object detectors, behaving like small classifiers of much more abstract content of the image. The end of the network is usually still connected to a few fully connected layers that merge these sub-classifiers into a final classification output or into any other kind of output that is targeted. Such an artificial neural network architecture is called a Convolutional Neural Network (CNN).\\
	
\begin{figure}[!t]
\vspace{-0.5cm}
	\hspace{-0.6cm}
	\begin{minipage}[!t]{1.05\textwidth}
	\centering
	\includegraphics[width=1.0\hsize]{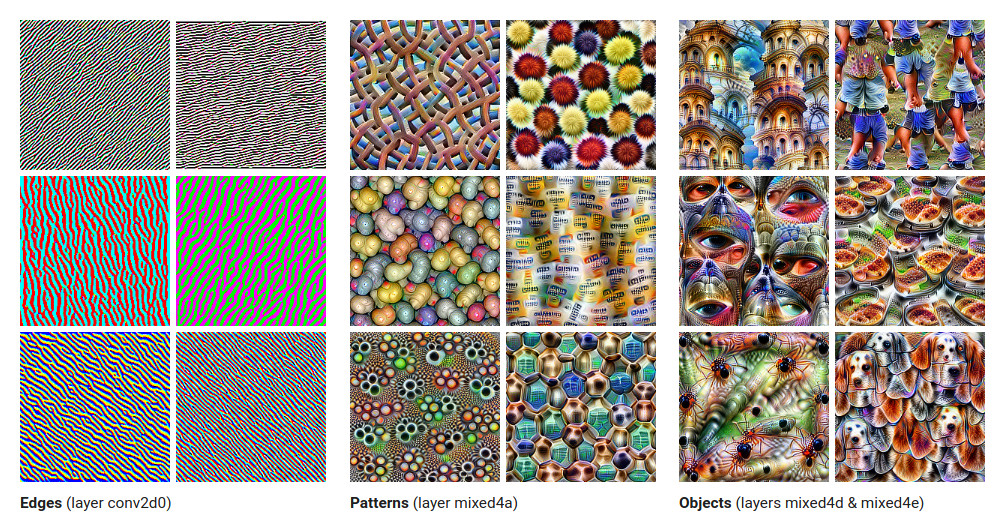}
	\end{minipage}
	\vspace{-0.2cm}
	\caption[CNN filters representation-level examples]{Example of mock images generated to maximize the activation of specific filters in the network. {\it Left}: using filters form an early convolution layer. {\it Middle}: using filters from a deeper layer. {\it Right}: using filters close to the output of the network. {\it Image from \href{https://distill.pub/2017/feature-visualization/}{Distil}}.}
	\vspace{-0.2cm}
	\label{features_visual}

	\end{figure}
	
While it is difficult to look directly at the filters themselves because they are usually very small, there are optimization techniques that can be used to construct mock input images that maximize the activation of a specific filter. While such methods are beyond the scope of this thesis, they still provide a didactic illustration of the previous multi-level representation in CNNs. Figure~\ref{features_visual} shows an example from the \href{https://distill.pub/2017/feature-visualization/}{Distil} online CNN visualization tool. The filters become more and more precise to identify specific sub-parts of the images, with the apparent repetition of the pattern in the image being just a construction effect. The filters become more and more precise to identify specific sub-parts of the images. We note that the apparent repetition of the pattern in the image is just a construction effect in the sens that the filters selected to produce the image are usually small ($3\times 3$ or $5 \times 5$) weight matrices, so that the complex pattern is constructed from the non-linear combination of all the previous layer filters that contribute to the looked at filterup to input image that maximizes the looked at filter.\\
	
	\subsubsection{Pooling layer}
	\label{pool_layer}

	\begin{figure}[!t]
	\centering
	\includegraphics[width=0.85\hsize]{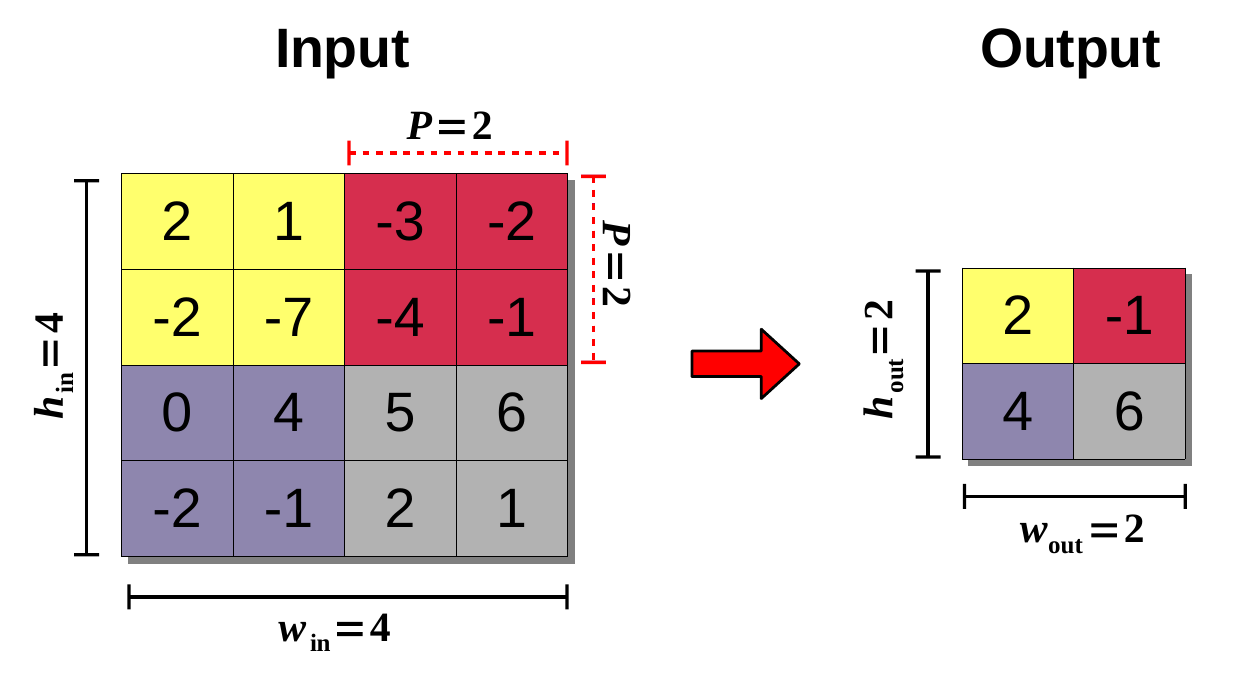}
	\caption[Illustration of the Max-Pooling operation]{Illustration of the Max-Pooling operation with $P_o=2$ on a single layer image, colored by sub-regions.}
	\label{pool_op}
	\vspace{-1.5cm}
	\end{figure}

	Convolutional layers are often used in combination with another new type of layer that reduces the dimensionality of the output images (or activation maps). This layer performs a so-called pooling operation, and is therefore named a pooling layer. While there are different types of pooling operations, the most commonly used is the Max-Pooling. Considering a pooling size $P_o$ it will decompose each depth channel of the image in non-overlaping sub-regions of $P_o \times P_o$ pixels and produce an output image that is composed only of the maximum value of each of these sub-regions. It is common to use $P_o = 2$, which results in an image that is half the size in both dimensions, conserving only a quarter of the pixels of the original image. This operation is performed for each depth channel of the image, which can be the activation maps of a previous layer, so it conserves its depth. We note that a pooling layer does not use weights or activation in any way, so it does not increase the number of learned parameters in the network. This operation is illustrated in Figure~\ref{pool_op} using a different color for each sub-region. Among the most common max-pooling alternatives we can cite the average pooling, max-out layers, L2-norm pooling, global pooling, fractional pooling, etc. \citep[][...]{LeCun-98b, Scherer_2012,Gulcehre_2013, Graham_2014}\\

The aim of this operation is to reduce dimensionality by selecting only the dominant pixels in the input image. This can lead to a very convenient speed up of the learning process with generally small or negligeable degradation of the prediction capacity of a CNN network. Using a pooling layer just after a convolutional one conserves an important overlap of the convolution filter with a small stride. This way the convolved image has a much better resolution and the max-pooling operation conserves only the most relevant information by selecting only one pixel. Such construction has long proven to be a very efficient architecture as exposed in Section~\ref{cnn_architectures}. While pooling layers were widely used a few years ago with a pooling after each convolution, they tend to be less common in modern architectures, with a pooling layer only every several convolution layers. Moreover, the pooling layers has shown to be replaceable by carefully designed convolution that performs equivalent dimensionality reduction \citep{Springenberg_2014}.\\

	\subsubsection{Learning the convolutional filters}
	\label{conv_layer_learn}
   
   \begin{figure}[!t]
	\centering
	\includegraphics[width=0.85\hsize]{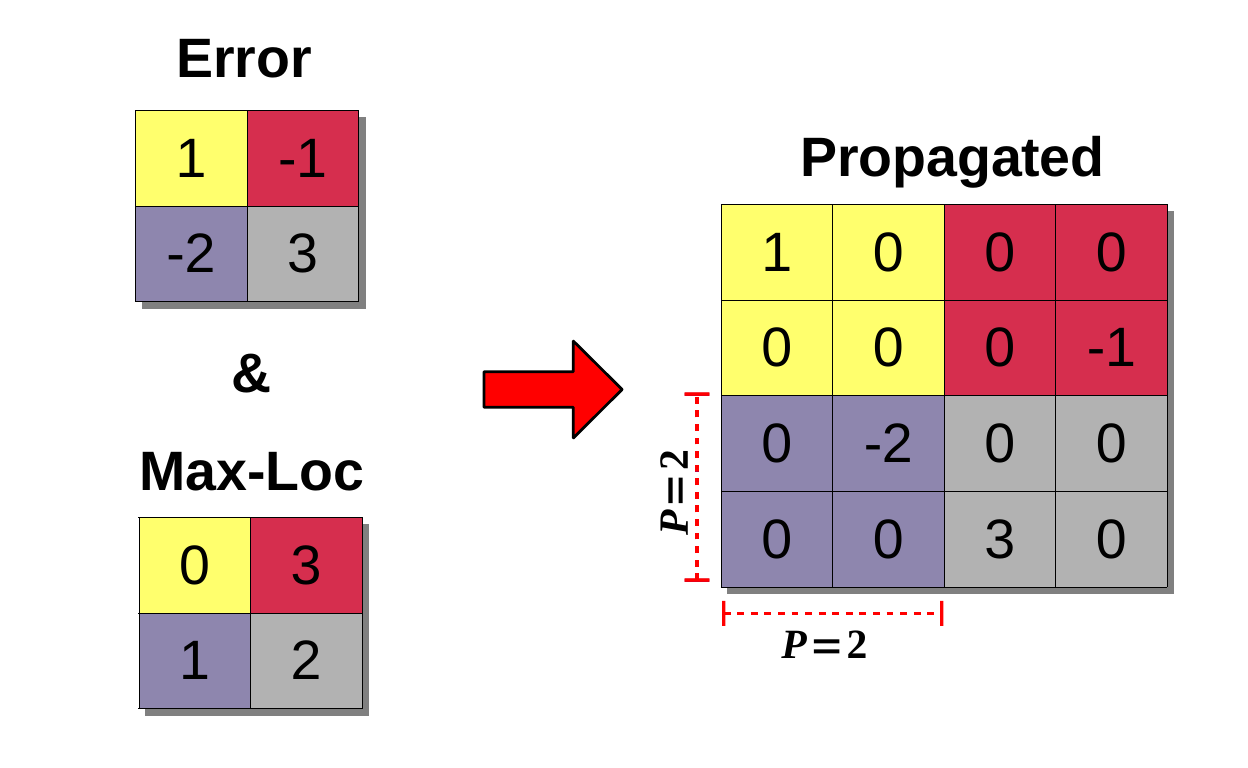}
	\caption[Illustration of the Max-Pooling error backpropagation]{Illustration of the Max-Pooling error backpropagation with $P_o=2$ on a single layer image, colored by sub-regions. The "Max-Loc" matrix gives the location where the maximum value were prior to the Max-Pooling step, counted from zero. The used maximum location are those of Figure~\ref{pool_op} for consistency.}
	\label{pool_op_back}
	\vspace{-1.0cm}
	\end{figure}
	
	In the previous sections we have expressed the convolution operation by using already appropriate weight filters. While it is possible to use only pre-defined filters like it was the case in the first few decades of machine learning image applications, the true objective with this architecture is to learn these filters \citep{lecun-98}. We note that, in echo to our first discussion in Section~\ref{mlp_backprop}, this is the boundary where many agree on the definition of "deep learning". The "deep" attribute here does not only represents the depth of the network as a stack of layers but refers to the fact that both the link between filters and the filters themselves are learned during the training process. Here, we describe all the elements that are necessary to propagate the error through a convolutional architecture. We note that this description is frequently missing in many deep learning courses or presentations (e.g Stanford \href{https://cs231n.github.io/convolutional-networks/}{CS231} or \citet{Bishop:2006:PRM:1162264}) while it is often the most difficult part of a CNN construction.\\

First, the simplest operation to propagate is the pooling. Considering that the error has been propagated using the classical rules for fully connected layers as described in Section~\ref{mlp_backprop}, the propagation produces an error volume for the pooling layer that has the size of its output. The error then has to be propagated to the input image with the appropriate error being associated to the input neurons (or pixels) that was the maximum value of each sub-region. In practice it means that one must conserve the memory of the position of the neuron that contained the maximum value. The error is reported to this element location and all the other elements involved in the associated pooling sub-region get their error set to zero. We illustrate this procedure in the typical case of a $P_o=2$ pooling size in Figure~\ref{pool_op_back}, where the input image was a single layer of size $4\times 4$, which is equivalent to the size of the propagated error. Identically to the forward pooling operation, the depth channels are independent.\\

\vspace{-0.2cm}
For the error propagation in the convolutional layers, we need to define an operation that is called a transposed convolution \citep{Dumoulin_2016}. It works in a very symmetrical way as the convolution, each pixel of an input image being multiplied by all the filter values to produce an identical number of elements in an output image. This operation is repeated for each pixel in the input image according to the stride. Pixels next to each other in the input image will often have an overlap of their projected field of contribution in the output image, depending on the stride. In this case the contribution from each input weighted by the filter are summed on overlapping output pixels. With this operation it is possible to propagate an error "image" of the size of the forward convolution output to produce an error table that has the size of the input image used by the forward convolution. This way each error pixel propagates its value to all the original input positions that were involved in its activation, accordingly weighted by the filter element used by each original input pixel. This operation is the exact equivalent of the Equation~\ref{eq_update_full_network}, or the simplified version from Equation~\ref{eq_deltah}, for each sub-region of the considered operation input image. Indeed, each error is scaled by each weight that were involved in the activation of the forward output neurons and the error is propagated the corresponding forward input neurons.\\

\begin{figure}[!t]
	\hspace{-0.9cm}
	\begin{minipage}{1.0\textwidth}
	\centering
	\includegraphics[width=1.0\hsize]{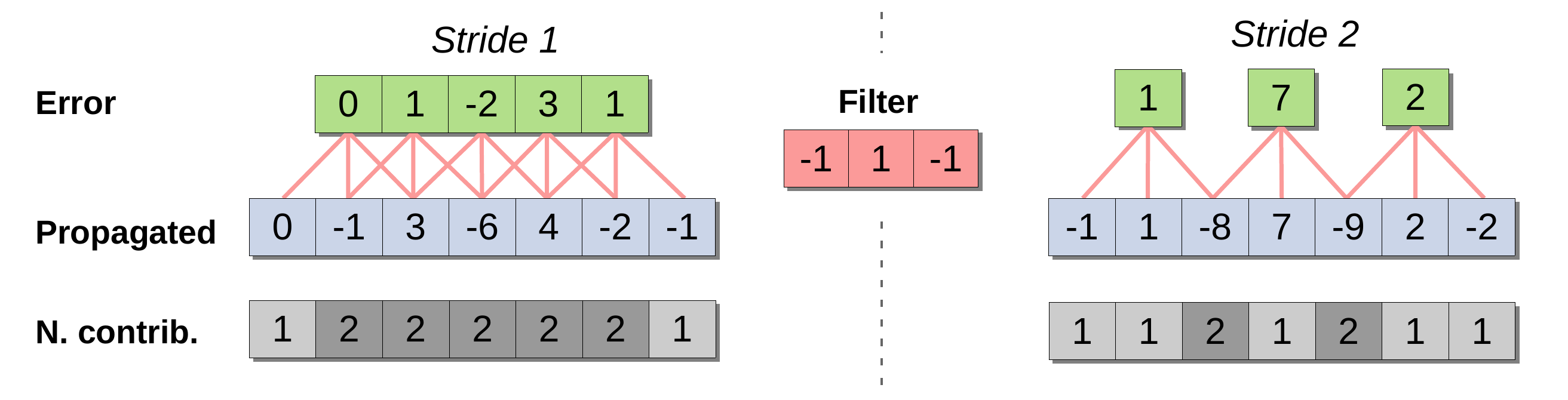}
	\end{minipage}
	\caption[Illustration of a 1D transposed convolution operation]{Illustration of a 1D transposed convolution operation using a 3-element filter (in red). The error is in green, the propagation result in blue and the grayscale table represents the number of error elements that contribute to each propagated pixel. Two examples are given for $S=1$ and $S=2$ that result in propagated errors with identical sizes.}
	\label{im_proc_1d_conv_back}
	\end{figure}

\vspace{-0.2cm}
We illustrate this transposed convolution operation in a one dimensional case in Figure~\ref{im_proc_1d_conv_back} that is the equivalent of the propagation of the case presented in Figure~\ref{im_proc_1d_conv}. We note that the stride in this image is the one that was used in the forward convolution. For the $S=2$ case an equivalent of the transposed convolution should be used, but it requires an addition that we explain below, in the 2D case. Interestingly, in this figure, the contribution pattern is identical to the one that was used for the forward convolution but this time it represents the overlapping of each pixel contribution.\\
	
\begin{figure}[!t]
	\centering
	\includegraphics[width=1.0\hsize]{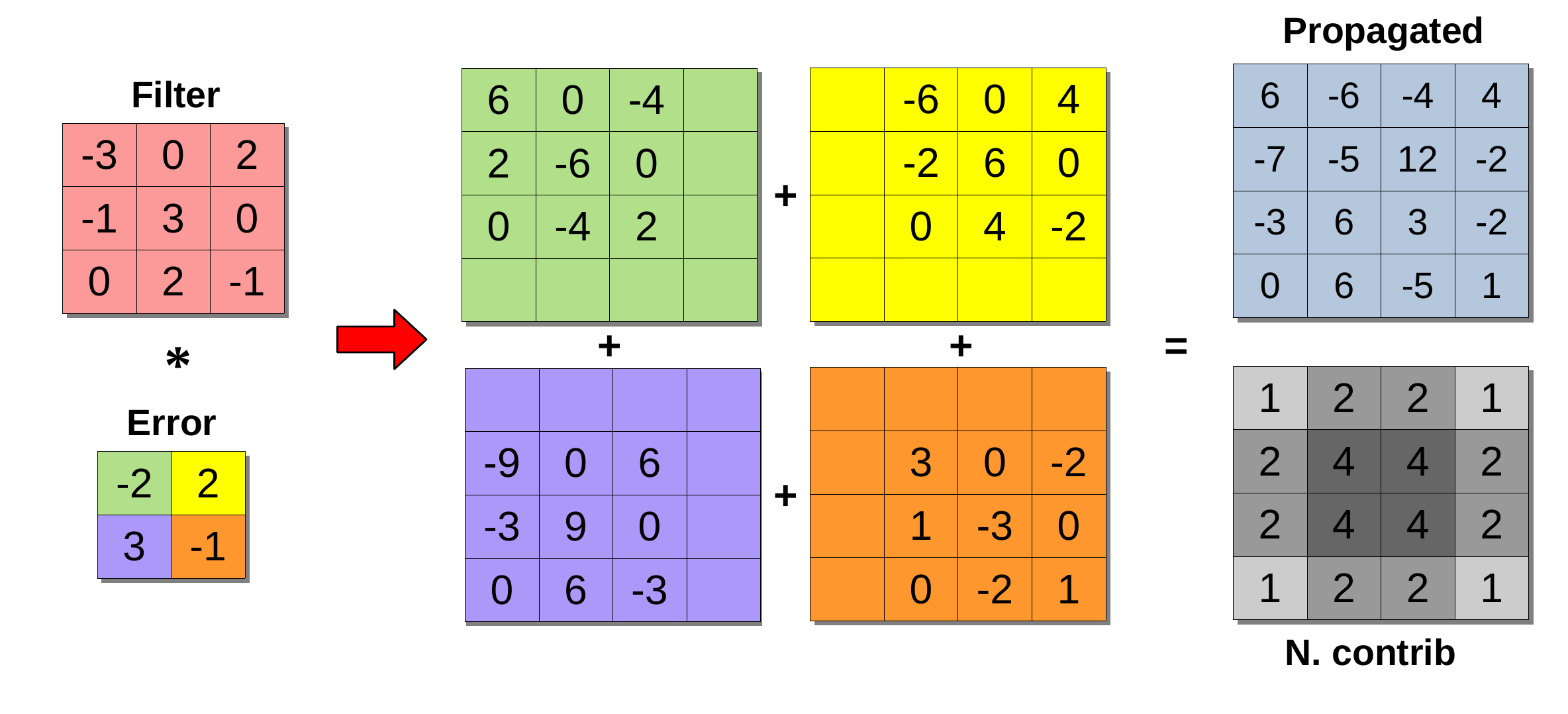}
	\caption[Illustration of a 2D transposed convolution operation]{Illustration of a 2D transposed convolution operation on a $2\times 2$ error using a $3\times 3$ filter (in red). Each error pixel has a different color and the individual contribution to the operation is shown for each of them in the corresponding color. The resulting $4 \times 4$ propagated error is in blue and the grayscale table show the number of contribution for each pixel.}
	\label{im_proc_2d_conv_back}
	\end{figure}

\vspace{-0.2cm}
We illustrate a 2D transposed convolution operation in Figure~\ref{im_proc_2d_conv_back}. We consider an original convolution operation that was performed on a $4\times 4$ input image using a $3\times 3$ filter producing a $2\times 2$ output image. Then the transposed convolution operation combines the error computed in this output with the same filter than the one used in the forward in order to propagate it into an input error that has the same size of the original input. To ease the understanding of the operation, the contribution from each error pixel has been represented independently and colored accordingly. These contributions are summed to produce the shown propagated error along with the associated contribution pattern \footnote{Useful animated illustrations of the transposed convolution can be found on the \href{https://github.com/vdumoulin/conv_arithmetic}{GitHub} page associated with the paper from \citep{Dumoulin_2016}}. \\

We stress that the transposed convolution is not a true "deconvolution" as it is often called. While the transposed convolution performs exactly the action we need, that is to distribute the error over all the original input neurons that were responsible of each output activation, a deconvolution in the sense of signal processing would reconstruct the convolution original input. This is not the same operation as the transposed convolution. Another way to depict a transposed convolution is through a so-called fractionally-stride convolution \citep{Dumoulin_2016}, which is discussed in the next paragraph.\\

While the transposed convolution performs the wanted operation it is often re-expressed as a regular convolution. There are two main motivations for this: (i) the homogeneity of the operations to perform, that already led to replace pooling and dense layers by convolutional ones \citep{Springenberg_2014}, (ii) and the fact that the convolution can efficiently be encoded as a matrix multiplication operation as we describe in Section~\ref{gpu_cnn}.\\

\begin{figure*}[!t]
	\centering
	\includegraphics[width=0.90\hsize]{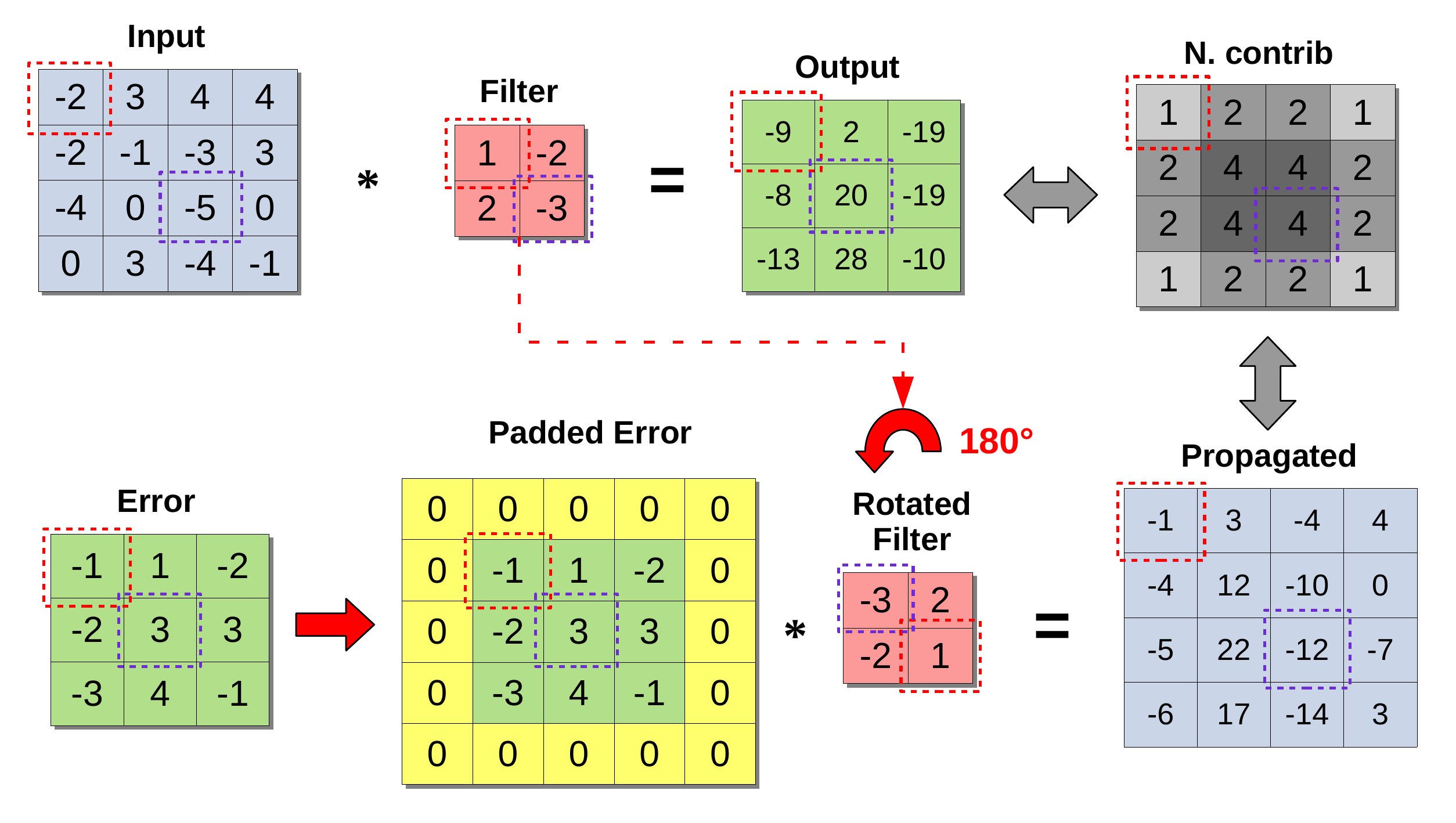}
	\caption[Illustration of the complete forward and backpropagation process for $S=1$]{Illustration of the complete forward and backpropagation process for $4\times 4$ input images using a filter of size $f_s=2$ (in red) and a stride $S=1$. The input and propagated error are in blue, the output and its error are in green. The yellow outline is the external zero-padding $P = 1$. The grayscale table represents the number of contributions from each input pixel and is equivalent to the number of contributing elements from the error. The colored dashed squares highlight specific input-weight pairs to help following pixel paths in the operation.}
	\label{im_proc_2d_conv_back_rot}

	\vspace{0.3cm}
	
	\centering
	\includegraphics[width=0.90\hsize]{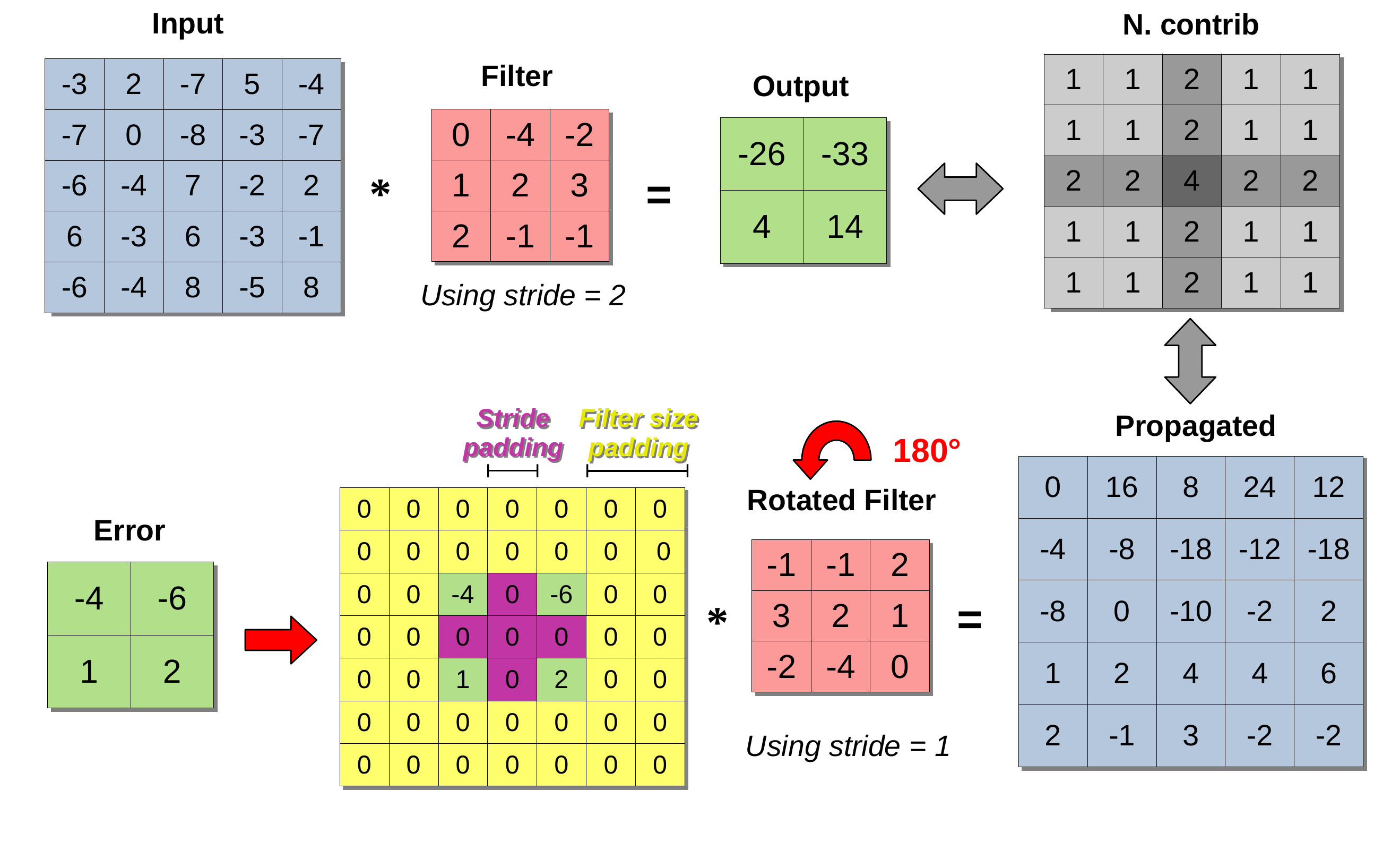}
	\caption[Illustration of the complete forward and backpropagation process for $S=2$]{Illustration of the complete forward and backpropagation process for $5 \times 5$ input image using a filter of size $f_s =3 $ (in red) and a stride $S=2$. The input and propagated error are in blue, the output and its error are in green. The yellow outline is the external zero-padding $P_o = 2$ and the purple internal zero-padding represents $P_s = 1$. The grayscale table represents the number of contributions from each input pixel and is equivalent to the number of contributing elements from the error.}
	\label{im_proc_2d_conv_back_stride}
	
\end{figure*}

This alternative formulation of the back-propagation transposed convolution takes the same output error image as its own input but with a scaling transformation so that a regular convolution operation produces an output image that has the size of the one from the transpose convolution. In order to replicate the correct propagation, this convolution has to reproduce the input size of the forward convolution operation. It means that it usually has to "upscale" its input image, here the output error to be propagated. This is possible by adding zero padding around the image similarly to the one that is used to preserve the size between the input and the output in the forward convolution. To avoid the filters to be applied to zero values only, the padding must be smaller than the filter size (see Fig.~\ref{im_proc_2d_conv}). The quantity of padding to add in a $S=1$ case is defined as $P' = f_s - P - 1$ where $P'$ is the padding of the propagation convolution and $P$ is the padding of the forward convolution. The change of size between the input and output image of the forward operation can then be reversed. However, to reproduce the transposed convolution operation, the weight filter must be rotated by $180^\circ$ when using this propagation convolution. This way, each weight in the filter is applied to the appropriate output value to reconstruct the correct propagated error. This full operation including the filter rotation is often referred to as a full convolution \citep{Dumoulin_2016}. We illustrate this operation in Figure~\ref{im_proc_2d_conv_back_rot} that contains both the forward convolution and the associated error propagation using the full padded convolution. The $4\times 4$ input image is convolved by a $f_s = 2$ filter with a stride $S=1$ and no padding, resulting in a $3\times 3$ output image, here pictured with no activation function. To illustrate the propagation convolution we used an arbitrary error image that has the same size of the output, on which a $P'=1$ padding is apply. Two specific input-weight pairs are highlighted using dashed colored squares, to help verifying that the proper association is kept between the forward and the backward operations.\\

In the previous example we addressed the full convolution in a case with $S=1$ but the use of a larger stride in the original convolution requires additional transformations of the error image. Indeed, it is necessary to control the overlapping pattern as well as the resizing process. This is done by using an additional internal zero-padding (or stride padding) that will enlarge the the error image. This padding is simply defined as $P_s = S - 1$ and is applied between each input image pixel in both dimensions. We illustrate the case of a $S=2$ convolution and the associated full convolution for error propagation in Figure~\ref{im_proc_2d_conv_back_stride}. In this example a $5\times 5$ input image is convolved by a $3 \times 3$ filter using a stride $S=2$, which results in a $2\times 2$ output image. The full convolution for the error propagation, then uses an external padding $P_o = 2$ and an internal padding $P_s = 1$ on the $2\times 2$ error that is propagated into a $5\times 5$ error input image by using the weights from the rotated filter applied to the corresponding output. As before the contribution pattern is reproduced in the forward and back-propagation operations.\\

Finally, there are usually more than one input and output depth channels $d_i$ and $d_o$ in a convolutional layer. In the forward convolution each filter presents a depth that is equivalent to the number of input depth channels $d_i$. The output depth is then the number of filters in the layer $n_f = d_o$ forming an individual filter volume $V = f_s\times f_s\times d_i$ inside a larger volume of $V\times d_o$. The backpropagation convolution must produce a propagated-error volume with the same depth $d_i$ as the input layer. To achieve this, the error propagation convolution must reorganized the filters so that it creates a new volume where a given depth $d_i$ for all the $d_o$ filters are associated in a new individual filter volume $V' = f_s \times f_s \times d_o$ inside a larger volume of $V'\times d_i$. The new filter volume then contains a number of filters corresponding to the the number of depth channel of the input image of the forward convolution, and each of this filter contain as much depth channel than the output of the forward convolution. An illustration of this process using the matrix formalism in presented later in Figure~\ref{matricial_form_fig}.\\

\clearpage
\subsection{Convolutional networks parameters}
\label{cnn_parameters}

	\subsubsection{Convolutional Neural Network architectures}
	\label{cnn_architectures}
	
	\vspace{-0.2cm}
	Convolutional layers can be stacked on top of each others (Sect.~\ref{convolution_stacking}) and pooling layers can be added to reduce the dimensionality of the intermediate activation maps (Sect.~\ref{pool_layer}). To achieve this, many questions need to be addressed: How to choose the appropriate detailed architecture? How many layers are necessary? With how many filters in each? What should be the size of the filters? Is a pooling layer necessary after each convolution, etc ? The general answer to all these questions is: it depends. Some extreme boundaries are easy to estimate: like not having larger filters than the input image, avoiding to reduce the dimensionality too quickly, or avoid having a stride that is larger than the filter size which would cause some pixels to never be scanned. It remains, however, difficult to provide proper general advice. The ANN community has long adopted a "trial and error" approach to find the most effective network architectures, but the number of possible combinations is huge and continues to rise exponentially with the addition of ever new ANN features and layer types.\\
	
\vspace{-0.2cm}
For these reasons, the ML community is mainly moving forward by organizing contests on always more difficult tasks or by comparing the success of various architectures on freely accessible datasets. This leads to a proliferation of algorithms and architectures to be tested and the most successful ones then spread to the rest of the user base. This principle is going even further these days with many architectures being so hard to train and unstable for a long training time that it is advised to get a pre-trained network for very general purpose classification, and then continue the training with completely different input images to adapt it to a new application.\\

\vspace{-0.2cm}
For general considerations, it has been observed that architectures that have many convolutional layers with many small filters were much more efficient than fewer layers with larger filters \citep{vggnet-2014}. While a large filter scans over a large region at once, the same large region can be scanned using successive layers with small filters. For example two layers of $3 \times 3$ filters with a stride of $S=1$ scan up to a $5 \times 5$ area of the original image. Adding an extra layer with identically sized filters results in a $7 \times 7$ area. The advantage of this type of architecture is that it decompose large scale pattern into a non-linear combination of several small patterns, increasing the diversity of objects that can be identified for a same given number of filter. This is exactly comparable to the difference between one MLP layer with much more neurons against the same number of neuron distributed over multiple MLP layers. Additionally, a complex $7 \times 7$ pattern can be decomposed into pieces that might be useful for another type of objects in the same dataset, reducing the global quantity of weights in the network. Most common filter sizes are $3 \times 3$ or $5\times 5$.\\

\begin{sidewaysfigure}
	\hspace{-0.8cm}
	\begin{minipage}{1.05\textwidth}
	\centering
	\includegraphics[width=\hsize]{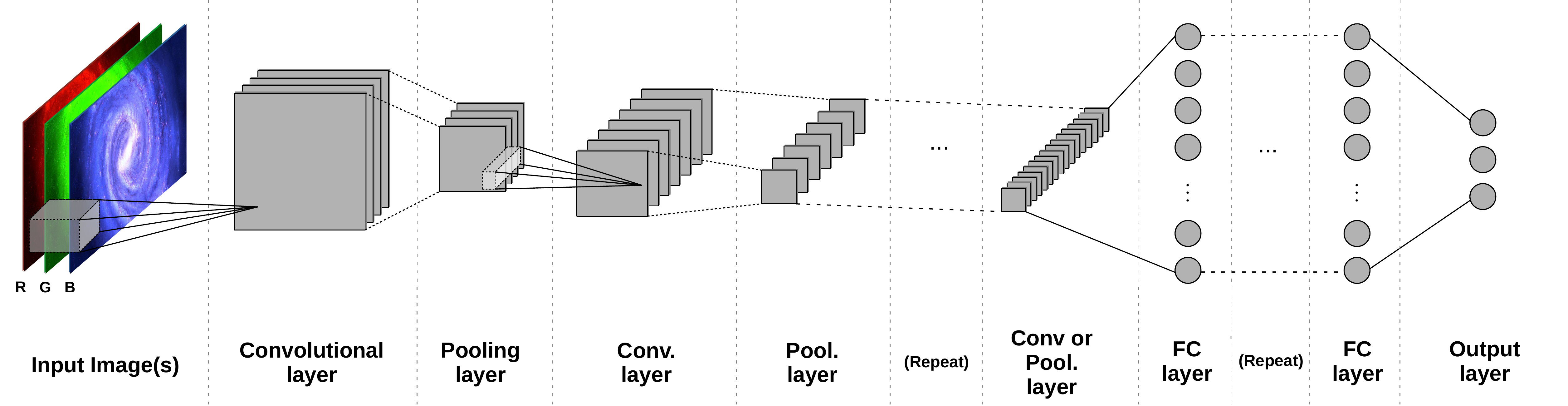}
	\end{minipage}
	\caption[Illustration of a typical CNN architecture]{Illustration of a typical CNN architecture with subsequent convolutional layers with regular pooling. In this representation the gray slices represent activation maps, the filters are not represented. The end of the network is composed of several fully connected layers with gray circles representing regular individual neurons.}
	\label{cnn_network}
	\end{sidewaysfigure}

\vspace{-0.2cm}
Other common practices are for example to use adequate zero-padding in order to conserve the image size in the convolutional layers. The dimension reduction is then completely endorsed by the pooling layers. Small strides are more common, generally with a simple $S=1$ value. In some cases larger filters are used in conjunction with a larger stride when the number of parameters in the network is problematic, but only on the first layers \citet{alexnet_2012}. This leads to a common practice that is to start with layers that have few filters in order to reduce the number of activations and that conserve the input image size. The subsequent layers are made denser with more filters, and the activation maps are pooled to reduce the dimensionality. Most networks finish their convolution part with a layer that contains many filters that produce very small activation maps. The objective being to list all the necessary sub-patterns from the input images that are necessary to perform the end classification. Ultimately, the last convolution layer is fully connected to a smaller series of MLP layers with, as before, each pixel of all the activation maps considered as an independent input feature. It is also common to perform a last convolution operation with many filters of the size of the last activation maps to produce a large mono-dimensional activation before adding the fully connected layers, which results in the exact same number of parameters than the previous solution. We illustrate a very generic architecture of this type in Figure~\ref{cnn_network}, where only the activation maps are represented, but their size and number is characteristic of the filters used at each layer.\\

\vspace{-0.2cm}
There are many other kinds of layers and tuning with for example non linear architectures, networks with outputs larger than their input, recurrent CNN, residual CNN... We can noticeably mention the widely used "depthwise separable convolution" that performs a regular convolution layer per layer and then recombines them using a $1 \times 1$ convolution. Some networks also work with tensor images that have more than two spatially coherent dimensions, and can still have several multiple depth channels, needing tensor weight filters and subsequent tensor representation of the network. However, these advanced techniques are for now irrelevant for our applications.\\ 

\vspace{-0.2cm}
Finally, we list here some famous CNN architectures that illustrate the previous general considerations. We note that, because the size of the activation maps, and therefore the depth of the network, depends on the size of the input map, each architecture is given with its typical input image size. We note that all these architectures have won one or several image classification contests that are described in the reference paper for each of them, leading to their wide adoption in the ANN community. To ease the architecture description we use the following naming system: the input volume is denoted I along with its dimensions (width, height, depth) as I-W.H.D, a convolutional filter is denoted C with the number of filters N followed by the filter size dimension $f_s$ and the stride $S$ as C-N.$f_s$.$S$ (the stride $S$ is omitted in the case of $S=1$), a pooling layer is simply denoted P along with its $P_o$ values as P-$P_o$, and dense layers are denote D followed by the number of neurons $n$ as D-$n$.
\begin{itemize}[leftmargin=0.5cm]
\setlength\itemsep{0.01cm}
\item \textbf{LeNet:} a "classical" simple CNN architecture from \citet{lecun-98}, from which the most known revision is the LeNet-5: [I-32.32, C-6.5, P-2, C-16.5, P-2, C-5.120, D-84, D10].
\item \textbf{AlexNet:} a much more recent and larger architecture from \citet{alexnet_2012}, but that remains modest enough to be usable on most modern individual computers. It noticeably uses larger filters and stride in the first layer to strongly reduce the image size. Its architecture is [I-224.224, C-96.11.4, P-2, C-256.5, 2$\times$[C-384.3], C-256.3, P-2, 2$\times$[D-4096], D-1000].
\item \textbf{VGGNet:} that made the demonstration that a very deep network with only small filters can achieve top-tier performances. This network architecture from \citet{vggnet-2014} is still widely used today due to it simplicity of implementation and very good performance even on modern problems. It is made of chunks of identical convolutional layers. The architecture can be described as [I-224.224.3, 2$\times$ [C-64.3], P-2, 2$\times$ [C-128.3], P-2, 3$\times$ [C-256.3], P-2, 3$\times$ [C-512.3], P-2, 3$\times$ [C-512.3], P-2, D-4096, D-4096, D-4096, D-1000].
\item \textbf{Inception:} a much more recent approach to CNN that is composed of several "blocks" of parallel networks that are concatenated after a number of layers \citep{inception_2014}. The number of continuous convolutional layers in the first version is 22 and it goes above 70 for more recent iterations of this network architecture \citep{inception_2016}, which we do not represent here because of its complexity. This category of networks can only be trained using powerful computing clusters but are capable of solving very diverse tasks efficiently using the very same architecture. However, the details of such an architecture is beyond the objectives of this thesis.
\end{itemize}

	\subsubsection{Weight initialization and bias value}
	\label{cnn_weight_init}
	
	As we exposed in Section~\ref{weight_init}, the method used to initialize the weights at the beginning of the training process can have a strong impact on the stability and convergence ability of ANN. We also discussed that it is strongly linked to the choice of activation function, bias value, and even to the network architecture. Overall, the ReLU activation function is much more reliable than the sigmoid activation for similarly deep networks. In the case of a very deep network like the one we described in the previous section, the ReLU activation quickly becomes the only suitable solution. In such a case, the choice of an appropriate weight initialization is crucial because an inapproriate choice can completely prevent very deep networks to converge. The main objective remains to conserve the weight small enough to preserve precision and stability but also large enough to quickly obtain very different behaviors of each neuron in the network. The performance comparison between the various weight initialization methods is mainly empirical. Still, the main objective of each method is to get as close to the same initial weight variance over all the network layers, independently of their size.\\
	
In the VGG network for example, they used the Xavier initialization (also named Glorot depending on the reference to the name of forename of the author) that is a uniform distribution scaled according to the size of the layers by $\sqrt{1/n_{l-1}}$, where $n_{l-1}$ is the size of the previous layer \citep{glorot_understanding_2010}. In the same paper they also propose what is now called the normalized Xavier initialization that scales the uniform distribution by:
\begin{equation}
\sqrt{\frac{6}{n_{l-1} + n_{l}}}.
\end{equation}
They claim that this initialization works better with layers that are unevenly sized. These initializations can be generalized to be used with a normal distribution instead. Using a zero mean and standard deviation of one, the Xavier initialization is scale by:
\begin{equation}
\sqrt{\frac{2}{n_{l-1} + n_{l}}}.
\label{eq_xavier_normal}
\end{equation}
We note that {\bf this is the weight initialization currently used by default in our CIANNA framework}.\\

Another initialization that is frequently used is the He initialization that follows the same idea but with a wider variance by scaling the uniform or normal distribution by $\sqrt{6/n_{l-1}}$ and $\sqrt{2/n_{l-1}}$, respectively \citep{he_delving_2015}. This initialization is claimed to be more efficient on very deep ReLU activated networks, while in practice it appears that both He and Xavier initialization are commonly used in such applications.\\

Regarding the bias value, there are many approaches using ReLU in convolutional layers that differ regarding their implementation. If the bias value itself changes during the learning phase, then it is often initialized to zero at the beginning of the training. The other approach that uses a constant bias value and an adaptive weight can emulate the same behavior with a weight set to zero for the bias at the initialization. It is common to rather use a small positive value in order to allow the ReLU to not start in an inactive state. {\bf Our approach to this problem has been to use a bias value of $\bm{ 0.1}$} with an associated weight that is randomly generated following the previously described rules for every neuron that uses a ReLU activation.
	
\clearpage
	\subsubsection{Additional regularization: Dropout and momentum}
	\label{dropout_sect}
	
In the context of modern ANN, regularization denotes any technique that is used to prevent overfitting and to even the generalization of the network between training data points. These methods aimed at gaining a better representation of the larger scale network prediction \citep{Goodfellow_2016}. Most of the modern CNN architectures use a regularization technique called dropout \citep{dropout_2014}. It consists in randomly removing a given proportion $d_r$ of neurons at each training step, taking into account that the $d_r$ proportion can be different for each layer. A dropout of $d_r = 0.6$ means that $60\%$ of neurons are dropped. It noticeably prevents overfitting by forcing the network weights to adapt in a more general way by preventing given weights to become the only overspecialized representation of a training dataset specificity. This type of regularization enables the network to work with smaller datasets without overfitting.\\
	
	\vspace{-0.2cm}
Usually, this technique is used only on dense layers at the end of the CNN architecture. This presents a very interesting side effect that is to better specialize the filters themselves, especially if dropout is used in combination with a momentum (Sect.~\ref{sect_momentum}). This way each filter really becomes responsible for one robust pattern as presented in Section~\ref{convolution_stacking}. Without dropout, multiple filters can share the responsibility for a pattern that could otherwise be represented by just one of them, making the interpretation of a filter more complex. Despite having less neurons to compute, using dropout usually makes the training process require much more epochs to converge due to the multiple suitable combination of weights that must be learned to account for the random shutdown of neurons. However the improved robustness of the network representation is often considered worth the slower training.\\

\vspace{-0.2cm}
Usually, the first dense layers after the convolution part present a high dropout rate with up to $d_r = 0.9$ that is consecutively reduced for layers closer to the output with a value depending on the usage. A value around $d_r=0.5$ is often adopted for the last dense layer in classification cases but smaller values of $0.2$ or $0.1$ are sometimes adopted on suitable applications like regression cases with great efficiency \citep{Gal_2015}. Naturally, the output layer does not have dropout since it encodes the problem prediction. One side effect of the use of a high dropout value is that the number of neurons must be significantly raised. An common heuristic is to have an "active" number of neurons that is the equivalent of the same network without dropout, but it might be insufficient is some cases. Still, having $n/d_r$ neurons where $n$ is the number of neuron necessary for a non-dropout model is a useful minimal approach. We also note that a high momentum ($\alpha >= 0.9$) is often advised to minimize dropout undesirable effects like gradients canceling each other and an increased noise in the global gradient descent. This can also be efficiently combined to a larger learning rate.\\

\vspace{-0.2cm}
Finally, once the network has been trained using dropout, there are two opposite approaches to compute the prediction of the network. The first one, that is recommended most of the time, is to scale down all the weights by the $d_r$ factor and then to use all the network weights without removing neurons. This way the global sum of the weights remains scaled similarly as during the training process but all representations are averaged. This is usually the most efficient way to preform the prediction. The other approach is to perform a Monte-Carlo estimation of the output by performing several predictions with a different dropout activation for each layer. This method requires several predictions of the network before being close to the average predicted by weight scaling but is equivalent to sampling a prediction probability distribution, just like a Bayesian approach (see Sect.~\ref{dropout_error}) We stress that we only scratched the surface of the complexity and capacity of dropout for many artificial neuron based ML applications, more details and examples being presented in the reference paper from \citet{dropout_2014}.
	
	\subsubsection{Implications for GPU formalism}
	\label{gpu_cnn}
	
We exposed in Section~\ref{matrix_formal} the advantages of a matrix formalism to speed up the network operations. Following the same objective, we discuss here a possible approach to express the convolutional layer operations as matrix multiplications. The element-wise multiplication between a filter and several sub-regions of an image is already very close to the operations performed in a matrix multiplication. Some methods also take advantage of its underlying SIMD structure to construct GPU kernels that perform the convolution operation directly. However, it is often more efficient to use the already strongly optimized linear algebra libraries.\\

	\begin{figure}[!t]
	\hspace{-1.6cm}
	\begin{minipage}{1.2\textwidth}
	\centering
	\includegraphics[width=1.0\hsize]{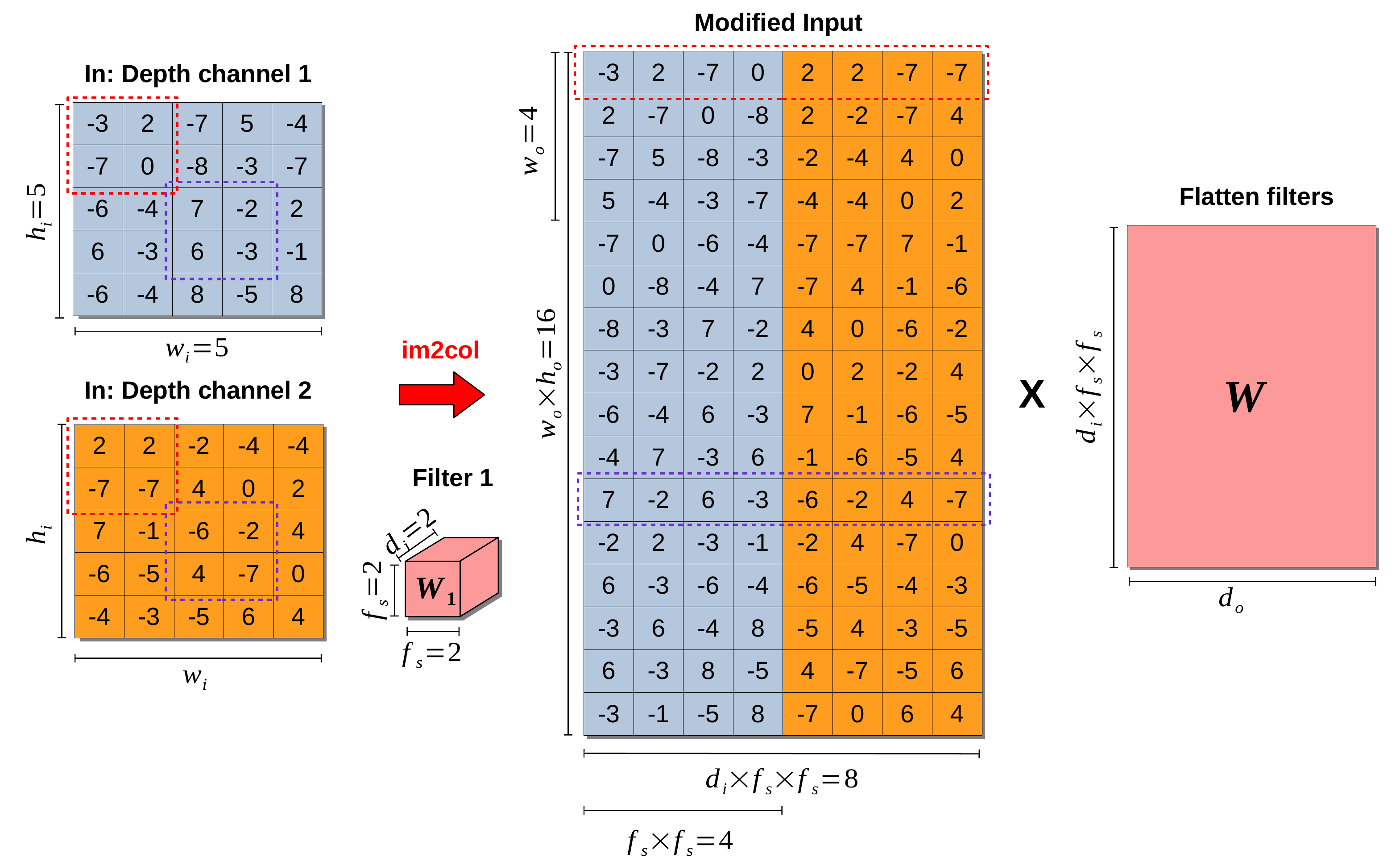}
	\end{minipage}
	\caption[Illustration of the im2col operation]{Illustration of the im2col (here, rather im2row) operation performed on an depth $d=2$ input image of size $w_i = h_i = 5$ using filters with $f_s=2$ represented in red. Elements from input depth channel 1 and 2 are colored in blue and orange, respectively. The expanded input present $w_o \times h_o \times$ columns which correspond to the flatten dimension of one input depth channel, and $d\times f_s \times f_s$ to correspond to a flatten sub-region. $W$ represents the flatten filter matrix, with $d_o$ independent filters, that is multiplied by the expanded input to produce the activation maps. The red and blue dashed rectangles highlight two specific sub-regions that go through the conversion.}
	\label{im2col_fig}
	\end{figure}

\begin{sidewaysfigure}
\centering
\includegraphics[width=0.82\hsize, height=0.75\paperwidth]{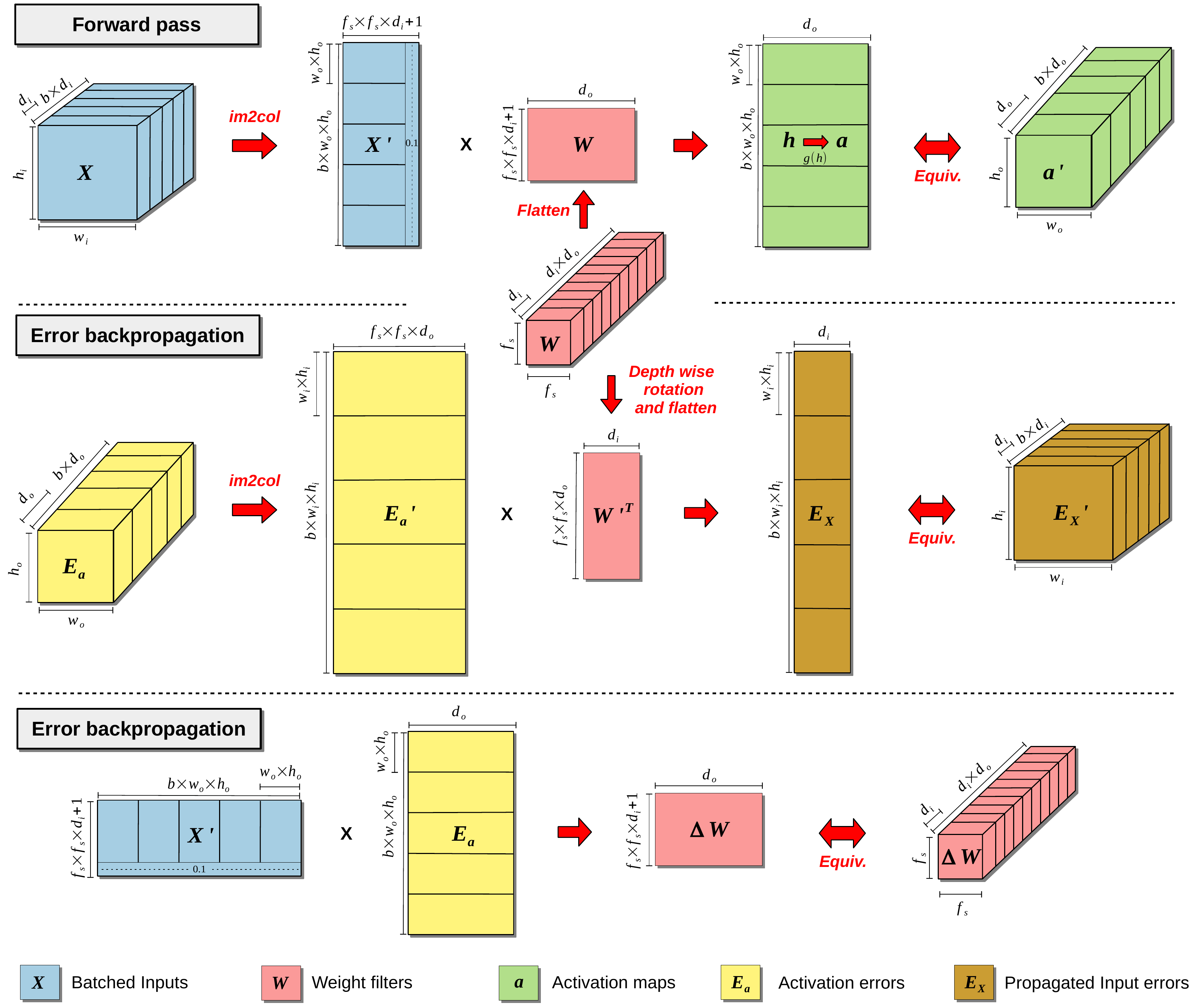}
\caption[Graphical representation of a CNN matrix batch training]{Graphical representation of the matrix batch operation for one convolutional layer. The large red arrows indicate the order of the operations. Large $\times$ symbols are matrix multiplications. The matrix sizes are as follows: $b$ is the batch size, $w_i$, $h_i$ and $d_i$ the width, height and depth of the input image, respectively, $w_o$, $h_o$ are the width and height of the activation maps, $d_o$ is the number of filters and activation maps, and $f_s$ is the filter size in both spatial dimensions.}
\label{matricial_form_fig}
\end{sidewaysfigure}

The most widely adopted approach is based on a representation of the weight filters as a matrix with columns representing each filter flattened ($f_s\times f_s \times d_i$) and with as many columns as the number of filters ($d_o$) in the layer. To correspond to this weight matrix, the input matrix must be composed of rows that represent all the elements of each sub-region flattened accordingly. Besides, multiple images from the same batch can be concatenated in the same expanded representation resulting in a $w_o \times h_o \times b$ number of rows. It is then possible to perform all the convolution operations of the batch using one single matrix operation \citep{Chellapilla_2006}. Using this representation, each sub-region of the input is multiplied by each filter of the layer and produces an output matrix where each column is the flattened activation maps for one filter. This conversion operation is often referred as "im2col" that stands for all the operations of this type, even if the operation we describe might be more suited by the "im2row" name. We illustrate this conversion in Figure~\ref{im2col_fig} where an input image with two depth channels of $5 \times 5$ pixels is converted into an expanded version that corresponds to filters of size $f_s = 2$ with the same number of dimensions. Two dashed areas highlight specific sub-regions that go through the conversion.\\

While it is relatively easy to construct the weight matrix in the right format and conserve it during training, it is much more difficult to convert the input in the right form. The first problem is that this representation uses much more memory than the regular image form due to the redundancy of elements caused by the overlap between the sub-regions. This is a common algorithmic trade off that consists in increasing the memory usage in order to improve computational performances. But, even if it is possible in some cases to keep the expanded form in memory for simple problems, most of the time the images must be converted dynamically to lower the memory footprint. The second problem is that the output format of the activation maps correspond to regular flattened images. This means that they must also be converted in the expanded form if the next layer is a convolutional one, and successively for the all network convolutional layers, increasing even more the memory issue.\\

However, even with a very poor conversion performance the large single matrix operation is so much faster than a raw computation of the convolution that it is almost always worth the additional cost of the conversion. There are many reasons for the raw convolution to be slow, the main one being that the input elements are poorly arranged in memory for this operation, inducing regular cache-misses \footnote{Cache-miss refers to the CPU being forced to load data from the host memory rather than from the cache. Memory copy being made by blocks, operations on continuous data in memory allows the CPU to read the subsequent data from the cache loaded from previous access.}. This is completely solved with the expanded matrix representation that fully uses the matrix multiplication optimization of cache and redundancy. Again, we insist on the fact that in almost all cases it is faster to use this approach anyway, but it makes the conversion operation the most probable bottleneck of the whole network, then any improvement in the conversion computational cost results in a large global network performance increase. There are many implementations of this function that make use of various hardware capabilities, with many of them being open source while others are protected inside closed frameworks. This function is so important in modern networks that it is the object of several researches and publications; a nice empirical method comparison can be found in \citet{Anderson_2017}. Still, we propose our own im2col implementation that takes the form of a CUDA kernel . While we kept a simple approach, we minimized the number of memory operations with only one read per input pixel and just the absolutely necessary number of affectations. Despite this, the kernel remains completely memory bound, which indicates that it is as optimized as it can be without using advanced shared GPU memory management and low-level cache operations. We emphasize that, depending on the network architecture, our im2col implementation is efficient enough not to be the computational bottleneck with most of the computation time spent in matrix multiplication even on a high-end GPU.\\

Using this approach we now have a method to convert convolutional layers into efficient matrix operations. Each subsequent layer can then use the same approach to construct the full network. The pooling operations are just a SIMD operation on all sub regions and do not require any additional treatment. It is possible to use the matrix formalism for fully connected layers described in Section~\ref{matrix_formal} for the end of a CNN, when necessary. Regarding the error backpropagation, we exposed that it can be expressed as a classical convolution, meaning that the same im2col routine can be used with minimal adjustments to propagate the error in a convolutional layer. We illustrate this matrix formalism in Figure~\ref{matricial_form_fig} that presents the forward, the backpropagation and the weight update for a convolutional layer that consists in the multiplication of the expanded input and the expanded output error. The dimensions of each element are provided in the figure following the same notations as in the previous sections. The figure also shows the addition of a bias value in the expanded image form, as before to minimize the number of kernel launches that must be performed. Still, unlike the matrix formalism we presented for fully connected layers (Fig.~\ref{matricial_form_fig}), this representation is not exhaustive: to limit the complexity of the figure, we omitted many small adjustments that are needed to achieve a computationally efficient matrix operation.\\

	\subsubsection{Example of a classical image classification}
	\label{mnist_example}
	
\begin{figure}[!t]
	\centering
	\includegraphics[width=0.8\hsize]{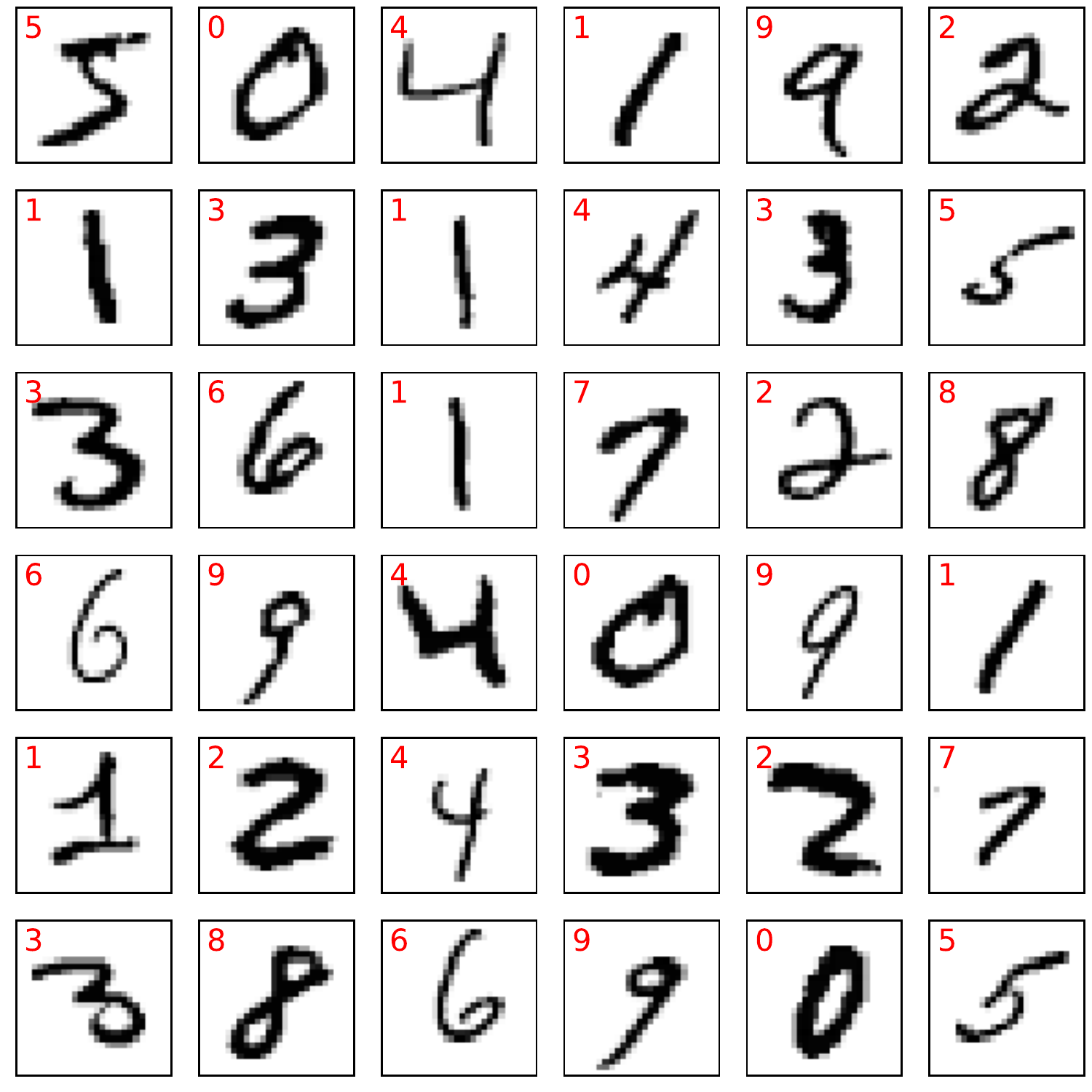}
	\caption[Excerpt of the first 36 images of the MNIST dataset]{Excerpt of the first 36 images of the MNIST dataset. Each image has a size of $28 \times 28$ pixels and is encoded using grayscale with integer values in the range 0-255. The corresponding targets are shown in red for each image.}
	\label{mnist_digit_fig}
	\end{figure}

	To illustrate the classification capacity of a CNN architecture and connect the exposed theoretical aspects to a real simple example, we will use a very well known dataset named \href{http://yann.lecun.com/exdb/mnist/}{MNIST} (Modified NIST's Special Database) \citep{Lecun-95, lecun-98}. It consists of a set of handwritten digits from 500 different writers expressed as $28 \times 28$ grayscale images (0-255 pixel values) positioned to their respective centers of mass. It is freely accessible in the form of a 60000-image training dataset and a 10000-image test set. We stress that although the digits are not perfectly equally represented, the proportions are sufficiently balanced to not cause any issue. Figure~\ref{mnist_digit_fig} shows the first 36 images of the training dataset with the corresponding labels. It has been the support of some of the first CNN architectures that automated the filter selection by training them as the rest of the network. It has then been used by many others to test the efficiency of architectures or even non-ANN classification methods due to its free access and balanced difficulty. The results for various architectures are listed on the dataset website along with the associated publications for each of them \citep[for example][]{lecun-98, Belongie_2002, Ciresan_2012} with best results near $0.23\%$ error rate to be compared to a human performance estimated at $0.2\%$ error rate. For comparison, a single-layer linear ANN gets only a $12\%$ error rate on this dataset. These days, MNIST remains a widely adopted benchmark set for CNN applications and is used for many pedagogical illustrations.\\
	
For this example we used our framework CIANNA to construct a simple CNN that is vastly inspired by the LeNET-5 from \citet{lecun-98} slightly described in Section~\ref{cnn_architectures}, and that achieved a $0.95\%$ error rate on this dataset. We note that, for our application, we did not use any pre-processing on the input data, like dataset augmentation, image distortion or deskewing and just used the raw MNIST training dataset. Firstly, we used a convolutional layer that is composed of 6 filters of size $f_s=5$ with a stride of $S=1$ and padding of $P=2$, directly followed by a max-pooling of size $P_o = 2$. It results in a set of $14\times 14$ activation maps that uses leaky ReLU with $\lambda = 0.01$, and goes through a second convolutional and pooling combination using the same parameters but with 16 filters, resulting in $7\times 7$ activation maps. A last convolutional layer is added without pooling using 48 filters of size $f_s=3$ and $P=1$. It produces $5\times 5$ activation maps. This output is flattened into a $5\times 5 \times 48 = 1200$ input vector that is connected to two fully connected layers with $n = 1024$ , $d_r = 0.5$ and $n = 256$, $d_r = 0.2$, respectively, that both use the same ReLU activation than the convolutional layers. The network ends with a fully connected layer of $o=10$ using Softmax activation with a cross-entropy error (Sect.~\ref{proba_class_intro}). The network is trained using mini-batches of size $b=64$, a learning rate of $2\times 10^{-4}$ that slowly decays to $1 \times 10^{-4}$, a momentum of $\alpha=0.9$, for a total of 40 epochs. \\

	\begin{table}[!t]
	\small
	\centering
	\caption{Confusion matrix for the MNIST prediction using our CNN implementation.}
	\begin{minipage}[!t]{1.2\textwidth}
	\hspace{-1.6cm}
	\begin{tabularx}{1.0\hsize}{r l |*{10}{m}| r }
	\multicolumn{2}{c}{}& \multicolumn{10}{c}{\textbf{Predicted}}&\\
	\cmidrule[\heavyrulewidth](lr){2-13}
	\parbox[l]{0.2cm}{\multirow{13}{*}{\rotatebox[origin=c]{90}{\textbf{Actual}}}} & Class & C0 & C1 & C2 & C3 & C4 & C5 & C6 & C7 & C8 & C9 & Recall \\
	\cmidrule(lr){2-13}
	& C0 &  976 &    0 &    1 &    0 &    0 &    0 &    1 &    1 &    1 &    0 &	 99.6\%  \\
	& C1 &    0 & 1132 &    1 &    0 &    1 &    0 &    0 &    0 &    1 &    0 &	 99.7\%  \\
	& C2 &    1 &    1 & 1027 &    0 &    1 &    0 &    0 &    1 &    1 &    0 &	 99.5\%  \\
	& C3 &    0 &    0 &    1 & 1004 &    0 &    3 &    0 &    1 &    1 &    0 &	 99.4\%  \\
	& C4 &    0 &    0 &    1 &    0 &  972 &    0 &    1 &    0 &    1 &    7 &	 99.0\%  \\
	& C5 &    0 &    0 &    0 &    4 &    0 &  886 &    1 &    0 &    0 &    1 &	 99.3\%  \\
	& C6 &    3 &    2 &    0 &    0 &    1 &    2 &  949 &    0 &    1 &    0 &	 99.1\%  \\
	& C7 &    0 &    2 &    3 &    0 &    0 &    0 &    0 & 1020 &    1 &    2 &	 99.2\%  \\
	& C8 &    0 &    0 &    1 &    1 &    0 &    1 &    1 &    1 &  968 &    1 &	 99.4\%  \\
	& C9 &    0 &    0 &    0 &    0 &    3 &    1 &    0 &    4 &    0 & 1001 &	 99.2\%  \\

	\cmidrule(lr){2-13}
	& Precision &     99.6\%  &  99.6\%  &  99.2\%  &  99.5\%  &  99.4\%  &  99.2\%  &  99.6\%  &  99.2\%  &  99.3\%  &  98.9\%  &  99.35\%  \\
	\cmidrule[\heavyrulewidth](lr){2-13}
	\end{tabularx}
	\end{minipage}
	\vspace{-0.1cm}
	\label{mnist_confmat}
\end{table}

With this simple network we reached a $0.65\%$ error rate on the test set at around epoch 30, corresponding to a $\sim 3$ minutes training running on a Nvidia \href{https://www.techpowerup.com/gpu-specs/quadro-p2000-mobile.c3202}{Quadro P2000} mobile. The corresponding confusion matrix (Sect.\ref{class_balance}) is shown in Table~\ref{mnist_confmat} that shows the global accuracy of $99.35\%$. It also highlights some specific confusion between digits that are more alike, for example between C4 and C9. More importantly for this manuscript, this result demonstrates the overall effectiveness of our implementation and choice of optimization as it is very competitive to similarly deep state-of-the art network implementations. For example, the same network architecture declared using the Keras framework with the much more advanced ADAM gradient descent optimization \citep{kingma2014} does not achieve better accuracy results. We illustrate the use of the CIANNA interfaces on this example in Appendix~\ref{cianna_app}, where we also use this example to make a performance comparison between CIANNA and Keras (TensorFlow).

\subsection{Use of the dropout to estimate the uncertainty in a regression case}
\label{dropout_error}

An interesting side-usage of the dropout (Sect.~\ref{dropout_sect}) in ANN is to provide an uncertainty measurement of the prediction. Indeed, a network trained using dropout can be used to make several predictions with different random selections of neurons performing a Monte-Carlo estimate of the network prediction \citep{dropout_2014}. In fact, the dropout of neurons in dense layers forces the network to learn a probability distribution of the output, each random selection being responsible for a sub-set of this distribution. Many applications that need this feature use Bayesian Neural Networks \citep{MacKay_92} that achieve such task using modern variational inference \citep{Titsias_2014}, but they often have an important supplementary computational cost. However, it has been demonstrated that a regular ANN with dropout can achieve similar predictions while being much more efficient in terms of computational performances \citep{Blundell_2015, Gal_2015}. \\

We illustrate this capacity here using the simple one-dimensional example from Section~\ref{regression_expl}. To better show the uncertainty measurement we slightly raised the added noise dispersion around our original function to $\sigma = 0.15$. The network we used here is composed of two hidden layers with a leaky-ReLU activation that contains $n=64$ and $n=48$ neurons using a dropout rate of $d_r = 0.6$ and $d_r=0.5$, respectively. Having larger layers than these values with a higher dropout rate has shown to degrade the global prediction performance. In contrast, smaller layers were able to get a similar global prediction but the uncertainty appeared to be underestimated in these cases. As expected, too large layers with a too small dropout rate led to overtraining. On this example the optimal learning rate was $\eta = 0.002$ using a batch size of $b_s = 64$ and a momentum of $0.8$.\\

\begin{figure}[!t]
	\centering
	\includegraphics[width=0.95\hsize]{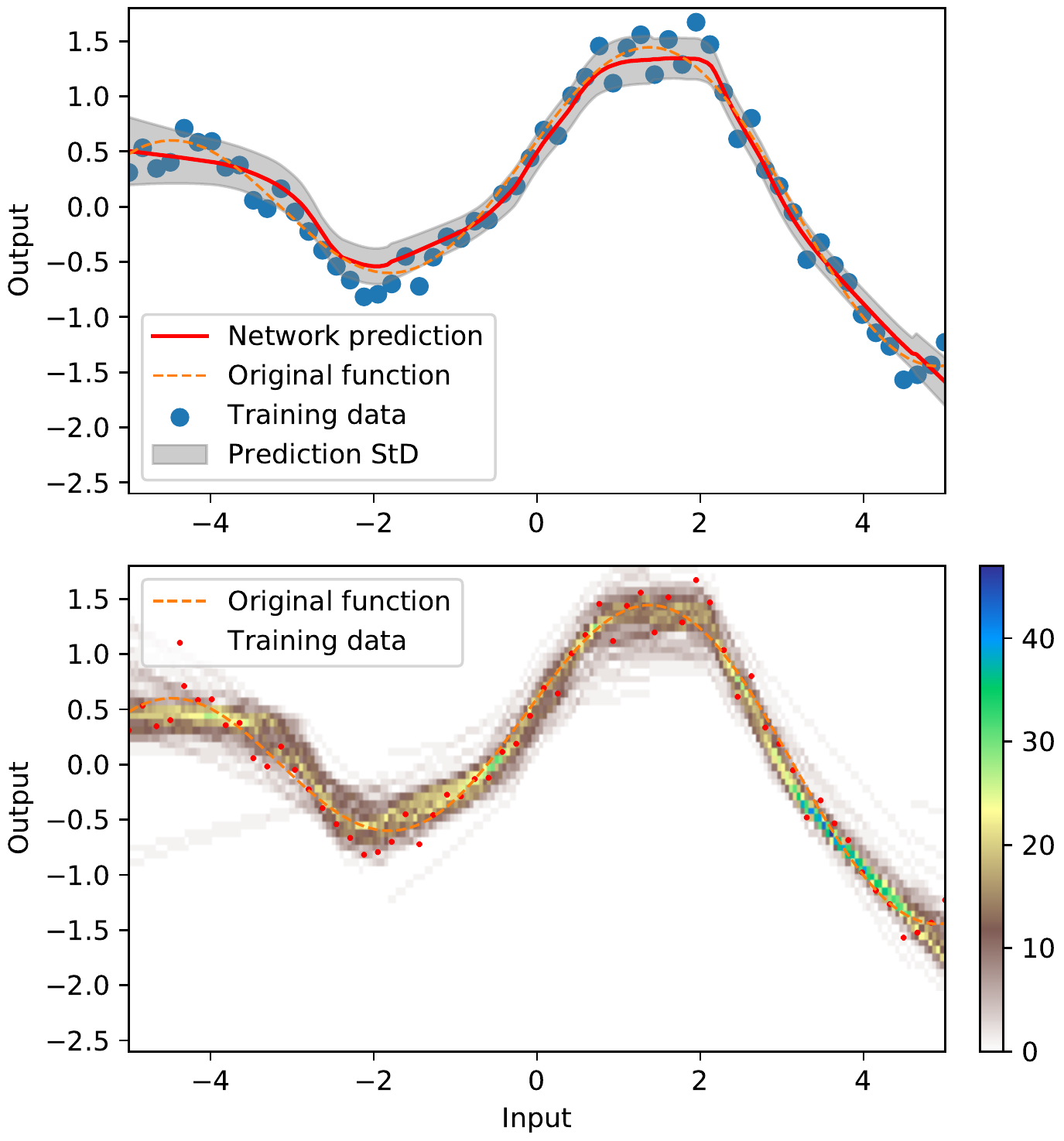}
	\caption[Error prediction using dropout in a 1D regression case]{Error prediction using dropout in a 1D regression example (from Sect.~\ref{regression_expl}) using a two fully-connected hidden layer network. {\it Top:} Average of 100 predictions using dropout. The gray area shows the uncertainty computed as the standard deviation of all the predictions at a given abscissa. {\it Bottom:} 2D histogram of the 100 predictions. Each input value correspond to a vertical histogram of prediction values.}
	\label{drop_error_regre}
	\end{figure}

The results are presented in Figure~\ref{drop_error_regre} corresponding to 100 predictions using the training network with random neuron exclusion. The figure shows two different representations, the first one using the mean and the standard deviation from all the predictions of each input point in our test set, and the second one by drawing a histogram of the predicted values for each input point. The two representations reveal that the global prediction mostly follows the original function in a good confidence interval. The points near the limits of the training interval ($X > 4$ and $X < -4$) are less constrained, like in the regular case due to the combination of few training points and quick local-changes in the original function at these places, just like a regular boundary effect. It is also visible that the network confidence interval is more narrow in regions that have a steeper slope. The histogram representation is very informative since it directly represents the probability distribution but it is less well suited for visualisation of higher dimension regressions.

\clearpage
\section{Extinction profile reconstruction for one line of sight}
\label{galmap_problem_description}

In this section we describe how the discussed CNN formalism can be used to reconstruct an extinction profile based on the comparison between observed and modeled data (see Sect.~\ref{extinction_with_bgm_intro}). We explain our choice of observed quantities for this comparison and describe various processing that had to be performed in order to make the BGM predictions as realistic as possible. We also detail our methodology to construct mock extinction profiles that are sufficiently representative of the interstellar dust distribution to train our network. We then describe various effects either from the previous construction or from the network architecture itself than can have effects on the prediction. Finally, we combine these elements to perform a first CNN prediction using one LOS and discuss its generalization capacity to neighboring lines of sight in the Galactic plane.

\etocsettocstyle{\subsubsection*{\vspace{-1cm}}}{}
\localtableofcontents

\vspace{0.5cm}

\subsection[Construction of a simulated 2MASS CMD using the BGM]{Construction of a simulated 2MASS color-magnitude diagram using the Besançon Galaxy model}
\label{cmds_construction_section}
	\subsubsection{Choice of BGM representation and observed quantity}
	\label{choice_of_BGM_cmd}

We stated in Section \ref{extinction_with_bgm_intro} that we aim at {\bf using the Besançon Galaxy Model to reconstruct extinction profiles by comparison to observed quantities} that are also predicted by the model. We also explained that, because the BGM is a statistical stellar synthesis model, we have to use observations in a statistical form as well. In a first step, for the sake of simplicity, we wanted to use solely 2MASS data. We thus take advantage of the potentially large distance range permitted by the relatively low extinction in the near IR, and of the possibility of a direct comparison to the other work we are involved in \citep{Marshall_2020}. There are 3 bands in the 2MASS survey that can also be predicted by the BGM, namely J ($1.235 \mu m$), H ($1.662 \mu m$) and K ($2.159 \mu m$). To constrain the extinction, it is better to have one color and one magnitude to be sensitive to both the reddening and the brightness decrease that it induces. To maximize the leverage in the color dimension (i.e. the difference in extinction between two wavelengths) we chose to use a [J-K]-[K] CMD, that we already illustrated in Figures~\ref{expl_observed_cmds} and \ref{obs_model_ext_comparison}.\\

\begin{figure*}[!t]
\hspace{-1.7cm}
\begin{minipage}{1.2\textwidth}
	\centering
	\begin{subfigure}[!t]{0.47\textwidth}
	\caption*{\bf Giant stars (Class III)}
	\includegraphics[width=1.0\hsize]{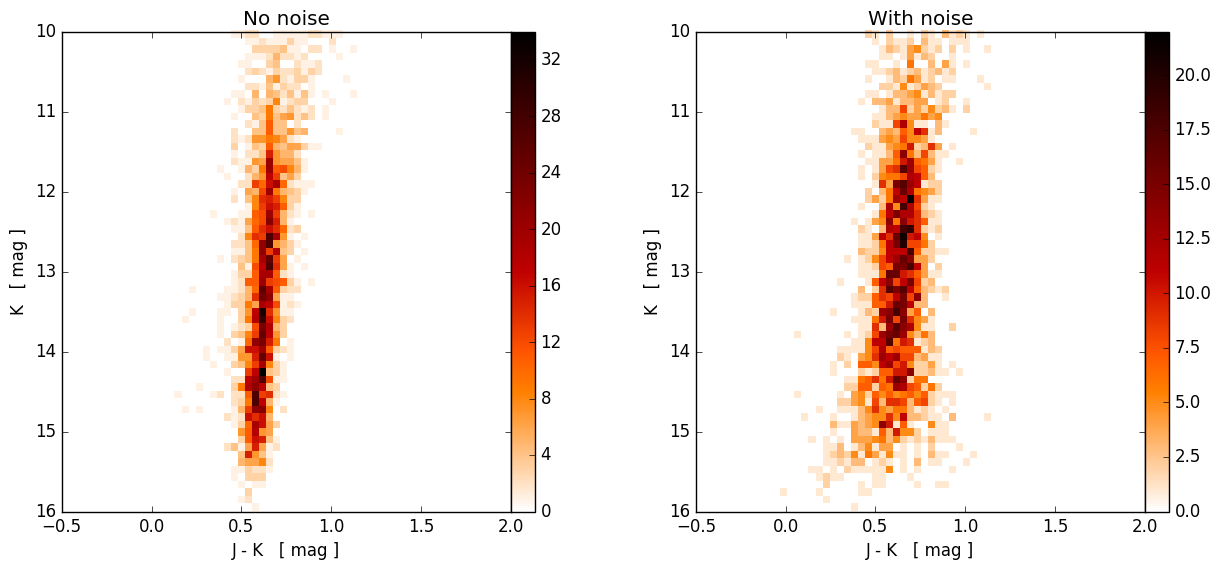}
	\end{subfigure}
	\hspace{0.3cm}
	\begin{subfigure}[!t]{0.47\textwidth}
	\caption*{\bf Main sequence stars (Class V)}
	\includegraphics[width=1.0\hsize]{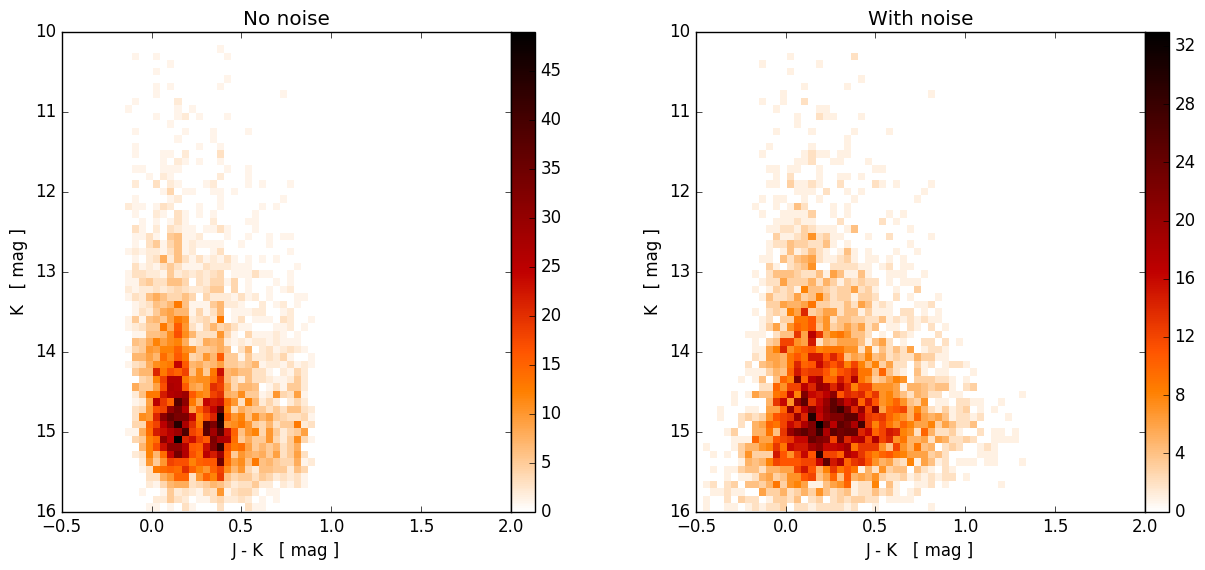}
	\end{subfigure}\\
	\vspace{0.3cm}
	\begin{subfigure}[!t]{0.47\textwidth}
	\caption*{\bf All modeled stars}
	\includegraphics[width=1.0\hsize]{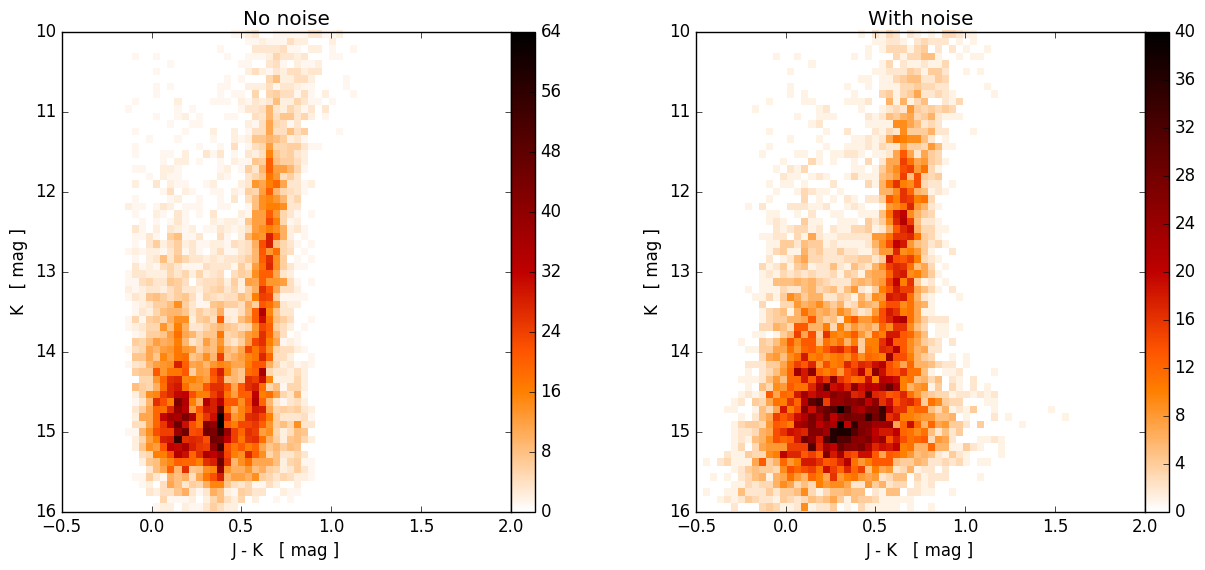}
	\end{subfigure}
	\end{minipage}
	\caption[Simulated 2MASS CMDs for giant and main sequence stars]{Color-magnitude diagram simulated by the BGM for mock 2MASS data, without extinction, in the direction $l = 280$ deg, $b = 0$ deg. The contributions of giant stars and main sequence stars are shown separately and together. The raw BGM values are show in the left frame of each case. The right frame of each case shows the same data after adding simulated 2MASS noise.}
	\vspace{-0.5cm}
	\label{stellar_types_in_CMD}
\end{figure*}

\begin{figure*}[!t]
\hspace{-2.1cm}
	\begin{minipage}{1.22\textwidth}
	\centering
	\begin{subfigure}[!t]{0.41\textwidth}
	\caption*{$d = 0 \rightarrow 1$ kpc}
	\includegraphics[width=1.0\hsize]{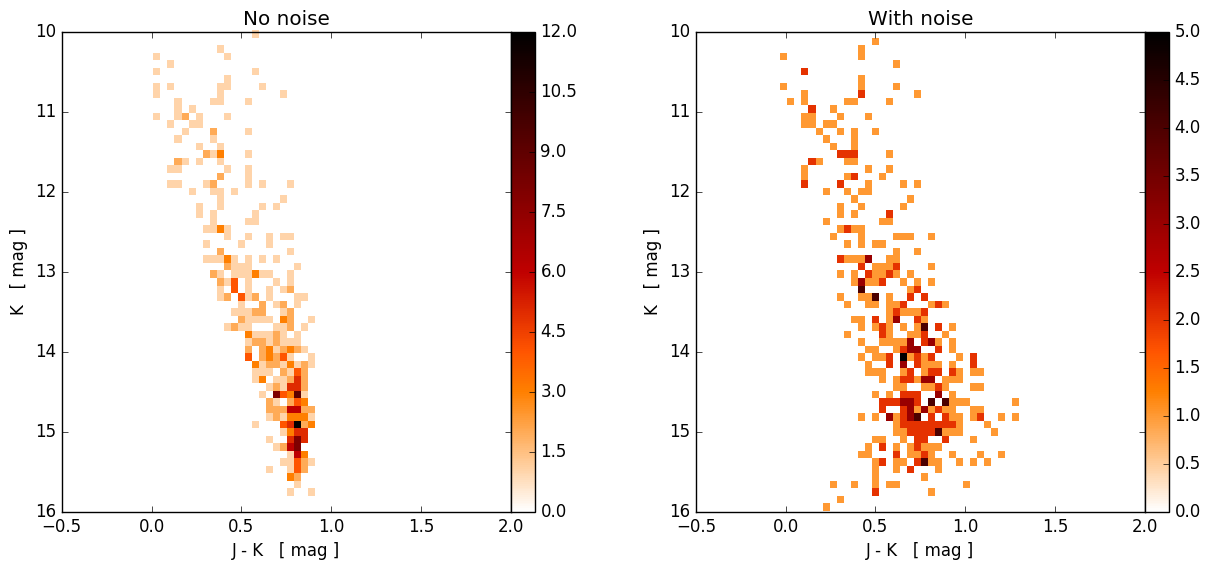}
	\end{subfigure}
	\hspace{1cm}
	\begin{subfigure}[!t]{0.41\textwidth}
	\caption*{ $d = 5 \rightarrow 6$ kpc}
	\includegraphics[width=1.0\hsize]{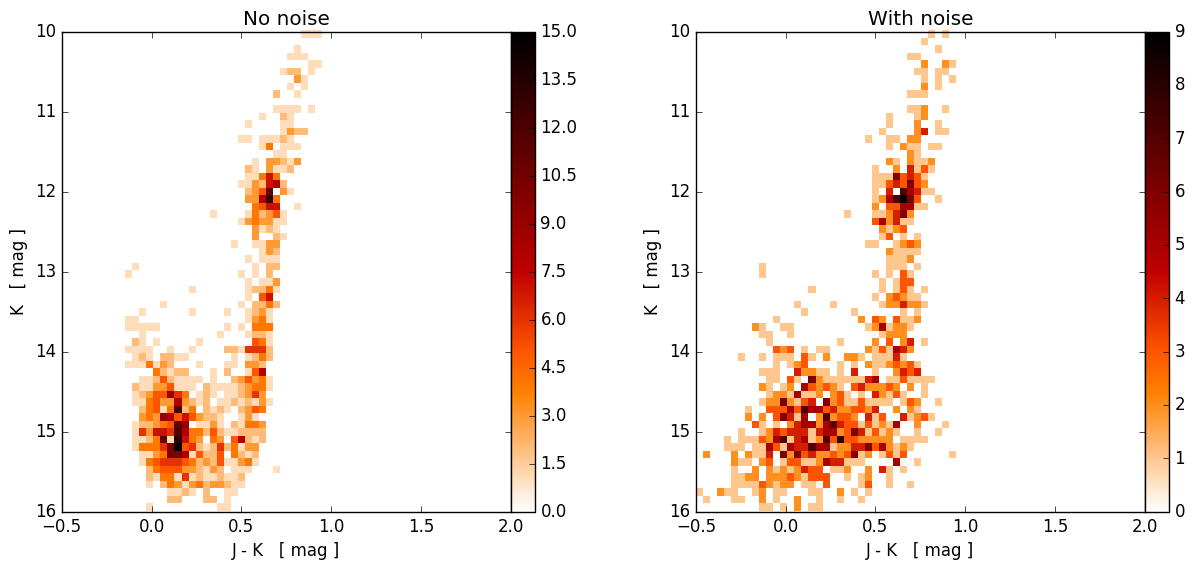}
	\end{subfigure}\\
	\begin{subfigure}[!t]{0.41\textwidth}
	\caption*{ $d = 1 \rightarrow 2$ kpc}
	\includegraphics[width=1.0\hsize]{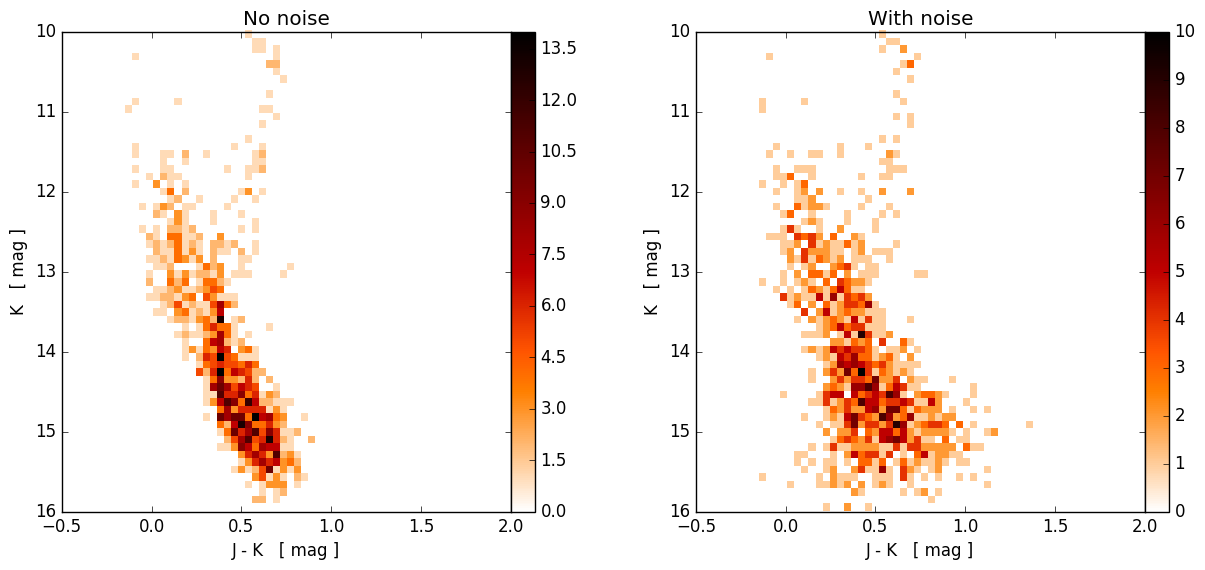}
	\end{subfigure}
	\hspace{1cm}
	\begin{subfigure}[!t]{0.41\textwidth}
	\caption*{ $d = 6 \rightarrow 7$ kpc}
	\includegraphics[width=1.0\hsize]{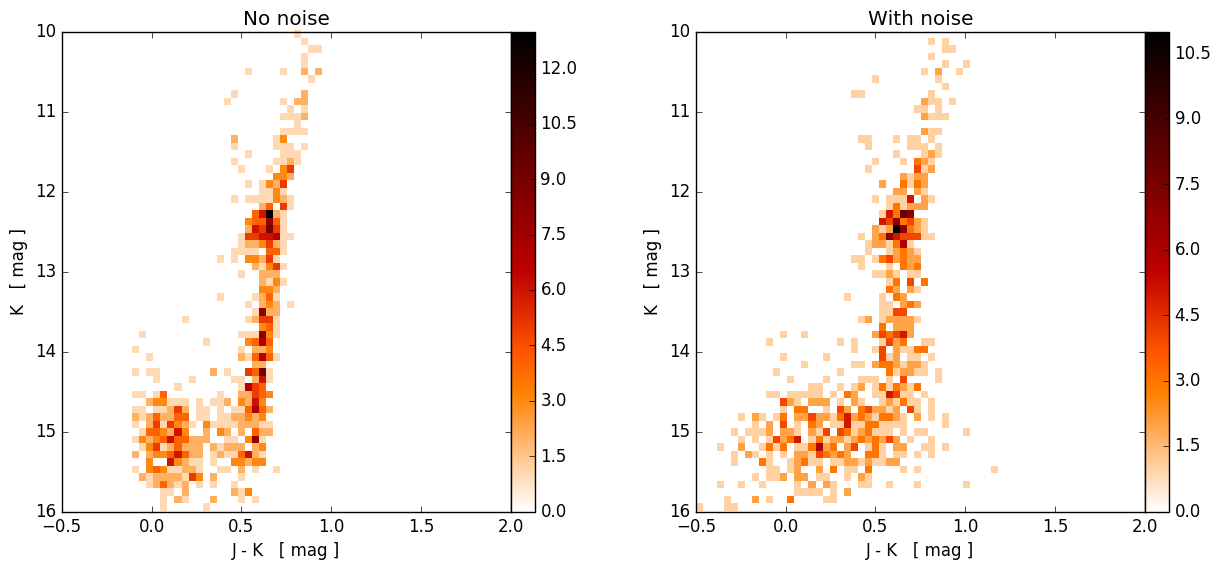}
	\end{subfigure}\\
	\begin{subfigure}[!t]{0.41\textwidth}
	\caption*{ $d = 2 \rightarrow 3$ kpc}
	\includegraphics[width=1.0\hsize]{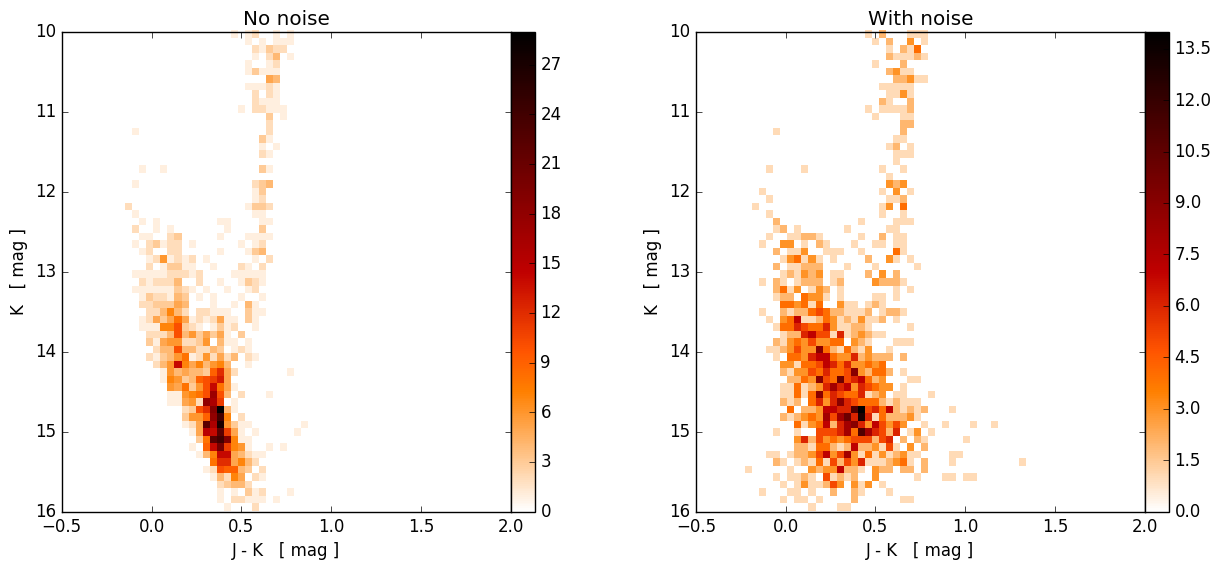}
	\end{subfigure}
	\hspace{1cm}
	\begin{subfigure}[!t]{0.41\textwidth}
	\caption*{ $d = 7 \rightarrow 8$ kpc}
	\includegraphics[width=1.0\hsize]{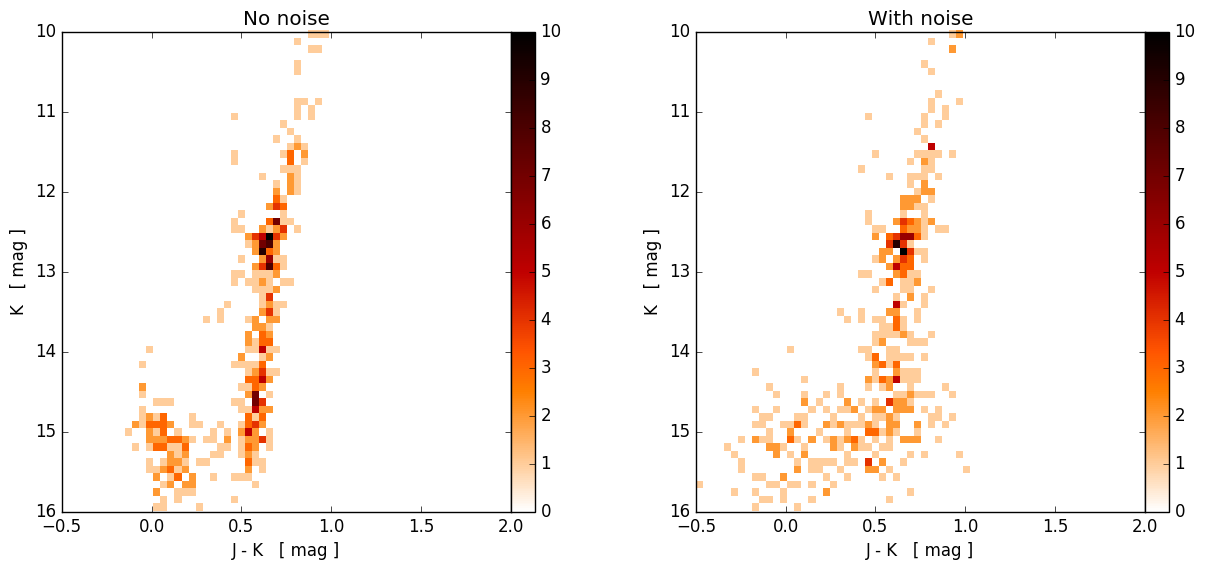}
	\end{subfigure}\\
	\begin{subfigure}[!t]{0.41\textwidth}
	\caption*{ $d = 3 \rightarrow 4$ kpc}
	\includegraphics[width=1.0\hsize]{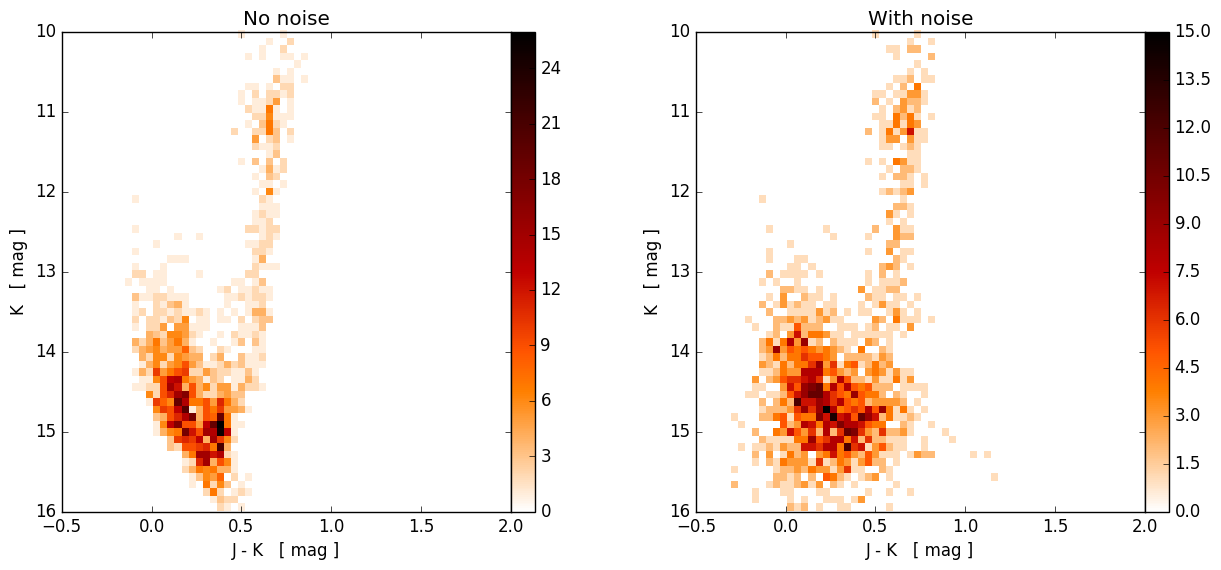}
	\end{subfigure}
	\hspace{1cm}
	\begin{subfigure}[!t]{0.41\textwidth}
	\caption*{ $d = 8 \rightarrow 9$ kpc}
	\includegraphics[width=1.0\hsize]{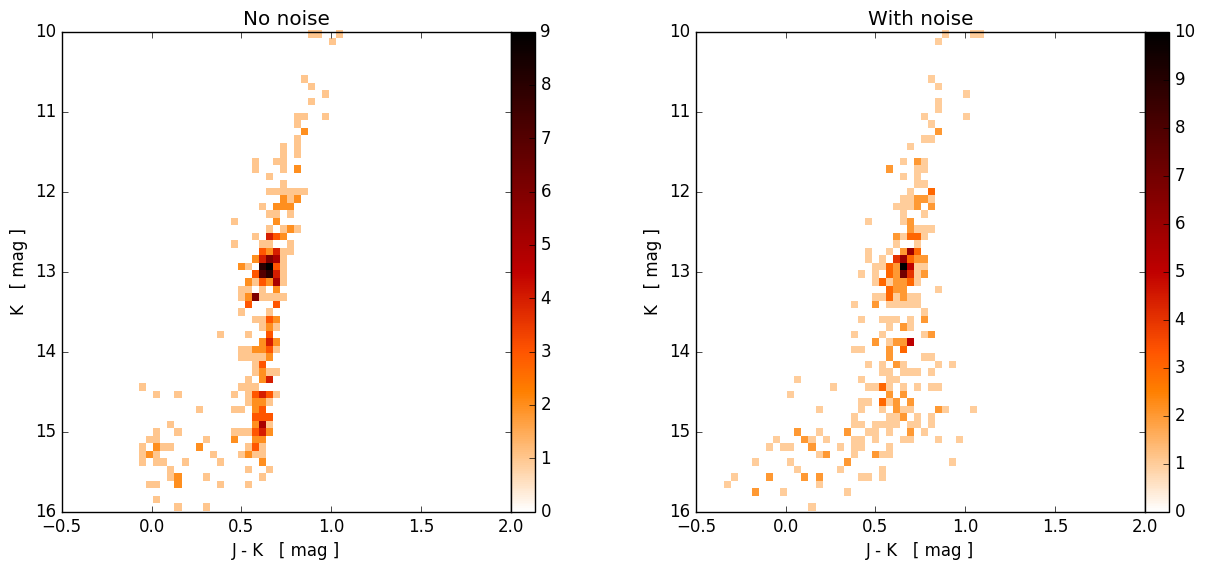}
	\end{subfigure}\\
	\begin{subfigure}[!t]{0.41\textwidth}
	\caption*{ $d = 4 \rightarrow 5$ kpc}
	\includegraphics[width=1.0\hsize]{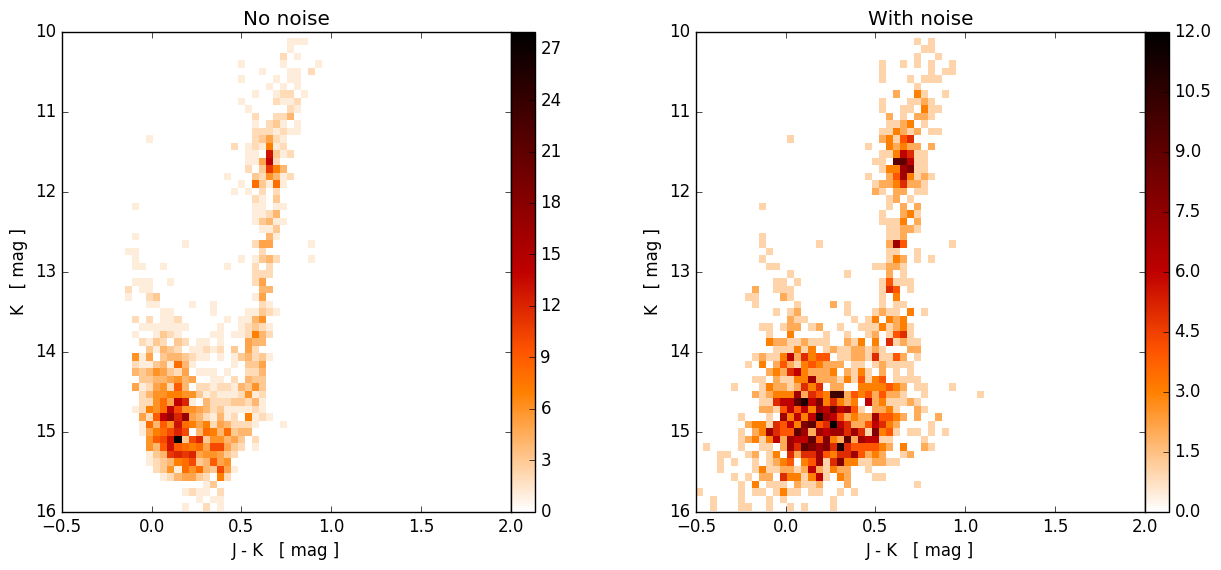}
	\end{subfigure}
	\hspace{1cm}
	\begin{subfigure}[!t]{0.41\textwidth}
	\caption*{ $d = 9 \rightarrow 10$ kpc}
	\includegraphics[width=1.0\hsize]{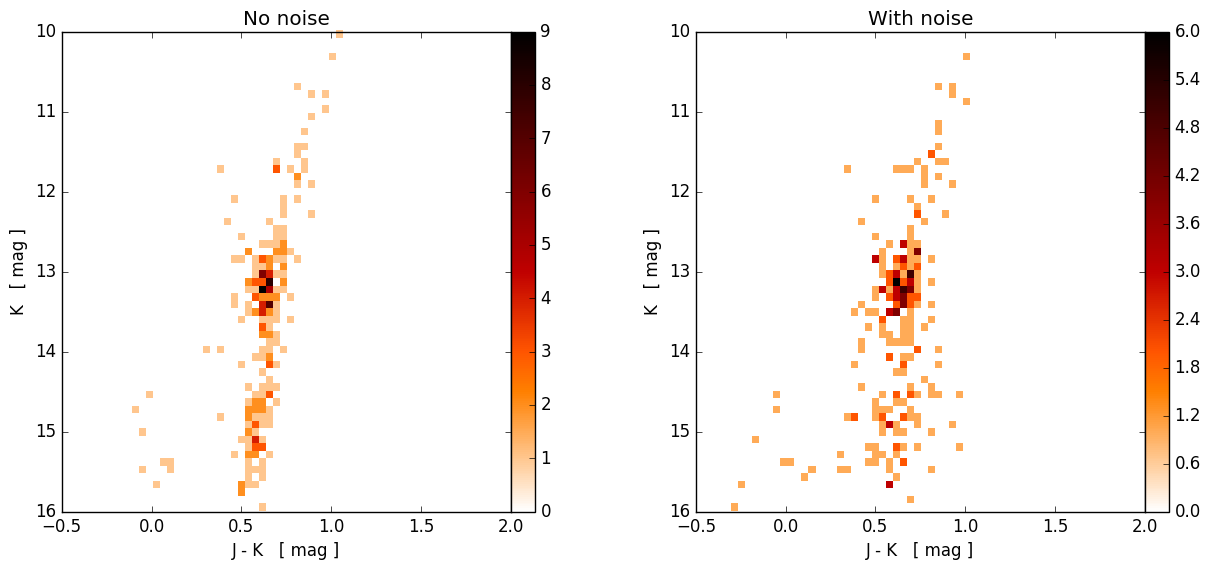}
	\end{subfigure}
	\end{minipage}
	\caption[Effect of distance on a 2MASS CMD]{Effect of distance on a 2MASS CMD. The data are the same as in Fig.~\ref{stellar_types_in_CMD}, but split in 1 kpc bins of distance.}
	\label{distance_slices_CMD}
\end{figure*}

\newpage
We highlight that the various stellar populations distribute differently in this diagram. This is an important aspect to understand which classes of stars are the most useful to the profile reconstruction. For example, without extinction, the giant stars mainly align vertically around a color of [J-K]$\simeq$0.7 forming a vertical continuous distribution in almost all our K magnitude range. The rest of the diagram mostly corresponds to main sequence stars with the highest density being due to the relatively low-mass star population centered around [J-K] = 0.25 and [K]$\simeq 15$ mag. These two populations are represented separately in Figure~\ref{stellar_types_in_CMD} for a modeled 2MASS CMD without extinction, with and without simulated noise (the details on the noise modeling are given in Section~\ref{ext_profile_and_cmd_realism}). Additionally, the respective proportions of each of them varies as a function of their location in the Milky Way, as parametrized by the BGM. This extinction free diagram is also shaped by the effect of distance to the Sun. Figure~\ref{distance_slices_CMD} shows the corresponding CMD for slices of distance for an example LOS at $l = 280$ deg, $b = 0$ deg. It shows that the primary effect of increasing distance (i.e. a vertical shift) also leads to a strong increase in the ratio of giant to main sequence stars, since the latter are to faint to be detected by 2MASS beyond $\sim 8$ kpc. These properties highlight again that the amount of information contained in a simple 2D image is considerable, and that powerful statistical techniques are needed to disentangle this complex information.\\

This choice of using a multi-faceted CMD as input for our method makes clear that, due to the properties exposed in Section~\ref{image_process_section}, the CNN architecture should be more suitable than a classical ANN for this task (see Sect.~\ref{input_output_cnn_dim}). Indeed, a CMD can be considered as an image with the pixel value encoding the quantity in each bin when using a sufficient numerical range, and in the comparison we mostly want to estimate the translation of specific patterns for each stellar population. In the present study we used CMDs of $64\times 64$ pixels with $-0.5<[J-K]<6.1$ and $10<[K]<16$. This choice of resolution and limits is discussed in Section~\ref{input_output_cnn_dim}. However, for a CNN architecture to properly predict on real data we have to assess to what extent the BGM prediction realistically represents an observation. This is of major importance because ANN can easily be biased by systematic differences in the training sample or by non-representative proportions (as seen in Sects.~\ref{class_balance}, \ref{training_test_datasets}, or \ref{detailed_feature_space_analysis_ON}), for example by populating a part of the training CMDs that is never present in real observations, or in the opposite case if the training data lack constrains on parts of the CMDs that contain information in the observations.
	
	\subsubsection{Reproducing realistic observations: uncertainty and magnitude cuts}
	\label{ext_profile_and_cmd_realism}
	
\begin{table}
	\centering
	\caption{Uncertainty fitting free parameters for all 2MASS bands}
	\vspace{-0.1cm}
	\def\arraystretch{1.1}
	\begin{tabularx}{0.80\hsize}{l @{\hskip 0.05\hsize} @{\hskip 0.05\hsize}*{3}{Y}}
	\toprule
	& a & b & c\\
	\toprule
	J				& $7.253 \times 10^{-8}$ & $8.590 \times 10^{-1}$ & $2.258 \times 10^{-2}$\\
	H				& $1.807 \times 10^{-8}$ & $9.894 \times 10^{-1}$ & $2.802 \times 10^{-2}$\\
	$\mathrm{K_s}$	& $2.242 \times 10^{-7}$ & $8.768 \times 10^{-1}$ & $2.044 \times 10^{-2}$\\
	\bottomrule
	\end{tabularx}
	\label{table_2MASS_uncertainty_fitting}
\end{table}
	
On the one hand, an astronomical observed data can be altered in several ways during the acquisition process. Observational instruments have limits to their sensitivity inducing incompleteness, measurement uncertainty, and can even have systematic biases. Usually, these effects are well documented for each instrument which makes it possible to take them into account in the data analysis. On the other hand, a model also have biases or incompleteness of other types that are often difficult to assess, even for models that are constrained by observations. Our objective here is to make the training CMDs and the observed ones as alike as possible. There are two main properties that must be evaluated: the measurement photometric uncertainty, and the magnitude detection limit cut of the telescope. In our approach, these quantities are estimated individually. Anticipating on Section~\ref{2mass_single_los} we note that we will focus on the Milky Way disk between galactic latitudes $|b|<5$ deg and galactic longitudes $257 < l < 303$ deg centered on $l=280$ deg.\\

\begin{figure}[!t]
	\centering
	\includegraphics[width=0.99\textwidth]{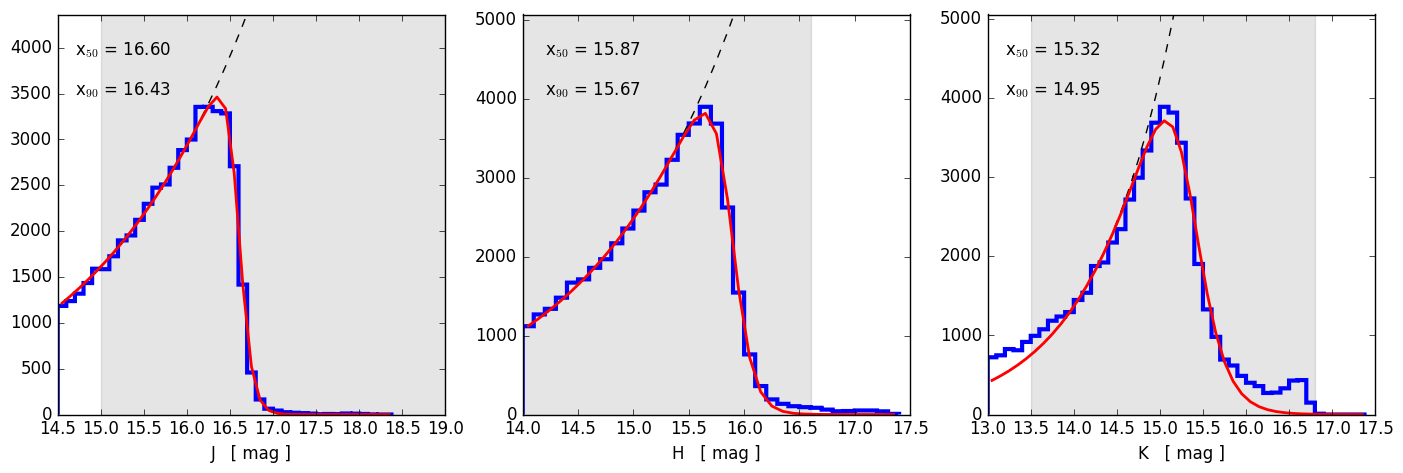}
	\caption[Fitting of the cut in magnitude for the three 2MASS bands]{Fitting of the cut in magnitude for the three 2MASS bands. The blue histograms show the observed distribution, the fitted models are in red. The gray area shows the range of magnitude values included in the fit.}
	\label{2MASS_cut_fitting}
	\vspace{-0.4cm}
\end{figure}

Regarding the magnitude cuts, we fitted data from the 2MASS point source catalog. For this, we downloaded the stars from a 1 $\mathrm{deg^2}$ region, centered on $l=280$ deg and $b=0$ deg. We excluded the stars for which one or more of the J, H, $\mathrm{K_s}$ bands was missing. We then fitted the magnitude histogram for each band individually using the following analytical formula:
\begin{equation}
f(x) = 0.5 a x^\alpha {\cal S}(x) \quad {\rm with} \quad {\cal S}(x) = 1 + \tanh \Bigg( b \times \frac{x_{50}-x}{x_{50}-x_{90}} \Bigg)
\label{magnitude_cut_fit}
\end{equation}
where $a$ and $\alpha$ are free parameters that correspond to the first part of the function following a power law, used as a simple model for the underlying star distribution. The constant $b=\arctanh ( 0.8 ) $ is fixed and $x_{50}$ and $x_{90}$ are free parameters that correspond to the abscissa values at half and $90\%$ increase of the selection function ${\cal S}$. To use this selection in our mock CMDs, for each star we drew a value randomly from the selection function according to the star magnitude for each involved in the CMD. Therefore the shape of the selection cut is reproduced statistically. Figure~\ref{2MASS_cut_fitting} shows the observed star distribution histograms (in blue) and the best fit obtained for each band. \\

\begin{figure*}[!t]
	\hspace{-1.9cm}
	\begin{minipage}{1.23\textwidth}
	\centering
	\begin{subfigure}{0.32\textwidth}
	\includegraphics[width=\textwidth]{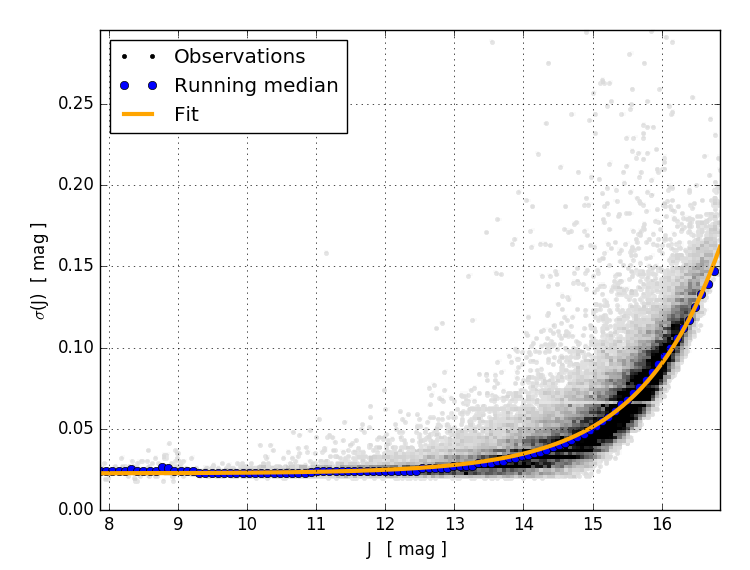}
	\end{subfigure}
	\begin{subfigure}{0.32\textwidth}
	\includegraphics[width=\textwidth]{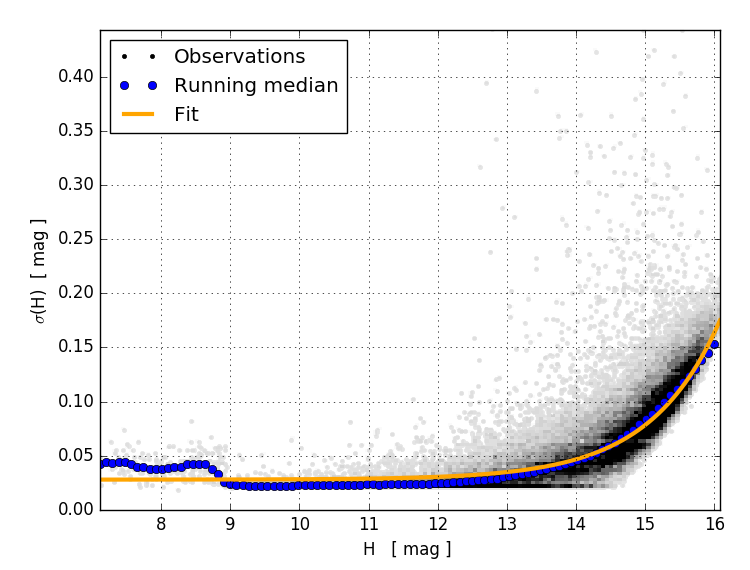}
	\end{subfigure}
	\begin{subfigure}{0.32\textwidth}
	\includegraphics[width=\textwidth]{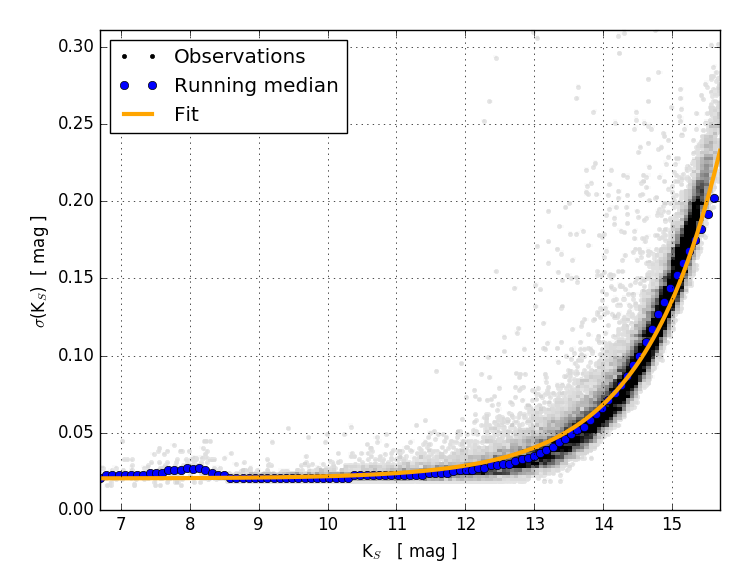}
	\end{subfigure}
	\end{minipage}
	\caption[Fit of 2MASS uncertainties]{Fit of 2MASS uncertainties. The gray dots are 2MASS stars, the gray scale representing the star density in the diagram. The running median (blue dots) is fitted by an exponential model (orange line).}
	\label{fig_2MASS_uncertainty_fitting}
	\vspace{-0.4cm}
\end{figure*}

To evaluate the photometric uncertainty for each 2MASS band we used the same 1 $\mathrm{deg^2}$ region centered on $l=280$ deg, $b=0$ deg. Similar, to the cut fitting, we excluded stars for which at least one of the J, H, $\mathrm{K_s}$ bands was missing but also the stars that do not have all the respective uncertainties. For each band we represented the corresponding magnitude-uncertainty diagram. Figure~\ref{fig_2MASS_uncertainty_fitting} shows that the stars mostly distribute following an exponential law. Following the example of \citet{Robin_2003}, we fitted the distribution with using the following form:
\begin{equation}
\sigma(x) = a  \exp\big (b \, x\big) + c
\label{uncertainty_fit}
\end{equation}
where $x$ is the magnitude of the fitted band and $a$, $b$, and $c$ are free parameters. To overcome the fact that the number of outliers greatly increases toward the higher magnitudes, we first computed the running median (RM) of the distribution, that has the advantage of being robust against outliers. In practice, the RM was evaluated for 100 magnitude bins of 0.3 mag evenly distributed between the 0.1 and 99.9 percentiles of the magnitude interval. The RM has then been fitted using Equation~\ref{uncertainty_fit} without weighting in order to prevent the less represented magnitude bins from being less constrained due to their smaller proportion. The corresponding RM values and our fit results are illustrated in Figure~\ref{fig_2MASS_uncertainty_fitting}. Table~\ref{table_2MASS_uncertainty_fitting} summarizes the free parameter values obtained for the three 2MASS bands. To compute mock photometric errors, the $\sigma$ value of each star was computed using the best fit $a$, $b$, and $c$ values, and was used to draw a random Gaussian error that was added to the errorless magnitude computed by the BGM.\\

\clearpage
	\subsubsection{Simple extinction effect on the diagram}
	\label{simple_extinction_effect_cmd}
	
\begin{figure*}[!t]
	\hspace{-1.4cm}
	\begin{minipage}{1.15\textwidth}
	\centering
	\begin{subfigure}{0.48\textwidth}
	\caption*{\bf \small No extinction}
	\includegraphics[width=\textwidth]{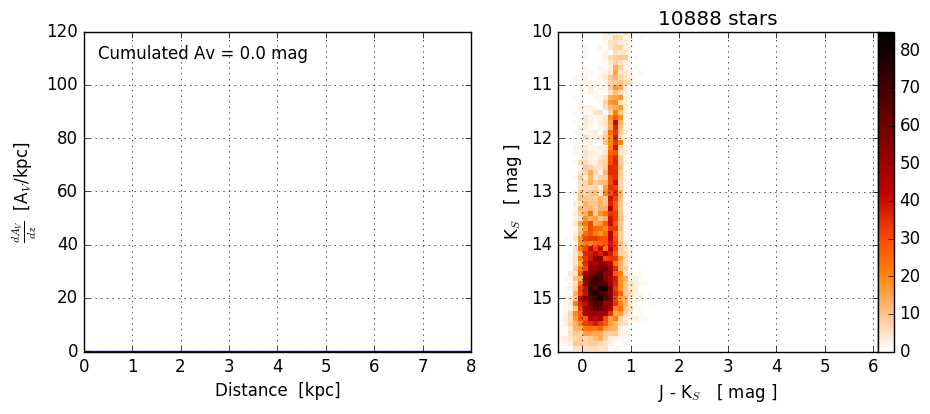}
	\end{subfigure}
	\hspace{0.4cm}
	\begin{subfigure}{0.48\textwidth}
	\caption*{\bf \small 1 Cloud, $\bm{ A_{V} = 3}$ mag, $\bm{d = 2}$ kpc}
	\includegraphics[width=\textwidth]{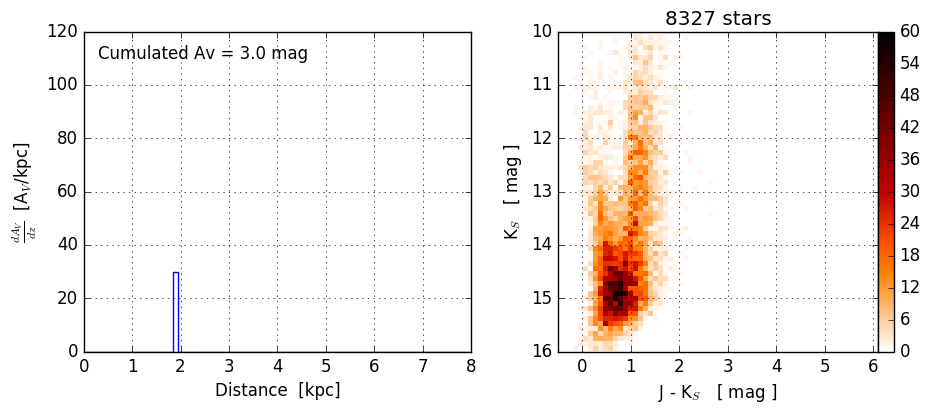}
	\end{subfigure}\\
	\vspace{0.6cm}
	\begin{subfigure}{0.48\textwidth}
	\caption*{\bf \small 1 Cloud, $\bm{A_{V} = 10}$ mag, $\bm{d = 2}$ kpc}
	\includegraphics[width=\textwidth]{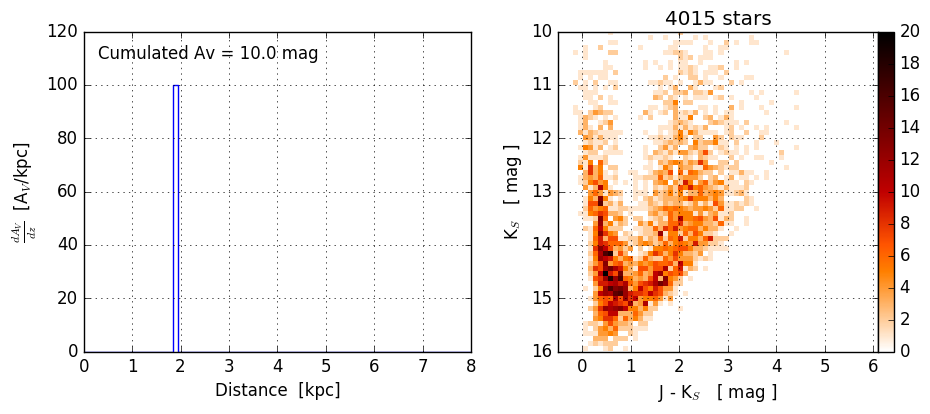}
	\end{subfigure}
	\hspace{0.4cm}
	\begin{subfigure}{0.48\textwidth}
	\caption*{\bf \small 1 Cloud, $\bm{A_{V} = 10}$ mag, $\bm{d = 4}$ kpc}
	\includegraphics[width=\textwidth]{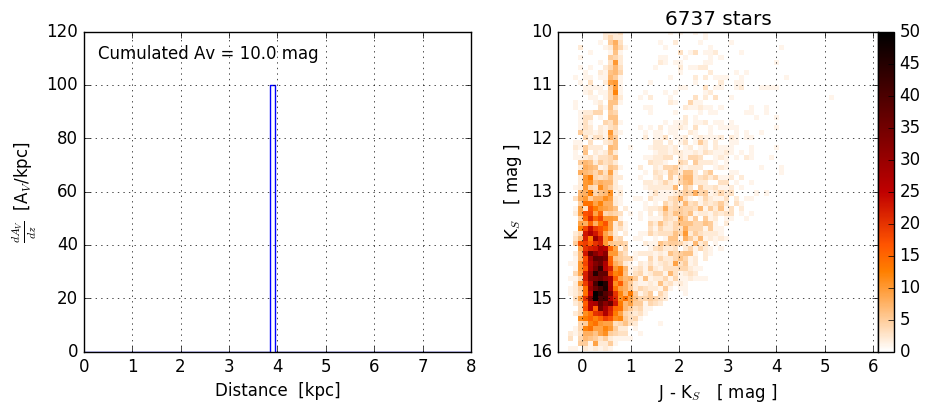}
	\end{subfigure}\\
	\vspace{0.6cm}
	\begin{subfigure}{0.48\textwidth}
	\caption*{\bf \small 2 Clouds, $\bm{A_{V} = 1.5, 6 }$ mag, $\bm{d = 1, 6}$ kpc}
	\includegraphics[width=\textwidth]{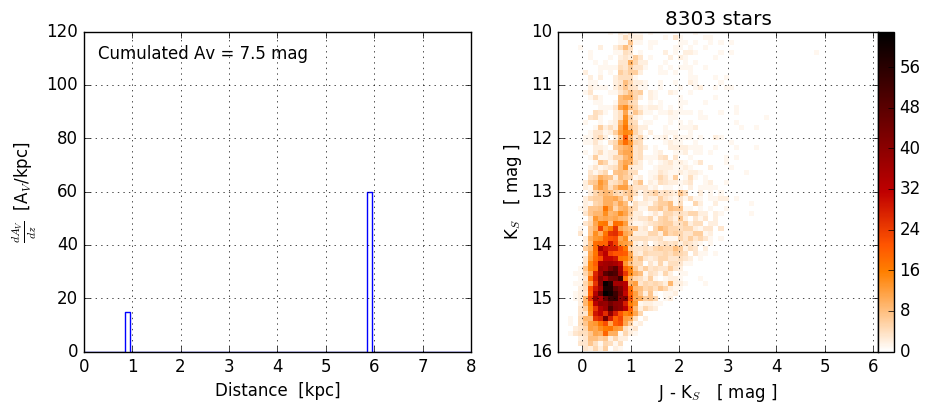}
	\end{subfigure}
	\hspace{0.4cm}
	\begin{subfigure}{0.48\textwidth}
	\caption*{\bf \small Uniform ext - 2 Clouds, $\bm{A_{V} = 1.5, 6 }$ mag, $\bm{d = 1, 6}$ kpc}
	\includegraphics[width=\textwidth]{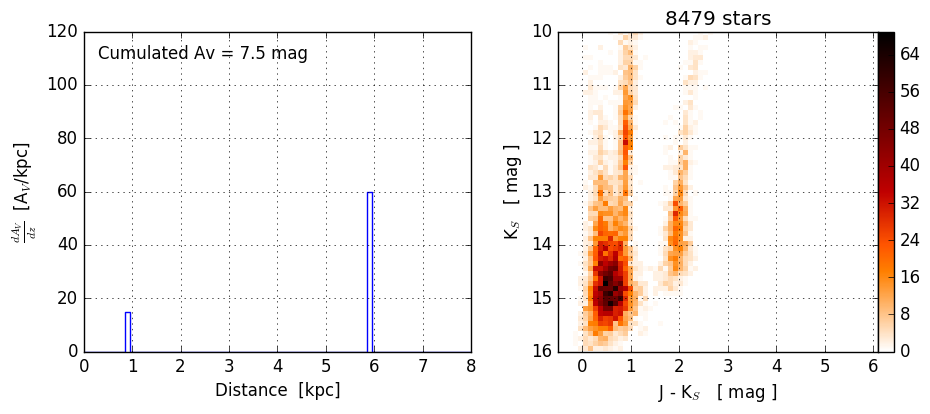}
	\end{subfigure}
	\end{minipage}
	\caption[Effect of individual clouds on the 2MASS CMD]{Effect of individual clouds on the 2MASS [J-K]-[K] CMD. The extinction is modeled as a log-normal distribution, except in the bottom-right panel where a uniform extinction is used.}
	\label{simple_ext_examples}
\end{figure*}

	Now that we have described the effects of observational noise and selection on the CMDs, we can build a realistic CMD without extinction. From this "bare" CMD we illustrate the effect of rudimentary extinction profiles in order to better understand what is the information that the network will be tasked to extract. As we described in Section~\ref{extinction_with_bgm_intro}, a star that is present in a given CMD pixel at the beginning will translate toward a higher color and a lower magnitude. If we consider the case with a single point-like cloud on the LOS, all stars that are in front of the cloud do not move at all, while all others are translated with respect to the extinction quantity of the cloud, as illustrated in Figure~\ref{simple_ext_examples}. In the same figure we also illustrate a simple two-cloud example in which the stars that are behind the clouds are affected by the cumulative extinction. The individual cloud extinction effect is especially visible in the vertical branch of the giant stars.\\

We note that with the typical angular resolution of our 3D extinction maps (15 arcmin per pixel), the sub-beam distribution of extinction is not uniform. For example, at 5 kpc, a 15 arcmin beam covers a physical area of $\sim 20$\,pc, which is enough to contain a whole molecular cloud, with its complex substructures (filaments, clumps, cavities, ...). To take this effect into account, we followed \citet{Marshall_2020} and modeled this so-called fractal structure of the ISM with a log-normal probability density function:
\begin{equation}
   f(A_V) = \frac{1}{\sigma A_V \sqrt{2\pi}} \exp \left( - \frac{\log^2 (A_V)}{2\sigma^2} \right)
   \label{eq:lognorm_ext}
\end{equation}
where $A_V$ is the cumulative extinction from the star to the observer as obtained from our extinction profile. This value can be considered as the mean extinction in the beam. The constant $\sigma$ characterizes the width of the distribution. We adopted a value $\sigma = 0.4$, typical of the values estimated by \citet{Kainulainen_2009} in a score of nearby molecular clouds using near-IR 2D extinction maps derived from 2MASS data. In practice, we used this probability density function to randomly draw the actual value of $A_V$ of each star. The bottom frames of Figure~\ref{simple_ext_examples} show a comparison of the produced CMD using the same extinction profile but with a log-normal and uniform extinction distribution, respectively.\\
	
We showed in Fig.~\ref{obs_model_ext_comparison} that extinction corresponds to a shift in CMD diagrams. This provides a very important insight about the expected resolution of the predicted extinction. Indeed, due to the position of the CMD and to the translation direction, it is challenging to retrieve more than 30 discrete values for the height of an extinction bin in the profile. In addition, the distance at which the extinction must be placed in the profile corresponds to the fraction between the number of reddened stars and the number of not reddened stars. Consequently the best achievable distance resolution is related to the number of stars per CMD bins, which is of a few tens of stars per bin for the example LOS. We elaborate more on consequences of the choice of CMD resolution in Section~\ref{input_output_cnn_dim}\\

Despite the relative easy understanding of the extinction effect on individual stars in this CMD it is striking from Figures~\ref{stellar_types_in_CMD}, \ref{distance_slices_CMD} and \ref{simple_ext_examples}, that the combination of the different stellar classes, the spatial variations of the Galactic structures, the observation noise, and sub-beam extinction distribution is a very complex problem for a 2D CMD representation. This justifies the need for a highly non-linear method that would be able to automatically extract all these correlations from the intricate CMD, just like our CNN formalism.
	
\clearpage
\subsection{Creating realistic extinction profiles for training}
\label{GRF_profiles_section}

We remind the reader that the objective of the study is to reconstruct the underlying dust distribution of each LOS, represented by a differential extinction profile. In this section {\bf we prepare a training sample to train an ANN} to perform this task. We used the BGM to generate many realistic star distributions on given lines of sight. Then, applying a given extinction profile to this star distribution we construct a mock extincted CMD. The network will then take this CMD as its input and will learn to predict the extinction profile that was used by taking it as its target.\\

To be capable of predicting extinction profiles from observed CMD, the ANN must have been trained using realistic extinction profile examples. It means that we have to find a prescription to construct sufficiently realistic example to train the network. One approach could be to use simulations of the interstellar medium \citep[e.g.][]{padoan_2017} but it would require very large hardware facilities considering the fact that we are interested in large distances at the Milky Way scale and due to the number of examples that will be necessary to train a relatively large ANN architecture (Sects.~\ref{cnn_hyperparameters}, \ref{2mass_single_los_training_and_test_set_prediction}, \ref{2mass_multi_los_training_and_test_set_prediction}, \ref{Gaia-2MASS_single_los_training_and_test_set_prediction}, and \ref{Gaia_2mass_multi_los_training_and_test_set_prediction}). Another approach would be to use previous 3D extinction maps of other methods as a prior for our training profiles but it would be difficult to assess the induced bias in our own results. Instead, we adopted a lower level approach that consists in creating mock training profiles from Gaussian Random Fields \citep[GRF, e.g.][]{Sale_2014} and to tune general construction parameters to correspond to our need.\\

\subsubsection{Gaussian Random Fields}

We succinctly describe here the necessary elements to construct 1D Gaussian Random Fields (GRFs) that are then used to construct our profiles. Details on the method formalism that we depict here can be found in the appendix B of \citet{Sale_2014}. The objective here is to construct a realistic dust density profile, for which the logarithm can be approximated by a GRF. In the present method we first start by assuming that the log of density $\rho$ (i.e. of differential extinction) has a power law $|FT(\log \rho^2(k))| \propto k^{-2\beta}$, where $FT$ is the Fourier transform, and $k$ is the spatial frequency. Each point in the Fourier space then receives a complex magnitude drawn randomly from a Gaussian probability distribution of standard deviation $k^{-\beta}$ that corresponds to the square root of its value from the power law. A complex phase is added to each point randomly in the $0 < \phi < 2\pi$ range. Applying the inverse Fourier transform to this space generates a GRF that follows the desired power spectrum. The final profile of differential extinction is obtained by exponentiating the GRF following $\frac{dA_V}{dz}(z) = \exp( \sigma \mathrm{GRF} )$. Two parameters can be tuned in this formalism in order to control the properties of the constructed profile, (i) a $\beta$ parameter that controls whether the peaks of the predicted density are narrow and frequent or more sparse and large, and (ii) a $\sigma$ parameter that acts as a posterior scaling of the profile, being responsible for the contrast between the peaks and the lows in the profile. Example of the generated GRF profiles are in Figure~\ref{grf_param_illustrations} for two different sets of $\beta$, $\sigma$ parameters.\\

\newpage
\subsubsection{GRF generated profile}

\begin{figure}[!t]
	\centering
	\includegraphics[width=0.9\hsize]{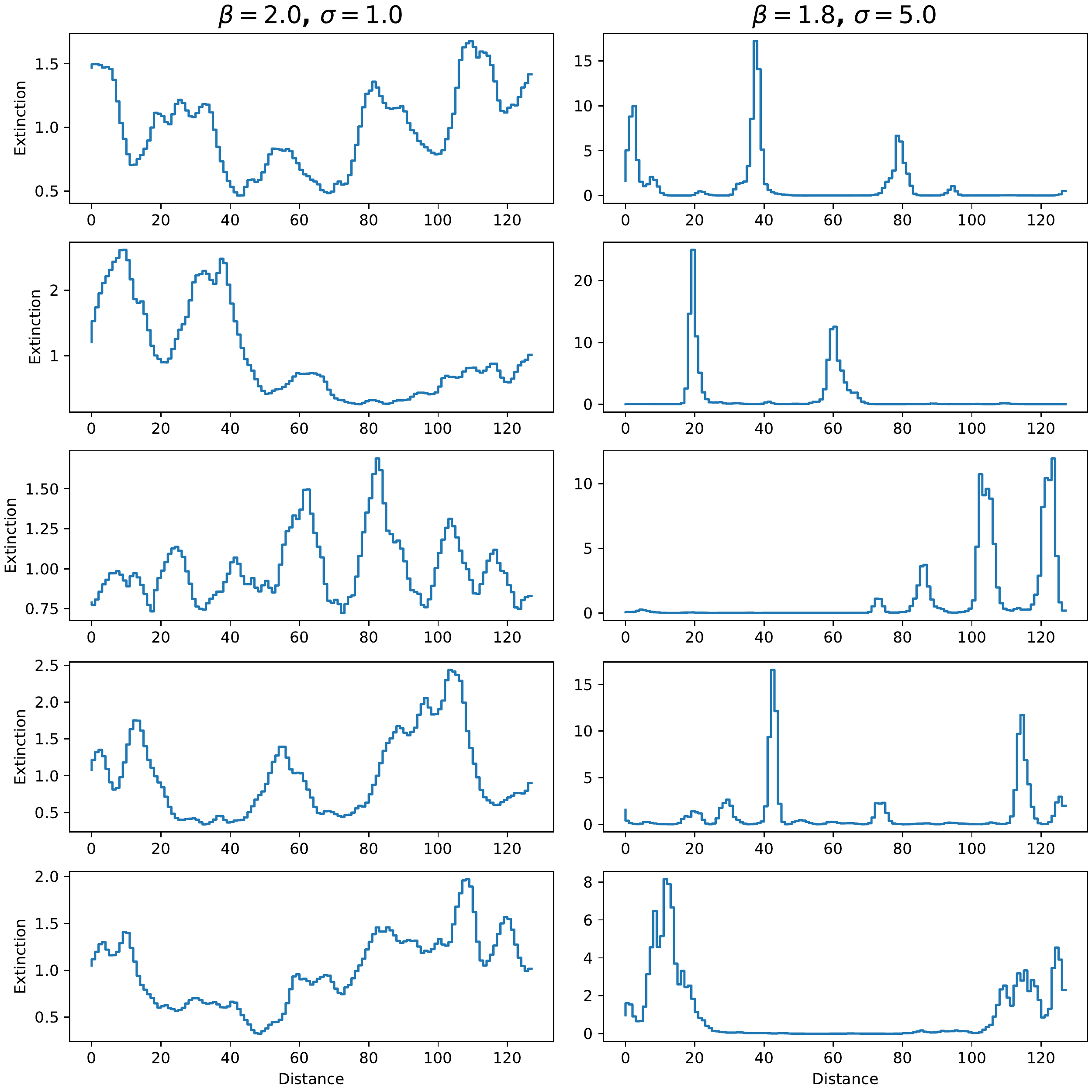}
	\caption[Examples of GRF realizations]{Examples of Gaussian random field realizations with the two sets of adopted parameters, as indicated above each column. We emphasize that the ordinate scales are different in each example.}
	\label{grf_param_illustrations}
\end{figure}

Using this formalism we tested many combinations of $\beta$ and $\sigma$ to assess which one was suitable for our application. One important point to notice here is that, providing the network with a target that contains too much details in comparison to the amount of information that is contained in the inputs is counterproductive. Indeed, the fine-grained error that would be produced from the target output comparison in such a case would not find any input information that correlates with this error. This would induce a non meaningful correction to all the weights of active neurons, globally adding noise to the network preventing it from properly converging to the information effectively accessible. Therefore, the training profile realism must also be limited. We noticeably tested training profiles that contains several narrow structures close to each others, and the network was enable to reconstruct them only finding smoother structures. We also tested to create training datasets that were based on random $\beta$ and $\sigma$ values within given ranges, but the diversity was too large for the network to converge without over-sized datasets. \\

\begin{figure}[!t]
	\centering
	\includegraphics[width=0.7\hsize]{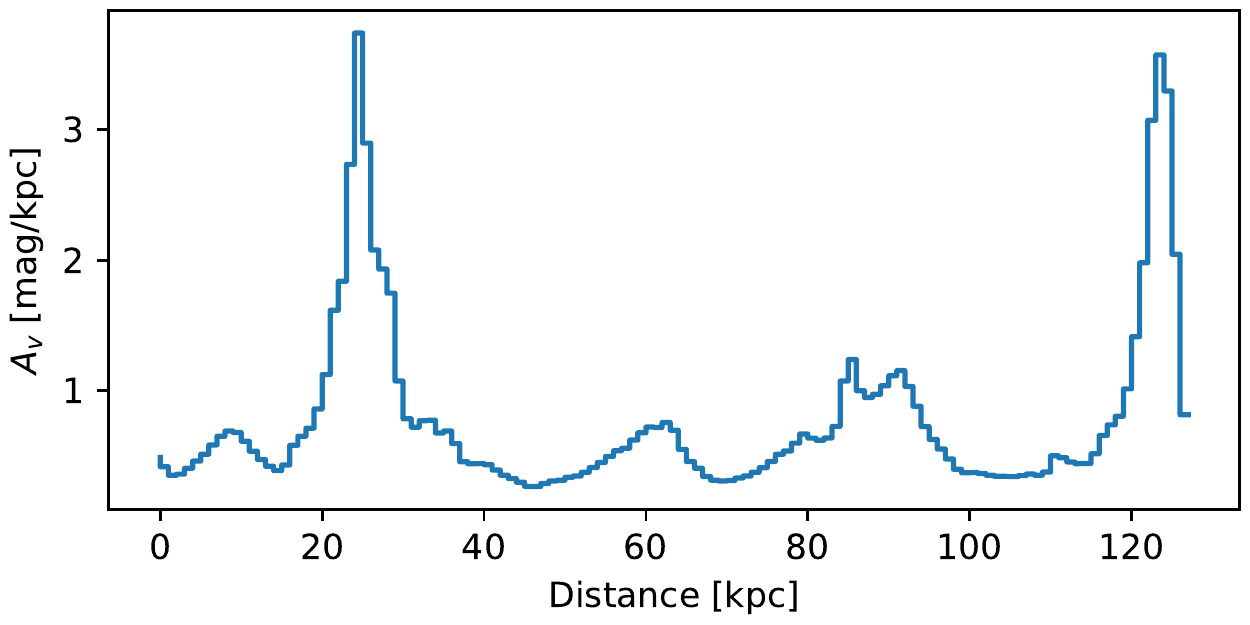}
	\caption[Example of a combined Gaussian random field profile]{Example of a profile obtained by summing two Gaussian random field profiles, one with $\beta=2.0$ and $\sigma=1.0$, the other with $\beta=1.8$ and $\sigma=5.0$, weighted with a random value randomly drawn between 0 and 1. }
	\label{grf_fractional_sum}
\end{figure}

By looking at the predictions from other maps and also at the prediction capacity of the network on several tested parameters we settled on an approach that makes a combination of two individual GRF profiles. We generated a profile with $\beta = 2.0$ and $\sigma = 1.0$ to obtain large structures (low spatial frequencies) along the LOS, in order to represent the more diffuse ISM. The second profile was generated with $\beta = 1.8$ and $\sigma = 5.0$ to represent more compact structures (higher spatial frequencies), representative of molecular complexes. We illustrate a few profiles from each of these two sets of parameter in Figure~\ref{grf_param_illustrations}. To construct a single training profile, a realization of each of these two types of GRF was summed using a random fraction between 0 and 1 for each of them. We illustrate a typical result of this combination in Figure~\ref{grf_fractional_sum}. The profile is then scaled based on its maximum value using a random $A_{V,\rm max}$ between $10^{-2}$ and 100 mag/kpc. To avoid having profiles with a very strong total extinction, which could occur with the GRF, we excluded any profile for which the total cumulative extinction is higher than $A_{V,\rm cumul} > 50$ mag. The latter effect is visible in several of our profiles in our result section (e.g. in Figures~\ref{single_los_2MASS_test_profiles_prediction} and \ref{multi_los_2MASS_test_profiles_prediction}). The exact values of our profile generation were tuned to maximize statistical similarities between our profiles and the prediction from other extinction maps, by looking for example at the maximum peak extinction distribution, or at the integrated extinction distribution. Tuning the parameters of the GRF profiles in order to modify their statistical properties is very similar to our training dataset rebalacing from Section~\ref{training_test_datasets}. More profiles generated using the same approach are visible in subsequent figures that illustrate the network predictions in the following sections (e.g. Figs.~\ref{single_los_2MASS_test_profiles_prediction} or \ref{multi_los_2MASS_test_profiles_prediction}).\\

\subsubsection{Profile star count limit and magnitude cap}
\label{zlim_subsection}

The obtained profiles are the ones that are effectively applied to the bare modeled CDMs to obtain the final mock CMDs of the training sample. However, after computing the mock CMDs, we performed two last transformations of the profiles before using them as targets, still following the idea that we should not have targets that are impossible to reproduce. The first transformation is motivated by the possibility that, despite our limit in $A_{V, \rm cumul}$, some profiles can have an sufficient cumulative extinction to completely screen the stars beyond a certain distance. In other words, it is possible that parts of the profile are not constrained at all, or not by a sufficient amount of stars. To account for this effect, we manually defined a star count limit $Z_{\rm lim}$ that was used to force the target profile to zero after a certain point. In practice, after the application of the full extinction profile to the star list, we searched the farthest distance beyond which remained only $Z_{\rm lim}$ stars. For each training profile every distance bin beyond this limit was set to zero, which is illustrated by the cut profiles in Figure~\ref{single_los_2MASS_test_profiles_prediction}. Interestingly, since the full profile and not the cut profile is used to compute the input CMD extinction, it only means that the network is trained to consider all the confused cases where there is not enough stars anymore to be zero, still conserving a fully realistic CMD. From a classification standpoint, it can be seen as making one large class that contains all the cases that are too difficult to discriminate and attribute the same target to all of them.\\

The second transformation followed a similar idea. We also capped the maximum extinction per bin in the target profile to $dA_V/dz = 50$ mag/kpc despite the $dA_V/dz = 100$ mag/kpc permitted by the GRF profile construction. Just like for the $Z_{\rm lim}$ cut, this modification is only made on the target profile while the CMD is still affected by the full profile. Again it just acts as an additional clustering of the cases with large extinction per bin, strongly stabilizing the training and improving the global network prediction. This choice is justified because we are more interested in the dust distribution rather than the exact extinction quantity at a first time. Additionally our various results showed that the cases where the predicted profile is saturated are rare (Sect.~\ref{2mass_maps_section}, e.g. Fig.~\ref{single_los_2mass_polar_plan}).\\

Finally, even if it is possible to generate profiles with a very small total extinction, we observed that adding some flat extinction profiles in the training sample significantly improved the prediction results. The might be due to the fact that the network has to see the original distribution of the stars in the bare CMD directly from the model without extinction. It helps better constraining the reference pixels for each star. We control the proportion of profiles that we manually set to zero using the $f_{\rm naked}$ parameter that is usually set at $0.01$ so that 1 out of 100 profiles is a flat zero extinction one. This might seems a large amount but we expect to have many predicted profiles with very faint extinction, especially if we try to perform prediction outside of the Galactic Place. Still, we exposed in our YSO application that a better representation of the most common case is a way of reducing false positive in the more rare classes. In other words, since ANN works by assessing differences between cases, having a strong constraint on what a null or faint profile looks like significantly reduces the noise in our prediction and increases the confidence one can have when there is a detection.\\

\newpage
\subsection{Tuning the method}

\subsubsection{Input and output dimensions}
\label{input_output_cnn_dim}

In the present section we describe some general network architecture properties that are shared by our different applications in Sections~\ref{2mass_maps_section} and \ref{gaia_2mass_ext_section}. First, it is important to note that the angular resolution in the plane of the sky will strongly affect how many stars will be present in our lines of sight, therefore affecting the proportions in the CMDs. For all our applications we adopted a pixel size of $0.25^\circ$, corresponding to a $0.2 \mathrm{deg^2}$ surface on the plane of the sky. In practice it means that for each pixel we built the forward sample from a query of the 2MASS or Gaia catalogs within an area of this size. \\

The input volume of our CNN is always composed of images of size $64\times 64$, representing different diagrams, in the case of 2MASS the [J-K]-[K] CMD. Considering the adopted LOS area, this CMD resolution has been identified as a proper balance between the number of pixels in the CMD and the number of stars per bin, following the considerations on how the extinction profile is encoded in the CMD (Section~\ref{simple_extinction_effect_cmd}). Indeed, in the case of a too low resolution (i.e. only few pixels in the CMD) the network is unable to assess properly both the original position of the stars and shift length in the diagram, leading to imprecise extinction quantity prediction at a given distance. In the opposite case of a too high resolution, the number of stars per bin becomes too small for the network to properly assess how many stars have moved due to extinction, making the distance estimate very inaccurate. We note that our input CMDs are systematically normalized by a simple scaling between 0 and 1 according to the maximum pixel value in the full training dataset, which works well with a CNN architecture.\\

Regarding the predicted profile resolution, we opted for 100 pc bins. We expect, somewhat naively, that it corresponds to the typical resolution one could expect around 2 to 3 kpc. Closer distance estimates could be slightly under resolved compared to the information we expect to be contained in the CMD. At larger distances, the profile will be clearly over resolved compared to what can be extracted from the CMD. In this case the network prediction is expected to be spread across several bins, following the distance uncertainty. We note that the same resolution was adopted by \citep[][in prep.]{Marshall_2020}. Testing progressive bin sizes would certainly be worth testing in the future. \\

We choose to have profiles of 128 bins allowing a maximum distance estimate of 12.8 kpc. We observed that having a maximum distance at 10 kpc induced boundary artifacts when there was indeed structures around this maximum distance. As we show in Section~\ref{los_combination}, our method manages to reconstruct extinction structures at distances around 10 kpc when looking at sufficiently populated line of sights. From a network perspective, these profiles are encoded using a set of 128 linearly activated neurons representing each bin in the profile. It means that each bin is independently activated and that any correlation between nearby bins is the result of the network training without any prior on the form of the output other than the list of training target profiles. The profile is numerically encoded using the differential extinction per bin $dA_V/dz$ in units of mag / 100 pc, which is similar to considering that the profile is an array of total extinction $A_V$ in each 100 pc-bin. These values are normalized using a constant division by 5, which corresponds the maximum possible value of a single bin. This leads to targets in the range 0 to 1, which works well for linearly activated neurons. We note that we did not force any output to be above 0, so it happened that some profiles present slightly negative predictions. When negative values were present in a map prediction, they were changed to 0.

\vspace{-0.2cm}
\subsubsection{Network architecture}
\label{cnn_architecture_test}

\vspace{-0.2cm}
We have described our input image and our output layer dimension for the targeted and predicted profiles. They define the boundary dimensions of the network between which we can arrange our internal network architecture. In the following paragraphs we compare the capacity of several network architectures based on their error on a modeled test dataset after training.\\

\vspace{-0.2cm}
We first attempted a fully connected architecture, considering each input pixel of the input CMD as an independent feature. We tested several variations with up to 5 layers of various size (up to 4096 neurons on each layer) and used several of the improvements from the Section~\ref{cnn_parameters} that are suitable for regular ANN, like the change in activation function for leaky ReLU, a better weight initialization, and the use of a dropout. While such an architecture is still suitable for the task to a certain extent, it never managed to get to a similar prediction quality than a carefully designed CNN architecture. It is interesting to note that the high number of weights induced by a fully connected architecture is not the limiting factor in our case since our best CNN network has a number of weights that is of the same order of magnitude.\\ 

\vspace{-0.2cm}
We then tried to use classical architectures (LeNet, AlexNet, VGG, ..., Sect.~\ref{cnn_architectures}) as an inspiration for our own convolutional layer construction. Despite many careful attempts almost no architecture with 3 or more convolutional layers was able to perform even as good as a fully connected one. We found that one solution to improve the performance is to have few convolutional layers that quickly increase their number of filters up to above 256 before the fully connected layers with a minimum amount of image size reduction. Such architecture barely outperformed the fully connected architecture in spite of a huge increase in computational time.\\

\vspace{-0.2cm}
From this observations we evaluated why the fully connected architecture was performing so well on this task. First, we remind that a convolutional layer extracts pattern in the images. At first glance it seems to be the operation that we want to perform here since we mostly want to detect an echo of a given pattern (typically, the giant branch) a several positions in the image. Additionally it is more indicated when a pattern to detect can be a different places in the image, which is not the case of our CMD for which the reference pattern is always at the same place for the stars that are not affected by extinction. Then the most common CNN architectures progressively reduce the image dimensionality and increase the number of filter, corresponding to the amount of feature. This architecture is well suited to differentiate between many very different objects that might share some sub-patterns. In our case this is not really the expected operation. In Sections~\ref{simple_extinction_effect_cmd} and \ref{input_output_cnn_dim}, we stated that the information about the profile is encoded in the CMD in the form of: (i) a shifting amount in pixel corresponding to the extinction quantity, and (ii) a ratio between pixels assessing how many stars have moved from their original position, corresponding to the distance. From this it is more evident than common CNN architectures that are mainly design around classification are not suitable for the task.\\

\vspace{-0.2cm}
Our approach to improve the results using the CNN formalism was then to assess what a convolutional layer can add to the fully connected architecture. Noticeably, it is efficient to find some patterns, to reduce noise in the original image, and to find a similar representation of the input with less dimensions. Each of these reasons is sufficient to justify the addition of at least one convolutional layer to the fully connected architecture. Indeed, even with as few as 4 filters of size $5\times 5$ with a stride of $1$ and then 3 fully connected layers similarly sized to the fully connected architecture, the prediction result was significantly improved. This attests that few filters are sufficient to average, denoise and strengthen the most important patterns in the CMD. We then explored various small improvements around this very simple architecture. \\

Following the notation introduced in Section~\ref{cnn_architectures} and including the dropout notation in a dense layer as D-n\_$d_r$ where $d_r$ is the dropout fraction, we list here a few of the architectures that we explored: 
\begin{enumerate}
\setlength\itemsep{0.2em}
\item {[I-64.64, C-4.5, $2\times$[D-3072\_0.1], D-2048, D128]}
\item {[I-64.64, C-8.5, P-2, $2\times$[D-3072\_0.1], D-2048, D128]}
\item {\bf [I-64.64, C-12.5, P-2, $\bm{2\times}$[D-3072\_0.1], D-2048, D128]}
\item {[I-64.64, C-8.5.2, P-2, $2\times$[D-3072\_0.1], D-2048, D128]}
\item {[I-64.64, C-6.5, C-8.5, $2\times$[D-3072\_0.1], D-2048, D128]}
\item {[I-64.64, C-6.5, C-8.5, P-2, $2\times$[D-3072\_0.1], D-2048, D128]}
\item {[I-64.64, C-6.5, P-2, C-8.5, P-2, $2\times$[D-3072\_0.1], D-2048, D128]}
\item {[I-64.64, C-6.5.2, C-8.5, P-2, $2\times$[D-3072\_0.1], D-2048, D128]}
\item {[I-64.64, C-12.3, P-2, C-24.3, P-2, $2\times$[D-3072\_0.1], D-2048, D128]}
\end{enumerate}
and many other variations, including modifications to the dense part.\\

We found that the architecture 3 in this list was the one that had the best balance between prediction quality, much better than the fully connected one, and computational efficiency. The first convolutional layer is used with a padding of two $P=2$ to conserve the input dimensionality and apply 12 filters of size $5\times 5$ with a stride of $S=1$. This leads to 12 activation maps that conserve the image resolution, and therefore preserve the shift quantity in the input CMD. Connecting these maps directly to the dense part of the network would induce a very large number of weights which would make the training much more difficult and the error convergence much more noisy as well as much slower. We had two choices for the reduction dimensionality, a Max-Pooling or a stride of 2 for the first convolution. We kept the first one since it was providing significantly better predictions. Adding more convolutional layers after this first construction almost always led to less good predictions and a significant increase in computational time. The end of the network is then made of two dense layers with $n=3072$ leaky-ReLU neurons with dropout. We discuss the choice of dropout rate in the following Section~\ref{cnn_hyperparameters}. There is then a last smaller dense layer with $n=2048$ leaky-ReLU without dropout, and then the output layer and its 128 linear activations corresponding to the extinction profile.

\subsubsection{Network hyperparameters}
\label{cnn_hyperparameters}

Our choice of hyperparameters regarding the selected architecture is the result of a meticulous manual exploration. Due to the number of hyperparameters to tune and to the time required to train a network on a given set (Sect. \ref{cnn_computational}), it was unrealistic to attempt an automated exploration of a large hyperparameter space, therefore it might exist a combination that works better than the one we adopted. We note that the adopted combination provides similar network training behavior in all the cases that are exposed in our results Sections~\ref{2mass_maps_section} and \ref{gaia_2mass_ext_section}, therefore the following description is valid for all the following applications. To better understand the choice of hyperparameters we already state here that our typical training dataset size will range between $5\times 10^5$ and $2\times 10^6$. We note that the few parameters that are not mentioned here follow the prescription from the corresponding section, usually following the CIANNA default values. \\

The network weights are initialized using the Xavier normal distribution (Sect.~\ref{cnn_weight_init}, equation~\ref{eq_xavier_normal}). We adopted a batch size of 32, which provided an appropriate balance between the computational time of a large enough batch (Sect.~\ref{gpus_prog}) and small weight updates to efficiently resolve the error in the weight space and converge in a reasonable number of epochs (Sect.~\ref{descent_schemes}). Interestingly, we observed that having momentum on this architecture was mostly preventing the network from reaching its optimal value while only slightly speeding up the error convergence. It is possible that the learning rate decay that we used was somehow redundant with the momentum effect, and decided to not use the latter. The training is then decomposed into 3 blocks of 50 epochs with their own learning rate. The blocks have individual learning rate prescription with an exponential decay that follows the equation:  
\begin{equation}
\eta(t) = \eta_{\rm min} + (\eta_{\rm max} - \eta_{\rm min}) \exp\big (-\tau t \big )
\label{eq_decay}
\end{equation}
where $\eta$ is the learning rate as a function of the epoch, $\eta_{\rm max}$ is the starting learning rate at the beginning of the block, $\eta_{\rm max}$ is the asymptotic minimum value, $\tau$ is the decay rate and, $t$ the current epoch number. We note that each block count its own epoch from zero. The first block starts with $\eta_{\rm max} = 0.002$ and decreases exponentially toward $\eta_{\rm min} = 0.001$. The second and third blocks are identical, starting at $\eta_{\rm max} = 0.0015$ and aiming at $\eta_{\rm min} = 0.001$. All the blocks have the same decay rate of $\tau = 0.005$. Most of our networks converge between the epoch 50 and 100. We kept the third block in case, similarly to a simulated annealing technique, the sudden increase in learning rate at the beginning of the third block gets the weights out of a local minima. In practice we found it to be rare, since the network properly converged almost all the time before epoch 100.\\

Regarding the dropout rate in the dense layer at the end of the network, we observed that having at least a small dropout is absolutely necessary on this architecture. Without dropout, repeating the same training several time could lead to inconsistent predictions on observed data and very noisy predictions, even if the valid and test dataset errors where very similar between the different training. We note that this is not solely due to a possible overfitting induced by potentially too large dense layers, since we observed the same behavior even with much smaller ones. Still, the dropout solves this issue very nicely and allowed us to estimate uncertainties on our predicted profiles. In practice, we adopted a dropout rate of $dr = 0.1$ for the two first dense layers afters the convolutional part as in \citet{Gal_2015}, the last layer being smaller and free of dropout since we observed that it evened the prediction between several training without a significant impact on the uncertainty predictions. However, in order to assess if the network produces uncertainties that are representative of a true underlying dispersion, we should have make several training, slowly increasing the dropout value to find the point where the dispersion do not increase anymore. Larger dropout would imply to resize the layers accordingly, which would increasing the raw computation time of these larger layers and would significantly increase the number of epochs required to converge (Sect.~\ref{dropout_sect}). Such an exploration was not compatible with the time given to the present study, but it would be an interesting future development in order to have accurate prediction uncertainties. Still, we note that an identical value of dropout rate was used to reproduce accurate posterior-error measurement on a regression case in \citet{Gal_2015} even more efficiently than other usual methods. In any case our dropout rate is useful to overcome undesirable effects on the prediction and provide at least a first estimate of the uncertainty morphology in large predictions (Sect.~\ref{2mass_single_los}).\\
\vspace{-0.6cm}

\newpage
\subsubsection{Computational aspects}
\label{cnn_computational}

\vspace{-0.2cm}
All our network trainings were performed on Deep-Learning dedicated GPU cluster nodes from the "Université de Franche-Comté" Mesocenter. These nodes contain a total of 7 Nvidia Tesla V100 GPU, which was at this point the most powerful Nvidia professional GPU for servers. Each of these card contains 5120 CUDA cores clocked at 1.38 GHz corresponding to 14.13 TFLOPS in FP32. We had access to two sub-models of V100 with either 16 or 32 GB of dedicated HBM2 memory that uses a 4096 bit interface to reach a bandwidth of $\sim$900 GB/s. Using one of these monstrous GPUs, our CIANNA framework managed to train at a rate of $\sim$7000 examples/s. Increasing the batch size, it would be possible to achieve up to $\sim$18000 examples/s for training but at the cost of more epochs needed and a less good overall error convergence. We note that these GPUs are equiped with first generation Tensor Cores, but that we did not used them since we did not had the time to add Mixed-Precision training in CIANNA.\\

\vspace{-0.1cm}
Depending on the training dataset size, our training process required between 2h and 8h to converge, and had a memory usage between 6 and 190 GB. The timescale and memory usage match with most CNN applications, considering for example that the training in the AlexNet \citep{alexnet_2012} study needed 6 days on an old generation $\sim$3 TFLOPS total GPU machine. Our persistent memory usage is also very high since testing different dataset constructions and storing the network weights all uncompressed we would accumulate several TB of data production. While there are still a lot of improvements possible from a numerical performance and memory usage standpoint, we highlight that this is already the result of large optimizations and meticulous choices of relevant data. This study would certainly not have been possible on a similarly sized hardware infrastructure without all the time we invested in tuning our own CNN framework to be optimized for the task and our careful dataset construction.\\

\vspace{-0.1cm}
Finally, while our training process is very computationally intensive we stress that using the trained network to produce large scale maps with dropout uncertainty is a matter of minutes using a mid-range laptop GPU Nvidia P2000-mobile with 768 CUDA cores clocked at 1.6 GHz. Still, most of our map predictions using our trained networks was made on a server belonging to the UFC Computational Physics Master that is equipped with a much more recent Nvidia Quadro RTX-5000 GPU. Overall, the network architecture is very computationally efficient and the time to train the network is only a consequence of the complexity of the problem to solve, which requires to constrain a large number of parameters. Otherwise, the prediction of a large scale map from a single path without dropout is always a matter of second on any GPU. This opens the possibility to distribute the trained network along with the map itself, the first one being much lighter with only 560 MB while a reasonable sized map with the full prediction probability distribution weights several GB. This would allow anyone to quickly reconstruct a map with an individual control of the probability distribution sampling or on the resolution.\\

\vspace{-0.1cm}
For the present work we accumulated more than 1000 GPU hours on the UFC Mesocenter. While the comparison with CPU hour is not straightforward since our code takes advantage of GPUs architecture specificity, we can roughly estimate the conversion between GPU and CPU hours. A Tesla V100 is estimated at 14.13 TFLOPS, while the CPUs on the same machine are Intel 4110 at 2.1 GHz, which convert roughly to 67 GFLOPs per core in single precision. Using the ratio between the two raw-compute power our GPU hours count converts to $\sim$210000 CPU hours. We note that for subsequent studies we plan to make a proposal to the \href{http://www.idris.fr/annonces/annonce-jean-zay-eng.html}{Jean Zay} GENCI super computer as stated in Section~\ref{gpus_variety}, which has a dedicated entry program for AI projects, granting access to large GPU-nodes equipped with several Tesla V100 GPUs.

\newpage
\section{2MASS only extinction maps}
\label{2mass_maps_section}

In this section we describe the results we obtained using our CNN architecture solely on 2MASS data. We first present results in a large zone from the generalization of a training on a single LOS, and in a second step we show how several lines of sight can be combined into a single training. We also illustrate the effect of some parameters of our training dataset like the $Z_{\rm lim}$ value on our network prediction. Because the results are mostly arranged in an linearly increasing complexity order, we perform most of the analysis of each case after the presentation of the results.

\etocsettocstyle{\subsubsection*{\vspace{-1cm}}}{}
\localtableofcontents

\subsection{Training with one line of sight}
\label{2mass_single_los}

\subsubsection{Network training and test set prediction}
\label{2mass_single_los_training_and_test_set_prediction}

The simplest approach we can elaborate to train a CNN on 2MASS CMDs is the one that has been described along with the dataset construction in Section~\ref{galmap_problem_description}). For this first application, we considered only one LOS. Still, we will show that a training on a single LOS can be generalized to a relatively large galactic longitude range. We selected the LOS $l = 280$ deg, $b = 0$ deg in galactic coordinates because this region approximately corresponds to the observable tangent of the Sagittarius-Carina galactic arm. In this region we expect to have a significant diversity of extinction distributions in a relatively narrow galactic longitude window.\\

Using solely the central LOS value, we generated a training sample of various extincted CMDs. For this we first produced several BGM realizations because they are always slightly different considering that it represents stellar populations statistically (Sect.~\ref{BGM_sect}). For each of these realizations, we generated 100 (i.e. $1/f_{\rm naked}$, Sect.~\ref{zlim_subsection}) mock extinction profiles following our GRF recipe (Sect.~\ref{GRF_profiles_section}) that are applied to the BGM star distance distribution. We then used the magnitude limit cuts to exclude too faint stars and added the modeled photometric errors, both following the description from Section~\ref{ext_profile_and_cmd_realism}. The extincted CDMs are generated from the extincted star lists and defined as the network inputs. Finally, the realistic list of stars for each example is used to modify the corresponding extinction profile based on a $Z_{\rm lim} = 100$ value, and the extinction profiles are capped (Sect.~\ref{zlim_subsection}). The modified profile is then defined as the network target. \\

For our first application using a single LOS we generated $5\times 10^5$ examples of 2MASS CMD-profile pairs meaning that the training sample is based on $1000$ BGM realizations. Also, a part of the profiles are considered as flat with no extinction to better constrains the bare CMDs distribution, using $f_{\rm naked} = 0.1$ as described in Section~\ref{zlim_subsection}. This dataset is separated into a training dataset that contains $94\%$ (470000) of the examples, a valid dataset with $5\%$ (25000), and the remaining $1\%$ (5000) are our test dataset (see Sect.~\ref{sect_overtraining}). The training is performed using the architecture we highlighted in Section~\ref{cnn_architecture_test} with only one convolutional layer with few filters followed by three dense layer that include a small dropout. The network hyperparameters are described in Section~\ref{cnn_hyperparameters}. It took around 60 epochs for the network to converge corresponding to a few hours on a Tesla V100 GPU, then the prediction is mostly stable up to epoch 100 where the network distinctly began to overtrain.\\

\begin{figure}[!t]
	\hspace{-0.9cm}
	\vspace{-0.2cm}
	\begin{minipage}{1.06\textwidth}
	\centering
	\includegraphics[width=1.0\hsize]{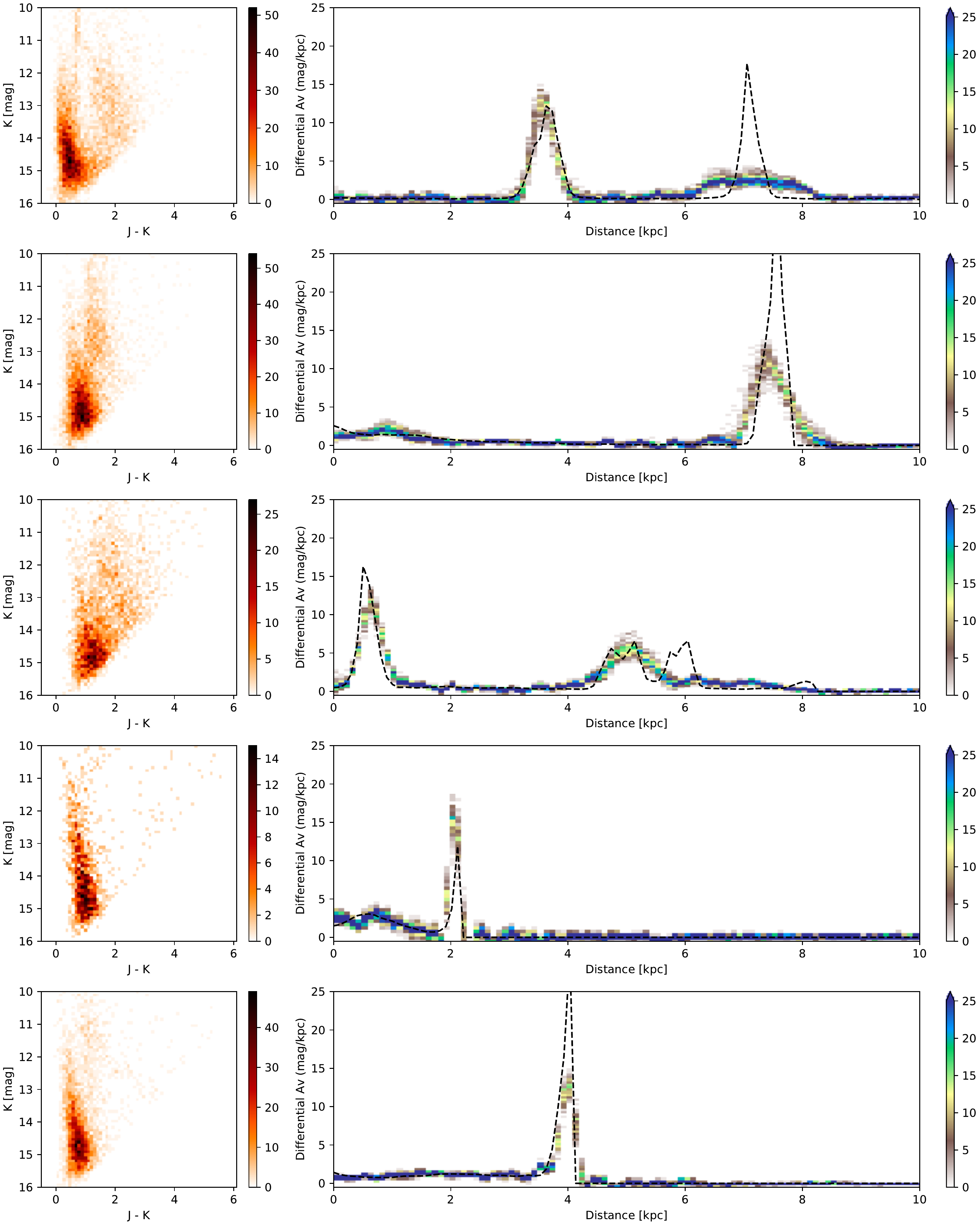}
	\end{minipage}
	\caption[Single 2MASS LOS training prediction on the corresponding test set]{Excerpt of a few objects from the test dataset of the 2MASS single LOS training. {\it Left:} View of the CMD for which the prediction is made. {\it Right:} View of corresponding profile. The dashed line shows the target of the network that accounts for the $Z_{\rm lim}$ maximum distance limit. The network prediction is presented in the form of a vertical histogram prediction for each distance bin corresponding to 100 random dropout predictions.}
	\label{single_los_2MASS_test_profiles_prediction}
\end{figure}

\vspace{-0.3cm}
A global error on the test dataset is not a very visual representation of the network capacity to reproduce the target, so we extracted a few extinction profile predictions of the network on the test dataset that we represent in Figure~\ref{single_los_2MASS_test_profiles_prediction}. This figure shows the CMD that is used as input for each case on the left column, and it compares the target for this case to the network prediction using a sample of 100 random dropout predictions to construct the prediction probability distribution. From the figure it is striking that the network greatly succeeds in localizing the extinction peaks, but it is slightly less accurate to reproduce the maximum extinction amount. We also observe that some structures are not reconstructed properly after a first extinction peak, for example in frames 1 and 3. The frame 3 illustrates a case where the network manages to reconstruct a relatively low second extinction peak after a first one, even at a large distance $d \simeq 5$ kpc. In all the cases the network mostly succeeds to localize the $Z_{\rm lim}$ maximum distance and appropriately predicts zero for larger distances. On the other hand, high extinction peaks are not always as nicely represented, as illustrated in frame 2, especially at large distance. We especially notice that in the case of a relatively strong first extinction peak, the network has much more difficulties to predict a second one that is still not cut by the $Z_{\rm lim}$ limit distance, as visible in frame 1. In such a case, we observe that the network still localizes an extinction increase at the appropriate location but with an underestimated extinction value. However, as our following results will show, our map will mostly predict extinction that lies in a range where our test examples are properly reconstructed. Indeed, the most difficult cases are in fact less realistic or would be very uncommon. We kept them in the network training in order to ensure that we have a large enough feature space coverage, in an attempt to obtain a sufficient diversity to avoid leaving unusual observed CMDs completely unconstrained.\\

\vspace{-0.7cm}
\subsubsection{Generalizing over a Galactic Place portion}

\vspace{-0.1cm}
From this trained network we were able to make predictions on real observed data. For this, we used observed CMDs in place of the mock ones as the input of our network. While our network was trained using solely mock CMDs that correspond to the $l=280$ deg, $b = 0$ deg it is still possible to construct a map using close enough lines of sight. In practice we constructed our maps using cone queries with $0.25$ deg radius on the 2MASS PSC, corresponding to the same $0.2 \mathrm{deg^2}$ solid angle that was used to build our mock CMDs from the BGM. In order to be sure that our observed and mock CMDs are constructed from similar stellar distributions we removed every 2MASS star that lacks one or more band detection since this is the approach followed in our mock CMD construction. Our map is then considered as a grid of 0.2 deg sized square pixels so that it is close to follow the Nyquist criterion. Our map range is as follow, $257 < l < 303$ deg and $ -5 < b < 5$ deg. Since each pixel of the map is a LOS with an extinction profile prediction, our map is a 3D volume of $230\times 50\times 128$ bins of differential extinction values that represent a large squared-cone field of view.\\

\subsubsection{Integrated view of the plane of the sky}

\begin{figure*}[!t]
\hspace{-1.6cm}
\begin{minipage}{1.25\textwidth}
	\centering
	\vspace{0.79cm}
	\begin{subfigure}[!t]{1.0\textwidth}
	\includegraphics[width=1.0\hsize]{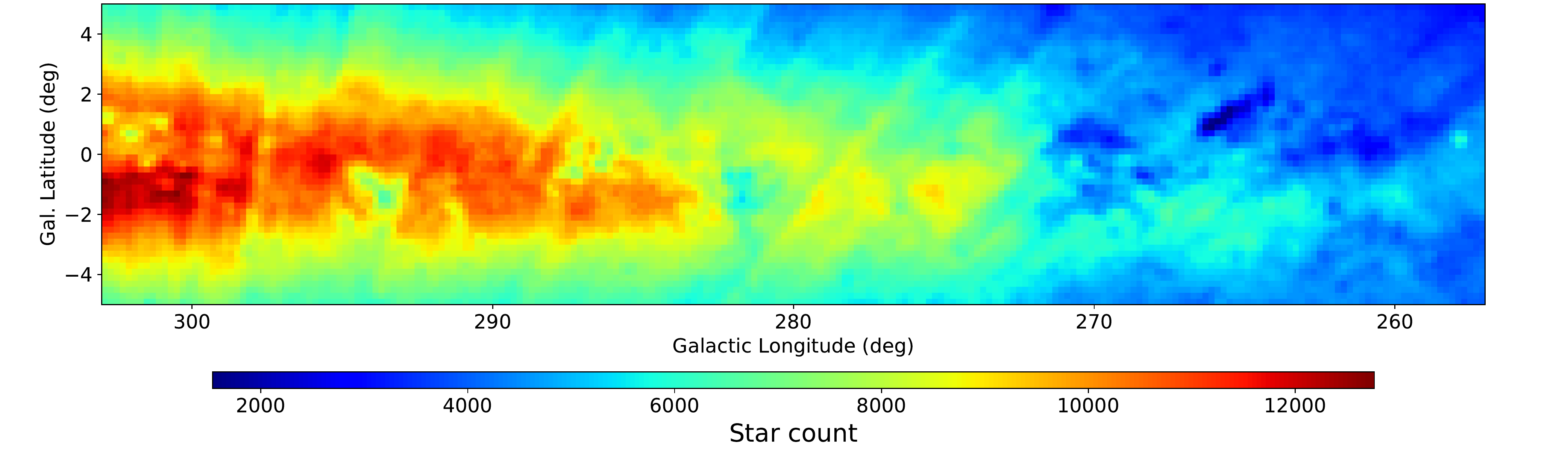}
	\end{subfigure}
	\end{minipage}
	\caption[2MASS star count]{Map of star counts in the 2MASS [J-K]-[K] CMDs for each pixel on the same grid as in Fig.~\ref{single_los_2mass_integrated_result}.}
	\label{2mass_sky_star_count}
	\vspace{-1.2cm}
\end{figure*}

\begin{figure*}[!t]
\hspace{-1.5cm}
\begin{minipage}{1.22\textwidth}
	\centering
	\begin{subfigure}[!t]{1.0\textwidth}
	\caption*{\bf \large Planck dust opacity $\bm{\tau_{353}}$}
	\includegraphics[width=1.0\hsize]{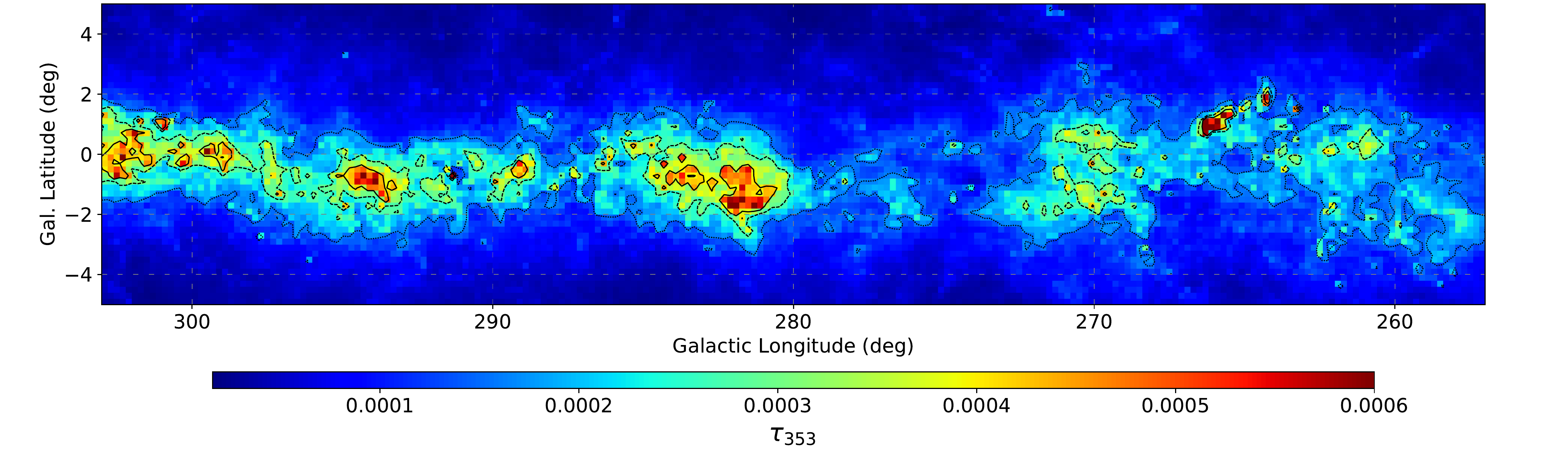}
	\end{subfigure}\\
	\vspace{0.6cm}
	\begin{subfigure}[!t]{1.0\textwidth}
	\caption*{\bf \large Single position training - $\bm{Z_{\rm lim}=50}$}
	\includegraphics[width=1.0\hsize]{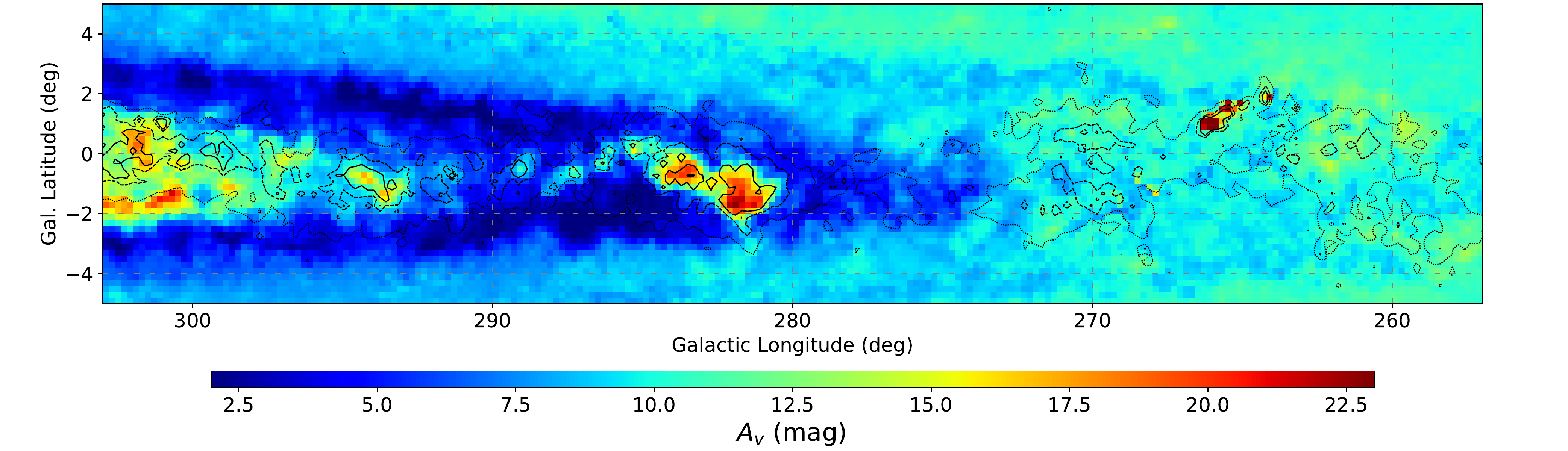}
	\end{subfigure}\\
	\vspace{0.6cm}
	\begin{subfigure}[!t]{1.0\textwidth}
	\caption*{\bf \large Single position training - $\bm{Z_{\rm lim}=100}$}
	\includegraphics[width=1.0\hsize]{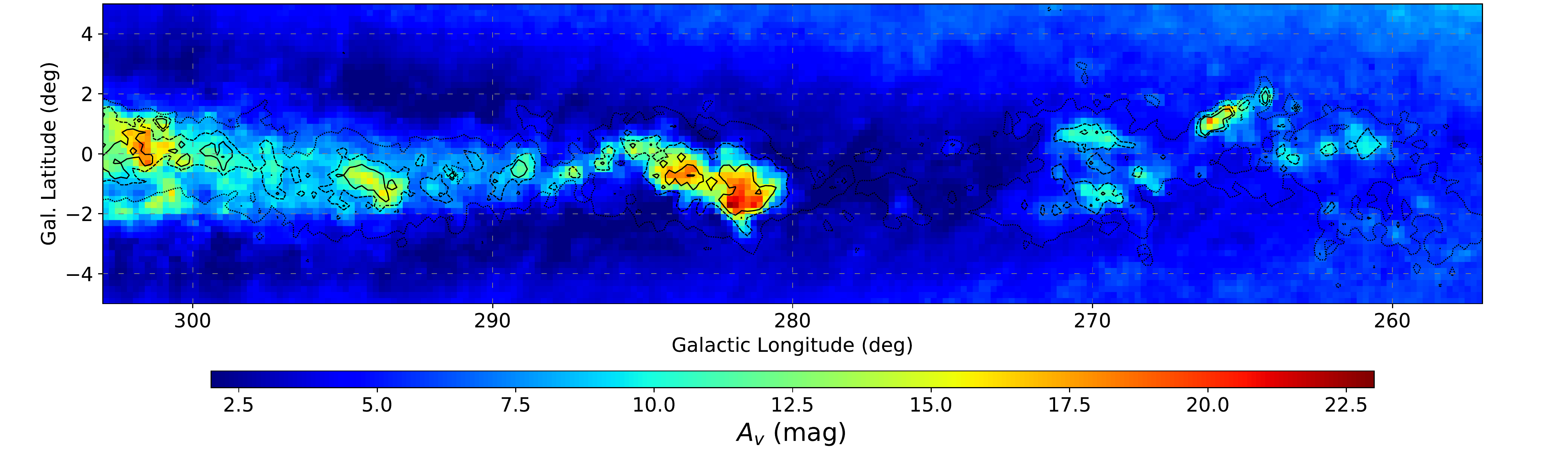}
	\end{subfigure}
	\end{minipage}
	\caption[Integrated view of the plane of the sky for the 2MASS single LOS training]{Comparison of the 2MASS single-LOS training results in a plane of the sky view using galactic coordinates. {\it Top:} Observed Planck dust opacity at 353 GHz. {\it Middle:} Integrated extinction over the whole LOS for each pixel corresponding to the 2MASS single-LOS training with $Z_{\rm lim}=50$. {\it Bottom:} Integrated extinction for the same case but with $Z_{\rm lim}=100$. The contours of the Planck map at $\tau_{353}=0.00016$, $0.00028$, and $0.0004$ are reproduced in the other two frames to ease the comparison.}
	\label{single_los_2mass_integrated_result}
\end{figure*}

From this predicted map we first reconstructed the integrated extinction as observed in the plane of the sky. With this test it is possible to verify that our method predicts realistic total extinction in the various line of sights, and more importantly to verify that the observed Galactic Plane morphology is properly reconstructed. We chose to use the Planck optical depth $\tau_{353}$ that is derived from dust emission as a proxy for the dust distribution morphology. Figure~\ref{single_los_2mass_integrated_result} shows the comparison between the Planck sky map and our CNN predictions for $Z_{\rm lim} = 50$ and $Z_{\rm lim} = 100$. The morphology of the Planck map mostly follows the Galactic Plane with structures mostly contained in the $|b| < 2$ deg interval.\\

The most striking result of this figure is that there is a large difference induced by the choice of the $Z_{\rm lim}$ value especially for the lines of sight that contain less stars, i.e. at higher latitudes outside the Plane, and at lower galactic longitudes away from the Galactic Center. When the drop in star count compared to the reference training LOS is not related to extinction, the network still predicts an important amount of extinction. The star count distribution is illustrated by Figure~\ref{2mass_sky_star_count}. We observe that this star count distribution strongly (anti-)correlates with our inappropriate extinction predictions. The $Z_{\rm lim} = 100$ value mitigates this effect, but we could not remove it completely. It probably cannot be done using a single LOS training.\\

In spite of the star count variations across the map, our $Z_{\rm lim} = 100$ result already reconstructs the Planck map morphology very well. Many of the strongest $\tau_{353}$ structures are also predicted as strong extinction LOS by our CNN and the contours of these structures are accurately followed most of the time. We stress that generalizing over a $\pm 23^\circ$ longitude range from a single training is a very challenging task since the corresponding CMD variations are important. It means that the CNN architecture manages to identify the parts of the CMD that are most relevant for the extinction, and probably also that the part that it learned to ignore happened to be the one that changes the most between galactic coordinate positions. We highlight that there is no strong constraint on the integrated extinction value in our CNN profile prediction, each output neuron being completely independent there is no error propagation corresponding to a total extinction error on the profile. Therefore, having a proper predicted integrated morphology is already a sign that the reconstructed profile is likely to be realistic.\\

\vspace{-1cm}
\subsubsection{Face-on view}
\label{face_on_view}

\begin{figure*}[!t]
\vspace{-0.5cm}
\hspace{-0.5cm}
	\begin{minipage}{1.05\textwidth}
	\centering
	\begin{subfigure}[!t]{0.47\textwidth}
	\caption*{\bf \large $\bm{Z_{\rm lim}=50}$}
	\includegraphics[width=1.0\hsize]{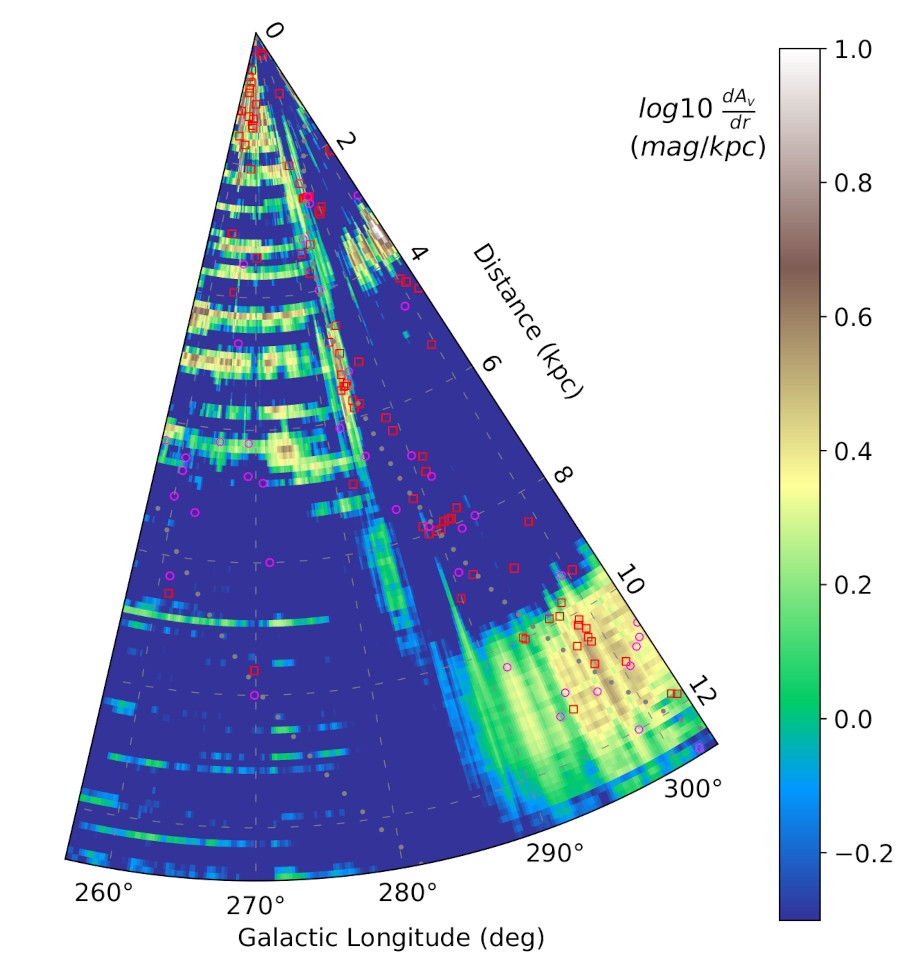}
	\end{subfigure}
	\vspace{0.5cm}
	\hspace{0.4cm}
	\begin{subfigure}[!t]{0.47\textwidth}
	\caption*{\bf \large $\bm{Z_{\rm lim}=100}$}
	\includegraphics[width=1.0\hsize]{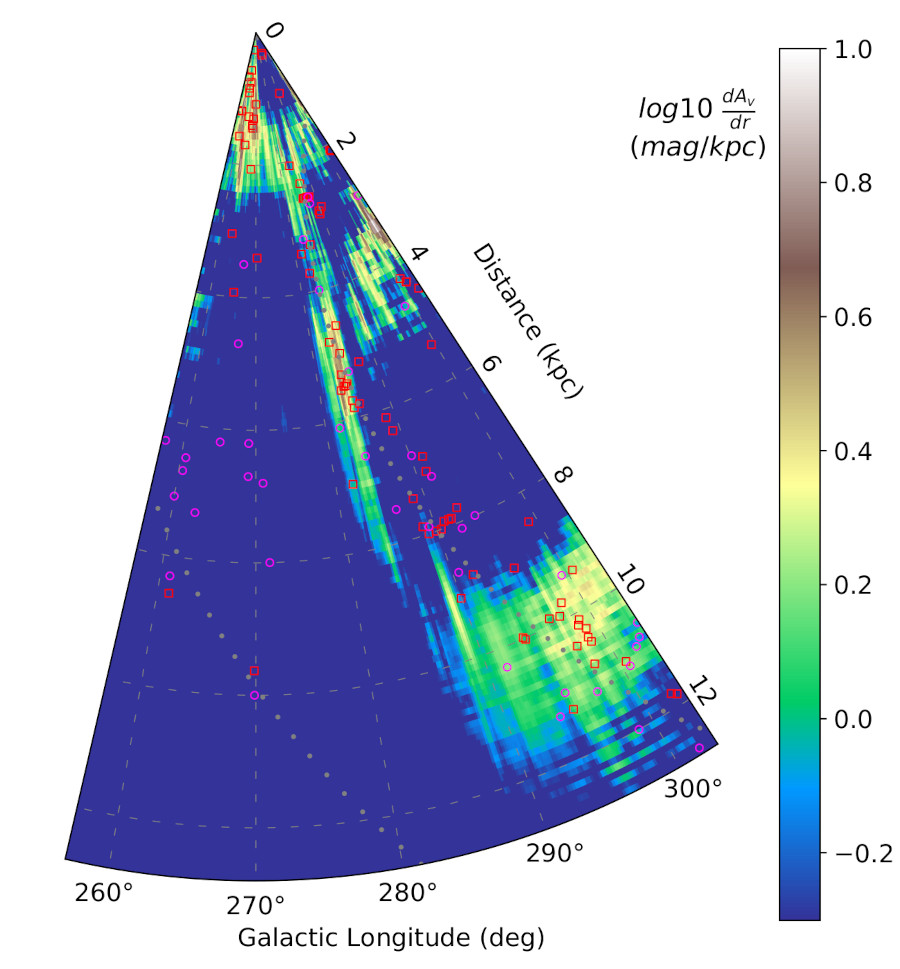}
	\end{subfigure}\\
	\begin{subfigure}[!t]{0.47\textwidth}
	\caption*{\bf \large $\bm{Z_{\rm lim}=50}$}
	\includegraphics[width=1.0\hsize]{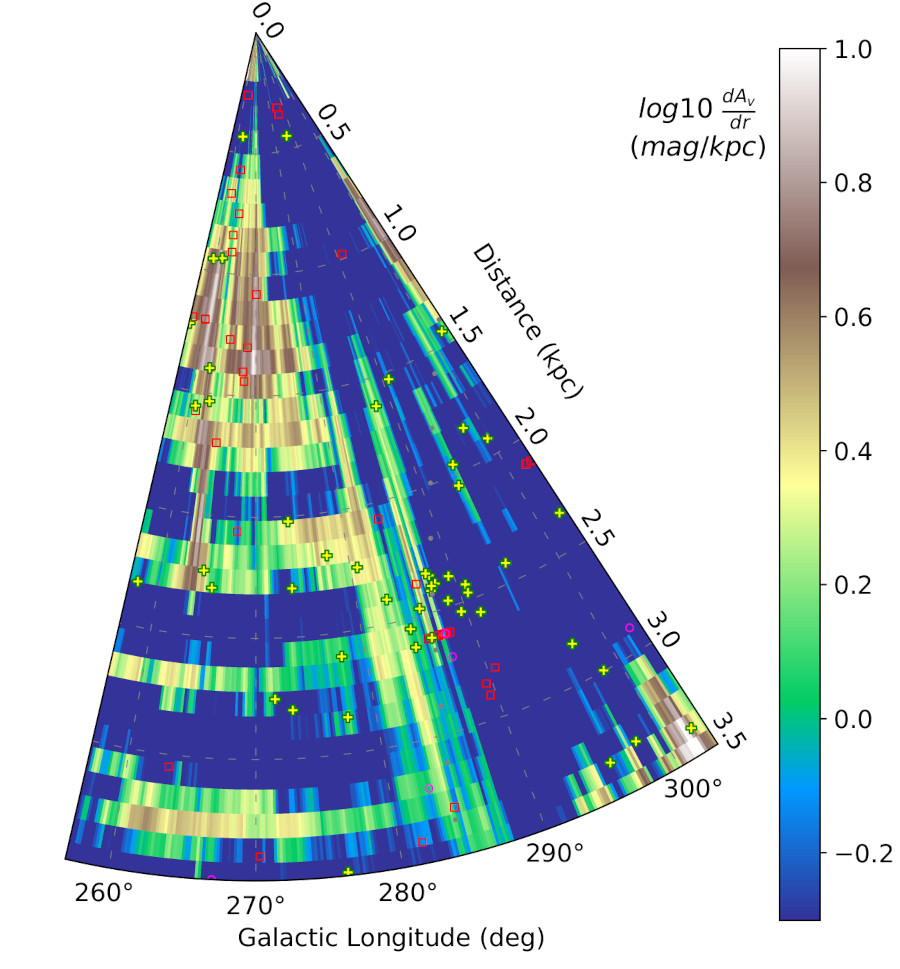}
	\end{subfigure}
	\vspace{0.5cm}
	\hspace{0.4cm}
	\begin{subfigure}[!t]{0.47\textwidth}
	\caption*{\bf \large $\bm{Z_{\rm lim}=100}$}
	\includegraphics[width=1.0\hsize]{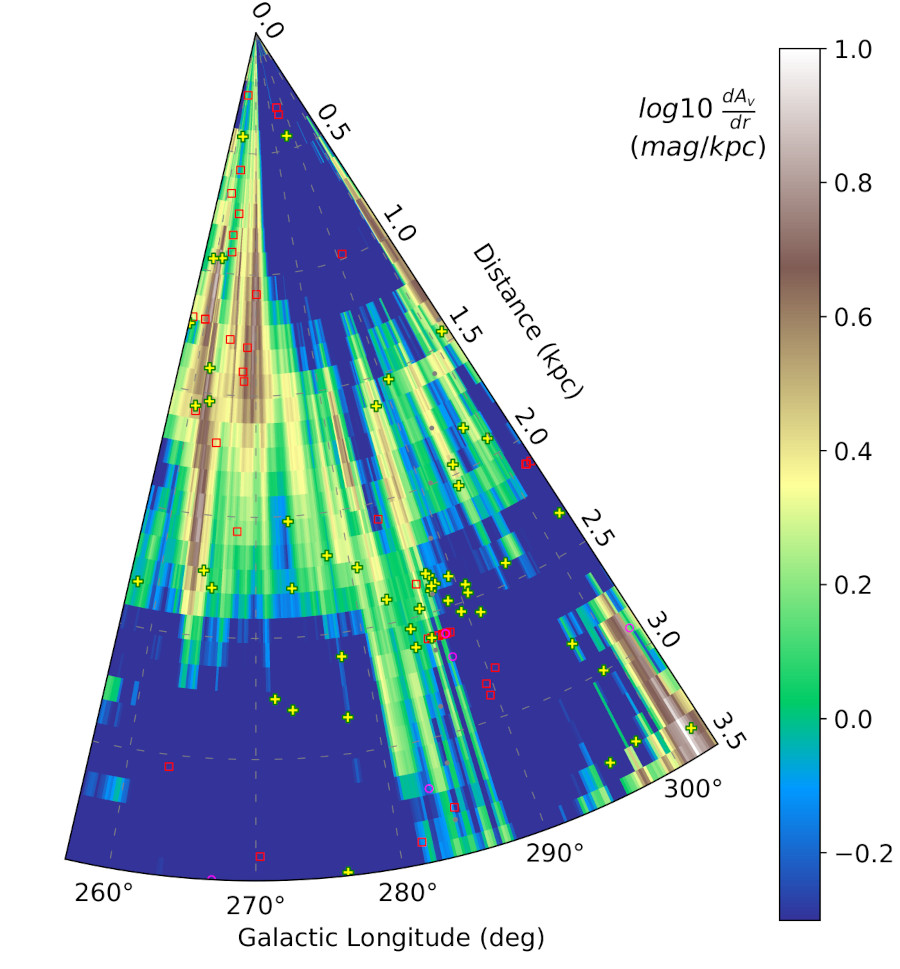}
	\end{subfigure}
	\end{minipage}
	\caption[Face-on view for the 2MASS single-LOS training]{Face-on view of the Galactic Plane $|b| < 1$ deg in polar galactic-longitude distance coordinates for the predicted Carina-arm region using our 2MASS single-LOS training. {\it Left column:} Network prediction with $Z_{\rm lim}=50$. {\it Right column:} Network prediction with $Z_{\rm lim}=100$. {\it Top row:} Full distance prediction. {\it Bottom row:} Zoom on the $d < 3.5$ kpc prediction. The HII regions and GMCs compiled by \citet{hou_and_han_2014} are displayed as red open squares and purple open circles, respectively. The yellow pluses are open clusters from the catalog by \citet{Cantat-Gaudin_2018}. In the top row, simple spiral arm models from \citep{Marshall_phd_2006} are represented by gray dots for comparison.}
	\label{single_los_2mass_polar_plan}
\end{figure*}

To visualize the distance prediction, we built face-on views of the Galactic Plane. To do so we had to select a latitude thickness in which we average our differential extinction cube. In order to be comparable to the other maps described in Section~\ref{ext_properties_part3} we used a slice of $|b| < 1$ deg. Due to the remark we made on Figure~\ref{single_los_2mass_integrated_result} that most of the structures truly lie in the $|b|<2$ deg range for our region, we compared the integrated map obtained with $|b| < 1$ deg and $|b| < 2$ deg, and found only marginal differences, so in the following we only discuss the one with $|b| < 1$ deg. We present our first distance prediction results in the Figure~\ref{single_los_2mass_polar_plan} that presents the polar face-on view of the portion of disk between $257 < l < 303$ deg. In this figure we show the full distance prediction up to 12.8 kpc and a zoom on the nearby predictions closer than 3.5 kpc, for both the $Z_{\rm lim} = 50$ and 100 values. On all the figures that follow this representation, we also display other observational constraints on the Milky Way morphology.\\

We added HII and GMC regions (red squares and purple circles) from \citet{hou_and_han_2014}, that are expected to trace the Galactic spiral arm structure. The distance to these regions was either compiled from various previous studies or estimated by the authors using the kinematic method based on a rotation curve of the Milky Way resulting in relatively heterogeneous sample with large uncertainties that are not provided in their fully accessible catalog. Still, from their catalog, the Carina arm appears very clearly, suggesting that the distance estimates are relatively reliable in this region of the Milky Way, although they could be subject to the same systematic error. The dispersion of their points suggests uncertainties of the order of a few hundred parsecs, but it is difficult to disentangle the genuine scatter of interstellar structures from the distance uncertainties. \\

We also added the Gaia stellar cluster catalog (Yellow crosses with green border) from \citet{Cantat-Gaudin_2018} that are much more accurately positioned in distance thanks to the Gaia parallaxes, but that are less likely to be directly related to dust distribution because it mostly includes relatively evolved stars (see our YSO-Gaia cross match in Sect.~\ref{3d_yso_gaia}). Additionally, there are only few of these clusters close to the Galactic Place that are found at large distances. We therefore only display these data points in our short-distance view. We note that for both catalogs, only the regions inside the $|b|<1$ deg range are displayed. \\

We also display a very simple elliptical arm model represented by light gray dots simply constrained by a distance and a pitch angle, and that is parameterized to represent the Carina arm (we used $r_0 = 5.0$\,kpc, $p = 14.5^\odot$) and the end of the Perseus arm at high distance (we used $r_0 = 3.8$\,kpc, $p = 14.5^\odot$) in the low galactic longitude part of our map, both assuming a sun distance to the galactic center of 8.4 kpc \citep{Marshall_phd_2006}. These are really simplistic arm equations just constrained by a distance and a pitch angle, therefore they are not adjusted on any data and just provide a global insight on arms shape in such representation. Still, we see that with our adopted parameters the mock Carina arm mostly follows the \citet{hou_and_han_2014} regions. \\

\begin{figure*}[!t]
	\begin{minipage}{0.98\textwidth}
	\centering
	\begin{subfigure}[!t]{1.0\textwidth}
	\caption*{\bf \large $\bm{Z_{\rm lim}=50}$}
	\includegraphics[width=1.0\hsize]{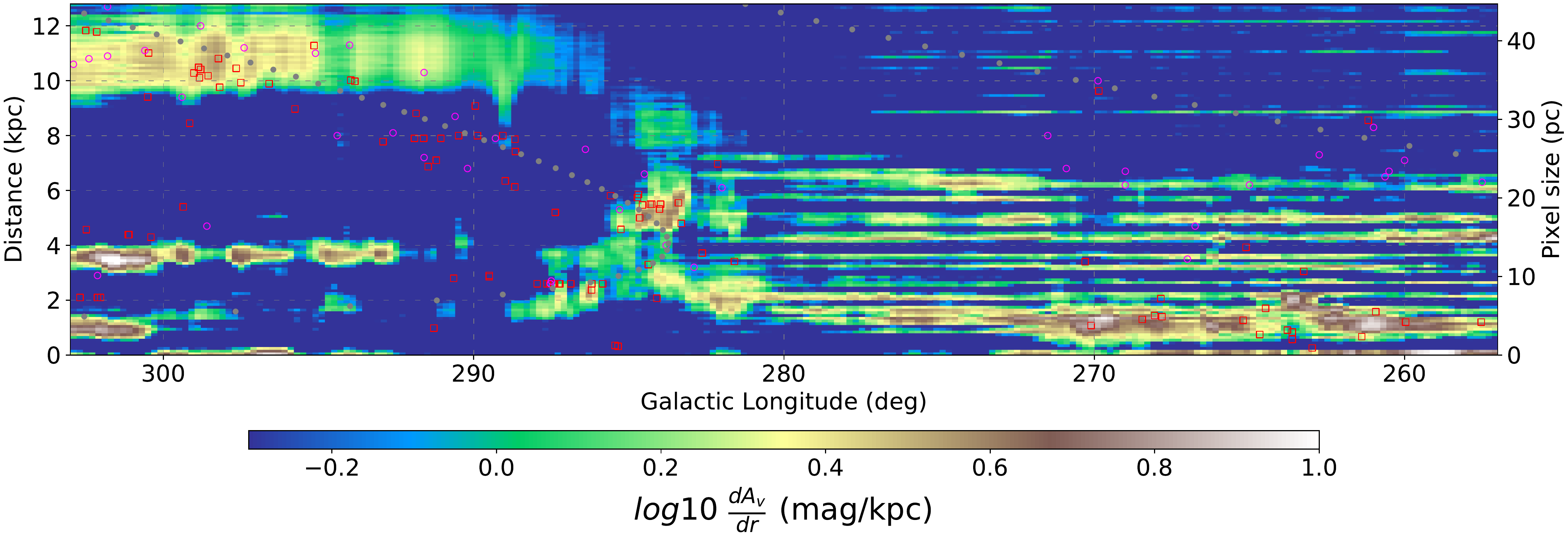}
	\end{subfigure}\\
	\vspace{0.6cm}
	\begin{subfigure}[!t]{0.98\textwidth}
	\caption*{\bf \large $\bm{Z_{\rm lim}=100}$}
	\includegraphics[width=1.0\hsize]{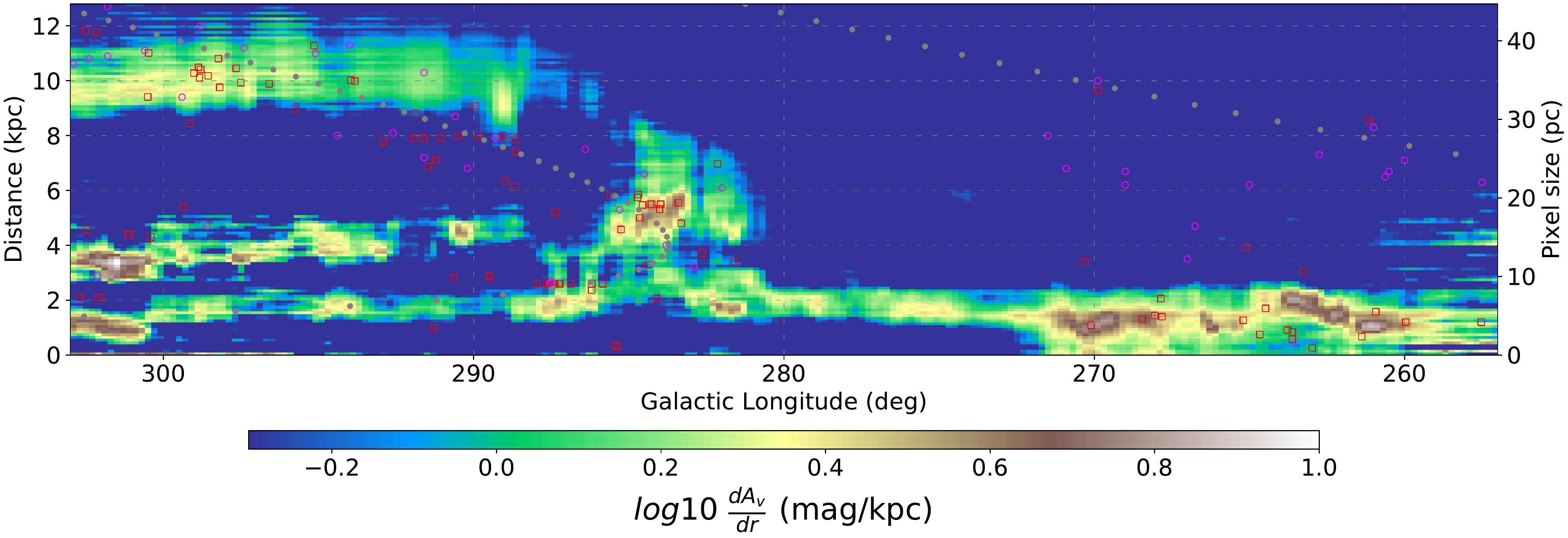}
	\end{subfigure}
	\end{minipage}
	\caption[Cartesian face-on view for the 2MASS single-LOS training]{Face on view of the Galactic Plane $|b| < 1$ deg in cartesian galactic-longitude distance coordinates for the predicted Carina-arm region using our 2MASS single-LOS training. The axis on the right border of each frame corresponds to the pixel height as a function of the distance induced by the conic shape of our LOS. The symbols are the same as in figure ~\ref{single_los_2mass_polar_plan}. {\it Top:} Network prediction with $Z_{\rm lim}=50$. {\it Bottom:} Network prediction with $Z_{\rm lim}=100$.}
	\label{single_los_2mass_cart_plan}
	\vspace{-0.3cm}
\end{figure*}

Finally, we note that the polar view is not the most representative of the data cube the CNN works with, especially at short distances where the first bins of all our lines of sight overlap. As an alternative, we provide a Cartesian view of the the same Galactic Place view in Figure~\ref{single_los_2mass_cart_plan}, that can be used to complement the previous Figure~\ref{single_los_2mass_integrated_result} and ease the prediction interpretation.\\

\newpage
From Figures \ref{single_los_2mass_polar_plan} and \ref{single_los_2mass_cart_plan}, we observe that the $Z_{\rm lim}=50$ case is much more noisy and presents quasi-periodic artifacts in the $l < 275$ deg part corresponding to the case where the latitude artifact joins the plane in Figure~\ref{single_los_2mass_integrated_result}. In contrast the $Z_{\rm lim} = 100$ case is much more realistic, and is devoid of these periodic artifacts in the low longitude region, with a convincing nearby extinction distribution. The other striking observation is how the larger longitude results are compatible with an arm shape. The most convincing part is the tangent structure found just above $l = 280$ deg at around 6 kpc with a clear structure interruption along the longitude axis. The large blurry prediction at between $285 < l < 303$ deg at 10 kpc is likely to be dominated by an artifact. The galactic disk morphology induce that the number of stars is rising quickly with the observational longitude and distance roughly following the shape of this structure. This is a surprising effect that the network tend to compensate higher star count for which it is not constrained by an excess of extinction at high distances.\\

These maps also illustrate that the network manages to reconstruct structures that are very coherent between lines of sight. We remind that there is no prescription of LOS correlation between adjacent pixels, and that the CNN still reconstructs convincing structures that are sometime coherent for more that 10 adjacent pixels following the longitude axis. \\

There is also a notable separation between two continuous structures at distances of $\sim$2 and $\sim$4 kpc in the large longitude part ($290 < l < 303$ deg), which is more visible in Figure~\ref{single_los_2mass_cart_plan}. The arm model seems to follow the closer structure but there are no HII region or GMC to confirm that this separation is real. Still, the group of Gaia clusters and HII region that is present at $l \simeq 287$ and $d = 2.5$ kpc, visible in the close distance representation of Figure~\ref{single_los_2mass_polar_plan}, suggests that the distance to the local maximum of extinction at $l \simeq 287$ and $d = 2.5$ kpc is under-estimated by $\sim 500$ pc in our map.\\

\begin{figure*}[!t]
\hspace{-0.8cm}
	\begin{minipage}{1.08\textwidth}
	\centering
	\begin{subfigure}[!t]{0.47\textwidth}
	\includegraphics[width=1.0\hsize]{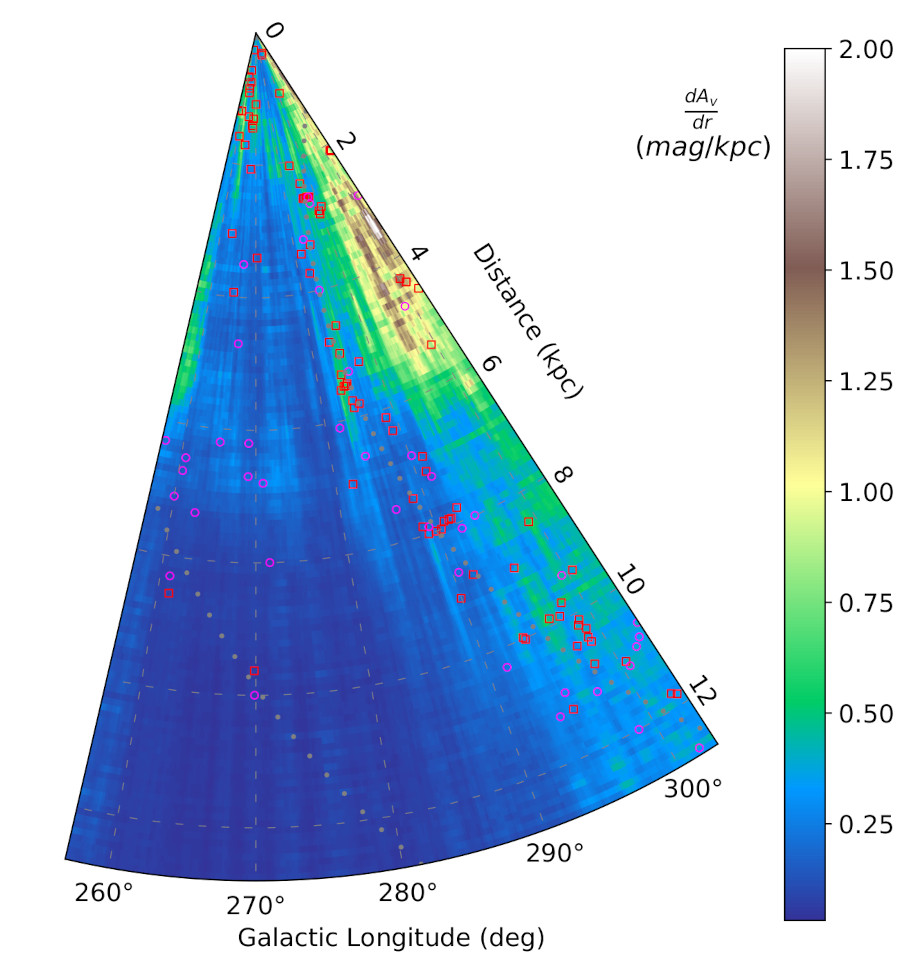}
	\end{subfigure}
	\hspace{0.5cm}
	\begin{subfigure}[!t]{0.47\textwidth}
	\includegraphics[width=1.0\hsize]{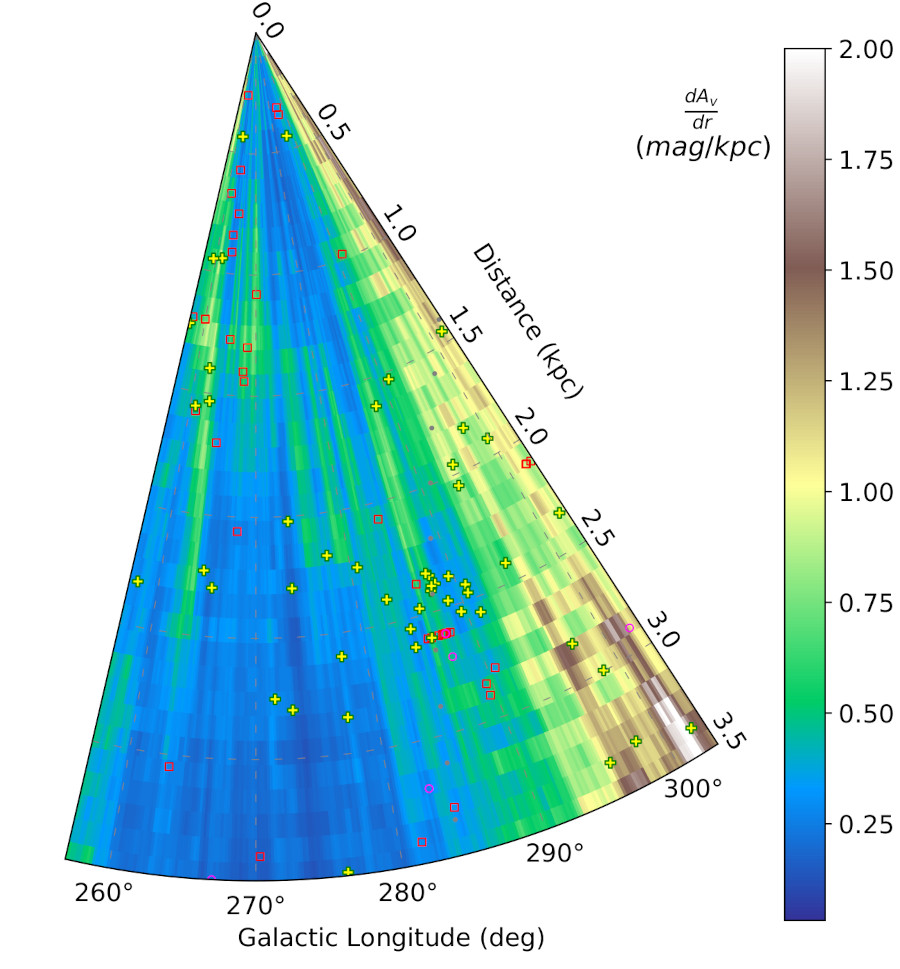}
	\end{subfigure}
	\end{minipage}
	\caption[Face-on view for the standard deviation of the 2MASS single-LOS training]{Face-on view of the Galactic Plane $|b| < 1$ deg in polar galactic-longitude distance coordinates for the standard deviation predicted in the Carina arm region using our 2MASS single-LOS training. The symbols are the same as in figure ~\ref{single_los_2mass_polar_plan}. {\it Left:} Full distance prediction. {\it Right:} Zoom in on the d < 3.5 kpc prediction region.}
	\label{std_single_los_2mass_polar_plan}
\end{figure*}

One of the use of our network prediction density probability from dropout (Sect.~\ref{2mass_single_los_training_and_test_set_prediction}) is to assess which structure could be less realistic according to the network own uncertainty. Figure~\ref{std_single_los_2mass_polar_plan} shows the same face-on view as before but representing the averaged standard deviation of individual prediction at each distance. This figure reveals that the large-longitude short-distance region discussed in the previous paragraph is the less well constrained, with the uncertainty maximum being reached for the secondary structure at $l \simeq 301$, $d = 3.5$ kpc. This is consistent with the fact that it is $\sim 23$ deg away from our training LOS, and that the star count starts to rise quickly at these galactic longitudes. It is likely that a similar issue occurs for the other boundary of the map at $l = 257$ due to the lower star count, but it is partly compensated by the $Z_{\rm lim} = 100$ parameter, still causing the network to loose the higher distance information. \\

Overall, these results on a single LOS training are surprisingly capable of a convincing generalization on a large galactic longitude window once tuned appropriately. However, many of the limitations we exposed in the present section should be resolved if we were able to train the network on several lines of sight at different galactic coordinates.\\

\newpage
\subsection{Combination of several lines of sight in the same training}
\label{los_combination}

\subsubsection{Sampling in galactic longitude}

There are a few methods that could be used to construct a map that is capable of taking into account the presented LOS. We saw in the previous Section~\ref{face_on_view} that the network prediction in the vicinity of the training LOS appears to be reconstructed properly within several degrees. The simplest approach would then be to sample the plane of the sky with a few reference LOS and to train an individual network on each of them. The map would then be made of several tiles centered on each reference LOS, each tile corresponding to an independent network training. We tested this method as follows.\\

\vspace{-0.25cm}
Considering that we are mainly interested in the Galactic Place and that we are reasonably free of latitude artifacts in the range $|b| <1$ deg we only sampled the galactic longitude axis with a training every $5^\circ$ centered on the $l = 280$ deg, $b = 0$ deg position, ending up with 9 LOS from $l = 260$ deg to $l = 300$ deg. While this approach provides interesting results that solved some of the issues we had with the single training generalization (Fig.~\ref{single_los_2mass_polar_plan}) it creates strong discontinuities at the junction between adjacent tiles, resulting in a very patchy prediction. Additionally, we saw that we needed up to $5\times 10^5$ training examples for a single LOS training, so 9 individual training were very expensive in terms of memory and training time.\\

\vspace{-0.5cm}
\subsubsection{Multiple line of sights in a single training}
\label{multi_los_CMD_constrution}

We found a more suitable approach in the idea that there should be redundant information between the different training. It is straightforward that convolutional filters that were found to be useful for one LOS is very likely to be useful for another LOS. This consideration can be generalized to the whole network architecture. One possible solution would have been to perform a single training on the central LOS and then use it as a pre-trained starting point for all the other LOS trainings. This would have significantly reduced the training time and possibly reduced a part of the tiling effects, considering that all networks have mostly similar weights. However, this solution is still not as appealing as a single training that capable of predicting the whole map at once. For this to be possible we made a few changes in our network input.\\

\vspace{-0.25cm}
For a single network to work on multiple lines of sight, it must be provided with the CMD-profile pairs, but also with some information about the reference bare CMD. In our single training this was done by including a $f_{\rm naked}$ proportion of bare CMDs in the training corresponding to a flat profile so that the network could learn the reference statistically. In the present approach we chose to change the form of the input by adding an input depth channel containing the bare BGM realization CMD that was used to compute the extincted one. This way the network is provided in a single input with both the reference CDM of the given LOS and the corresponding extinct one. We kept our $f_{\rm naked} = 0.1$ value corresponding to cases with the bare CDM in both input depth channels, so that the network can still associate this reference to a flat profile. We note that the bare CDM is presented with the same magnitude limit cut and uncertainties. Using a completely non-processed CMD as reference, the network was not able to extract the link between a reference CMD and a CMD with almost no extinction. With this approach it is possible to construct a single training that learns from various lines of sight at the same time. Because making BGM realizations requires a significant computational time, we still used the 9 LOS sampling described in the previous section, but here they were merged into one single training dataset. We highlight that in this approach we double the input dimensions which results in twice larger dataset memory usage. 

\subsubsection{Dataset construction, architecture effect and training}
\label{2mass_multi_los_training_and_test_set_prediction}

Considering that there will be redundancy of a non-negligible part of the information from different lines of sight we were able to reduce the number of examples to $2 \times 10^5$ for each reference LOS. This still results in a $1.8 \times 10^6$ dataset that has twice the number of pixels per object compared to Section~\ref{2mass_single_los}, requiring a careful choice of numerical range for each of our pixels to reduce storage footprint that could reach 50 GB easily. From this dataset we used the same 0.94, 0.05, 0.01 proportions for the training, valid and test dataset, respectively, than in the previous single LOS training. We took care of applying these proportions to each reference LOS set of $2 \times 10^5$ examples before merging them. Additionally, each input depth channel is scaled separately by looking for the pixel with the highest star count in the whole dataset. This way, both our reference and observed CMD fall in a 0 to 1 range by conserving the proportions between various example on a given diagram and also ensuring that both the diagrams have similar actual pixel values, just like we normalized every input feature individually in Section~\ref{input_norm}.\\

We highlight that despite this change in input size, the first convolutional layers of the network produce an activation volume with the same size as before. The only change in network parameters is that the filters of this convolutional layer get an extra depth channel, which is an insignificant increase in regards of the more than $50\times 10^6$ weights in our network architecture. We note that despite this change in input dimension and a much more general context, the CNN architecture described in Section~\ref{cnn_architecture_test} remained the one with the lowest error on the test dataset from all the other architectures we tested. Since only the first convolutional layer is changed, the network mostly conserves its training speed in terms of number of objects per seconds. However, the dataset is much larger than in the single LOS case so each epoch takes much more time. The typical number of epochs required to train then remains mostly unchanged meaning that we have provided a similar amount of information overall, considering both the increased generality of this case and the additional global statistic of the larger dataset. We note that using only $1 \times 10^5$ examples for each reference LOS was providing significantly less good results, still with very acceptable predictions. Therefore, the redundancy of information is effectively present but in a smaller amount than we expected. We could not try $3\times 10^5$ since it would exceed the maximum host RAM memory of the compute cluster we used for training (250 GB). Still, since the difference between $1 \times 10^5$ and $2 \times 10^5$ examples per LOS was only slightly improving the results, we do not expect important improvements for a $3\times 10^5$ sample. Overall, the training using this dataset with $2 \times 10^5$ examples per LOS requires between 8h and 12h to complete using the Tesla V100 GPU (Sect.~\ref{cnn_computational}) depending on the number of epochs required to converge.\\

\begin{figure}[!t]
	\begin{minipage}{1.00\textwidth}
	\centering
	\includegraphics[width=1.0\hsize]{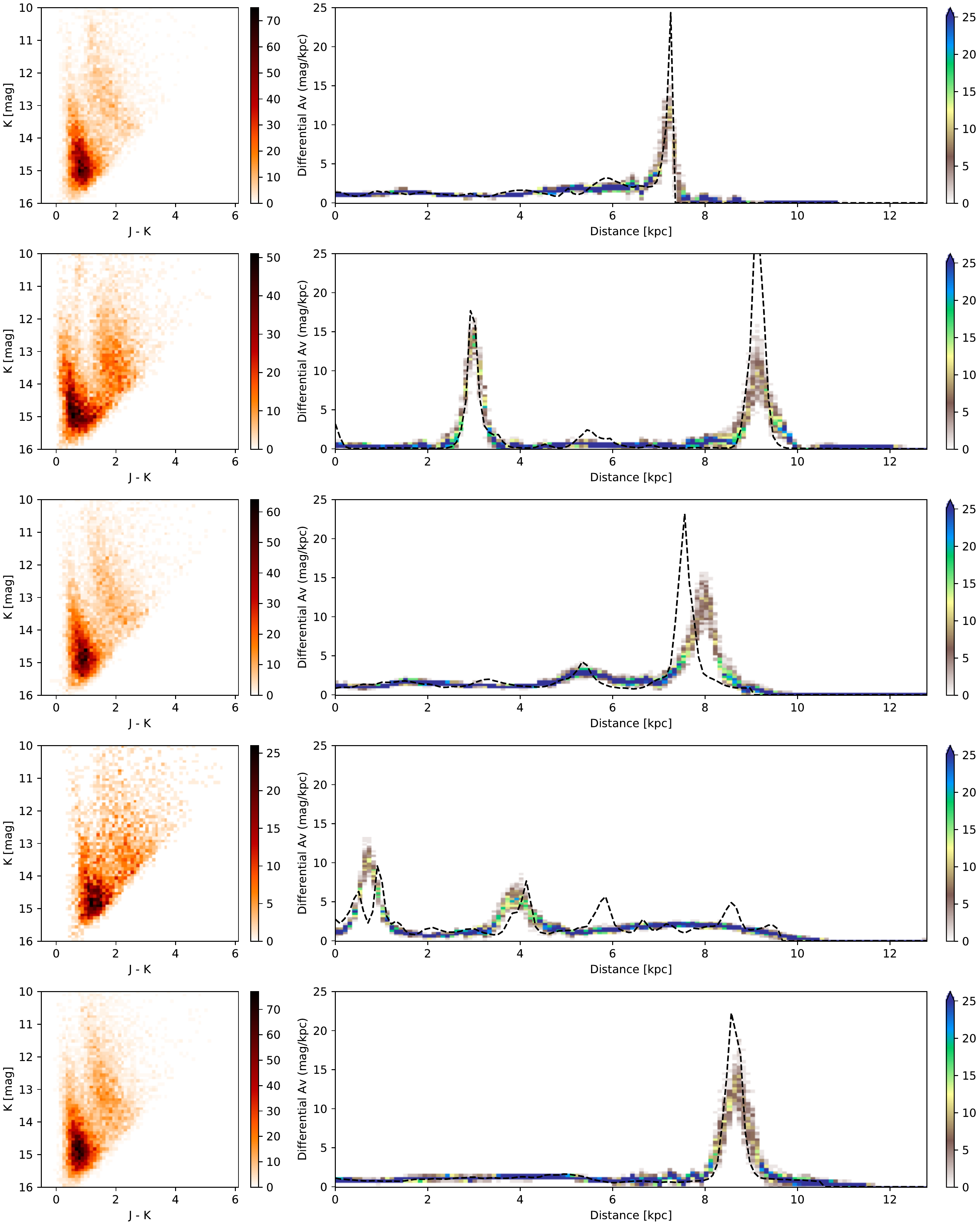}
	\end{minipage}
	\caption[2MASS Multiple-LOS training prediction on the corresponding test set]{Excerpt of a few objects from the test dataset of the 2MASS multiple-LOS training, for the $l=280$ deg, $b=0$ deg reference LOS. {\it Left:} View of the CMD for which the prediction is made. {\it Right:} Corresponding profile. The dashed line shows the target of the network that accounts for the $Z_{\rm lim}$ maximum distance limit. The network prediction is presented in the form of a vertical histogram prediction for each distance bin, corresponding to 100 random dropout predictions.}
	\label{multi_los_2MASS_test_profiles_prediction}
\end{figure}

Following the approach of Section~\ref{2mass_single_los_training_and_test_set_prediction} we present in Figure~\ref{multi_los_2MASS_test_profiles_prediction} a typical prediction of our multiple-LOS CNN training. All the predictions in the figure refer to the middle LOS $l=280$ deg, $b=0$ deg, so it can be compared to Figure~\ref{single_los_2MASS_test_profiles_prediction}. The reconstruction of the profiles is very similar to the one on the single LOS training, despite the number of example given for this specific LOS being reduced by 60\%. This confirms that our dual depth-channel input approach is suitable for the task, and that the network successfully shared a part of the information between several LOS to maintain a similar prediction capacity on individual one. This figure noticeably highlights a case in frame 2, where a first extinction structure with a non negligible total extinction is very well predicted and where the second structure at high distance ($\sim$9 kpc) is perfectly localized. The amount of extinction however is significantly underestimated for the second structure, but interestingly the dropout dispersion is maximal at the same location and can therefore be used to diagnose the poor reliability of this structure in the map.\\

\vspace{-0.9cm}
\subsubsection{Map results}
\label{sec:2MASS_main_result}

Performing a prediction on an observed CMD with this combined network is slightly more complicated than in the previous case. Indeed, additionally to the observed CMD we have to provide a reference bare CMD from the BGM. We could use different approaches for this independently of the reference CMD that was used for the training. One solution would be to perform a BGM realization for each pixel of our map. While it will obviously be at the cost of a significant computational time, it is indeed possible to have a bare BGM CMD at each pixel, although due to the statistical fluctuations of the realizations it may produce artifacts. To remove them it would be necessary to have several BGM realizations for each map pixel and then either average them or including a random reference choice in our 100 predictions that already account for the dropout, making this solution too much time and resource consuming. A much simpler approach consists in using a random reference CMD from the training dataset corresponding to the closest reference LOS. This would also induce a possible tiling effect but that would be much lighter than with a completely independent training on each reference LOS. We chose to use this approach with the addition that we constructed an average reference CMD from 10 of the bare reference CMD of each training reference LOS. From this the map is constructed by using the observed CMD and the closest averaged reference CMD as input. We note that an intermediate approach would be either to interpolate the reference CMD or to construct a sub grid of BGM references that has a better resolution than the training, but still is significantly less sampled than the map resolution.\\

\begin{figure*}
\hspace{-1.5cm}
	\begin{minipage}{1.15\textwidth}
	\centering
	\begin{subfigure}[!t]{1.0\textwidth}
	\includegraphics[width=1.0\hsize]{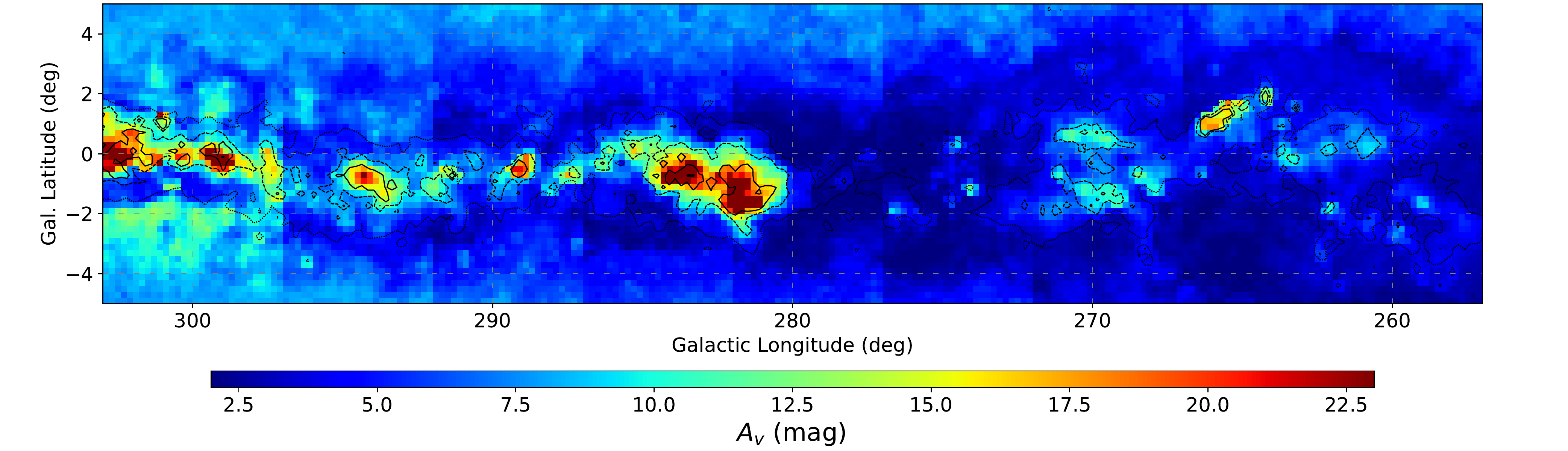}
	\end{subfigure}
	\end{minipage}
	\caption[Plane of the sky view for the 2MASS multiple-LOS training]{Integrated extinction for each pixel of the 2MASS multiple-LOS training prediction in a plane of the sky view using galactic coordinates. Contours are from Planck $\tau_{353}$.}
	\label{multi_los_2mass_polar_sky}
\end{figure*}

\begin{figure*}[!t]
\vspace{2cm}
\hspace{-1.5cm}
\begin{minipage}{1.15\textwidth}
	\centering
	\begin{subfigure}[!t]{1.0\textwidth}
	\caption*{\bf \large $\bm{0 \leqslant d < 4}$ kpc}
	\includegraphics[width=1.0\hsize]{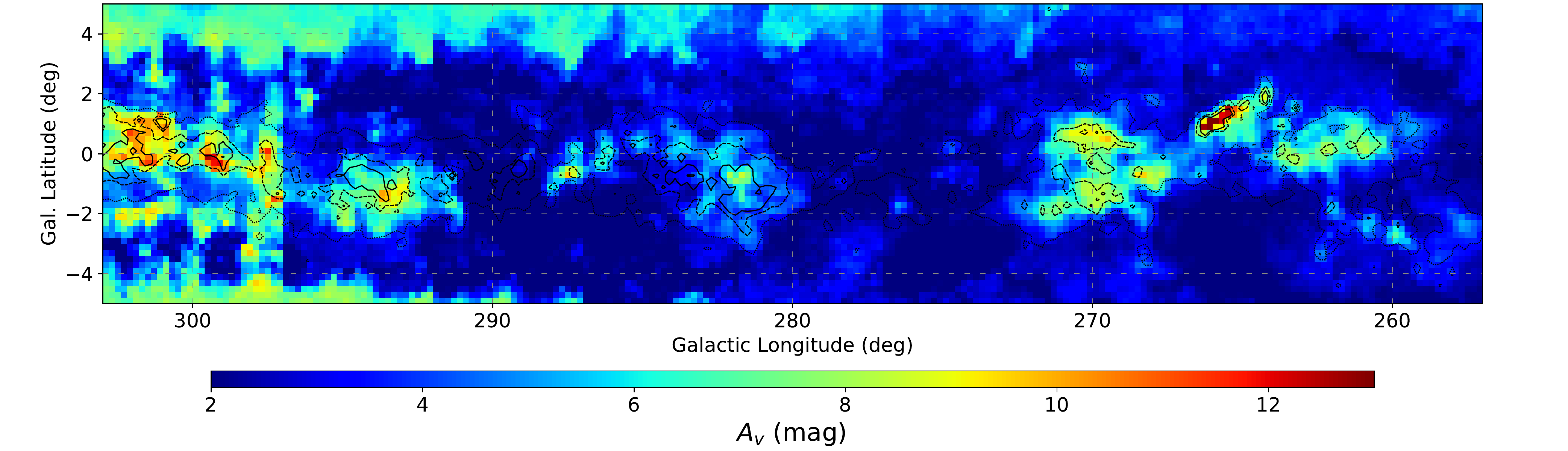}
	\end{subfigure}\\
	\vspace{0.8cm}
	\begin{subfigure}[!t]{1.0\textwidth}
	\caption*{\bf \large $\bm{4 < d \leqslant 7}$ kpc}
	\includegraphics[width=1.0\hsize]{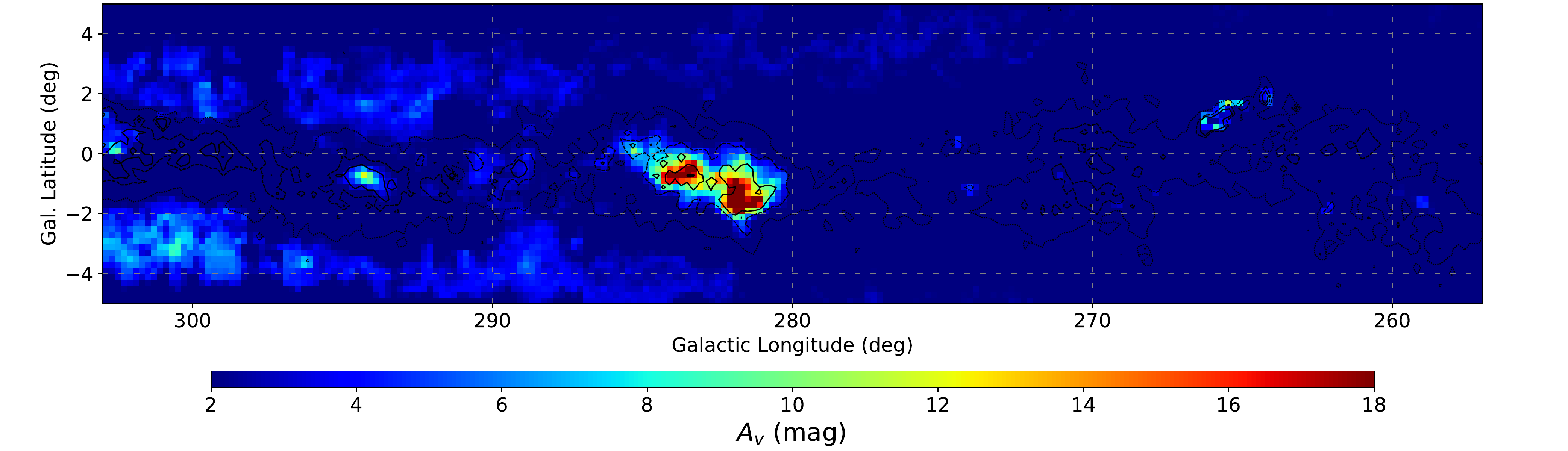}
	\end{subfigure}\\
	\vspace{0.8cm}
	\begin{subfigure}[!t]{1.0\textwidth}
	\caption*{\bf \large $\bm{ d \geqslant 7}$ kpc}
	\includegraphics[width=1.0\hsize]{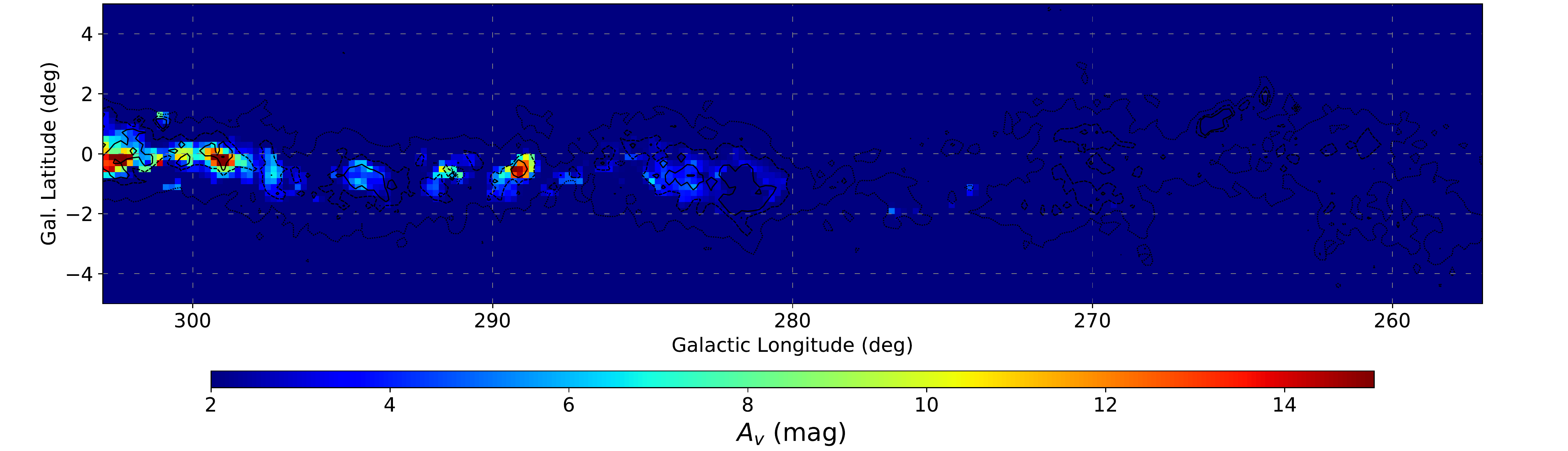}
	\end{subfigure}
	\end{minipage}
	\caption[Integrated view as a function of distance]{Integrated extinction for different distance range from the 2MASS multiple-LOS training prediction in a plane of the sky view using galactic coordinates. Contours are from Planck $\tau_{353}$.}
	\label{multiple_los_2mass_integrated_distance_results}
\end{figure*}

\begin{figure*}[!t]
\hspace{-0.8cm}
	\begin{minipage}{1.08\textwidth}
	\centering
	\begin{subfigure}[!t]{0.47\textwidth}
	\includegraphics[width=1.0\hsize]{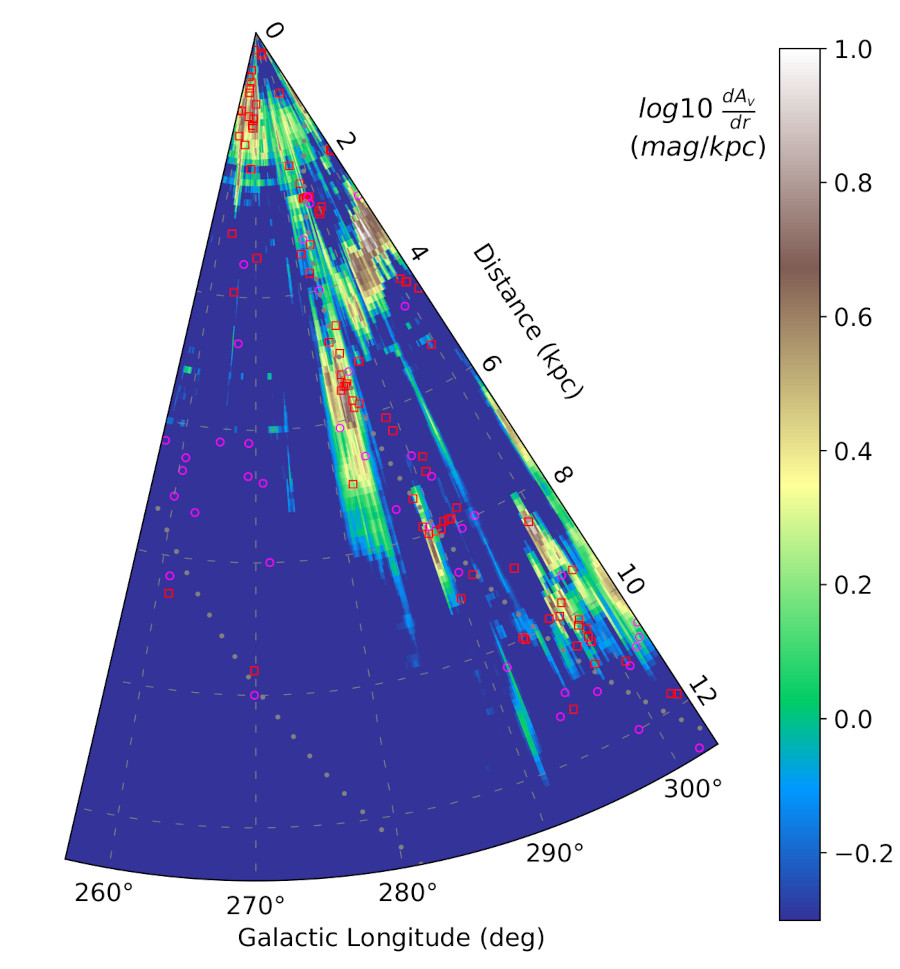}
	\end{subfigure}
	\hspace{0.5cm}
	\begin{subfigure}[!t]{0.47\textwidth}
	\includegraphics[width=1.0\hsize]{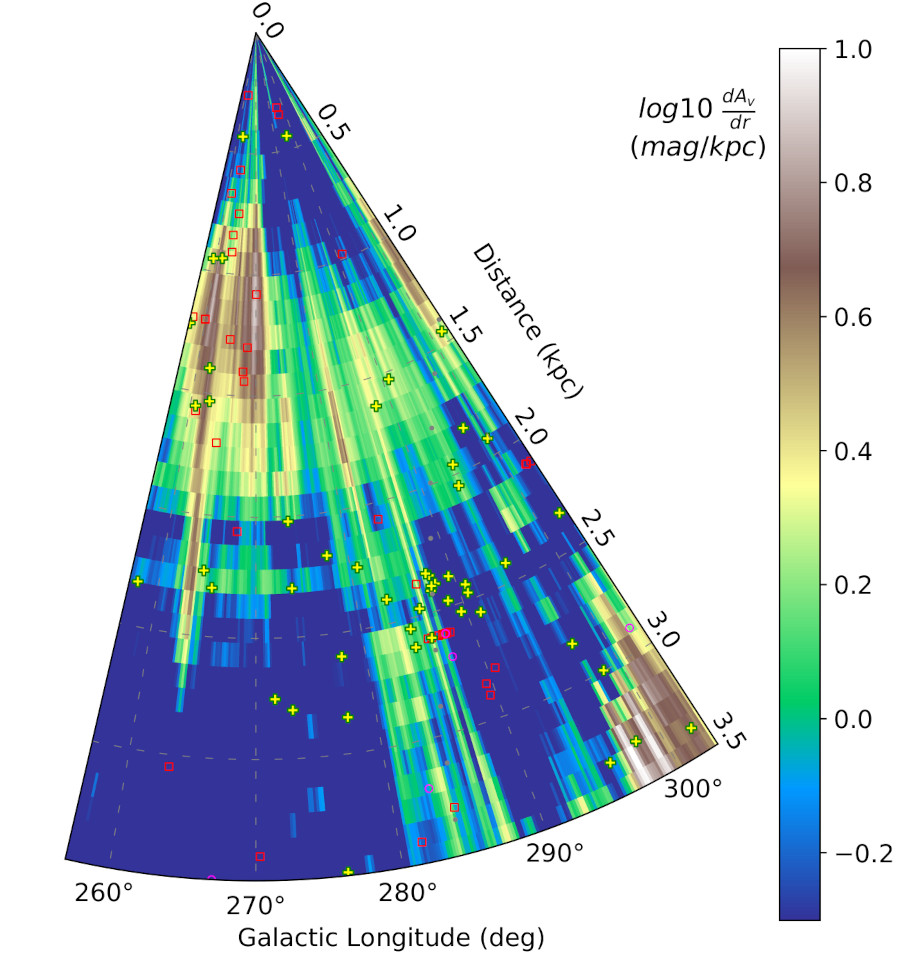}
	\end{subfigure}
	\end{minipage}
	\caption[Face-on view for the 2MASS multiple-LOS training]{Face-on view of the Galactic Place $|b| < 1$ deg in polar galactic-longitude distance coordinates for the predicted Carina arm region using our 2MASS multiple-LOS training. The symbols are the same as in figure ~\ref{single_los_2mass_polar_plan}. {\it Left:} Full distance prediction. {\it Right:} Zoom in on the $d < 3.5$ kpc prediction region. The displayed symbols are as in Fig.~\ref{single_los_2mass_polar_plan}.}
	\label{multi_los_2mass_polar_plan}
\end{figure*}

The results for this network prediction are presented in Figure~\ref{multi_los_2mass_polar_sky} that shows the plane of the sky map of integrated extinction, and in Figure~\ref{multi_los_2mass_polar_plan} that shows the face-on polar view of our distance prediction. For this result we also present in Figure~\ref{multiple_los_2mass_integrated_distance_results}, a decomposition of the integrated extinction corresponding to different distance range, $[0 \leqslant d < 4]$, $[4 \leqslant d < 7]$ and $[d \geqslant 7]$ kpc. From these figures it is visible that a tiling effect remains in the integrated extinction, especially for the large-longitude tile. However, it mainly impacts the highest latitudes and close distance range ($d < 4$ kpc) of the map, which are not in our Galactic Place slice represented in the face-on view. Interestingly, these results were made using $Z_{\rm lim}=50$ meaning that the effect of properly sampling the galactic longitude into several lines of sight already considerably reduces the very large amount of artifacts we had in the equivalent training using the same value of $Z_{\rm lim}$. The reason we chose not to use a larger $Z_{\rm lim}$ was that it reduced the map maximum distance estimation loosing most of the structures for $d > 8kpc$, which were recovered using a smaller $Z_{\rm lim}$ value. Another striking difference between this result and the single training one, even with high $Z_{\rm lim}$, is the contrast we obtain in high-extinction regions. For example the structure  around $l = 283$ deg and $b=-1$ deg contains a much higher integrated extinction than before. The few structures at low longitude are also much more convincing, following accurately several of the Planck map structures highlighted by the contours. The structure at $l = 266$ deg and between $1 < b < 2 $ deg is much better identified as containing much more extinction that the surrounding small regions, which was not the case in our single LOS training. In the large-longitude part (l > 297) the situation is more complicated. Even if we have a dedicated LOS for this region, the predictions appear to be less good. It is still much better than the single previous single LOS training but the structures less accurately follow the Planck morphology and are more prone to artifacts. Indeed, the large difference in star count with the previous one (see Fig.\ref{2mass_sky_star_count}) tends to indicate that either our longitude sampling is too coarse and more reference LOS should be used here, or the information is too different from the other lines of sight for the generalization on them to be useful here. This would be equivalent to consider that this LOS was trained solely on the $2\times 10^5$ examples which is insufficient to constrain a single LOS training. However, trying to voluntarily imbalance this training dataset would require much care and we preferred to delay such an approach to a future work.\\

\newpage
Regarding the details from the face-on view (Fig.~\ref{multi_los_2mass_polar_plan}), the extinction distribution in distance is quite similar to that obtained with our single-LOS training. Still, we stress again that this result is obtained using $Z_{\rm lim}=50$, which led to many artifacts in the single-LOS training, whereas they are totally absent here. This tends to confirm that there is effectively no strong extinction structures in the low longitude $l<280$ deg high distance d > 3 region, since a lower value of $Z_{\rm lim}$ increases the sensitivity to structures traced by a limited number of stars, and since we used multiple dedicated reference lines of sight for this large region. Still, the short-distance figure (Right frame) shows an interesting extinction dynamic in this region for $ 0.5 < d < 2.5$ kpc with regions mostly being in agreement with both the HII regions and the Gaia clusters. Overall the region at $l=283$ deg and $d=5.5$ kpc interpreted as the Carina arm tangent is still well resolved and is mostly in agreement with the HII regions. The regions at $d=10$ kpc and $295 < l < 303$ deg show a good match with many HII regions and a much more detailed structure than the diffuse structure obtained in this area with our single-LOS training. For this reason, although we found the single-LOS result of this area suspicious, we are quite confident with that of the multiple-LOS training. We note that this is typically the kind of structures that are removed if the $Z_{\rm lim}$ parameter is increased, since they are at large distances and are most likely constrained by a relatively small number of stars in the CMD. Finally, the problematic structure at $l \simeq 300$ deg, $d = 4$ kpc is still present and has a stronger maximum differential extinction than for the single LOS training. The foreground for this large-longitude LOS still remains compatible with our arm shape. The result from this multiple-LOS training and presented in the two previous figures (\ref{multi_los_2mass_polar_sky} and \ref{multi_los_2mass_polar_plan}) constitutes our main reference result. {\bf For the rest of the present manuscript we refer to this result as our "main 2MASS result", to ease the comparison with our cases.}\\

\begin{figure*}[!t]
\hspace{-0.8cm}
	\begin{minipage}{1.08\textwidth}
	\centering
	\begin{subfigure}[!t]{0.47\textwidth}
	\includegraphics[width=1.0\hsize]{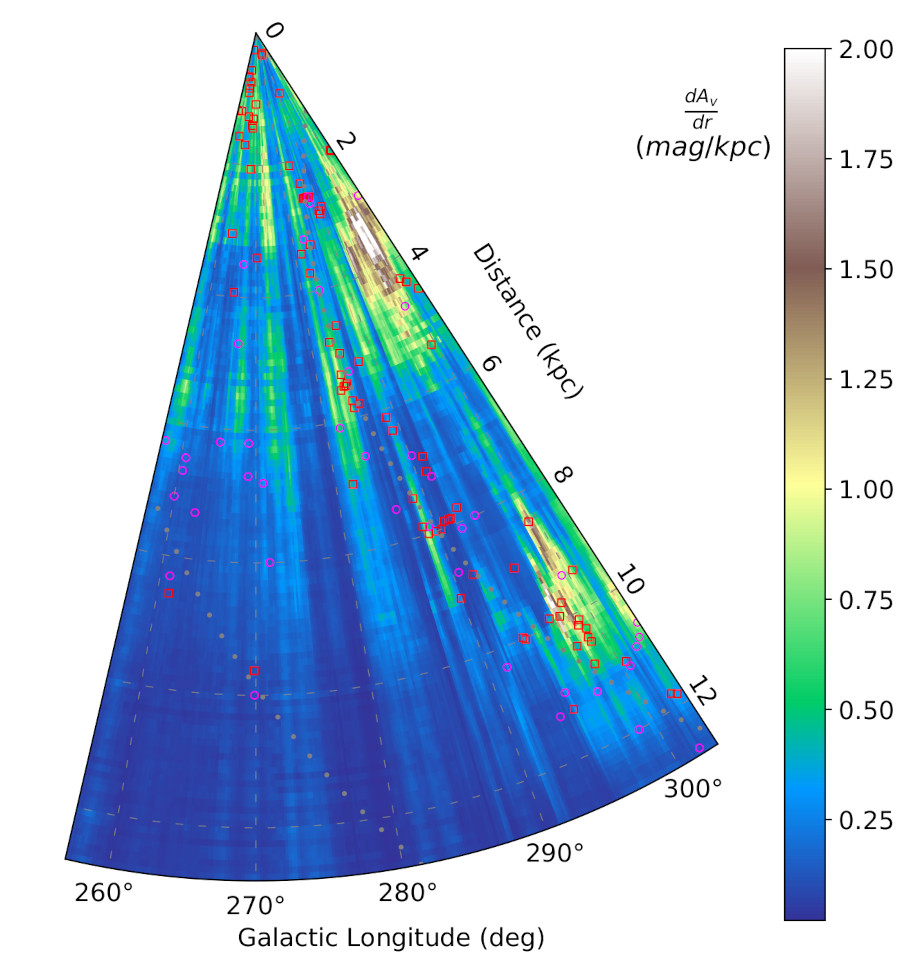}
	\end{subfigure}
	\hspace{0.5cm}
	\begin{subfigure}[!t]{0.47\textwidth}
	\includegraphics[width=1.0\hsize]{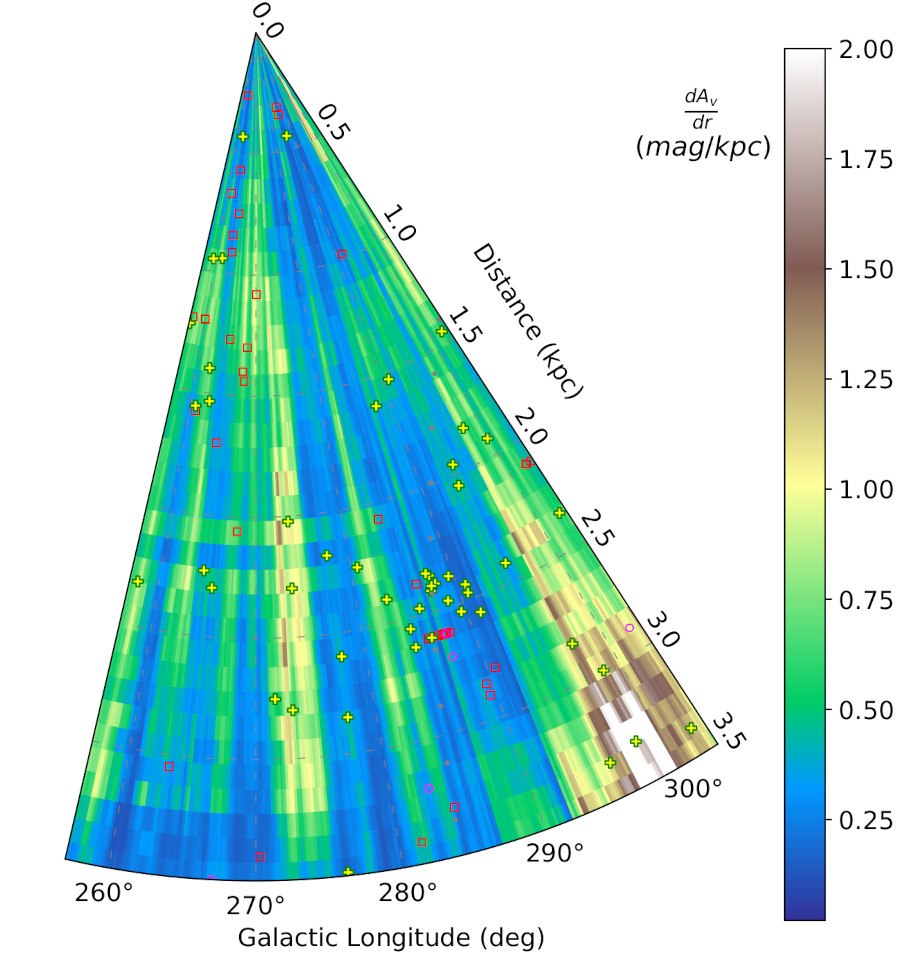}
	\end{subfigure}
	\end{minipage}
	\caption[Face-on view for the standard deviation of the 2MASS multiple-LOS training]{Face-on view of the Galactic Place $|b| < 1$ deg in polar galactic-longitude distance coordinates for the standard deviation predicted in the Carina arm region using our 2MASS multiple-LOS training. The symbols are the same as in figure ~\ref{single_los_2mass_polar_plan}. {\it Left:} Full distance prediction. {\it Right:} Zoom in on the d < 3.5 kpc prediction region.}
	\label{std_multi_los_2mass_polar_plan}
\end{figure*}

\newpage
Figure~\ref{std_multi_los_2mass_polar_plan} shows the averaged standard deviation of the prediction coming from the dropout using the face-on view. Overall this figure appears more contrasted than the equivalent one from the single-LOS training (Fig.~\ref{std_single_los_2mass_polar_plan}. Again it is mostly explained by the lower $Z_{\rm lim}$ value that allows the network to consider less stars as relevant for the extinction prediction, inducing that we recover structures that we potentially missed before, but also reduce the signal-to-noise ratio of the global map prediction. For this reason and because we conserved the same scale for comparison, we expect only the two regions with the highest dispersion to reflect a true underlying issue. These regions are both found in the large-longitude region that corresponds to the same problematic reference LOS. The region at $d = 10$ kpc that contains several small predictions lost by the higher $Z_{\rm lim}$ are in fact expected to be uncertain from this standpoint since they are at our detection limit. This does not raise much concern about the fact that they are representative of genuine physical structures, but rather that their differential extinction might be off the true value and that they are possibly too extended around their central position, similarly to what is illustrated in the second frame of Fig.~\ref{multi_los_2MASS_test_profiles_prediction}. In contrast, the region between $3 < d < 4$ kpc is much more likely to be an artifact since it has a very large dispersion while it is much closer and does not have a high extinction foreground. We expect this behavior for a significantly unconstrained region, or for an input that lies off the part of the feature space that was effectively constrained during training. Without additional investigation of this region we were unable to draw firm conclusion about the reality of this strong extinction structure.\\

\begin{figure}[!t]
	\hspace{-1.2cm}
	\begin{minipage}{1.10\textwidth}
	\centering
	\includegraphics[width=1.0\hsize]{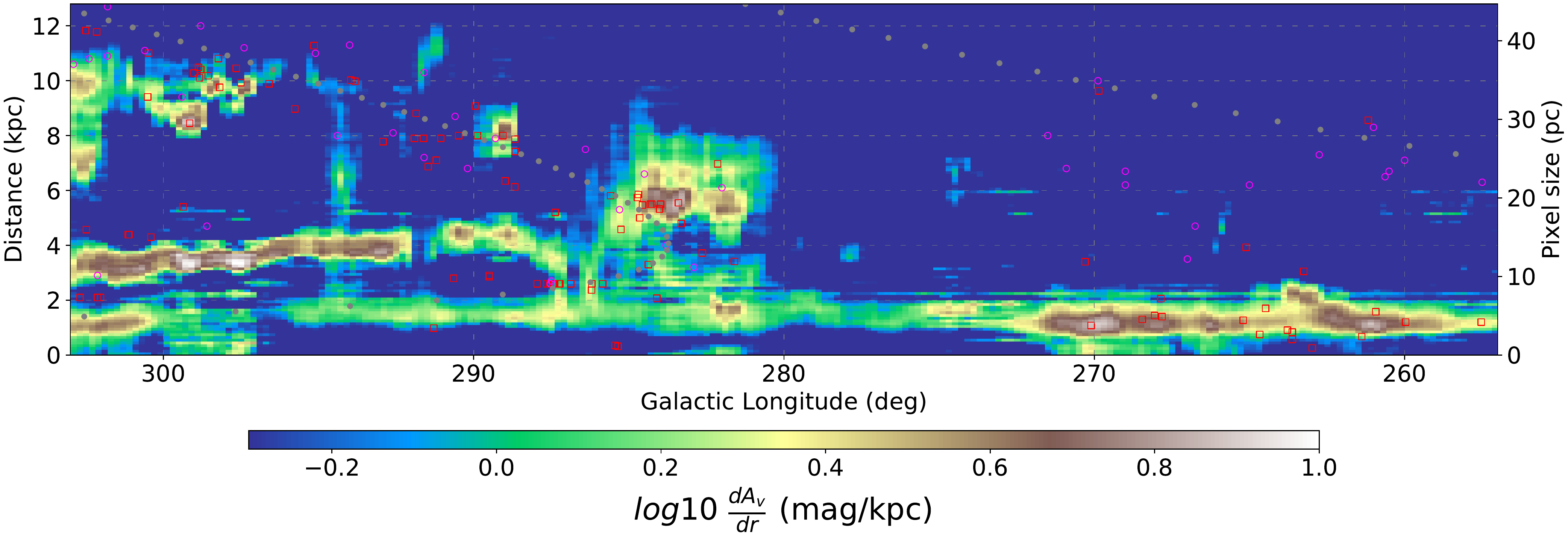}
	\end{minipage}
	\caption[Cartesian face-on view for the 2MASS multiple-LOS training]{Face-on view of the Galactic Place $|b| < 1$ deg in cartesian galactic-longitude distance coordinates for the predicted Carina arm region using our 2MASS multiple-LOS training. The axis on the right border corresponds to the pixel height as a function of the distance induced by the conic shape of our LOS. The symbols are the same as in Figure ~\ref{single_los_2mass_polar_plan}.}
	\label{multi_los_2MASS_test_profiles_prediction}
\end{figure}

\newpage
\subsubsection{Effect of the galactic latitude}

In this section, we use a new multiple line of sights training to examine the effects of a latitude sampling. The general purpose of this test is to assess whether a multiple-LOS training based on a sampling both in longitude and latitude would be useful, since it would considerably increase the size of our training dataset. For this we performed a training again centered on $l=280$ deg, $b=0$ deg using 5 reference lines of sight between $-4 < b < 4$ deg, each LOS being used as a reference for a 2 deg slice in latitude. Each reference LOS is again provided with $2\times 10^5$ examples following the same construction described in Section \ref{multi_los_CMD_constrution}.\\

\begin{figure*}[!t]
\hspace{-1.5cm}
	\begin{minipage}{1.15\textwidth}
	\centering
	\begin{subfigure}[!t]{0.48\textwidth}
	\caption*{\bf \large Single LOS}
	\includegraphics[width=1.0\hsize]{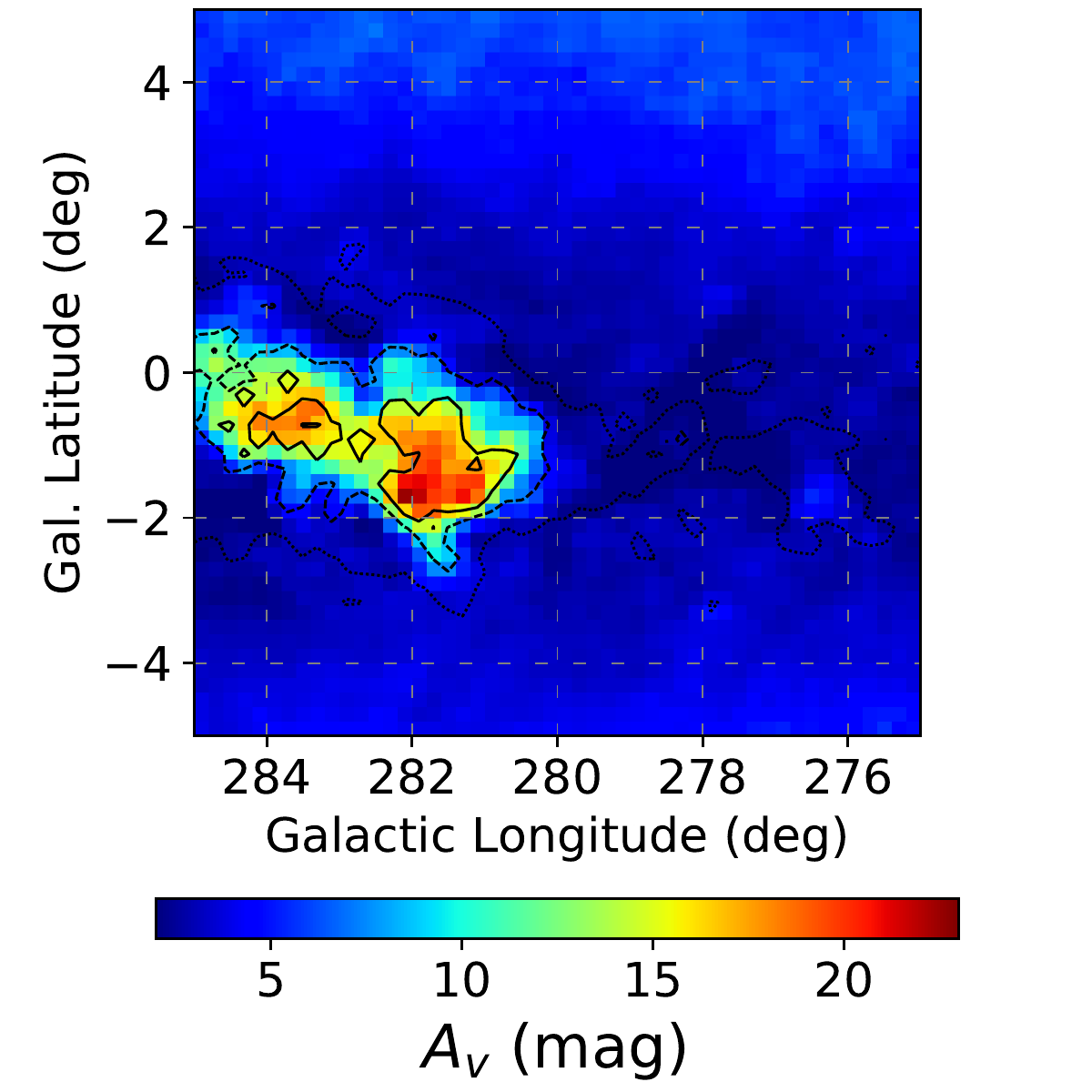}
	\end{subfigure}
	\hspace{0.3cm}
	\begin{subfigure}[!t]{0.48\textwidth}
	\caption*{\bf \large Multi LOS longitude}
	\includegraphics[width=1.0\hsize]{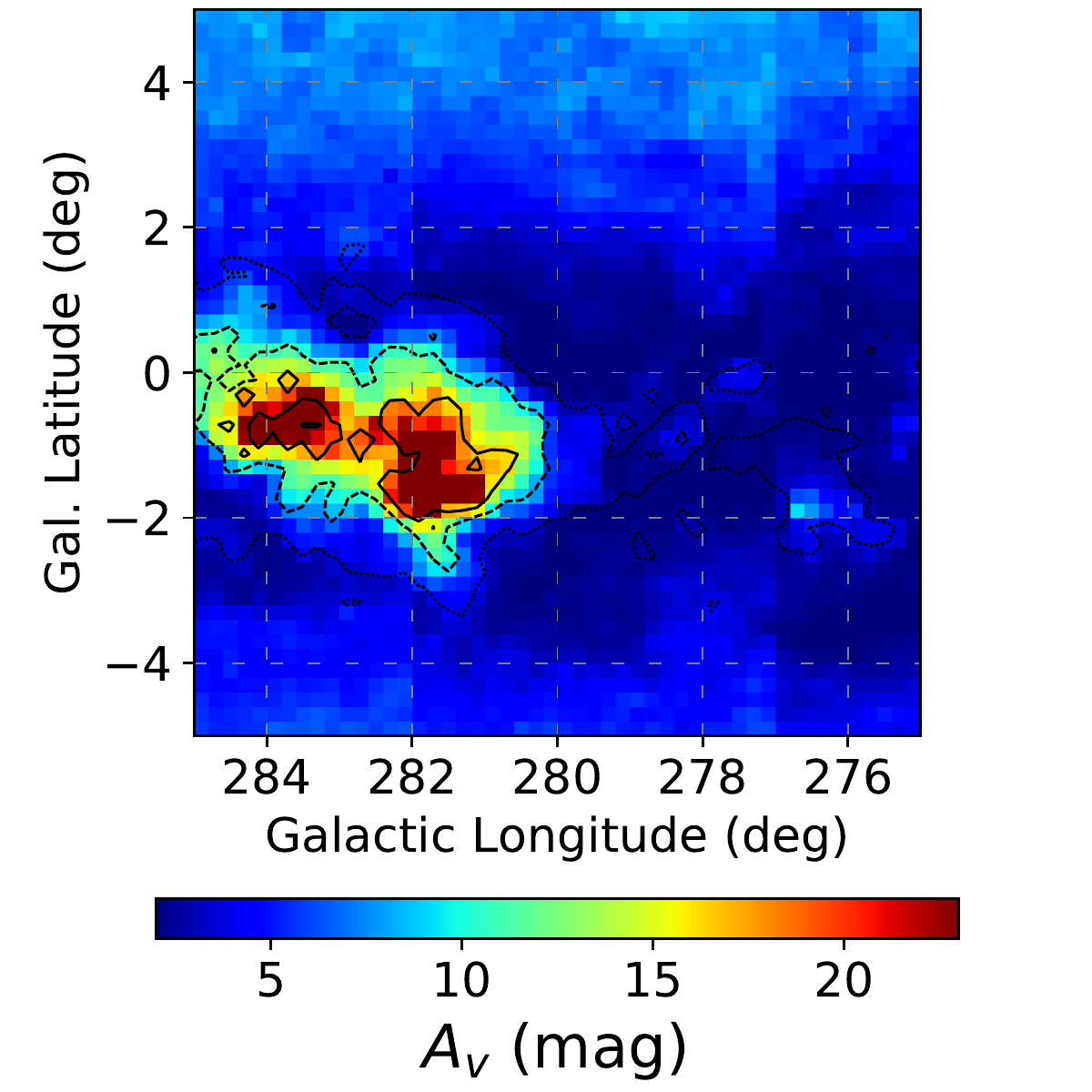}
	\end{subfigure}\\
	\vspace{0.8cm}
	\begin{subfigure}[!t]{0.48\textwidth}
	\caption*{\bf \large Multi LOS latitude}
	\includegraphics[width=1.0\hsize]{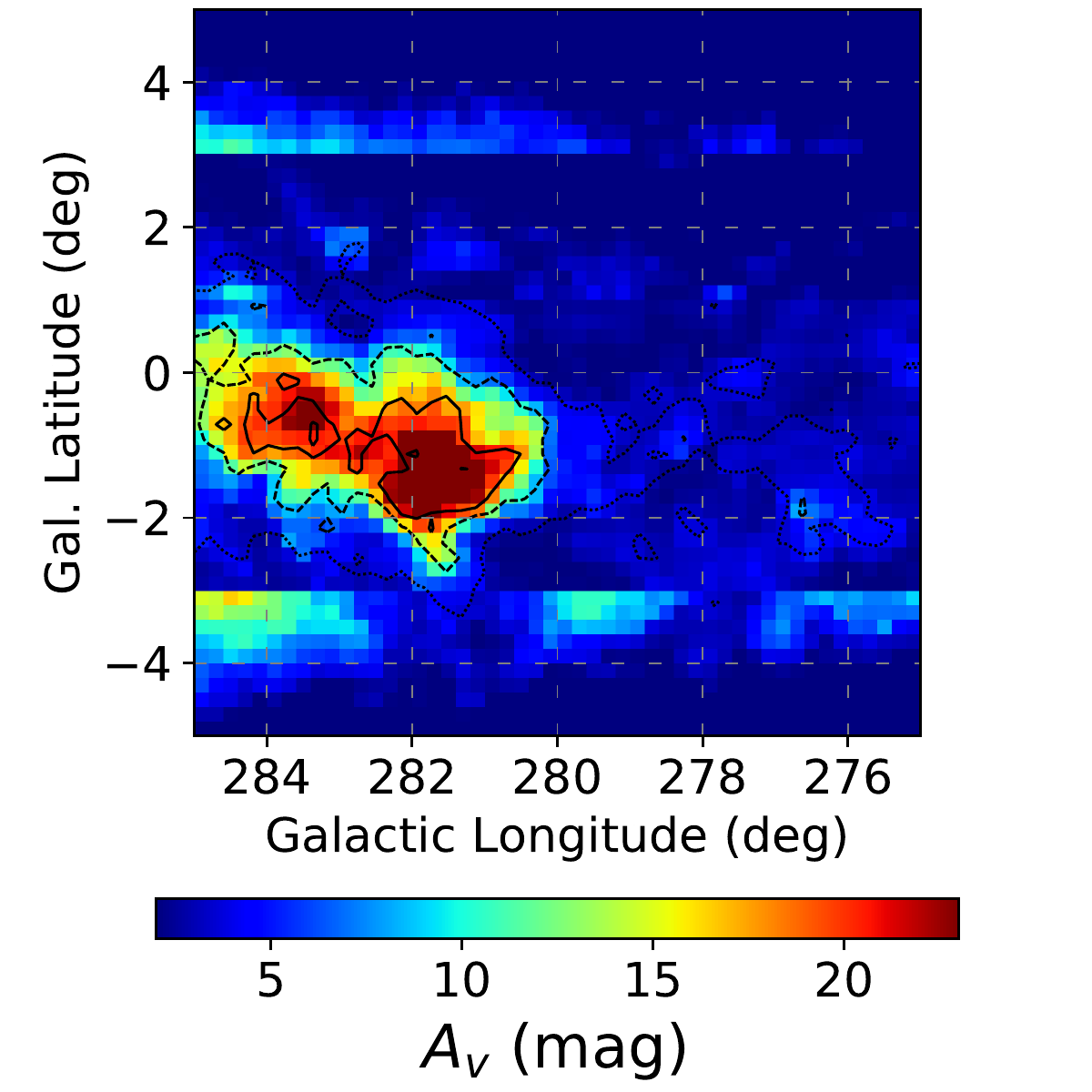}
	\end{subfigure}
	\hspace{0.3cm}
	\begin{subfigure}[!t]{0.48\textwidth}
	\caption*{\bf \large Planck $\tau_{353}$}
	\includegraphics[width=1.0\hsize]{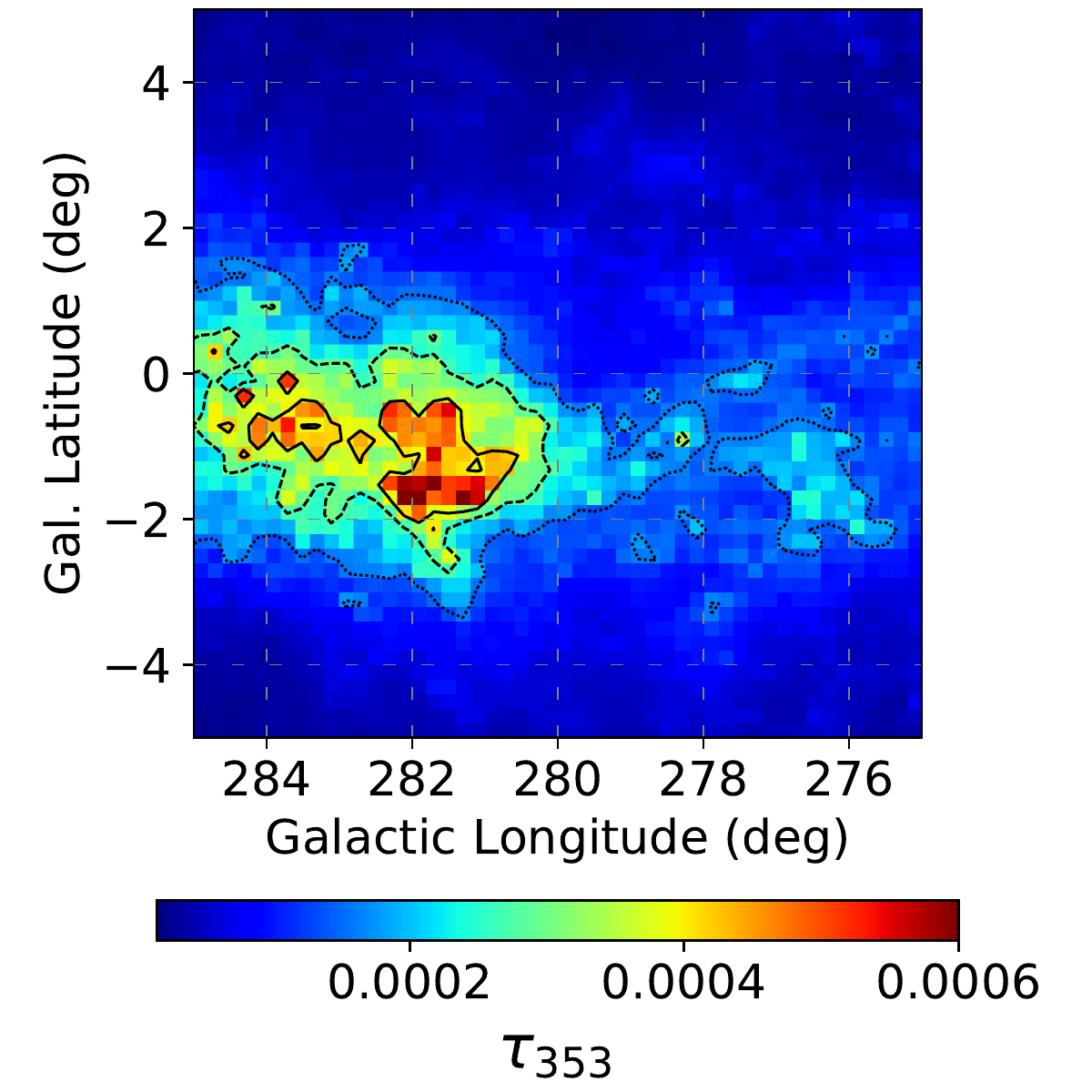}
	\end{subfigure}
	\end{minipage}
	\vspace{0.2cm}
	\caption[Plane-of-the-sky view for the 2MASS latitude sampling multiple-LOS training]{Comparison of the integrated extinction for different 2MASS training prediction in a plane of the sky view using galactic coordinates. All the predictions are cropped at the longitude prediction range of the 2MASS latitude sampling training. The contours are from Planck $\tau_{353}$. {\it Top-left:} 2MASS single LOS training result. {\it Top-right:} 2MASS multi LOS longitude sampling training result. {\it Bottom-left:} 2MASS multi LOS latitude sampling training result. {\it Bottom-Right:} Observed Planck dust opacity at 353.}
	\label{multi_los_2MASS_latitude}
\end{figure*}

Figure~\ref{multi_los_2MASS_latitude} shows a comparison of different integrated extinction map over the plane of the sky for that include this training in the bottom-left frame. From this figure, the three middle reference LOS $b = -2, 0, 2$ deg seems to have slightly improve the boundary of the structure in comparison to the two previous 2MASS training. Still the most striking effect is that there are significantly less latitude artifacts and extinction overall predicted in empty regions, especially considering that this training was made with $Z_{\rm lim}=50$. However there is a relatively strong tiling effect. This should come from the same effect we exposed for the multiple-LOS training in the previous section, where the discontinuity between tiles is stronger when the star count changes quickly. As we get farther from the Galactic Place this is likely to be the case here. Surprisingly, the network seems to react to an overall excess of stars compared to the reference LOS by overestimating the extinction, as revealed by the strong latitude gradient in the $+4$ and $-4$ deg tiles. Finally, we observed that the extinction visible for $|b|$ between 3 and 4 deg is mostly localized at large distance $d > 10$ kpc.\\

There are a few solutions to overcome this effect and still take advantage of the added information coming from multiple LOS in latitude. First, since most of the errors appear to be at large distance, we could simply raise the $Z_{\rm lim}$ value. Another solution would be to increase the number of LOS to better sample the latitude axis. This second solution would require again a very large amount of data for large maps. Another approach would be to refine only the spatial grid of the reference CMDs that are used during the forward step along with the observed one. Indeed, we expect that from several reference LOS the network should have found a continuous transformation of the CMD to account for the variations in star count. Therefore, changing only the reference CMD for the one corresponding to each pixel LOS when constructing the map should benefit from this automated CMD interpolation by the network. However, as we stated before, creating that many BGM realizations would be very compute intensive. As proposed before, a convenient solution would be to use a grid for the BGM CMDs used in the forward step with an intermediate resolution between that of the map and that used for training. In any case, this tiling effect should be fixed before attempting a multiple-LOS training on longitude and latitude simultaneously.

\clearpage
\subsection{Comparison with other 3D extinction maps}

\begin{sidewaysfigure}
\hspace{-0.6cm}
	\begin{minipage}{1.05\textwidth}
	\centering
	\begin{subfigure}[!t]{1.0\textwidth}
	\caption*{\bf \large \hspace{0.8cm} Planck dust opacity $\bm{\tau_{353}}$}
	\includegraphics[width=1.0\hsize]{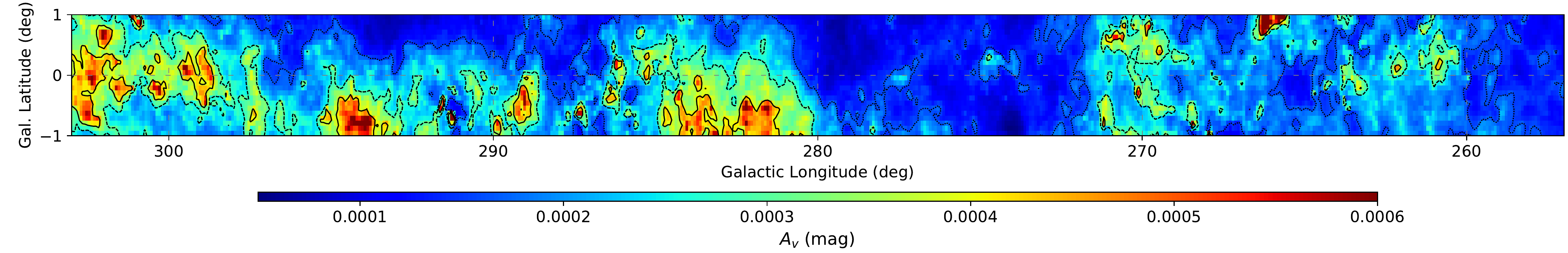}
	\end{subfigure}\\
	\vspace{0.4cm}
	\begin{subfigure}[!t]{1.0\textwidth}
	\caption*{\bf \large \hspace{0.8cm} Main 2MASS result}
	\includegraphics[width=1.0\hsize]{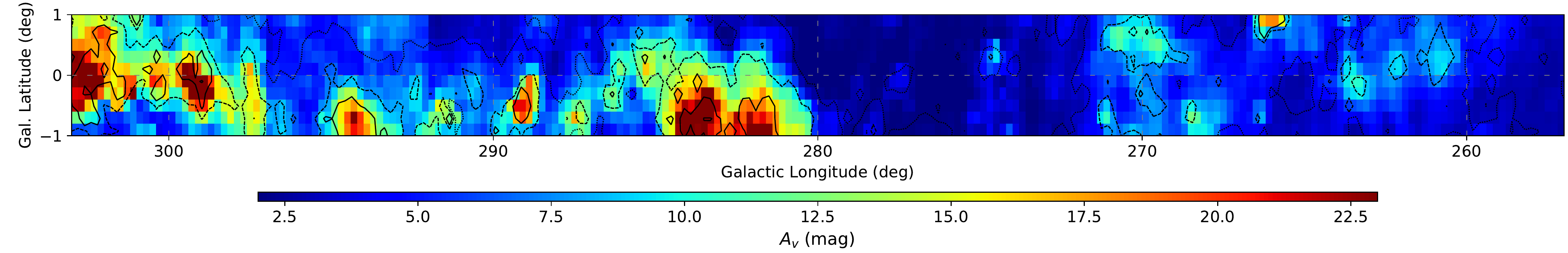}
	\end{subfigure}\\
	\vspace{0.4cm}
	\begin{subfigure}[!t]{1.0\textwidth}
	\caption*{\bf \large \hspace{0.8cm} Marshall+20}
	\includegraphics[width=1.0\hsize]{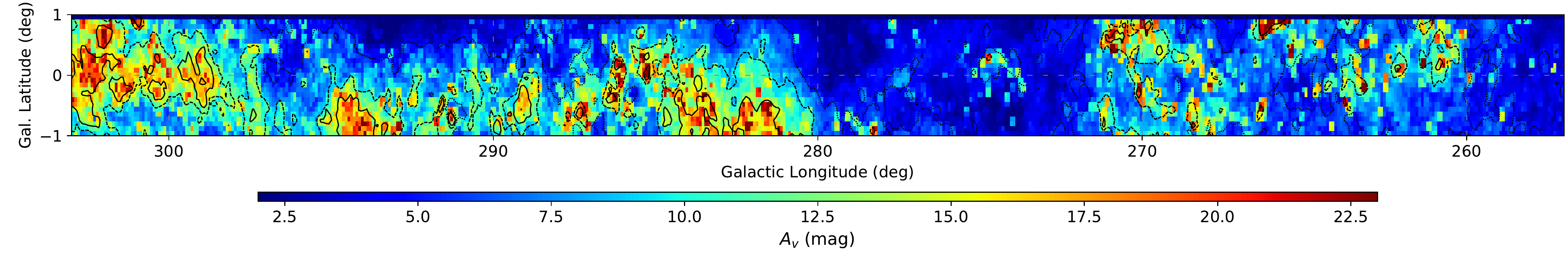}
	\end{subfigure}
	\end{minipage}
	\caption[Comparison of Galactic Place for the 2MASS multiple-LOS training with M20]{Comparison of the 2MASS multiple-LOS training results with Planck opacity and with the 3D extinction map by M20, in a plane of the sky view using galactic coordinates. All the maps are cropped according to the M20 latitude limits. {\it Top:} Observed Planck dust opacity at 353 GHz using the pixel resolution of M20. {\it Middle:} Integrated extinction over the whole LOS for each pixel corresponding to the 2MASS multiple-LOS training. {\it Bottom:} Integrated extinction over the whole LOS for each pixel for the M20 prediction.}
	\label{cornu_marshall_int_maps_comparison}
\end{sidewaysfigure}

In this section we discuss briefly the comparison of our main 2MASS result with 3D extinction maps obtained with different methods. The simplest one to perform is with \citet{Marshall_2020} (here after M20) since both our maps rely on the BGM, use the same survey as input, use a LOS approach and have the same distance bin resolution. Figure~\ref{cornu_marshall_int_maps_comparison} shows the comparison of our map with the M20 one. In this figure we added Planck data for which we used the angular resolution of the M20 map rather than our lower resolution (the Planck map with our resolution is in Fig.~\ref{single_los_2mass_integrated_result}). To ease the comparison we cropped our map to the $|b| < 1$ deg range of the M20 map. From this figure our prediction appears less noisy that but it might be affected by the resolution choice. Still, the prediction of larger structure regarding the radial axis could be explained by the typical structure width from our GRFs, pushing the network to avoid very narrow structures. It can also be an effect of the convolution and pooling steps of our network which tend to smooth the fine grained differences in the input CMD, while in the MCMC method of M20, these fine grained differences are likely to be responsible for the strong contrast between some adjacent pixels in the map. \\

\begin{figure*}[!t]
\vspace{-0.3cm}
	\begin{minipage}{1.0\textwidth}
	\centering
	\begin{subfigure}[!t]{0.47\textwidth}
	\caption*{\bf \large Main 2MASS result}
	\includegraphics[width=1.0\hsize]{images/run147_polar_plan_map_log_terrain_ep70.jpg}
	\end{subfigure}
	\hspace{0.4cm}
	\begin{subfigure}[!t]{0.47\textwidth}
	\caption*{\bf \large Marshall+20}
	\includegraphics[width=1.0\hsize]{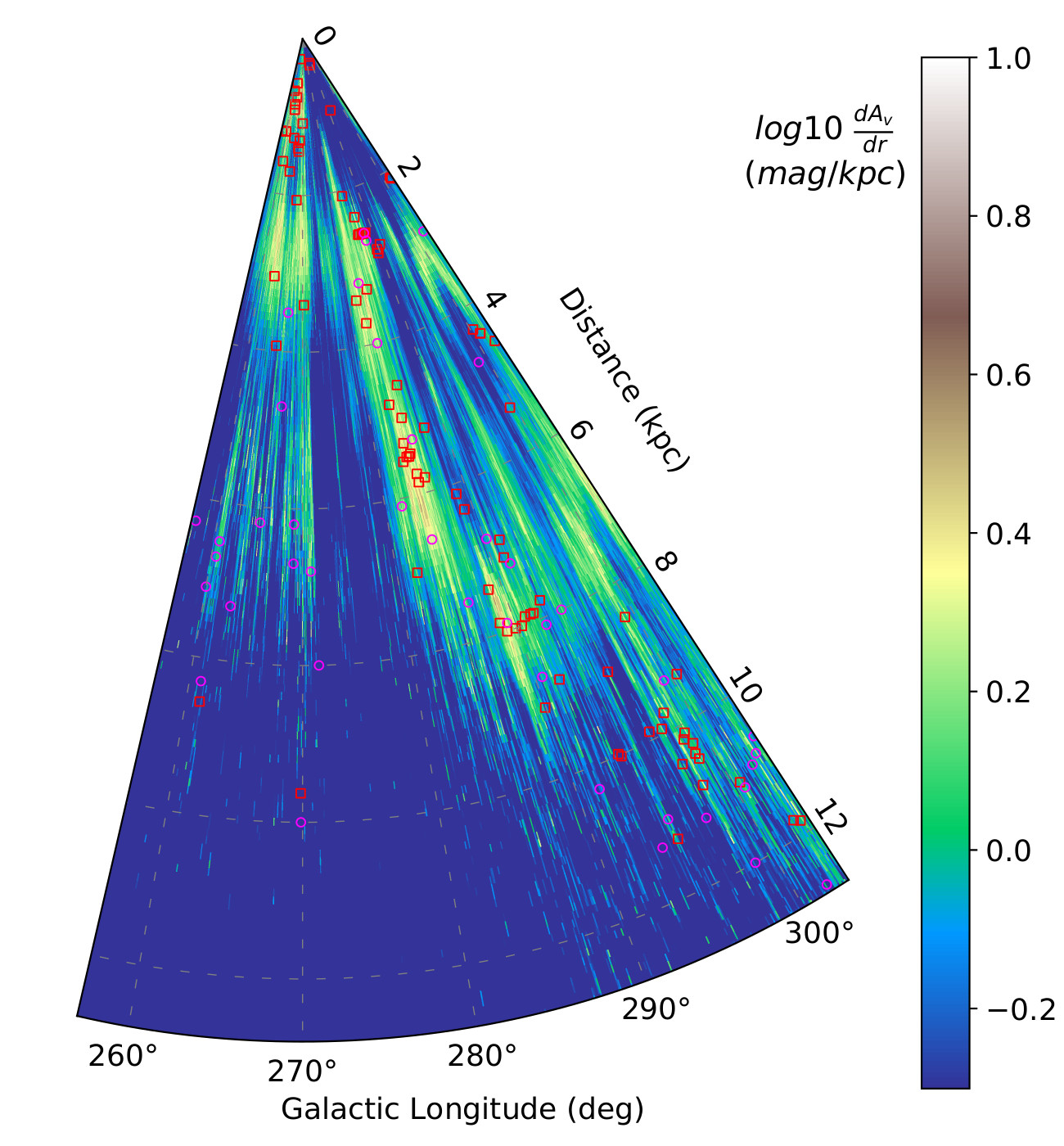}
	\end{subfigure}
	\end{minipage}
	\caption[Comparison of face-on views for the 2MASS multiple-LOS training with M20]{Comparison of the face-on view of the Galactic Place $|b| < 1$ deg in polar galactic-longitude distance coordinates for the Carina arm region. The symbols are the same as in figure ~\ref{single_los_2mass_polar_plan}. {\it Left:} Our main 2MASS multiple-LOS training result. {\it Right:} The M20 prediction using the same galactic slice construction.}
	\label{cornu_marshall_maps_comparison}
	\vspace{-0.5cm}
\end{figure*}

In Figure~\ref{cornu_marshall_maps_comparison} we compare our maps with the M20 map using the same face on representation using an identical plotting methodology from the M20 raw data cube. Both of the maps are averaged for a $|b| < 1$ deg slice. In this figure, the two maps are compared with the same color scale, the same colobar range and the same method to average on latitude pixels. At first glance, most of the important structures are reconstructed similarly. The Carina arm tangent is very similar in our two maps. However, in our prediction the low longitude part is free of structures that could either be artifacts in the M20 map or be missed by our method. The $d = 10$ kpc, $l \simeq 300$ deg group of structures we recover in our map are absent from the M20, or may correspond to the extinction detected near 7 - 8 kpc in the M20 map. The fact that we have a very clear extinction free area in our results at the same place, and that it matches very well the distribution of HII region by \citet{hou_and_han_2014} suggests that our map is more reliable in this area. The structure at $l \simeq 300$ deg, $3 < d < 4$ kpc has an equivalent in the M20 map but with a much lower extinction and a lesser extent along the longitude axis than our prediction. As discussed in Section~\ref{sec:2MASS_main_result}, we doubt that our prediction is accurate for this structure.\\

\begin{figure*}[!t]
\vspace{0.6cm}
\hspace{-0.4cm}
	\begin{minipage}{1.05\textwidth}
	\centering
	\begin{subfigure}[!t]{0.48\textwidth}
	\caption*{\bf \large Main 2MASS result}
	\includegraphics[width=1.0\hsize]{images/run147_polar_plan_map_log_terrain_close_ep70.jpg}
	\end{subfigure}
	\vspace{0.6cm}
	\hspace{0.1cm}
	\begin{subfigure}[!t]{0.48\textwidth}
	\caption*{\bf \large Marshall+20}
	\includegraphics[width=1.0\hsize]{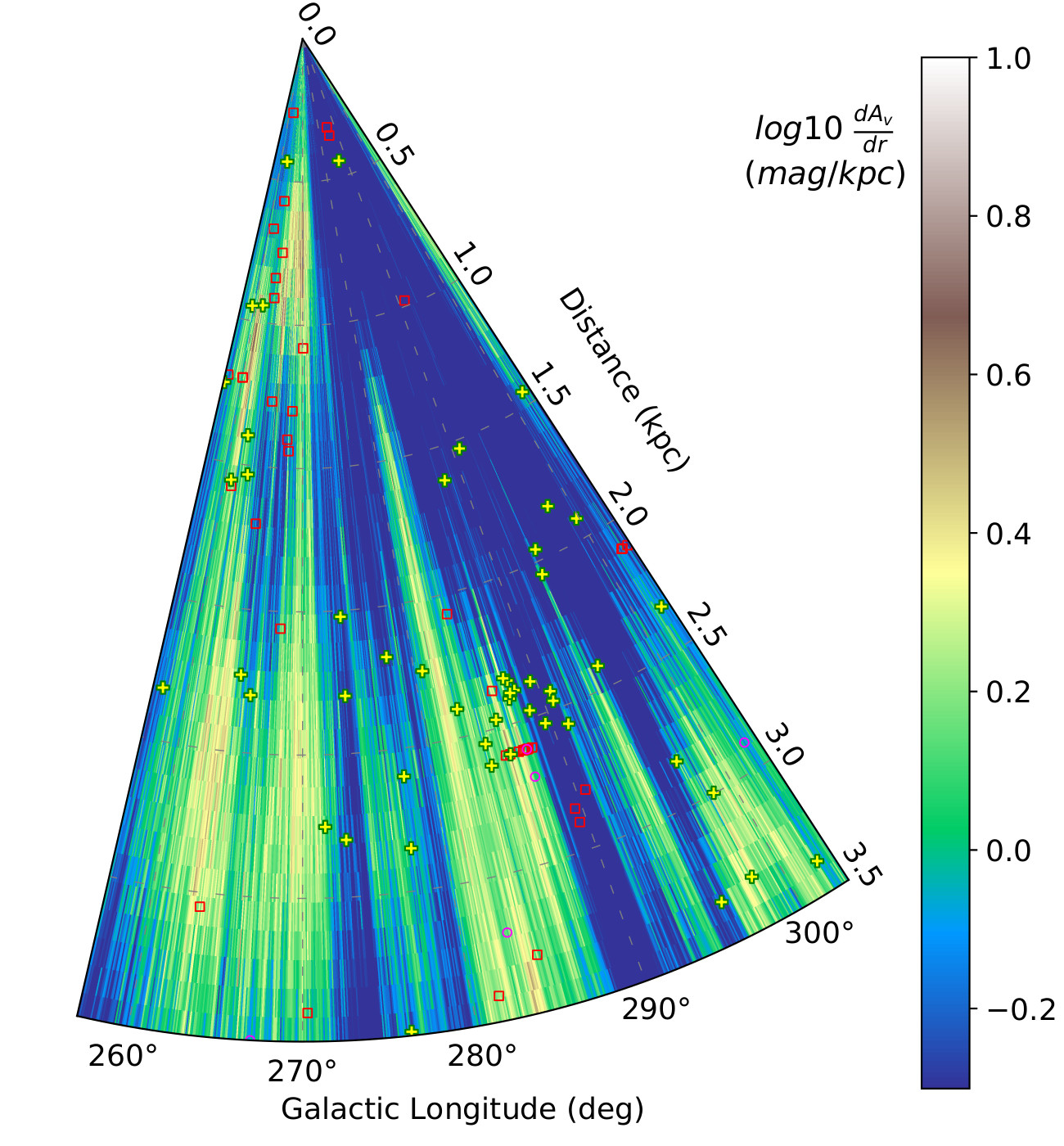}
	\end{subfigure}\\
	\begin{subfigure}[!t]{0.48\textwidth}
	\caption*{\bf \large Lallement+19}
	\includegraphics[width=1.0\hsize]{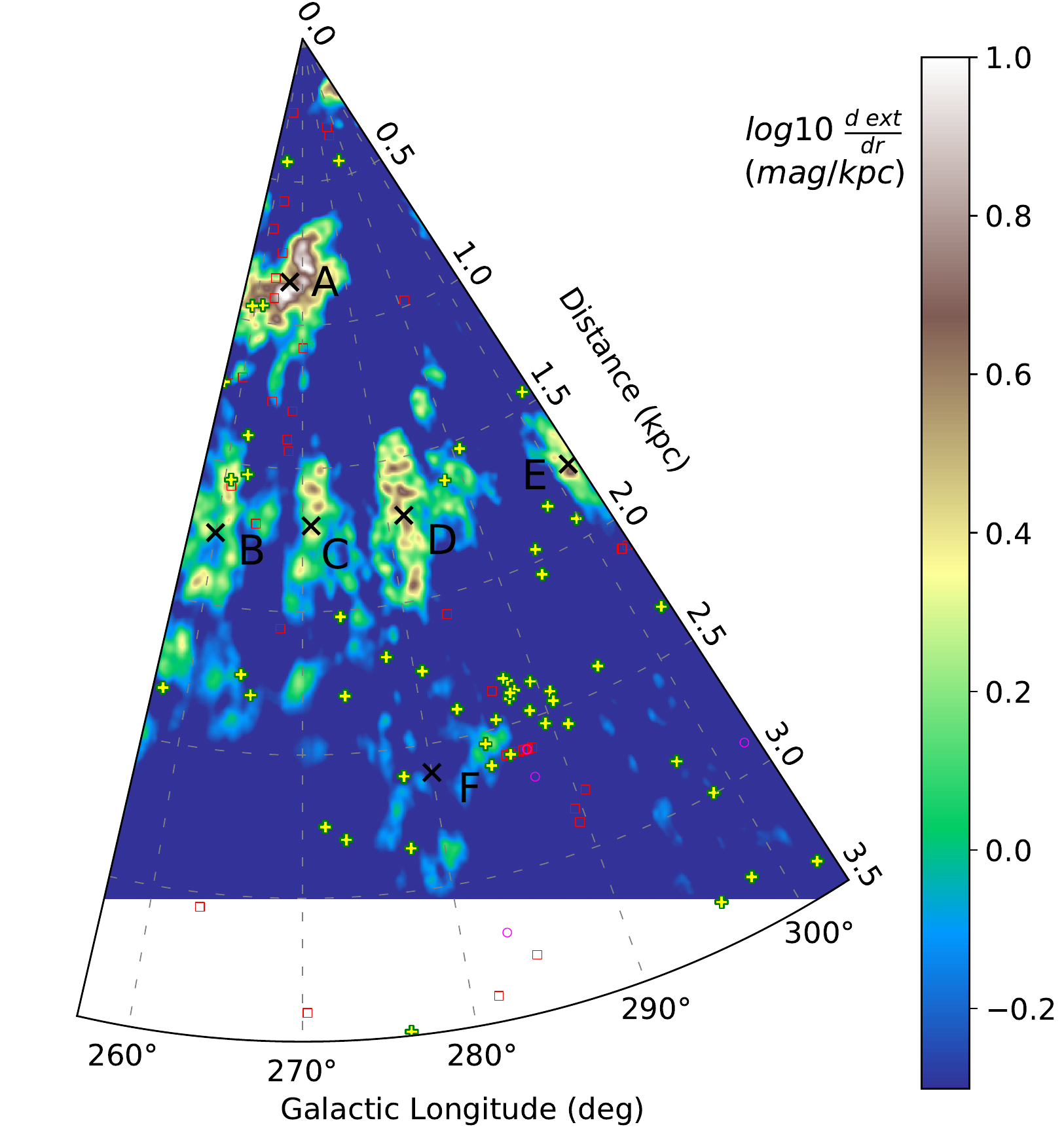}
	\end{subfigure}
	\vspace{0.6cm}
	\hspace{0.1cm}
	\begin{subfigure}[!t]{0.48\textwidth}
	\caption*{\bf \large Chen+19}
	\includegraphics[width=1.0\hsize]{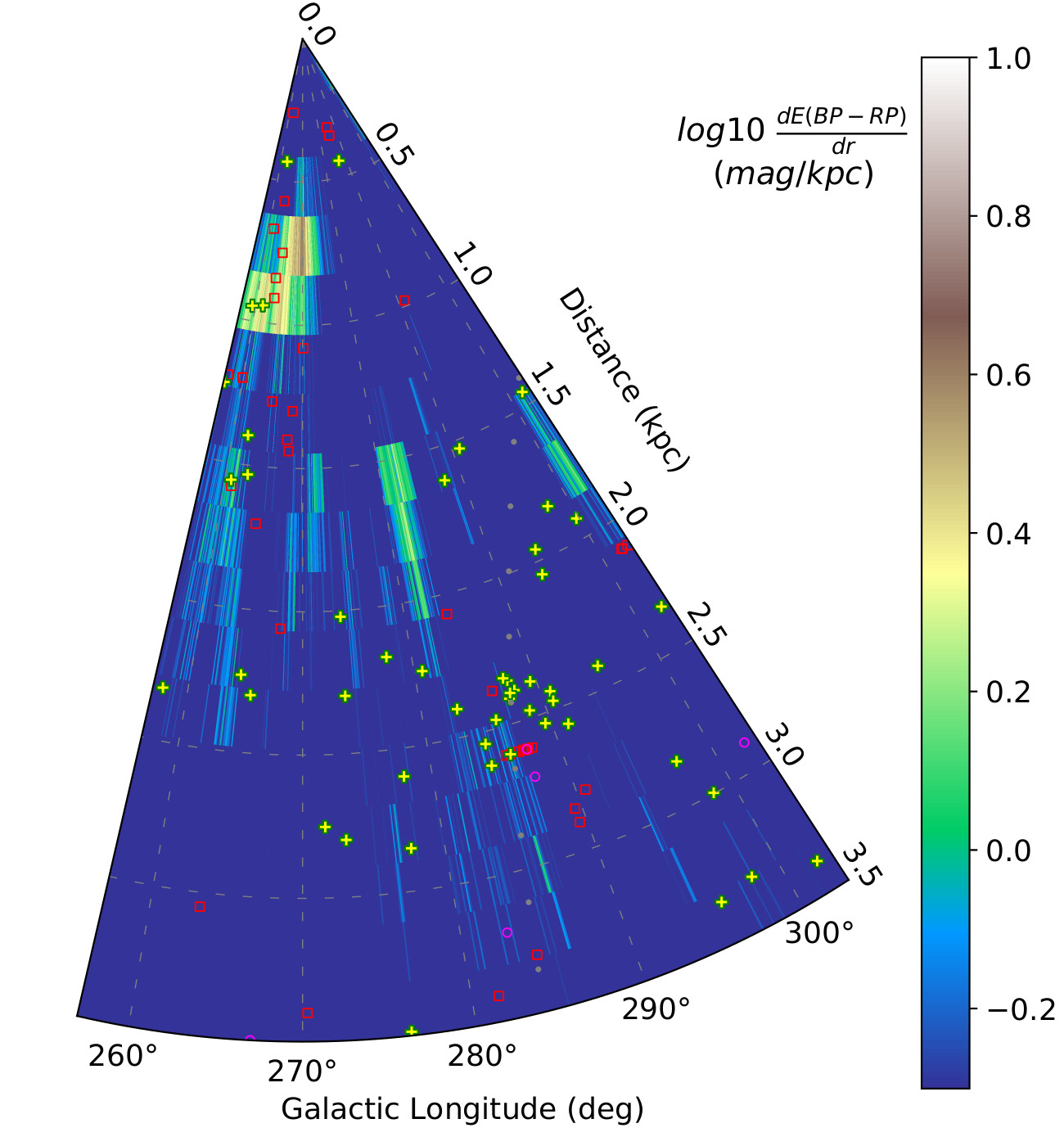}
	\end{subfigure}
	\end{minipage}
	\caption[Face-on view comparison at short distances for the 2MASS multiple-LOS training]{Short distance comparison of the face-on view of the Galactic Place in polar galactic-longitude distance coordinates for the Carina arm region using various maps. All the maps are limited to $d < 3.5$ kpc. The symbols are the same as in figure ~\ref{single_los_2mass_polar_plan}. {\it Top-left:} our 2MASS multiple-LOS training result for $|b|< 1$ deg. {\it Top-Right:} M20 prediction for $|b|< 1$ deg. {\it Bottom-left:} L19 prediction for $|z| < 35$ pc. {\it Bottom-right:} C19 prediction for $|b|< 1$ deg.}
	\label{cornu_marshall_lallement_close_maps_comparison}
\end{figure*}

We compare in Figure~\ref{cornu_marshall_lallement_close_maps_comparison} our short-distance prediction with M20, and also with \citet{Lallement_2019} (hereafter L19) and \citet{Chen_2019} (hereafter C19) since the distance range is now comparable. As before, all the maps are made from the accessible raw data cubes and therefore compared using the same color-scale and intervals. The L19 map does not use galactic coordinate but a cartesian frame. Our approach for this comparison was to average the Galactic Place in a constant $\pm 35$ pc slice around the Galactic Place, roughly corresponding to the height of our map at 2kpc. We note that both the L19 and C19 maps use Gaia DR2 parallaxes through a cross-match in their process of recovering the extinction distance, which might explain the global morphology match between the two. We also did not degrade the resolution of any map to correspond to our resolution since the very high resolution of the L19 provides interesting substructures that are worth discussing. From this comparison we see that at distance $d > 2.5$ kpc there is almost no extinction left in both L19 and C19, while our result and M20 predict a significant extinction passed this distance, with a similar morphology. The high-extinction structure at $l\simeq 300$ deg, $d=3.5$ kpc is the problematic one discussed in Section~\ref{sec:2MASS_main_result} and in the previous paragraph. \\

\newpage
To compare the sub structures in the close range comparison (Fig.~\ref{cornu_marshall_lallement_close_maps_comparison}), we will refer to the sub structures observed in the L19 map. We call A the structure at close $d = 0.8$ kpc and that is observed in the range $257 < l < 280$ deg, then the structures that are all between $1.5 < d < 2.0$ kpc are called B, C, D, E and correspond to the longitude 260, 272, 283, 302 respectively. The last, more blurry, structure around $l=280$ deg at a higher distance $2.5 < d < 3$ kpc is called F. We observe that the L19 map D structure has a counterpart in M20 and in our map as well. In the M20 map, this structure is blurred by the large uncertainties in distance and it cannot be disentangled from the arm tangent structure. Our method seems to dissociate this D structure from a more distant one that is compatible with the F structure from the L19 map and that is visible in the C19 maps as well. Regarding the region A, B, and C the morphology of the M20 map in the corresponding region seems compatible with that of L19 and C19, with a significant detection of the gap at $\simeq 265$ deg but with a higher distance prediction for the B and C structures that are behind A. In our result it seems that our CNN has rather smoothed and packed all these structures together, with a continuous structure between $d=0.8$ and $d=2.5$ kpc. A gap may also be present at $\simeq 265$ deg, although it is less clear than in the three other maps. We not that the structure E from L19 and that is also predicted by C19 does not have a counterpart in the M20 map and in our prediction. However, our prediction and the M20 map both reconstruct a closer structure at $l=303$ deg around $d=1$ kpc, the more distant one at the same longitude being the probable artifact we described before.\\

\vspace{-0.2cm}
From all these comparisons our method seems to be at least as efficient as the M20 approach that uses the same data, with the advantage of predicting less noisy maps with more compact structures and more resolved high distance structures. Our map still contains uncertainty on the distance estimate that can spread over several distance bins, but the finger of god effect is greatly reduced. Our CNN might still lack distance resolution at closer distance to match the overall morphology of the L19, which might be improved by adding Gaia data in our approach (Section~\ref{gaia_2mass_ext_section}).

\clearpage
\subsection{Addition of a second color-magnitude diagram}

\begin{figure*}[!t]
\vspace{-0.3cm}
	\centering
	\begin{subfigure}[!t]{0.47\textwidth}
	\includegraphics[width=1.0\hsize]{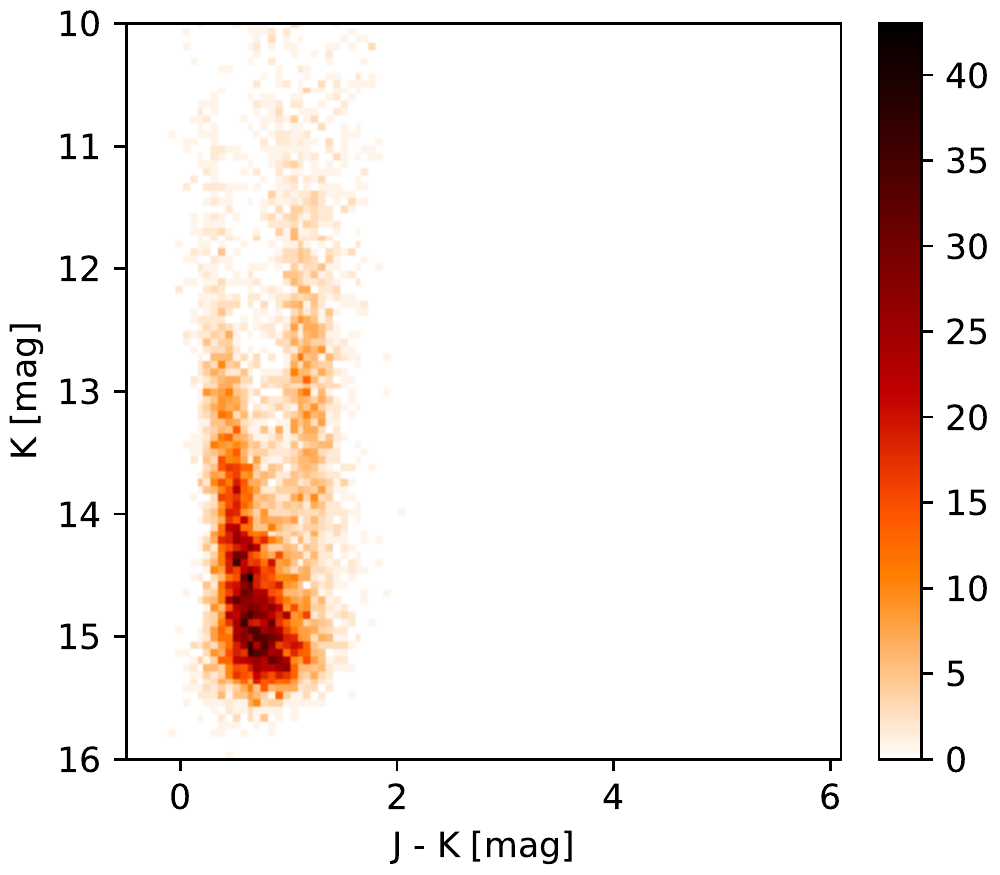}
	\end{subfigure}
	\hspace{0.5cm}
	\begin{subfigure}[!t]{0.47\textwidth}
	\includegraphics[width=1.0\hsize]{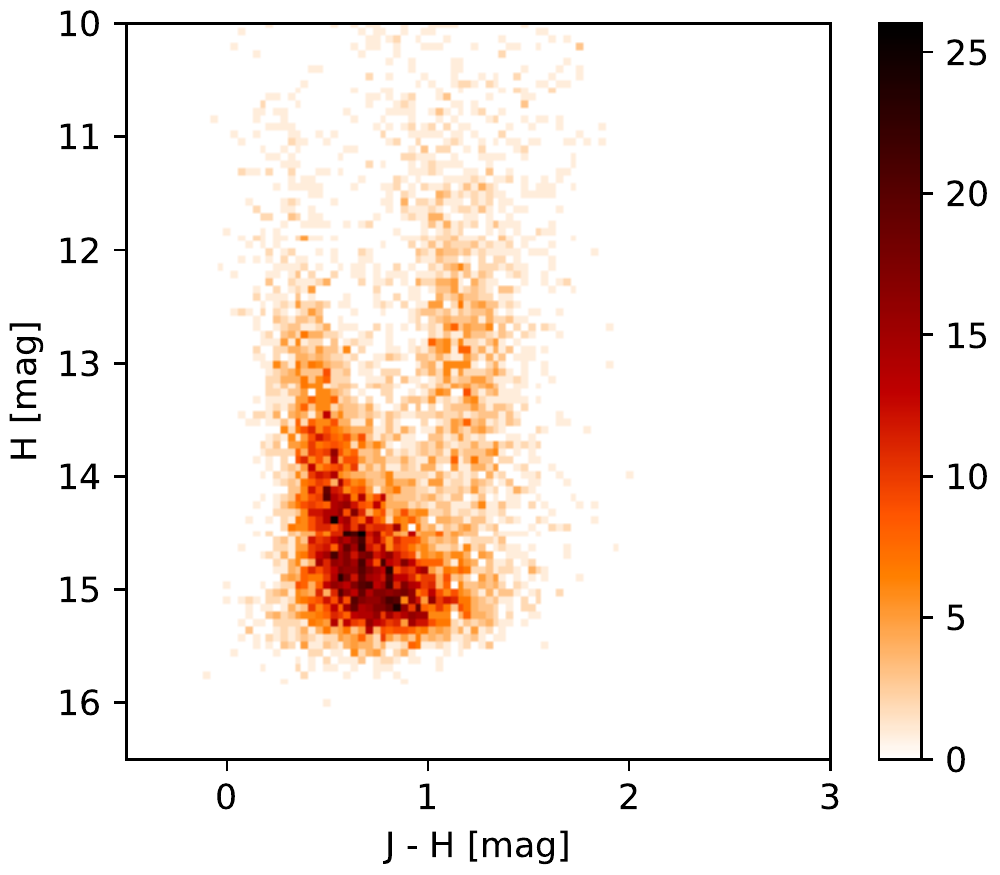}
	\end{subfigure}
	\caption[Dual 2MASS CDM illustration]{Illustration of the two different 2MASS CMDs [J-K]-[K] and [J-H]-[H] as observed by 2MASS for $l = 280$ deg, $b = 0$ deg with a cone query radius of $0.25\, \mathrm{deg}$.}
	\label{second_2mass_cmd comparison}
\end{figure*}

\begin{figure}[!t]
	\hspace{-0.9cm}
	\begin{minipage}{1.10\textwidth}
	\centering
	\includegraphics[width=1.0\hsize]{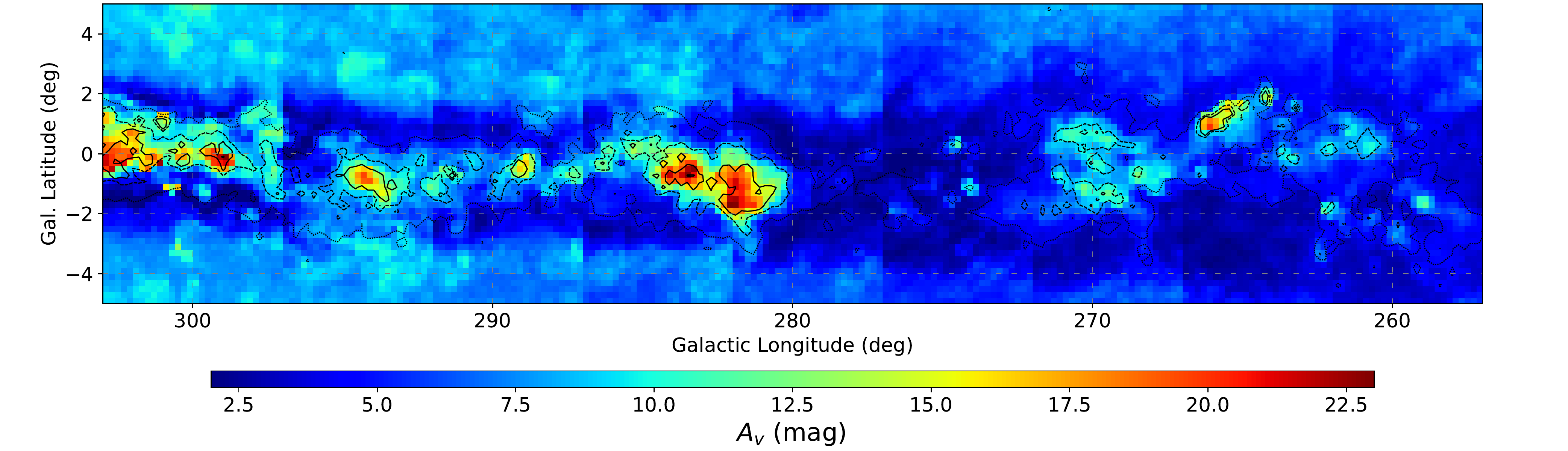}
	\end{minipage}
	\caption[Plane-of-the-sky view of the 2MASS dual-CMD multiple-LOS training]{Integrated extinction of the 2MASS dual-CMD multiple-LOS training prediction in a plane of the sky view using galactic coordinates. Contours are from Planck $\tau_{353}$.}
	\label{multi_los_dual_2MASS_int_map}
	\vspace{-0.2cm}
\end{figure}

Based on our multiple-LOS training, we demonstrated that it is possible to efficiently add a new input depth-channel without significantly increasing the network computational time and that it manages to infer information correlation between the two channels. While this approach was used to allow the network to distinguish different line of sigh references, it can be used to add more information in the form of additional images. Thus, we added a second 2MASS CMD representing [J-H]-[H] with the same $64 \times 64$ resolution. The J-H colors are lesser than the J-K colors, because of the lesser difference in wavelengths, so we reduced the color range of this diagram to better resolve the induced star shift. Figure~\ref{second_2mass_cmd comparison} shows a comparison of the two observed diagrams. For this training we directly used the multiple reference LOS approach conserving the same 9 reference LOS for training. Therefore, our input for each example is now a set of 4 CMD, two observed ones and two bare references, consequently increasing the dataset size. We conserved $f_{\rm naked} = 0.1$ and $Z_{\rm lim} = 50$ but the definition of the latter change slightly. It is still used to assess a maximum distance after which the target profile is set to zero (Sect.~\ref{zlim_subsection}), but this time it does so only after both CMDs have reached this limit individually. Each CMD input depth channel is normalized individually by looking for the highest pixel value in the all dataset corresponding to a given depth. The training time required for this is similar to our main 2MASS result, although the convergence requires more epochs overall.\\

\begin{figure*}[!t]
\hspace{-0.8cm}
	\begin{minipage}{1.10\textwidth}
	\centering
	\begin{subfigure}[!t]{0.48\textwidth}
	\includegraphics[width=1.0\hsize]{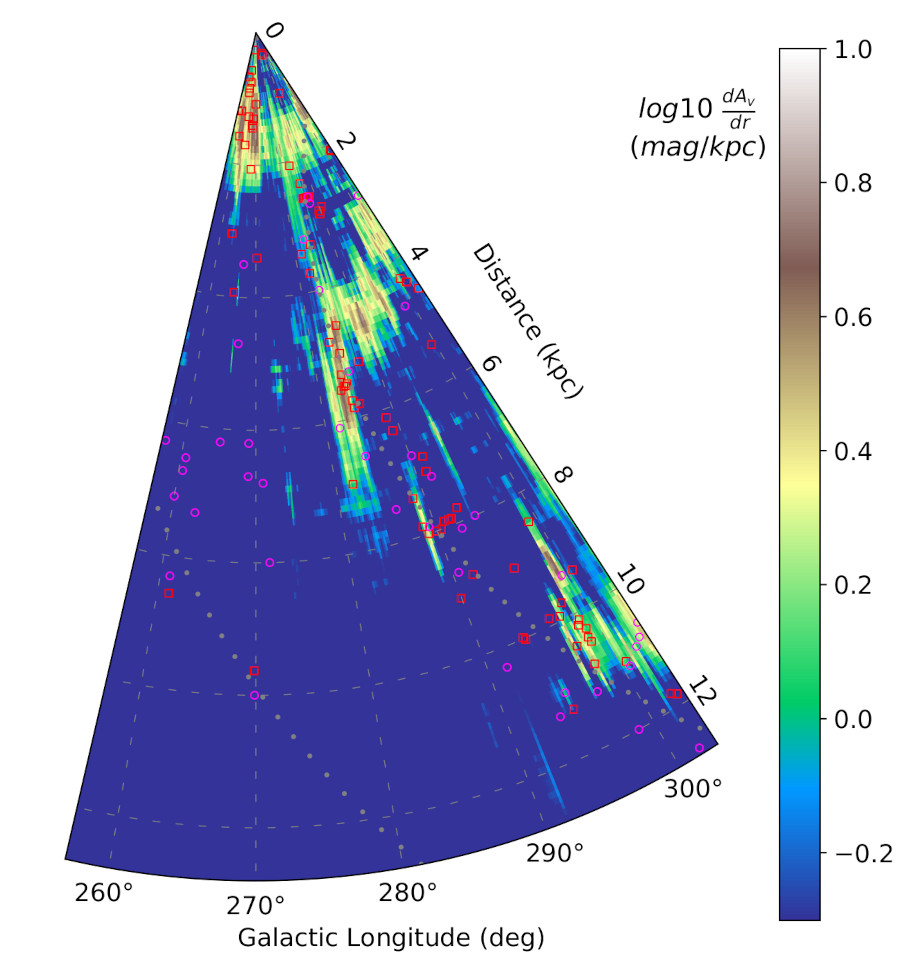}
	\end{subfigure}
	\hspace{0.5cm}
	\begin{subfigure}[!t]{0.48\textwidth}
	\includegraphics[width=1.0\hsize]{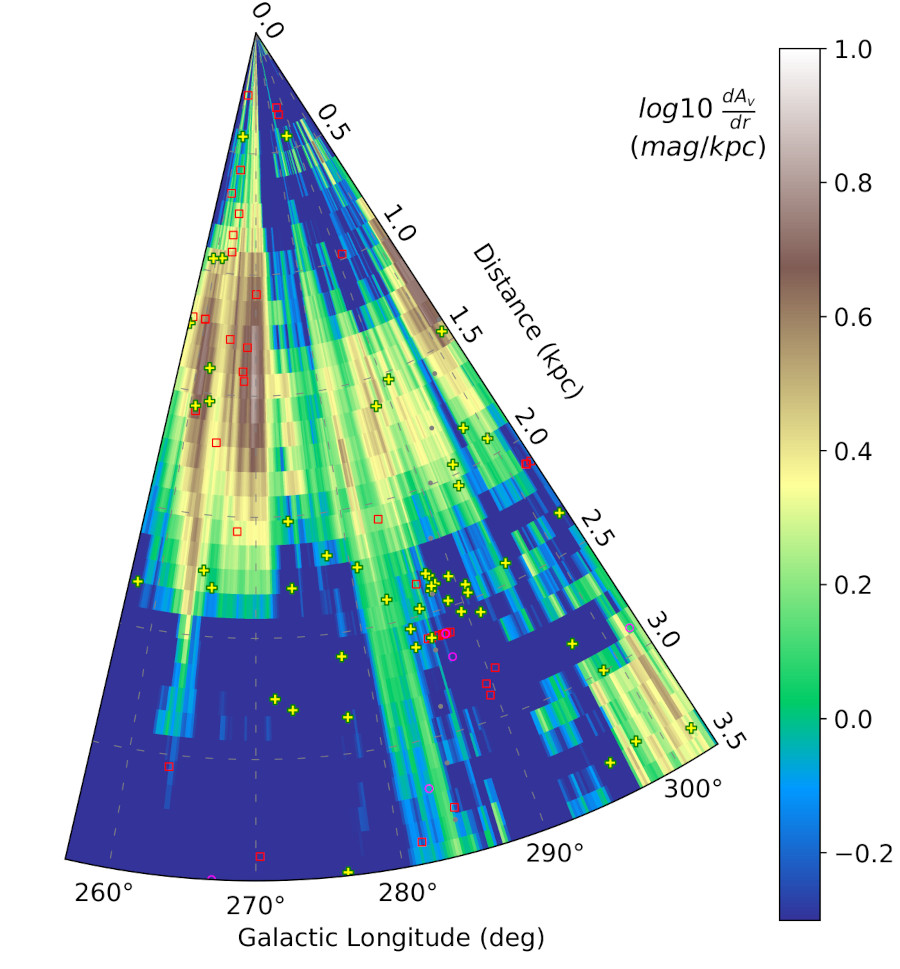}
	\end{subfigure}
	\end{minipage}
	\caption[Face-on view for the dual-CMD 2MASS multiple-LOS training]{Face-on view of the Galactic Place $|b| < 1$ deg in polar galactic-longitude distance coordinates for the predicted Carina arm region using our dual-CMD 2MASS multiple-LOS training. The symbols are the same as in figure ~\ref{single_los_2mass_polar_plan}. {\it Left:} Full distance prediction. {\it Right:} Zoom on the $d < 3.5$ kpc prediction.}
	\label{multi_los_dual_2MASS_polar_map}
	\vspace{3cm}
\end{figure*}

\begin{figure}[!t]
	\hspace{-0.5cm}
	\begin{minipage}{1.05\textwidth}
	\centering
	\includegraphics[width=1.0\hsize]{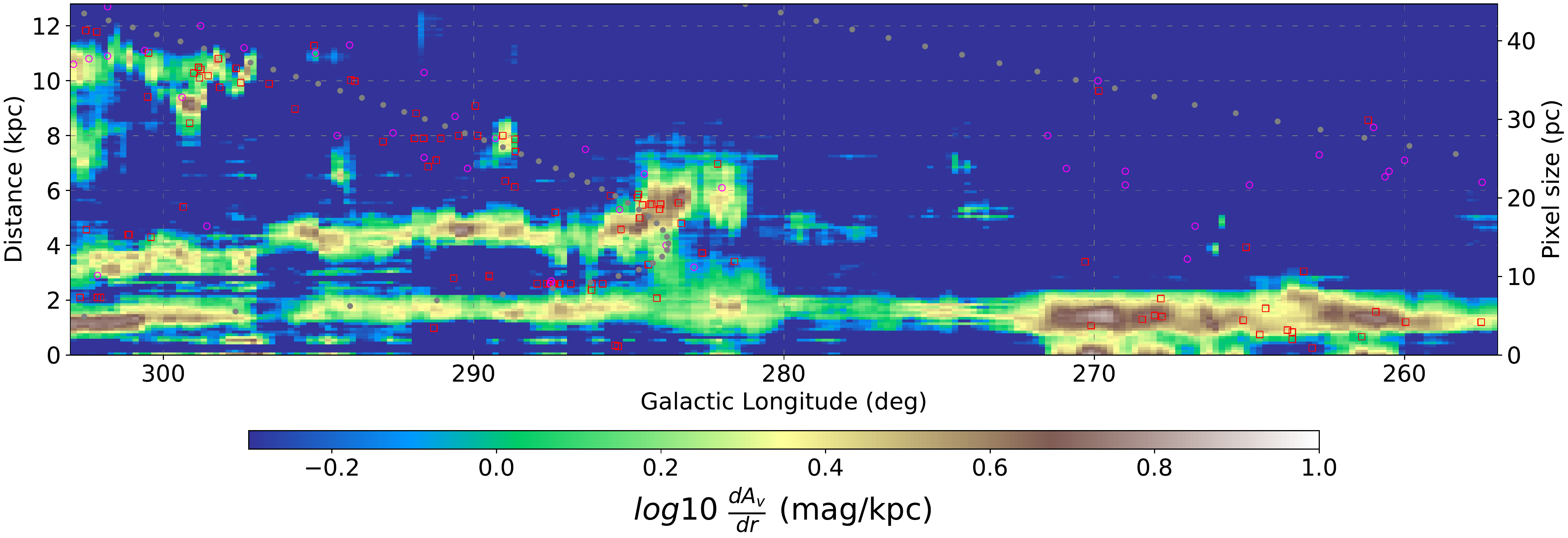}
	\end{minipage}
	\caption[Cartesian face-on view for the dual-CMD 2MASS multiple-LOS training]{Face on view of the Galactic Place $|b| < 1$ deg in cartesian galactic-longitude distance coordinates for the predicted Carina arm region using our dual-CMD 2MASS multiple-LOS training. The axis on the right border corresponds to the pixel height as a function of the distance induced by the conic shape of our LOS. The symbols are the same as in figure ~\ref{single_los_2mass_polar_plan}.}
	\label{multi_los_dual_2MASS_cart_map}
\end{figure}

\newpage
We illustrate our prediction in the plane of the sky in Figure~\ref{multi_los_dual_2MASS_int_map}. It mainly highlights that there are more latitude artifacts and a stronger tiling effect than in our main 2MASS result. Still, the Galactic Place in the $|b| < 2$ deg range is mainly free of this artifacts and accurately follows the Planck morphology contours. The most interesting difference with our main results is that the extinction quantity seems better constrained overall. We conserved mostly the same integrate extinction dynamic range in the map but there are much less saturated pixels and the transition between the high and low extinction regimes is smoother. This could be explained by the fact that, even if the second CMD did not improve the distance estimate (that is mostly induced by the amount of pixel shift), it has certainly improved the resolution of extinction value (that depends of the star ratio between pixels, Sect.~\ref{input_output_cnn_dim}). Indeed, two CMDs improve the global statistic but there are also more stars per pixel overall in the [J-H]-[H] CMD, definitely improving the extinction quantity resolution. However, the fact that this result presents more latitude artifacts might indicate that the $Z_{\rm lim} = 50$ value is not adapted to this case. The [J-H]-[H] CMD generally being more populated than the [J-K]-[K] one, it usually reaches the limit later, and it can still be used to infer the extinction profile up to greater distances. However, the color leverage being lesser in this diagram it tasks the network to predict more extinction with less information. A suitable solution to counter this effect would be to have a different $Z_{\rm lim}$ value for each CMD, but we did not had time for this test for now.\\

The face-on view for this training is presented in Figure~\ref{multi_los_dual_2MASS_polar_map} and the cartesian view of the same prediction is in Figure~\ref{multi_los_dual_2MASS_cart_map}. The lower longitude part $l < 280$ deg is roughly identical to our main 2MASS result with a bit more of very close $d < 0.4$ kpc foreground. In contrast, the high longitude part $l > 280$ deg presents a few differences that are more evident in Figure~\ref{multi_los_dual_2MASS_cart_map}. First there is much more continuity along the structure at $1 < d < 2$ kpc, and along that at $3 < d < 5$ kpc. Even if this view strongly stretches the short distances, it is visible that the extinction is more evenly distributed between these two structures than in our main result where the structure at $3 < d < 5$ kpc was much denser. This latter structure now has a much lower extinction value and looks much more alike the M20 prediction in Figures~\ref{cornu_marshall_maps_comparison} and~\ref{cornu_marshall_lallement_close_maps_comparison}. These elements support the idea that it is not an artifact, but a genuine interstellar structure, although it remains more extended in longitude in the present prediction than in M20. Interestingly, the network tends to predict a connection between the tangent and this secondary structure more than with the closer one that would better correspond to our arm model. The more distant group of structures around $d\simeq 10$ kpc is roughly identical with an order of variation similar to the one we would obtain from repeated training over the same dataset.\\

While these results are sufficiently improved to justify the addition of the second 2MASS CMD, we were not able to always include it in all our tests since it would lead to very large datasets that are hard to work with using our main hardware infrastructure. Therefore, in the following section this diagram is not used.\\

\clearpage

\section{Combined Gaia-2MASS extinction maps}
\label{gaia_2mass_ext_section}

In this section we generalize our approach of multiple CMDs as input by adding a Gaia data diagram. We present some specificity related to the Gaia dataset like the band and parallax error fittings. We also present the results from a CNN training on a single reference LOS case and from a multiple reference LOS one. We compare them to our main 2MASS result and to other extinctions maps. We finally discuss the current limitations of our approach with Gaia along with some possible adjustment that we considered.

\etocsettocstyle{\subsubsection*{\vspace{-1.2cm}}}{}
\localtableofcontents

\vspace{-0.3cm}

\subsection{Realistic Gaia diagram construction from the BGM}
\label{gaia_diag_constuction}

In the previous Section~\ref{2mass_maps_section}, we illustrated that the network architecture and training dataset construction using the BGM and GRF generated profiles (Sect.~\ref{galmap_problem_description}) was suitable to reconstruct large 3D extinction maps. We exposed that our CNN architecture allows to efficiently combine multiple diagrams as input depth channels allowing the network to generalize complex problem representations. One of the main advantages of this construction is that each diagram can theoretically be completely independent and that the network will automatically extract the relevant information contained in each of them regarding the task to perform. From these observations and theoretical elements it should be possible to add an independent diagram from Gaia without the necessity of cross matching the stars with 2MASS. For this to be possible we still had to follow the rules that allowed us to construct a suitable 2MASS CMD from the BGM, meaning that we had to select a statistical representation that accurately reproduced an observed quantity based on cuts and uncertainties (Sect.~\ref{ext_profile_and_cmd_realism}).\\

\begin{figure*}[!t]
\hspace{-0.6cm}
	\begin{minipage}{1.05\textwidth}
	\centering
	\begin{subfigure}[!t]{0.49\textwidth}
	\caption*{\bf No extinction}
	\includegraphics[width=1.0\hsize]{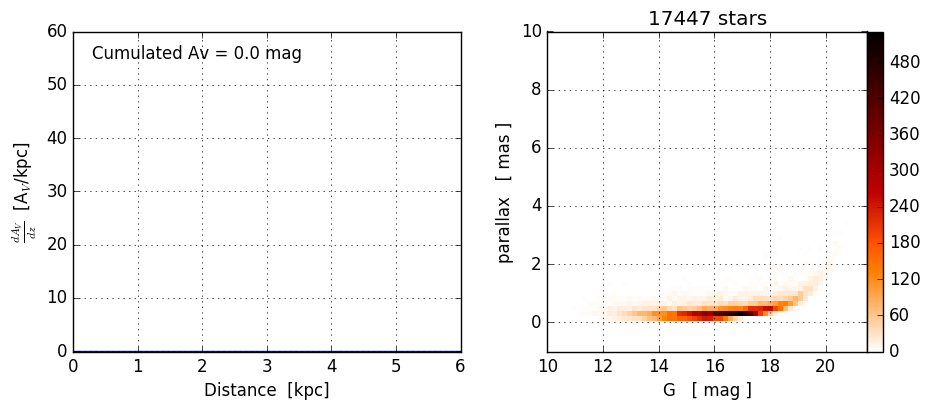}
	\end{subfigure}
	\vspace{0.6cm}
	\begin{subfigure}[!t]{0.49\textwidth}
	\caption*{\bf 1 Cloud, $\bm{A_v = 3}$ mag, $\bm{d = 0.5}$ kpc}
	\includegraphics[width=1.0\hsize]{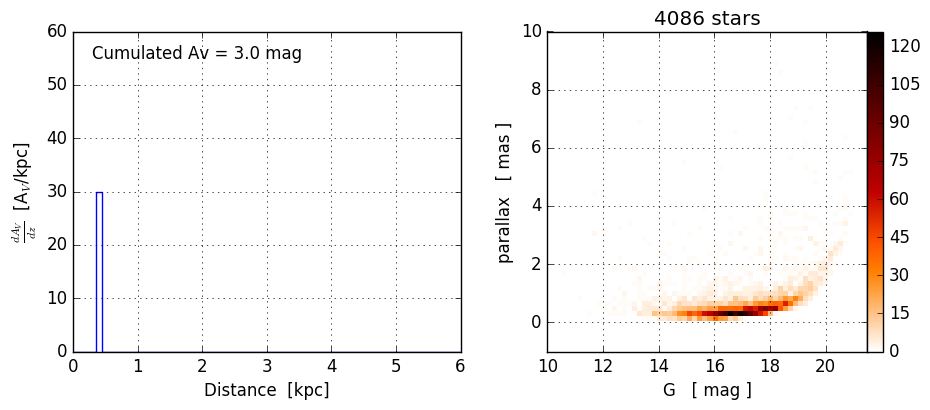}
	\end{subfigure}\\
	\begin{subfigure}[!t]{0.49\textwidth}
	\caption*{\bf 1 Cloud, $\bm{A_v = 5}$ mag, $\bm{d = 0.5}$ kpc}
	\includegraphics[width=1.0\hsize]{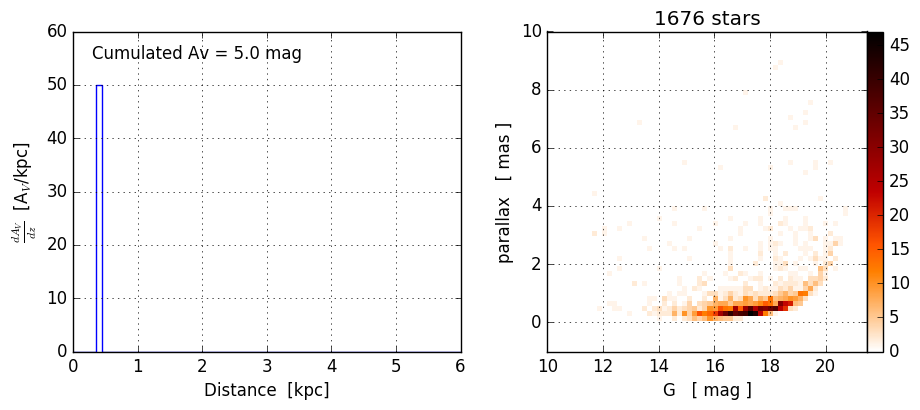}
	\end{subfigure}
	\begin{subfigure}[!t]{0.49\textwidth}
	\caption*{\bf 2 Clouds, $\bm{A_v = 5,3}$ mag, $\bm{d = 0.5,2}$ kpc}
	\includegraphics[width=1.0\hsize]{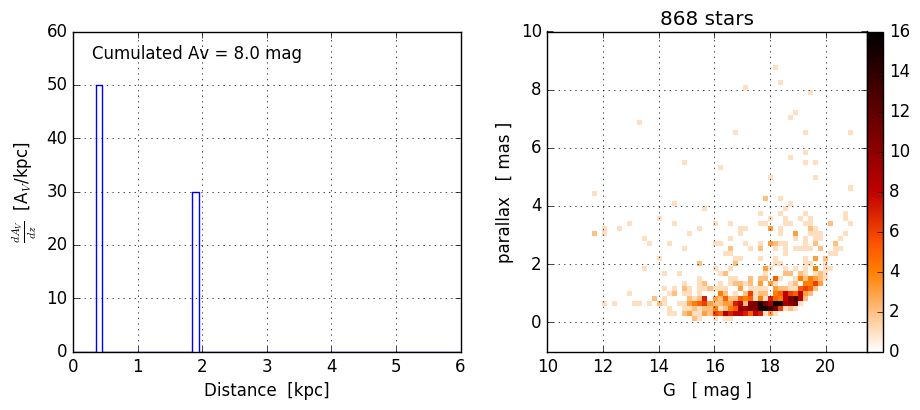}
	\end{subfigure}\\
	\begin{subfigure}[!t]{0.49\textwidth}
	\caption*{\bf 1 Cloud, $\bm{A_v = 3}$ mag, $\bm{d = 2}$ kpc}
	\includegraphics[width=1.0\hsize]{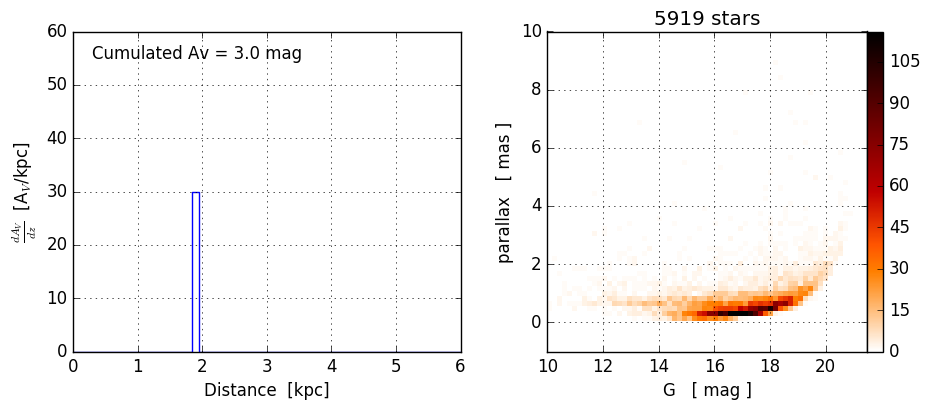}
	\end{subfigure}
	\begin{subfigure}[!t]{0.49\textwidth}
	\caption*{\bf Uniform ext - 1 Cloud, $\bm{A_v = 3}$ mag, $\bm{d = 2}$ kpc\\}
	\includegraphics[width=1.0\hsize]{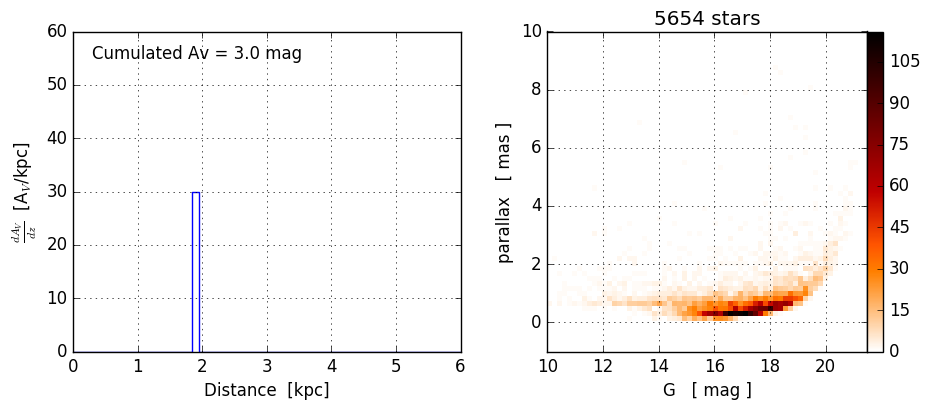}
	\end{subfigure}
	\end{minipage}
	\caption[Effect of individual clouds on the Gaia diagram]{Effect of individual clouds on the Gaia [G]-[$\varpi$] diagram. The extinction is modeled as a log-normal distribution, except in the bottom-right panel where a uniform extinction is used.}
	\label{Gaia_diag_dissection}
\end{figure*}

For this application we chose to use a [G]-[$\varpi$] diagram, where G is the photometric Gaia band at $\lambda_{\rm eff} = 0.623$\,$\mu m$, and $\varpi$ is the parallax measurement. We note that using directly the parallax instead of the distance removes the necessity of a possibly inaccurate or complex distance inversion \citep{bailer-jones_2018}. In this diagram the extinction effect will be to decrease the star luminosity (increasing the G magnitude value) of all stars after the corresponding distance. This should mainly result in platforming the continuous parallax distribution following the G magnitude axis. We illustrate the effect of extinction on this diagram in Figure~\ref{Gaia_diag_dissection}. We note that this effect is very strong due to the relatively short wavelength of the G band, implying greater extinction than with 2MASS, and that therefore the diagram will not provide any additional information for large distance estimates. Still, it should be useful to increase the close distance resolution, especially providing a better first extinction front position. It is also possible that it helps to better constrains the low extinction lines of sight.\\

\begin{figure}[!t]
	\centering
	\includegraphics[width=1.0\textwidth]{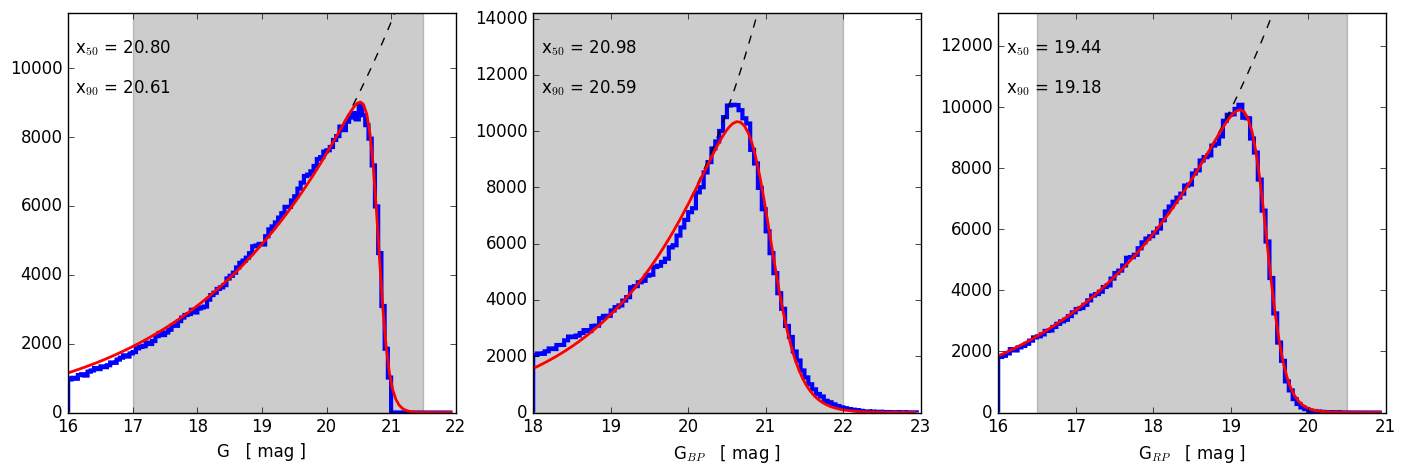}
	\caption[Fitting of the cut in magnitude for the three Gaia bands]{Fitting of the cut in magnitude for the three Gaia bands. The blue histograms show the observed distribution, the fitted models are in red. The gray area shows the range of magnitude values included in the fit.}
	\label{Gaia_cut_fitting}
\end{figure}

Like for 2MASS, we had to characterize the selection cut and the observational uncertainties for the Gaia data. For this we used the same approach than the one described in Section~\ref{ext_profile_and_cmd_realism} on the same $l=280$ deg, $b=0$ deg LOS with a 1 $deg^2$ radius. For the magnitude cut in the G, $\mathrm{G_{BP}}$ and $\mathrm{G_{RP}}$ bands we excluded the stars that lack one or more detection and followed the equation \ref{magnitude_cut_fit} to fit each magnitude histogram. Figure~\ref{Gaia_cut_fitting} shows the resulting fitted cuts in the three Gaia bands. For the parallax, we selected the stars based on the ratio of their parallax over parallax-error with $\varpi/\sigma(\varpi) > 3$, excluding all stars that miss any of these quantities. This last addition reduced the number of stars that is then closer to the order of 2MASS star count.\\

\begin{figure*}[!t]
\hspace{-0.4cm}
	\begin{minipage}{1.05\textwidth}
	\centering
	\begin{subfigure}{0.47\textwidth}
	\includegraphics[width=\textwidth]{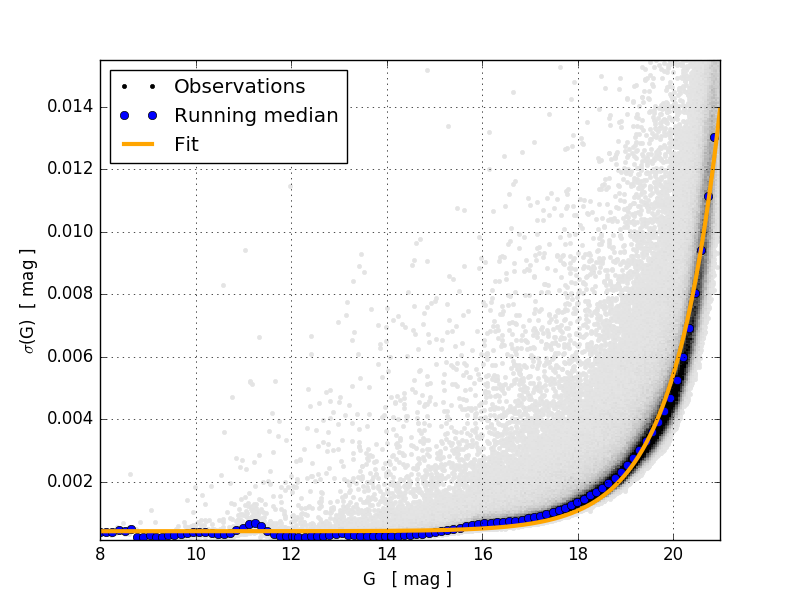}
	\end{subfigure}
	\begin{subfigure}{0.47\textwidth}
	\includegraphics[width=\textwidth]{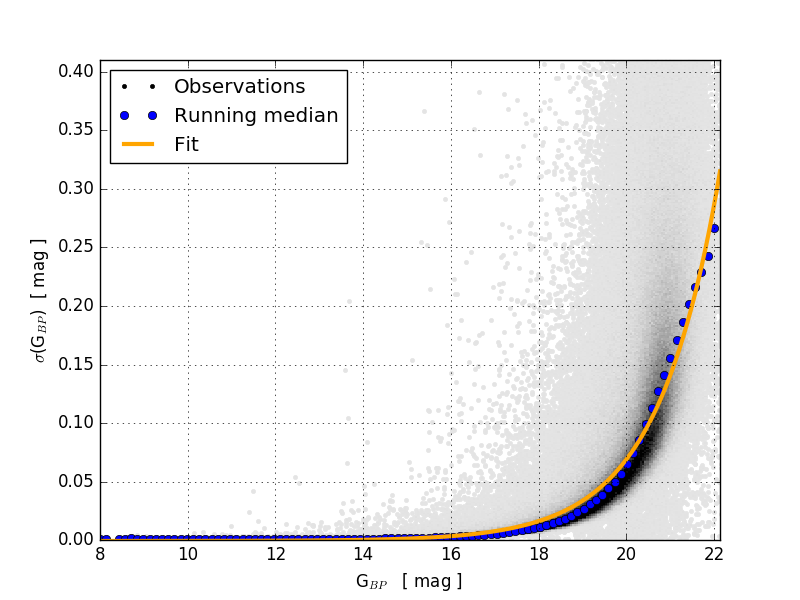}
	\end{subfigure}\\
	\begin{subfigure}{0.47\textwidth}
	\includegraphics[width=\textwidth]{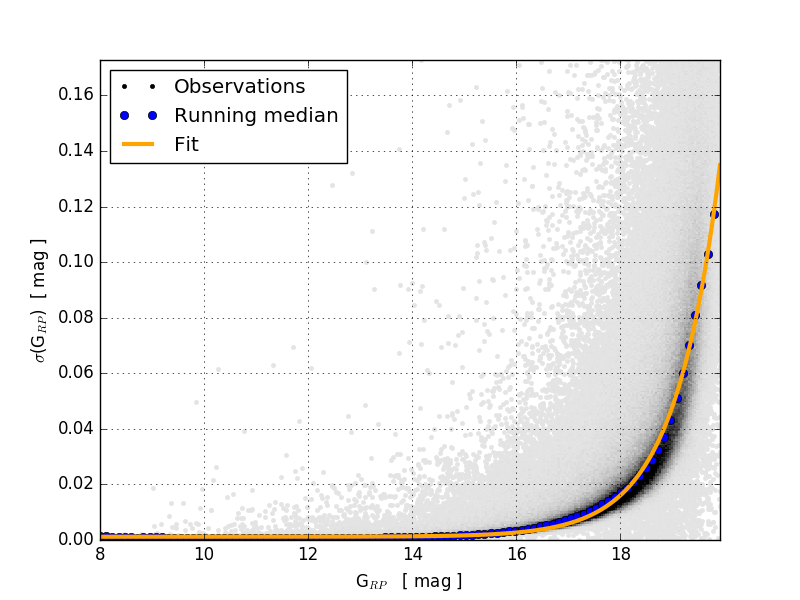}
	\end{subfigure}
	\begin{subfigure}{0.47\textwidth}
	\includegraphics[width=\textwidth]{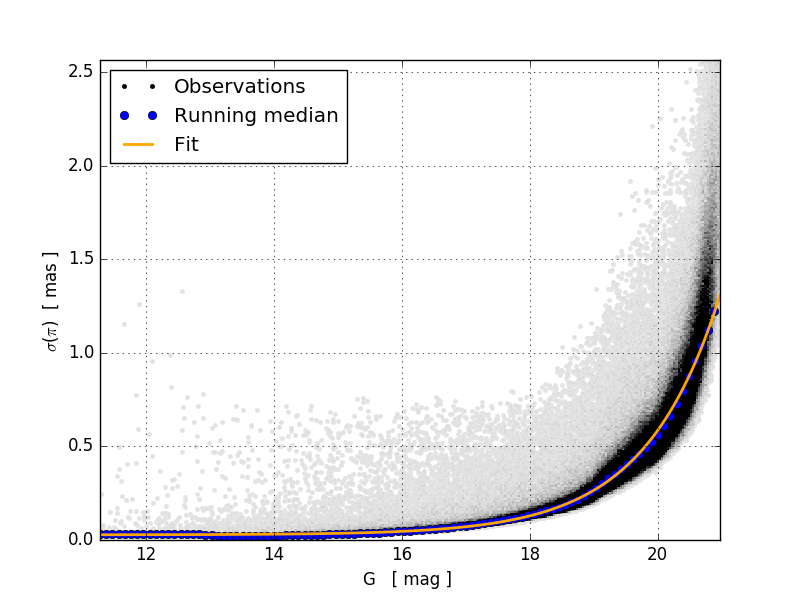}
	\end{subfigure}
	\end{minipage}
	\caption[Fit of Gaia uncertainties]{Fit of Gaia uncertainties. The gray dots are Gaia stars, the gray scale representing the star density in the diagram. The running median (blue dots) is fitted by an exponential model (orange line).}
	\label{fig_Gaia_uncertainty_fitting}
\end{figure*}

Regarding the uncertainties we kept only stars that present all the bands, all the uncertainties and both the parallax value and its uncertainty, which conserved around 80\% of the stars on the present LOS. For the 3 magnitude band uncertainties we followed the same procedure described in Section~\ref{ext_profile_and_cmd_realism} following Equation \ref{uncertainty_fit}. Regarding the parallax uncertainty, we observed that it has a higher correlation with the G band than with the parallax itself. So we decided to fit this uncertainty using the [G]-[$\sigma(\varpi)$] diagram. All the fits are illustrated in Figure~\ref{fig_Gaia_uncertainty_fitting} and the Table~\ref{table_Gaia_uncertainty_fitting} contains the corresponding best fit parameters for the Gaia magnitudes and the parallaxes. There is an important observed difference between the diagram with and without the added uncertainty which plays a major role to make the diagram realistic.\\

\begin{table}
	\centering
	\caption{Uncertainty best fit parameters for all Gaia bands and parallax}
	\vspace{-0.1cm}
	\def\arraystretch{1.1}
	\begin{tabularx}{0.80\hsize}{l @{\hskip 0.05\hsize} @{\hskip 0.05\hsize}*{3}{Y}}
	\toprule
	& a & b & c\\
	\toprule
	G				  & $7.950 \times 10^{-12}$ & $1.013 \times 10^{0}$  & $4.069 \times 10^{-4}$\\
	$\mathrm{G_{BP}}$ & $5.003 \times 10^{-8}$  & $7.074 \times 10^{-1}$ & $-8.295 \times 10^{-4}$\\
	$\mathrm{G_{RP}}$ & $1.621 \times 10^{-11}$ & $1.147 \times 10^{0}$  & $1.187 \times 10^{-3}$\\
	$\mathrm{\varpi}$ & $2.138 \times 10^{-8}$  & $8.539 \times 10^{-1}$  & $2.767 \times 10^{-2}$\\
	\bottomrule
	\end{tabularx}
	\label{table_Gaia_uncertainty_fitting}
\end{table}

\newpage
\vspace{-0.3cm}
\subsection{Training with one line of sight}
\label{Gaia-2MASS_single_los_training_and_test_set_prediction}

For comparison with the single LOS training using solely 2MASS from Section~\ref{2mass_single_los}, we present here a Gaia-2MASS training on a single LOS. For this we defined our input as a dual depth channel CMD, the first one being the same [J-K]-[K] CMD and the new [G]-[$\varpi$] diagram, both using the same $64 \times 64$ resolution. As for 2MASS only, the order of the operations is as follow: we started with raw BGM realizations, we generated composite GRF extinction profiles (Sect.~\ref{GRF_profiles_section}) that are used to extinct all stars from each list, then the fitted cuts are applied and the noise is added. From these extincted star lists we determined the limiting distances using $Z_{\rm lim} = 100$ and forced all profiles to zero after this point. Finally we cut every extinction peak above five in the targets (Sect.~\ref{zlim_subsection}). The created input-target pairs are then used to train the network accounting for a $f_{\rm naked} = 0.1$ value in order to regularly present the bare diagrams as inputs. Similarly to the single LOS training using 2MASS, we generated $5\times 10^5$ of these examples and kept the $0.94$, $0.05$, $0.01$ proportions for the training, valid, and test datasets. Each input depth channel is normalized into the 0 to 1 range according to the maximum pixel value in the full dataset. This step is even more important in this case because there are significantly more stars in the [G]-[$\varpi$] diagrams than in the [J-K]-[K] CMDs, therefore the normalization evens out the initial respective influence of the two diagram types. Training the network on this dataset is quick since it only contains two input depth channels and $5\times 10^5$ examples per epoch. The convergence is reached at a very similar epoch as for the 2MASS single-LOS training.\\

\begin{figure*}
\hspace{-0.6cm}
	\begin{minipage}{1.05\textwidth}
	\centering
	\begin{subfigure}[!t]{1.0\textwidth}
	\includegraphics[width=1.0\hsize]{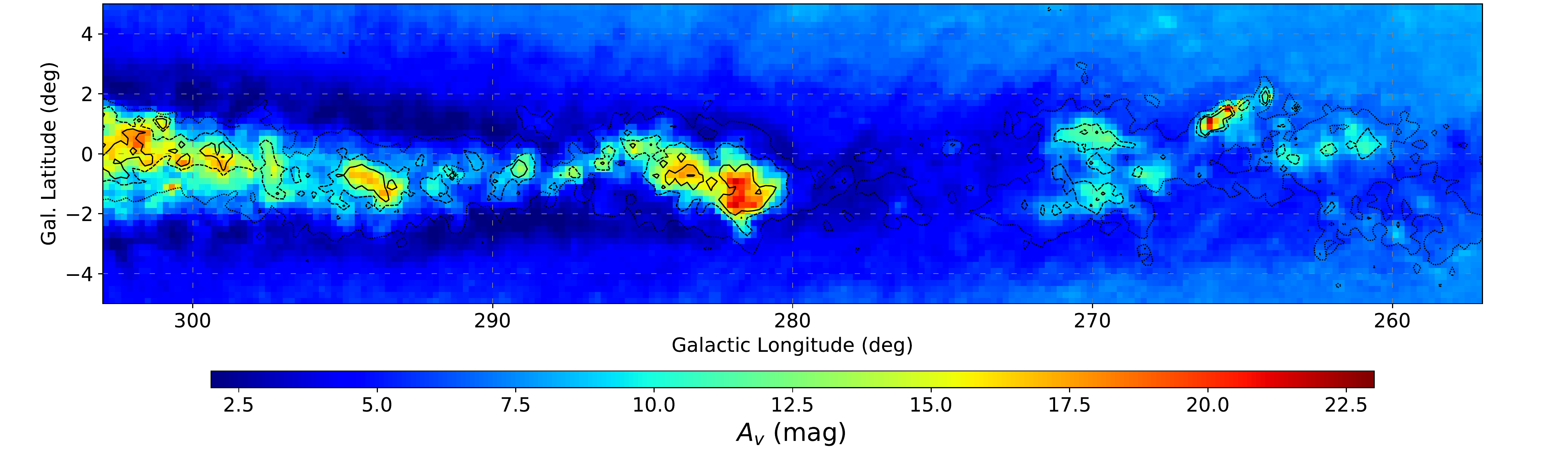}
	\end{subfigure}
	\end{minipage}
	\caption[Plane of the sky view for the Gaia-2MASS single-LOS training]{Integrated extinction for each pixel of the Gaia-2MASS single-LOS training prediction in a plane of the sky view using galactic coordinates. Contours are from Planck $\tau_{353}$.}
	\label{single_gaia_2mass_int_map}
\end{figure*}

\begin{figure*}[!t]
\hspace{-0.9cm}
	\begin{minipage}{1.1\textwidth}
	\centering
	\centering
	\begin{subfigure}[!t]{0.47\textwidth}
	\includegraphics[width=1.0\hsize]{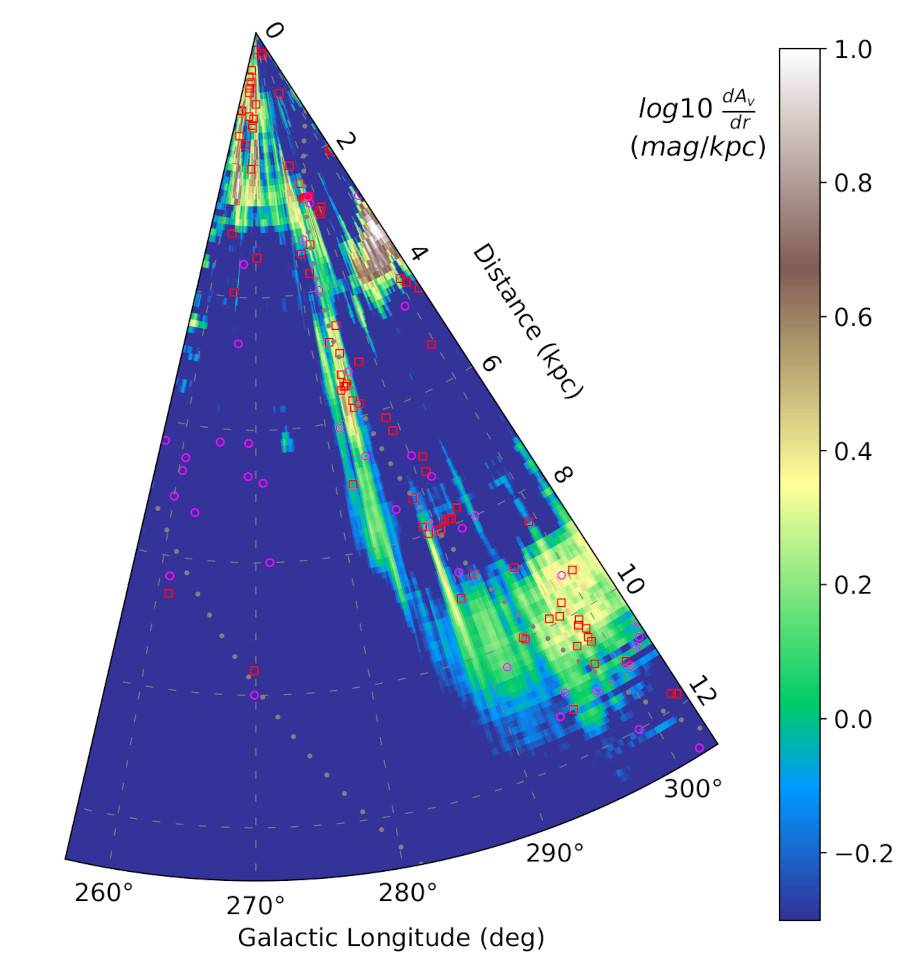}
	\end{subfigure}
	\hspace{0.3cm}
	\begin{subfigure}[!t]{0.47\textwidth}
	\includegraphics[width=1.0\hsize]{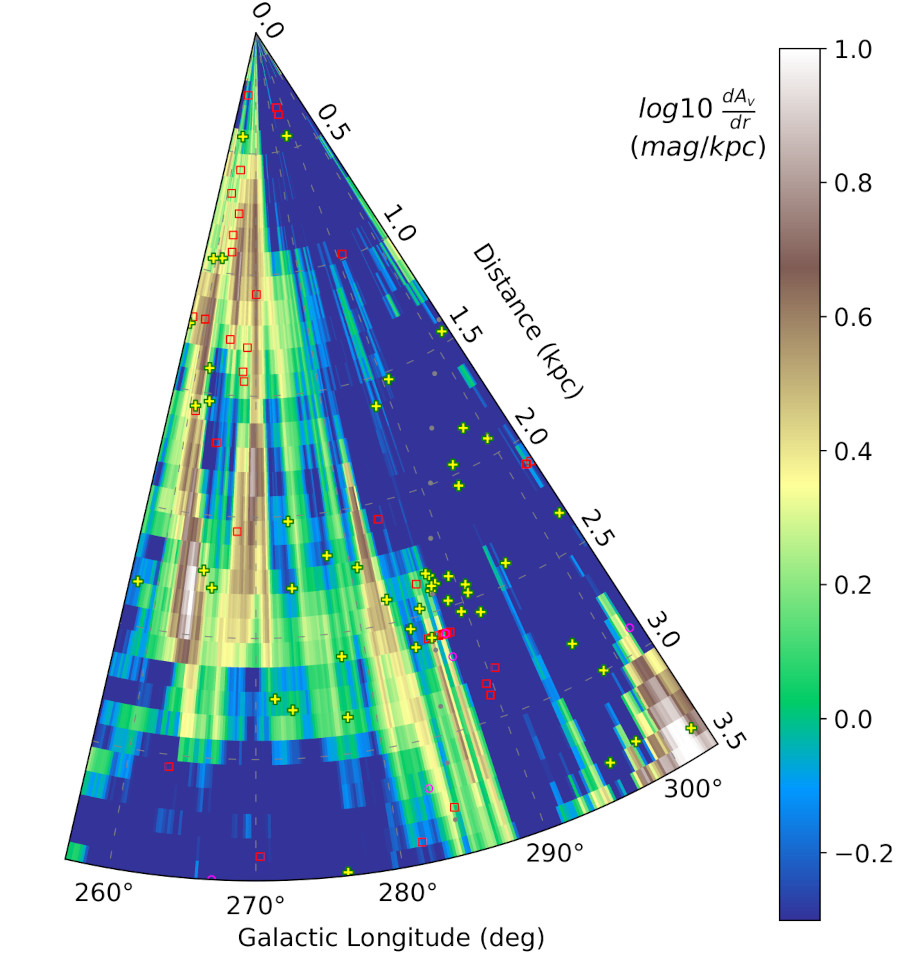}
	\end{subfigure}
	\end{minipage}
	\caption[Face-on view for the Gaia-2MASS single-LOS training]{Face-on view of the Galactic Place $|b| < 1$ deg in polar galactic-longitude distance coordinates for the predicted Carina arm region using our Gaia-2MASS single LOS training. The symbols are the same as in figure ~\ref{single_los_2mass_polar_plan}. {\it Left:} full distance prediction. {\it Right:} zoom in on the d < 3.5 kpc prediction region.}
	\label{single_gaia_2mass_polar_plan}
\end{figure*}

\begin{figure*}[!t]
	\centering
	\begin{subfigure}[!t]{0.98\textwidth}
	\includegraphics[width=1.0\hsize]{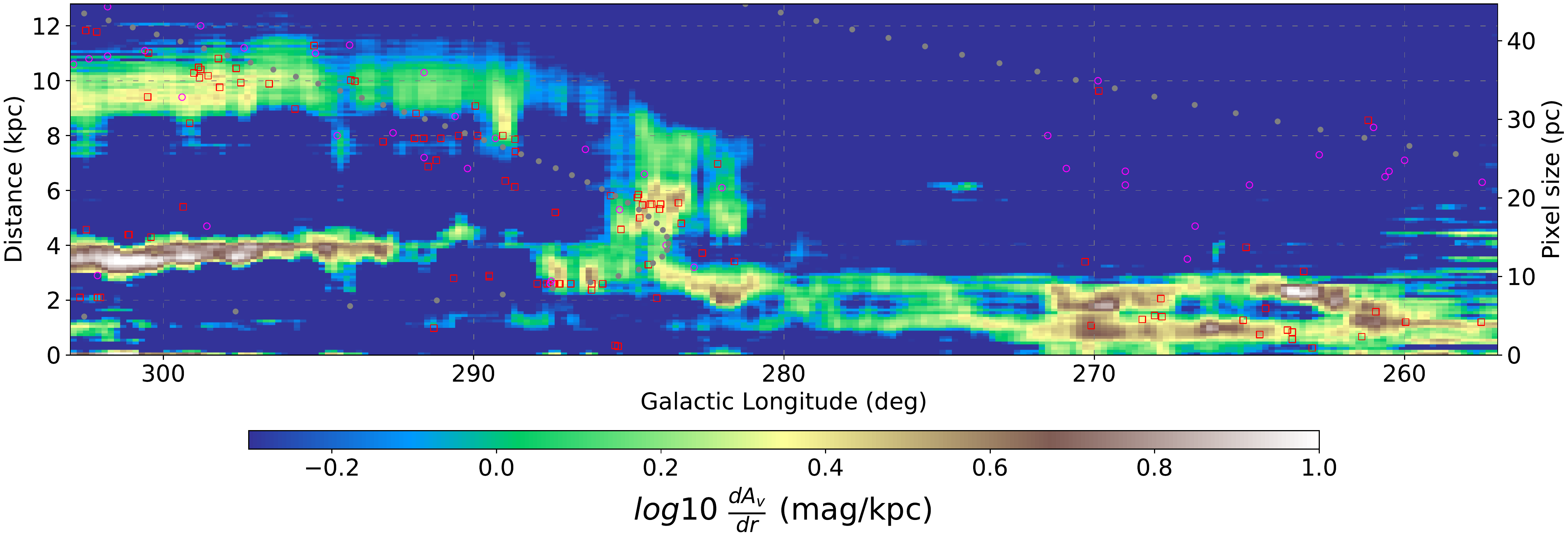}
	\end{subfigure}

	\caption[Cartesian face-on view of the Gaia-2MASS single-LOS training]{Face-on view of the Galactic Place $|b| < 1$ deg in cartesian galactic-longitude distance coordinates for the predicted Carina arm region using our gaia-2MASS single-LOS training. The axis on the right border corresponds to the pixel height as a function of the distance induced by the conic shape of our LOS. The symbols are the same as in figure ~\ref{single_los_2mass_polar_plan}.}
	\label{single_gaia_2mass_cart_plan}
\end{figure*}

The predictions from the previously described network training are presented in the following figures: the integrated extinction plane of the sky view is presented in Figure~\ref{single_gaia_2mass_int_map}, the face-on view of the Galactic Place corresponds to Figure~\ref{single_gaia_2mass_polar_plan} and the cartesian view of the same quantity is in Figure~\ref{single_gaia_2mass_cart_plan}. From these results we observe that the prediction is very similar to the one from the 2MASS only results. It mostly suffers from the same issues and presents the same strengths. However we note two major differences: (i) in the low longitude part $l < 280$ deg the prediction seems to better follow the L19 or C19 morphology from Figure~\ref{cornu_marshall_lallement_close_maps_comparison}, with a much clearer distinction between two groups of structures in distance, and (ii) in the high longitude part $ l > 290$ deg almost all the extinction is concentrated in the structure we interpreted as an artifact in previous sections in the region $ 295 < l < 303$ deg and $2.5 < d < 3.5$ kpc. There might be two explanations for the second point. One possibility is that the very large star count increase from Gaia, when following the longitude axis, pushes the network more quickly to perform prediction on non-constrained parts of the feature space. Indeed, it has never been presented such highly populated CMD in the training dataset. The other explanation would be that we did not manage to create a representative Gaia diagram and that this artifact is the result of a systematic difference between modeled and observed diagrams. We note that the high distance diffuse structure is predicted very similarly to the 2MASS single LOS training indicating that the network must have automatically identified that it can use 2MASS only for such high distance prediction independently of the foreground for which Gaia dominates. This might also come from a non sufficiently restrictive $Z_{\rm lim}$ but we already adopted a relatively large $Z_{\rm lim} = 100$ value, as further discussed for the next case.\\

\newpage
\subsection{Combined sampled training}
\label{Gaia_2mass_multi_los_training_and_test_set_prediction}

For the same reason as exposed in Section~\ref{los_combination}, we proceed to a single training using the same sampling of 9 LOS over the galactic longitude range $260 < l < 300$ deg. This should help take into account the specificity of each LOS and share the redundant information between them. The input is then adjusted to account for 4 individual diagrams at once, a 2MASS [J-K]-[K] extinct CMD and its bare reference, and a Gaia [G]-[$\varpi$] extinct diagram and its bare reference. The output is still a single LOS profile. We used the same previously defined parameters with $2\times 10^5$ examples per reference LOS, $Z_{\rm lim} = 100$ and kept $f_{\rm naked} = 0.1$ so that the network can associate the bare references to a flat-zero extinction profile. Each input depth channel is normalized as an independent feature from the maximum pixel value of each diagram in the whole dataset. This last step is of critical importance because a bare Gaia reference diagram can contain a very large number of stars that would completely dominate the network training error at the beginning otherwise.\\

\begin{figure*}[!t]
\hspace{-0.6cm}
	\begin{minipage}{1.05\textwidth}
	\centering
	\begin{subfigure}[!t]{1.0\textwidth}
	\includegraphics[width=1.0\hsize]{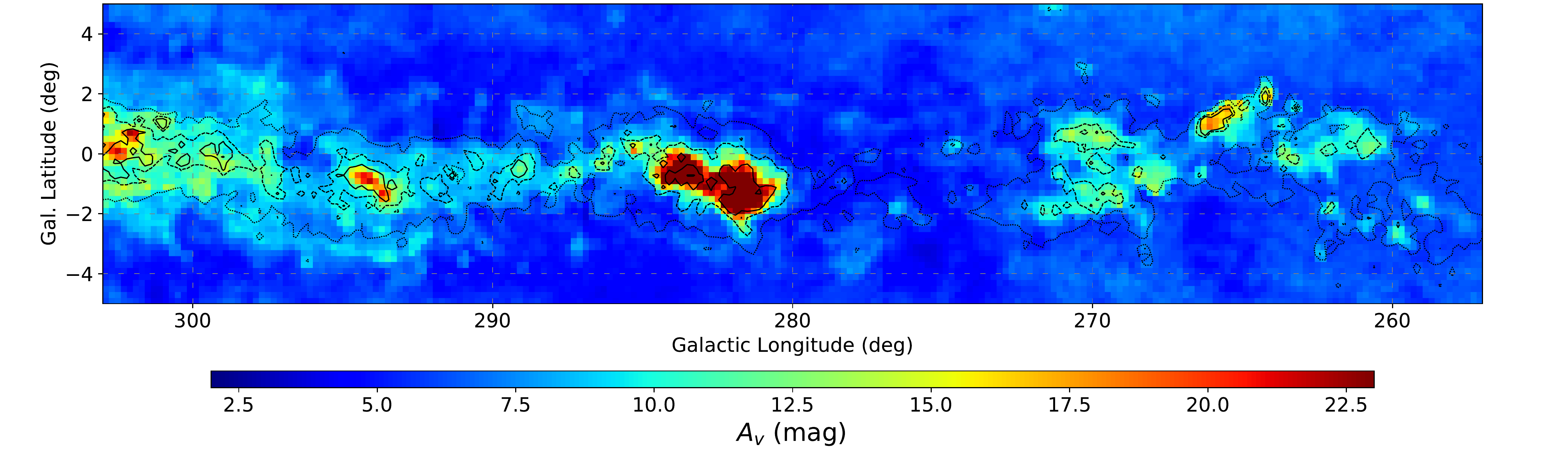}
	\end{subfigure}
	\end{minipage}
	\caption[Plane-of-the-sky view for the Gaia-2MASS multiple-LOS training]{Integrated extinction for each pixel of the Gaia-2MASS multiple-LOS training prediction in a plane of the sky view using galactic coordinates. Contours are from Planck $\tau_{353}$.}
	\label{multi_gaia_2mass_int_map}
\end{figure*}

The integrated extinction plane of the sky view is presented in Figure~\ref{multi_gaia_2mass_int_map}. It is visible that this map presents significantly less contrast in the low extinction regions, but the high-latitudes predictions seem more even than in our previous maps. The central structure presents a very high integrated extinction that is significantly higher than in several of our previous results but this high extinction is compatible with our main 2MASS result (Fig.~\ref{multi_los_2mass_polar_plan}). The low longitude part appears properly reconstructed, again with less contrast, which is also the case for the high longitude part but with an added diffuse latitude extinction.\\

\begin{figure*}[!t]
\hspace{-1.2cm}
	\begin{minipage}{1.12\textwidth}
	\centering
	\begin{subfigure}[!t]{0.48\textwidth}
	\includegraphics[width=1.0\hsize]{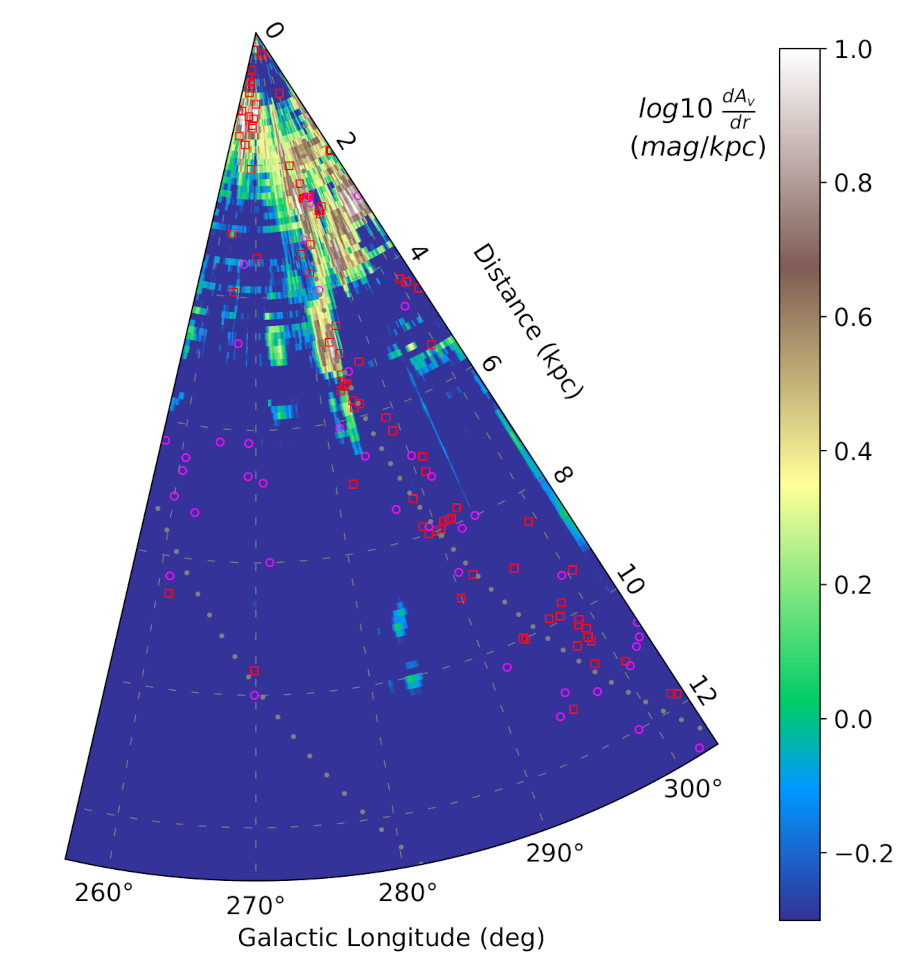}
	\end{subfigure}
	\hspace{0.3cm}
	\begin{subfigure}[!t]{0.48\textwidth}
	\includegraphics[width=1.0\hsize]{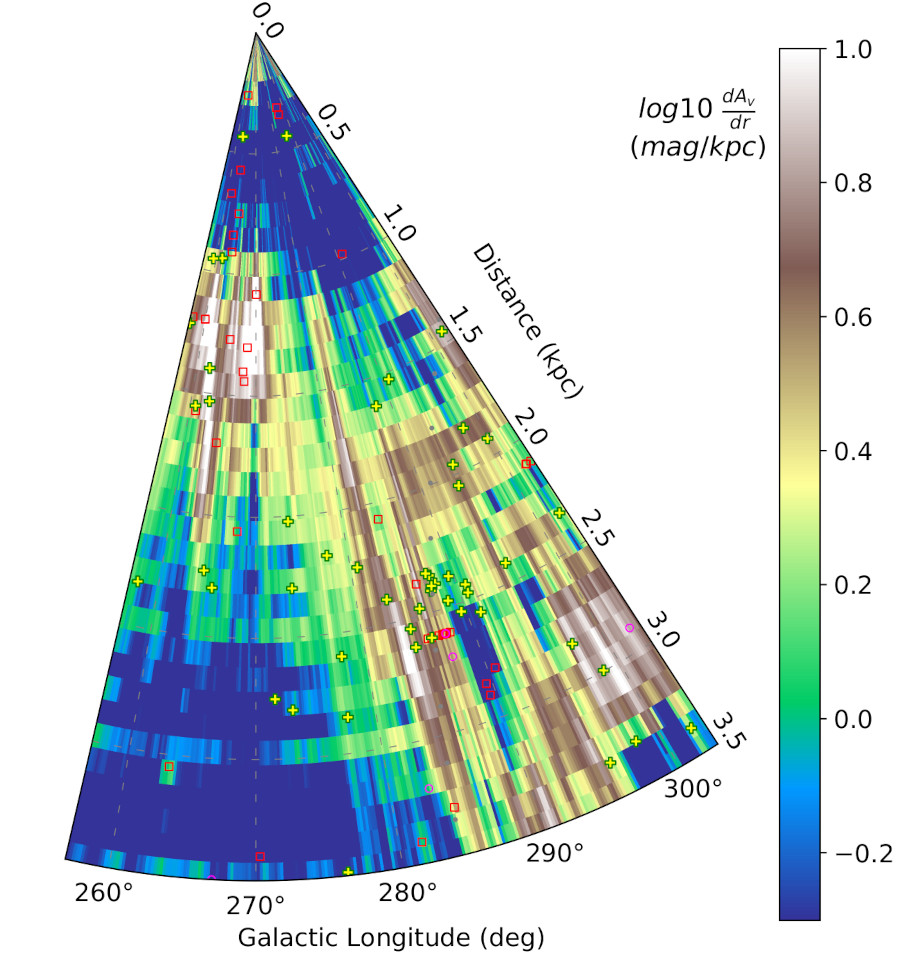}
	\end{subfigure}
	\end{minipage}
	\caption[Face-on view for the Gaia-2MASS multiple-LOS training]{Face-on view of the Galactic Place $|b| < 1$ deg in polar galactic-longitude distance coordinates for the predicted Carina arm region using our Gaia-2MASS multiple-LOS training. The symbols are the same as in figure ~\ref{single_los_2mass_polar_plan}. {\it Left:} Full distance prediction. {\it Right:} Zoom in on the $d < 3.5$ kpc prediction region.}
	\label{multi_gaia_2mass_polar_plan}
\end{figure*}

\begin{figure*}[!t]
	\begin{minipage}{1.0\textwidth}
	\centering
	\begin{subfigure}[!t]{1.0\textwidth}
	\includegraphics[width=1.0\hsize]{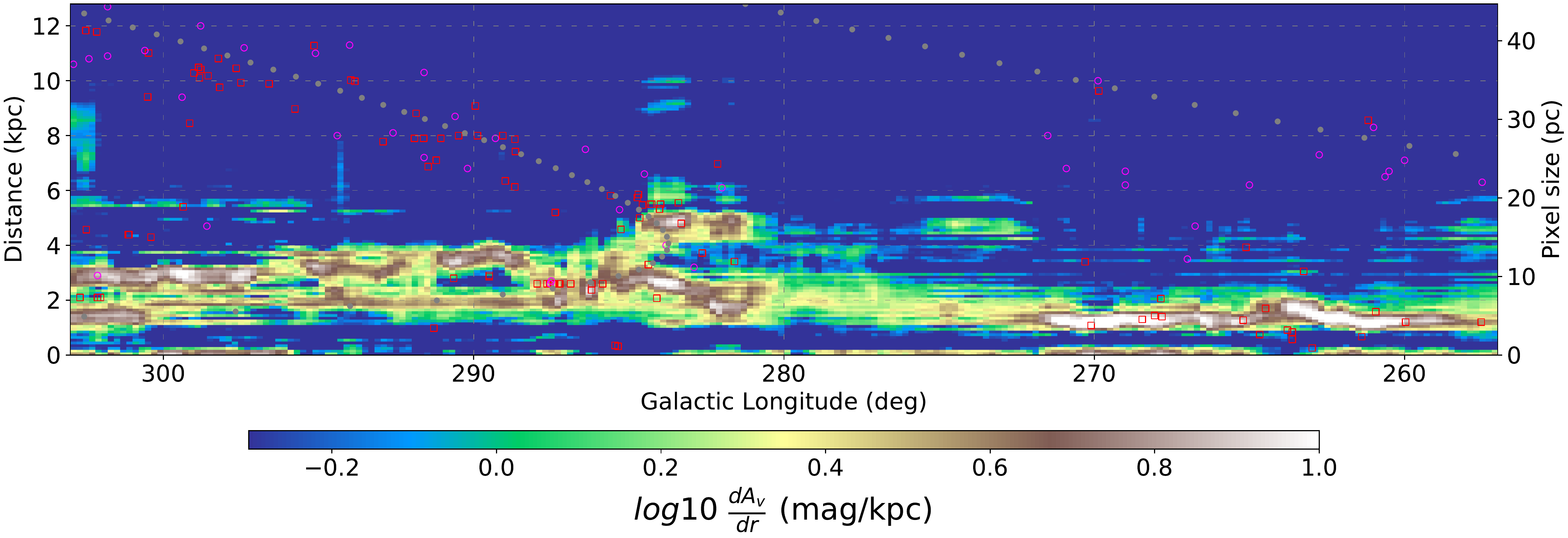}
	\end{subfigure}
	\end{minipage}
	\caption[Cartesian face-on view for the Gaia-2MASS multiple-LOS training]{Face-on view of the Galactic Place $|b| < 1$ deg in cartesian galactic-longitude distance coordinates for the predicted Carina arm region using our gaia-2MASS multiple-LOS training. The axis on the right border corresponds to the pixel height as a function of the distance induced by the conic shape of our LOS. The symbols are the same as in figure ~\ref{single_los_2mass_polar_plan}.}
	\label{multi_gaia_2mass_cart_plan}
\end{figure*}

\begin{figure*}[!t]
\hspace{-0.5cm}
	\begin{minipage}{1.05\textwidth}
	\centering
	\begin{subfigure}[!t]{0.48\textwidth}
	\caption*{\bf \large Main 2MASS result}
	\includegraphics[width=1.0\hsize]{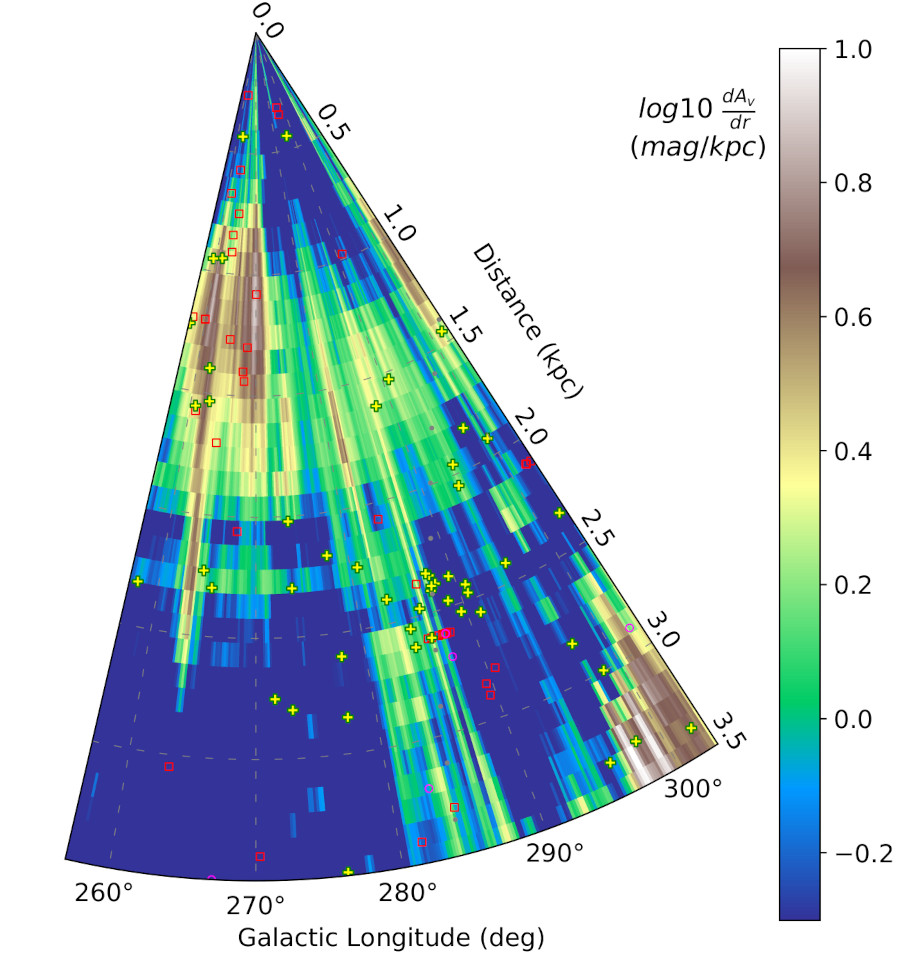}
	\end{subfigure}
	\vspace{0.5cm}
	\hspace{0.3cm}
	\begin{subfigure}[!t]{0.48\textwidth}
	\caption*{\bf \large 2MASS \& Gaia result}
	\includegraphics[width=1.0\hsize]{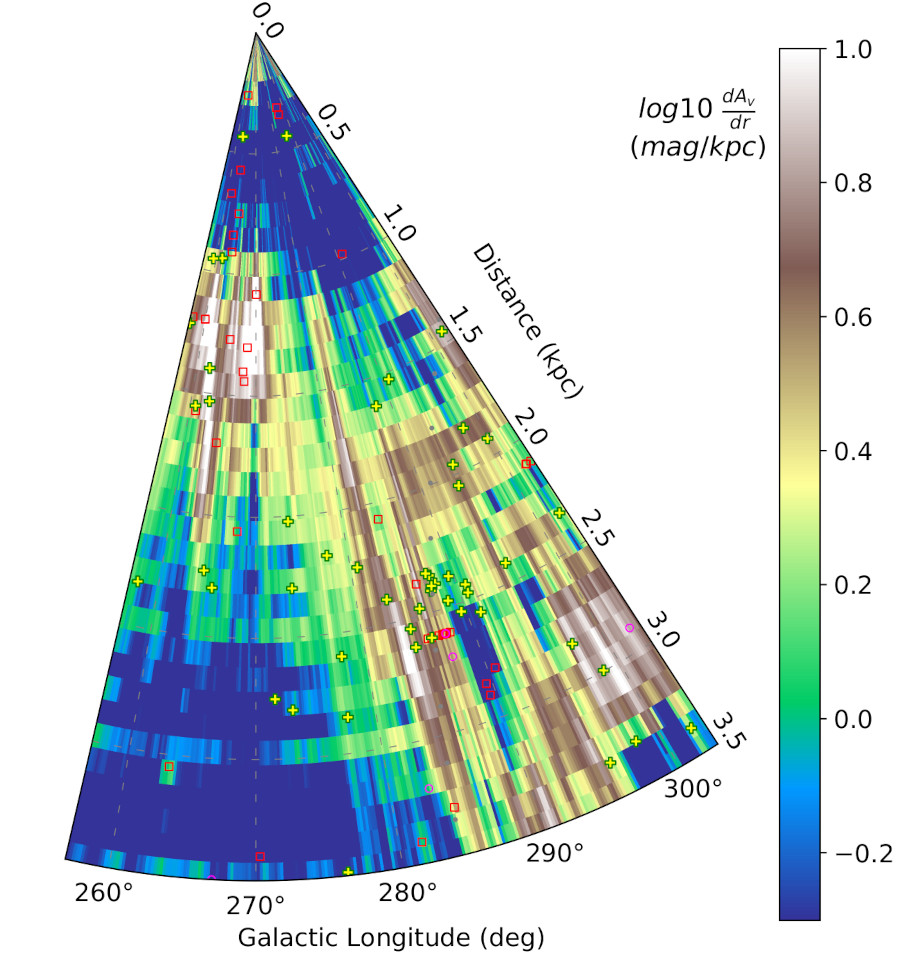}
	\end{subfigure}\\
	\begin{subfigure}[!t]{0.48\textwidth}
	\caption*{\bf \large Marshall+20}
	\includegraphics[width=1.0\hsize]{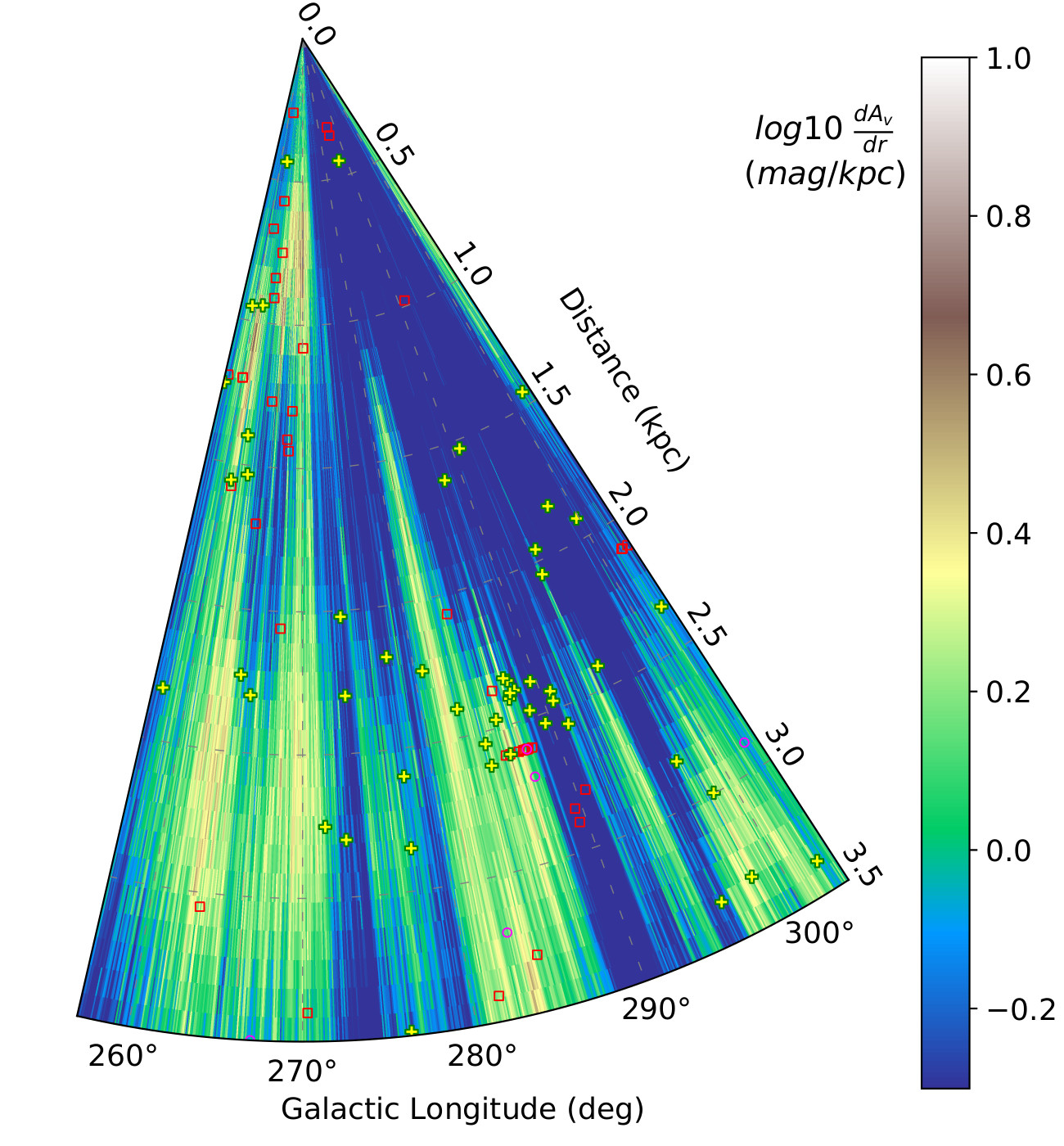}
	\end{subfigure}
	\vspace{0.5cm}
	\hspace{0.3cm}
	\begin{subfigure}[!t]{0.48\textwidth}
	\caption*{\bf \large Lallement+19}
	\includegraphics[width=1.0\hsize]{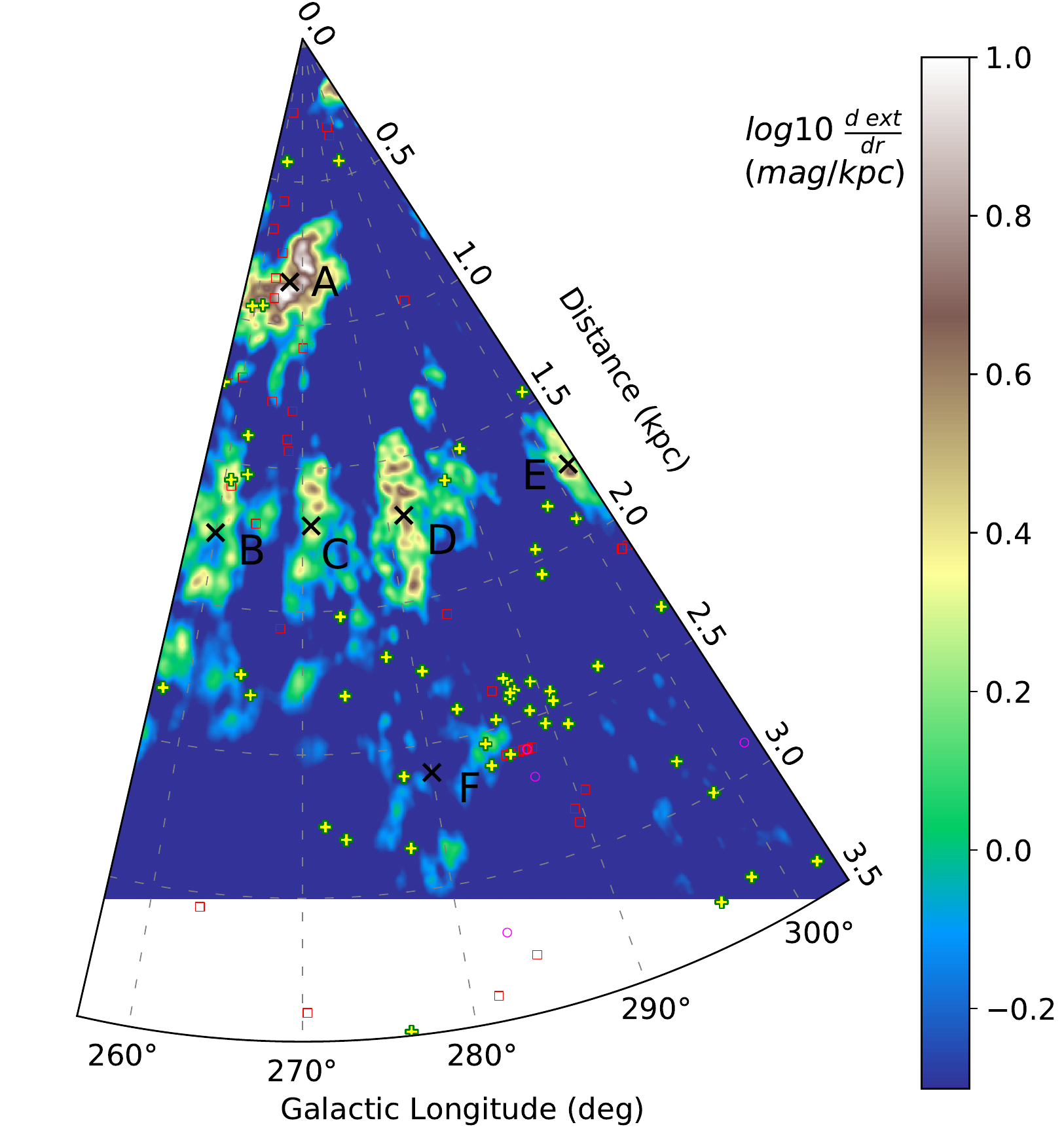}
	\end{subfigure}
	\end{minipage}
	\caption[Face-on view comparison at short distance for various maps]{Short distance comparison of the face-on view of the Galactic Place in polar galactic-longitude distance coordinates for the Carina arm region using various maps. All the maps are limited to $d < 3.5$ kpc. The symbols are the same as in figure ~\ref{single_los_2mass_polar_plan}. {\it Top-left:} Our 2MASS multiple-LOS training result for $|b|< 1$ deg. {\it Top-Right:} Our Gaia-2MASS multiple-LOS training result for $|b|< 1$ deg. {\it Top-left:} M20 prediction for $|b| < 1$ deg. {\it Bottom-right:} L19 prediction for $|z| < 35 $ pc.}
	\label{gaia_2mass_cornu_marshall_lallement_close_maps_comparison}
\end{figure*}

The usual face-on view of the Galactic Place slice is presented in Figure~\ref{multi_gaia_2mass_polar_plan} and the same quantity using a cartesian view is in Figure~\ref{multi_gaia_2mass_cart_plan}. We also provide a comparison of this result with our previous main 2MASS result along with the M20 and the L19 maps in a close distance view in Figure~\ref{gaia_2mass_cornu_marshall_lallement_close_maps_comparison}. These figures illustrate important issues we have with this result. First, from a simple prediction stability standpoint, the prediction seems more noisy with several continuous and quasi-periodic stripes with widths of 1 to 3 bins, and that are much more visible in Figure~\ref{multi_gaia_2mass_cart_plan}. These are obvious network artifacts, althouh most of them have low extinction values ($\lesssim 1$ mag/kpc). Regarding the low longitude part, the prediction roughly follows our main 2MASS result but with much larger absolute extinction and with a late first extinction estimate and more compact structure (see the short-distance comparison in Fig.~\ref{gaia_2mass_cornu_marshall_lallement_close_maps_comparison}). There is also a foreground extinction in the first 2 distance bins in this region. The high longitude region is similarly dominated by the likely artifact structure at $d=3$ kpc, that is now closer by about 0.5 kpc in comparison to the main 2MASS result. We also note that we do not detect any structure after $d > 6$ kpc anymore, which is unlikely to be due to the addition of Gaia since we did not remove any information from the 2MASS data, and that structures were present at those distances in the single LOS Gaia-2MASS result (Fig.~\ref{single_gaia_2mass_polar_plan}). At short distance in the middle longitude range $275<l<285$ deg, the prediction appears realistic. The first central extinction peak around region D at $ l \simeq 282$ and $d = 1.5$ kpc roughly corresponds to the L19 one for the same structure and our structures in this small area are compatible with the M20 prediction. Still, in the same middle longitude range, the secondary peak that could correspond to the arm tangent around $l = 282$ deg and $d = 6$ kpc from our main 2MASS result, is predicted with a shorter distance by at least 1 kpc, and is not as elongated than in the previous Gaia-2MASS single LOS result. Considering that the single LOS training had 2.5 times the number of training example for this centered LOS it is more likely that it has a better prediction here, and that with just $2 \times 10^2$ this LOS is underconstrained in the present result.\\

\newpage
We do not push much further the detailed analysis of this result since it presents clear signs of a very underconstrained training. There are several reasons that could explain this behavior. First, we observed that our prediction from a single-LOS training presented more plausible results than the multiple LOS training. Cases where adding information causes the network to degrade its prediction are likely to be a sign of non-realistic input examples. This assumption is strongly supported by the fact that this network performs significantly better with lower average error on its test set than an identical training with the same parameters but using 2MASS only CMDs. It means that the network better reproduces the training profiles, but is still unable to predict the observed quantity correctly, confirming an intrinsic difference between our training input and the observed one. Additionally, the fact that the central LOS prediction is also less well reconstructed illustrates that the information generalized from the other reference LOS is at best not used, and at worse badly affecting the central LOS prediction. The $Z_{\rm lim}$ maximum distance is unlikely to be affected by Gaia since most of the corresponding stars are extincted very early, therefore the high distance range should be constrained solely based on 2MASS. The fact that the prediction at these distances is degraded means that the addition of Gaia affects the 2MASS prediction, which should not be the case. Here, again the explanation might be in the existence of a systematic difference between our training Gaia diagram and the observed ones. We note that such effects have also been observed in our early-testing results using 2MASS only, and that usually managing to create more realistic training CMD solved the issues. The main difference here is that the Gaia diagram seem more affected by aspects of the construction that was negligible for the 2MASS data. Overall, the present inconsistency of our Gaia-2MASS multiple reference LOS training, is likely to come from a combination of all the limits we discuss in Section~\ref{extinction_maps_discussion}.\\

Independently of the previous discussion, we note that our choice of diagram for Gaia was not the most appropriate. Using a Gaia color [$\mathrm{G_{BP}}$ - $\mathrm{G_{RP}}$] would probably have worked better in place of the G band. This way the diagram may have contained more information thanks to the correlation between distance and extinction. Also, as we discussed for the 2MASS dual CMD case, it should be possible to add another Gaia diagram. Ideally, we aim at using a [$\mathrm{G_{BP}}$ - $\mathrm{G_{RP}}$]-[$\varpi$] diagram and a [$\mathrm{G_{BP}}$ - $\mathrm{G_{RP}}$]-[G] CMD all together with one or several 2MASS CMDs. This is a step-by-step and ongoing work where we try to combine the most of the two surveys.\\

\vspace{-1cm}
\section{Method discussion and conclusion}
\label{extinction_maps_discussion}

\subsection{Dataset construction limits and improvements}

\subsubsection{Magnitude cuts and uncertainty issues}

We highlighted in several section that the construction of realistic examples to train the network is the most critical aspect of this application. It is difficult to assess whether our training data are realistic enough since even if it is not the case, the network will mostly reconstruct a good prediction of the test dataset because it is based on the same construction than the training dataset. Therefore, it only provides information on how well the network is able to perform the task it was given, but not in any case if the prediction using observed inputs will be realistic. One can directly have a look at the diagrams to search for striking differences, but the use of a complex ANN method was selected exactly because the fine analysis of this diagram is difficult. Therefore, imperceptible differences could remain. Another solution is to look at the predicted map and its uncertainty and to compare it to other predictions to identify clear errors. According to the region where major differences are noticed, it is possible to better analyze the underlying prediction and input to diagnose the origin of the problem. In the end, it is only by trying to reproduce the acquisition scheme of the observed data that a realistic example can be built.\\

One of our main assumption, that we did not stress strongly before, is that we fit our magnitude cut limits and uncertainties solely on the $l = 280$ deg, $b = 0$ deg. However, it is well known that 2MASS and Gaia have variations of these values across the plane of the sky, mainly due to the variations in stellar confusion across the Milky Way \citep{skrutskie_two_2006, Evans_2018}. For this reason it should be more efficient to perform at least an uncertainty and limit fitting for each of the reference LOS used in multiple-LOS training. From preliminary tests we observed significant variations of the cut limit following the galactic longitude axis for 2MASS. However, since the magnitude cut limit mainly affects faint main sequence dwarf stars, it could explain that our 2MASS results are not that much affected. Indeed, the network most likely does not use these limit stars, as a small extinction is sufficient to shift those stars beyond the detection limit in the CMD, preventing the network from extracting any information from them. In contrast, it is very probable that for the Gaia diagram the network is significantly affected by this limit. Overall, it means that the Gaia and 2MASS mock diagrams differ from the observed ones, which could affect some results like the last combined Gaia-2MASS multiple-LOS maps.\\

\subsubsection{Modular $Z_{\rm lim}$ value}

Regarding the $Z_{\rm lim}$ value, we stressed that its value is important to assess the network prediction limit capacity and to avoid providing target profiles that are impossible to reconstruct from the input data information. A simple addition to improve results from multiple-CMD would be to have a different $Z_{\rm lim}$ for each. Especially, in the case of Gaia, since the number of stars is larger and because there is less extinction information contained in each of them, it should be justified to adopt a significantly higher $Z_{\rm lim}$ for Gaia than for 2MASS in the same training. Another improvement would be to have a different $Z_{\rm lim}$ for each reference CMD. This could be used to force the network to perform higher distance estimates in less crowded areas of the Milky Way, toward the anti-center for example. This would results in less precise predictions in these regions, but it could be better than no prediction at all. More populated areas would keep a high $Z_{\rm lim}$ to improve the prediction uncertainty since the number of stars will be sufficient to still reconstruct high distance structures. We note that we tried to use a relative $Z_{\rm lim}$ value so that each LOS has a $Z_{\rm lim}$ that correspond to $1\%$ of its total star count. We combined this solution with fixed $Z_{\rm lim}$ to avoid very low star counts LOS to be too noisy. However, for now we did not manage to improve the results using this approach. The other limitations of the study may remain the dominant issue, or we did not found an appropriate $Z_{\rm lim}$ recipe yet. Still, having the possibility to choose the map detection limit for each region would be a very useful addition to the map prediction, which would make it tunable as a function of the use case.\\

\subsubsection{Construction of realistic profiles}

Our profile construction using GRFs could certainly benefit from some refinements as well. Our current prescription induces two main problems. The first one is that there is a characteristic structure size range. Similarly to the effect of a prior in a Bayesian approach, our predictions might then be biased toward structures of the same size. This means that it might be necessary to increase the diversity of our training dataset, and therefore to also increase its size. In contrast, there is a significant part of our generated profiles that remains very unrealistic due to the intrinsic randomness of the GRF generation. Reducing the occurrence of these unrealistic examples would help reducing the training dataset size for an identical prediction capacity, and could enable the network to save many of its weights for the actually realistic profiles. While our GRF recipe could definitely be improved, we could also use other profile prescriptions. We could for example use profiles predicted from a variety of other extinction maps. In this case, we would only keep the realistic profiles independent of the studied LOS, so we could apply all theses profiles randomly to a large variety of CMDs from any region of the Milky Way. A similar solution would be to use simulations of the large scale Milky Way distribution to also recreate realistic profiles. One advantage would be that several simulation realizations could be used to significantly increase the training dataset size and diversity. Both of the previous approaches could also be used just to constrain the best parameters for our GRF construction. This would then allow us to generate as many profiles we want with the added property that it could create new realistic examples that were not seen in simulations or observations, but that have a similar statistical distribution. On the other hand, in this approach the predictions would inherit of the biases of the simulations or the other maps used as a model for our profiles. \\

Another approach would be to reverse the methodology of the present study. We could use a generative method that learns to reconstruct realistic CMDs based on a mock profile. We performed preliminary attempts in this direction by designing a Generative Adversarial Network (GAN) for this task \citep{Goodfellow_2014}. This is a double ANN setup where a network learns to predict a realistic extinction profile from a random vector, and a second network tries to distinguish extincted CMDs produced from the combination of a bare BGM CMD and the previous network profile prediction. This is then a "zero sum game" where the first network learns to fool the second one, while the second one tries to distinguish between true and fake extincted CMDs. This approach would allow us to have a GAN profile generator that was constrained to reconstruct realistic profiles based on observations. Still, GAN are long to train and usually do not have a transforming process between the generator prediction and the discriminator input as we described here. Getting to a working configuration of this approach would still require a lot of efforts.

\newpage
\subsubsection{The "perfect BGM model" assumption}
\vspace{-0.1cm}
One of our strongest assumptions is that the BGM prediction perfectly reproduces the observed LOS without extinction, at least statistically. Obviously the BGM itself has assumptions and limits, but one of the most important issues is that it predicts only the general shape of the Milky Way disk. In this model, there is no galactic arm, and no local stellar over-density. For example, the youngest stars are modeled with the same assumptions as the older stars, so that they are well mixed with the general population, while in observations they are more often grouped in open clusters. It means that the difference between our bare BGM and our observed quantity is certainly not solely due to the extinction. While it might be possible to construct an input quantity that would be sensitive to the presence of stellar clusters using Gaia, they are still missing in the training dataset for the network to learn to make the difference. A long term approach here would be to create star lists that may or may not contain star clusters using an additional construction recipe, that can again be constrained by observations or simulations. This would obviously be a considerable work to properly parameterize such datasets and it would very significantly increase the problem complexity and therefore the training dataset sizes and training times. It is not excluded that it would be possible to construct a very large network infrastructure that smartly combines all this information in a near future.

\vspace{-0.3cm}
\subsection{CNN method discussion}
\vspace{-0.1cm}
Regarding the CNN architecture itself, there is still room for improvement. The first modification to attempt in a near future will be to allow our network to work on Mixed precision datasets. This would enable us to significantly reduce the training dataset storage by at least a factor of two, and additionally it would strongly improve the network training speed without a significant prediction penalty regarding the expected precision of extinction maps. Still from a computational standpoint, allowing our framework to load data dynamically from the durable storage source would strongly reduce our RAM memory usage. The next step is then multi-GPU support. This way we could expect to work on much larger datasets and to add several input depth-channels from other studies to be all combined at once by our CNN, still without the necessity of a cross match.\\

Another modification would be to improve the network architecture itself. There are still recent CNN approaches that we did not attempt in the present construction like Residual Neural Networks \citep{He_ResidualNN_2015} or multi-path networks like in the Inception architecture \citep{inception_2016} that might improve the generalization capacity of the method. We note that we were surprised by the performance of a $1\times 1$-filter convolution in our architecture exploration, so a careful redesign of the network architecture with more of them could lead to prediction improvements. Finally, at a much larger time scale, we could consider to include spatial coherency in our input, by for example showing the input depth channel for the presented LOS but also for all the ones that are close to it. Our input could also be constructed from higher dimension depth channels, for example using 3D histograms like [G]-[H]-[K], or a similar one that includes the Gaia parallax. This way the network would be provided with a sampling of the star distance distribution and it would look for 3D coherent patterns in this volume. Finally, we can imagine a network that would be able to take large-scale Milky Way volumes as its input at once (instead of discrete LOS-profiles), allowing it to construct several realistic 3D Milky Way dust distributions as single examples and generalizing from it. Such an application would be a huge computational challenge, but it is becoming more and more realistic considering the important improvement in ANN dedicated hardware in the past few years and the present performance prediction for the upcoming new hardware technologies.

\subsection{Conclusion and perspectives}

\vspace{-0.0cm}
In the final part (Part III) of the manuscript, we presented a Convolutional Neural Networks methodology to reconstruct 3D extinction maps of the Milky Way, based on the Besançon Galaxy Model and applied to 2MASS and Gaia data, and that is suitable for large scale predictions. This study led to the following conclusion. A comparison between modeled extinction-free and observed extincted CMDs (or 2D histogram) contains a sufficient amount of information to reconstruct extinction profiles with a large distance range $d \gtrsim 10$ kpc prediction. Usual methods compare star list directly to avoid too intricate information in a CMD formalism for which it is necessary to construct a highly non-linear analysis method. \\

\vspace{-0.2cm}
A Convolutional Neural Network is a suitable method for this task. It is able to extract the information contained on the input CMD efficiently. The most efficient CNN architecture that was found is based on very few convolutional layers with a minimal dimensionality reduction of the input and then requires large dense layers to reconstruct extinction profiles accurately. The exposed architecture is computationally efficient but the training process requires large datasets of examples ($5\times 10^5$ to $2 \times 10^6$). This formalism can efficiently be generalized over a multiple LOS combination in a single training by adding reference LOS as input depth channel.\\

\vspace{-0.2cm}
The realism of the chosen input is of critical importance. Small differences between the training modeled examples and the observations lead to significant prediction artifacts in some cases. The magnitude limit cut and the uncertainty fitting of all used quantity from both our used survey has been identified as the most important step. The target profile realism is of similar importance. It must be, at the same time, diverse enough to properly constrain the feature space, and restricted enough to avoid too many unrealistic examples that slow the training process down and add noise to the predictions. The target profiles must also be modified so they do not correspond to unpredictable results regarding the information contained in the input volume.\\

\vspace{-0.2cm}
The CNN construction we propose is able to reconstruct large portions of the Milky Way Galactic Place by learning from a relatively sparse sampling in galactic longitude. The network predicts spatially coherent structures even without forcing any correlation between adjacent lines of sight. The distance dispersion of the prediction is smaller than in maps working from the same datasets and the prediction contains much less finger of gods artifacts.\\

\vspace{-0.2cm}
It has been exposed that the CNN architecture can efficiently combine several diagrams as input allowing to combine information from multiple surveys like Gaia and 2MASS without a cross match of the stars. Presently, the results from large scale application of such combination is limited by realism of our modeled diagrams especially in terms of magnitude cut limit and uncertainty fitting. The star count limit that the network considers as relevant for making a prediction also appears to be insufficiently described for the combined surveys prediction to conserve the distant 2MASS-identified structures.\\

\vspace{-0.2cm}
Our on-going works are focused on the improvement of the training dataset and construction and on the assessment of hyperparameters relative to reference LOS used during the training. The objective is to correct the Gaia-2MASS prediction to better balance the contribution from each survey, and then to predict a full Galactic Place map. Several important improvements are currently under study, such as adding other surveys, adding spatial coherence between lines of sight, changing the profile construction toward more realistic ones, network architecture improvements to speed up the training and increase the generalization capacity. The method description and first results on 2MASS will be soon published in the form of a letter to the journal Astronomy and Astrophysics \citep{cornu_montillaud_20b}.

\clearpage

\section{General conclusion}

Each of the previous parts has already been discussed in depth individually. In the present section we partly describe the timeline of the presented work. We remind the reader some of the global questioning with our approach to solve them and briefly the main results and limitations. We also provide more general insights on our approach and slightly discuss what could be the role of ML methods in astronomy during the coming years.\\

\vspace{-0.1cm}
In this study we presented the work that we have done toward the reconstruction of the large-scale 3D structure of the Milky Way. The objective of this work was to design a new methodology based on Machine Learning and that could efficiently use infrared surveys in combination with Gaia to infer this Galactic structure. From the beginning of this work, the underlying goal was to assess if these methods would be able to construct new 3D extinction maps. The main motivation for this was the present opposition between: (a) small distnace range but high resolution maps that use astrometry from optical surveys in combination with infrared ones, and (b) high distance range with lower resolution maps that use infrared surveys only. The rationale was based on the reputation of ML methods that should be able to combine heterogeneous datasets automatically. This property would permit the construction of extinction maps for which the most suitable survey, or a complex combination of them, is used as a function of the distance in an automated way. Additionally, the very large dataset usually involved in such tasks would ensure a proper training of the ML methods and their efficiency in handling a very high number of dimensions would open the possibility for the combination of several large surveys at once.\\

\vspace{-0.1cm}
It quickly became apparent that, independent to the intrinsic strengths of ML methods, they require the given task to be constructed in a very specific form. Replacing some steps of existing methods with ML fitting would have been possible and relatively easy, but not satisfying due to the fact that it would not make use of the full generalization potential of a ML approach. For this reason we decided to build a strong theoretical and practical knowledge on ML methods by designing our own framework from scratch. The choice of Artificial Neural Network, or Deep Learning, was motivated by their almost unmatched versatility and for their high computational performance. In order to build experience on their usage we approached a simpler problem, YSO classification, that would still be able to provide 3D spatial constrains on medium scale structures.\\

\vspace{-0.1cm}
As demonstrated in the present study, Young Stellar Objects trace dense clouds and can be used with Gaia to reconstruct their distance, elongation, and even global morphology in 3D. One difficulty of this approach is the lack of detected YSOs to enable robust predictions, and more importantly the fact that some regions that could contain YSOs are not investigated. We then decided to construct an identification that would focus on the deep infrared survey Spitzer. It provides the addition of granting constrains that can help to distinguish youngest Class I and older Class II YSOs, which are missing in many surveys, and its sensitivity also allows it to detect more YSOs overall, even in dense environment. The two limitations are that Spitzer does not cover the full sky but mostly the Galactic Place and that even if it detects more deeply embedded YSOs they are unlikely to have a counterpart with Gaia. Putting aside this limitation, we managed to design an ANN that is able to accurately reproduce the classification scheme from \citep{gutermuth_spitzer_2009} and that also provides an additional membership probability for each object. It allowed us to identify a few weaker points in the usual classification scheme, but more importantly we demonstrated that this probability can be used to select the most reliably identified YSOs, providing additional constraints for the 3D reconstruction of observed clouds. This approach should be able to provide large Spitzer census of YSO candidates in the near future that could be used to identify new regions of interest or even constrains more clouds morphology. We identified several possible improvements of the method mainly relying on further improving the training dataset, especially by reducing the reliance on the \citep{gutermuth_spitzer_2009} classification and rather use strong observational targets or simulations. This approach is also suitable to be applied to other all-sky surveys, or even to be used with a combination of surveys without cross-match.\\

\vspace{-0.1cm}
This first application allowed us to construct a much deeper understanding of ANN structure, behavior, limits, etc. This allowed us to identify more clearly a specific problem construction that would be able to work on the original objective of building a new 3D extinction map. After significant improvements of our framework and a careful problem description, we managed to use a Convolutional Neural Network to perform direct extinction profile reconstruction. The approach is based on the comparison of the Besançon Galaxy Model with observations from the 2MASS infrared survey. From a set of training mock profile examples, the CNN has proved capable of accurately predicting the position and size of typical extinction structures for individual lines of sight. We successfully generalized our approach in a manner that allows the network to be trained simultaneously on multiple lines of sight from different galactic longitudes. Thus, our network can be trained once and is then capable to generalize its prediction for large Galactic Place portions. From it we constructed large extinction maps that exhibit spatial coherence between adjacent lines of sight and that strongly reduce common elongation artifacts in distance. Our map based on 2MASS exhibits compact substructures up to 10 kpc and is in correct agreement with many other maps, and it also correlates with other tracers of high density structures like HII regions that are expected to trace the spiral arms. We also demonstrated that it is possible to combine Gaia and 2MASS data in a manner that does not require a cross-match. We observed that the network efficiently combines the information without requiring a significant increase in the number of parameters in the network and a negligible computation time increase. Our results are promising, particularly in some places of the map where it is clear that the information from the two surveys is efficiently combined. However, most of our results from this 2MASS-Gaia combination remain dominated by artifacts that are likely to come from insufficiently realistic training examples that still have to be refined. Since this artifacts are similar to the ones we had in our first attempts of 2MASS only predictions, we are confident in the fact that they will be solved and do not represent a fundamental limitation of either the data or the method.\\

\vspace{-0.1cm}
To conclude, Machine Learning methods are becoming an essential tool in astronomy, especially in regard of the future challenges from very-large and highly-dimensional surveys. In this work we successfully constructed ML approaches around the identified limitations that were observed in two different astronomical problems. These methods must be used properly in order to truly provide genuine improvements to various present works in astronomy. The easy accessibility of ML frameworks and the frequent publicized breakthroughs they permit often convey the idea that these methods are so efficient that they can be used even on poorly described problems. We exposed in this work that this is at the opposite of the proper approach, that consists in a fine identification of the parameter space to constrain the dataset reliability, the training and observed proportions, etc. These methods will solve many currently impossible problems in astronomy, but that they will first need to be progressively tamed by the community.




\newpage
\thispagestyle{empty}

\part*{\null}
\addcontentsline{toc}{part}{Appendix}
\begin{appendix}

\newpage
\thispagestyle{empty}
\hfill\\
\vspace{0.35\textheight}

\textbf{\LARGE Appendix}
\section{Detailed description of the CIANNA framework}
\label{cianna_app}

\clearpage
\null
\thispagestyle{empty}
\newpage
\thispagestyle{empty}
\etocsetnexttocdepth{4}
\etocsettocstyle{\subsection*{Detailed description of the CIANNA framework}}{}

\localtableofcontents{}

\clearpage
\section*{Detailed description of the CIANNA framework}
\vspace{-0.1cm}
In this section we describe our numerical framework CIANNA (Convolutional Interactive Artificial Neural Networks by/for Astrophysicists) in its present state. Our framework is evolving fast so some of the information presented here might be out of date at some point, but the main programming philosophy should remain the same. For up to date information please look for CIANNA on GitHub or similar code-hosting solution (currently at \href{https://github.com/Deyht/CIANNA}{github.com/Deyht/CIANNA}). All the results from the previous study were obtained using CIANNA or its precursors, from the simpler illustrative examples (Sect.~\ref{global_ann_section}) up to the advanced various CNN architectures (Sect.~\ref{cnn_global_section}). Despite being a very generalist framework that could be applied to any ANN application, the development was mostly driven by its ability to solve various astrophysical problems. For example we successfully used it for: regression, classification, clustering, other program acceleration, and dimensionality reduction, all in the context of an astrophysical application. This demonstrates an already satisfying maturity of the framework that is available as Open Source under the Apache 2 license. In the present appendix section we detail the overall development philosophy, the global programming scheme, some very important functions like \textit{im2col}, how the interface works on a concrete example, and finally the framework performance in comparison to the widely used Keras (TensorFlow) framework.\\

\vspace{-0.7cm}
\subsection{Global description}

\vspace{-0.0cm}
CIANNA is written using the C language (C99 revision), but it also contains a large amount of Nvidia CUDA code allowing for Nvidia GPU acceleration. Any application can be coded using a low-level C interface to access to the full potential of the framework. Additionally, we constructed a Python high-level interface that resembles the Keras one. This interface is suitable for the vast majority of applications and can be easily modified to suit many other needs (see Sect.~\ref{cianna_interface}). CIANNA allows the use of several computing methods through different implementations: a very simple no-dependency CPU, OpenMP, an OpenBLAS matrix formalism, and finally the CUDA GPU matrix acceleration. Overall, the framework has been built to permit subsequent additions in a modular way. However, each time a choice had to be made between modularity or ease of use against performance, we systematically privileged the latter. The rational behind this decision is that, with this framework, we wanted to get the finest possible control over the detailed compute behavior. A focus on modularity would have enclosed the detailed behavior into too many high-level considerations, which would have strongly complicated any application that would get out of the pre-determined range of use cases. This noticeably explains that there are some repetitions in the network in order to obtain a proper modularity over several independent fine grained implementations. Overall, the framework aims at providing basic elements for which we proposed several arrangements but that remain suitable to construct any non anticipated application with less efforts than with a very high level framework.\\

\vspace{-0.2cm}
In practical terms, CIANNA presently allows the user to construct arbitrarily deep convolutional or fully connected networks with a fine control over each layer property. In addition to the overall architecture construction, it is possible to control several aspects of the network including: learning rate with decay, weight initialization, activation function, gradient descent scheme, shuffle method, momentum, weight decay, dropout, etc. Interestingly, the framework being based on low-level functions it can be adapted to easily get out of the classical feedforward ANN with backpropagation scheme. For example we were able to construct several other applications from CIANNA, namely: Generative Adversarial Networks, object detection networks, semi-supervised clustering by K-means, Self Organizing Maps (SOM) and Radial Basis Functions. Even if we did not had time to test other constructions, we are confident that the framework could be adapted to other applications, for example to ANN trained from a Genetic Algorithm formalism.

\vspace{-0.2cm}
\subsection{CIANNA objects}

While CIANNA is coded with regular C that is not focused on high-level object programming, we still reproduced some simple object properties in the framework based on C data structures that are associated to specific sets of functions. This choice was mainly made due to strong pre-existing programming habit, still presenting a few advantages. It is easier to manipulate for programmers who are not used to the object formalism, and it provides full access to every element of our data structures at any place. This allows one to elaborate on the pre-existing CIANNA functions from completely new files without the need for modifying the existing code, or to create derivative objects. We list here the principal data-structures and briefly describe their role in the framework:\vspace{-0.6cm}\\

\begin{itemize}[leftmargin=0.5cm]
\setlength\itemsep{0.01cm}
\item \textbf{The network structure} allows the user to create several networks in the same instance of CIANNA. This structure contains all the properties of the networks that have an impact on the subsequent structures. For example it contains the choice of gradient descent scheme, the batch size if needed, the learning rate and momentum, but also all the information about the dataset to be processed, with their input and output dimensions. But more importantly it is the home structure of the list of layer objects that will be declared for this network. It also contains references to the training, test, and valid datasets that are associated to this network. Still, since this structure mostly contains references, it means that layers or datasets can technically be associated to several networks.

\item \textbf{The dataset structure} is the simplest one. It contains the memory location of the raw data already divided into the several batches. This structure then only contains the sizes of the data it handles, but also references their location that can be either on the host memory or the device (GPU) memory. The independence of this structure allows to create and manipulate any dataset before an association to a network in the rare case it is useful, but it also allows the creation of transition datasets for various internal operations of the framework.

\item \textbf{The layer structure} is certainly the most important one. This structure is highly modular since it must handle all the types of layers in the same object: Convolutional, pooling, dense, or any other kind of layers. Each layer is associated to a network structure but is constructed independently, so that a given layer could be used by different networks, for example to vary the dataset to be trained from. Each layer contains a reference to the previous layer or to the input dataset if it is the first one. It contains all the internal data that it needs, like the weights, but also its output and any memory set that would be needed for intermediate transformations between layers. It also usually contains all the equivalent memory required for the backpropagation. Interestingly, the layer structure contains placeholder references in the form of a per-type-of-layer structure and of activation function pointers that are referenced during layer initialization.

\item \textbf{The layer parameter structures} are as many as the number of layer types in the framework. Each of them is an independent structure that is able to store the information that is required regarding the layer type. For example the parameter structure associated to a convolutional layer contains the number of filters, the filter size, the padding, the stride, etc. All these structures can take the same void place-holder in a layer structure and are then converted back to their true type in any function that requires this information.

\end{itemize}

\newpage

\subsection{Description of the layers}

\vspace{-0.2cm}
The previous description of data structures already provides a lot of details on how the framework is constructed. Still we want to provide more details here by listing the typical major operations performed by each type of layer. Since it would be too long to distinguish the different approaches we only discuss the matrix formalism in the general case.\\

\vspace{-0.7cm}
\subsubsection{Dense layer}

\vspace{-0.1cm}
The dense layers follow the various prescriptions from Section~\ref{matrix_formal}. They are composed of 2D matrices that represent the batched input with a size depending on the previous layer output, the weight matrix sized accordingly to the input and to the number of the neurons in the layer, and finally the batched output depending on the number of neurons in the layer. All these elements have a counterpart used for the error propagation through the layer. In the case where the previous layer is dense as well, the input is not duplicated and the previous layer output is directly in the right form to be used as input in the present layer, assuming the use of the bias propagation trick described in Section ~\ref{gpus_prog}. In the case where the previous layer is a convolutional or pooling one, then a temporary matrix is made to re-arrange the data in the proper format. We also note that this is the only layer type that can use dropout in CIANNA for now.\\

\vspace{-0.1cm}
The dense layers are mainly characterized by two functions, that strongly resemble what was illustrated in Figure~\ref{matricial_form_fig}: the forward propagation and the back propagation. The forward function is mainly composed of a matrix multiplication between the weight matrix and the input followed by the use of the activation function associated to the layer applied on each element of the output. Each activated neuron then gets a chance of being set to zero according to the dropout rate. Therefore, the number of neurons dropped at each path is not constant. The position of the deactivated neurons is stored for the backpropagation phase. The layer does not actively transfer its output to the next layer, it is at the charge of the next layer to gather this output. The construction of the backpropagation function is very similar, starting by setting the propagated output error elements corresponding to the dropped neurons to zero. Then it continues with a large matrix multiplication between the transposed weight matrix and the remaining sparse output error. The result goes through the derivative of the activation function of the \textit{previous} layer and is then propagated to the previous layer output-error directly. Finally, the backpropagation function handles the weight update by multiplying the present layer output error by the transposed input it has received and stored from the forward phase. This creates an update matrix that is used to change the weights. This update matrix is kept in the layer memory and all subsequent update matrices use it to account for the momentum parameter.

\vspace{-0.2cm}
\subsubsection{Pooling layer}

\vspace{-0.1cm}
The pooling layers are by far the simplest ones. For now the pooling is systematically considered as a max pooling, but other kinds could be added simply. The present pooling layers are characterized only by the pooling size. They are characterized by forward and backward functions with a single internal operation for both (See Figures~\ref{pool_op} and ~\ref{pool_op_back}). The forward function performs the pooling operation by selecting the maximum input pixel values in each $P_o\times P_o$ area for each depth channel and stores the position of the maximum. The backpropagation function then takes the output error it has inherited from the previous layer propagation and propagates it in a matrix that contains zeros everywhere except for the position of the maximums of the forward phase that get the error of the associated output error pixel corresponding to their area. The pooling layer also handles the transformation of this propagated error by applying the derivative of the activation function of the previous layer to its own output.

\subsubsection{Convolutional layer}

The convolutional layers are the most complicated to handle. They are characterized by all the necessary parameters: filter size, stride, padding, and number of filters. They are composed of many 2D matrices, several of them being in fact flattened versions of 3D matrices. Presently we consider that a convolutional layer can only be preceded by a convolution or pooling layer. It should be possible to allow the network to grow convolutional layers from dense ones in order to have for example auto-encoders convolutional architectures, but it is currently not present in CIANNA. The flattened input usually considers that each depth channel is a continuous 1D array, and that these depths can be stacked after another. There are two arrangements here, (i) one where all the first depth channel images are subsequent in a 1D array and the second dimension is the number of input depths, (ii) and the second where all the depth channels of a given image are flattened in a 1D array and the second dimension is the batch size. The input image is usually considered in the second format, while the activation maps at the output of a convolutional layers are in the first one. The layer is then composed of the filter volume that depends both on the previous layer number of depth channels and on the number of filters of the present layers. Finally the output is in the form of a flatten volume from the different activation maps. As before, all these inputs have a duplicate form that is dedicated to the error backpropagation phase. We remind that a simple example of matrix formalism for a single convolution layer is given in Figure~\ref{matricial_form_fig}.\\

As for the previous layers, the convolutional ones are separated into a forward and a backpropagation function, but more importantly here is the im2col function that transforms the input in the appropriate form to be handled as a matrix multiplication. We dedicate the next Section ~\ref{im2col_function_algorithm} to this function and only refer to its use here for the rest of the description. The forward function of a convolutional layer actually starts with the im2col transformation that rearranges the input volume into a larger flattened volume that separates all the sub regions to which the filters must be applied. This volume is then transposed and multiplied by the flattened weight filter matrix to obtain all the activation maps with all their depth channels from the full batch at once. The resulting volume then goes through the activation function of the present layer. While the ReLU activation is strongly recommended for convolutional layers it is still possible to use other activation functions in CIANNA.\\

The backpropagation function is very similar. Based on the input and output shapes it evaluates the parameters that must be used for the transposed convolution like the external and internal padding or the stride. It then rotates and rearranges the filters following the prescription from Section~\ref{conv_layer_learn} so that each weight is effectively used to propagate each error pixel to the input pixels that were involved in its activation. The im2col includes the padding and stride parameters in its transformation so there is no need to have an intermediate transformation. Then the transformed error volume is transposed and multiplied by the rotated filter volume. The propagated error obtained then goes through the derivative of the activation function from the previous layer. As for the dense layer, the convolutional layer handles the weight update after the propagation by multiplying the im2col transformed input from the forward phase by the output error \textit{before} the im2col transformation. The obtained weight update matrix is considered with a momentum using its previous value and then used to change the weight filter volume.\\

\newpage

\subsection{Im2col function}
\label{im2col_function_algorithm}

\vspace{-0.1cm}
The im2col function is by far the most important one of the all framework when considering convolutional networks. We remind that a graphical representation of the procedure is presented in Figure~\ref{im2col_fig}. The development of this function has concentrated a lot of optimization efforts and the version we present here is already the 4th major version. This function is very important because it is a prerequisite of the matrix formalism for convolutional layers, which is usually much quicker than direct convolution implementation. The total time is then dominated by the im2col function itself as we discussed in Section~\ref{gpu_cnn}. Any improvement to this function then leads to strong overall performance improvements of the whole network. In our approach the im2col operation directly handles the transformation induced by the external and internal padding and by the stride and filter size choice. This way we avoid any non-necessary intermediate computation and kernel launch time when using the GPU acceleration. We note that the position of the zeros induced by the previous parameters do not move in the matrix expanded form. For this reason the full matrix is set to zero at the layer creation and all elements that must stay at zero are left untouched by our im2col implementation. We note that there is still room for improvement for this function, especially using advanced CUDA shared memory management and better thread block fine tuning. Our version currently minimizes the number of read and write in memory by only accessing once to all input and output pixels. The fact that our function remains memory bound with a minimal cache-miss in this situation is in fact a strong indication that we have constructed a computationally efficient implementation that is mainly limited by the GPU memory clock.\\

\vspace{-0.2cm}
We present our im2col implementation in Algorithm~\ref{im2col_algo} using the notations from Table~\ref{im2col_algo_parameters}. While this algorithm is fairly difficult to follow, we resume here the overall approach. The different loops allow to go through all the pixels of the original input volume. Then the objective is to find all the locations of the transformed volume that must contain a duplicate of this input pixel. We remind that this is due to the overlap of the different sub-regions in the image that occurs when $S < f_s$. With this approach there is only one reading per input pixel, which then stays on the cache, and only one affectation for each output pixel that must receive a value and no memory action for those that must stay to zero. To do so the algorithm searches all the possible filter positions around the running input pixel. This can be seen as overlapping a filter above the running pixel starting at the top-right corner position of the filter and then moving the filter around this point following the stride value searching for all filter placements that still contain the looked-at pixel. Each of the contribution of the current input pixel is then associated to all the transformed output pixel identified, taking into account the external and internal padding.\\

\vspace{-0.2cm}
This algorithm form is in fact slightly different than the concrete implementation in CIANNA. In fact our im2col approach was designed directly into a CUDA kernel. The conversion between the presented algorithm and the kernel is relatively easy and mainly consists in replacing the $i$, $d$ and $z$ loops by the indices of a 3D CUDA thread block. The rest of the differences should be marginal and was just made to improve the readability of the present algorithm. We note that, using CUDA blocks, it is very important to consider the memory arrangement to avoid cache-misses. Therefore, the $z$ loop that goes through contiguous pixels must absolutely be associated with the quickest CUDA block index. We also note that the present matrix is suitable for both forward and backpropagation by adjusting the $I_{\mathrm{shift}}$, $D_{\mathrm{pad}}$ and $T_{\mathrm{flat}}$ parameters that depict the arrangement of the images and their depth channels in memory. This function handles the conversion to correspond to a fractionally-stride convolution with internal padding and various strides. The choice of parameters that correspond to the adequate backpropagation is made automatically by the convolutional layer in the backpropagation function.

\begin{table}[!t]
	\centering
	\vspace{-0.6cm}
	\caption{Variable list for the im2col algorithm}
	\vspace{-0.3cm}
	\footnotesize
	\def\arraystretch{1.0}
	\begin{tabularx}{1.0\hsize}{l @{\hskip 0.06\hsize} l @{\hskip 0.06\hsize} l }
	\toprule
	\toprule
	{\small Symbol} & {\small Type} & {\small Description}\\
	\toprule
	$i$ & variable & Index of current image in the batch\\
	$d$ & variable & Index of depth channel in the input image\\
	$z$ & variable & Index of pixel in the input depth channel\\
	$w$ & variable & Pixel width coordinate in transformed image\\
	$h$ & variable & Pixel height coordinate in transformed image\\
	$x$ & variable & The corresponding region on the width axis\\
	$y$ & variable & The corresponding region on the height axis\\
	$p_w$ & variable & The width position inside the filter\\
	$p_h$ & variable & The height position inside the filter\\
	$pos$ & variable & Pixel 1-D flatten coordinate in transformed image\\
	$\mathrm{in}$  & pointer & Temporary position in the input image\\
	$\mathrm{mod}$ & pointer & Temporary position in the transformed image\\
	$B_{\mathrm{size}}$ & constant & Batch size\\
	$D$ & constant & Number of input depth channel\\
	$I_{\mathrm{flat}}$ & constant & Size of a flattened input depth channel, i.e $w_{\mathrm{in}} \times h_{\mathrm{in}}$\\
	$D_{\mathrm{pad}}$ & constant &  Separation between two depth channels in memory ($I_{\mathrm{flat}}$ or $B_{\mathrm{size}} I_{\mathrm{flat}}$)\\
	$P_{\mathrm{ext}}$ & constant & External padding\\
	$P_{\mathrm{int}}$ & constant & Internal padding\\
	$T_{\mathrm{flat}}$ & constant & Size of a flattened image in the transformed format, i.e $w_{\mathrm{out}} \times h_{\mathrm{out}} \times f_s^2D$\\
	$I_{\mathrm{shift}}$ & constant & Separation between two input images in memory ($I_{\mathrm{flat}}$ or $D I_{\mathrm{flat}}$)\\
	$f_s$ & constant & Filter 1D size\\
	$S$ & constant & Convolution stride \\
	$N_{\mathrm{area}}$ & constant & Number of regions in the image in one axis, i.e $w_{\mathrm{out}}$\\
	\bottomrule
	\bottomrule
	\end{tabularx}
	\label{im2col_algo_parameters}
	\vspace{-0.2cm}
\end{table}

\begin{algorithm}[!h]
\setstretch{1.0}
\SetInd{0.3cm}{0.6cm}
	\For{ $i \leftarrow 0$ \KwTo $B_{\mathrm{size}}$ }{
		\vspace{0.1cm}
		\For{ $d \leftarrow 0$ \KwTo $D$ }{
			\vspace{0.1cm}
			$\mathrm{in}\quad \leftarrow i\times I_{\mathrm{shift}} + d\times D_{\mathrm{pad}}$\\
			$\mathrm{mod}   \leftarrow  i\times T_{\mathrm{flat}} + d\times f_s^2$\\
			\For{ $z \leftarrow 0$ \KwTo $I_{\mathrm{flat}}$ }{
				\vspace{0.1cm}
				$w \leftarrow (z \div w_s) \times(1 + P_{\mathrm{int}}) + P_{\mathrm{ext}} $\\
				$h \, \leftarrow (z\ \, / \ w_s) \times(1 + P_{\mathrm{int}}) + P_{\mathrm{ext}} $\\
				$x \, \leftarrow w/S$\\
				\While{$(w-xS) < f_s$ and $ x \geqslant 0 $}{
					\vspace{0.1cm}
					$p_w \leftarrow w-xS$\\
					$y \ \  \,\leftarrow h/S$\\
					\While{$(h-yS) < f_s$ and $ y \geqslant 0$}{
						\vspace{0.1cm}
						$p_h \leftarrow h-yS$\\
						$pos \leftarrow x f_s^2 D + y N_{\mathrm{area}} f_s + p_h f_s + p_w$\\
						\If{$pos \geqslant 0$ and $pos < I_{\mathrm{flat}}$}{
							\textcolor{red}{$\bm{\mathrm{mod}[pos]  \leftarrow \mathrm{in}[z] }$}\\
							}
						$y \ \ \leftarrow y-1$\\
					}
					$x \ \  \,\leftarrow x-1$\\
				}
			}
		}
	}
	\caption{Im2col algorithm}
 \label{im2col_algo}
\end{algorithm}

\newpage

\subsection{Other important functions}

\vspace{-0.2cm}
We have reviewed the most important elements in the CIANNA framework. Around them are also various auxiliary functions like the activation functions, weight initializations, dataset shuffling, normalizations, dataset loading, etc. It is already possible to assemble all these elements using C programming to create almost any network structure. Still, we added a few higher lever functions that provide an easier network construction for the classical cases. Among them, there are several construction functions that declare the network, dataset and layer structures. The link between them is then automatized. Once assembled the network can be used with a global training and forward functions that take various network hyperparameters as argument.\\

\vspace{-0.2cm}
The simplest of these large scale functions is the one that performs a forward on a given dataset. It simply takes the dataset as input and forwards each pre-constructed batch through the network layers one after the other. The layer construction has already linked the respective layer inputs and outputs so that each of them only has to be called through its associated internal forward function. This global function is in fact built among another one that performs this task in addition to computing the error between the targets and the network outputs. In this case the error is only used as a monitoring information and not for training. It can noticeably display a confusion matrix on the provided dataset. This forward with error computation allows one to monitor the evolution of the error on the valid dataset during the training process.\\

\vspace{-0.2cm}
Finally, the highest-level function is the training one. This function includes all the necessary elements to perform the full network training during a given number of epochs. It handles the learning rate with decay, momentum, the frequency of control steps, the frequency of shuffle, etc. It means that it has access to all the required datasets, to the network structure, and to all the hyperparameters that are not considered as parameters of the function itself. If needed, it also manages a part of the data transfers between the CPU host memory and the GPU device memory. Overall, the function proceeds by forwarding the training dataset batches one after another through all the network layers, then computes an output error that is used for the backpropagation through each layer again. After a full epoch, the function shuffles the training dataset if necessary and it computes the error on the valid dataset using the previous error function. Interestingly, we also added a performance measurement inside this function that takes the form of the number of objects processed per second by the network.\\

\vspace{-0.7cm}
\subsection{Python and C interfaces}
\label{cianna_interface}

\begin{lstlisting}[float=!t, caption={Python interface},label=cianna_python_interf, language=Python, basicstyle=\fontsize{8pt}{8pt}\selectfont,frame=single]
import CIANNA as cnn

#Various parameters
image_width = 28
image_height = 28
image_depth = 1
input_dim = np.array([image_width,image_height,image_depth])
output_dim = 10
input_bias = 0.1
batch_size = 64

#Initialize the network
cnn.init_network(input_dim,output_dim ,input_bias,batch_size,'C_CUDA')
#Reading data into numpy fp32 arrays (data_train, target_train, ...)
[...]

#Creating dataset inside the CIANNA C framework from the python data
cnn.create_dataset("TRAIN", 60000, data_train, target_train)
cnn.create_dataset("VALID", 10000, data_valid, target_valid)
cnn.create_dataset("TEST", 10000, data_test, target_test)

#Create all the network layers
cnn.conv_create(f_size=5, nb_filters=6, stride=1, padding=2, activation="RELU")
cnn.pool_create(pool_size=2)
cnn.conv_create(f_size=5, nb_filters=16, stride=1, padding=2, activation="RELU")
cnn.pool_create(pool_size=2)
cnn.conv_create(f_size=3, nb_filters=48, stride=1, padding=1, activation="RELU")
cnn.dense_create(nb_neurons=1024, activation="RELU", drop_rate=0.5)
cnn.dense_create(nb_neurons=256, activation="RELU", drop_rate=0.2)
cnn.dense_create(10, activation="SOFTMAX")

#Train the network and control regularly on the valid dataset
cnn.train_network(nb_epoch=40, learning_rate=0.0002, end_learning_rate=0.0001, \
control_interv=1, momentum=0.9, decay=0.009, confmat=1)

#Forward on the test dataset
cnn.forward_network()
\end{lstlisting}

\begin{lstlisting}[float=!t, caption={C interface},label=cianna_c_interf, language=Python, basicstyle=\fontsize{8pt}{8pt}\selectfont,frame=single]
[...]

int input_dim[3] = {28,28,1};
int output_dim = 10, batch_size=64, net_id = 0;
float bias = 0.1;
network* c_net;

init_network(net_id, input_dim, output_dim, bias, batch_size, C_CUDA, /*dynamic_loading*/ 0);
c_net = networks[net_id];

c_net->train = create_dataset(c_net, 60000);
c_net->test = create_dataset(c_net, 10000);
c_net->valid = create_dataset(c_net, 10000);
//Reading data directly into the dataset structures referenced in the network
[...]

//Explicit cuda dataset conversion
cuda_convert_dataset(c_net, &c_net->train);
cuda_convert_dataset(c_net, &c_net->test);
cuda_convert_dataset(c_net, &c_net->valid);

conv_create(c_net, NULL, 5, 6, 1, 2, RELU, NULL);
pool_create(c_net, c_net->net_layers[c_net->nb_layers-1], 2);
conv_create(c_net, c_net->net_layers[c_net->nb_layers-1], 5, 16, 1, 2, RELU, NULL);
pool_create(c_net, c_net->net_layers[c_net->nb_layers-1], 2);
conv_create(c_net, c_net->net_layers[c_net->nb_layers-1], 3, 48, 1, 1, RELU, NULL);
dense_create(c_net, c_net->net_layers[c_net->nb_layers-1], 1024, RELU, 0.5, NULL);
dense_create(c_net, c_net->net_layers[c_net->nb_layers-1], 256, RELU, 0.2, NULL);
dense_create(c_net, c_net->net_layers[c_net->nb_layers-1], 
	c_net->output_dim, SOFTMAX, 0.0, NULL);

train_network(c_net, /*epochs*/ 40, /*control_interv*/ 1, /*learning_rate*/ 0.0002, 
	/*end_learning_rate*/0.0001, /*momentum*/ 0.9, /*decay*/ 0.009, 
	/*confmat*/ 1, /*save_net_every*/1, /*shuffle_gpu*/ 1, /*shuffle_every*/1);
	
forward_testset(c_net, c_net->epoch, /*dropout_realisations*/ 1);
\end{lstlisting}

\vspace{-0.2cm}
Now that we have covered most of the internal elements of CIANNA, we present here the two interfaces, in C and Python, on a practical example. For this section and the following one about the network performance we used the example from Section~\ref{mnist_example} on MNIST. We illustrate the two interfaces in Listings~\ref{cianna_python_interf} and~\ref{cianna_c_interf} for Python and C, respectively. There are many tunable parameters in the CIANNA interface so we only discuss the global approach here since this section does not stand in place of the full instructions that will be provided with the source code. Both interfaces require the same list of actions that resemble the Keras approach. First, the network must be created. This is done using a function that takes as arguments the size of the input and output of the network and the batch size, which are all required to properly arrange memory in the dataset constructions. Then the data can be loaded directly into the dataset structure in the C interface, while for the Python one it requires the user to create Numpy arrays that are then converted into the right format. The network construction is made through individual layer creation functions for which the order is important to automatize the connection between the subsequent layers. It is then possible to call the train function on the constructed network. Once trained it is possible to forward the test dataset through the network to construct predictions.\\

\begin{lstlisting}[float=!t, caption={Typical CIANNA training monitor},label=cianna_output, language=sh, basicstyle=\fontsize{6pt}{6pt}\selectfont,frame=single, linewidth=1.1\textwidth, xleftmargin=-1.2cm]
############################################################
Importing CIANNA python module V-p.0.4, by D.Cornu
############################################################

Network have been initialized with : 
Input dimensions: 28x28x1 
Output dimension: 10 
Batch size: 64 
Using CUDA compute methode

Reading inputs ... Done !
Reading targets ... Done !
Setting training set
input dim :784,Creating dataset with size 60000 (nb_batch = 938) ... Done !
Converting dataset to GPU device (CUDA)
Setting valid set
input dim :784,Creating dataset with size 10000 (nb_batch = 157) ... Done !
Converting dataset to GPU device (CUDA)
Setting testing test
input dim :784,Creating dataset with size 10000 (nb_batch = 157) ... Done !
Converting dataset to GPU device (CUDA)

Layer output: 28 28
L:1 - Convolutional layer created:
 Input: 28x28x1, Filters: 5x5x6, Output: 28x28x6 
 Activation: (RELU), Stride: 1, padding: 2
L:2 - Pooling layer layer created:
 Input: 28x28x6, P. size: 2, Output: 14x14x6 
Layer output: 14 14
L:3 - Convolutional layer created:
 Input: 14x14x6, Filters: 5x5x16, Output: 14x14x16 
 Activation: (RELU), Stride: 1, padding: 2
L:4 - Pooling layer layer created:
 Input: 14x14x16, P. size: 2, Output: 7x7x16 
Layer output: 7 7
L:5 - Convolutional layer created:
 Input: 7x7x16, Filters: 3x3x48, Output: 7x7x48 
 Activation: (RELU), Stride: 1, padding: 1
L:6 - Dense layer created:
 Input: 2353, Nb. Neurons: 1024 
 Activation: (RELU), Dropout: 0.500000
L:7 - Dense layer created:
 Input: 1025, Nb. Neurons: 256 
 Activation: (RELU), Dropout: 0.200000
L:8 - Dense layer created:
 Input: 257, Nb. Neurons: 10 
 Activation: (SMAX), Dropout: 0.000000
 
Epoch: 1
Net. training perf.: 11215.43 items/s
Learning rate : 0.0002
Confusion matrix load
Net. forward perf.: 27050.57 items/s
Cumulated error: 	 0.0758373
                                                                                                                              Recall
                 969 |         1 |         2 |         0 |         0 |         0 |         2 |         1 |         2 |         3 |	 98.88%
                   0 |      1128 |         3 |         0 |         0 |         0 |         1 |         0 |         3 |         0 |	 99.38%
                   6 |         1 |      1012 |         3 |         1 |         0 |         0 |         7 |         2 |         0 |	 98.06%
                   1 |         0 |         4 |       981 |         0 |        12 |         0 |         5 |         5 |         2 |	 97.13%
                   0 |         0 |         3 |         0 |       953 |         0 |         6 |         1 |         0 |        19 |	 97.05%
                   2 |         0 |         1 |         8 |         0 |       878 |         2 |         0 |         0 |         1 |	 98.43%
                  10 |         4 |         1 |         0 |         1 |         8 |       933 |         0 |         1 |         0 |	 97.39%
                   0 |         3 |         6 |         4 |         0 |         0 |         0 |      1009 |         1 |         5 |	 98.15%
                   6 |         0 |         5 |         3 |         3 |         4 |         0 |         3 |       946 |         4 |	 97.13%
                   4 |         6 |         1 |         6 |         5 |         3 |         0 |        18 |         3 |       963 |	 95.44%
Precision :    97.09%      98.69%      97.50%      97.61%      98.96%      97.02%      98.83%      96.65%      98.23%      96.59%  
Correct : 97.719994%

Epoch: 2
Net. training perf.: 11232.12 items/s
Learning rate : 0.000199104
Confusion matrix load
Net. forward perf.: 27467.02 items/s
Cumulated error: 	 0.0570341
                                                                                                                              Recall
                 974 |         1 |         0 |         0 |         0 |         0 |         2 |         1 |         2 |         0 |	 99.39%
                   1 |      1124 |         4 |         1 |         0 |         0 |         0 |         1 |         3 |         1 |	 99.03%
                   1 |         1 |      1020 |         0 |         0 |         0 |         0 |         4 |         6 |         0 |	 98.84%
                   0 |         0 |         2 |       994 |         0 |         4 |         0 |         5 |         5 |         0 |	 98.42%
                   0 |         0 |         1 |         0 |       970 |         0 |         1 |         1 |         2 |         7 |	 98.78%
                   2 |         0 |         0 |         5 |         0 |       879 |         2 |         0 |         4 |         0 |	 98.54%
                   8 |         4 |         0 |         0 |         4 |         4 |       936 |         0 |         2 |         0 |	 97.70%
                   0 |         3 |        12 |         6 |         0 |         0 |         0 |       998 |         3 |         6 |	 97.08%
                   4 |         0 |         4 |         3 |         1 |         2 |         1 |         1 |       956 |         2 |	 98.15%
                   3 |         2 |         0 |         8 |         4 |         3 |         0 |         7 |         9 |       973 |	 96.43%
Precision :    98.09%      99.03%      97.79%      97.74%      99.08%      98.54%      99.36%      98.04%      96.37%      98.38%  
Correct : 98.239998%

Epoch: 3
Net. training perf.: 11355.42 items/s
Learning rate : 0.000198216
Confusion matrix load
Net. forward perf.: 27464.00 items/s
Cumulated error: 	 0.0463487
                                                                                                                              Recall
                 973 |         0 |         0 |         0 |         0 |         2 |         2 |         1 |         2 |         0 |	 99.29%
                   1 |      1116 |         3 |         1 |         0 |         0 |         5 |         4 |         5 |         0 |	 98.33%
                   1 |         0 |      1025 |         2 |         0 |         0 |         0 |         2 |         2 |         0 |	 99.32%
                   0 |         0 |         1 |       998 |         0 |         4 |         0 |         4 |         3 |         0 |	 98.81%
                   0 |         0 |         2 |         0 |       970 |         0 |         2 |         1 |         2 |         5 |	 98.78%
                   2 |         0 |         0 |         3 |         0 |       884 |         1 |         0 |         2 |         0 |	 99.10%
                   6 |         2 |         0 |         0 |         4 |         6 |       939 |         0 |         1 |         0 |	 98.02%
                   0 |         0 |         7 |         4 |         0 |         0 |         0 |      1017 |         0 |         0 |	 98.93%
                   1 |         0 |         5 |         3 |         0 |         6 |         3 |         1 |       954 |         1 |	 97.95%
                   3 |         3 |         0 |         7 |         2 |         3 |         0 |         6 |         5 |       980 |	 97.13%
Precision :    98.58%      99.55%      98.27%      98.04%      99.39%      97.68%      98.63%      98.17%      97.75%      99.39%  
Correct : 98.559998%


[...]


\end{lstlisting}

\vspace{-0.7cm}
Now that the general procedure is described we list here a few details on the interfaces. First, it is visible that many of the network hyperparameters are expressed in the example we selected. It is possible to custom simple parameters like the batch size or the learning rate, momentum, etc; but also layer dependent parameters like the stride, padding, number of filters, number of neurons, dropout, etc; and more importantly it is possible to select the activation function for each layer. In the Python interface there are many optional parameters that are not all expressed here, while in our classic C interface all the parameters are required every time. This exposes some details of the underlying C structure of CIANNA. It is important to note that the Python interface only calls the equivalent functions from the C interface in a simplified way. Therefore, the C interface provides much more control over the network behavior. For example it is possible to design two networks that share some layers, or to have much more datasets that can be switched to emulate transfer learning. A more practical example that we actually made was a Generative Adversarial Network, which we were able to design using a succession of networks to associate the generative and the discriminative parts and train them separately or simultaneously. Finally, we note that both the interfaces allow for what we call "dynamical loading" on GPU, which consists in loading each batch individually on the GPU memory while keeping the all dataset on the host memory. While this approach is expected to add memory loading overhead during the training, we observed that this overhead mostly overlaps with other actions performed by the framework like kernel launch latency. In addition, this allows to delegate the shuffle task to the CPU host memory that is much more efficient than the GPU for this and that can therefore be done concurrently to GPU computations. For these reasons, the performance hit induced by the dynamic loading is most of the time negligible while allowing to handle much larger datasets that would not fit entirely inside the GPU memory. We also highlight that the network training and forward functions can be serialized to construct training blocks or to construct more complex evolution of the training hyperparameters or dataset. Finally, the CIANNA framework can save the network state regularly during the training, and consequently it is able to reload any previously trained network. This allows to pursue training from a given point, to make predictions using a saved network, or to perform transfer learning from any network.\\

We illustrate in Listing~\ref{cianna_output} a typical console output of the network construction and training for 3 epochs using the Python interface on the MNIST example. This constitutes the CIANNA log file that can be saved during training and that shows several hyperparameters as well as the network layer structure. This output was made using the option that displays the confusion matrix from the test dataset at each control step, here at every epoch. This allows one to have a detailed view of the network prediction for all the classes at each epoch. While the network appears to have already a very good prediction at the first epoch with a global accuracy of 97.72\%, we remind that the best accuracy achieved by this network is around 99.35\% as exposed in Section~\ref{mnist_example}. With this log it is also possible to monitor the network error on the validation dataset expressed as the "cumulated error" at each epoch, which correspond to the average of the cross-entropy error computed on the prediction of each validation object in this case. On this example it is visible that the error slowly decreases with the values 0.0758, 0.0570, and 0.0463. Because this error is measured from a given dropout selection and not from a scaling of the weights, it is expected that it will present significant oscillations during training, the best classification result on this example being obtained at an error around 0.036. Still, it is usually easy to spot overtraining by looking for a global increase of this error over several control steps.

\clearpage
\subsection{Performance comparison}

\vspace{-0.2cm}
Now that we have presented how CIANNA can be used, we proceed to an evaluation of its compute performance. To have a reference we will declare the exact same network on an identical dataset with CIANNA and with a Keras implementation that relies on TensorFlow, both using GPU acceleration. All the measurements were made on our Nvidia P2000 mobile since it is the system on which we have the finest control over the software versions and hardware properties. Even if we usually use the latest CUDA 10.2 version with CIANNA, the present TensorFlow stable release is limited to CUDA 10.1, we then compiled our framework with this version as well. We remind that the specifications of the P2000 mobile are: 768 CUDA cores clocked at 1544 MHz (for our version), with 4GB of GDDR5 dedicated memory clocked at 1502 MHz (6008 Mhz effective) and interfaced with a 128 bit connection to the host memory, achieving 96 GB/s of bandwidth. The host system uses a Xeon E-2176M, and both frameworks use only a single CPU thread to drive the GPU on a single core that sustains a sturdy 4.2 GHz clock under load. The reference network used for the comparison is again the one from the MNIST dataset, since it is one of our network with the most convolutional layers.\\

\vspace{-0.1cm}
First we observed that both frameworks achieved very similar prediction quality on a identical architecture, even if the ADAM error gradient optimization is automatically used in the Keras version. The only advantage that this advanced gradient optimization provides is that it is more resilient to changes in the network learning rate, momentum, and batch size, and more generally of all elements that affect the size of the weight updates. In practical terms it means that the network is more likely to converge properly in a larger range of values for these parameters than with our present very naive learning rate decay implementation. In terms of number of epochs to converge, CIANNA has usually a lesser accuracy during the first few epochs on which the ADAM optimized Keras version is more efficient, but still the two frameworks reach their best prediction at a similar epoch.\\

\vspace{-0.1cm}
In terms of compute performance, it is difficult to separate some network construction elements from the training function itself. For this reason we only excluded the initial data loading and kept the network initialization, the layers creation and the data conversion into the comparison, even if the times for these operations are marginal against the training time. Both framework implementations were executed several times in a row to account for possible overheating of the system that would lead to thermal throttling in one run and not in the other.\\

\vspace{-0.1cm}
We present the performance comparison results in Table~\ref{cianna_vs_keras} for 4 network architectures that are variations around the one we used originally for MNIST. We used the same architecture description as in Section~\ref{cnn_architecture_test}. Before analyzing the results we note that: (i) all the convolutional layers use an external padding that preserves the image size between the input and the output, but that is not displayed to increase the readability, (ii) Keras performance metric are the time in second for an epoch, or the time in millisecond for a batch, both not being very accurate to compute the number of items used per second during training, which is the metric used in CIANNA. From these results we observed that CIANNA is most of the time faster for the considered architectures. For the dense-only network (N°2) CIANNA is even twice faster than Keras. The other architectures tend to confirm a general trend of CIANNA being much faster for dense layers, and Keras being significantly faster for the convolutional layers. The latter effect is expected considering that Keras/TensorFlow relies on the cuDNN closed framework from Nvidia that is strongly optimized for convolution, using dedicated kernels that was specifically designed for the task. \\

We also noticed a few interesting behavior during training. First, the Keras framework fully utilizes the GPU memory while CIANNA only uses a third of it, and does not suffer of important performance impact when using the dynamic load that allows to use only a few hundred MB on the GPU. Concurrently, we observed that the Nvidia monitoring tools report a $50$ to $70\%$ GPU utilization when using Keras, while the GPU utilization is always saturated at $100\%$ with CIANNA. We note that many behaviors of the frameworks can be affected by the present P2000 architecture and what are its inherent bottlenecks. For example we noticed that CIANNA as well does shows a $\sim 75\%$ GPU utilization on the V100 GPU. Still, this difference highlights that Keras must be mostly memory bound, while CIANNA remains compute bound for at list a significant subset of its operations. For this reason we expect that much deeper convolutional architectures will certainly train faster with Keras than CIANNA. Still, our results here are sufficient to predict that CIANNA will not be far behind in computational time, and that it will be as good or better on large networks with a more balanced architecture.

\begin{table}[!t]
	\vspace{-0.5cm}
	\caption{Performance comparison CIANNA vs Keras/TensorFlow}
	\vspace{-0.1cm}
	\small
	\def\arraystretch{1.6}
	\hspace{-0.6cm}
	\begin{tabularx}{1.1\hsize}{ l  @{\hskip 0.06\hsize} *{2}{Y} c@{\hskip 0.04\hsize} *{2}{Y}}
	\toprule
	\toprule
	\multirow{2}{*}{\textbf{Network architecture}} & \multicolumn{2}{c}{\textbf{CIANNA}} & & \multicolumn{2}{c}{\textbf{Keras}}\\
	\cmidrule[1pt](r){2-3} \cmidrule[1pt](r){5-6}
	 & Time [s] & Object/s & & Time (s) & Object/s\\
	\toprule
	\makecell[cl]{1. I-28.28, C-6.5, P-2, C-16.5, P-2, C-48.3,\\ D-1024\_0.5, D-256\_0.2, D10} & 226 & 11360 & & 241 & $\sim$9900\\
	\midrule
	\makecell[cl]{2. D-1024\_0.5, D-1024\_0.5, D-256\_0.2, D10\\ \hfill } & 63 & 41000 & & 144 & $\sim$18700\\
	\midrule
	\makecell[cl]{3. I-28.28, C-6.5, P-2, C-16.5, P-2, C-48.3,\\ D-256\_0.2, D10} & 185 & 14000 & & 208 & $\sim$12800\\
	\midrule
	\makecell[cl]{4. I-28.28, C-6.5, P-2, C-16.5, P-2, C-48.3,\\ C-96.3, D-1024\_0.5, D-256\_0.2, D10} & 339 & 7540 & & 324 & $\sim$7800\\
	\bottomrule
	\bottomrule
	\end{tabularx}
	\label{cianna_vs_keras}
	\vspace{-0.2cm}
\end{table}

\subsection{Future improvements}

Presently, the objectives for the development of CIANNA are the addition of the "mixed precision" support to be used with Nvidia tensor cores, which should lead to a huge performance boost of the framework when using modern GPUs. We also want to include multi GPU support, at least on a single cluster node. Overall, we aim at improving the memory loading scheme since we were limited by the host memory usage in our extinction map application. Just like we added dynamic load to the GPU we would like to serialize the data loading from the permanent memory into the host RAM memory with a minimum performance impact overall and allow the possibility of dynamic data augmentation that is especially useful when working with images. Another major improvement would be to generalize CIANNA to handle inputs that present more than 2D spatial coherency, so we could have several 3D cubes as input "channels". We would also like to increase the diversity of weight initialization, activation functions and layers. We would like to create more high-level functions to handle some cases like GAN, Genetically trained ANN, semi-supervised learning including clustering, etc. Finally, we are looking forward to automatize more complex layer connections like in Recurrent Neural Network, or Residual Neural Networks, or even blocks of layers like in the Inception architecture.

\end{appendix}

\clearpage
\listoffigures
\addcontentsline{toc}{part}{\listfigurename}

\clearpage
\listoftables
\addcontentsline{toc}{part}{\listtablename}

\clearpage
\addcontentsline{toc}{part}{References}

\newgeometry{left=2.0cm,right=2.0cm,top=2.8cm,bottom=2.8cm}
\setlength{\bibsep}{3pt}
\bibliographystyle{aa}
\bibliography{these}

\clearpage

\thispagestyle{empty}
\newgeometry{left=2.4cm,right=2.4cm,top=1.8cm,bottom=2.2cm}

\begin{center}
\textbf{Abstract / Résumé}
\end{center}

\fontsize{10pt}{9pt}\selectfont

Large-scale structure in the Milky Way (MW) is, observationally, not well constrained. Studying the morphology of other galaxies is straightforward but the observation of our home galaxy is made difficult by our internal viewpoint. Stellar confusion and screening by interstellar matter are strong observational limitations to assess the underlying 3D structure of the MW. At the same time, very large-scale astronomical surveys are made available and are expected to allow new studies to overcome the previous limitations. The Gaia survey that contains around 1.6 billion star distances is the new flagship of MW structure and stellar population analyses, and can be combined with other large-scale infrared (IR) surveys to provide unprecedented long distance measurements inside the Galactic Plane. Concurrently, the past two decades have seen an explosion of the use of Machine Learning (ML) methods that are also increasingly employed in astronomy. With these methods it is possible to automate complex problem solving and efficient extraction of statistical information from very large datasets.\\

In the present work we first describe our construction of a ML classifier to improve a widely adopted classification scheme for Young Stellar Object (YSO) candidates. Born in dense interstellar environments, these young stars have not yet had time to significantly move away from their formation site and therefore can be used as a probe of the densest structures in the interstellar medium. The combination of YSO identification and Gaia distance measurements enables the reconstruction of dense cloud structures in 3D. Our ML classifier is based on Artificial Neural Networks (ANN) and uses IR data from the Spitzer Space Telescope to reconstruct the YSO classification automatically from given examples. We extensively explore dataset constructions and the effect of imbalanced classes in order to optimize our ANN prediction and to provide reliable estimates of its accuracy for each class. Our method is suitable for large-scale YSO candidate identification and provides a membership probability for each object. This probability can be used to select the most reliable objects for subsequent applications like cloud structure reconstruction.\\

In the second part, we present a new method for reconstructing the 3D extinction distribution of the MW and that is based on Convolutional Neural Networks (CNN). With this approach it is possible to efficiently predict individual line of sight extinction profiles using IR data from the 2MASS survey. The CNN is trained using a large-scale Galactic model, the Besançon Galaxy Model, and learns to infer the extinction distance distribution by comparing results of the model with observed data. This method has been employed to reconstruct a large Galactic Plane portion toward the Carina arm and has demonstrated competitive predictions with other state-of-the-art 3D extinction maps. Our results are noticeably predicting spatially coherent structures and significantly reduced artifacts that are frequent in maps using similar datasets. We show that this method is able to resolve distant structures up to 10 kpc with a formal resolution of 100 pc. Our CNN was found to be capable of combining 2MASS and Gaia datasets without the necessity of a cross match. This allows the network to use relevant information from each dataset depending on the distance in an automated fashion. The results from this combined prediction are encouraging and open the possibility for future full Galactic Plane prediction using a larger combination of various datasets.

\vspace{0.5cm}
\begin{center}
\textbf{\footnotesize Titre en français : Modélisation de la Voie Lactée en 3D par \textsc{machine learning}\\ avec les données infrarouges et Gaia}\\
\end{center}

La structure à grande échelle de la Voie-Lactée (VL) n'est actuellement toujours pas parfaitement contrainte. Contrairement aux autres galaxies, il est difficile d'observer directement sa structure du fait de notre appartenance à celle-ci. La confusion entre les étoiles et l'occultation de la lumière par le milieu interstellaire (MIS) sont les principales sources de difficulté qui empêchent la reconstruction de la structure sous-jacente de la VL. Par ailleurs, de plus en plus de relevés astronomiques de grande ampleur sont disponibles et permettent de surmonter ces difficultés. Le relevé Gaia et ses 1.6 milliards mesures de distances aux étoiles est le nouvel outil de prédilection pour l’étude de la structure de la VL et l’analyse des populations stellaires. Ces nouvelles données peuvent être combinées avec d’autres grands relevés infrarouges (IR) afin d’effectuer des mesures à des distances jusque-là inégalées. Par ailleurs, le nombre d’applications reposant sur des méthodes d’apprentissage machine (AM) s’est envolé ces vingt dernières années et celles-ci sont de plus en plus employées en astronomie. Ces méthodes sont capables d’automatiser la résolution de problèmes complexes ou encore d’extraire efficacement des statistiques sur de grands jeux de données.\\

Dans cette étude, nous commençons par décrire la construction d’un outil de classification par AM utilisé pour améliorer les méthodes classiques de classification des Jeunes Objets Stellaires (JOS). Comme les étoiles naissent dans un environnement interstellaire dense, il est possible d’utiliser les plus jeunes d’entre elles, qui n’ont pas encore eu le temps de s’éloigner de leur lieux de formation, afin d’identifier les structures denses du MIS. La combinaison des JOS et des distances mesurées par Gaia permet alors de reconstruire la structure 3D des nuages denses. Notre méthode de classification par AM est basée sur les réseaux de neurones artificiels et se sert des données du télescope spatial Spitzer pour reconstruire automatiquement la classification des JOS sur la base d’une liste d’exemples. Nous détaillons la construction des jeux de données associés ainsi que l’effet du déséquilibre entre les classes, ce qui permet d’optimiser les prédictions du réseau et d’estimer la précision associée. Cette méthode est capable d’identifier des JOS dans de très grands relevés tout en fournissant une probabilité d’appartenance pour chacun des objets testés. Celle-ci peut alors être utilisée pour retenir les objets les plus fiables afin de reconstruire la structure des nuages.\\

Dans une seconde partie, nous présentons une méthode permettant de reconstruire la distribution 3D de l’ extinction dans la VL et reposant sur des réseaux de neurones convolutifs. Cette approche permet de prédire des profils d’extinction sur la base de données IR provenant du relevé 2MASS. Ce réseau est entraîné à l’aide du modèle de la Galaxie de Besançon afin de reproduire la distribution en distance de l’extinction à grande échelle en s’appuyant sur la comparaison entre le modèle et les données observées. Nous avons ainsi reconstruit une grande portion du plan Galactique dans la région du bras de la Carène, et avons montré que notre prédiction est compétitive avec d’autres cartes d’extinction 3D qui font référence. Nos résultats sont notamment capables de prédire des structures spatialement cohérentes, et parviennent à réduire les artefacts fréquents dits ``doigts de Dieu''. Cette méthode est parvenue à résoudre des structures distantes jusqu’à 10 kpc avec une résolution formelle de 100 pc. Notre réseau est également capable de combiner les données 2MASS et Gaia sans avoir recours à une identification croisée. Cela permet d’utiliser automatiquement le jeu de données le plus pertinent en fonction de la distance. Les résultats de cette prédiction combinée sont encourageants et ouvrent la voie à de nouvelles reconstructions du plan Galactique en combinant davantage de jeux de données.

\end{document}